\documentclass[a4paper,11pt]{report}
\pdfoutput=1
\usepackage{mystyle}
\usepackage[utf8]{inputenc}
\usepackage[T1]{fontenc}
\usepackage[english,french]{babel}
\usepackage{layout}
\usepackage[top=3cm, bottom=3cm, left=3cm, right=3cm]{geometry}
\usepackage{setspace}
\usepackage{lmodern}

\usepackage{sectsty}

\usepackage{enumitem}
\usepackage{soul}
\usepackage{eurosym}
\usepackage{url}

\usepackage{verbatim}
\usepackage{moreverb}
\usepackage{listings}
\usepackage{fancyhdr}
\usepackage{afterpage}

\usepackage{wrapfig}
\usepackage{pdfpages}

\usepackage{mathrsfs}
\usepackage{cancel}

\usepackage{makeidx}
\usepackage{titlesec}
\usepackage{stmaryrd}
\usepackage{sectsty}

\usepackage{array}
\usepackage{booktabs}
\usepackage{multirow}
\usepackage{diagbox}

\usepackage{amsmath}
\usepackage{amssymb}
\usepackage{epsfig}
\usepackage{graphicx}
\usepackage[dvipsnames]{xcolor}

\usepackage[nottoc,notlot,notlof]{tocbibind}

\newcommand{\ket}[1]{|\,#1\,\rangle}
\newcommand{\bra}[1]{\langle\, #1\,|}
\newcommand{\scp}[2]{\langle\,#1\,|\,#2\,\rangle} 

\renewcommand{\bar}[1]{\overline{\mathstrut #1}}
\renewcommand{\vec}[1]{\overrightarrow{\mathstrut #1}}

\newcommand\blankpage{
    \null
    \thispagestyle{empty}
    \addtocounter{page}{-1}
    \newpage
    }

\partfont{\color{RedViolet}}
\chapterfont{\color{RedViolet}}
\sectionfont{\color{RedViolet}}
\subsectionfont{\color{RedViolet}}
\subsubsectionfont{\color{RedViolet}}

\begin{document}

\selectlanguage{english}

\pagestyle{plain}

\includepdf[pages={-}]{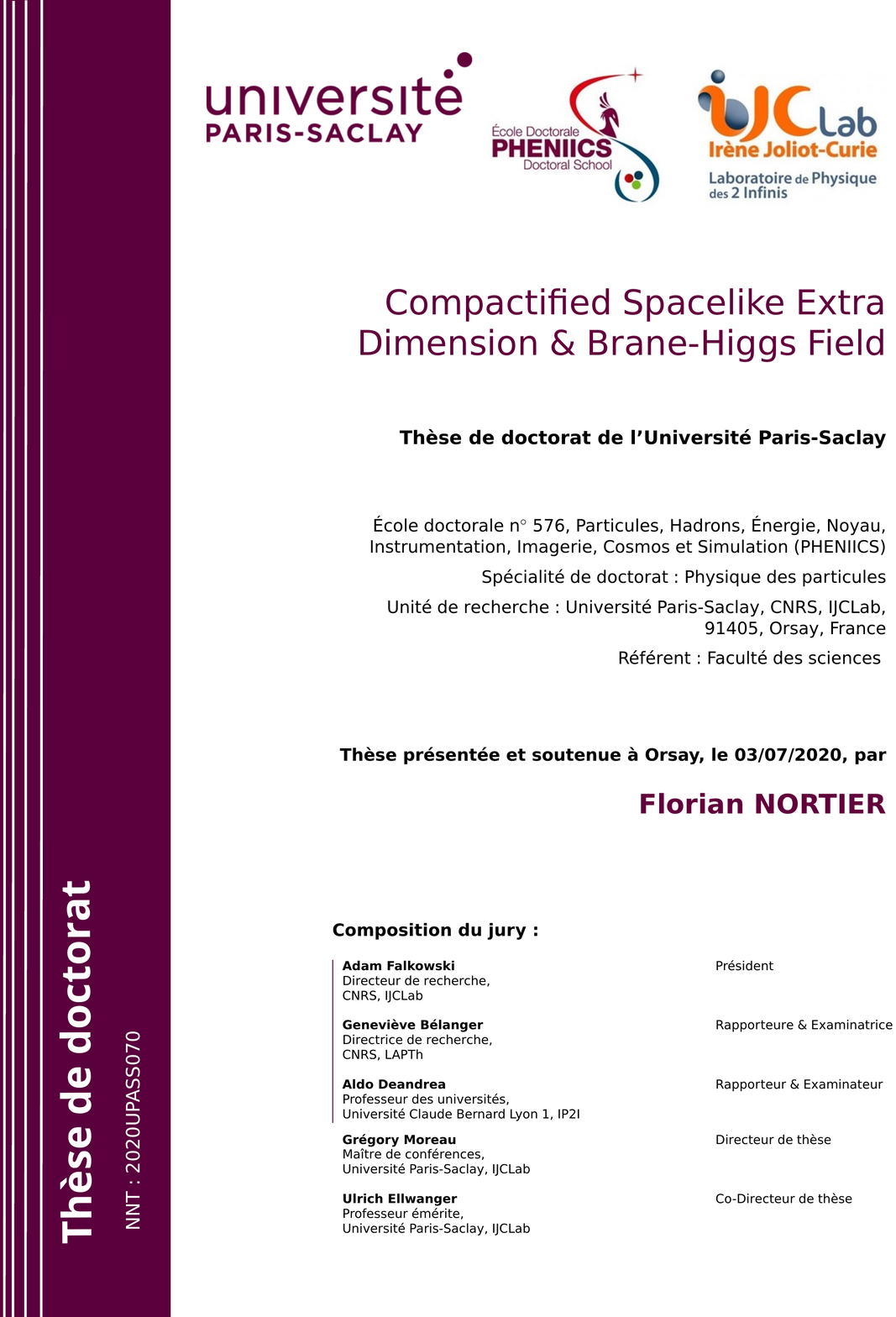}
\blankpage

\chapter*{Acknowledgments}
\addcontentsline{toc}{chapter}{Acknowledgments}


\selectlanguage{french}

{\fontfamily{pzc}\selectfont \Large{
Je tiens tout d'abord à remercier mes parents et grands-parents, qui m'ont soutenu tout au long de mes études.\\

Merci à Grégory Moreau, mon directeur de thèse, pour m'avoir donné l'occasion de faire ma thèse sur le sujet passionnant des dimensions spatiales supplémentaires, ce qui m'a permis de rencontrer des personnes que j'estime beaucoup. Merci pour son temps consacré à nos discussions sur mon sujet de thèse.\\

Merci à Ulrich Ellwanger, mon co-directeur de thèse, pour toute son aide durant ma dernière année de thèse. Sa relecture minutieuse de mon manuscrit, de mon article sur la compactification en étoile/rose, et de mes diapositives de soutenance m'a été très précieuse. Un grand merci pour tous ses conseils et sa patience.\\

Merci aux membres de mon jury de thèse : Geneviève Bélanger, Aldo Deandrea et Adam Falkowski, pour avoir accepté d'examiner ma thèse.\\

Merci à Bartjan van Tent et Gatien Verley, mes parrains de thèse, pour leur écoute et leurs conseils.\\

Merci à Andrei Angelescu et Ruifeng Leng pour leur collaboration scientifique durant cette thèse.\\

Merci à Emilian Dudas, Jérémie Quevillon, Danièle Steer et Robin Zegers pour toutes nos discussions scientifiques.\\

Merci à mes amis proches : Timothy Anson, Imène Belahcène, Hermès Bélusca-Maïto, Lydia Chabane, Charles Delporte, Ma\'ira Dutra, Timothée Grégoire, Dounia Helis, Giulia Isabella, Mathieux Lamoureux, Amaury Micheli, Thomas Montandon, Elie Mounzer et Mart\'in Novoa Brunet, pour leur soutien dans les moments difficiles, pour les bons moments passés ensemble en dehors du travail, pour toutes les pauses partagées pour ceux qui étaient avec moi au laboratoire, et pour toutes nos discussions scientifiques.\\

Merci à Sébastien Descotes-Genon, directeur du LPT, pour sa disponibilité et ses conseils.\\

Merci à Sarah Même pour toute son aide durant l'année qu'elle a passée au laboratoire, et pour toutes les pauses café partagées ensemble. Merci à elle pour son sens des relations humaines et du devoir.\\

Merci à tous les doctorants, post-doctorants, visiteurs et stagiaires avec qui j'ai sympathisé durant mon séjour au laboratoire : Giorgio Arcadi, Nicol\'as Bernal, Sjoerd Bouma, Renaud Boussarie, Nicolas Delporte, Emilien Dockes, Sylvain Fichet, Florentin Jaffredo, Gabriel Jung, Simone La Cesa, Antoine Lehebel, Natalie Macdonald, Xavier Mercado Imaz, Florent Michel, Marta Moscati, Sara Ouassidi, Mathias Pierre, Timothé Poulain, Jérémie Quarroz, Matias Rodriguez Vazquez, Olcyr Sumensari, Luiz Vale Silva et Hadrien Vroylandt.\\

Merci aux chercheurs du laboratoire avec qui j'ai pu discuter, ainsi que pour l'aide qu'ils m'ont apportée : Asmaa Abada, Damir Becirevic, Benoît Blossier, Philippe Boucaud, Christos Charmousis, Sébastien Descotes-Genon, Adam Falkowski, Jean-Pierre Leroy, Yann Mambrini, Samuel Wallon. Une pensée particulière pour Renaud Parentani, qui nous a quitté bien trop tôt, et avec qui j'ai souvent discuté à la pause café du midi et au restaurant du personnel. C'était un homme et physicien remarquable.\\

Merci à Marie-Agnès Poulet pour toute son aide au laboratoire. Merci également au reste du personnel administratif et informatique : Olivier Branc-Foissac, Odile Heckenauer, Yvette Mamilonne, Philippe Molle et Jocelyne Raux.\\

Merci à mes anciens encadrants de stage en laboratoire, qui m'ont initié à la recherche : Elena Gonzalez Ferreiro, Jean-Philippe Lansberg, Stanislas Rohart et Samuel Wallon.\\

Merci à mes anciens professeurs de physique et mathématiques de lycée et de CPGE, qui ont largement contribué à ma réussite. En particulier, je voudrais citer : Aymeric Autin, Claude Brisemure, Véronique Gadiou, Dominique Pageon et Jean-Marie Rivette, qui m'apportent encore aujourd'hui leur soutien. Merci aussi à tous les enseignants dont j'ai croisé la route au Magistère de Physique Fondamentale d'Orsay.\\

Merci aux représentants de l'école doctorale PHENIICS et de la Faculté des Sciences d'Orsay.\\

Enfin, merci à Rajaa Jourdy pour ses précieux conseils qui m'ont aidé à surmonter les difficultés de la vie et de la thèse.\\

Encore merci à tous,\\

\begin{flushright}
Florian Nortier
\end{flushright}
}}

\selectlanguage{english}

\chapter*{List of Publications}
\addcontentsline{toc}{chapter}{List of Publications}

Publications during the preparation of this PhD thesis:
\begin{itemize}
\item the article~\cite{Angelescu:2019viv} was published in a peer-reviewed journal:\\
A.~Angelescu, R.~Leng, G.~Moreau and F.~Nortier, \textit{Beyond brane-Higgs regularization: Clarifying the method and model}, Phys.~Rev.~D.~\textbf{101}~(2020)~075048, arXiv:1912.12954.
\item the preprint~\cite{Nortier:2020lbs} was published on ArXiv:\\
F.~Nortier, \textit{Large Star/Rose Extra Dimension with Small Leaves/Petals}, arXiv:2001.07102.
\end{itemize}

\renewcommand{\contentsname}{Contents}
\tableofcontents
\addcontentsline{toc}{chapter}{Contents}

\listoffigures
\addcontentsline{toc}{chapter}{List of Figures}

\listoftables
\addcontentsline{toc}{chapter}{List of Tables}


\chapter*{Introduction}
\addcontentsline{toc}{chapter}{Introduction}

The Standard Model (SM) of particle physics \cite{Donoghue2014}, based on Quantum Field Theory (QFT) \cite{Peskin:1995ev, Weinberg:1995mt, Weinberg:1996kr}, is well established nowadays, thanks to an undeniable experimental success. All its elementary particle content is discovered, and their properties are in good accordance with the theoretical predictions \cite{Tanabashi:2018oca}.

In particular, the ElectroWeak (EW) sector is described by the Glashow-Salam-Weinberg (GSW) model \cite{Glashow:1961tr,Weinberg:1967tq,Salam:1968rm}, based on the spontaneous symmetry breaking of a gauge theory. For that propose, one uses the Brout-Englert-Guralnik-Hagen-Higgs-Kibble mechanism \cite{Higgs:1964pj,Higgs:1964ia,Englert:1964et,Guralnik:1964eu,Higgs:1966ev,Kibble:1967sv} where a weakly coupled scalar field acquires a Vacuum Expectation Value (VEV). The fluctuations of the field around its VEV describe a spin 0 particle, the Higgs boson. It is an elementary scalar field which is not protected by a symmetry: its mass squared is quadratically sensitive to every new mass scale above $\Lambda_{EW} \sim 100$ GeV \cite{Wilson:1970ag,Susskind:1978ms,tHooft:1979rat}, so such a light scalar is not technically natural. In the modern Effective Field Theory (EFT) point of view, when an UltraViolet (UV) theory contains an unprotected elementary scalar of mass $m$ in addition to particles at a mass scale $M>m$, once one integrates out the particles of mass $M$, one needs to fine tune the parameters $m$ and $M$ of the UV theory to have a light scalar in the spectrum of the EFT (see Ref.~\cite{Cohen:2019wxr} for more details). In the case of the Higgs boson of mass $m_h \simeq 125$ GeV, if a new mass scale exists above $\mathcal{O}(1)$ TeV, then there is a fine tuning in the UV theory. There are several motivations to believe that new scales exist well above $\Lambda_{EW}$:

First of all, the SM does not give a description of gravity which includes a backreaction of matter and energy on the spacetime geometry in general relativity \cite{Carroll:2004st}. Einstein's theory is classical and is not able to describe the quantum fluctuations of spacetime. When one considers small fluctuations around a background metric, one gets a field of spin 2 particles, the gravitons, propagating on a 4D classical background. One can quantize the gravitons coupled to the SM as a non-perturbatively renormalizable QFT. Nevertheless, near the 4D Planck scale $\Lambda_P^{(4)} \simeq 2.4 \times 10^{18}$ GeV, one cannot consider small quantum fluctuations around a classical metric any more: one needs a new theory of quantum spacetime geometry. It is expected that new degrees of freedom appear in the UV completion near $\Lambda_P^{(4)}$ which contribute via loop corrections to $m_h$ and induce an incredible fine tuning to have this Higgs boson light.

Besides, in particle physics, one observes oscillations of neutrinos which implies that at least two of them are massive. In the SM, neutrinos are massless so one needs to add new physics to give them a mass. In the fermion sector of the SM, one observes a large mass hierarchy between the neutrino mass scale and the top mass (see Fig.~\ref{SM}). Many high energy theoreticians suspect that this hierarchy could be the result of an unknown mechanism at high scale, which could also explain the particular textures of the mixing matrices in the quark and neutrino sectors and the origin of the three fermion generations. Moreover, the absence of CP violation in strong interactions, described by Quantum ChromoDynamics (QCD) \cite{GellMann:1964nj, Zweig:1964pd, Zweig:1964jf, Fritzsch:1973pi, Gross:1973id}, is also a natural question since a CP violating topological term is authorized by the symmetries of the SM. Besides, the SM of cosmology ($\Lambda$CDM) needs additional ingredients which are not present in the SM of particle physics: a Cold Dark Matter (CDM) sector\footnote{Notice that warm dark matter is also possible.}, an inflaton sector, and a new CP violating sector for the baryogenesis. Beyond $\Lambda$CDM, if the role of dark energy is not entirely played by the cosmological constant $\Lambda$, one needs to introduce a new exotic ingredient to explain the observed accelerated expansion of the Universe. In the abundant literature concerning Beyond Standard Model (BSM) physics, many models whose goal is to solve one of these questions require a new mass scale above $\Lambda_{EW}$. The naturalness of the EW scale, also called gauge hierarchy problem, is thus a serious problem which needs to be explained by some BSM scenario.

In QFT, spacetime is a classical background where fields propagate. In general relativity, spacetime is dynamic and the theory describes how classical sources backreact on the geometry. In both theories, the number of spacetime dimensions is not determined by first principles. Possibly, in a complete theory of gravity which describes Planckian physics, the dimensionality of spacetime can be determined by the dynamics or the consistency of the theory as in string theory \cite{Becker:2007zj}. In absence of a particular UV completion of gravity, one is free to build models adding an arbitrary number of spacetime dimensions compactified on some geometries with a given topology. The criteria are that one should be consistent with all the observations indicating that our Universe appears four-dimensional in current experiments. A higher-dimensional field theory with compactified extra dimensions can be rewritten as a 4D theory by a procedure which is called Kaluza-Klein (KK) dimensional reduction. A higher-dimensional field gives a tower of 4D fields: the KK modes. On the one hand, compactified timelike extra dimensions seem to lead to physically inconsistent theories because the KK excitations of the fields propagating in the extra dimensions are tachyons, which imply violation of physically reasonable conditions like causality and unitarity \cite{Yndurain:1990fq,Dvali:1999hn}. On the other hand, spacelike extra dimensions have a long history in fundamental physics, since the pioneer works by G.~Nordström \cite{Nordstrom:1914fi}, T.~Kaluza \cite{Kaluza:1921tu} and O.~Klein \cite{Klein:1926tv,Klein:1938jm}, because they allow to build consistent EFTs. They are the subject of the present PhD thesis where we will consider only one timelike dimension. The beginning of the new millenium was marked by an explosion of the number of scientific publications on the subject of spacelike extra dimension after the appearance of some key articles with the goal of solving the gauge hierarchy problem for some of them.

The scenario proposed by Arkani-Hamed, Dimopoulos and Dvali (ADD) \cite{ArkaniHamed:1998rs, Antoniadis:1998ig, ArkaniHamed:1998nn}, with compactified Large Extra Dimensions (LEDs) and the SM fields localized on a 3D wall (3-brane), allows to have a higher-dimensional Planck scale at a few TeV. The 4D Planck scale $\Lambda_P^{(4)}$ is just an effective scale which is not related to masses of new degrees of freedom. This model allows also to generate small Dirac masses for the neutrinos where the right-handed neutrinos are KK modes of a gauge singlet field propagating into the whole spacetime (the bulk) \cite{Dienes:1998sb, ArkaniHamed:1998vp, Dvali:1999cn}. In the simplest version with toroidal compactification, and with less than seven LEDs motivated by superstring/M theory, the compactification radii are large compared to the higher-dimensional Planck length: the gauge hierarchy disappears at the price of introducing a geometrical hierarchy to stabilize. The ADD proposition is thus just a reformulation of the gauge hierarchy problem instead of a solution of it.

The most popular way to overcome the ADD geometrical hierarchy problem is to use a single warped extra dimension as proposed by Randall and Sundrum \cite{Randall:1999ee}: the RS1 model. The SM fields are localized at the boundary of a slice of an AdS$_5$ spacetime where the warp factor redshifts the scale at which gravity becomes strongly coupled from $10^{18}$ GeV to the TeV scale. Quickly, it was realized that only the Higgs field needs to be localized at the boundary and that gauge bosons and fermions can propagate into the bulk \cite{Davoudiasl:1999tf, Pomarol:1999ad, Grossman:1999ra, Chang:1999nh, Gherghetta:2000qt, Davoudiasl:2000wi}. The zero modes of the 5D fields are identified with the SM particles. The fermion zero modes can be quasi-localized near one of the two boundaries thanks to 5D Dirac masses. The wave function of a heavy (light) SM fermion has then a big (small) overlap with the boundary localized Higgs field. Without hierarchies in the 5D masses and Yukawa couplings, one can generate the flavor mass hierarchy observed in Nature \cite{Gherghetta:2000qt, Huber:2000ie}. In order to study the phenomenology of this model, it is then crucial to have a field theoretical treatment of 5D fermions in an extra dimension with spacetime boundaries which can accommodate couplings to a brane localized Higgs field.

Most of the authors use a perturbative approach \cite{Goertz:2008vr, Barcelo:2014kha}, which we call 4D approach, where the KK spectrum and wave functions of the 5D fermion fields are worked out without the brane localized interactions. They use \textit{ad hoc} Dirichlet boundary conditions on the wave functions of one of the two chiralities of the KK modes in order to have a chiral theory at the zero mode level. After that, they treat the Yukawa interactions with the brane localized Higgs field VEV as a perturbation by truncating the KK tower and bi-diagonalize the mass matrix. An alternative method is to treat directly the brane localized mass terms when one solves the equations for the wave functions: the 5D method. Many authors were puzzled by an apparent discontinuity in the KK wave functions at the Higgs field position, and they introduce a regularization method by smoothing or shifting away from the boundary the brane localized Higgs boson \cite{Csaki:2003sh, Csaki:2005vy, Grojean:2007zz, Casagrande:2008hr, Azatov:2009na, Casagrande:2010si, Azatov:2010pf, Goertz:2011hj, Carena:2012fk,  Malm:2013jia, Hahn:2013nza, Malm:2014gha, Barcelo:2014kha}. It seems puzzling that one cannot treat the Higgs boson at the boundary without this regularization procedure.

In this thesis, we begin by developping a consistent field theoretical method to treat 5D fermions coupled to a brane localized Higgs field without regularization \cite{Angelescu:2019viv}. For that purpose, we have to introduce Henningson-Sfetsos boundary terms \cite{Henningson:1998cd, Mueck:1998iz, Arutyunov:1998ve, Henneaux:1998ch, Contino:2004vy, vonGersdorff:2004eq, vonGersdorff:2004cg, Angelescu:2019viv} at both sides of a brane. One can then solve the problem treating fields as functions or distributions. After that, we apply different methods (function/distribution fields, 4D/5D calculations, etc) to various brane localized terms (kinetic terms, Majorana masses, etc), as well as 
generalizing to several classified models (flat/warped dimensions, intervalle/orbifold, etc).

The end of the PhD is dedicated to revisit the LED possibility by compactifying an extra dimension on a star/rose graph with $N \gg 1$ identical leaves/petals of length/circumference $\ell$ \cite{Kim:2005aa}. The SM fields are localized on a 3-brane at the central vertex of the graph. One can then have a 5D Planck scale at a few TeV and choose for example $\ell \sim 1/\Lambda_{EW}$. The large hierarchy between the 4D Planck scale $\Lambda_P^{(4)}$ and $\Lambda_{EW}$ is then reformulated as a large $N$. As $N$ is a radiatively stable quantity fixed by the spacetime geometry, which is just a classical brackground for the EFT, this solves the technical naturalness problem of the Higgs boson mass $m_h$. The question of why $N$ is very large is postponed until a Planckian theory of gravity. Subsequently, we use our previous results on 5D fermions to build a model of small neutrino masses, where the right-handed neutrinos are the KK modes of a gauge singlet fermion propagating into the LED and coupled to the 4D left-handed neutrino through the 4D Higgs field, both localized at the central vertex.

The manuscript of this PhD thesis entitled ``Compactified Spacelike Extra Dimension \& Brane-Higgs Field'' is organized as follows:
\begin{itemize}
\item Part~\ref{State_Art} is written in French and gives the state of the art of the most popular models with spacelike extra dimensions whose purpose is to tackle the gauge and/or flavor hierarchies problems:
\begin{itemize}
\item Chapter~\ref{Higgs_boson_gauge_hierarchy} is a short review of the SM of particle physics and of the motivations for BSM model buildings, insisting on the gauge hierarchy problem.
\item Chapter~\ref{univers_branaires} is a review of the historical models with spacelike extra dimensions and in particular the ones which solve or reformulate the gauge hierarchy problem.
\item Chapter~\ref{tranche_AdS5} is an introduction to the models with the SM Higgs field localized at the boundary of a slice of an AdS$_5$ spacetime with bulk fermion and gauge fields.
\end{itemize}
\item Part~\ref{research_work} contains the main research work made during this PhD thesis:
\begin{itemize}
\item Chapter~\ref{1_4D perturbative approach} describes the method to treat 5D fermions coupled to a boundary localized Higgs field with a compactification on an interval.
\item Chapter~\ref{Applications} contains a generalization of the method of Chapter~\ref{1_4D perturbative approach} towards a compactification on 5D orbifolds, to a Higgs field localized on a brane away from a boundary, and to other brane localized terms.
\item Chapter~\ref{large_star_rose_ED_small_leaves_petals} describes our model of LED compactified on a star/rose graph with a large number of weak scale leaves.
\end{itemize}
\item Our conventions are given in Appendix~\ref{conventions}.
\item A summary in French of the chapters in English is given in Appendix~\ref{summary_french}.
\item The acronyms used in this manuscript are listed in a Glossary p.~\pageref{Glossary}.
\end{itemize}

\part{State of the Art}
\label{State_Art}

\selectlanguage{french}

\chapter{Du champ de Higgs électrofaible à la hiérarchie de jauge}
\label{Higgs_boson_gauge_hierarchy}

\section{Modèle standard de la physique des particules}
\label{SM_part_phys}
\begin{figure}[h]
\begin{center}
\includegraphics[width=15cm]{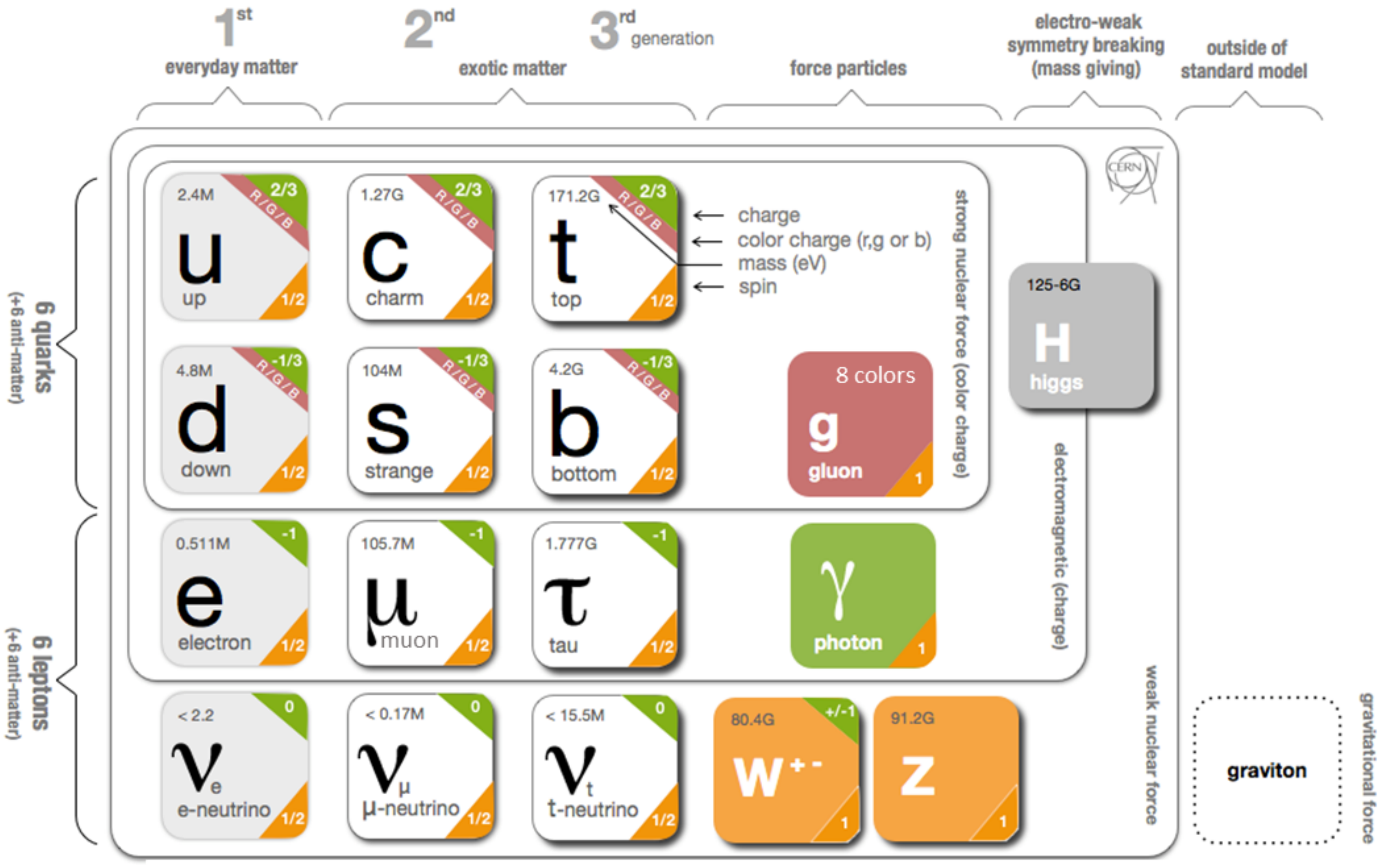}
\end{center}
\caption[Modèle standard de la physique des particules]{(Adaptée de la Réf.~\cite{CERN}.) Vue schématique du modèle standard de la physique des particules après brisure de symétrie EW.
}
\label{SM}
\end{figure}

Le SM de la physique des particules décrit les particules élémentaires de la matière et leurs interactions. Il est construit dans le cadre de la QFT\footnote{Pour une revue sur le SM de la physique des particules et la QFT, voir par exemple les Réfs.~\cite{Itzykson:1980rh, Halzen:1984mc, Cheng:1985bj, Ryder:1985wq, Bailin:1986wt, Griffiths:1987tj, Pokorski:1987ed, Sterman:1994ce, Peskin:1995ev, Weinberg:1995mt, Weinberg:1996kr, Aitchison:2003tq, Aitchison:2004cs, Zee:2003mt, Banks:2008tpa, Quigg:2013ufa, Schwartz:2013pla, Thomson:2013zua}.} sur une théorie de jauge renormalisable \cite{tHooft:1971akt, tHooft:1971qjg, tHooft:1972tcz} et unitaire \cite{LlewellynSmith:1973yud, Bell:1973ex, Cornwall:1973tb, Cornwall:1974km} décrivant les interactions EW et fortes.

La théorie GSW des interactions EWs \cite{Glashow:1961tr, Weinberg:1967tq, Salam:1968rm} décrit les interactions électromagnétiques \cite{Dirac:1927dy, Heisenberg:1929xj, Tomonaga:1946zz, Schwinger:1948iu, Feynman:1949hz} et faibles \cite{Fermi:1934sk, Fermi:1934hr, Feynman:1958ty} entre les quarks et les leptons. C'est une théorie de Yang-Mills \cite{Yang:1954ek} basée sur les groupes de symétrie d'isospin $SU(2)_W$ et d'hypercharge $U(1)_Y$ faibles. Ajoutée à la QCD \cite{GellMann:1964nj, Zweig:1964pd, Zweig:1964jf, Fritzsch:1973pi, Gross:1973id}, la théorie de l'interaction forte entre les quarks colorés fondée sur le groupe de jauge $SU(3)_C$, on obtient une description unifiée des interactions subatomiques. Le SM contient trois types de champs : les fermions, les bosons de jauge et le boson de Higgs (\textit{c.f.} la Fig.~\ref{SM}).

\subsection{Zoologie des fermions}
\label{fermions}
Les champs de matière sont des fermions de spin $1/2$ dans la représentation fondamentale du groupe de jauge $SU(3)_C \times SU(2)_W \times U(1)_Y$. Il y a trois générations séquentielles, \textit{i.e.} trois répliques ayant les mêmes nombres quantiques mais de masses différentes. Les fermions de chiralité gauche sont des isodoublets faibles\footnote{On note d'ailleurs plus communément le groupe de jauge d'isospin faible $SU(2)_L$ plutôt que $SU(2)_W$. Cependant, dans le cadre de modèles avec des fermions se propageant dans des dimensions spatiales supplémentaires, des quarks vectoriels apparaissent, dont la chiralité droite est un isodoublet de $SU(2)_L$, ce qui peut prêter à confusion.} : $T^3(f) = \pm 1/2$ (la troisième composante d'isospin faible). On distingue les leptons:
\begin{equation}
L_1 =
\begin{pmatrix}
\nu_e \\
e^-
\end{pmatrix}_L,
\ \ \ 
L_2 =
\begin{pmatrix}
\nu_\mu \\
\mu^-
\end{pmatrix}_L,
\ \ \ 
L_3 =
\begin{pmatrix}
\nu_\tau \\
\tau^-
\end{pmatrix}_L,
\nonumber
\end{equation}
et les quarks:
\begin{equation}
Q_1 =
\begin{pmatrix}
u \\
d
\end{pmatrix}_L,
\ \ \ 
L_2 =
\begin{pmatrix}
c \\
s
\end{pmatrix}_L,
\ \ \ 
L_3 =
\begin{pmatrix}
t \\
b
\end{pmatrix}_L,
\nonumber
\end{equation}
où on a utilisé les notations de la Fig.~\ref{SM}. Les fermions de chiralité droite sont des isosingulets faibles : $T^3(f) = 0$. Il y en a un pour chaque élément des isodoublets, sauf pour les neutrinos\footnote{On sait aujourd'hui que les neutrinos sont des particules massives via le phénomène d'oscillation. Pour générer cette masse, il est souvent nécessaire d'ajouter des neutrinos de chiralité droite (\textit{c.f.} les Réfs.~\cite{Zuber:2004nz, Giunti:2007ry}, par exemple, pour une revue). Dans le SM, les neutrinos sont considérés sans masse.} :
\begin{equation}
e_{1R} = e^-_R, \ \ \ 
e_{2R} = \mu^-_R, \ \ \ 
e_{3R} = \tau^-_R,
\nonumber
\end{equation}
\begin{equation}
u_{1R} = u_R, \ \ \ 
u_{2R} = c_R, \ \ \ 
u_{3R} = t_R,
\nonumber
\end{equation}
\begin{equation}
d_{1R} = d_R, \ \ \ 
d_{2R} = s_R, \ \ \ 
d_{3R} = b_R.
\nonumber
\end{equation}
L'hypercharge faible est définie à partir de $T^3(f)$ et de la charge électrique $Q(f)$ (en unité de charge électrique élémentaire $+e$):
\begin{equation}
Y(f) = 2 \left[ Q(f) - T^3(f) \right],
\end{equation}
\begin{equation}
Y(L_i) = -1, \ \ \ Y(e_{iR}) = -2, \ \ \ Y(Q_i) = \dfrac{1}{3}, \ \ \ Y(u_{iR}) = \dfrac{4}{3}, \ \ \ Y(d_{iR}) = - \dfrac{2}{3}.
\end{equation}
La théorie GSW assigne donc les fermions de chiralité gauche et droite à des représentations différentes de $SU(2)_W \times U(1)_Y$ : on parle de théorie chirale, par opposition à une théorie vectorielle comme la QCD. En effet, les deux chiralités de quarks (leptons) sont des triplets (singulets) de couleur sous $SU(3)_C$. On notera la relation :
\begin{equation}
\sum_\psi Y(\psi) = \sum_\psi Q(\psi) = 0,
\end{equation}
où on somme sur tous les champs d'une génération de fermions (chaque couleur étant comptée comme un champ différent). Ceci assure l'annulation des anomalies chirales \cite{Adler:1969er, Jackiw1972} pour chaque génération, préservant ainsi les symétries de jauge au niveau quantique (dans les diagrammes à boucle).

\subsection{Zoologie des bosons de jauge}
\label{jauge}
Les champs de jauge décrivent des bosons de spin-1, médiateurs des interactions. Le nombre de bosons par interaction est égal au nombre de générateurs des groupes de symétrie associés.

Dans le secteur EW, on a les champs $B_\mu$ et $W^{1,2,3}_\mu$, correspondants respectivement aux générateurs $Y$ et $T^a$ ($a = 1,2,3)$ des groupes $U(1)_Y$ et $SU(2)_W$. Les générateurs $T^a$ sont reliés aux matrices de Pauli tels que
\begin{equation}
T^a = \dfrac{\sigma^a}{2},
\end{equation}
vérifiant
\begin{equation}
\text{Tr}(\sigma_{a}\sigma_{b})=2\delta_{ab}.
\end{equation}
Les relations de commutation entre générateurs sont :
\begin{equation}
[T^a, T^b] = \epsilon_{abc} T^c \ \ \ \text{et} \ \ \ [Y,Y]=0,
\end{equation}
où $\epsilon_{abc}$ est le tenseur de Levi-Civita totalement antisymétrique.

Dans le secteur de l'interaction nucléaire forte, on a un octet de gluons $G^{1, \ldots, 8}$ associés aux huit générateurs du groupe $SU(3)_C$, dont les relations de commutation sont :
\begin{equation}
\left[\dfrac{\lambda^A}{2},\dfrac{\lambda^B}{2}\right]=i f_{ABC}\dfrac{\lambda^C}{2},
\end{equation}
avec les matrices de Gell-Mann ($A = 1, \ldots, 8$):
\begin{align}
	\lambda_{1} &=
	\begin{pmatrix}
	0 & 1 & 0 \\
	1 & 0 & 0 \\
	0 & 0 & 0
	\end{pmatrix},
	\phantom{aaa}
	\lambda_{2}=
	\begin{pmatrix}
	0 & -i & 0 \\
	i & 0 & 0 \\
	0 & 0 & 0
	\end{pmatrix},
	\phantom{aaa}
	\lambda_{3}=
	\begin{pmatrix}
	1 & 0 & 0 \\
	0 & -1 & 0 \\
	0 & 0 & 0
	\end{pmatrix},\nonumber\\ \nonumber \\
	\lambda_{4}&=
	\begin{pmatrix}
	0 & 0 & 1 \\
	0 & 0 & 0 \\
	1 & 0 & 0
	\end{pmatrix},
	\phantom{aaa}
	\lambda_{5} =
	\begin{pmatrix}
	0 & 0 & -i \\
	0 & 0 & 0 \\
	i & 0 & 0
	\end{pmatrix},
	\phantom{aaa}
	\lambda_{6}=
	\begin{pmatrix}
	0 & 0 & 0 \\
	0 & 0 & 1 \\
	0 & 1 & 0
	\end{pmatrix},\nonumber\\ \nonumber \\
	\lambda_{7}&=
	\begin{pmatrix}
	0 & 0 & 0 \\
	0 & 0 & -i \\
	0 & i & 0
	\end{pmatrix},
	\phantom{aaa}
	\lambda_{8}=\frac{1}{\sqrt{3}}
	\begin{pmatrix}
	1 & 0 & 0 \\
	0 & 1 & 0 \\
	0 & 0 & -2
	\end{pmatrix},
	\end{align}
vérifiant
\begin{equation}
\text{Tr}(\lambda_{A}\lambda_{B})=2\delta_{AB}.
\end{equation}
Les constantes de structure non-nulles de $SU(3)_C$ sont :
\begin{equation}
f_{123}=1, \nonumber
\end{equation}
\begin{equation}
f_{147}=-f_{156}=f_{246}=f_{257}=f_{345}=-f_{367}=\frac{1}{2},
\end{equation}
\begin{equation}
f_{458}=f_{678}=\frac{\sqrt{3}}{2}. \nonumber
\end{equation}

Les tenseurs des champs pour chaque interaction s'écrivent :
\begin{align}
B_{\mu \nu} &= \partial_\mu B_\nu - \partial_\nu B_\mu, \nonumber \\ \nonumber \\
W^a_{\mu \nu} &= \partial_\mu W^a_\nu - \partial_\nu W^a_\mu + g_w \; \epsilon_{abc} W^b_\mu W^c_\nu, \\ \nonumber \\
G^A_{\mu \nu} &= \partial_\mu G^A_\nu - \partial_\nu G^A_\mu + g_c \; f_{ABC} G^B_\mu G^C_\nu, \nonumber
\label{tenseur}
\end{align}
où $g_c$, $g_w$ et $g_y$ sont respectivement les constantes de couplage des groupes $SU(3)_C$, $SU(2)_W$ et $U(1)_Y$. On remarquera la présence des constantes de structures dans l'expression des tenseurs de champs associés aux groupes de jauge non-abéliens. Cela se traduit physiquement par une auto-interaction des gluons et bosons $W$, \textit{i.e.} des couplages triples et quartiques dans les lagrangiens.

Un champ fermionique $\psi$ est couplé de manière minimale aux champs de jauge par la dérivée covariante :
\begin{equation}
D_\mu \psi = \left( \partial_\mu - i g_c \dfrac{\lambda^A}{2} G_{A \mu} - i g_w \dfrac{\sigma^a}{2} W_{a \mu} - i g_y \dfrac{Y(\psi)}{2} B_\mu \right) \psi.
\end{equation}
Dans une théorie de Yang-Mills, imposer l'invariance d'un lagrangien de Dirac sous une symétrie de jauge entraine de manière automatique le couplage du fermion au(x) boson(s) de jauge.

Le lagrangien du SM, invariant sous $SU(3)_C \times SU(2)_W \times U(1)_Y$, que l'on peut écrire à ce stade, est :
\begin{align}
\mathcal{L}_{SM} &= - \dfrac{1}{4} G^A_{\mu \nu} G_A^{\mu \nu} - \dfrac{1}{4} W^a_{\mu \nu} W_a^{\mu \nu} - \dfrac{1}{4} B_{\mu \nu} B^{\mu \nu} \nonumber \\ \\
&+ i L_i^\dagger D_\mu \bar{\sigma}^\mu L_i + i e_{iR}^\dagger D_\mu \sigma^\mu e_{iR} + i Q_i^\dagger D_\mu \bar{\sigma}^\mu Q_i + i u_{iR}^\dagger D_\mu \sigma^\mu u_{iR} + i d_{iR}^\dagger D_\mu \sigma^\mu d_{iR}, \nonumber
\end{align}
où toutes les particules sont de masses nulles, ce qui est correct pour les gluons. En revanche, on mesure une masse non-nulle pour les fermions et les bosons de l'interaction faible. De tels termes de masse correspondent à :
\begin{equation}
\dfrac{1}{2} m_W^2 W^a_\mu W_a^\mu \ \ \ \text{et} \ \ \ \dfrac{1}{2} m_B^2 B_\mu B^\mu
\label{mass_gauge}
\end{equation}
pour les bosons de jauge EWs, et
\begin{equation}
-m_\psi \bar{\psi} \psi = -m_\psi \left( \psi_R^\dagger \psi_L + \psi_L^\dagger \psi_R \right)
\label{mass_ferm}
\end{equation}
pour un fermion $\psi$. Or, si on écrit les transformations des champs sont le groupe $SU(2)_L \times U(1)_Y$, on obtient
\begin{align}
\psi_L(x) &\mapsto \exp \left( \displaystyle{i \alpha_a(x) \dfrac{\sigma^a}{2} + i \beta(x) Y(\psi)} \right) \psi_L(x) \nonumber \\ \nonumber \\
\psi_R(x) &\mapsto \exp \left( \displaystyle{i \beta(x) Y(\psi)} \right) \psi_R(x) \nonumber \\ \\
\vec{W}_\mu(x) &\mapsto \vec{W}_\mu(x) - \dfrac{1}{g_w} \partial_\mu \vec{\alpha}(x) - \vec{\alpha}(x) \times \vec{W}_\mu(x) \nonumber \\ \nonumber \\
B_\mu(x) &\mapsto B_\mu(x) - \dfrac{1}{g_y} \partial_\mu \beta(x). \nonumber
\end{align}
On en conclut que les termes de masses dans les \'Eqs.~\eqref{mass_gauge} et \eqref{mass_ferm} sont manifestement non-invariants sous ces transformations : de tels termes sont interdits dans une théorie de jauge. Il faut donc rajouter un ingrédient au modèle.

\subsection{Champ de Higgs}
\label{Higgs}
\begin{figure}[h]
\begin{center}
\includegraphics[width=15cm]{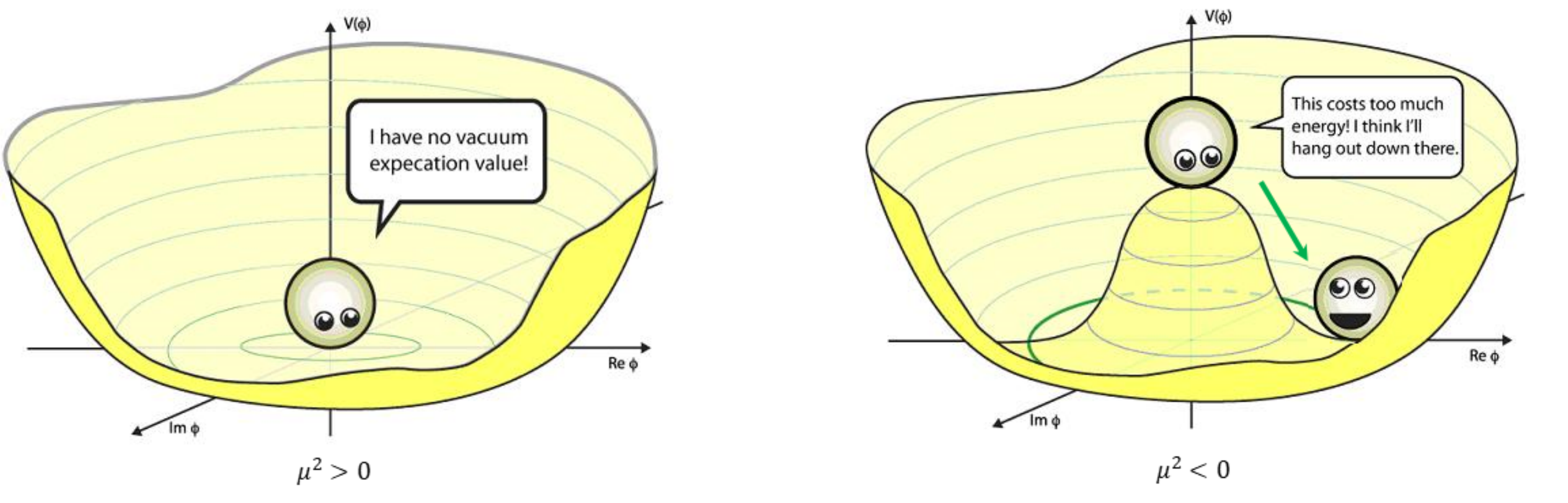}
\end{center}
\caption[Mécanisme de Higgs]{(Adaptée de la Réf.~\cite{QD}.) Schéma du potentiel $V(\phi)$ d'un champ scalaire complexe $\phi$ brisant (à droite) ou non (à gauche) une symétrie de jauge U(1), selon le signe du terme de masse $\mu^2$. Dans le cas de la brisure de symétrie, les fluctuations le long du cercle de minima correspondent au boson de Nambu-Goldstone \og mangé \fg{}  par le boson de jauge, qui obtient ainsi sa masse. Les fluctuations, dans la direction radiale du potentiel, décrivent une particule de spin-0 massive : le boson de Higgs.}
\label{H_potential}
\end{figure}

Afin de donner une masse aux fermions et bosons de l'interaction faible, le SM a recourt au mécanisme de Brout-Englert-Guralnik-Hagen-Higgs-Kibble (plus communément appelé mécanisme de Higgs) de brisure spontanée de symétrie~\cite{Higgs:1964pj, Higgs:1964ia, Englert:1964et, Guralnik:1964eu, Higgs:1966ev, Kibble:1967sv}. Le but est de générer une masse pour les bosons $W^\pm$ et $Z$, tout en préservant une masse nulle pour le photon. Pour cela, on introduit un isodoublet faible de champs scalaires complexes
\begin{equation}
H =
\begin{pmatrix}
H^+ \\
H^0
\end{pmatrix},\ \ \ 
Y(H) = 1,
\end{equation}
tel que sa composante électriquement neutre $H^0$ développe une VEV non-nulle. Le potentiel scalaire $V(H)$ dans le lagrangien
\begin{equation}
\mathcal{L}_H = (D_\mu H)^\dagger (D^\mu H) - V(H), \ \ \ V(H) = \mu^2 H^\dagger H + \lambda (H^\dagger H)^2,
\label{L_H}
\end{equation}
doit alors avoir une instabilité tachyonique $\mu^2 < 0$ et être stabilisé par un terme quartique $\lambda (H^\dagger H)^2$ avec $\lambda > 0$. Ainsi, le minimum du potentiel est obtenu pour (\textit{c.f.} la Fig.~\ref{H_potential})
\begin{equation}
\bra{0} H \ket{0} = \dfrac{1}{\sqrt{2}}
\begin{pmatrix}
0 \\
v
\end{pmatrix}
\ \ \ \text{avec} \ \ \ 
v = \sqrt{-\dfrac{\mu^2}{\lambda}}.
\end{equation}
Si on prend à la place $\mu^2 > 0$, $\lambda > 0$ alors $\bra{0} H \ket{0} = 0$ (\textit{c.f.} la Fig.~\ref{H_potential}) et on a simplement le lagrangien de Klein-Gordon ordinaire d'une particule de spin-0 avec un terme d'interaction quartique. Notons au passage que le cas $\mu^2 < 0$, $\lambda < 0$ correspond à un potentiel tachyonique, et donc non-physique, et que si $\mu^2 > 0$, $\lambda < 0$ alors $\bra{0} H \ket{0} = 0$ est un minimum métastable qui peut se désintégrer en un vide tachyonique.

\newpage

En revenant au premier cas qui nous intéresse, la composante chargée $H^+$ ne doit pas acquérir de VEV afin de préserver la symétrie de jauge $U(1)_{EM}$ à l'origine de l'interaction électromagnétique. Le champ $H(x)$ peut alors s'écrire, au premier ordre, en fonction de quatre champs réels $\theta_{1, 2, 3}(x)$ et $h(x)$ :
\begin{equation}
H(x) =
\begin{pmatrix}
\theta_2(x) + i \theta_1(x) \\
\dfrac{1}{\sqrt{2}} (v + h(x)) - i \theta_3(x)
\end{pmatrix}
= \dfrac{1}{\sqrt{2}} \exp\left({\displaystyle{i \dfrac{\theta_a(x)}{v} \dfrac{\sigma^a}{2}}}\right)
\begin{pmatrix}
0 \\
v + h(x)
\end{pmatrix}.
\end{equation}
On peut alors effectuer une transformation de jauge sous $SU(2)_W$, afin de se placer dans la jauge dite unitaire, telle que
\begin{equation}
H(x) \mapsto \exp\left({\displaystyle{-i \dfrac{\theta_a(x)}{v} \dfrac{\sigma^a}{2}}}\right) H(x) = \dfrac{1}{\sqrt{2}}
\begin{pmatrix}
0 \\
v + h(x)
\end{pmatrix}.
\end{equation}
En écrivant cette expression de $H(x)$ dans le lagrangien~\eqref{L_H}, on obtient :
\begin{equation}
|(D_\mu H)|^2 = \dfrac{1}{2} (\partial_\mu h)^2 + \dfrac{g_w^2}{8} (v + h)^2 |W_\mu^1 + i W_\mu^2|^2 + \dfrac{1}{8} (v + h)^2 |g_w W_\mu^3 - g_y B_\mu|^2.
\label{L_DH}
\end{equation}
Définissons les champs $W^\pm_\mu$, $Z_\mu$, $A_\mu$ des bosons $W^\pm$, $Z$ et du photon :
\begin{align}
W^\pm &= \dfrac{1}{\sqrt{2}} (W_\mu^1 \mp i W_\mu^2),
\nonumber \\ \nonumber \\
Z_\mu &= \dfrac{1}{\sqrt{g_w^2 + g_y^2}} (g_w W_\mu^3 - g_y B_\mu),
\\ \nonumber \\
A_\mu &= \dfrac{1}{\sqrt{g_w^2 + g_y^2}} (g_w W_\mu^3 + g_y B_\mu),
\nonumber
\end{align}
que l'on peut écrire sous forme matricielle :
\begin{equation}
\begin{pmatrix}
Z_\mu \\
A_\mu
\end{pmatrix}
=
\begin{pmatrix}
\cos \theta_w & - \sin \theta_w \\
\sin \theta_w & \cos \theta_w
\end{pmatrix}
\begin{pmatrix}
W^3_\mu \\
B_\mu
\end{pmatrix},
\end{equation}
en introduisant l'angle de mélange faible $\theta_w$, tel que
\begin{equation}
\cos \theta_w = \dfrac{g_w}{\sqrt{g_w^2 + g_y^2}},
\ \ \ 
\sin \theta_w = \dfrac{g_y}{\sqrt{g_w^2 + g_y^2}}.
\end{equation}
Les termes de masse dans \'Eq.~\eqref{L_DH} deviennent
\begin{equation}
m_W^2 W^+_\mu W^{- \mu} + \dfrac{1}{2} m_Z^2 Z_\mu Z^\mu + \dfrac{1}{2} m_A^2 A_\mu A^\mu
\end{equation}
avec
\begin{equation}
m_W = \dfrac{v}{2} g_w,
\ \ \ 
m_Z = \dfrac{v}{2} \sqrt{g_w^2 + g_y^2},
\ \ \ 
m_A = 0.
\end{equation}
\`A la fin, on obtient bien que les trois bosons de l'interaction faible sont massifs et que le photon reste sans masse : c'est la brisure spontanée de symétrie $SU(2)_W \times U(1)_Y \rightarrow U(1)_{EM}$. En fait, trois des quatre degrés de liberté de l'isodoublet $H$ (les bosons de Nambu-Goldstone~\cite{Nambu:1960xd, Nambu:1961tp, Nambu:1961fr, Goldstone:1961eq, Goldstone:1962es}) sont \og mangés \fg{}  par les bosons $W^\pm$ et $Z$ qui acquièrent, de ce fait, une polarisation longitudinale, et donc une masse. Le degré de liberté restant, $h(x)$, décrit une particule massive de spin-0 : le boson de Higgs, qui constitue ainsi la signature du mécanisme éponyme. En faisant la correspondance entre la théorie GSW et celle de Fermi, on peut relier la masse des bosons $W^\pm$ à la constante de Fermi $G_F$, ce qui permet de dériver la valeur de la VEV dans le SM :
\begin{equation}
M_W = \dfrac{v}{2} g_w = \sqrt{\dfrac{\sqrt{2} g_w^2}{8 G_F}} \Rightarrow v = \sqrt{\dfrac{1}{\sqrt{2}G_F}} \simeq 246 \ \text{GeV},
\end{equation}
qui définit l'échelle EW.

Après brisure de symétrie EW (EWSB -- \textit{ElectroWeak Symmetry Breaking}), on peut écrire la dérivée covariante en fonction des états propres de masse. Introduisons les matrices :
\begin{equation}
\sigma^\pm = \dfrac{1}{2} (\sigma^1 \pm i \sigma^2).
\end{equation}
Pour un fermion $\psi$, on a :
\begin{align}
D_\mu \psi &= \left( \partial_\mu -i \dfrac{g_w}{\sqrt{2}} \left[ W^+_\mu \sigma^+ + W^-_\mu \sigma^- \right] - i \dfrac{g_w}{\cos \theta_w} Z_\mu \left[ T^3(\psi) - \sin^2 \theta_w \; Q(\psi) \right] \right. \nonumber \\
&\left. - ie A_\mu \; Q(\psi) \right) \psi,
\end{align}
où on a défini la charge électrique :
\begin{equation}
e = \dfrac{g_w g_y}{\sqrt{g_w^2 + g_y^2}} = g_w \sin \theta_w.
\end{equation}

Le mécanisme de Higgs permet aussi de générer les masses des fermions du SM. Pour une génération, on introduit le lagrangien
\begin{equation}
\mathcal{L}_{H\psi} = - y_e L^\dagger H e_R - y_d Q^\dagger H d_R - y_u Q^\dagger \widetilde{H} u_R + \text{c.h.},
\end{equation}
invariant sous $SU(2)_W \times U(1)_Y$, où $\widetilde{H} = i \sigma^2 H^*$ avec $Y(\widetilde{H}) = -1$. Dans le cas de l'électron, on a par exemple :
\begin{equation}
\mathcal{L}_{He} = - \dfrac{y_e}{\sqrt{2}}(v + h) e_L^\dagger e_R + \text{c.h.}
\end{equation}
Le couplage à la VEV constitue un terme de masse, on obtient :
\begin{equation}
m_e = \dfrac{v}{\sqrt{2}} y_e,
\ \ \ 
m_u = \dfrac{v}{\sqrt{2}} y_u,
\ \ \ 
m_d = \dfrac{v}{\sqrt{2}} y_d.
\end{equation}
Dans le cas de trois générations, on peut écrire des couplages de Yukawa impliquant deux fermions de générations différentes. Ceci induit des mélanges de saveur dans le secteur des quarks. La matrice, permettant de passer des états propres de l'interaction faible (avec \og ' \fg{}) aux états propres de masse (sans \og ' \fg{}), est celle de Cabibbo-Kobayashi-Maskawa (CKM) \cite{Cabibbo:1963yz, Kobayashi:1973fv} :
\begin{equation}
\begin{pmatrix}
d' \\
s' \\
b'
\end{pmatrix}
=
\begin{pmatrix}
V_{ud} & V_{us} & V_{ub} \\
V_{cd} & V_{cs} & V_{cb} \\
V_{td} & V_{ts} & V_{tb}
\end{pmatrix}
\begin{pmatrix}
d \\
s \\
b
\end{pmatrix},
\end{equation}
dont les éléments sont en général complexes. On peut absorber certaines phases en redéfinissant les champs de quarks. Au final, il en reste une : la phase de violation de $CP$, qui vient s'ajouter aux trois angles de mélange.

\begin{figure}[h!]
\begin{center}
\includegraphics[width=15cm]{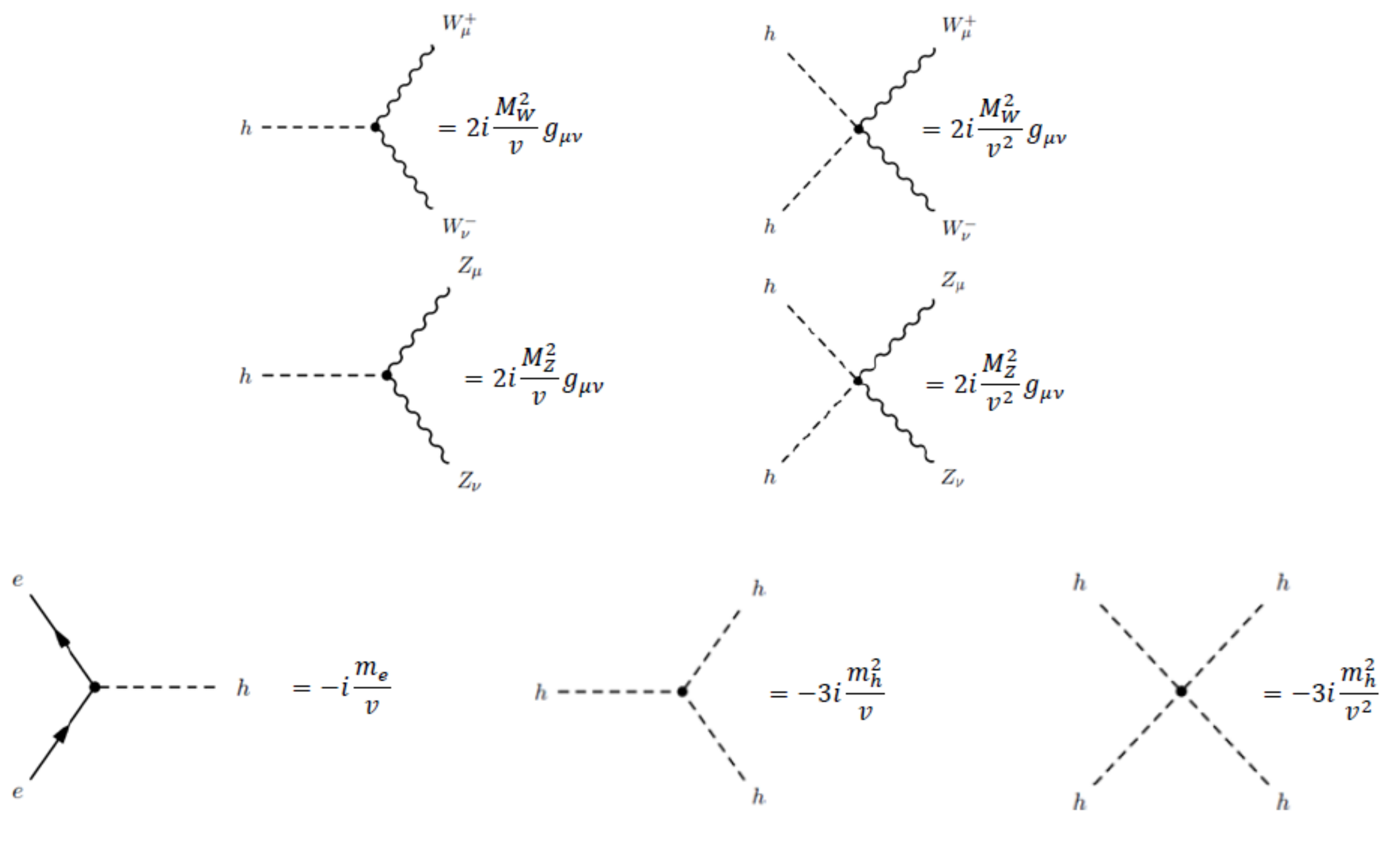}
\end{center}
\caption[Règles de Feynman pour le boson de Higgs]{(Adaptée de la Réf.~\cite{Logan:2014jla}.) Règles de Feynman impliquant le boson de Higgs dans le SM de la physique des particules.}
\label{H_Feyrules}
\end{figure}

Venons en maintenant au boson de Higgs lui-même. En développant $H$ autour de sa VEV dans l'\'Eq.~\eqref{L_H}, et en ne gardant que les termes impliquant $h$ seul, on a :
\begin{equation}
\mathcal{L}_h = \dfrac{1}{2} (\partial_\mu h)^2 - \lambda v^2 h^2 - \lambda v h^3 - \dfrac{\lambda}{4} h^4,
\end{equation}
d'où on lit la masse du boson de Higgs :
\begin{equation}
m_h = \sqrt{2 \lambda} v = \sqrt{2} i \mu,
\end{equation}
en rappelant que $\mu$ est imaginaire pur dans nos conventions. On notera la présence des couplages triple et quartique de $h$. Quant aux termes couplant ce dernier aux bosons de jauge de l'interaction faible ($V = W^\pm, Z$) et aux fermions, ils donnent
\begin{equation}
m_{V}^2 \left( 1 + \dfrac{h}{v} \right)^2,
\ \ \ 
- m_\psi \left( 1 + \dfrac{h}{v} \right).
\end{equation}
Les règles de Feynman, pour le boson de Higgs, sont répertoriées sur la Fig.~\ref{H_Feyrules}.\\

\section{Au-delà du modèle standard de la physique des particules}
Dans cette section, nous allons rapidement passer en revue les principales motivations pour élargir le SM et construire une théorie plus fondamentale des interactions entre les particules élémentaires.

\subsection{Lacunes du modèle standard}

\subsubsection{Masses des neutrinos}
L'observation des oscillations des trois saveurs de neutrinos connues implique qu'au moins deux d'entre eux aient une masse \cite{Wang:2015rma}. Or, de tels termes de masse n'existent pas dans le SM. De nombreux mécanismes ont été proposés, impliquant souvent de nouvelles échelles physiques à haute énergie. La solution la plus naturelle, du point de vue du SM, serait d'ajouter deux ou trois champs de neutrinos de chiralité droite, singulets sous le groupe de jauge du SM, pour pouvoir écrire des termes de Yukawa (impliquant les neutrinos de chiralité droite, le champ de Higgs et les isodoublets faibles contenant les neutrinos de chiralité gauche) qui seraient à l'origine des termes de masse pour les neutrinos après EWSB. Il est alors possible d'ajouter des termes de masse de Majorana pour les neutrinos droits, puisqu'ils sont invariants sous le groupe de jauge du SM et renormalisables. Dans ce cas, les neutrinos seraient des particules de Majorana. Le fait que les neutrinos soient des fermions de Majorana ou de Dirac reste aujourd'hui une question ouverte en physique des particules.

\subsubsection{Interaction gravitationnelle}
La relativité générale\footnote{Voir les Réfs.~\cite{Weinberg:1972kfs, Hawking:1973uf, Misner:1974qy, Wald:1984rg, Schutz:1985jx, Hartle:2003yu, Carroll:2004st}, par exemple, pour une revue.} d'Einstein~\cite{Einstein:1914bt, Einstein:1914bw, Einstein:1914bx, Einstein:1915by, Einstein:1915ca, Hilbert:1915tx, Einstein:1916vd, Einstein:1916cc, Einstein:1916cd} est la théorie de l'espace, du temps et de la gravitation. C'est l'outil central pour comprendre les phénomènes astrophysiques extrêmes comme les trous noirs, les pulsars, les quasars, la fin de vie d'une étoile, la naissance et l'évolution de l'Univers. La relativité générale est une théorie classique, qui n'est pas miscible avec le SM : la gravité est la seule interaction fondamentale connue qui n'a pas encore de formulation quantique achevée, principalement parce qu'une quantification naïve aboutit à une théorie qui n'est pas perturbativement renormalisable. La relativité générale est donc une EFT\footnote{Pour une revue introduisant le concept d'EFT, \textit{c.f.} les Réfs.~\cite{Georgi:1994qn, Manohar:1996cq, Rothstein:2003mp, Burgess:2007pt, Bain:2013imf, Gripaios:2015qya, Manohar:2018aog}} qui cesse d'être valide à une certaine échelle d'énergie de coupure\footnote{Les revues~\cite{Donoghue:1995cz, Burgess:2003jk, Donoghue:2017pgk} donnent une introduction à l'étude de la gravité basée sur les EFTs.} $M_{QG}$, correspondant à l'échelle de masse des degrés de liberté UVs les plus légers de la gravité.

La relativité générale a une formulation lagrangienne : l'action d'Einstein-Hilbert, invariante sous les difféomorphismes et les transformations locales de Lorentz, s'écrit :
\begin{equation}
S_{EH} = \int d^4x \sqrt{|g|} \left( - \dfrac{M_P^2}{2}  R - \Lambda_c + \mathcal{L}_{SM} \right),
\label{action_Einstein-Hilbert}
\end{equation}
où $g$ est le déterminant de la métrique $g_{\mu \nu}$, $R$ est la courbure scalaire de Ricci, $\Lambda_c$ est la constante cosmologique, $\mathcal{L}_{SM}$ est le lagrangien du SM, et $M_P$ est la masse de Planck\footnote{La définition de $M_P$ adoptée ici est généralement celle de la masse de Planck réduite. Il y a de nombreuses définitions de la masse de Planck dans la littérature, la plus commune diffère de la notre par un facteur $\sqrt{8 \pi}$.},
\begin{equation}
M_P = \sqrt{\dfrac{1}{8 \pi G_N}} \simeq 2,4 \times 10^{18} \ \text{GeV},
\end{equation}
où $G_N$ est la constante de Newton. On définit aussi la longueur de Planck,
\begin{equation}
\ell_P = \dfrac{1}{M_P} \simeq 8,0 \times 10^{-35} \ \text{m} \, .
\end{equation}
En l'absence d'un achèvement UV explicite, on peut \textit{a priori} ajouter à la relativité générale une infinité d'opérateurs non-renormalisables invariants sous les difféomorphismes et les transformations locales de Lorentz. Dans le régime où la relativité générale est applicable, il est possible d'étudier les fluctuations quantiques linéaires $h_{\mu \nu}$ autour d'une métrique de fond classique $\bar{g}_{\mu \nu}$ :
\begin{equation}
g_{\mu \nu} = \bar{g}_{\mu \nu} + \dfrac{1}{2 M_P} h_{\mu \nu},
\label{graviton}
\end{equation}
où $\bar{g}_{\mu \nu}$ est obtenue en résolvant les équations d'Einstein. Les fluctuations $h_{\mu \nu}$ se comportent comme une particule de spin-2 de masse nulle, le graviton, se propageant sur un espace-temps classique. Il est alors possible de quantifier la relativité générale linéarisée, mais la théorie reste non-renormalisable.

Une théorie non-renormalisable devient fortement couplée à une certaine échelle d'énergie, où :
\begin{itemize}
\item tous les opérateurs, quelque soit leur dimension, contribuent de manière égale à l'arbre,
\item les contributions quantiques (à boucle), associées à une interaction, donnent toutes une contribution égale à celle à l'arbre, ce qui signifie que l'expansion perturbative est perdue.
\end{itemize}
Le procédé d'analyse dimensionnelle naïve \cite{Weinberg:1978kz, Manohar:1983md, Cohen:1997rt} permet d'estimer l'échelle d'énergie $\Lambda_{NDA}$ à partir de laquelle une théorie effective devient non-perturbative, du fait de son caractère non-renormalisable, et en l'absence des nouveaux degrés de liberté de l'achèvement UV. Si ce dernier est une théorie fortement couplée, alors les nouveaux degrés de liberté les plus légers ont leur masse de l'ordre de $\Lambda_{NDA}$. Au contraire, si c'est une théorie faiblement couplée, alors les nouveaux degrés de liberté les plus légers ont, en général, leur masse bien inférieure à $\Lambda_{NDA}$ \cite{Giudice:2016yja}. Il est donc impossible de déterminer l'échelle d'énergie de coupure d'une théorie effective sans en connaître l'achèvement UV. Au mieux, on peut estimer l'échelle de coupure maximale $\Lambda_{NDA}$, grâce à l'analyse dimensionnelle naïve. En relativité générale linéarisée, le couplage du graviton est
\begin{equation}
g_h = \dfrac{E}{M_P} \, ,
\end{equation}
où $E$ est l'échelle d'énergie typique du processus étudié. Chaque boucle apporte un facteur $g_h^2/\ell_4$, où $l_4 = 16 \pi^2$ est le facteur de boucle à 4D, donc la relativité générale linéarisée devient fortement couplée quand $g_h^2 = \ell_4$. Ceci correspond à l'échelle d'énergie \cite{Giudice:2003tu}
\begin{equation}
\Lambda_{NDA} = \sqrt{l_4} \, M_P \sim 3 \times 10^{19} \ \text{GeV} \, .
\end{equation}
Cependant, si la gravité est décrite dans l'UV par une théorie faiblement couplée, \textit{i.e.} si l'un de ses couplages $g_{QG}^2 \ll l_4$, on s'attend à avoir $M_{QG} \ll \Lambda_{NDA}$ puisqu'une analyse dimensionnelle donne, pour une théorie à 4D,
\begin{equation}
M_{QG} = \dfrac{g_{QG}}{\sqrt{l_4}} \, \Lambda_{NDA} \, .
\label{M_QG_M_P}
\end{equation}

\begin{figure}[h!]
\begin{center}
\includegraphics[width=15cm]{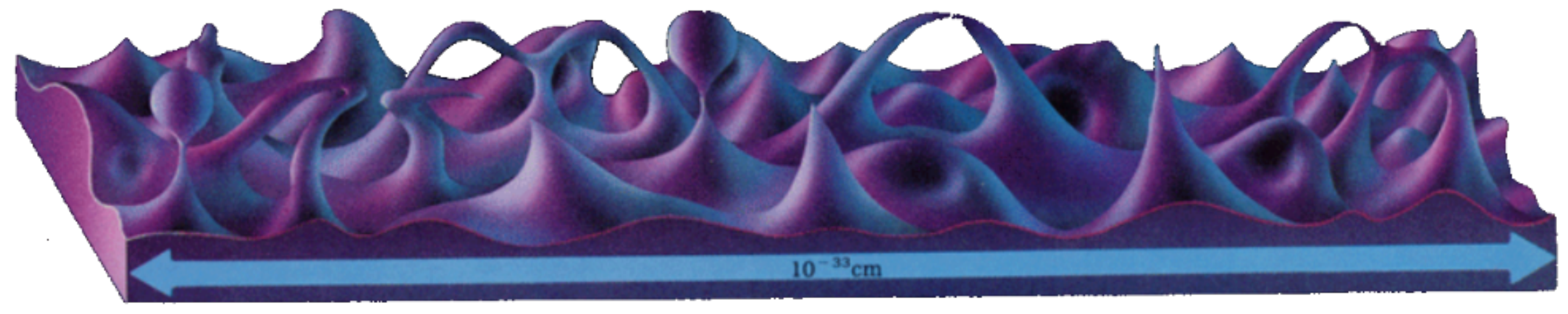}
\end{center}
\caption[Espace-temps quantique]{(Adaptée de la Réf.~\cite{siteW3}.) Vue d'artiste de l'espace-temps à l'échelle de Planck, dominé par des effets gravitationnels non-perturbatifs. On peut imaginer que l'espace-temps est alors une \og mousse \fg{} de trous noirs, trous de ver, bébés univers et instantons gravitationnels.}
\label{quantumfoam}
\end{figure}

On vient de voir que la relativité générale linéarisée peut être quantifiée en tant que théorie effective. Qu'en est-il des effets non-linéaires ? À ce jour, il n'y a pas de cadre théorique clair pour quantifier la relativité générale en incluant les effets non-linéaires. La physique des trous noirs permet d'intuiter \cite{Han:2002yy} qu'ils doivent apparaître à une énergie de l'ordre de $M_P$, où \'Eq.~\eqref{graviton} n'est plus applicable. On s'attend à ce qu'à une échelle d'énergie de l'ordre de $M_P$ ou, de manière équivalente, à une échelle de distance de l'ordre de $\ell_P$, les effets de gravité non-linéaires et non-perturbatifs deviennent dominants. En s'inspirant de l'étude de la théorie de la gravité quantique euclidienne \cite{EQG}, on peut imaginer, de manière très spéculative, que l'espace-temps quantique est une \og mousse \fg{} \cite{Hawking:1988ae, Hawking:1995ag} de trous noirs, trous de ver, bébés univers, instantons gravitationnels, et autres objets exotiques (\textit{c.f.} la Fig.~\ref{quantumfoam}), dont le cadre théorique cohérent, notamment l'achèvement UV, reste largement à construire (\textit{c.f.} la Réf.~\cite{Hebecker:2018ofv} pour une revue récente). Comme $M_P < \Lambda_{NDA}$, $M_{QG} \lesssim M_P$. Si $M_{QG} \ll M_P$, les degrés de liberté UV de la gravité, même faiblement couplés, peuvent altérer significativement le comportement UV de la gravité, notamment l'échelle à laquelle les effets non-linéaires deviennent importants, qui peut être très différente de celle spéculée à partir de la relativité générale.\\

\begin{figure}[h]
\begin{center}
\includegraphics[height=10cm]{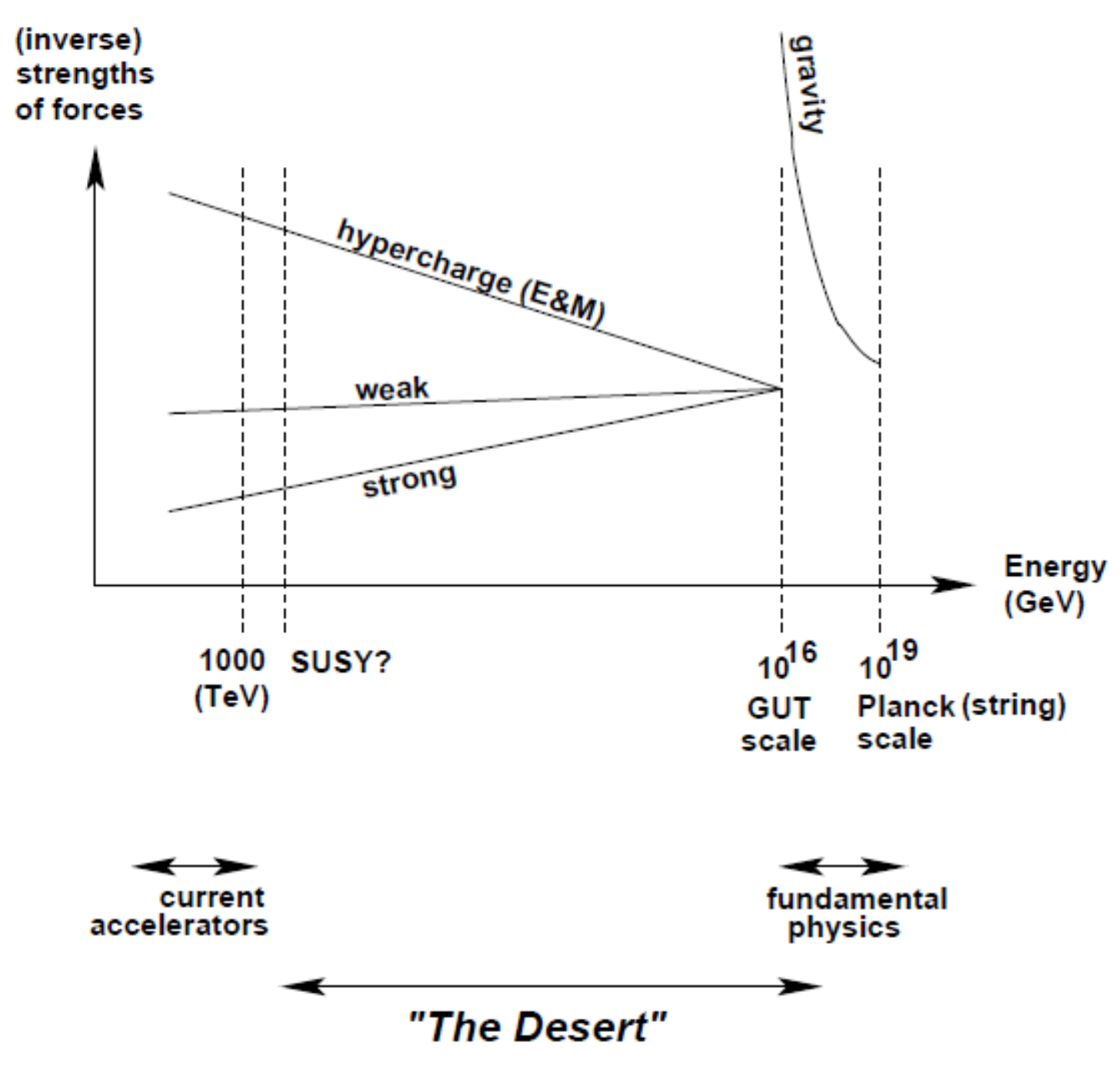}
\end{center}
\caption[Paradigme standard]{(Adaptée de la Réf.~\cite{Dienes:2002hg}.) Vue schématique du paradigme standard.}
\label{desert}
\end{figure}

Encore aujourd'hui, beaucoup de physiciens admettent le \og paradigme standard \fg{}, qui repose sur deux hypothèses fortes \cite{Dvali:2001gx} :
\begin{itemize}
\item $M_{QG} \sim M_P$, la théorie UV de la gravité est alors fortement couplée. Dans ce cas, rien ne se passe concernant la gravité entre l'échelle du TeV, actuellement scrutée au LHC, et $M_P$ (\textit{c.f.} la Fig.~\ref{desert}).
\item $M_{QG}$ est supposée être aussi l'énergie de coupure des modèles de physique des particules décrivant les interactions fondamentales dans le cadre de la QFT (le SM et ses extensions).
\end{itemize}
Le \og désert \fg{} physique entre l'échelle EW et $M_P$, qui en résulte, est appelé la hiérarchie de jauge. Souvent, on inclut dans le paradigme standard la solution la plus populaire à ce que l'on appelle le problème de hiérarchie de jauge (\textit{c.f.} Section~\ref{gauge_hier}) : la SUperSymétrie (SUSY) \cite{Weinberg:2000cr}. Celle-ci est une éventuelle symétrie de l'espace-temps reliant bosons et fermions. Elle permet aux couplages de jauge du SM de s'unifier à une énergie autour de $M_{GUT} \sim 10^{16} \ \text{GeV}$. On parle de théories de grande unification (GUTs -- \textit{Grand Unified Theories}). La hiérarchie de jauge désigne alors l'écart en énergie entre l'échelle de brisure de la SUSY, $M_{SUSY} \sim 1 \ \rm{TeV}$, et $M_{GUT} \sim 10^{16} \ \text{GeV}$. Dans ce manuscrit, on s'intéressera à des modèles où la SUSY n'est pas forcément nécessaire, et on n'inclura donc pas la SUSY et les GUTs dans notre définition du paradigme standard.

Les hypothèses du paradigme standard sont très discutables. La première a été remise en cause par les Réfs.~\cite{ArkaniHamed:1998rs, Antoniadis:1998ig, ArkaniHamed:1998nn}. Les auteurs supposent la présence de dimensions spatiales supplémentaires compactes, dont le rayon de compactification $R$ est supérieure à $\ell_P$. Ceci implique que $M_P$ n'est qu'une échelle effective, différente de l'échelle de Planck réelle $M_*$ :
\begin{equation}
M_P^2 = (2 \pi R)^n \, M_*^{n+2} \, .
\end{equation}
Le même résultat peut être obtenu si on rajoute, au SM, un secteur caché (champs neutres sous les interactions du SM) avec un grand nombre de degrés de liberté \cite{Dvali:2007hz, Dvali:2007wp} :
\begin{equation}
M_P = \sqrt{N} \, M_* \, ,
\end{equation}
où $N$ est le nombre d'espèce de particules du modèle. Les Réfs.~\cite{Sundrum:1997js, Dvali:2000hr, Dvali:2001gx, Sundrum:2003jq} abandonnèrent la deuxième hypothèse pour justifier la petitesse de la constante cosmologique mesurée par rapport aux échelles de la physique des particules testées aux collisionneurs. $M_{QG}$ peut alors être très inférieure à l'échelle d'énergie où le SM cesse d'être une description correcte de la Nature. Les effets de la théorie UV de la gravité doivent alors être \og doux \fg{}, \textit{i.e.} qu'ils n'augmentent pas le couplage effectif du SM à la gravité, celle-ci peut alors continuer à être négligée dans l'étude des processus en collisionneur.

\subsubsection{Modèle standard de la cosmologie}
\begin{figure}[h]
\begin{center}
\includegraphics[width=15cm]{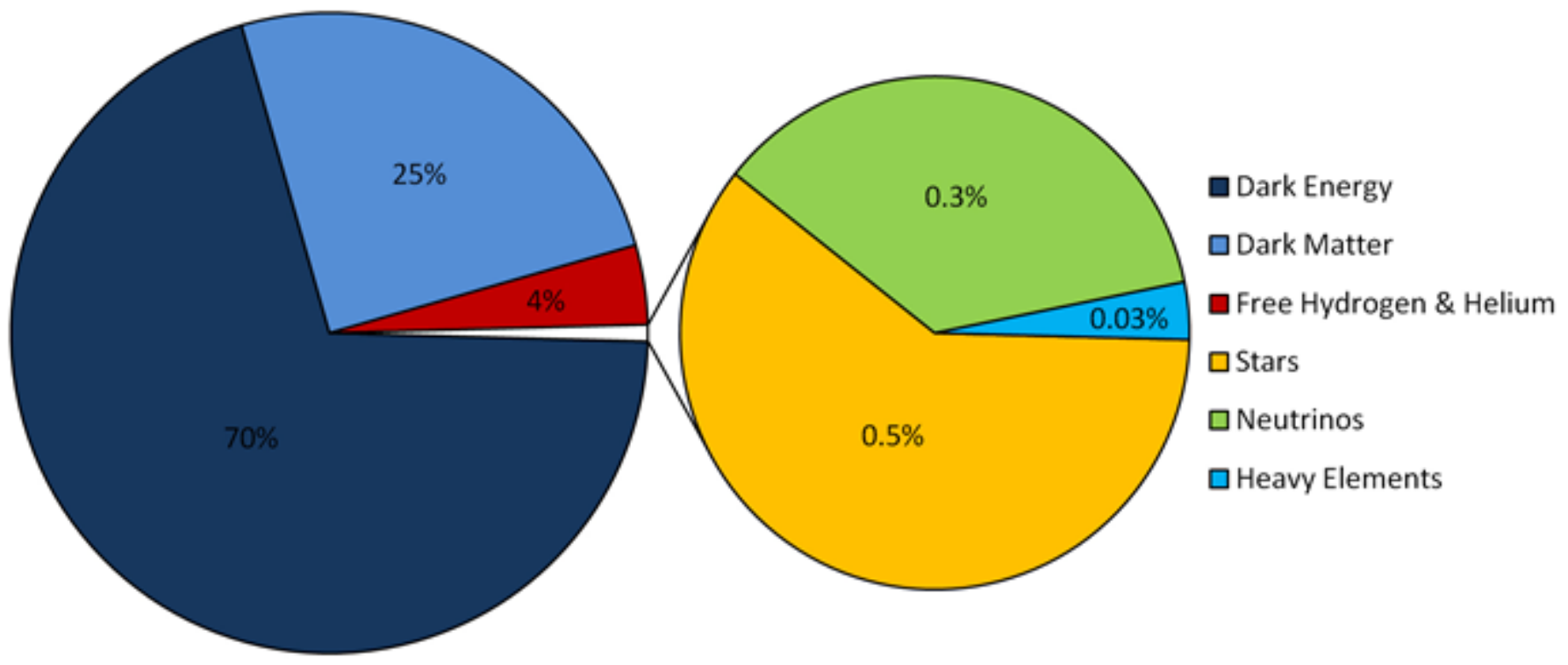}
\end{center}
\caption[Contenu en énergie de l'Univers]{(Adaptée de la Réf.~\cite{siteW4}.) Contenu en énergie de l'Univers.}
\label{LCDM}
\end{figure}

Le SM de la cosmologie, le $\Lambda$CDM, suppose l'existence d'autres champs de particules pour reproduire les observations. Ainsi, environ $70\%$ du contenu en énergie de l'Univers est constitué d'énergie noire pour expliquer l'accélération de son expansion. Dans le $\Lambda$CDM, ce rôle est tenu par la constante cosmologique $\Lambda_c$. Environ $25\%$ est constitué de matière noire froide (CDM), et seulement environ $5\%$ est constitué de matière du SM de la physique des particules. Le contenu en énergie de l'Univers est résumé par la Fig.~\ref{LCDM}

\begin{figure}[h]
\begin{center}
\includegraphics[width=15cm]{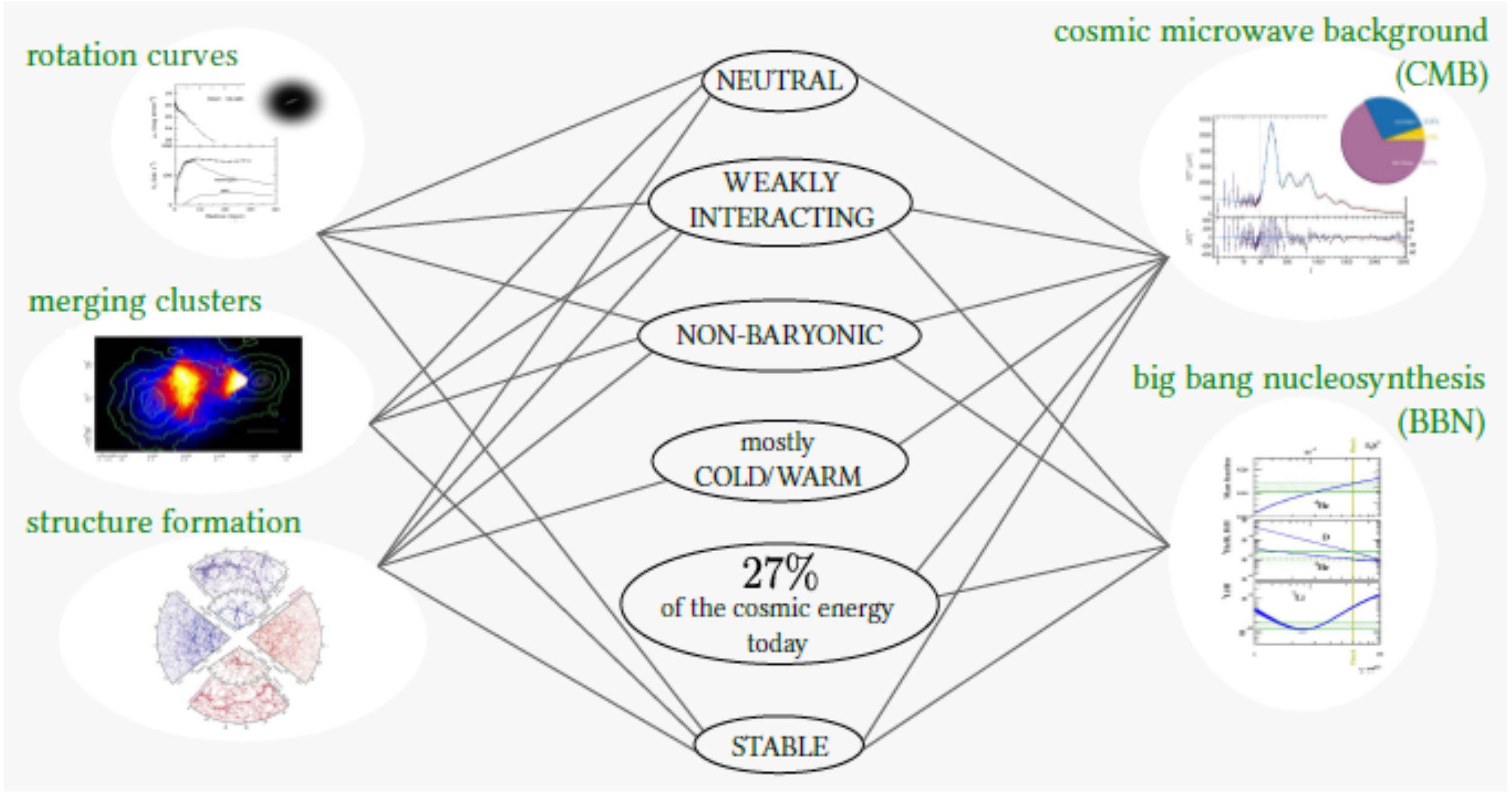}
\end{center}
\caption[Propriétés de la matière noire]{(Adaptée de la Réf.~\cite{Dutra}.) Résumé des propriétés de la matière noire et de leurs évidences observationnelles.}
\label{CDM}
\end{figure}

L'hypothèse de nouvelles particules constituant la CDM est aujourd'hui la meilleure pour expliquer toutes les évidences observationnelles de masse manquante dans l'Univers : les courbes de rotation des galaxies \cite{Rubin:1970zza, Faber:1979pp, Kent:1987zz, Sofue:2000jx, Bertone:2004pz}, la collision d'amas \cite{Clowe:2006eq}, la formation des structures \cite{Springel:2006vs}, le fond diffus cosmologique (CMB -- \textit{Cosmic Microwave Background}) \cite{Akrami:2018odb}, et la nucléosynthèse primordiale (BBN -- \textit{Big Bang Nucleosynthesis}) \cite{Coc:2017pxv}. Les propriétés de la matière noire et les évidences observationnelles pour celles-ci sont résumées par la Fig.~\ref{CDM}.

Le $\Lambda$CDM fait également l'hypothèse d'une époque de l'Univers primordial où celui-ci est en expansion rapide : l'inflation \cite{Guth:1980zm}. Cette phase est provoquée par la rétroaction sur la métrique d'au moins un champ scalaire : l'inflaton.

\newpage

\subsection{Naturalité de l'échelle électrofaible}
\label{gauge_hier}
\begin{figure}[h]
\begin{center}
\includegraphics[height=2.5cm]{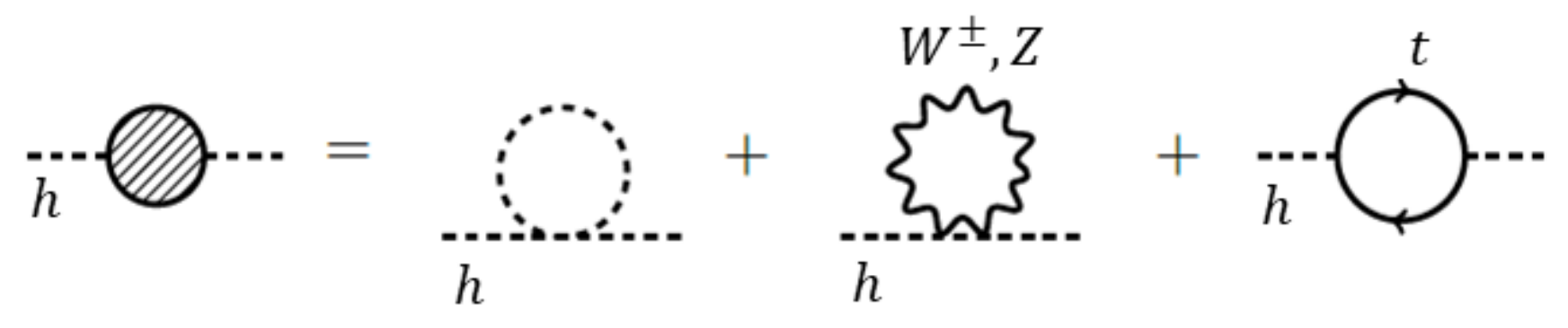}
\end{center}
\caption[Corrections à une boucle à la masse du boson de Higgs]{(Adaptée de la Réf.~\cite{Csaki:2016kln}.) Diagrammes de Feynman des corrections à une boucle à la masse du boson de Higgs.}
\label{rad_H}
\end{figure}

On vient de voir que le SM n'est pas la théorie ultime de la physique. C'est donc une EFT, certes renormalisable et unitaire, mais qui cesse d'être valide à une échelle de coupure $\Lambda_{UV}$. Intéressons nous aux corrections radiatives à une boucle à la masse du boson de Higgs, générées principalement par le quark top, le boson de Higgs lui-même, les bosons $W^\pm$ et $Z$ (\textit{c.f.} la Fig.~\ref{rad_H}). En coupant les moments dans l'intégrale de boucle à l'échelle de coupure $\Lambda_{UV}$, on obtient :
\begin{equation}
\delta m_h^2 = m_h^2 - (m_h^0)^2 = \dfrac{\Lambda_{UV}^2}{32 \pi^2} \left[ 6 \lambda + \dfrac{9 g_w^2 + 3 g_y^2}{4} - y_t^2 \right],
\label{hierarchie}
\end{equation}
où on a gardé que les contributions dominantes. $m_h$ est la masse physique du boson de Higgs, $m_h^0$ est sa masse \og nue \fg{} et $\delta m_h$ est la contribution radiative. Si $\Lambda_{UV} > 10 \ \text{TeV}$, alors les corrections radiatives sont plus grandes que la masse elle-même : la masse du boson de Higgs est quadratiquement sensible à toute échelle de nouvelle physique au-dessus de l'échelle EW. S'il n'y a pas de physique BSM à une échelle d'énergie inférieure à $M_P$, alors $\Lambda_{UV} \sim M_P$ et $(m_h^0)^2$ doit être finement ajustée pour compenser la contribution radiative sur une trentaine d'ordre de grandeur ! C'est le problème de hiérarchie de jauge, inhérent au caractère d'EFT du SM, et caractéristique des champs scalaires élémentaires, comme le boson de Higgs. En effet, les corrections radiatives aux masses des fermions et bosons de jauge croissent logarithmiquement avec $\Lambda_{UV}$, et sont proportionnelles aux masses des particules elles-mêmes :
\begin{align}
\delta m_\psi &\sim m_\psi \; \text{ln} \left( \dfrac{\Lambda_{UV}}{m_\psi} \right), \nonumber \\ \\
\delta m_V &\sim m_V^2 \; \text{ln} \left( \dfrac{\Lambda_{UV}}{m_V} \right). \nonumber
\end{align}
Les corrections de boucles sont donc supprimées par le petit paramètre à l'arbre. Ce phénomène est relié au critère de naturalité technique de t'Hooft~\cite{tHooft:1979rat, tHooft:1980xss} : un paramètre est naturellement petit si, lorsqu'il est mis à zéro, la théorie a une symétrie supplémentaire. Pour les fermions et les bosons de jauge, ce sont respectivement les symétries chirale et de jauge.

\begin{figure}[h!]
\begin{center}
\includegraphics[height=2.5cm]{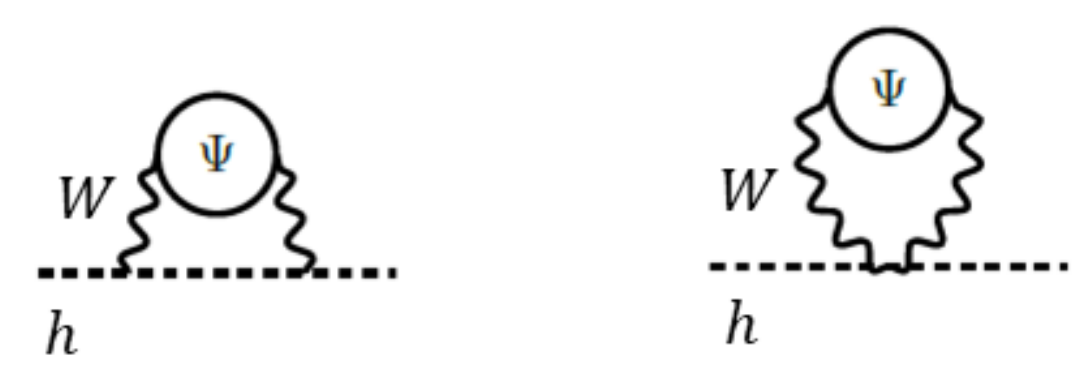}
\end{center}
\caption[Corrections à deux boucles d'un fermion lourd à la masse du boson de Higgs]{(Adaptée de la Réf.~\cite{Csaki:2016kln}.) Diagrammes de Feynman des corrections à deux boucles à la masse du boson de Higgs impliquant un fermion lourd $\Psi$, couplant indirectement au boson de Higgs via les interactions de jauge.}
\label{rad_H_2}
\end{figure}

Il est important de rappeler que le problème de hiérarchie est indépendant du schéma de renormalisation. On trouve parfois dans la littérature que la régularisation dimensionnelle fait disparaître la divergence quadratique en $\Lambda_{UV}$, remplacée par un pôle en $1/\epsilon$ correspondant à une divergence logarithmique. Ce n'est pas le bon argument : le problème de hiérarchie n'est pas relié à la divergence quadratique mais à le sensibilité quadratique aux nouvelles échelles d'énergie. Par exemple, supposons l'existence d'un nouveau scalaire lourd $S$ de masse $m_S$ couplant au boson de Higgs par un terme $-\lambda_S |H|^2|S|^2$. La contribution à une boucle de la particule $S$ est
\begin{equation}
\delta m_h^2 = \dfrac{\lambda_S}{16 \pi^2} \left[ \Lambda_{UV}^2 - 2 m_S^2 \; \text{ln} \left( \dfrac{\Lambda_{UV}}{m_S} \right) + \ldots \right].
\end{equation}
Si $\Lambda_{UV}$ n'est pas, cette fois, une échelle BSM mais juste un régulateur (que l'on peut faire disparaître par une régularisation dimensionnelle), la seule échelle de nouvelle physique est $m_S$. On voit que la masse du boson de Higgs est quadratiquement sensible à $m_S$, et ce peu importe le schéma de régularisation choisi pour le calcul de la boucle. Ceci reste vrai même si les états BSM ne couplent pas directement au boson de Higgs mais interagissent seulement avec les autres champs du SM. Prenons par exemple une paire de fermions lourds $\Psi$, chargés sous le groupe de jauge du SM mais ne couplant pas directement à $H$.
Ils contribuent aux corrections radiatives à $m_h$ via des diagrammes à deux boucles (\textit{c.f.} la Fig.~\ref{rad_H_2}) :
\begin{equation}
\delta m_h^2 \sim \left( \dfrac{g_w^2}{16 \pi^2} \right)^2 \left[ a \Lambda_{UV}^2 + 48 m_\Psi^2 \; \text{ln} \left( \dfrac{\Lambda_{UV}}{m_\Psi} \right) + \ldots \right],
\end{equation}
où l'on retrouve une dépendance en $m_\Psi^2$. \\

La valeur de $m_h$ est donc naturellement à l'échelle de la théorie fondamentale. Du point de vue du groupe de renormalisation Wilsonien, $m_h$ est le seul paramètre relevant du SM, donc son importance augmente au cours du flot vers l'infrarouge (IR -- \textit{InfraRed}). Il est donc difficile (ajustement fin) de trouver une trajectoire du groupe de renormalisation qui donne la valeur de $m_h$ mesurée dans l'IR. Si le boson de Higgs est réellement une particule élémentaire qui n'est pas protégée par une symétrie, comme c'est le cas dans le SM, alors toute échelle de nouvelle physique doit être inférieure à 10 TeV, afin que la valeur mesurée de $m_h$ soit techniquement naturelle. Le boson de Higgs, comme toute autre particule, couple à la gravité. Si les effets de la théorie UV de la gravité ne sont pas doux, on s'attend à ce que $\delta m_h^2$ soit dominée par $M_{QG}^2$. $m_h$ est alors techniquement naturelle si $M_{QG} < 10$ TeV et donc, d'après \'Eq.~ \eqref{M_QG_M_P}, si $g_{QG} < 10^{-15}$ pour une théorie à 4D et en l'absence d'un secteur caché comprenant un grand nombre de degrés de liberté. On peut ainsi se demander pourquoi la théorie UV de la gravité serait si faiblement couplée mais, sans théorie explicite, il n'est pas possible de dire si c'est une reformulation du problème de hiérarchie de jauge ou non.

Si l'échelle de gravité est très grande devant l'échelle EW, la manière conservative (du point de vue de la structure du SM) la plus simple pour résoudre le problème de hiérarchie de jauge est de supposer l'existence de la SUSY. Dans les modèles les plus simples, toutes les particules du SM ont un partenaire de même masse et couplages dont le spin diffère d'une demie unité. Comme les boucles de fermions et de bosons ont un signe opposé, les corrections radiatives s'annulent. Ces particules n'ayant pas été découvertes, la SUSY doit être spontanément brisée à une échelle $M_{SUSY}$. Les particules supersymétriques sont alors plus lourdes que leurs partenaires du SM. L'échelle $M_{SUSY}$ doit être autour du TeV pour stabiliser la masse du boson de Higgs contre les corrections radiatives.

\chapter{La voie des Univers branaires}
\label{univers_branaires}

\section{Préambule historique}
\subsection{Modèles à la Kaluza-Klein}
La première apparition d'une dimension spatiale supplémentaire, dans la littérature scientifique, est due à Nordström~\cite{Nordstrom:1914fi}. En 1914 (un an avant la publication de la relativité générale par Einstein), il étend la théorie de l'électromagnétisme à 5D, en utilisant un 5-vecteur,
\begin{equation}
A_M =
\begin{pmatrix}
A_\mu \\
\phi
\end{pmatrix},
\end{equation}
le 4-vecteur $A_\mu$ et le scalaire $\phi$ décrivant respectivement les interactions électromagnétiques et gravitationnelles. Ainsi, il proposa une théorie unifiée de ces deux interactions fondamentales.

En 1921, s'appuyant sur la relativité générale, Kaluza~\cite{Kaluza:1921tu} repris cette idée avec cette fois une théorie tensorielle de la gravité. La métrique à 5D se décompose comme
\begin{equation}
g_{MN} =
\left(
\begin{array}{ccccc|c}
& & & & & \\
& & & & & \\
& & g_{\mu \nu} & & & A_{\mu} \\
& & & & & \\
& & & & & \\
\hline
& & & & & \\
& & A_{\mu}^{T} & & & \phi\\
& & & & &
\end{array}
\right),
\label{graviton_KK}
\end{equation}
où le scalaire $\phi$, quelque peu embarrassant alors pour Kaluza, ajouté à la métrique $g_{\mu \nu}$ est compris aujourd'hui comme constituant une théorie tenseur-scalaire de la gravité à la Brans-Dicke. Ainsi, dans le modèle de Nordström, la gravité n'est qu'un effet électromagnétique dans une dimension spatiale supplémentaire alors que, chez Kaluza, c'est l'interaction électromagnétique qui est un effet gravitationnel dans celle-ci. Nordström, comme Kaluza, firent l'hypothèse drastique que les champs ne dépendent pas de la coordonnée de la dimension spatiale supplémentaire, afin d'expliquer la non-observabilité de cette dernière.

\begin{figure}[h]
\begin{center}
\includegraphics[width=15cm]{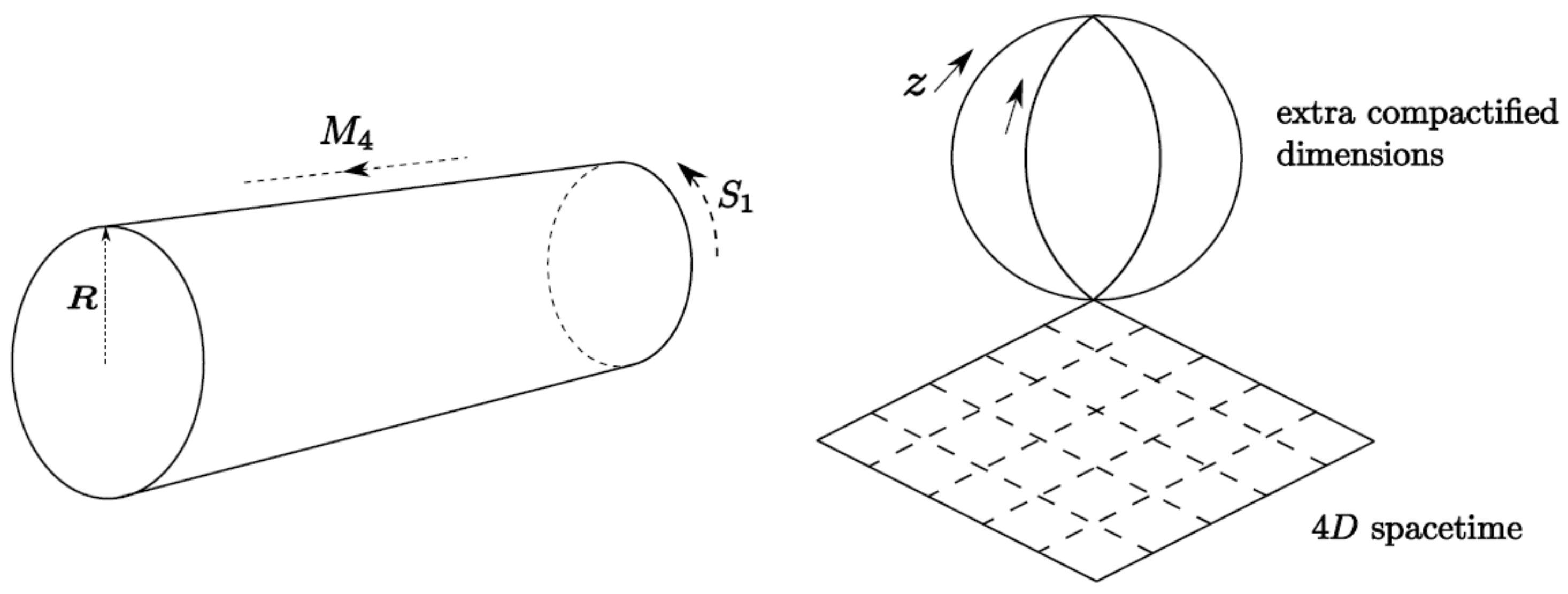}
\end{center}
\caption[Dimension spatiale supplémentaire compactifiée]{(Adaptée de la Réf.~\cite{Shifman:2009df}.) À gauche, schéma de l'espace-temps de Kaluza-Klein, où $M_4$ est l'espace-temps de Minkowski à 4D et $S^1$ est la dimension spatiale supplémentaire compactifiée sur un cercle. À droite, généralisation à deux dimensions spatiales supplémentaires compactifiées sur une sphère.}
\label{KK}
\end{figure}

Pour remédier à ce problème, Klein~\cite{Klein:1926tv} proposa en 1926, après l'avènement de la mécanique quantique, de compactifier la dimension spatiale supplémentaire sur un cercle $S^1$ dont le rayon $R$ serait très petit (\textit{c.f.} la Fig.~\ref{KK}). Les champs sur le cercle admettent alors une décomposition en modes de Fourier, étiquetés par l'entier $n$, dont le moment est quantifié en $n/R$ ; on parle aujourd'hui de décomposition de KK. Par exemple, pour un champ scalaire $\phi (x^\mu,y)$, dépendant des coordonnées $x^\mu$ de l'espace-temps de Minkowski à 4D et de la coordonnée $y$ d'une dimension spatiale supplémentaire, on a
\begin{equation}
\phi (x^\mu,y) = \sum_n \phi_n (x^\mu) f_n(y),
\end{equation}
où $\phi_n (x^\mu)$ et $f_n(y)$ sont respectivement les champs et les fonctions d'onde des modes de KK dans la dimension spatiale supplémentaire, étiquetés par l'entier $n$. En prenant un rayon suffisamment petit, les modes $n>0$ sont hors de portée des expériences. $U(1)$ étant le groupe d'isométries de $S^1$, on comprend bien le lien intime entre la géométrie de la dimension spatiale supplémentaire et la nature de l'interaction induite. Dans les années 30, Pauli~\cite{Pauli:1933gc} et Klein~\cite{Klein:1938jm} généralisèrent l'idée au cas de deux dimensions spatiales supplémentaires compactifiées sur une sphère $S^2$, dont le groupe d'isométries est $SU(2)$ (\textit{c.f.} la Fig.~\ref{KK}). Cette fois-ci, le modèle unifie la gravité avec une interaction basée sur un groupe de jauge non-abélien : ils découvrirent alors la première théorie de Yang-Mills.

Le principe des théories de KK est simple et attrayant : compactifier des dimensions spatiales supplémentaires sur une variété dont le groupe d'isométries générées par les vecteurs de Killing se manifeste comme une théorie de jauge à 4D. En 1975, Cho et Freund~\cite{Cho:1975sf} présentèrent la dérivation complète des théories gravitationnelles et de Yang-Mills à partir d'une théorie de dimension supérieure. Néanmoins, cette voie a un problème de taille : l'espace-temps à 4D usuel est nécessairement courbe, rejetant la solution de type Minkowski. Pour y remédier, les tentatives qui suivirent se focalisèrent sur la réalisation d'une compactification spontanée des dimensions spatiales supplémentaires, proposée par Cremmer et Scherk~\cite{Cremmer:1976zc}, en incluant des scalaires et champs de jauge additionnels, mais abandonnant ainsi l'idée fondatrice de KK d'une \og théorie du tout \fg{} purement gravitationnelle.

Motivé par le développement de la SUSY et de son application à la gravité, la supergravité\footnote{Pour une revue sur la SUGRA, \textit{c.f.} la Réf.~\cite{Freedman:2012zz}.} (SUGRA -- \textit{SUperGRAvity}), Witten~\cite{Witten:1981me} démontra en 1981 que le nombre minimal de dimensions spatiales supplémentaires pour réaliser une théorie de KK, incluant le SM, est sept, ce qui correspond également au nombre maximal de dimensions spatiales supplémentaires qu'une théorie SUGRA cohérente peut avoir, laissant entrevoir le rêve que la théorie du tout pourrait être la SUGRA à 11D. Cependant, une telle théorie n'est pas renormalisable, et Witten~\cite{Witten:1981me} montra qu'aucune compactification de la variété à 7D ne peut aboutir à une théorie chirale comme le secteur EW du SM. Ceci enterra définitivement la voie des scenarii de KK pour construire une théorie ultime de la physique.

\subsection{Théories des supercordes}

\begin{figure}[h]
\begin{center}
\includegraphics[height=8cm]{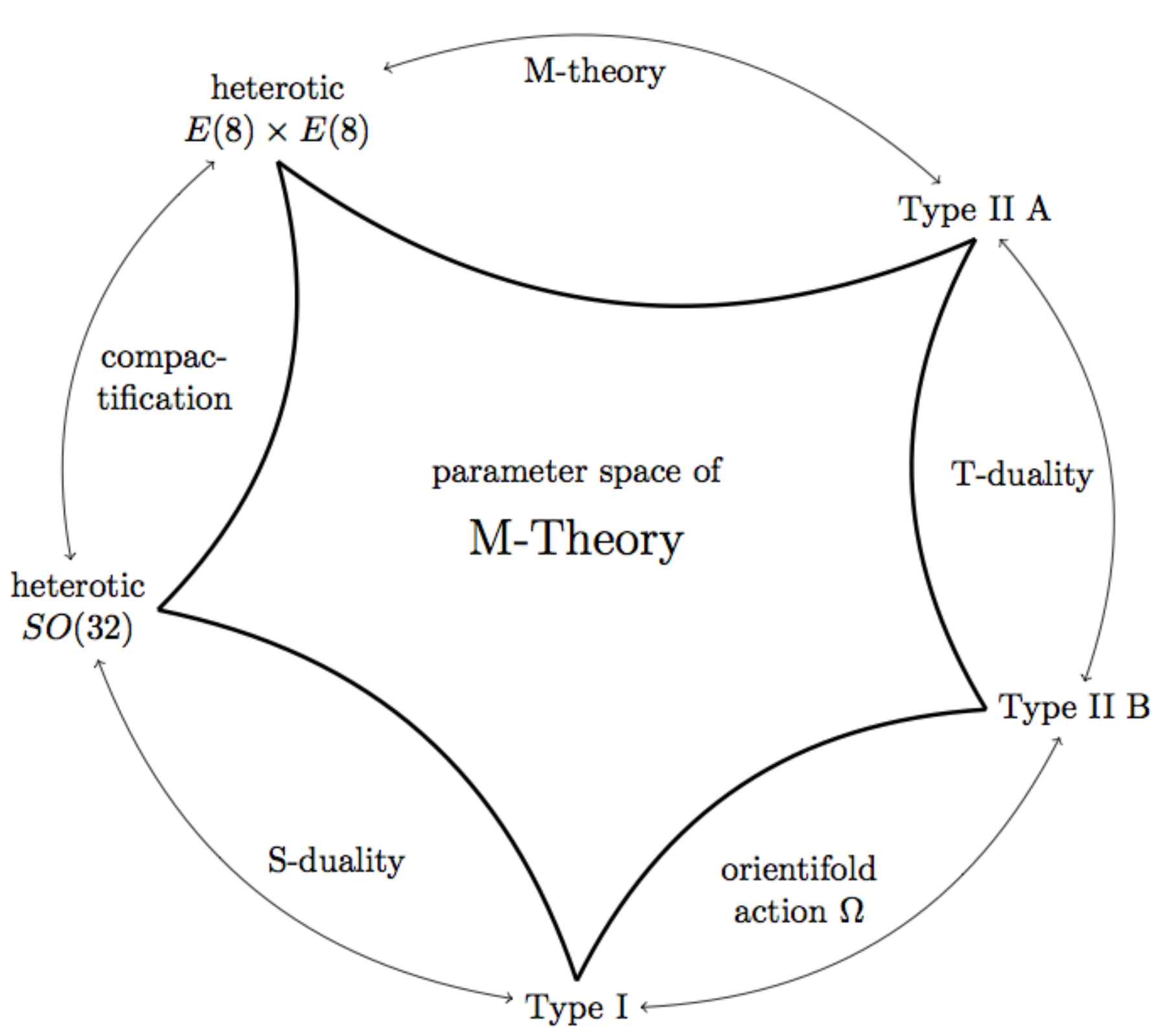}
\end{center}
\caption[Les cinq théories des supercordes et la théorie M]{(Adaptée de la Réf.~\cite{siteM}.) Vue schématique des cinq théories des supercordes reliées entre elles via des relations de dualité. On pense que toutes ces théories ne sont que des cas limites d'une théorie plus fondamentale à 11D : la théorie M.}
\label{M-theory}
\end{figure}

En 1974, Scherk et Schwarz~\cite{Scherk:1974ca} proposèrent que les particules élémentaires de la matière ne seraient pas des objets ponctuels, mais de petites cordes vibrantes de taille $1/M_s$, où l'échelle de masse $M_s$ des premières excitations des cordes est souvent prise proche de $M_P$ (paradigme standard). Le spectre de la théorie des cordes prévoyant une particule sans masse de spin-2, la gravité est donc naturellement incluse dans le modèle. En outre, il a été démontré que la gravité à grande distance se comporte comme le prévoit la relativité générale, et que c'est à petites distances que les effets cordistes apparaissent. Dans une théorie des cordes\footnote{Pour une revue sur les théories des cordes, voir par exemple les Réfs.~\cite{Green:1987sp, Green:1987mn, Polchinski:1998rq, Polchinski:1998rr, Johnson:2003gi, Zwiebach:2004tj, Becker:2007zj, Ibanez:2012zz}.}, les grands moments euclidiens $p_E$ dans les intégrales de boucle sont coupés par des facteurs $\text{e}^{- (p_E / M_P)^2}$, impliquant l'absence de divergences UV dans la théorie : cette dernière n'a donc pas besoin d'être renormalisée. Une telle théorie est donc une bonne prétendante au titre de théorie quantique de la gravitation dans l'UV. Pour introduire des fermions en théorie des cordes, et s'assurer de l'absence de tachyons dans le spectre, il faut avoir recours à la SUSY. De plus, la quantification de la supercorde requiert un espace temps à 10D. On connaît aujourd'hui cinq théories des supercordes consistantes, reliées par des relations de dualité, dont on pense qu'elles ne sont que des cas limites d'une théorie plus fondamentale à 11D : la théorie M (\textit{c.f.} la Fig.~\ref{M-theory}). C'est aujourd'hui, dans ce cadre théorique, qu'est reformulé le fantasme d'unification des interactions fondamentales et de la matière : un unique objet fondamental, la supercorde, est à l'origine de toutes les particules élémentaires connues.

\subsection{Branes}
Les $p$-branes sont des membranes de dimension $p$ apparaissant souvent dans les modèles extra-dimensionnels. Elles ont une densité d'énergie appelée tension, et sont capables de piéger certains champs à leur surface, les empêchant ainsi de se propager dans tout l'espace compactifié. Cela ouvre la porte à une myriade de nouvelles possibilités de construction de modèles. Par essence, ce sont des objets effectifs dont la formation et la description microscopique peut avoir deux origines connues à ce jour : les D$p$-branes en théorie des supercordes ou les défauts topologiques en théorie des champs.

\subsubsection{D-branes}
\begin{figure}[h]
\begin{center}
\includegraphics[height=8cm]{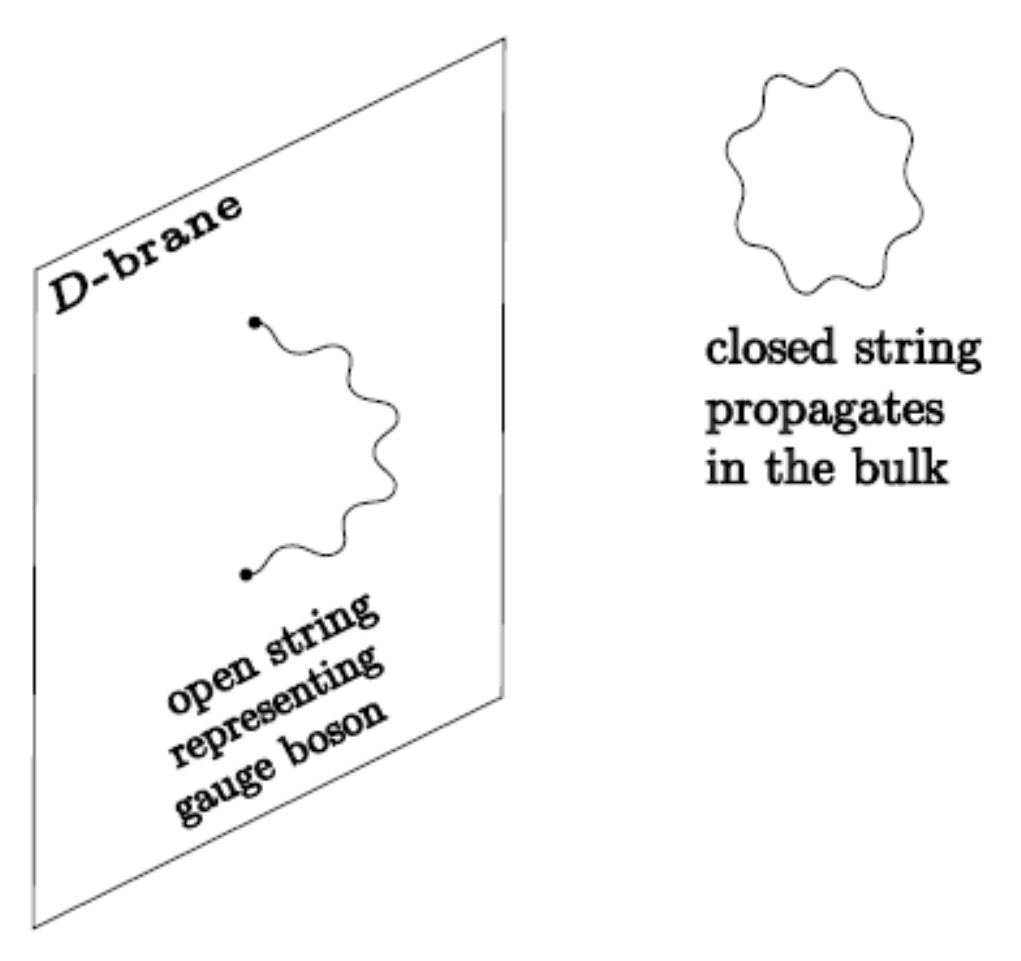}
\end{center}
\caption[D-branes en théorie des cordes]{(Adaptée de la Réf.~\cite{Shifman:2009df}.) Dans certains modèles de supercordes, les bosons de jauge sont des cordes ouvertes attachées à une pile de D-branes, alors que les gravitons sont des cordes fermées se propageant dans le bulk.}
\label{D-brane}
\end{figure}

En 1995, Polchinski~\cite{Polchinski:1995mt} découvrit que dans les théories des supercordes de type I et II apparaissent des membranes solitoniques de dimension $p$ appelées $p$-branes de Dirichlet, ou D$p$-branes. Les cordes ouvertes sont attachées par leurs extrémités à des piles de D$p$-branes, et y développent des charges de Chan-Paton~\cite{Paton:1969je} : elles y sont chargées sous un groupe de jauge dont le rang est déterminé par le nombres de D$p$-branes de la pile. On se retrouve alors avec une membrane sur laquelle sont piégés des champs de jauge et des fermions (\textit{c.f.} la Fig.~\ref{D-brane}). Quant aux cordes fermées, comme le graviton, elles n'ont pas d'extrémités où s'attacher aux D$p$-branes et sont donc libres de se propager dans tout l'espace-temps, appelé le bulk. Dans la vision de la relativité générale, la gravité est une propriété géométrique de l'espace-temps, qui se propage donc dans toutes les dimensions, ce qui rejoint la description cordiste.

\subsubsection{Défauts topologiques}
\begin{figure}[h]
\begin{center}
\includegraphics[width=15cm]{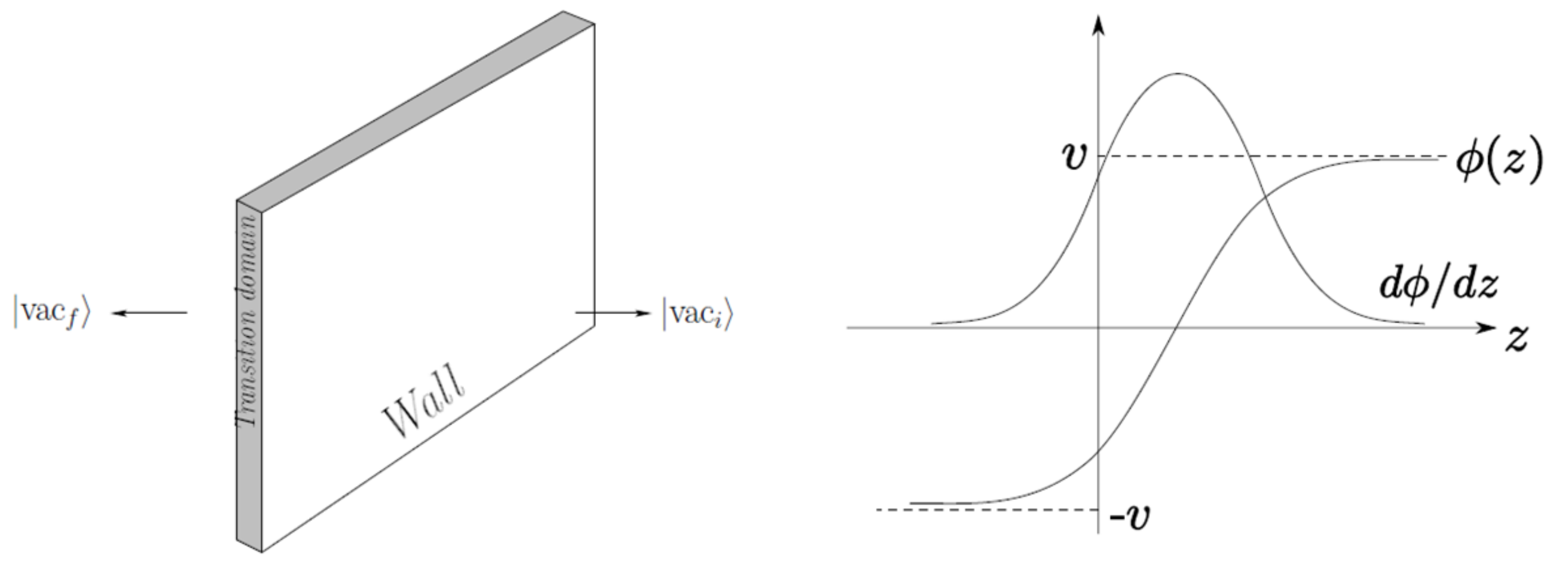}
\end{center}
\caption[Mur de domaine]{(Adaptée de la Réf.~\cite{Shifman:2009df}.) À gauche, un mur de domaine séparant deux vides distincts dégénérés. À droite, le profil du champ $\phi(z)$  modélisant le mur de domaine. Sa dérivée détermine la localisation du mode zéro.}
\label{domain_wall}
\end{figure}

Dans les années 80, Akama~\cite{Akama:1982jy}, Rubakov et Shaposhnikov~\cite{Rubakov:1983bb}, ainsi que Visser~\cite{Visser:1985qm}, proposèrent l'idée selon laquelle les champs du SM seraient piégés au sein d'un défaut topologique dans le bulk, avec une ou plusieurs dimensions spatiales supplémentaires non nécessairement compactifiées. Supposons, par exemple, l'existence d'une dimension spatiale supplémentaire de coordonnée $z$, et que la théorie a plusieurs vides discrets dégénérés, correspondant à des valeurs différentes d'un paramètre d'ordre. Appelons deux de ces vides I et II. Il existe une configuration statique des champs, un mur de domaine, qui divise le bulk en deux compartiments : dans l'un, le système est dans le vide I, alors que, dans l'autre, il est dans le vide II (\textit{c.f.} la Fig.~\ref{domain_wall}). Le mur de domaine représente une région transitoire topologiquement stable. Son épaisseur $\epsilon$ dépend des détails microscopiques de la théorie. À des distances très grandes devant $\epsilon$, il peut être vu comme une membrane 3D.

Les excitations de la configuration en champs du mur de domaine peuvent être classées en deux catégories. Certaines sont localisées sur le mur, leur extension spatiale le long de la direction perpendiculaire au mur étant du même ordre que $\epsilon$. Elles sont souvent associées aux modes zéro de la décomposition en modes 4D : ce sont des particules sans masse se propageant uniquement le long de la surface du mur. Les autres excitations, dont les masses les plus légères sont de l'ordre de $1/\epsilon$, sont délocalisées et peuvent s'échapper dans le bulk. En supposant que toute la matière de notre Univers est composée de modes zéro, ces derniers sont donc confinés à la surface du mur et sont perçus comme constituant un monde à 4D. Afin de découvrir la dimension spatiale supplémentaire, un observateur devra avoir accès à des énergies supérieures à $1/\epsilon$.

La principale distinction, entre les scénarii de KK et la localisation sur un défaut topologique, est l'échelle de masse des modes excités $m_E$. Dans les théories de KK, cette dernière est reliée au rayon de compactification $m_{E} \sim 1/R$ alors que, dans le cas des murs de domaine, elle est reliée à l'épaisseur du mur $m_{E} \sim 1/\epsilon$, et la dimension spatiale supplémentaire est possiblement infinie.

L'existence d'au moins un mode zéro peut être démontrée. La théorie avec une dimension spatiale supplémentaire infinie est invariante sous les translations 4D. La présence du mur de domaine brise spontanément l'invariance par translation selon la direction $z$ : la physique devient dépendante de la distance au mur le long de la dimension spatiale supplémentaire. On a donc l'existence d'un boson de Nambu-Goldstone de spin-0 confiné à la surface du mur. La fonction d'onde du mode zéro dans la dimension spatiale supplémentaire est donnée par la dérivée par rapport à $z$ de celui du paramètre d'ordre $\phi(z)$ (\textit{c.f.} la Fig.~\ref{domain_wall}). Ainsi, la localisation d'un boson de spin-0 est possible. Supposons que la théorie ait une symétrie globale sous le groupe $G$, qui reste non-brisée dans les vides I et II. Supposons ensuite que $G$ soit brisé spontanément en $H$ sur le mur. Alors les bosons de Nambu-Goldstone, correspondant aux générateurs brisés, sont localisés sur celui-ci.

Qu'en est-il de la localisation des fermions de spin-1/2 ? Ces derniers peuvent être couplés au champ scalaire $\phi$ modélisant le mur. Le nombre de modes zéro est donné par le théorème de l'index de Jackiw-Rebbi~\cite{Jackiw:1975fn}. L'épaisseur de la fonction d'onde du mode zéro est de l'ordre de l'inverse de la masse du fermion dans le bulk. Si le bulk est à 6D et le défaut topologique une hypercorde cosmique abélienne, Frère, Libanov et Troitsky proposèrent en 2000 \cite{Libanov:2000uf, Frere:2000dc} de générer les trois générations de fermions du SM à partir d'une seule génération à 6D grace au nombre quantique de vorticité de l'hypercorde.

\begin{figure}[h]
\begin{center}
\includegraphics[height=8cm]{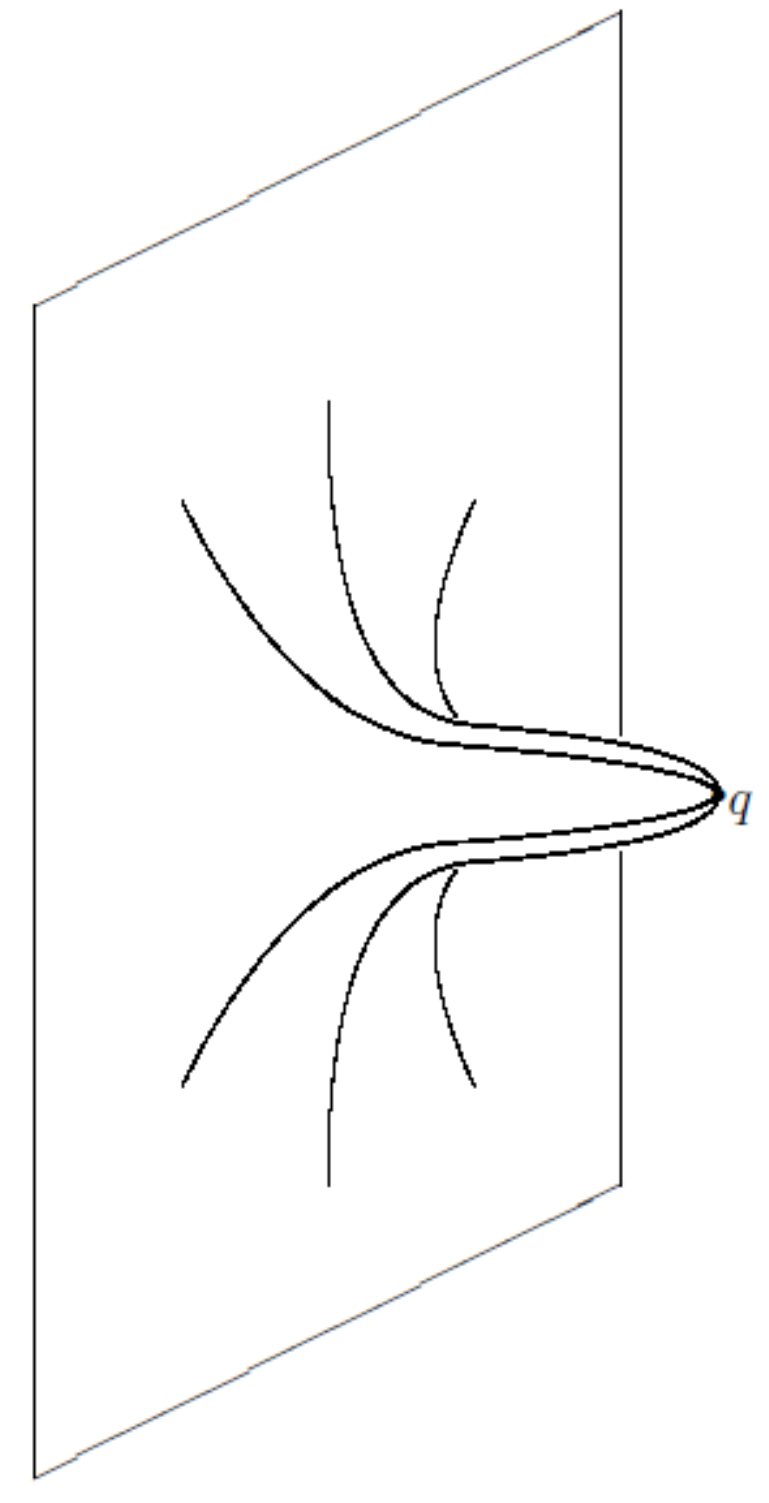}
\end{center}
\caption[Charge électrique reliée à une brane par un tube de champ]{(Adaptée de la Réf.~\cite{Rubakov:2001kp}.) Une charge $q$, que l'on éloigne de la brane, est connectée à celle-ci par un tube de champ.}
\label{flux_tube}
\end{figure}

Quant aux champs de jauge, il est notoirement difficile de les localiser sur un mur de domaine tout en préservant l'universalité des couplages aux fermions. En effet, dans le SM par exemple, les différentes saveurs de quarks couplent de la même manière aux gluons. En 1996, Dvali et Shifman~\cite{Dvali:1996xe} proposèrent un mécanisme réalisant cette tâche. L'idée est d'avoir une théorie de jauge dans une phase de confinement dans le bulk, et une phase déconfinée sur le mur. De cette manière, le champ chromo-électrique d'une charge résidant sur le mur ne peut pas pénétrer dans le bulk. Un modèle dual  est le suivant : un supraconducteur inhomogène avec une phase non-supraconductrice sur un plan. Des monopoles magnétiques placés loin du plan vont ressentir le confinement, alors que ceux sur le plan interagissent via la loi de Coulomb 2D. L'universalité de la charge est préservée dans ce modèle. Si celle-ci est déplacée du mur vers le bulk, un vortex la connectant au mur va se former (\textit{c.f.} la Fig.~\ref{flux_tube}). Le champ de jauge induit par cette charge sur le mur à grande distance est alors indépendant de sa position dans la dimension spatiale supplémentaire, et est identique à l'interaction générée par une charge placée sur le mur. Une reformulation plus économe de ce mécanisme a été proposée en 2010 par Ohta et Sakai~\cite{Ohta:2010fu}, en introduisant une perméabilité diélectrique dépendante de $z$ pour les champs de jauge, \textit{i.e.} un couplage de jauge dépendant de la position dans la dimension spatiale supplémentaire. Cette difficulté de localisation des bosons de jauge disparaît avec deux dimensions spatiales supplémentaires : Oda~\cite{Oda:2000zj, Oda:2000zc} montra en 2000 que les modes zéro sont localisés grâce à la gravité près d'une hypercorde (3D), dans un bulk à 6D, en préservant un couplage universel aux fermions.

\subsection{Modèles phénoménologiques}
À la fin des années 90 et au début du nouveau millénaire, de nouveaux paradigmes provoquèrent un gigantesque engouement pour les dimensions spatiales supplémentaires en physique des particules et en cosmologie, ouvrant la voie à une approche \og du bas vers le haut \fg{} pour construire des modèles extra-dimensionnels.

La première est due à Arkani-Hamed, Dimopoulos et Dvali~\cite{ArkaniHamed:1998rs, Antoniadis:1998ig, ArkaniHamed:1998nn} (ADD) en 1998. Ils proposèrent de confiner le SM à une 3-brane, avec des dimensions spatiales supplémentaires compactifiées. L'idée phare de leur modèle est de supposer le rayon de compactification $R \gg \ell_P$, \textit{i.e.} entre le millimètre et le femtomètre, en fonction du nombre de dimensions spatiales supplémentaires. Ainsi, l'échelle de gravité peut être abaissée à l'échelle du TeV\footnote{L'idée d'abaisser l'échelle de la gravité était apparue quelques années auparavant en théorie des supercordes \cite{Antoniadis:1990ew, Antoniadis:1993jp, Witten:1995ex, Horava:1995qa, Lykken:1996fj, Banks:1996ss}.}, faisant disparaître la hiérarchie de jauge et le problème qui y est inhérent. Cependant, cette dernière y est remplacée par une hiérarchie géométrique : la taille des dimensions spatiales supplémentaires considérées est très grande devant la distance caractéristique de la gravité.

Pour remédier à ce problème, Randall et Sundrum~\cite{Randall:1999ee} (RS) publièrent, en 1999, un modèle avec une dimension spatiale supplémentaire courbe, de taille naturelle, compactifiée sur un intervalle. Le SM est toujours sur une 3-brane à une extrémité, mais l'échelle de gravité dépend de la position dans la dimension spatiale supplémentaire, évoluant d'un bout à l'autre de l'intervalle de $M_P$ jusqu'à l'échelle du TeV (là où est positionnée la 3-brane du SM). Cette fois, la hiérarchie de jauge apparente à 4D est expliquée par un effet de courbure dans la dimension spatiale supplémentaire, et l'échelle EW est stabilisée par une échelle de coupure au TeV sur la 3-brane : c'est le modèle de RS1.

Dans la foulée, la même année, Randall et Sundrum~\cite{Randall:1999vf} montrèrent qu'une 3-brane, où est localisé le SM, avec une dimension spatiale supplémentaire infinie, courbe l'espace dans celle-ci de manière à localiser le mode zéro du graviton près de la 3-brane. Il n'y a donc pas toujours besoin de compactifier les dimensions spatiales supplémentaires pour retrouver la gravité newtonienne à 4D dans le régime accessible actuellement par l'expérience : c'est le modèle de RS2.

En 2000, Dvali, Gabadadze et Porrati (DGP) \cite{Dvali:2000hr} proposèrent un modèle où le SM est localisé sur une 3-brane plongée dans un espace-temps de Minkowski infini à 5D. Un terme de courbure scalaire à 4D localisé sur la 3-brane implique que la gravité apparaît 4D à courte distance. À grandes distances, par contre, elle devient 5D, donnant lieu à de nouvelles solutions pour la cosmologie, notamment une réinterprétation de l'origine de l'accélération de l'expansion de l'Univers.

Toujours la même année, Appelquist, Cheng et Dobrescu \cite{Appelquist:2000nn} étudièrent un modèle avec une dimension spatiale supplémentaire plate, compactifiée sur un intervalle dont la longueur est de l'ordre d'un dixième d'attomètre, dans laquelle tous les champs du SM se propagent : on parle de dimension spatiale supplémentaire universelle (UED -- \textit{Universal Extra Dimension}). Si les 3-branes situées à chaque bord de l'UED sont identiques (mêmes couplages localisés) alors le modèle possède une symétrie sous la réflexion par rapport au milieu de l'intervalle. Ceci implique que la parité de KK $(-1)^n$, où $n$ étiquette les niveaux de KK, est conservée. La KK-particule $n=1$ la plus légère est stable, on parle de LKP (\textit{Lightest Kaluza-Klein Particle}). Celle-ci est alors une bonne candidate pour la matière noire.

En 2004, Cremades, Ib\'anez et Marchesano \cite{Cremades:2004wa} étudièrent des modèles avec 2, 4 ou 6 dimensions spatiales supplémentaires plates, compactifiées sur des tores magnétisés. Si des fermions se propagent dans les dimensions supplémentaires, le champ magnétique génère $N$ modes zéro chiraux dégénérés pour les tours de KK. On peut ainsi générer à la fois la chiralité et la structure en trois générations du SM. Ces modèles effectifs sont censés capturer l'essentiel de la physique à basse énergie de modèles UVs semi-réalistes en théorie des supercordes.

On a pris ici le parti de citer les modèles historiques dont l'impact est si important qu'ils ont chacun ouvert leur axe de recherche en physique des particules et/ou en cosmologie. La plupart des modèles qui suivirent en sont des variantes et des raffinements. L'explosion du nombre d'articles dans la littérature sur les dimensions spatiales supplémentaires ne permet pas d'en faire une liste exhaustive, et de nombreux mécanismes très intéressants ne sont pas cités ici.

\section{Mode d'emploi pour construire un modèle extra-dimensionnel}
\subsection{Compactification de dimensions spatiales supplémentaires}
Comme nous l'avons vu dans la section précédente, il est possible que des dimensions spatiales supplémentaires soient indétectables pour des expériences de basse énergie, si elles sont enroulées en un espace compact de petit volume. Nous allons revenir ici plus en détail sur cette idée.

\subsubsection{Compactification sur une variété lisse}
\begin{figure}[h]
\begin{center}
\includegraphics[height=2.5cm]{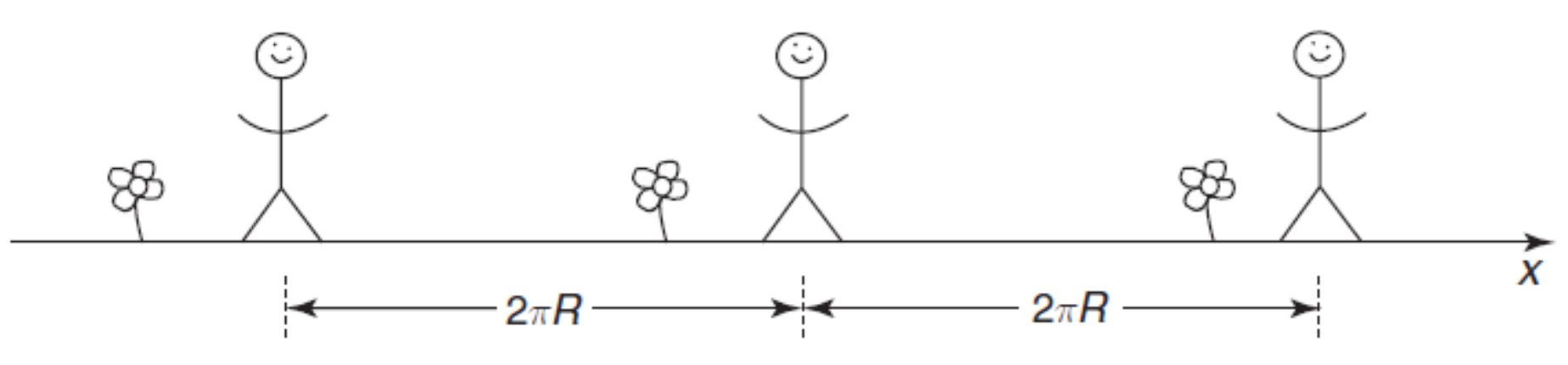}
\end{center}
\caption[Un monde unidimensionnel qui se répète tous les $2 \pi R$]{(Adaptée de la Réf.~\cite{Zwiebach:2004tj}.) Un monde unidimensionnel qui se répète tous les $2 \pi R$.}
\label{circle1}
\end{figure}

\begin{figure}[h]
\begin{center}
\includegraphics[height=2.5cm]{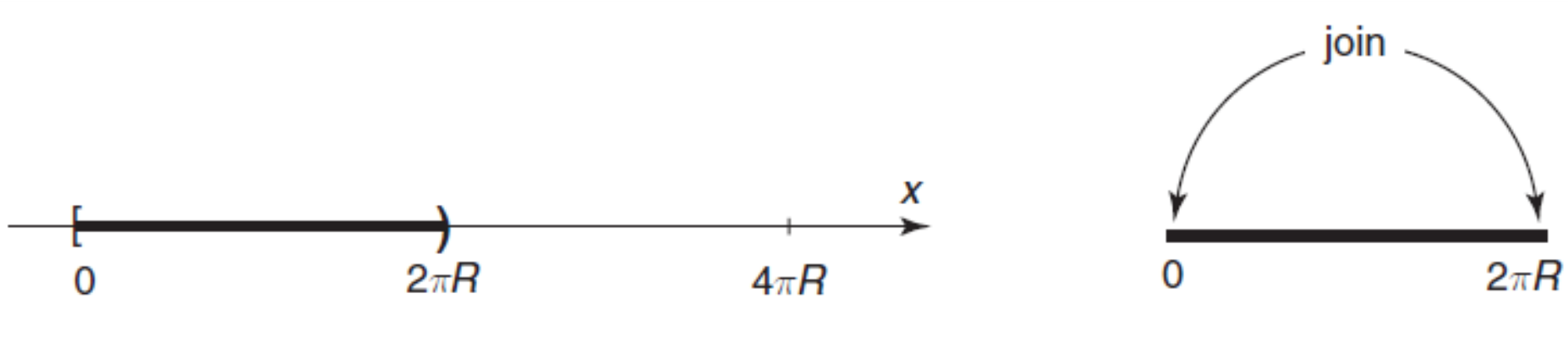}
\end{center}
\caption[Domaine fondamental]{(Adaptée de la Réf.~\cite{Zwiebach:2004tj}.) L'intervalle $[0, 2 \pi R)$ est le domaine fondamental pour la droite avec l'identification \eqref{compact_circle}. L'espace généré est le cercle de rayon $R$.}
\label{circle2}
\end{figure}

Considérons un monde unidimensionnel, la droite réelle $\mathbb{R}$, paramétrisé par la coordonnée $x$. Pour chaque point $P$ le long de cette droite, il y a un unique nombre réel $x(P)$ appelé la coordonnée $x$ du point $P$. Un bon système de coordonnée, pour cette droite infinie, satisfait deux conditions :
\begin{itemize}
\item Deux points distincts $P_1 \neq P_2$ ont des coordonnées différentes $x(P_1) \neq x(P_2)$.
\item L'affectation des coordonnées aux points est continue : deux points voisins ont des coordonnées presque égales.
\end{itemize}
Choisissons une origine à cette droite, et utilisons la distance depuis cette origine comme bonne coordonnée. Imaginons maintenant un observateur se déplaçant dans cette ligne-monde et remarquant l'étrange caractéristique suivante : la même scène se répète à chaque fois qu'il se déplace d'une distance $2 \pi R$. S'il rencontre une autre personne en $x=0$, il voit des clones de celle-ci en $x = 2 \pi R n$, avec $n \in \mathbb{Z}$ (\textit{c.f.} la Fig.~\ref{circle1}). Il n'y a pas moyen de distinguer une droite avec cette étrange propriété d'un cercle $S^1$ de rayon $R$. Dans ce dernier cas, il n'y a pas de clones mais seulement la même personne que l'observateur rencontre à chaque tour du cercle. Pour exprimer cela mathématiquement, il suffit de penser le cercle comme une droite où tous les points, dont les coordonnées diffèrent de $2\pi R$, correspondent au même point physique. Cela revient à définir la classe d'équivalence suivante :
\begin{equation}
P_1 \sim P_2 \ \ \  \Longleftrightarrow \ \ \ x(P_1) = x(P_2) + 2 \pi R n, \ \ \ n \in \mathbb{Z}.
\label{compact_circle}
\end{equation}
L'intervalle $[0, 2 \pi R)$ est appelé le domaine fondamental de l'identification \eqref{compact_circle}. C'est un sous-espace qui doit satisfaire deux conditions :
\begin{itemize}
\item Deux points dans le domaine fondamental ne sont pas identifiés.
\item Tout point de l'espace entier est dans le domaine fondamental ou est relié par une identification à un point de ce dernier.
\end{itemize}
Pour construire l'espace généré par cette identification, on prend l'adhérence du domaine fondamental, ici $[0, 2 \pi R]$, et on effectue l'identification de ses bords, ici $x=0$ et $x = 2 \pi R$, ce qui donne le cercle $S^1$ de rayon $R$ (\textit{c.f.} la Fig.~\ref{circle2}). On dit que l'identification \eqref{compact_circle} a compactifié la dimension. Cela peut sembler un moyen bien compliqué de décrire un cercle. Cependant, pour $S^1$, une coordonnée est soit discontinue, soit multivaluée. Définir la théorie des champs sur $\mathbb{R}$, avec la classe d'équivalence \eqref{compact_circle}, permet d'utiliser un bon système de coordonnée, lequel est multivalué sur $S^1$. De plus, en théorie des champs, une telle identification revient à imposer une symétrie pour l'action, ce qui rend le problème plus simple en pratique.

Voyons maintenant comment construire de manière générale une théorie des champs avec des dimensions spatiales supplémentaires compactifiées sur une variété lisse. Considérons une théorie à $(d+1)$D, avec $k = d-3$ dimensions spatiales supplémentaires et une action définie comme
\begin{equation}
S_{d+1} = \int d^{d+1} \mathcal{L}_{d+1}[\phi(z)],
\end{equation}
où on a le lagrangien $\mathcal{L}_{d+1}$ du champ $\phi(z)$. On dit que la théorie est compactifiée sur $M_4 \times C$, où $M_4$ est l'espace-temps de Minkowski à 4D, et $C$ est un espace compact à $k$D, si les coordonnées de l'espace à $(d+1)$D peuvent être séparées comme $z^M = (x^\mu, y^m)$ ($M \in \llbracket 0, d \rrbracket$, $\mu \in \llbracket 0, 3 \rrbracket$, $m \in \llbracket 1, k \rrbracket$). Le lagrangien à 4D est obtenu après intégration sur les coordonnées $y^m$ de $C$ :
\begin{equation}
\mathcal{L}_4 = \int d^ky \; \mathcal{L}_{d+1}[\phi(x^\mu, y^m)].
\end{equation}
En général, on peut écrire $C = M/G$, où $M$ est une variété non-compacte, et $G$ est un groupe discret agissant librement sur $M$ par les opérateurs $\tau_g : M \rightarrow M$, pour $g \in G$. $M$ est appelé l'espace de recouvrement de $C$. On dit que $G$ agit librement sur $M$ si seulement $\tau_\iota$ a des points fixes dans $M$, où $\iota$ est l'identité dans $G$. Les opérateurs $\tau_g$ constituent une représentation de $G$. $C$ est donc construit par identification des points $y$ et $\tau_g(y)$ qui appartiennent à la même orbite ; on définit alors la relation d'équivalence :
\begin{equation}
y \sim \tau_g(y).
\end{equation}
Après identification, la physique ne dépend pas des points individuels de $M$ mais seulement des orbites (chaque point de $C$ correspond à une unique orbite), ainsi
\begin{equation}
\mathcal{L}_{d+1}[\phi(x, y)] = \mathcal{L}_{d+1}[\phi(x, \tau_g(y))],
\end{equation}
dont une condition nécessaire et suffisante pour le champ est
\begin{equation}
\phi(x, \tau_g(y)) = T_g \; \phi(x, y) \, ,
\end{equation}
puisque
\begin{equation}
\mathcal{L}_{d+1}[T_g \, \phi] = \mathcal{L}_{d+1}[\phi] \, ,
\end{equation}
où $T_g$ est un élément de la représentation du groupe $G$ agissant dans l'espace des champs. Si $T_g = I$, où $I$ est l'identité, on parle de compactification ordinaire. Si $T_g \neq I$, on parle de compactification de Scherk-Schwarz \cite{Scherk:1978ta, Scherk:1979zr, Cremmer:1979uq}, et $T_g$ est appelé un twist de Scherk-Schwarz. Pour les compactifications ordinaires et de Scherk-Schwarz, les champs sont mono-valués sur l'espace de recouvrement $M$. Pour la compactification ordinaire, les champs sont mono-valués sur l'espace compact $C$. Par contre, pour la compactification de Scherk-Schwarz, les champs sont multi-valués sur $C$.

Dans l'exemple de la compactification de $\mathbb{R}$ sur $S^1$ : $d=4$ ($k=1$), $M = \mathbb{R}$, $G = 2 \pi R \mathbb{Z}$ et $C = S^1$. Le $n$-ième élément du groupe $2 \pi R \mathbb{Z}$ est représenté par $\tau_n$, tel que
\begin{equation}
\forall y \in \mathbb{R}, \; \tau_n(y) = y + 2 \pi R n.
\end{equation}
L'identification \eqref{compact_circle} revient donc à faire l'identification $y \sim \tau_n(y)$. Le groupe $2 \pi R \mathbb{Z}$ a une infinité d'éléments mais ils peuvent tous être obtenus à partir d'un seul générateur : la translation de $2 \pi R$ représentée par $\tau_1$. Le twist $T$ correspondant est défini par :
\begin{equation}
\phi(x, y + 2 \pi R) = T \; \phi(x,y),
\label{twist}
\end{equation}
et les twists des autres éléments de $2 \pi R \mathbb{Z}$ sont donnés par $T_n = T^n$.

\subsubsection{Compactification sur un orbifold}
\label{compact_orbifold}
\begin{figure}[h]
\begin{center}
\includegraphics[height=2cm]{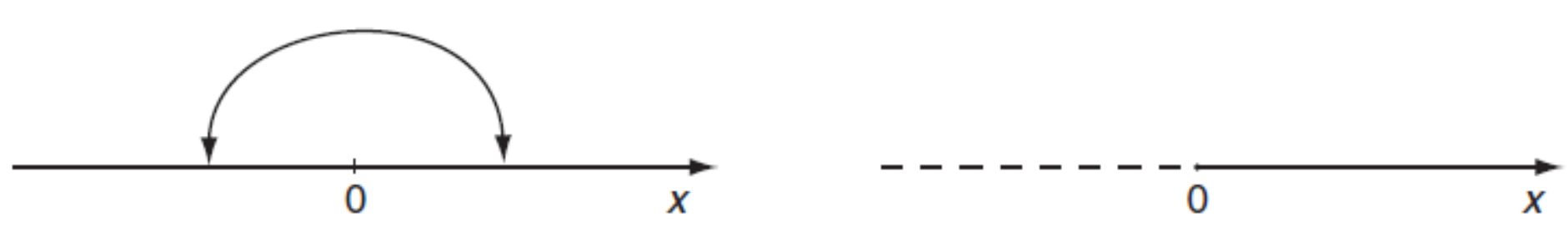}
\end{center}
\caption[Orbifold $\mathbb{R}/\mathbb{Z}_2$]{(Adaptée de la Réf.~\cite{Zwiebach:2004tj}.) L'identification $x \sim -x$ sur la droite donne une demi-droite : c'est l'orbifold $\mathbb{R}/\mathbb{Z}_2$.}
\label{demi-droite}
\end{figure}

On peut également construire des identifications qui ont des points fixes \cite{Candelas:1985en, Witten:1985xc, Dixon:1985jw, Dixon:1986jc}, \textit{i.e.} des points reliés à eux-mêmes par celles-ci. Prenons l'exemple de la droite réelle, paramétrisée par la coordonnée $x$, sur laquelle on effectue l'identification $x \sim -x$. Un domaine fondamental est la demi-droite $x \geqslant 0$. Notons que le point au bord $x=0$ doit être inclus dans le domaine fondamental, lequel est identifié avec lui-même : c'est l'exemple le plus simple d'un orbifold, un espace obtenu par des identifications qui ont des points fixes. Ces derniers sont des points singuliers : la demi-droite $x \geqslant 0$ est une variété unidimensionnelle conventionnelle pour $x>0$, alors que le point $x=0$ est un point d'arrêt. Cet orbifold est noté $\mathbb{R}/\mathbb{Z}_2$, où $\mathbb{Z}_2$ est la transformation : $x \mapsto -x$ (\textit{c.f.} la Fig.~\ref{demi-droite}).

On peut définir la compactification sur un orbifold de la manière suivante. Soit $C$ une variété compacte et $H$ un groupe discret représenté par les opérateurs $\zeta_h : C \rightarrow C$, pour $h \in  H$, agissant non-librement sur $C$. On définit alors la classe d'équivalence suivante :
\begin{equation}
y \sim \zeta_h(y).
\end{equation}
La condition nécessaire et suffisante est que les champs définis en deux points identifiés diffèrent par une transformation $Z_h$, élément de la représentation de $H$ dans l'espace des champs :
\begin{equation}
\phi(x, \zeta_h(y)) = Z_h \; \phi(x, y).
\end{equation}
Le fait que $H$ agisse non-librement sur $C$ signifie que certaines transformations $\zeta_h(y)$ ont des points fixes dans $C$. L'espace quotient $O = C/H$ n'est pas une variété lisse mais a des singularités aux points fixes : c'est un orbifold.

\begin{figure}[h]
\begin{center}
\includegraphics[height=3.5cm]{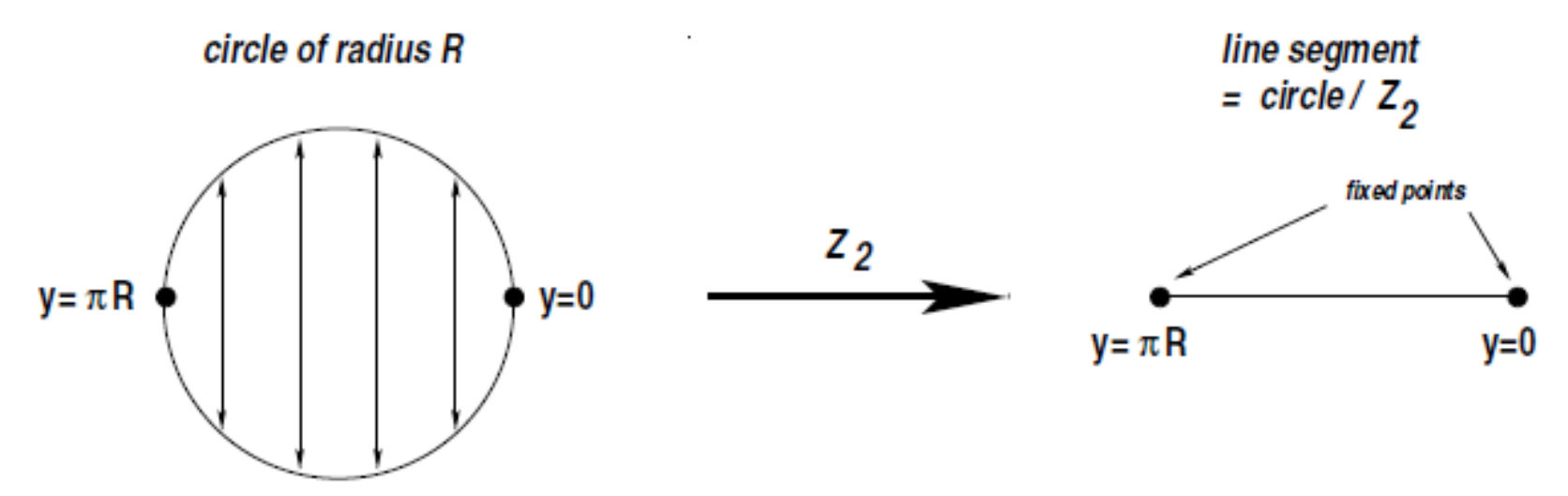}
\end{center}
\caption[Orbifold $S^1/\mathbb{Z}_2$]{(Adaptée de la Réf.~\cite{Dienes:2002hg}.) Le cercle $S^1$ de rayon $R$ est soumis à une identification sous $\mathbb{Z}_2$, devenant l'orbifold $S^1/\mathbb{Z}_2$ avec deux points fixes en $y=0$ et $y= 2 \pi R$.}
\label{intervalle}
\end{figure}

L'exemple le plus simple d'orbifold compact est l'intervalle. Dans ce cas, $k=1$, $C=S^1$ de rayon $R$ et $H = \mathbb{Z}_2$. L'action du seul élément non-trivial de $\mathbb{Z}_2$ (l'inversion) est représentée par $\zeta_0$, telle que
\begin{equation}
\zeta_0(y) = -y,
\end{equation}
ce qui satisfait $\zeta_0^2 = 1$. La condition sur les champs est
\begin{equation}
\phi(x, -y) = Z_0 \; \phi(x, y),
\end{equation}
où $Z_0^2 = I_2$, et $I_2$ est la matrice identité $2 \times 2$. Dans l'espace des champs, $Z_0$ est une matrice diagonalisable, de valeurs propres $\pm 1$. L'orbifold $S^1/\mathbb{Z}_2$ est une variété avec des bords en $y=0$ et $y=\pi R$, les points fixes (\textit{c.f.} la Fig.~\ref{intervalle}). La compactification sur $S^1$ permet aussi de définir un twist de Scherk-Schwarz $T$ avec l'\'Eq.~\eqref{twist}. $Z_0$ et $T$ doivent satisfaire une condition de consistance. Considérons un point de coordonnée $y$ dans le domaine fondamental de $S^1$. Appliquons d'abord une réflexion par rapport à $y=0$ et ensuite une translation de $2 \pi R$, ce qui correspond à $y \mapsto 2 \pi R - y$. C'est équivalent à considérer d'abord une translation de $- 2 \pi R$ et une réflexion par rapport à $y=0$. Ceci implique la condition :
\begin{equation}
\tau_1 \zeta_0 = \zeta_0 \tau_1^{-1},
\end{equation}
ce qui équivaut dans l'espace des champs à :
\begin{equation}
T Z_0 = Z_0 T^{-1} \ \ \ \Leftrightarrow \ \ \ Z_0 T Z_0 = T^{-1}.
\label{twist_orbi}
\end{equation}
Pour un twist non-trivial $T \neq I$, on peut toujours trouver une combinaison de $T$ et $Z_0$ qui agit comme une autre réflexion :
\begin{equation}
\zeta_L = \tau_1 \zeta_0 : y \mapsto 2 \pi R - y.
\end{equation}
C'est en fait une réflexion par rapport à $y=\pi R$ : si $y = \pi R -x$, $\zeta_L$ effectue la transformation $y \mapsto \pi R + x$, donc $x \mapsto -x$. La condition de consistance \eqref{twist_orbi} appliquée à $Z_L = T Z_0$ donne
\begin{equation}
Z_L^2 = (T Z_0)^2 = (T Z_0) (Z_0 T^{-1}) = I_2.
\end{equation}
La description de l'orbifold $S^1 / \mathbb{Z}_2$, avec des twists de Scherk-Schwarz non-triviaux, est donnée par deux parités $\mathbb{Z}_2$ et $\mathbb{Z}_2'$ par rapport aux points fixes respectifs $y=0$ et $y=L$, qui ne commutent pas nécessairement. On parle d'orbifold $S^1 / \left( \mathbb{Z}_2 \times \mathbb{Z}_2' \right)$.

Il n'est cependant pas obligatoire de décrire l'intervalle avec un orbifold. On peut très bien construire une théorie des champs consistante directement sur l'intervalle $I = [0, L]$ où $L = \pi R$ \cite{Hebecker:2001jb, Csaki:2003dt, Csaki:2003sh}. Les Réfs.~\cite{Lalak:2001fd, Csaki:2005vy} comparent les descriptions orbifold et intervalle.

\subsection{Réduction dimensionnelle et décomposition de Kaluza-Klein}
\begin{figure}[h]
\begin{center}
\includegraphics[height=10cm]{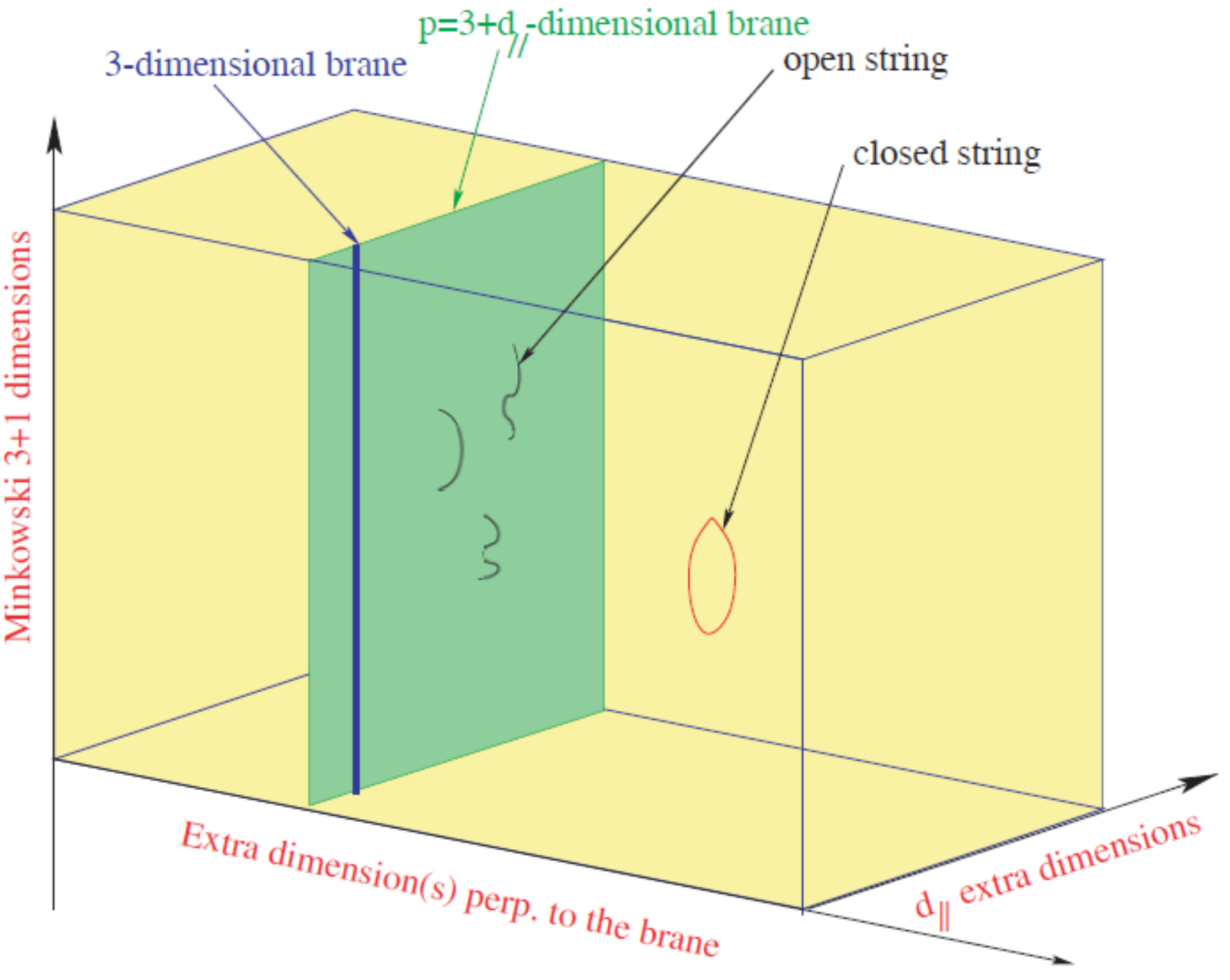}
\end{center}
\caption[Univers branaire]{(Adaptée de la Réf.~\cite{Antoniadis:2010zzc}.) Vue schématique d'un modèle d'Univers branaire issu d'un modèle de théorie des supercordes à haute énergie. Au-delà des trois dimensions spatiales connues (en bleu), il peut y avoir des dimensions spatiales supplémentaires le long de notre $p$-brane (en vert) où les extrémités de cordes ouvertes sont attachées (en noir), et aussi des dimensions spatiales supplémentaires transverses (en jaune) où la gravité décrite par des cordes fermées (en rouge) se propage.}
\label{bulk}
\end{figure}

Pour construire une théorie avec des dimensions spatiales supplémentaires, on commence par choisir : la géométrie de l'espace compactifié, la position des éventuelles $p$-branes, le contenu en champs (localisés sur une $p$-brane ou libres de se propager dans le bulk), \textit{c.f.} la Fig.~\ref{bulk}.
 
On écrit l'action du bulk à $(d+1)$D en termes des champs dans les différentes représentations du groupe de Lorentz à $(d+1)$D $SO(d,1)$. On effectue la même démarche avec l'action des $p$-branes. On compactifie ensuite les dimensions spatiales supplémentaires sur la géométrie choisie, et on résout les équations d'Einstein (si on désire tenir compte de la rétroaction de la tension des $p$-branes sur la métrique) avec un ansatz assurant un espace-temps de Minkowski pour les quatre dimensions usuelles (ou de de Sitter si on fait de la cosmologie). 

L'étape suivante est la réduction dimensionnelle à 4D. $SO(d,1)$ est plus grand que $SO(3,1)$, pour $d>3$, chaque représentation de $SO(d,1)$ étant plus grande que celle correspondante de $SO(3,1)$. Ceci implique qu'une seule représentation de $SO(d,1)$ se décompose en plusieurs représentations différentes de $SO(3,1)$. Cependant, différentes représentations de $SO(3,1)$ correspondent à des particules de spins différents. Ainsi, on voit que certaines particules de spins différents en 4D peuvent être vues comme différentes composantes d'une seule particule dans un espace-temps de plus haute dimensionalité, avec un seul spin associé à $SO(d,1)$. C'est l'idée à la base de la théorie de KK : l'\'Eq.~\eqref{graviton_KK} montre que le graviton à 5D se décompose en graviton, graviphoton et graviscalaire à 4D. Lors de la construction du modèle, on décompose donc les champs à $d+1$D en représentations irréductibles de $SO(3,1)$. Ensuite, on effectue une décomposition de KK des champs, qui dépendent toujours des coordonnées des dimensions spatiales supplémentaires, en une somme infinie de produits de champs à 4D et de leurs fonctions d'onde de KK, fonctions propres de l'opérateur de Laplace dans l'espace compactifié. L'action effective à 4D est alors obtenue en intégrant sur les coordonnées de l'espace compactifié, l'orthonormalisation des fonctions d'onde de KK étant imposée afin d'avoir des termes cinétiques canoniques à 4D.

Il y a ainsi deux signatures expérimentales de l'existence de dimensions spatiales supplémentaires :
\begin{itemize}
\item L'apparition d'une multitudes d'états de KK, appelée tour de KK, pour le graviton et certaines ou toutes les particules connues du SM.
\item L'apparition de nouveaux champs de spins différents se combinant avec les champs connus pour former des représentations irréductibles de $SO(d,1)$.
\end{itemize}
La première implication signifie qu'il y aurait une tour de KK-gravitons (qui ont toujours accès à toutes les dimensions spatiales supplémentaires), et possiblement des tours pour les leptons, les quarks et les bosons de jauge (à part s'ils sont piégés sur une 3-brane), dont le mode zéro correspond à la particule du SM. Tous les modes d'une tour de KK ont les mêmes nombres quantiques (spins, charges de jauge, ...). Pour la deuxième implication, les champs de spins différents issus d'une même représentation de $SO(d,1)$ ont les mêmes charges de jauge, et ont possiblement une tour de KK associée (parfois, les degrés de liberté d'une tour de KK peuvent être \og mangés \fg{}  par une autre tour : les modes de celle-ci acquièrent ainsi leur masse, d'une manière analogue au mécanisme de Higgs).

\section{Univers branaires et hiérarchie de jauge}
Dans cette section, nous allons passer en revue les différents modèles extra-dimensionnels historiques, résolvant le problème de hiérarchie de jauge, où le SM est localisé sur une 3-brane. Pour des revues sur le sujet, voir les Réfs.~\cite{Rubakov:2001kp, Dienes:2002hg, Hewett:2002hv, Gabadadze:2003ii, Rattazzi:2003ea, Csaki:2004ay, PerezLorenzana:2005iv, Hewett:2005uc, Kribs:2006mq, Shifman:2009df, Rizzo:2010zf, Cheng:2010pt}.

\subsection{Théorie effective pour une 3-brane}
\label{EFT_brane}
\begin{figure}[h]
\begin{center}
\includegraphics[height=8cm]{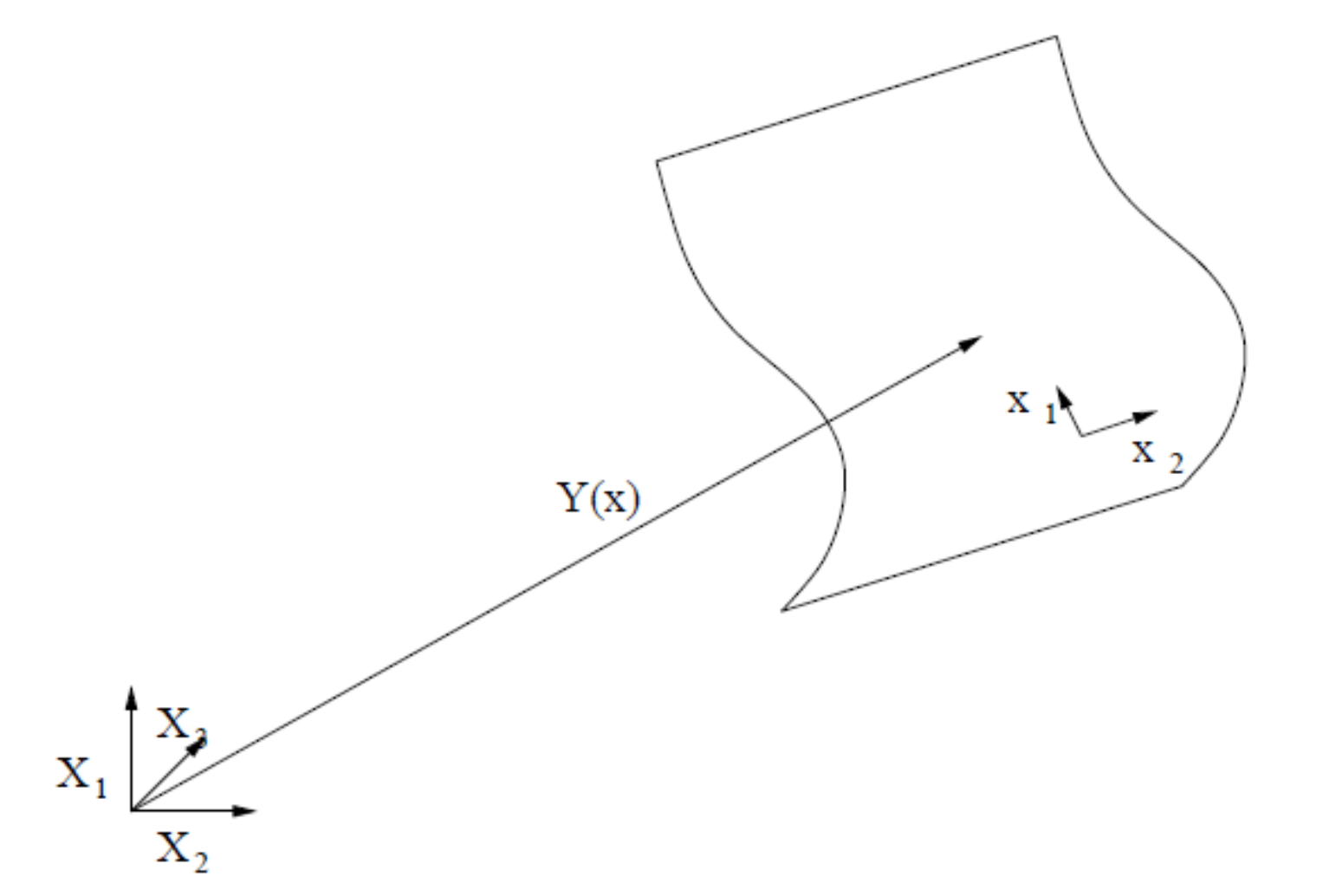}
\end{center}
\caption[Paramétrisation de la position de la brane dans le bulk]{(Adaptée de la Réf.~\cite{Csaki:2004ay}.) Paramétrisation de la position de la brane dans le bulk.}
\label{brane}
\end{figure}

Commençons par poser les bases de la description effective d'un Univers confiné à une 3-brane~\cite{Sundrum:1998sj, Sundrum:1998ns}. Considérons une 3-brane dans un espace temps à $(d+1)$D. Les coordonnées dans le bulk, $X^M$, sont étiquetées par $M=0, 1, \cdots, d$ et celles sur la $p$-brane, $x^\mu$, par $\mu=0,1,2,3$. Les coordonnées du point $x$ sur la 3-brane sont $Y^M(x)$, dont la dépendance en $x$ rappelle que les $Y^M(x)$ sont des champs (\textit{c.f.} la Fig.~\ref{brane}). La métrique dans le bulk est $G^{MN}(X)$, associée au vielbein à $(d+1)$D $E^A_M(X)$, où $A/ \alpha$ est l'indice de Lorentz de l'espace tangent au point $X/x$. L'EFT que l'on veut construire doit décrire de petites fluctuations des champs autour d'un état du vide. On peut supposer que la 3-brane est plate, plongée dans un bulk lui-aussi plat. Le vide est alors :
\begin{equation}
E^A_M(X) = \delta^A_M, \ \ \ G_{MN}(X) = \eta_{MN}, \ \ \ Y^M(x) = \delta^M_\mu x^\mu.
\label{vide}
\end{equation}

Si on considère la distance entre deux points de la 3-brane, séparés par une distance infinitésimale, un observateur sur celle-ci ou dans le bulk mesure la même distance, \textit{i.e.}
\begin{align}
ds^2 &= G_{MN}(X(x)) dY^M dY^N \nonumber \\
&= G_{MN}(Y(x)) \partial_\mu Y^M \partial_\nu Y^N dx^\mu dx^\nu,
\end{align}
ce qui donne la métrique induite sur la 3-brane :
\begin{equation}
g_{\mu \nu}(x) = G_{MN}(Y(x)) \partial_\mu Y^M \partial_\nu Y^N.
\end{equation}
On peut alors écrire l'action générale pour des champs localisés sur la 3-brane. Les symétries qui nous guident sont celles des transformations générales de coordonnées dans l'espace des $X$ et des $x$. Pour décrire des fermions, on doit définir le vielbein induit sur la 3-brane :
\begin{equation}
e^\alpha_\mu = R^\alpha_A E^A_M(X) \partial_\mu Y^M,
\end{equation}
où $R^\alpha_A$ est une représentation de $SO(d,1)$.

Sur la 3-brane, sont localisés des champs typiques du contenu du SM : un scalaire $\phi$, deux spineurs de Weyl $\psi_{L/R}$ qui se couplent à $\phi$ via un couplage de Yukawa $y$, un boson de jauge $A_\mu$ qui se couple à $\psi$ et $\phi$ avec la constante de couplage $g$. L'action effective, décrivant ces champs sur la 3-brane et leurs interactions, est donnée par
\begin{align}
S_{brane} = \int d^4x \sqrt{|g|} \left( -f^4 \right. + \left. g^{\mu \nu} D_\mu \phi^\dagger D_\nu \psi - \dfrac{1}{4} g^{\mu \rho} g^{\nu \sigma} F_{\mu \nu}  F_{\rho \sigma} \right. \nonumber \\
+ \left. \left[ i \psi^\dagger_L e^\mu_\alpha \bar{\sigma}^\alpha D_\mu \psi_L + i \psi^\dagger_R e^\mu_\alpha \sigma^\alpha D_\mu \psi_R - y \; \psi_R^\dagger \phi \psi_L + \text{c.h.} \right] + \ldots \right),
\label{3-brane_action}
\end{align}
où $F_{\mu \nu} = \partial_\mu A_\nu - \partial_\nu A_\mu$, $D_\mu = \partial_\mu - i g A_\mu$, et l'ellipse désigne des termes de dimensions supérieures. Le terme $-f^4$ est la tension de la 3-brane, \textit{i.e.} sa densité d'énergie, qui est déterminée par la théorie UV donnant une description microscopique de la 3-brane. Elle doit être exactement compensée par un ajustement fin, et ainsi pouvoir travailler avec le vide minkowskien de l'\'Eq.~\eqref{vide}. L'action \eqref{3-brane_action} s'ajoute à celle d'Einstein-Hilbert décrivant la gravité dans le bulk :
\begin{equation}
-\int d^{d+1}X \sqrt{|G|} \left( \dfrac{M_*^{d-1}}{2} R + \Lambda_B \right),
\end{equation}
où $M_*$ et $\Lambda_B$ sont respectivement la masse de Planck et la constante cosmologique dans le bulk.

À cause de l'invariance sous la reparamétrisation à 4D de la 3-brane, on doit choisir des conditions de fixation de jauge, afin d'éliminer les composantes non-physiques de $Y^M(x)$. Il y a quatre coordonnées sur la brane, donc on a besoin des quatre conditions :
\begin{equation}
Y^\mu(x) = x^\mu,
\end{equation}
qui fixent complètement la jauge. Seules les composantes le long de la dimension spatiale supplémentaire, $Y^m(x)$, demeurent, correspondant aux degrés de liberté des fluctuations de la 3-brane autour de son état du vide. Après réduction dimensionnelle, ils se manifestent comme des particules de spin-0 : les branons. Pour obtenir les champs $Y^m(x)$, avec une normalisation canonique des termes cinétiques, on effectue l'expansion du terme dominant dans l'action de la 3-brane, ce qui se réduit à simplement la tension de cette dernière :
\begin{equation}
\int d^4x \sqrt{|g|} \left( -f^4 + \cdots \right),
\end{equation}
où la dépendance à l'ordre dominant en $Y$ de la métrique induite est
\begin{equation}
g_{\mu \nu} = \eta_{\mu \nu} + \partial_\mu Y^m \partial_\nu Y_m + \cdots
\end{equation}
et
\begin{equation}
\sqrt{|g|} = 1 + \dfrac{1}{2} \partial_\mu Y^m \partial^\mu Y_m + \cdots
\end{equation}
Ainsi, le terme dominant de l'action est
\begin{equation}
\int d^4x \left( -f^4 + \dfrac{f^4}{2} \partial_\mu Y_m \partial^\mu Y_m \right).
\end{equation}
On peut alors définir le champ avec une normalisation canonique :
\begin{equation}
Z^m = f^2 Y^m.
\end{equation}
Dans le cas d'une brane de tension $f^4<0$, le terme cinétique est négatif et les branons sont des fantômes d'Ostrogradsky. La brane est donc instable et se désagrège. La solution est d'éliminer les degrés de liberté des branons, \textit{i.e.} empêcher la brane de bouger en la positionnant, par exemple, à un point fixe d'un orbifold.

\subsection{Modèles d'ADD}
\subsubsection{Compactification sur une variété toroïdale}
\begin{figure}[h]
\begin{center}
\includegraphics[height=6cm]{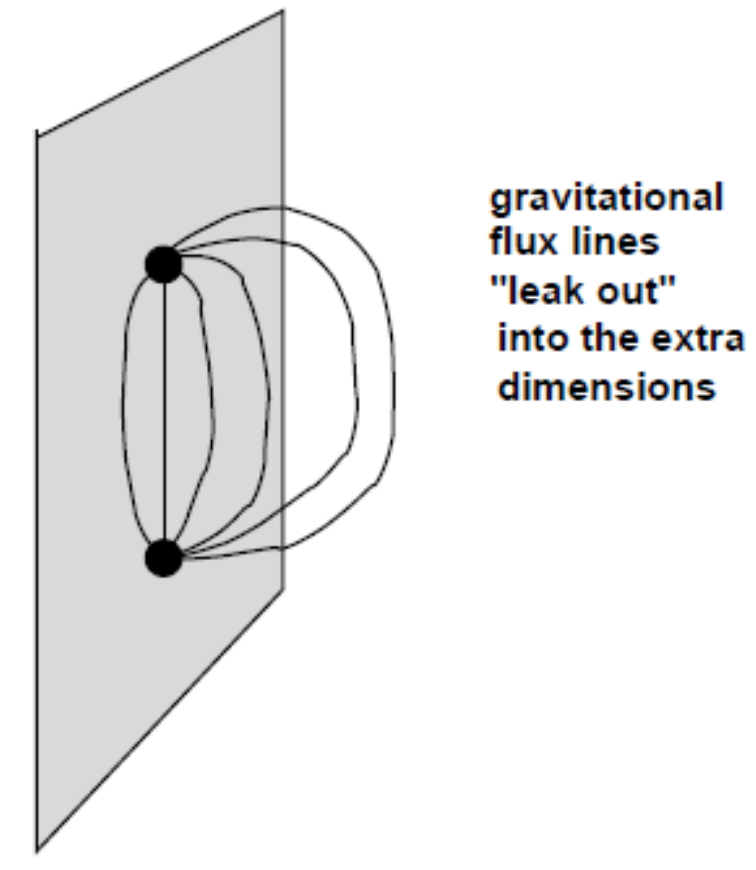}
\end{center}
\caption[Lignes du champ gravitationnel dans les modèles d'ADD]{(Adaptée de la Réf.~\cite{Dienes:2002hg}.) S'il existe des dimensions spatiales supplémentaires transverses à notre 3-brane, les lignes du champ gravitationnel, entre deux masses sur cette dernière, s'échappent dans l'espace compactifié, expliquant ainsi la faiblesse de la gravité comparée aux autres interactions fondamentales qui n'ont pas accès aux dimensions spatiales supplémentaires.}
\label{ADD}
\end{figure}

La nouvelle classe de modèles BSM, proposée par Arkani-Hamed, Dimopoulos et Dvali (ADD) \cite{ArkaniHamed:1998rs, Antoniadis:1998ig, ArkaniHamed:1998nn}, constitue la renaissance des dimensions spatiales supplémentaires en physique fondamentale. L'idée de base est de ramener l'échelle de la gravité au TeV, afin de faire disparaître la hiérarchie de jauge et le désert du paradigme standard. Les champs du SM sont confinés à une 3-brane, et l'espace compactifié doit avoir un volume suffisamment grand pour diluer la gravité, qui apparaît de faible intensité pour un observateur sur la 3-brane (\textit{c.f.} la Fig.~\ref{ADD}).

Dans sa version la plus simple, le modèle d'ADD repose sur les hypothèses suivantes :
\begin{itemize}
\item Il existe $k$ dimensions spatiales supplémentaires de rayon de compactification commun $R$, l'espace compactifié étant un tore à $k$D de volume $V_k = L^k$, où $L=2 \pi R$.
\item Les champs fermioniques, de jauge et de Higgs, sont piégés sur une 3-brane dans le bulk.
\item La constante cosmologique est nulle dans le bulk, et la tension la 3-brane compense exactement l'énergie du vide créée par les champs qui y sont localisés : l'espace-temps est minkowskien dans le bulk et sur la 3-brane.
\item La 3-brane est rigide, \textit{i.e.} la brane brise explicitement l'invariance par translation dans le bulk et les branons sont massifs et découplent de l'EFT: on peut donc négliger les fluctuations de la position de la brane dans le bulk : $Y^m = 0$.
\end{itemize}
L'action $S$ de ce modèle est constituée de deux parties :
\begin{equation}
S = S_{bulk} + S_{brane},
\end{equation}
où $S_{brane}$ est l'action du SM localisé sur la 3-brane (les indices de Lorentz étant contractés avec la métrique induite sur la 3-brane) et l'action dans le bulk $S_{bulk}$ est l'action d'Einstein-Hilbert à $(4+k)$D :
\begin{equation}
S_{bulk} = - \dfrac{M_*^{k+2}}{2} \int d^{4+k}x \sqrt{|g^{(4+k)}|} R^{(4+k)},
\label{S_ADD_bulk}
\end{equation}
où $M_*$ est l'échelle la masse de Planck à $(4+k)$D, et $g^{(4+k)}$, $R^{(4+k)}$ sont respectivement le déterminant de la métrique et le scalaire de courbure de Ricci à $(4+k)$D, avec
\begin{equation}
ds^2 = g^{(4+k)}_{MN} dx^M dx^N.
\end{equation}
Sous nos hypothèses, en développant la métrique autour de son état du vide, on a :
\begin{equation}
ds^2 = \left( \eta_{\mu \nu} + \hat{h}_{\mu \nu} \right) dx^\mu dx^\nu  - R^2 \sum_k d\theta_{(k)}^2,
\end{equation}
où on a utilisé un système de coordonnées cylindriques. On a écrit seulement les fluctuations $\hat{h}_{\mu \nu}$ de la partie à 4D car on veut déterminer comment l'action à 4D est incluse dans celle à $(4+k)$D. On peut alors calculer les quantités :
\begin{equation}
\sqrt{|g^{(4+k)}|} = R^k \sqrt{|g^{(4)}|}, \ \ \ R^{(4+k)} = R^{(4)},
\end{equation}
et ainsi, l'action de l'\'Eq.~\eqref{S_ADD_bulk} donne :
\begin{equation}
S_{bulk} = - \dfrac{M_*^{k+2}}{2} V_k \int d^4x \sqrt{|g^{(4)}|} R^{(4)}.
\end{equation}
Si on définit l'échelle de Planck du paradigme standard comme
\begin{equation}
M_P^2 = M_*^{k+2} V_k = M_*^{k+2} L^k,
\label{M_P_ADD}
\end{equation}
on retrouve l'action d'Einstein-Hilbert habituelle à 4D \eqref{action_Einstein-Hilbert} avec une constante cosmologique nulle. Si $L \sim 1/M_*$, alors $M_* \sim M_P$. Dans ce cas, la masse de Planck de la théorie extra-dimensionnelle est la même que celle de la relativité générale à 4D, ce qui a été longtemps supposé en théorie des supercordes. En revanche, si $L > 1/M_*$, $M_* < M_P$. $M_P$ n'est alors plus une échelle fondamentale, mais une échelle effective dérivée de la géométrie du système.

On peut linéariser la relativité générale :
\begin{equation}
g_{MN}^{(4+k)} = \eta_{MN}^{(4+k)} + \dfrac{1}{2M_*^{(k/2)+1}} h_{MN} \, .
\end{equation}
Dans les scénarii d'ADD, l'échelle d'énergie, à partir de laquelle la relativité générale linéarisée devient fortement couplée, est alors
\begin{equation}
\Lambda_{NDA} = \ell_{4+k}^{1/(2+k)} \, M_* \, ,
\label{grav_quant_strong}
\end{equation}
où $\ell_{4+k}$ est le facteur de boucle à $(4+k)$D,
\begin{equation}
\ell_{4+k} = (4 \pi)^{(4+k)/2} \, \Gamma \left( \dfrac{4+k}{2} \right) \, .
\end{equation}
Les effets gravitationnels non-linéaires apparaissent à l'échelle $M_* < \Lambda_{NDA}$. Il est important de rappeler que l'échelle, à laquelle apparaissent les nouveaux degrés de liberté de la théorie UV de la gravité, peut tout de même être très inférieure à $M_*$, si l'achèvement UV est une théorie faiblement couplée :
\begin{equation}
g_{QG}^2 = \ell_{4+k} \, \left( \dfrac{M_{QG}}{\Lambda_{NDA}} \right)^{2+k} \, ,
\end{equation}
comme discuté dans le troisième article d'ADD \cite{ArkaniHamed:1998nn}. Cependant, on suppose généralement dans la littérature que $M_{QG} \sim M_*$ est l'échelle de coupure du SM (ou de l'une de ses extensions). On choisit alors le volume de l'espace compactifié suffisamment grand pour ramener la masse de Planck près de l'échelle EW (\'Eq.~\eqref{M_P_ADD}). La théorie UV de la gravité n'a donc plus forcément besoin d'être très faiblement couplée pour résoudre le problème de hiérarchie avec un secteur de Higgs non-protégé par une symétrie. 

La réduction dimensionnelle du graviton à $(d+1)$D et la décomposition de KK donnent, pour les modes zéro, le graviton à 4D, des graviphotons et graviscalaires, dont le radion : un scalaire décrivant les fluctuations de la taille de l'espace compact. Dans le modèle d'ADD, l'espace-temps est plat, donc la décomposition de KK pour le graviton est une simple décomposition de Fourier, et la masse du $n$-ième niveau de KK est $m_n = n/R$. La présence d'une tour de KK-gravitons modifie la loi Newton de la gravité à 4D. Le potentiel gravitationnel, entre deux corps de masse $m_1$ et $m_2$ distants de $r$, est alors
\begin{equation}
V(r) = - G_N^{(4)} \dfrac{m_1 m_2}{r} \left( 1 + \sum_n g_n \text{e}^{-m_n r} \right) \, ,
\end{equation}
où $G_N^{(4)}$ est la constante gravitationnelle de Newton usuelle à 4D. Pour $r > R$, on peut ne retenir que le premier mode, pour lequel $g_1 = \dfrac{8}{3}k$ \cite{Adelberger:2003zx}. La déviation à la loi de Newton dépend donc de $R$, qui se calcule à partir de l'échelle de Planck dans le bulk $M_*$ avec l'\'Eq.~\eqref{M_P_ADD} :
\begin{equation}
R = \dfrac{1}{2 \pi} \, \left( \dfrac{M_P^2}{M_*^{k+2}} \right)^{1/k}.
\end{equation}
On peut prendre, par exemple, $M_* = 10 \ \text{TeV}$, pour résoudre le problème de hiérarchie de jauge. La longueur de Planck associée est $\ell_* = 1/M_* = 1,9 \times 10^{-20} \ \text{m}$. On obtient le tableau suivant :
\begin{center}
\begin{tabular}{c|l|l|l}
$k$ & $L$ & $L/\ell_*$ & $m_1$ \\
\hline 
1 & $1,2$ Gm & $5,9 \times 10^{28}$ & $1,1$ feV \\ 
2 & $4,7$ $\mu$m & $2,4 \times 10^{14}$ & $0,26$ eV \\ 
3 & $76$ pm & $3,9 \times 10^9$ & $16$ keV \\ 
4 & $0,30$ pm & $1,6 \times 10^7$ & $4,0$ MeV \\
5 & $11$ fm & $5,7 \times 10^5$ & $0,11$ GeV \\ 
6 & $1,2$ fm & $6,2 \times 10^4$ & $1,0$ GeV \\
7 & $0,25$ fm & $1,2 \times 10^4$ & $4,9$ GeV
\end{tabular} 
\end{center}
La plupart des études du modèle d'ADD considèrent $k \in \llbracket 1-7 \rrbracket$, motivé par un achèvement UV en théorie des supercordes \cite{Antoniadis:1998ig} ou en théorie M. On remarque que l'on a $L \gg \ell_*$. On a donc fait disparaître la hiérarchie de jauge au prix d'en faire apparaître une autre, cette fois géométrique. Il faut alors chercher un mécanisme de stabilisation de la taille des dimensions spatiales supplémentaires, ce qui est très difficile dans le modèle d'ADD\footnote{Néanmoins, certaines compactifications en théorie des supercordes peuvent autoriser de grands volumes \cite{Balasubramanian:2005zx, Conlon:2005ki, Cicoli:2011yy}. Cependant, pour une échelle de corde au TeV, certains modules sont si légers qu'ils médient de nouvelles forces à très longue portée, exclues par les limites expérimentales actuelles.} \cite{ArkaniHamed:1998kx, ArkaniHamed:1999dz, Mazumdar:2001ya, Carroll:2001ih, Albrecht:2001cp, Antoniadis:2002gw, Peloso:2003nv}. Le cas $k=1$ est clairement exclu par le succès de la gravité newtonienne dans l'étude du système solaire. Pour $k>1$, il faut tester la gravité à des échelles submillimétriques avec des expériences de Cavendish. La limite actuelle pour $k=2$ est $R<37 \ \mu \text{m}$ \cite{Kapner:2006si} et donc $M_* > 1,4 \ \text{TeV}$.\\

On peut définir le tenseur énergie-impulsion à partir de l'action du SM localisé sur la 3-brane:
\begin{equation}
T^{\mu \nu} = \dfrac{2}{\sqrt{|g|}} \dfrac{\delta S_{SM}}{\delta g_{\mu \nu}},
\end{equation}
où on a utilisé la métrique induite sur la 3-brane $g_{\mu \nu}$. Le champ de graviton est alors défini comme une fluctuation linéaire de la métrique de fond minkowskienne :
\begin{equation}
g_{\mu \nu} = \eta_{\mu \nu} + \dfrac{1}{2 M_*^{(k/2)+1}} h_{\mu \nu}.
\end{equation}
L'interaction entre le graviton et le SM s'écrit alors
\begin{equation}
S_{int} = \int d^4x \; \dfrac{1}{2 M_*^{(k/2)+1}} T^{\mu \nu} h_{\mu \nu},
\label{int_grav_SM}
\end{equation}
où $h_{\mu \nu}$ est la superposition des états propres de KK. En effet, la décomposition de KK du graviton s'écrit
\begin{equation}
h_{\mu \nu}(x^\mu, y^m) = \sum_{K_1 = - \infty}^{\infty} \cdots \sum_{K_k = - \infty}^{\infty} \dfrac{1}{\sqrt{V_k}} h_{\mu \nu}^{\vec{K}}(x^\mu) \exp \left( i \dfrac{\vec{K} \cdot \vec{y}}{R} \right),
\end{equation}
où les coordonnées des dimensions spatiales supplémentaires sont notées $y^m$. Après insertion dans l'action d'interaction graviton/SM, on obtient
\begin{equation}
S_{int} = \sum_{\vec{K}} \int d^4x \; \dfrac{1}{2 M_P} T^{\mu \nu} h_{\mu \nu}^{\vec{K}},
\end{equation}
où on a utilisé la relation \eqref{M_P_ADD} entre l'échelle de Planck à 4D et celle à $(4+k)$D. Chaque KK-graviton couple individuellement aux champs du SM avec un couplage en $1/M_P$. Néanmoins, la taille des dimensions spatiales supplémentaires étant grande, ils forment un quasi-continuum, et les processus inclusifs, où l'on somme sur tous les états, ont leurs sections efficaces supprimées seulement par une puissance de $M_*$. 

\begin{figure}[h]
\begin{center}
\includegraphics[height=4cm]{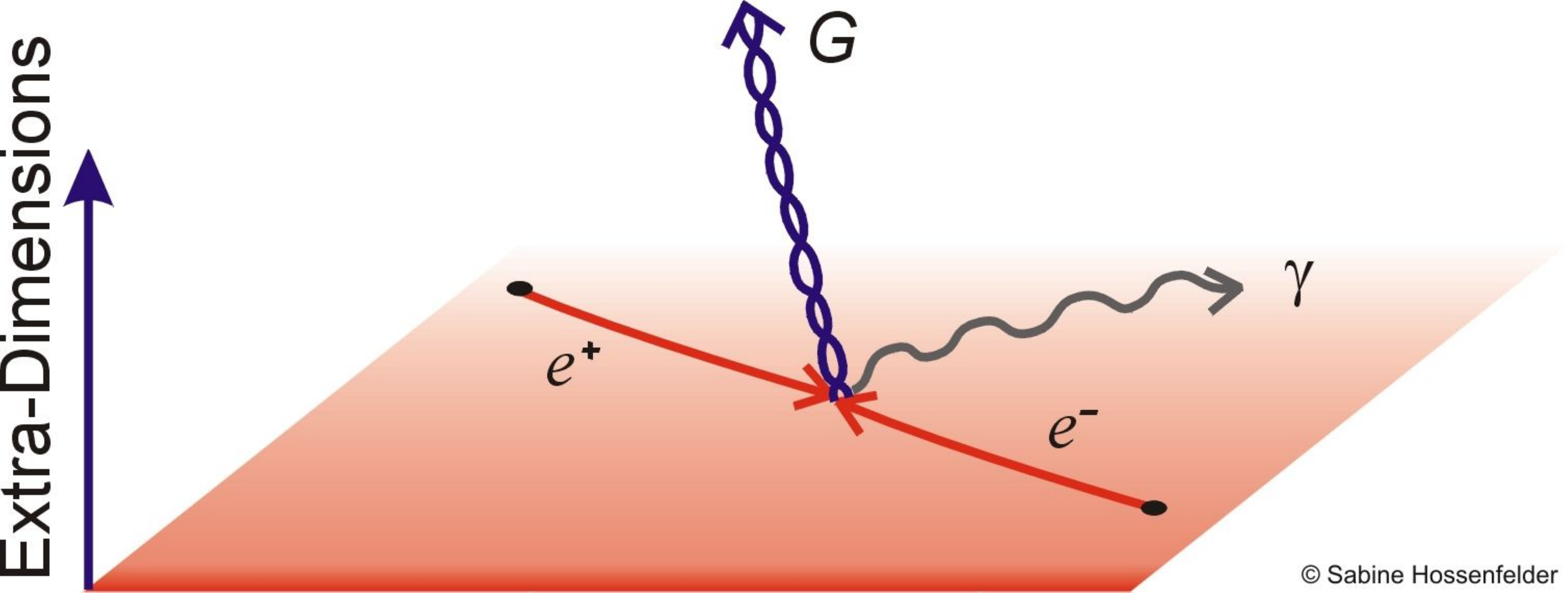}
\end{center}
\caption[\'Emission d'un graviton dans le bulk]{(Adaptée de la Réf.~\cite{siteW2}.) Une collision électron-positron sur la 3-brane, où sont localisés les champs du SM, peut produire un photon qui se propage sur la 3-brane et un graviton qui est libre de s'échapper dans le bulk.}
\label{bulk_graviton}
\end{figure}

Dans les collisionneurs de particules, on peut produire des gravitons qui s'échappent dans le bulk (\textit{c.f.} la Fig.~\ref{bulk_graviton}). Après réduction de KK, l'émission de gravitons dans le bulk correspond à la production de KK-gravitons. Leur largeur de désintégration est de l'ordre de $\Gamma \sim \dfrac{m_n^3}{M_P^2}$. Ils ont donc un temps de vie extrêmement long et, une fois produits, ils ne se désintègreront pas dans le détecteur. Ce sont donc des particules quasi-stables interagissant très faiblement. Leur signature est donc de l'énergie transverse manquante ($\cancel{E}_T$). On peut citer les processus typiques :
\begin{equation}
e^+ + e^- \rightarrow \gamma + \cancel{E}_T
\end{equation}
ou encore
\begin{equation}
q + \bar{q} \rightarrow j + \cancel{E}_T
\end{equation}
où $j$ désigne un jet. On définit souvent l'échelle $\widehat{M}_*$, telle que $\widehat{M}_*^{k+2} = (2 \pi)^k M_*^{k+2}$, considérée comme l'échelle de coupure du modèle dans la littérature expérimentale. Dans un collisionneur hadronique comme le LHC, les canaux les plus prometteurs sont ceux dont les états finaux sont $j + \cancel{E}_T$ ou $\gamma + \cancel{E}_T$. Les bornes les plus sévères proviennent de l'analyse du canal $j + \cancel{E}_T$ \cite{Sirunyan:2017jix} : $\widehat{M}_* > 9,9 \ (5,3) \ \text{TeV}$ pour $k=2 \ (6)$. D'autres contraintes provenant d'échanges de gravitons virtuels existent, mais dépendent de la manière de traiter la coupure UV de l'EFT, et sont donc plus discutables.

Les KK-gravitons légers peuvent être produits en grande quantité dans les étoiles, emportant ainsi de l'énergie. Il faut alors que la luminosité des gravitons émis soit suffisamment faible pour assurer un bon accord entre les modèles stellaires et les observations, donnant une borne inférieure sur $M_*$. La plus sévère provient de la supernova SN 1987a \cite{Hanhart:2001fx} : $M_* > 11 \ (0.8) \ \text{TeV}$ pour $k = 2 \ (3)$. Après une explosion de supernova, la plupart des gravitons restent piégés par la gravité dans l'étoile à neutron résultante. Il faut que celle-ci ne soit pas excessivement chauffée par les désintégrations des KK-gravitons en photons, on obtient ainsi \cite{Hannestad:2003yd} $\widehat{M}_* > 701 \ (25.5) \ \text{TeV}$ pour $k = 2 \ (3)$. On a aussi des bornes venant de la cosmologie \cite{Hall:1999mk}. Pour éviter l'effondrement gravitationnel de l'Univers par les KK-gravitons reliques, il faut $\widehat{M}_* > 7 \ \text{TeV}$ pour $k=2$. La désintégration des KK-gravitons reliques en photons donne également une contribution au fond diffus cosmologique, on a alors la borne $\widehat{M}_* > 100 \ \text{TeV}$ pour $k=2$. Les bornes, provenant des processus de désintégration des KK-gravitons en photons, peuvent être réduites si ceux-ci se désintègrent principalement en particules d'un secteur caché, par exemple des champs localisés sur une autre brane \cite{ArkaniHamed:1998nn}.

Notons que l'on a supposé les dimensions spatiales supplémentaires compactifiées sur un tore carré. De manière générale, une variété compactifiée est décrite par des modules de volume et des modules de forme. Par exemple, pour $k=2$, le module de forme est l'angle $\theta$ entre les deux directions de compactification. Dans la discussion précédente, le tore carré correspond à $\theta = \pi / 2$. La Réf.~\cite{Dienes:2001wu} a mis en évidence que le spectre des KK-gravitons dépend fortement de $\theta$. Dans certains cas, il est possible de maintenir le rapport entre $M_*$ et $M_P$ tout en augmentant l'écart entre les masses de deux KK-gravitons. En faisant varier $\theta$, on peut aussi induire des croisements entre niveaux de KK ou interpoler entre des modèles avec différents $k$.

\subsubsection{Compactification sur une variété hyperbolique}
Une variété hyperbolique à $k$ dimension $H_k$ est une variété avec une courbure constante négative. On définit la variété compacte hyperbolique $H_k/\Gamma$ où $\Gamma$ est un sous-groupe discret des isométries de $H_k$. On obtient ainsi un espace de genre (nombre de \og trou \fg{}) $g>1$, caractérisé par deux longueurs $R_c$ et $L_\Gamma$. $R_c$ est le rayon de courbure, c'est une propriété locale, alors que $L_\Gamma \sim R_c \; \text{ln}(g)$ est reliée au volume et est une propriété globale. On peut alors considérer un modèle d'ADD en remplaçant le tore $T^k$ par $H_k/\Gamma$ \cite{Kaloper:2000jb}. Les variétés compactes hyperboliques ont une propriété importante donnée par le théorème de rigidité de Mostow-Prasad : pour $k>2$, une fois que l'on a fixé $R_c$ et le volume, il n'y a pas d'autre module. De plus, le volume de la variété, en unité de $R_c$, ne peut pas être changé tout en maintenant l'homogénéité de la géométrie. Ainsi, la stabilisation d'un tel espace compactifié se réduit au problème de stabilisation du seul module $R_c$, \textit{i.e.} le radion. La topologie des variétés compactes hyperboliques est en général compliquée, on peut néanmoins estimer leur volume pour $L_\Gamma \gg R_c/2$ :
\begin{equation}
V_k^{CHM} \sim R_c^k \exp \left[ (k-1) \dfrac{L_\Gamma}{R_c} \right],
\end{equation}
avec toujours $M_P^2 = M_*^{k+2} V_k^{CHM}$. Pour $k=3$, $M_* = 1 \ \text{TeV}$ et $R_c \sim \ell_*$, on a $L_\Gamma \sim 35 \ell_*$ et donc $g \sim \text{e}^{35}$, on parle alors d' \og Univers gruyère \fg{} \cite{Orlando:2010kx}. En conséquence, l'une des caractéristiques intéressantes de la compactification sur une variété compacte hyperbolique est de générer une hiérarchie exponentielle entre $M_*$ et $M_P$, mais sans réintroduire une nouvelle hiérarchie géométrique trop grande entre $L_\Gamma$ et $\ell_*$.

Après réduction dimensionnelle, on peut montrer que le spectre gravitationnel contient un mode zéro pour le graviton, le radion (pas d'autres graviscalaires sans masse pour $d>2$), mais pas de graviphoton (pas d'isométries pour $H_k/\Gamma$). Concernant les KK-gravitons, KK-graviphotons et KK-graviscalaires, l'échelle de KK $m_{KK}$ et l'écart d'énergie entre les modes dépendent de la topologie précise de la variété compacte hyperbolique. On peut cependant estimer que $m_{KK} \geqslant R_c^{-1}$, ce qui relâche la plupart des contraintes expérimentales existantes pour le modèle d'ADD compactifié sur un tore \cite{Kaloper:2000jb}.

\subsection{Modèle de RS1}
\label{Modele_RS1}
\subsubsection{Principe}
Randall et Sundrum proposèrent une autre classe de modèles résolvant le problème de hiérarchie de jauge avec des dimensions spatiales supplémentaires de petite taille. Cette fois, le mécanisme utilise une propriété étonnante des espaces courbes : l'échelle de coupure d'une EFT dépend de la position de l'observateur dans l'espace compactifié. Dans les modèles d'ADD, on néglige la rétroaction gravitationnelle de la tension des branes sur la métrique, contrairement aux modèles de RS1. Dans le modèle originel \cite{Randall:1999ee}, on a une dimension spatiale supplémentaire compactifiée sur l'orbifold $S^1/\mathbb{Z}_2$ de rayon $R$, topologiquement équivalent à un intervalle de longueur $L = \pi R$. Deux 3-branes sont localisés aux points fixes de l'orbifold, \textit{i.e.} aux bords de l'intervalle en $y=0, L$, de tensions respectives $V_{UV}$ et $V_{IR}$. On peut contre-balancer l'effet des tensions branaires par une constante cosmologique $\Lambda_B$ dans le bulk. Après intégration sur la dimension spatiale supplémentaire, la théorie effective à 4D a alors une constante cosmologique nulle. En chaque point de la dimension spatiale supplémentaire, on impose que la métrique induite à 4D soit minkowskienne, et que les composantes de la métrique à 5D dépendent seulement de la coordonnée $y$ de la dimension spatiale supplémentaire. L'ansatz général pour cette métrique est alors
\begin{equation}
ds^2 = \text{e}^{-A(y)} \eta_{\mu \nu} dx^\mu dx^\nu - dy^2,
\label{RS_metric}
\end{equation}
où $\text{e}^{-A(y)}$ est appelé le facteur de déformation.

\begin{figure}[h]
\begin{center}
\includegraphics[height=6cm]{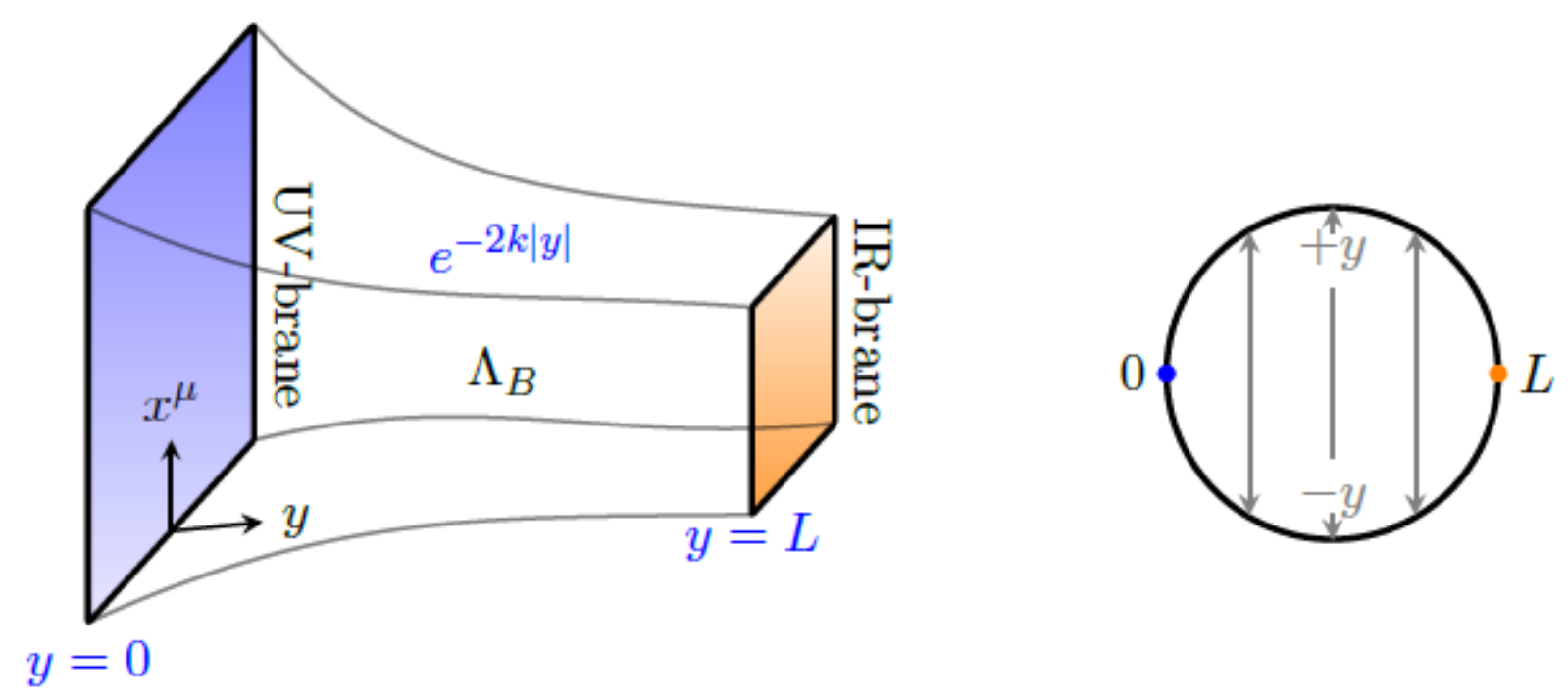}
\end{center}
\caption[Modèle de RS1]{(Adaptée de la Réf.~\cite{Ahmed:2015vde}.) Schéma de la géométrie du modèle de RS1. À gauche, tranche AdS$_5$ bornée par deux 3-branes. À droite, construction de l'orbifold $S^1/\mathbb{Z}_2$.}
\label{RS1}
\end{figure}

Dans la description dite de \og l'intervalle \fg{} \cite{Chamblin:1999ya, Lalak:2001fd, Carena:2005gq}, l'action du système s'écrit
\begin{equation}
S = S_{EH} + S_{UV} + S_{IR} + S_{GH} \, ,
\end{equation}
avec l'action d'Einstein-Hilbert du bulk à 5D,
\begin{equation}
S_{EH} = - \int d^4x \int_0^L dy \; \sqrt{|g|} \left( \dfrac{M_*^3}{2} R + \Lambda_B \right) \, ,
\end{equation}
l'action de la brane UV,
\begin{equation}
S_{UV} = - \int_{UV-brane} d^4x \, \sqrt{|g_{UV}|} \, V_{UV} \, ,
\end{equation}
l'action de la brane IR,
\begin{equation}
S_{IR} = - \int_{IR-brane} d^4x \, \sqrt{|g_{IR}|} \, V_{IR} \, ,
\end{equation}
et le terme de Gibbons-Hawking \cite{Gibbons:1976ue, Chamblin:1999ya, Lalak:2001fd, Carena:2005gq} sur les bords de l'intervalle,
\begin{equation}
S_{GH} = - \int_{UV-brane} d^4x \, \sqrt{|g_{UV}|} \, M_*^3 \, K - \int_{IR-brane} d^4x \, \sqrt{|g_{IR}|} \, M_*^3 \, K \, ,
\end{equation}
où $K$ est la courbure extrinsèque, et $g_{UV/IR}$ est le déterminant de la métrique induite sur la brane UV/IR. En résolvant les équations d'Einstein, on trouve alors que
\begin{equation}
A(y) = 2ky \ \ \ \text{avec} \ \ \ \Lambda_B = -6 M_*^3 k^2 < 0 \ \ \ \text{et} \ \ \ V_{UV} = - V_{IR} = - \dfrac{\Lambda_B}{k} > 0,
\label{fine_tune_RS}
\end{equation}
où $k$ est la courbure de l'espace. L'espace à 5D est donc une tranche Anti-de Sitter à 5D (AdS$_5$) reposant sur deux conditions d'ajustement fin : l'une pour la valeur de la constante cosmologique à 4D et l'autre imposant que les tensions des 3-branes doivent être exactement opposées. On remarquera que $V_{IR} < 0$, mais cela ne pose pas de problème d'instabilité ici, car la 3-brane est positionnée à un point fixe d'un orbifold (\textit{c.f.} la Fig.~\ref{RS1}), donc il n'y a pas de branons-fantômes associés.

Dans la littérature, on suppose aussi généralement, de manière implicite, que l'achèvement UV de la gravité est une théorie fortement couplée. La relativité générale linéarisée devient non-perturbative à l'échelle
\begin{equation}
\Lambda_{NDA} = \ell_5^{1/3} \, M_* \, ,
\label{strong_grav_RS}
\end{equation}
où $M_*$ et $\ell_5 = 24 \pi^3$ sont respectivement l'échelle de Planck réduite et le facteur de boucle à 5D. Quant aux effets gravitationnels non-linéaires, on estime qu'ils devraient apparaître à l'échelle $M_* < \Lambda_{NDA}$, donc $M_*$ est supposée être l'échelle fondamentale du modèle.

Dans le modèle de RS1 originel, les champs du SM sont localisés sur la brane IR, son action est :
\begin{equation}
S_{SM} = \int d^4x \int_0^L dy \; \delta(y-L) \sqrt{|g|} \left[ g^{\mu \nu} (D_\mu H)^\dagger D_\nu H - \lambda (H^\dagger H - v^2)^2 + \cdots \right],
\end{equation}
où on a écrit les termes impliquant seulement le champ de Higgs $H$ dont $v$ est la VEV, naturellement à l'échelle de gravité fondamentale $M_*$, mais suffisamment inférieure à celle-ci pour être encore dans le régime de validité de la théorie des champs : $v \sim 0.1 \; M_*$. En insérant l'expression de la métrique (\'Eqs.~\eqref{RS_metric} et \eqref{fine_tune_RS}), l'action du SM devient
\begin{equation}
S_{SM} = \int d^4x \left[ \eta^{\mu \nu} (D_\mu \widetilde{H})^\dagger D_\nu \widetilde{H} - \lambda (\widetilde{H}^\dagger \widetilde{H} - (\text{e}^{-kL} v)^2)^2 + \cdots \right],
\end{equation}
où les champs ont été canoniquement renormalisés :
\begin{equation}
\widetilde{H} = \text{e}^{-kL} H \ \ \ \text{pour un scalaire},
\label{renorm_1}
\end{equation}
\begin{equation}
\widetilde{A}_\mu = A_\mu \ \ \ \text{pour un boson de jauge},
\label{renorm_2}
\end{equation}
\begin{equation}
\widetilde{\psi} = \text{e}^{-(3/2)kL} \psi \ \ \ \text{pour un fermion}.
\label{renorm_3}
\end{equation}
Après avoir intégré sur la dimension spatiale supplémentaire, la VEV du Higgs dans la théorie effective à 4D $\widetilde{v}$ est alors naturellement à une échelle d'énergie plus basse que le paramètre de la théorie extra-dimensionnelle $v$:
\begin{equation}
\widetilde{v} = \text{e}^{-kL} v.
\end{equation}
Comme on ne veut pas réintroduire de hiérarchie géométrique, on s'attend à ce que $M_* \sim M_P$ et donc pour $kL \sim 35$, on a
\begin{equation}
\widetilde{v} \sim 0.1 \; \text{e}^{-kL} M_* \sim 0.1 \ \text{TeV}.
\end{equation}
On retrouve ainsi l'échelle d'énergie de la VEV du Higgs dans le SM ! Cependant, pour résoudre le problème de hiérarchie de jauge, la valeur de la VEV doit être stabilisée par rapport aux corrections radiatives. Le moyen le plus simple de le vérifier est de régulariser la théorie à la Pauli-Villars en introduisant des particules fantômes de masse $\Lambda_{UV}$, l'échelle de coupure de la théorie sur la brane UV. Lors de la renormalisation des champs, la masse de ces particules, \textit{i.e.} l'échelle de coupure, est naturellement abaissée comme la VEV du Higgs, assurant ainsi la stabilité de cette dernière, d'après l'\'Eq.~\eqref{hierarchie}. Ainsi, l'échelle de coupure $\Lambda$ dans une tranche AdS$_5$ dépend de la position dans la dimension spatiale supplémentaire \cite{Goldberger:2002cz} :
\begin{equation}
\Lambda(y) = \text{e}^{-ky} \Lambda_{UV},
\end{equation}
où on suppose $\Lambda_{UV} \sim M_*$.

Pour relier la masse de Planck réduite à 4D $M_P$ à celle dans le bulk $M_*$, on va procéder comme pour le modèle d'ADD. Considérons la fluctuation $\hat{h}_{\mu \nu}$ de la métrique associée au graviton à 4D :
\begin{equation}
ds^2 = \text{e}^{-2ky} \left( \eta_{\mu \nu} + \hat{h}_{\mu \nu} \right) dx^\mu dx^\nu - dy^2.
\end{equation}
Le terme donnant l'action d'Einstein-Hilbert à 4D est alors
\begin{equation}
- \dfrac{M_*^3}{2} \int d^4x \int_0^L dy \sqrt{|g^{(4)}|} \text{e}^{-2ky} R^{(4)},
\end{equation}
où on identifie la masse de Planck réduite
\begin{equation}
M_P^2 = M_*^3 \int d^4x \int_{0}^{L} dy \; \text{e}^{-2ky} = \dfrac{M_*^3}{2k} \left( 1 - \text{e}^{-2kL} \right).
\label{M*_MP}
\end{equation}
Pour $kL \sim 35$ et $k \sim M_*$, on a bien $M_* \sim M_P$. Plus précisément, la métrique du modèle de RS1 est obtenue par résolution des équations d'Einstein, on a donc la contrainte $k<M_*$. On peut alors estimer une borne supérieure sur $k/M_P \lesssim 0.1$.

\subsubsection{Secteur gravitationnel}
Le graviton à 5D donne, en principe, après réduction dimensionnelle à 4D, une tour de gravitons, de graviphotons et de graviscalaires. La fluctuation de la métrique du modèle de RS1 s'écrit
\begin{equation}
ds^2 = \text{e}^{-2ky} g_{\mu \nu} dx^\mu dx^\nu + A_\mu dx^\mu dy - b^2 dy^2,
\end{equation}
où $g_{\mu \nu}$ paramétrise les fluctuations du graviton, $A_\mu$ celles du graviphoton et $b$ celles du graviscalaire. La compactification intervalle/orbifold\footnote{À cause de la symétrie $\mathbb{Z}_2$ inhérente à l'orbifold, $ds^2$ doit être invariant sous celle-ci : $A_\mu$ est donc impair sous $\mathbb{Z}_2$ et ne peut avoir de mode zéro.} implique que la tour de graviphotons n'a pas de mode zéro. Les états excités des tours de graviphotons et de graviscalaires sont absorbés par ceux de la tour de gravitons, donnant ainsi leur masse aux KK-gravitons, de manière similaire au mécanisme de Higgs. Finalement, dans une jauge physique, il reste le graviton sans masse, sa tour de KK-gravitons massifs et un graviscalaire sans masse.

La décomposition de KK, pour le champ $h_{\mu \nu}(x^\mu, y)$ du graviton dans le modèle de RS1, est donnée par
\begin{equation}
h_{\mu \nu}(x^\mu, y) = \sum_{n=0}^{\infty} h_{\mu \nu}^n (x^\mu) f_h^n (y),
\end{equation}
où les champs $h_{\mu \nu}^n (x^\mu)$ décrivent les KK-gravitons, et les $f_h^n (y)$ sont leurs fonctions d'onde de KK normalisées telles que
\begin{equation}
\dfrac{1}{L} \int_0^L dy \; f_h^n (y) f_h^m (y) = \delta^{nm}.
\end{equation}
La résolution des équations d'Euler-Lagrange, obtenues à partir de la variation de l'action, permet d'obtenir une forme analytique pour ces fonctions d'onde de KK. Le mode zéro est ainsi quasi-localisé près de la 3-brane en $y=0$ avec une fonction d'onde de KK
\begin{equation}
f_h^0 (y) = \text{e}^{-ky}.
\end{equation}
On a alors un modèle où la gravité est quasi-localisée dans la dimension spatiale supplémentaire, ce qui permet de reformuler la résolution du problème de hiérarchie de jauge dans le modèle de RS1 : le graviton est quasi-localisé sur la 3-brane, dite UV, où la gravité est forte, et un observateur positionné sur l'autre 3-brane, dite IR, ressent seulement la queue de la fonction d'onde de KK du graviton ; l'intensité de la gravité y est exponentiellement supprimée. Pour les modes de KK, leurs fonctions d'onde sont des fonctions de Bessel quasi-localisées près de la brane IR, et leurs masses sont données par
\begin{equation}
m_j = x_j \; m_{KK},
\end{equation}
avec l'échelle de KK $m_{KK} = k \text{e}^{-kL}$, et
\begin{equation}
x_j \simeq \left\{
\begin{array}{l c l}
3,8 & & j = 1 \, , \\
7,0 & & j = 2 \, , \\
10,2 & & j = 3 \, , \\
16,5 & & j = 4 \, , \\
& \vdots &
\end{array}
\right.
\end{equation}
Une prédiction importante du modèle de RS1 est donc l'apparition d'une tour de KK-gravitons à une échelle $m_{KK} \sim 1 \ \text{TeV}$, accessible au LHC, et non à l'échelle de l'inverse de la longueur propre de la dimension spatiale supplémentaire (proche de l'échelle de Planck avec une dimension spatiale supplémentaire de petite taille). De plus, la quasi-localisation des KK-gravitons près de la brane IR, où est confinée la matière, induit un couplage avec celle-ci exponentiellement augmenté par rapport à celui du mode zéro, localisé près de la brane UV ; le lagrangien d'interaction est de la forme :
\begin{equation}
- \dfrac{1}{M_P} T^{\mu \nu} h_{\mu \nu}^0 - \dfrac{1}{M_P \text{e}^{-kL}} T^{\mu \nu} \sum_{n=1}^\infty h_{\mu \nu}^n.
\end{equation}
On peut alors rechercher ces KK-gravitons sous la forme de résonances individuelles au LHC. En supposant un rapport $k/M_P = 0.1$, les bornes inférieures les plus sévères, sur la masse du premier mode de KK, proviennent des canaux en dilepton \cite{Khachatryan:2016zqb} ($m_1 > 3,1 \ \text{TeV}$) et diphoton \cite{Khachatryan:2016yec} ($m_1 > 3,85 \ \text{TeV}$). Dans le scénario du chapitre suivant, où les bosons de jauge et les fermions se propagent dans la dimension spatiale supplémentaire, les limites précédentes ne s'appliquent plus : les couplages des KK-gravitons au $e e/ \mu \mu / \gamma \gamma$ sont supprimés. Le premier mode de KK du champ de graviton est alors plus lourd que les premiers modes de KK des autres champs se propageant dans la dimension spatiale supplémentaire.

Intéressons nous maintenant au graviscalaire. C'est en fait le champ qui décrit les fluctuations de la longueur $L = \pi R$ de la dimension spatiale supplémentaire, d'où son nom de radion. Dans le modèle de RS1 que l'on a présenté jusqu'alors, $L$ n'est pas fixée dynamiquement : on a juste choisi la valeur qui nous arrangeait pour résoudre le problème de hiérarchie de jauge. L'absence d'un potentiel jouant ce rôle entraîne que le radion est sans masse. Il génère ainsi une nouvelle interaction gravitationnelle scalaire de portée infinie qui modifie la loi de Newton, ce qui est expérimentalement exclu. Le mécanisme de Goldberger-Wise \cite{Goldberger:1999uk} permet de stabiliser dynamiquement la dimension spatiale supplémentaire, et ainsi de donner une masse au radion. Il faut alors ajouter un champ scalaire dans le bulk avec un terme de masse, cela engendre un potentiel non-trivial pour $L$ : la masse dans le bulk tend à réduire $L$. En ajoutant des potentiels scalaires, chacun localisé sur une des deux 3-branes, et ayant des minima différents, le scalaire acquiert un profil non-trivial le long de la dimension spatiale supplémentaire. Le terme cinétique tend alors à avoir ses dérivées petites et donc à augmenter $L$, pour minimiser l'énergie cinétique dans la direction de la dimension spatiale supplémentaire. Il y a ainsi apparition d'un minimum qui stabilise $L$ avec des valeurs naturelles pour les paramètres du lagrangien : le radion acquiert une masse inférieure à celle des KK-gravitons \cite{Csaki:2000zn, Tanaka:2000er}. On peut approximer le facteur de déformation de la métrique \eqref{RS_metric} par sa valeur sans la rétroaction du scalaire : $A(y) = 2ky$. Des deux conditions d'ajustement fin \eqref{fine_tune_RS}, il ne reste que celle associée à la constante cosmologique : le modèle de RS1 fait disparaître l'ajustement fin de la masse du boson de Higgs, sans en introduire un autre que celui de la constante cosmologique, déjà présent dans le SM couplé à la gravité. Du fait de l'ajout d'un champ scalaire dans le bulk, le radion acquiert une tour de KK dont les masses des modes sont du même ordre de grandeur que celles des KK-gravitons. Le radion peut se mélanger avec le boson de Higgs \cite{Giudice:2000av}, altérant potentiellement la phénoménologie standarde de ce dernier \cite{Csaki:2000zn, Dominici:2002jv, Gunion:2003px}. Des recherches expérimentales du radion au LHC ont été effectuées \cite{Khachatryan:2015yea, Khachatryan:2016cfa}, via sa production en fusion de gluons et sa désintégration en dihiggs, excluant sa masse entre 300 et 1100 GeV, et entre 1150 et 1550 GeV, pour une constante de désintégration de 1 TeV.

\subsection{Problèmes liés à une échelle de gravité au TeV}
\label{Problemes_quantique_TeV}
\begin{figure}[h]
\begin{center}
\includegraphics[height=4.5cm]{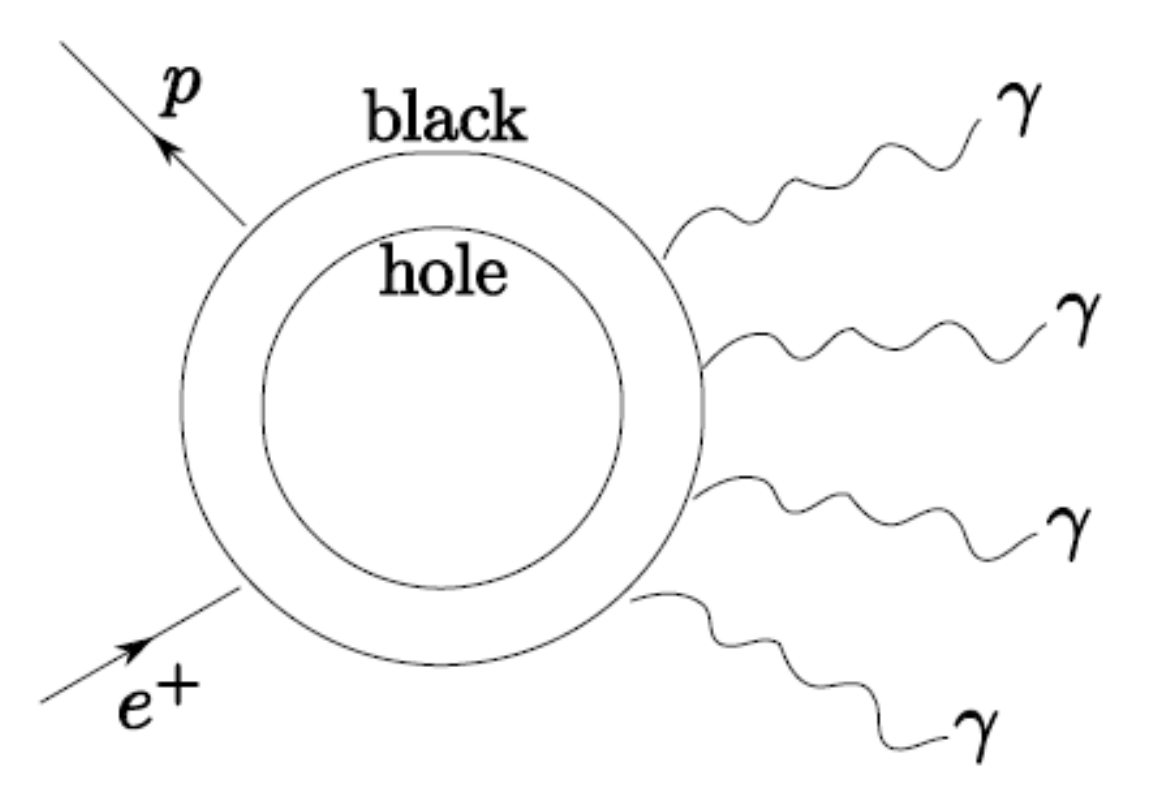}
\end{center}
\caption[Désintégration d'un proton par un trou noir]{(Adaptée de la Réf.~\cite{Shifman:2009df}.) Désintégration d'un proton par un trou noir. Le nombre baryonique est violé alors que la charge électrique est conservée.}
\label{QBH}
\end{figure}

\begin{figure}[h]
\begin{center}
\includegraphics[height=5cm]{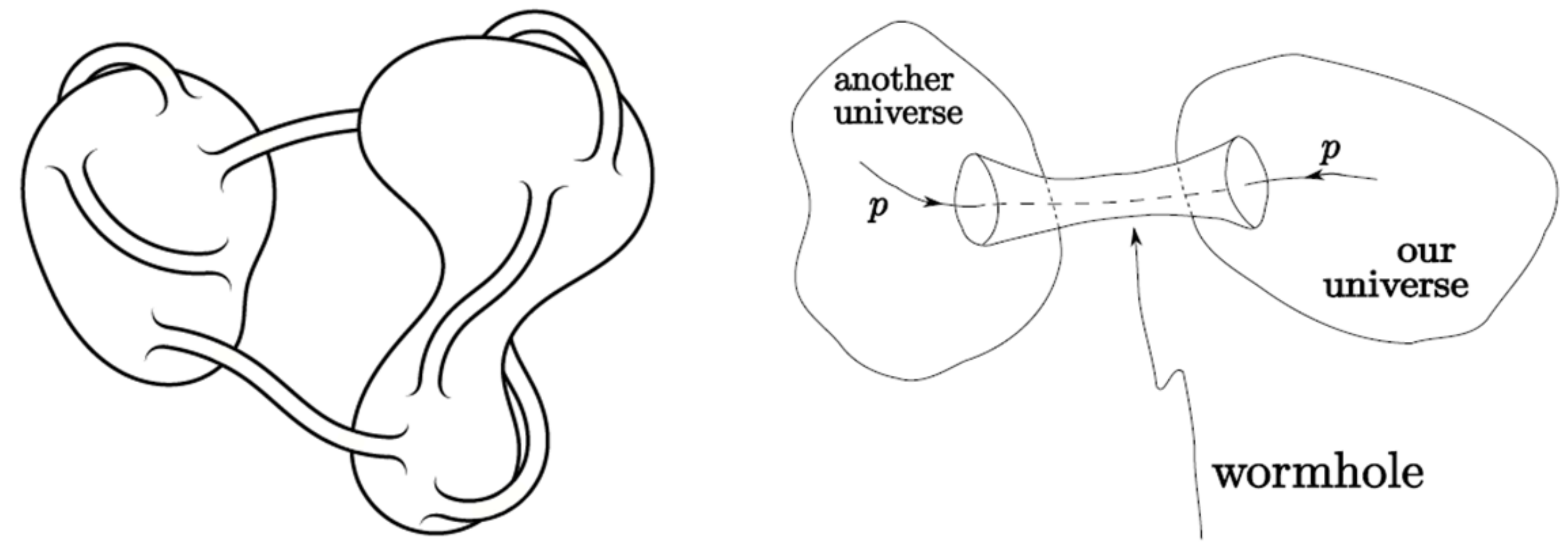}
\end{center}
\caption[Trous de ver]{À gauche (adaptée de la Réf.~\cite{Hebecker:2018ofv}), un trou de ver, en gravité quantique euclidienne, relie deux régions d'un même Univers ou constitue un pont entre deux Univers. À droite (adaptée de la Réf.~\cite{Shifman:2009df}), un proton s'échappe de notre Univers par un trou de ver virtuel.}
\label{QWH}
\end{figure}

\begin{figure}[h]
\begin{center}
\includegraphics[height=4cm]{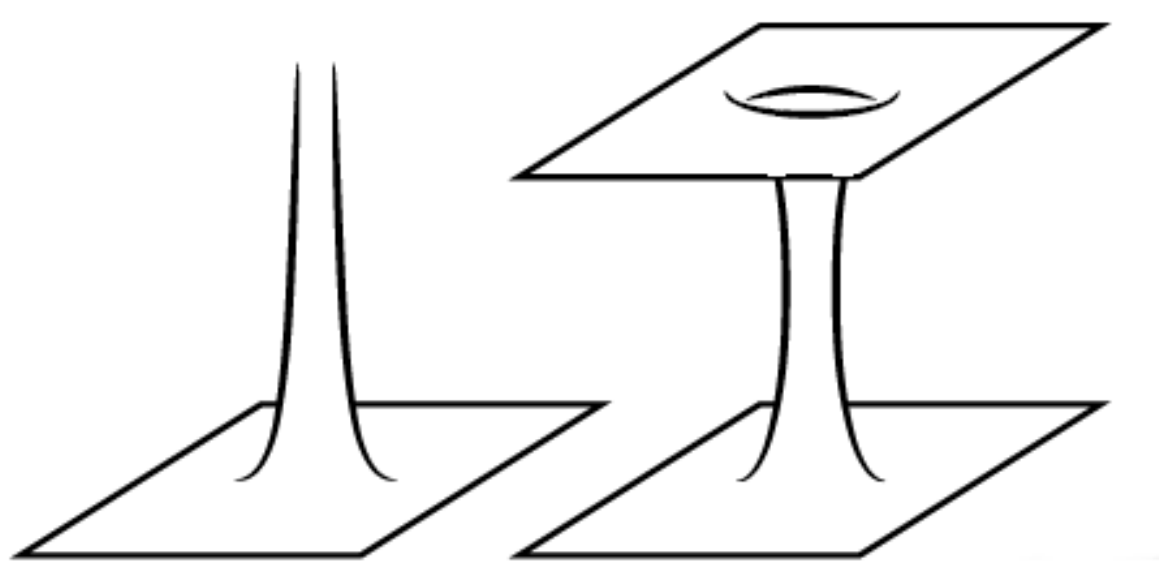}
\end{center}
\caption[Trous de ver et bébés Univers]{(Adaptée de la Réf.~\cite{Hebecker:2018ofv}.) À droite, trou de ver connectant deux Univers. À gauche, nucléation d'un bébé Univers (ou semi-trou de ver).}
\label{BU}
\end{figure}

\begin{figure}[h]
\begin{center}
\includegraphics[width=9cm]{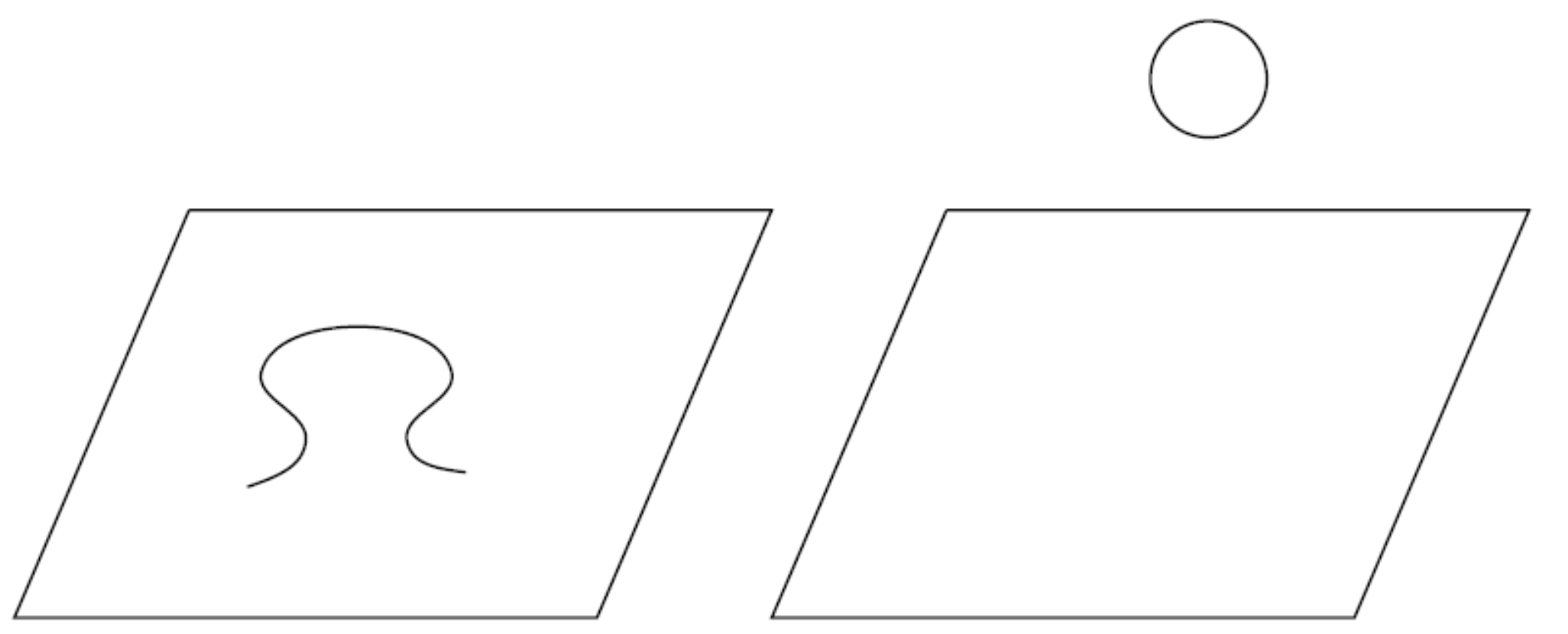}
\end{center}
\caption[Nucléation d'une bébé brane]{(Adaptée de la Réf.~\cite{Dvali:1999gf}.) Nucléation d'une bébé brane.}
\label{BB1}
\end{figure}

\begin{figure}[h!]
\begin{center}
\includegraphics[width=9cm]{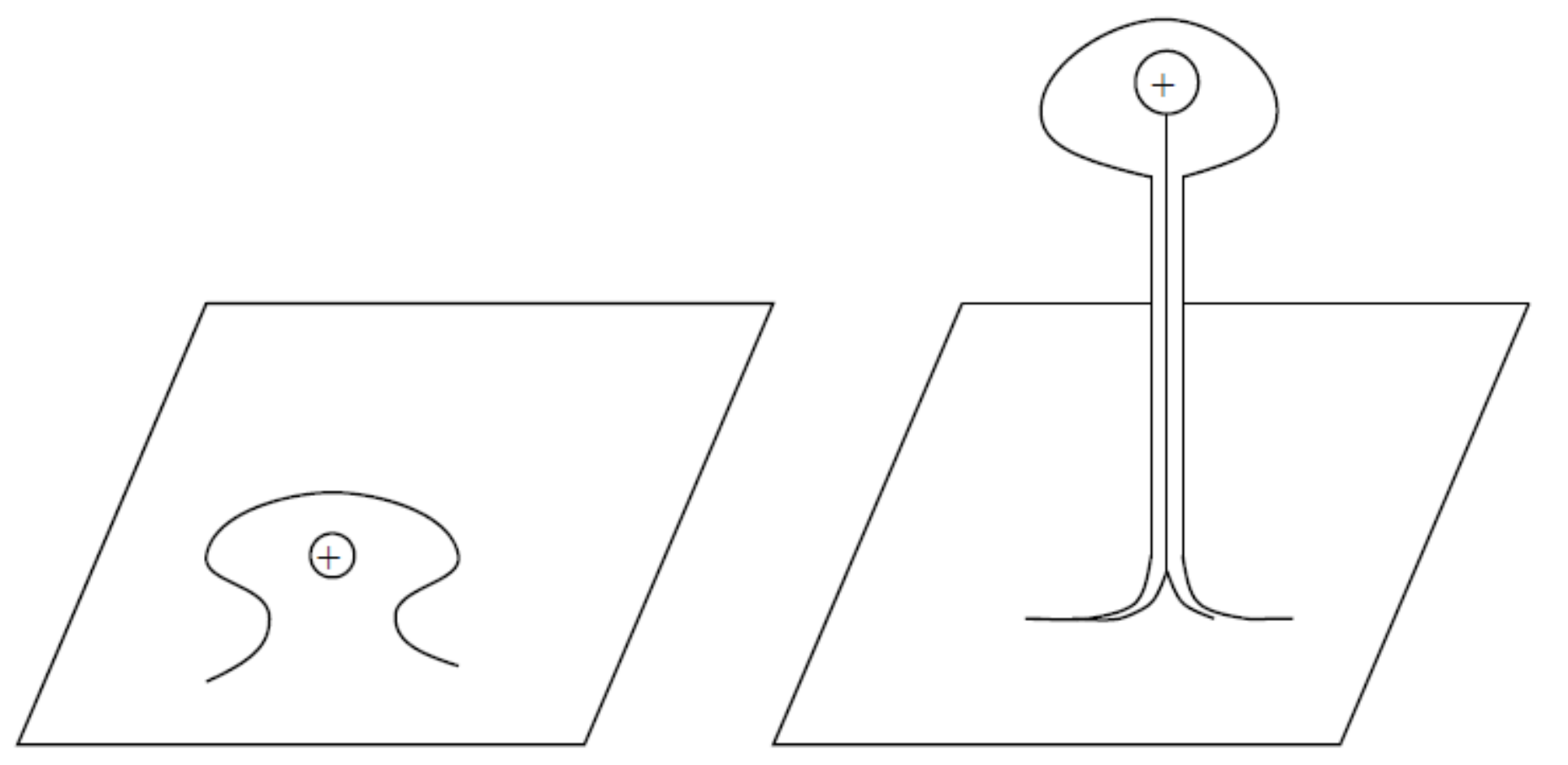}
\end{center}
\caption[Tube de champ reliant une charge locale, emportée par une bébé brane, à la brane mère]{(Adaptée de la Réf.~\cite{Dvali:1999gf}.) Tube de champ reliant une charge locale, emportée par une bébé brane, à la brane mère.}
\label{BB2}
\end{figure}

Les modèles qui résolvent le problème de hiérarchie de jauge, avec des dimensions spatiales supplémentaires et les champs du SM sur une 3-brane, ramènent l'échelle de coupure, et donc de gravité, à l'échelle du TeV. Or, des phénomènes gravitationnels non-perturbatifs, comme les trous noirs, les instantons gravitationnels, les trous de ver, et la nucléation de bébés univers ou de bébés branes sont connus pour ne pas conserver les charges associées à des symétries globales. Il est donc couramment admis qu'une théorie UV de la gravité, reproduisant la relativité générale à basse énergie, ne peut pas avoir de symétrie globale : c'est l'un des critères des \og Marécages \footnote{On appelle \og Marécages \fg{} l'ensemble des EFTs qui ne sont pas consistantes avec un achèvement UV décrivant la gravité.} \fg{} \cite{Brennan:2017rbf, Palti:2019pca}. Par contre, la gravité conserve les symétries de jauge. Prenons l'exemple de la conservation du nombre baryonique $B$ (charge globale associée à la symétrie accidentelle $U(1)_B$) et de la charge électrique $Q$ (charge locale) dans le SM.
\begin{itemize}
\item Un trou noir n'a pas de charge globale, c'est le théorème de calvitie \cite{Ruffini:1971bza, Hartle:1971qq, Bekenstein:1971hc, Bekenstein:1972ky, Bekenstein:1972ny, Teitelboim:1972pk, Teitelboim:1972ps, Teitelboim:1972qx}. À l'inverse, il peut avoir une charge de jauge \cite{Calmet:2008dg, Gingrich:2009hj}. Comme l'illustre la Fig.~\ref{QBH}, un trou noir virtuel peut absorber un proton et s'évaporer par radiation de Hawking \cite{Zeldovich:1976vw, Zeldovich:1976vq, Adams:2000za}. Cependant, les modèles microscopiques de trous noirs, où ces derniers sont décrits comme des condensats de Bose-Einstein de gravitons \cite{Dvali:2011aa, Dvali:2012en}, prédisent que les trous noirs peuvent avoir des charges globales \cite{Dvali:2012rt, Kuhnel:2015qaa}. Dans ces modèles, les arguments impliquant le théorème de calvitie, basés sur la relativité générale, sont alors valables dans la limite semi-classique idéale, que ne vérifie jamais un trou noir dans la Nature.
\item Un trou de ver est constitué d'une gorge terminée par une bouche à chaque extrémité. Il forme ainsi un pont entre deux régions du même Univers ou entre deux Univers différents (\textit{c.f.} la Fig.~\ref{QWH}). Un trou de ver virtuel peut transporter une charge globale d'une région de l'espace à une autre beaucoup plus éloignée, ou dans un autre Univers \cite{Hawking:1988ae, Coleman:1988cy, Coleman:1988tj, Giddings:1988wv, Abbott:1989jw, Giddings:1989bq, Coleman:1989zu, Hebecker:2018ofv}, comme le montre la Fig.~\ref{QWH}.
\item Un bébé Univers correspond à un trou de ver coupé au niveau de la gorge (\textit{c.f.} Fig.~\ref{BU}). Autrement dit, il n'y a pas encore d'Univers où ressortir. La gorge en formation correspond à la nucléation du bébé Univers. Une charge globale peut être emmenée par le bébé Univers en formation \cite{Giddings:1988wv, Hebecker:2018ofv}. 
\item Comme on l'a vu à la Section~\ref{EFT_brane}, si la 3-brane sur laquelle est localisé le SM n'est pas un point singulier de l'espace compactifié, elle peut fluctuer. On appelle branons les degrés de liberté de fluctuation. La Réf.~\cite{Dvali:1999gf} a mis en évidence, qu'à une énergie comparable à la tension de la 3-brane, appelée brane mère, celle-ci peut se contorsionner jusqu'à créer une bébé brane indépendante (\textit{c.f.} Fig.~\ref{BB1}). Lors de sa formation, la bébé brane peut capturer des particules de la brane mère, et les emmener avec elle lorsqu'elle s'éloigne dans le bulk. Si ces particules ont des charges globales, comme $B$, un observateur sur la brane mère verra une disparition de ces charges, donc une non-conservation de $B$. On pourrait craindre qu'il en soit de même pour les charges locales, comme $Q$. Cependant, une particule de charge Q, emportée par la bébé brane, va créer un tube de champ qui la relie à sa position initiale sur la brane mère (\textit{c.f.} Fig.~\ref{BB2}). Si l'énergie est suffisante, ce tube va se briser, en générant une paire particule-antiparticule, telle que l'antiparticule ait une charge $-Q$. La bébé brane va alors emporter la paire électriquement neutre, formée de la particule initiale et de l'antiparticule créée. Quant à la particule créée de charge $Q$, elle va rester sur la brane mère, assurant la conservation de la charge électrique. On peut généraliser ce raisonnement au cas de charges locales associées à des symétries non-abéliennes.
\end{itemize}
Tous ces phénomènes seraient interprétés dans une expérience comme une désintégration du proton. On doit alors inclure, dans l'EFT extra-dimensionnelle, des opérateurs de contact à quatre fermions sur la 3-brane du SM de type $\dfrac{C}{\Lambda^2}\bar{\psi}\psi\bar{\psi}\psi$, où $C$ est le coefficient adimensionné de l'opérateur, et $\Lambda$ est l'échelle de coupure sur la 3-brane ramenée au TeV (échelle fondamentale dans les modèles à la ADD, ou échelle de la gravité sur la brane IR due à l'effet du facteur de déformation dans le modèle de RS1). Le même constat peut être fait avec les opérateurs violant le nombre leptonique $L$, ou induisant des courants neutres de changement de saveur (FCNCs -- \textit{Flavor Changing Neutral Currents}). Par ailleurs, on sait expérimentalement que :
\begin{align}
\Lambda_{\cancel{B}} &\gtrsim 10^{15} \ \text{GeV}, \nonumber \\
\Lambda_{\cancel{L}} &\gtrsim 10^{14} \ \text{GeV}, \nonumber \\
\Lambda_{FCNC, \cancel{CP}} &\gtrsim 10^{3}-10^5 \ \text{GeV}. \label{suppress_op}
\end{align}
Pour un modèle où l'échelle de gravité est au TeV, on a alors deux cas possibles :
\begin{itemize}
\item Ces processus sont rares dans la théorie UV de la gravité, et donc le coefficient, devant les opérateurs effectifs, est supprimé : $C \ll 1$, il n'y a alors pas de problème.
\item Dans le cas contraire, $C \sim 1$ et il faut donc trouver des mécanismes supplémentaires pour supprimer ces opérateurs dangereux.
\end{itemize}

\begin{figure}[h!]
\begin{center}
\includegraphics[height=2.5cm]{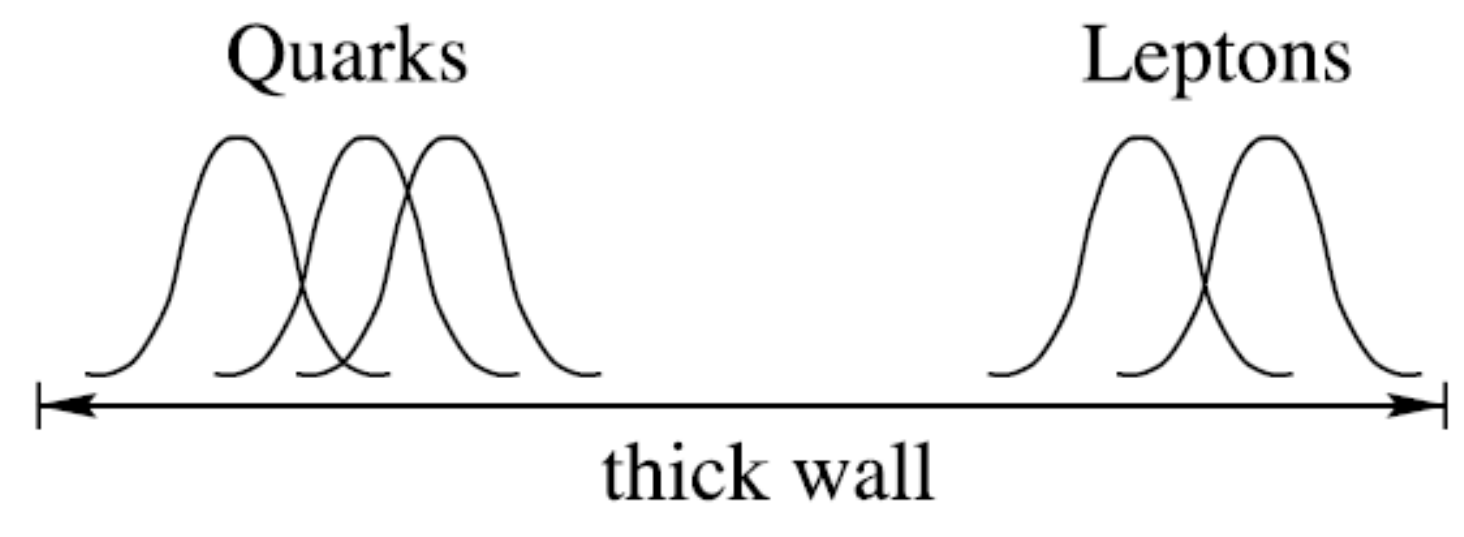}
\end{center}
\caption[Quasi-localisation des fermions dans une brane épaisse]{(Adaptée de la Réf.~\cite{Shifman:2009df}.) 3-brane épaisse où les fonctions d'onde des quarks et des leptons sont localisés en des points différents de l'épaisseur de la brane.}
\label{AS}
\end{figure}

Pour les symétries $U(1)_B$ et $U(1)_L$, on peut, par exemple, les jauger (la gravité conserve les charges de jauge). On obtient une théorie avec des anomalies de jauge qui peuvent être annulées par le mécanisme de Green-Schwarz \cite{Green:1984sg, Sagnotti:1992qw, Ibanez:1998qp, Poppitz:1998dj, Anastasopoulos:2006cz}, motivé par les constructions de la théorie de jauge du SM avec des D-branes \cite{Antoniadis:2000ena, Ibanez:2001nd, Antoniadis:2002qm} : les bosons de jauge acquièrent une masse à l'échelle des supercordes tout en préservant les symétries globales $U(1)_B$ et $U(1)_L$. Dans d'autres modèles, on jauge les symétries de saveur approximatives du SM, $SU(3)^5 \equiv SU(3)_{Q_L} \times SU(3)_{u_R} \times SU(3)_{d_R} \times SU(3)_{L_L} \times SU(3)_{e_R}$, pour supprimer les opérateurs de FCNCs \cite{Berezhiani:1998wt, ArkaniHamed:1998sj, ArkaniHamed:1999yy, Rattazzi:2000hs, Delaunay:2010dw, Eshel:2011wz}. Ces nouvelles symétries de jauge sont alors brisées sur des branes distantes de celle du SM, et la brisure est communiquée à cette dernière par des champs scalaires dans le bulk : les flavons.
Alternativement, si la 3-brane du SM est épaisse (un mur de domaine, par exemple), on peut construire un modèle où les fonctions d'onde des modes 4D des fermions sont des gaussiennes localisées en des points différents de la dimension spatiale supplémentaire pour les leptons et les quarks (\textit{c.f.} la Fig.~\ref{AS}). Le faible recouvrement, entre les gaussiennes des quarks et des leptons, supprime ainsi les opérateurs de contact de désintégration du proton \cite{ArkaniHamed:1999dc}.

\chapter{Bosons de jauge et fermions dans une tranche AdS$_5$}
\label{tranche_AdS5}

\begin{figure}[h]
\begin{center}
\includegraphics[width=10cm]{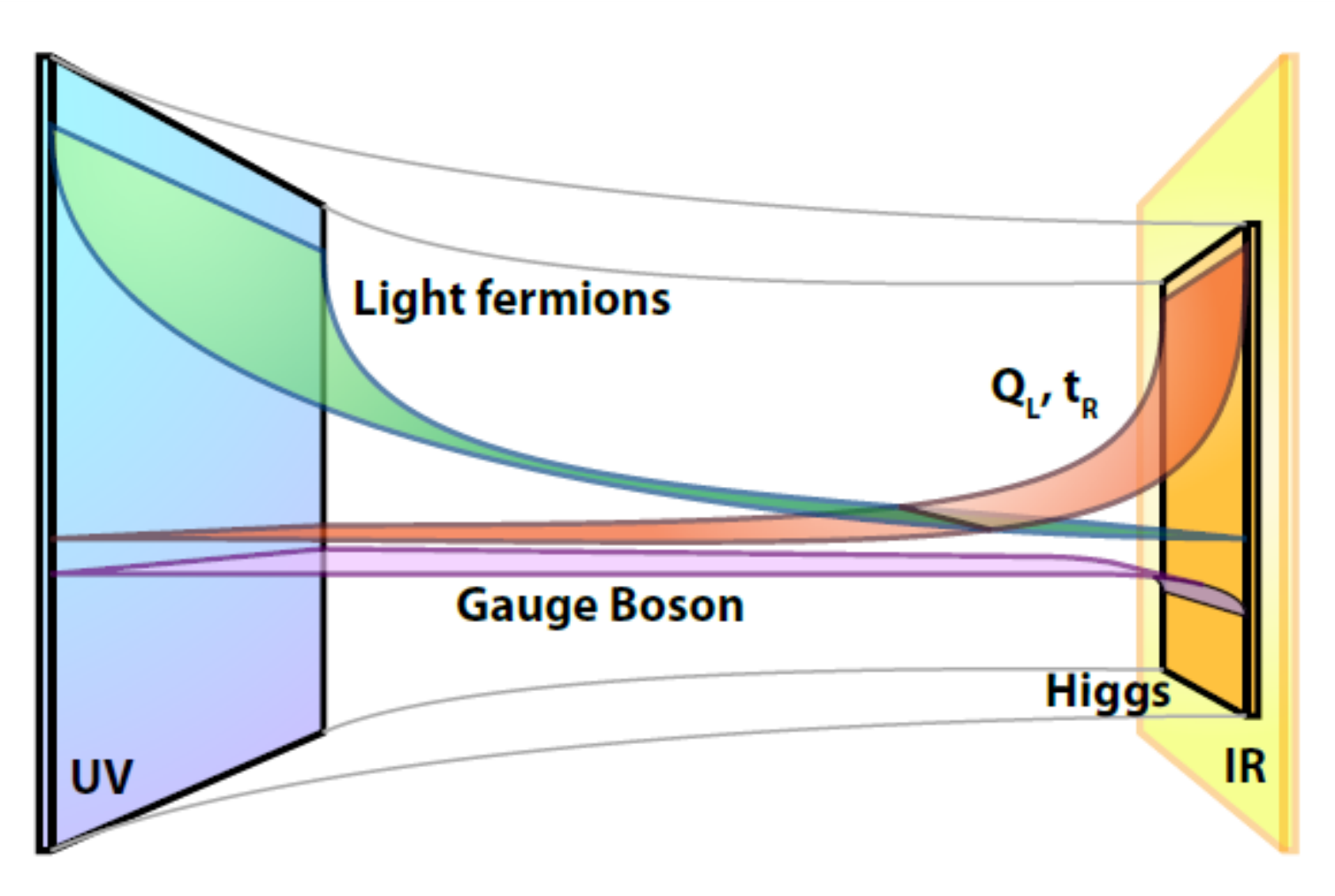}
\end{center}
\caption[Bosons de jauge et fermions dans une tranche AdS$_5$]{(Adaptée de la Réf.~\cite{Csaki:2016kln}.) Vue schématique du modèle de RS1 avec les bosons de jauge et les fermions dans le bulk. Le champ de Higgs est localisé sur la brane IR. Les modes zéro des bosons de jauge sans masse ont des fonctions d'onde de KK constantes le long de la dimension spatiale supplémentaire, assurant ainsi l'universalité des couplages avec les modes zéro des fermions. Ces derniers sont quasi-localisés différemment dans le bulk selon s'ils sont légers (près de la brane UV : faible recouvrement avec le champ de Higgs), ou lourds (près de la brane IR : fort recouvrement avec le champ de Higgs).}
\label{bulk_RS}
\end{figure}

Afin de résoudre les éventuels problèmes liés à une échelle de gravité au TeV sur la brane IR du modèle de RS1, on a vu au chapitre précédent que l'on peut construire des modèles en jaugeant $U(1)_B \times U(1)_L$ et les symétries de saveur approximatives du SM, $SU(3)^5$. Cependant, jauger toutes les symétries de saveur peut paraître très artificiel. Il a très vite été réalisé que, pour résoudre le problème de hiérarchie de jauge, seul le champ de Higgs nécessite d'être localisé sur la brane IR, alors que les bosons de jauge et les fermions peuvent se propager dans le bulk \cite{Davoudiasl:1999tf, Pomarol:1999ad, Grossman:1999ra, Chang:1999nh, Gherghetta:2000qt, Davoudiasl:2000wi} (\textit{c.f.} la Fig.~\ref{bulk_RS}). La brane UV assure alors une échelle de coupure suffisamment élevée pour supprimer les opérateurs de contact générant des FCNCs. On peut aussi utiliser la liberté sur la quasi-localisation des modes zéro des fermions pour générer la hiérarchie de masse du SM \cite{Gherghetta:2000qt, Huber:2000ie}. Pour une introduction à cette thématique, voir les Réfs.~\cite{Csaki:2005vy, Gherghetta:2006ha, Gherghetta:2010cj, Ponton:2012bi}.

\section{Bosons de jauge et fermions dans le bulk}
Nous allons ici décrire, en théorie des champs, des bosons de jauge et des fermions se propageant dans le bulk du modèle de RS1. On utilisera la description de la dimension spatiale supplémentaire sur un intervalle $I=[0,L]$ \cite{Csaki:2003dt, Csaki:2003sh}, équivalente à celle sur l'orbifold $S^1/\mathbb{Z}_2$. On suppose que ces champs ont une rétroaction gravitationnelle négligeable sur la géométrie. La métrique est alors :
\begin{equation}
ds^2 = \text{e}^{-2ky} \eta_{\mu \nu} dx^\mu dx^\nu  - dy^2.
\label{RS1_met}
\end{equation}

\subsection{Champs de jauge}
Considérons un champ de jauge $A_M$ dans le bulk du modèle de RS1 \cite{Davoudiasl:1999tf, Pomarol:1999ad, Csaki:2003dt}. Il a une dimension de masse $[A_M] = 1$. On va chercher à décrire le champ libre, et les interactions sont traitées en théorie des perturbations comme usuellement en QFT. On peut alors se concentrer sur une théorie de jauge abélienne, puisque le caractère non-abélien se traduit par une auto-interaction des bosons de jauge. L'action s'écrit
\begin{equation}
S_A = - \dfrac{1}{4 g_5^2} \int d^4x \int_0^L dy \, \sqrt{|g|} \, g^{MP} g^{NS} F_{MN} F_{PS},
\label{action_A}
\end{equation}
où $g_5$ est le couplage de jauge à 5D et $F_{MN} = \partial_M A_N - \partial_N A_M$. On travaille dans la jauge où $A_4=0$ et $\partial_\mu A^\mu = 0$. L'application du principe de Hamilton requiert l'annulation séparée des variations de l'action dans le bulk et sur chaque 3-brane. On obtient ainsi les équations d'Euler-Lagrange dans le bulk :
\begin{equation}
\partial_\mu F^{\mu \nu} - \partial_y \left( \text{e}^{-2ky} \partial_y A^\nu \right) = 0,
\label{eq_5D_A}
\end{equation}
et les conditions aux bords :
\begin{equation}
\left. \delta A^\mu \; \partial_y A_\mu \right|_{y=0,L} = 0.
\end{equation}
Ces dernières sont satisfaites si l'on suppose les champs arbitraires sur les 3-branes et si on les y laisse varier : $\left. \delta A^\mu \right|_{y=0,L} \neq 0$. On parle de conditions aux bords naturelles, qui imposent ainsi des conditions aux bords de Neumann (N) pour le champ : $\left. \partial_y A_\mu \right|_{y=0,L} = 0$.

Pour construire l'EFT à 4D, on effectue une décomposition de KK :
\begin{equation}
A_\mu (x^\mu, y) = \dfrac{1}{\sqrt{L}} \sum_n A_\mu^n (x^\mu) f_A^n(y),
\label{KK_A}
\end{equation}
où les $A_\mu^n (x^\mu)$ sont les modes de KK de masse $m_A^n$ satisfaisant l'équation de Proca :
\begin{equation}
\partial^\mu F_{\mu \nu} + (m_A^n)^2 A_\nu^n = 0.
\label{eq_Proca}
\end{equation}
Les fonctions d'onde $f_A^n(y)$ des modes de KK sont orthonormalisées de manière à avoir des termes cinétiques canoniques dans le lagrangien à 4D :
\begin{equation}
\dfrac{1}{L} \int_0^L dy \; f_A^n(y) f_A^m(y) = \delta^{nm}.
\label{orthonorm_A}
\end{equation}
En insérant la décomposition \eqref{KK_A} et l'\'Eq.~\eqref{eq_Proca} dans l'\'Eq.~\eqref{eq_5D_A}, on obtient des équations de Sturm-Liouville pour les fonctions d'onde de KK :
\begin{equation}
\partial_y \left( \text{e}^{-2ky} \partial_y f_A^n \right) + \left( m_A^n \right)^2 f_A^n = 0.
\end{equation}

Pour le mode zéro, $m_A^0 = 0$, la solution générale est donnée par
\begin{equation}
f_A^0(y) = A + B \text{e}^{2ky}
\end{equation}
où $A$ et $B$ sont des constantes à déterminer. Pour des conditions aux bords de Neumann sur chaque 3-branes (NN), $A = 1$ et $B = 0$, la solution est donc
\begin{equation}
f^0_A(y) = 1.
\label{profil_0_A}
\end{equation}
Le mode zéro d'un boson de jauge sans masse a une fonction d'onde de KK constante le long de la dimension spatiale supplémentaire, il n'est donc pas localisé dans le bulk. En insérant la décomposition de KK \eqref{KK_A} et la fonction d'onde de KK du mode zéro \eqref{profil_0_A} dans l'action \eqref{action_A}, on peut identifier le couplage de jauge à 4D :
\begin{equation}
g_4 = \dfrac{g_5}{\sqrt{L}}.
\label{couplage_jauge}
\end{equation}

Pour les modes de KK avec $m_A^n \neq 0$, la solution générale pour les fonctions d'onde de KK est donnée par
\begin{equation}
f_A^n(y) = N_A^n \; \text{e}^{ky} \left[ J_1 \left( \dfrac{m_A^n}{k \text{e}^{-ky}} \right) + b_A^n \; Y_1 \left( \dfrac{m_A^n}{k \text{e}^{-ky}} \right) \right],
\end{equation}
où $b_A^n$, $N_A^n$ et $m_A^n$ sont des constantes que l'on peut déterminer en imposant les conditions aux bords (NN) et en utilisant les conditions d'orthonormalisation \eqref{orthonorm_A}. On peut alors aussi dériver le spectre de masse :
\begin{equation}
\forall n \in \mathbb{N}^*, \; m_A^n = x_n \; m_{KK},
\end{equation}
avec l'échelle de KK, $m_{KK} = k \text{e}^{-kL}$, et
\begin{equation}
x_j \simeq \left\{
\begin{array}{l c l}
2,45 & & n = 1 \, , \\
5,56 & & n = 2 \, , \\
8,70 & & n = 3 \, , \\
11,83 & & n = 4 \, , \\
& \vdots &
\end{array}
\right.
\end{equation}
On peut approximer le spectre de masse avec ($kL \gg 1$)
\begin{equation}
x_j \simeq \left( n - \dfrac{1}{4} \right) \pi.
\end{equation}
L'étude des fonctions d'onde des modes de KK montre que ces derniers sont quasi-localisés près de la brane IR.

\subsection{Fermions de Dirac}
Intéressons nous maintenant au cas d'un fermion libre de spin 1/2, $\Psi$, dans le bulk du modèle de RS1 \cite{Grossman:1999ra, Gherghetta:2000qt}. Il a une dimension de masse $[\Psi] = 2$. L'action s'écrit :
\begin{equation}
S_\Psi = \int d^4x \int_0^L dy \, \sqrt{|g|} \left[ \dfrac{i}{2} \left( \bar{\Psi} e_A^M \Gamma^A \nabla_M \Psi - \bar{\nabla_M \Psi} e_A^M \Gamma^A \Psi \right) - ck \; \bar{\Psi} \Psi \right]
\label{L_Psi}
\end{equation}
où $\bar{\Psi} = \Psi^\dagger \Gamma^0$ et $\bar{\nabla_M \Psi} = (\nabla_M \Psi)^\dagger \Gamma^0$. $\Gamma^M = (\gamma^\mu, i \gamma^5)$ sont les matrices de Dirac à 5D. $c$ paramétrise la masse du champ fermionique dans le bulk, il est naturellement d'ordre $\mathcal{O}(1)$. Pour l'espace AdS$_5$, on a le fünfbein $e_A^M = (\text{e}^{ky} \delta^\mu_\alpha, 1)$, et la connexion de spin :
\begin{equation}
\omega_M = \left( \dfrac{i}{2} k \text{e}^{-ky} \gamma_\mu \gamma^5, 0 \right).
\end{equation}
La dérivée covariante est définie par $\nabla_M = \partial_M + \omega_M$. On notera que la contribution de la connexion de spin s'annule dans l'action \eqref{L_Psi}, on peut alors faire la substitution $\nabla_M \rightarrow \partial_M$. En prévision de la réduction dimensionnelle à 4D, on décompose le champ $\Psi$ en spineurs de Weyl:
\begin{equation}
\Psi =
\begin{pmatrix}
\psi_L \\
\psi_R
\end{pmatrix}.
\end{equation}
On peut appliquer le principe de Hamilton, où on annule toujours séparément les variations de l'action dans le bulk et sur les bords. On obtient ainsi les équations d'Euler-Lagrange pour les champs $\psi_L$ et $\psi_R$:
\begin{align}
\text{e}^{ky} i \bar{\sigma}^\mu \partial_\mu \psi_L - \partial_y \psi_R + (2-c)k \; \psi_R &= 0, \nonumber \\
\text{e}^{ky} i \sigma^\mu \partial_\mu \psi_R + \partial_y \psi_L - (2+c)k \; \psi_L &= 0,
\end{align}
et les conditions aux bords :
\begin{equation}
\left. \delta \psi_L^\dagger \psi_R \right|_{y=0,L} = 0 \ \ \ \text{et} \ \ \ \left. \delta \psi_R^\dagger \psi_L \right|_{y=0,L} = 0.
\label{eq_Psi_5D}
\end{equation}
La méthode habituelle, pour satisfaire ces conditions aux bords sur l'une des 3-branes, est d'y imposer une condition de Dirichlet (D) pour l'un des champs $\psi_C$, avec $C \equiv L \ \text{ou} \ R$, ce qui implique $\delta \psi_C = 0$.

L'étape suivante de la réduction dimensionnelle est d'effectuer la décomposition de KK:
\begin{equation}
\psi_C (x^\mu, y) = \dfrac{\text{e}^{(3/2)ky}}{\sqrt{L}} \sum_n \psi_C^n(x^\mu) f_C^n(y),
\label{KK_Psi}
\end{equation}
où les $\psi_C^n(x^\mu)$ sont les modes de KK de masse $m_\Psi^n$ satisfaisant les équations de Dirac-Weyl :
\begin{align}
i \bar{\sigma}^\mu \partial_\mu \psi_L^n (x^\mu) &= m_\Psi^n \; \psi_R^n(x^\mu), \nonumber \\
i \sigma^\mu \partial_\mu \psi_R^n (x^\mu) &= m_\Psi^n \; \psi_L^n(x^\mu).
\label{eq_Dirac}
\end{align}
Les fonctions d'onde $f_C^n(y)$ des modes de KK sont orthonormalisées, tel que :
\begin{equation}
\dfrac{1}{L} \int_0^L dy \; f_C^n(y) f_C^m(y) = \delta^{nm}.
\label{orthonorm_Psi}
\end{equation}
Le facteur $\text{e}^{(3/2)ky}$ dans la décomposition de KK \eqref{KK_Psi} permet d'avoir une normalisation canonique des termes cinétiques pour les modes de KK, ainsi qu'une interprétation physique des fonctions d'onde $f_C^n(y)$ en terme de localisation des KK-fermions dans la dimension spatiale supplémentaire. En utilisant les \'Eqs.~\eqref{eq_Dirac} et \eqref{KK_Psi} dans l'\'Eq.~\eqref{eq_Psi_5D}, on trouve les équations pour les fonctions d'onde de KK :
\begin{align}
\partial_y f_L^n + m_\Psi^n \text{e}^{ky} f_R^n - k \left( \dfrac{1}{2} + c \right) f_L^n &= 0, \nonumber \\
\partial_y f_R^n - m_\Psi^n \text{e}^{ky} f_L^n - k \left( \dfrac{1}{2} - c \right) f_R^n &= 0.
\label{eq_profils_Psi}
\end{align}

\begin{figure}[h]
\begin{center}
\includegraphics[height=6cm]{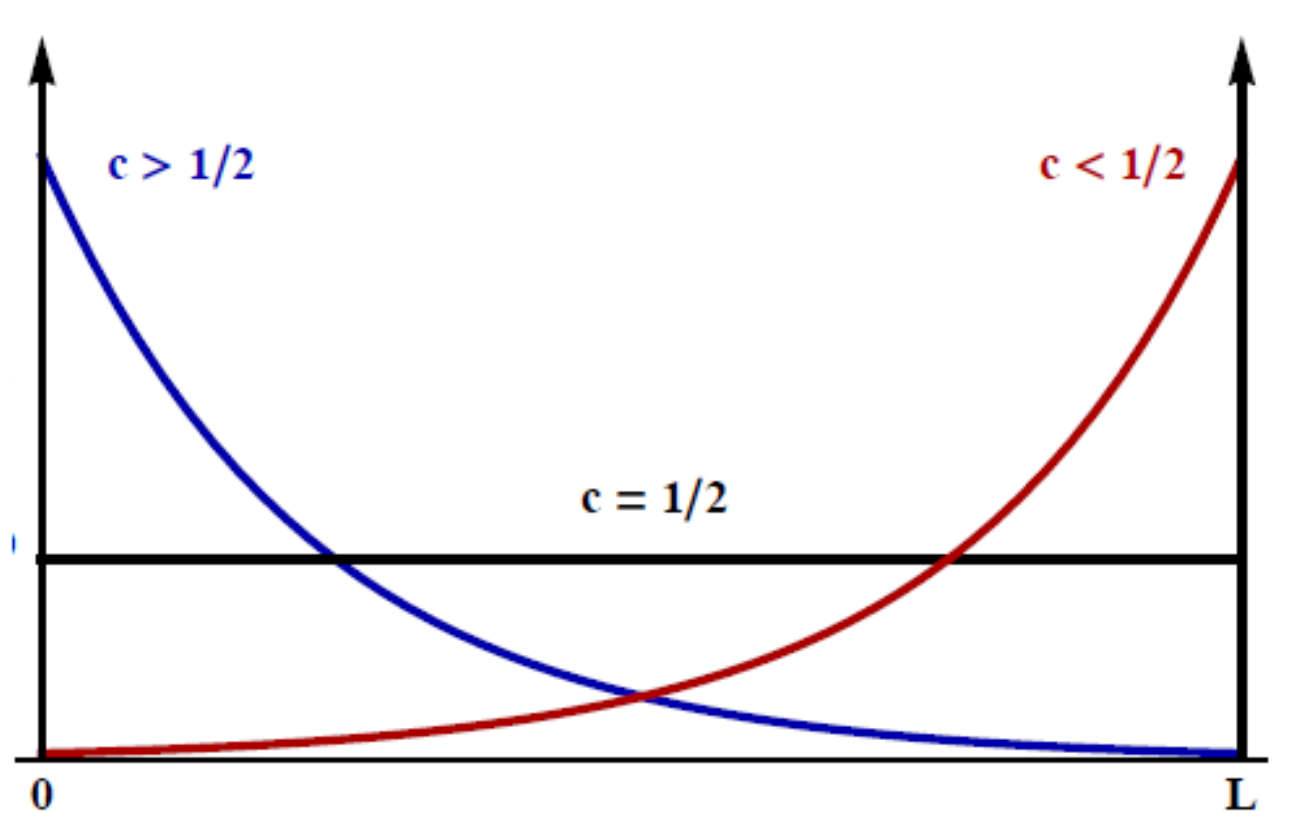}
\end{center}
\caption[Quasi-localisation d'un mode zéro fermionique]{(Adaptée de la Réf.~\cite{Ponton:2012bi}.) Schéma de la quasi-localisation d'un mode zéro de chiralité droite, en fonction du paramètre $c$.}
\label{fermion_local}
\end{figure}

Pour le mode zéro, $m_0 = 0$, les équations pour les fonctions d'onde de KK $f_L^n$ et $f_R^n$ sont découplées. La solution générale est donnée par
\begin{align}
f_L^0 (y) &= A_L \; \exp \left( \left[ \dfrac{1}{2} + c \right] ky \right), \nonumber \\
f_R^0 (y) &= A_R \; \exp \left( \left[ \dfrac{1}{2} - c \right] ky \right),
\label{profil_0_Psi}
\end{align}
où $A_L$ et $A_R$ sont des constantes à déterminer. Imposer une condition de Dirichlet sur $\psi_L$ ou $\psi_R$, à l'une des deux extrémités de l'intervalle, se transpose sur les fonctions d'onde après décomposition de KK \eqref{KK_Psi}. Avec la forme des fonctions d'onde de KK \eqref{profil_0_Psi}, une condition (D) pour $\psi_L$ ($\psi_R$) revient à avoir $f_L^0(y) = 0$ ($f_R^0(y) = 0$) et donc à éliminer le mode zéro correspondant. Un champ 5D $\Psi$ auquel on impose des conditions aux bords de Dirichlet à chaque bord (DD) sur $\psi_L$ ($\psi_R$) aura donc un seul mode zéro $\psi_R^0$ ($\psi_L^0$) : la théorie est chirale au niveau des modes zéro. Le coefficient de normalisation de sa fonction d'onde de KK est donné par les \'Eqs.~\eqref{orthonorm_Psi} et \eqref{profil_0_Psi}:
\begin{equation}
A_L = \sqrt{\dfrac{kL (1+2c)}{\text{e}^{kL (1+2c)} -1}}, \ \ \ 
A_R = \sqrt{\dfrac{kL (1-2c)}{\text{e}^{kL (1-2c)} -1}}.
\label{norm_coeff_Psi}
\end{equation}
D'après les \'Eqs.~\eqref{profil_0_Psi}, un mode zéro de chiralité gauche (droite) est quasi-localisé près de la brane UV si $c < -1/2$ ($c > 1/2$) ou près de la brane IR si $c > -1/2$ ($c < 1/2$). Il n'est pas localisé dans le bulk dans le cas particulier où $c=-1/2$ ($c=1/2$). La quasi-localisation du mode zéro est illustrée sur la Fig.~\ref{fermion_local}.

Pour les modes de KK avec $m_\Psi^n \neq 0$, les équations des fonctions d'onde \eqref{eq_profils_Psi} peuvent être découplées en les dérivant. On obtient alors des équations de Sturm-Liouville dont les solutions ont la forme :
\begin{align}
f_L^n (y) &= N_L^n \; \text{e}^{ky} \left[ J_{c - (1/2)} \left( \dfrac{m_\Psi^n}{k \text{e}^{-ky}} \right) + b_L^n \; Y_{c - (1/2)} \left( \dfrac{m_\Psi^n}{k \text{e}^{-ky}} \right) \right], \nonumber \\
f_R^n (y) &= N_R^n \; \text{e}^{ky} \left[ J_{c + (1/2)} \left( \dfrac{m_\Psi^n}{k \text{e}^{-ky}} \right) + b_R^n \; Y_{c + (1/2)} \left( \dfrac{m_\Psi^n}{k \text{e}^{-ky}} \right) \right],
\end{align}
où $b_C^n$, $N_C^n$ et $m_\Psi^n$ sont des constantes à déterminer en faisant un choix de conditions aux bords et en utilisant les conditions d'orthonormalisation \eqref{orthonorm_Psi}. En se focalisant sur les conditions aux bords induisant l'existence d'un mode zéro identifiable avec l'une des deux chiralités d'un fermion du SM, le spectre de masse peut être approximé par ($kL \gg 1$) :
\begin{equation}
\forall n \in \mathbb{N}^*, \; m_\Psi^n \simeq \left( n - \dfrac{1}{4} + \dfrac{|\alpha|}{2} \right) \pi \; m_{KK},
\end{equation}
où $\alpha = c - \dfrac{1}{2}$ $\left( \alpha = c + \dfrac{1}{2} \right)$ pour $\psi_L$ ($\psi_R$) ayant des conditions aux bords (DD). Comme pour les bosons de jauge, les modes de KK sont quasi-localisés près de la brane IR.

\section{Le Modèle Standard dans le bulk}
\begin{figure}[h]
\begin{center}
\includegraphics[height=6cm]{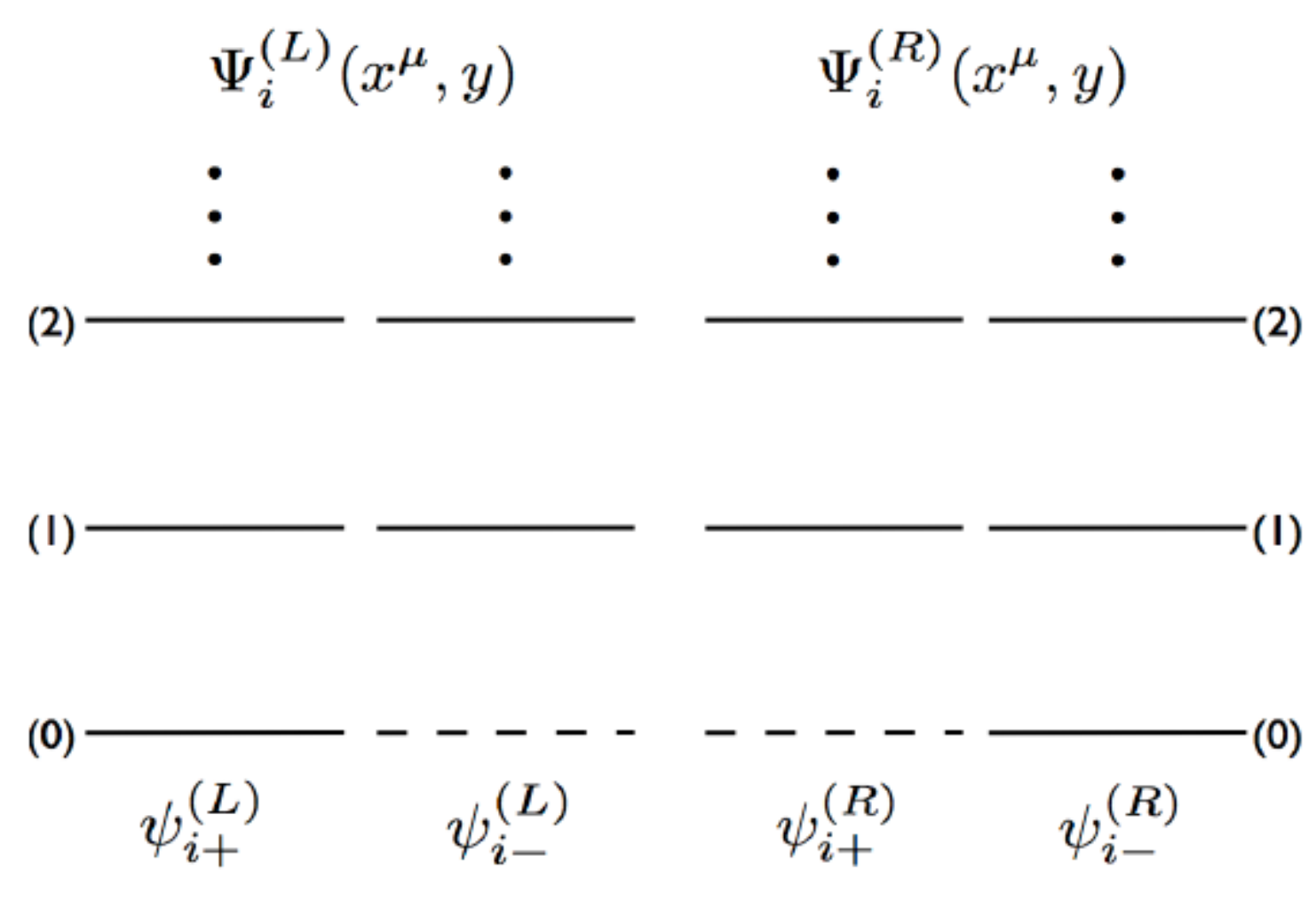}
\end{center}
\caption[Identification des spineurs de Weyl du SM avec les modes zéro des spineurs de Dirac 5D]{(Adaptée de la Réf.~\cite{Gherghetta:2010cj}.) Identification des spineurs de Weyl du SM avec les modes zéro des spineurs de Dirac 5D. Les lignes en pointillés indiquent l'absence d'un mode zéro.}
\label{fermion_KK_tower}
\end{figure}

Dans la suite de ce chapitre, on va considérer une tranche AdS$_5$, où le champ de Higgs est localisé sur la brane IR, et où les champs fermioniques et de jauge du SM sont promus à des champs 5D. Les bosons de jauge observés sont alors identifiés avec les modes zéro de champs de jauge 5D avec des conditions aux bords (NN). Le secteur EW du SM est une théorie chirale : les fermions de chiralité gauche et droite se transforment différemment sous le groupe de jauge. Or, on a vu que le mode zéro d'un spineur de Dirac 5D est de chiralité gauche ou droite. Ainsi, pour chaque spineur de Weyl $\psi_i$ du SM, on introduit un spineur de Dirac 5D $\Psi_i$ dans la même représentation du groupe de jauge (où $i$ est l'index de saveur). Les conditions aux bords sont alors choisies pour que le spineur de Weyl de chiralité gauche $\psi_{i,L}$ (droite $\psi_{i,R}$) soit identifié avec le mode zéro du spineur dans le bulk $\Psi_i^{(L)}$ ($\Psi_i^{(R)}$), \textit{c.f.} la Fig.~\ref{fermion_KK_tower}.

\subsection{Hiérarchie des couplages de Yukawa}
L'action des couplages de Yukawa, entre les fermions dans le bulk et le champ de Higgs localisé sur la brane IR, s'écrit :
\begin{align}
\int d^4x \int_0^L dy \; \sqrt{|g|} \; \delta(y-L) \; \dfrac{Y_{ij}}{k} \; \bar{\Psi}_i^{(L)} (x^\mu, y) H(x^\mu) \Psi_j^{(R)} (x^\mu, y) + \text{c.h.} \nonumber \\
= \int d^4x \; y_{ij} \; \bar{\widetilde{\psi}}_{i,L}^{(0)}(x^\mu) \widetilde{H}(x^\mu) \widetilde{\psi}_{j,R}^{(0)}(x^\mu) + \text{c.h.} + \text{couplages des KK-particules},
\end{align}
où $Y_{ij}$ est le couplage de Yukawa à 5D adimensionné car on a introduit l'échelle naturelle du modèle $1/k$, et $y_{ij}$ est le couplage de Yukawa effectif 4D. Pour passer de la première à la deuxième ligne, on a effectué une décomposition de KK \eqref{KK_Psi} dans l'action et on a renormalisé le Higgs (\textit{c.f.} l'\'Eq.~\eqref{renorm_1}), $y_{ij}$ est alors défini par :
\begin{equation}
y_{ij} = \int_0^L dy \; \delta(y-L) \; \sqrt{|g|} \; \dfrac{\text{e}^{k(3y+L)}}{L} f_L^0(y) f_R^0(y).
\label{Yuk_4D}
\end{equation}

Avec des couplages à 5D et des paramètres de masse dans le bulk naturels, $Y_{ij} \simeq 1$ et $c_i \sim \mathcal{O}(1)$, on va voir que l'on peut générer une hiérarchie dans les couplages à 4D \cite{Gherghetta:2000qt, Huber:2000ie} sans invoquer de symétries de saveur. Prenons un modèle simplifié où $c_{i,R} = - c_{i,L} \equiv c_i$, sans mélange de saveur. Les \'Eqs.~\eqref{profil_0_Psi}, \eqref{norm_coeff_Psi} et \eqref{Yuk_4D} donnent :
\begin{equation}
y_{ii} = Y_{ii} \; \dfrac{1-2c_i}{\text{e}^{kL(1-2c_i)} -1} \text{e}^{kL(1-2c_i)}.
\end{equation}
Quand $c_i > 1/2$, le couplage de Yukawa à 4D dépend exponentiellement du paramètre de masse dans le bulk $c_i$, on a approximativement ($kL \gg 1$) :
\begin{equation}
y_{ii} \simeq Y_{ii} \left( 2 c_i - 1 \right) \text{e}^{kL (1-2c_i)},
\label{y_light_ferm}
\end{equation}
consistant avec des fermions quasi-localisés près de la brane UV, dont les fonctions d'onde de KK ont un faible recouvrement avec la brane IR où est localisé le champ de Higgs. On peut alors reproduire, par exemple, le couplage de Yukawa de l'électron dans le SM, $y_e \sim 10^{-6}$, avec $c_e \simeq 0,68$. À l'inverse, quand $c_i < 1/2$, les fermions sont quasi-localisés près de la brane IR sans suppression exponentielle :
\begin{equation}
y_{ii} \simeq Y_{ii} \left( 1 - 2 c_i \right).
\end{equation}
Dans ce cas, le couplage de Yukawa du quark top dans le SM, $y_t \sim 1$, est obtenu pour $c_t \simeq 0$. Les couplages des autres fermions sont générés pour $c_t \lesssim c_i \lesssim c_e$. On donne une illustration de la quasi-localisation des modes zéro des champs fermioniques à la Fig.~\ref{bulk_RS}.

Il est important de noter que les couplages de Yukawa, entre les fermions dans le bulk et le champ de Higgs, induisent un mélange des modes zéro avec les modes excités : la base de KK dans la décomposition \eqref{KK_A} n'est pas celle des états propres de masse après EWSB. N'utiliser que les modes zéro, pour obtenir les masses des fermions du SM, reste néanmoins une bonne approximation à l'ordre $v^2/m_{KK}^2$. Les Réfs.~\cite{Huber:2003tu, Casagrande:2008hr} donnent une analyse détaillée des paramètres $c_i$ où la hiérarchie de masse des fermions et la matrice CKM peuvent être naturellement expliquées par le recouvrement des fonctions d'onde de KK.

\subsection{Suppression des opérateurs de contact induisant des FCNCs}
Dans le modèle de RS1 originel, que l'on a décrit à la Section~\ref{Modele_RS1}, les fermions sont localisés, comme le champ de Higgs, sur la brane IR. Comme expliqué à la Section~\ref{Problemes_quantique_TeV}, des opérateurs de contact, potentiellement générés par des effets gravitationnels, sont affectés par le facteur de courbure sur la brane IR lors de la renormalisation des champs des \'Eqs.~\eqref{renorm_1}, \eqref{renorm_2} et \eqref{renorm_3}. Ainsi, les opérateurs, induisant la désintégration du proton et des FCNCs, ne sont supprimés que par l'échelle de coupure sur la brane :
\begin{align}
\int d^4x \int_0^L dy \; \delta(y-L) \; \sqrt{|g|} \; \dfrac{1}{\Lambda_5^2} \bar{\psi}_i(x^\mu) \psi_j(x^\mu) \bar{\psi}_k(x^\mu) \psi_l(x^\mu) \nonumber \\
= \int d^4x \; \dfrac{1}{(\Lambda_5 \text{e}^{-kL})^2} \bar{\widetilde{\psi}}_i(x^\mu) \widetilde{\psi}_j(x^\mu) \bar{\widetilde{\psi}}_k(x^\mu) \widetilde{\psi}_l(x^\mu),
\end{align}
où $\Lambda_5$ est l'échelle à laquelle la gravité devient fortement couplée sur la brane UV. De la même manière, pour l'opérateur générant des masses pour les neutrinos, on a
\begin{align}
\int d^4x \int_0^L dy \; \delta(y-L) \; \sqrt{|g|} \; \dfrac{1}{\Lambda_5} \psi^T_{L,i}(x^\mu) C_5 \psi_{L,i}(x^\mu) H(x^\mu) H(x^\mu) \nonumber \\
= \int d^4x \; \dfrac{1}{\Lambda_5 \text{e}^{-kL}} \widetilde{\psi}^T_{L,i}(x^\mu) C_5 \widetilde{\psi}_{L,i}(x^\mu) \widetilde{H}(x^\mu) \widetilde{H}(x^\mu).
\end{align}
Les contraintes expérimentales de l'\'Eq.~\eqref{suppress_op} sur les échelles de suppression de ces opérateurs excluent le modèle de RS1 comme solution au problème de hiérarchie de jauge, si les coefficients de ces opérateurs ne sont pas naturellement très supprimés dans la théorie UV de la gravité. 

Pour éviter de jauger les symétries de saveur dans le bulk, comme évoqué à la Section~\ref{Problemes_quantique_TeV}, on autorise les fermions et les bosons de jauge à se propager dans la dimension spatiale supplémentaire \cite{Gherghetta:2000qt, Huber:2000ie}. Cette fois, les opérateurs de contact, générant les FCNCs et la désintégration du proton, sont de la forme :
\begin{align}
\int d^4x \int_0^L dy \; \sqrt{|g|} \; \dfrac{1}{\Lambda_5^3} \bar{\Psi}_i(x^\mu,y) \Psi_j(x^\mu,y) \bar{\Psi}_k(x^\mu,y) \Psi_l(x^\mu,y) \nonumber \\
= \int d^4x \; \dfrac{1}{\Lambda_4^2} \bar{\psi}_i^0(x^\mu) \psi_j^0(x^\mu) \bar{\psi}_k^0(x^\mu) \psi_l^0(x^\mu) + \cdots,
\end{align}
où on a utilisé la décomposition de KK \eqref{KK_Psi}. En utilisant les Eqs.~\eqref{profil_0_Psi} et \eqref{norm_coeff_Psi}, l'échelle effective de suppression à 4D est donnée par :
\begin{equation}
\dfrac{1}{\Lambda_4^2} = \dfrac{k}{\Lambda_5^3} \; N_i N_j N_k N_l \; \dfrac{\text{e}^{(4 - c_i - c_j - c_k - c_l)kL} - 1}{4 - c_i - c_j - c_k - c_l} \ \ \ \text{avec} \ \ \ N_i = \sqrt{\dfrac{1-2c_i}{\text{e}^{kL (1-2c_i)} -1}}.
\end{equation}
Pour des valeurs $1/2 \lesssim c_i \lesssim 1$, $1 \ \text{TeV} \lesssim \Lambda_4 \lesssim M_P$. On peut vérifier que les opérateurs générant des FCNCs sont suffisamment supprimés pour être en accord avec les contraintes expérimentales de l'\'Eq.~\eqref{suppress_op}, tout en générant la hiérarchie de masse des fermions. Pour la troisième génération de quarks, on s'attend, par contre, à observer des déviations par rapport aux prédictions du SM, car les processus de FCNCs sont moins supprimés. Quant à l'opérateur générant la désintégration du proton, il nécessite $c_i \gtrsim 1$. Les masses des fermions sont alors trop faibles pour être compatibles avec le SM. Pour finir, intéressons nous à l'opérateur de contact générant des masses de Majorana pour les neutrinos :
\begin{align}
\int d^4x \int_0^L dy \; \delta(y-L) \sqrt{|g|} \; \dfrac{1}{\Lambda_5^2} \psi^T_{L,i}(x^\mu,y) C_5 \psi_{L,i}(x^\mu,y) H(x^\mu) H(x^\mu) \nonumber \\
= \int d^4x \; \dfrac{1}{\Lambda_4} \psi^{0T}_{L,i}(x^\mu) C_5 \psi^0_{L,i}(x^\mu) H(x^\mu) H(x^\mu) + \cdots,
\end{align}
où
\begin{equation}
\dfrac{1}{\Lambda_4} = \dfrac{k}{\Lambda_5^2} \; N_i^2 \; \text{e}^{2 (1-c_i) kL}
\end{equation}
Ceci, combiné à l'\'Eq.~\eqref{y_light_ferm} donnant les couplages de Yukawa pour les doublets d'isospin gauche des leptons, donne une masse trop grande aux neutrinos (\'Eq.~\eqref{suppress_op}). La symétrie $U(1)_B \times U(1)_L$ doit donc quand même être jaugée afin d'interdire ces opérateurs de contact.

\subsection{Couplages de jauge}
Les modes zéro des fermions étant quasi-localisés différemment le long de la dimension spatiale supplémentaire, on pourrait penser que l'universalité des couplages des fermions aux bosons de jauge est perdue, excluant ainsi le modèle extra-dimensionnel. Considérons l'action du couplage entre un boson de jauge (associé à une symétrie $U(1)$ pour simplifier) et un champ fermionique se propageant dans le bulk :
\begin{align}
\int d^4x \int_0^L dy \; \sqrt{|g|} \; g_5 \; \bar{\Psi}_i(x^\nu, y) e_\alpha^\mu \Gamma^\alpha A_\mu (x^\nu,y) \Psi_i(x^\nu,y) \nonumber \\
= \sum_{C,n,m} \int d^4x \; g_4^{0nm} \bar{\psi}_{C,i}^n (x^\nu) \gamma^\mu A_\mu^0 (x^\nu) \psi_{C,i}^m (x^\nu) + \cdots,
\end{align}
où on a inséré les décompositions de KK \eqref{KK_A} et \eqref{KK_Psi}. Le couplage de jauge à 4D est alors défini par :
\begin{equation}
g_4^{0nm} = \dfrac{g_5}{L \sqrt{L}} \int_0^L dy \; f_A^0(y) f_{C,i}^n(y) f_{C,i}^m(y)
\end{equation}
La fonction d'onde de KK du boson de jauge étant constante le long de la dimension spatiale supplémentaire, on peut utiliser la relation d'orthonormalisation des fonctions d'onde de KK fermioniques \eqref{orthonorm_Psi} pour trouver que :
\begin{equation}
g_4^{0nm} = \dfrac{g_5}{\sqrt{L}} \delta^{nm},
\end{equation}
en accord avec l'\'Eq.~\eqref{couplage_jauge}. En conclusion, le couplage des KK-fermions aux bosons de jauge est universel, et le modèle est sauf.

Cependant, il est important de remarquer que ce résultat n'est valable que pour les couplages aux bosons de jauge non affectés par la VEV du Higgs sur la brane IR. Pour les bosons $W^\pm$ et $Z$, les conditions aux bords ne sont pas de simples conditions de Neumann, mais impliquent la VEV du Higgs. Les fonctions d'onde de KK des modes zéro sont déformées (par rapport au cas sans VEV) près de la brane IR, affectant les couplages aux fermions lourds (localisés près de la brane du Higgs) comme les quarks $t$ et $b$. On s'attend donc à avoir des déviations par rapport aux couplages du SM, induisant ainsi des contraintes sur le modèle (voir, par exemple, la Réf.~\cite{Casagrande:2008hr} pour une discussion détaillée).

\subsection{\'Echelle de coupure du modèle}
Il est important d'insister sur le fait qu'un modèle extra-dimensionnel doit être traité dans le cadre d'une EFT car, comme on l'a vu dans les paragraphes précédents, la plupart des couplages renormalisables du SM ont, lorsque les champs sont promus à se propager dans le bulk, une dimension de masse négative et sont donc non-renormalisables. On peut estimer l'échelle d'énergie à partir de laquelle un couplage non-renormalisable devient non-perturbatif par analyse dimensionnelle naïve \cite{Chacko:1999hg, Ponton:2012bi}. À 5D, un couplage de jauge $g_5$ et un couplage de Yukawa  $Y_5$ (entre un scalaire localisé sur une 3-brane et un fermion dans le bulk) ont une dimension de masse $-1$. Ils deviennent non-perturbatifs quand \cite{Ponton:2012bi}
\begin{equation}
g_5^2 = \dfrac{l_5}{N \Lambda}, \ \ \ Y_5 = \sqrt{\dfrac{N}{l_4}} \dfrac{l_5}{N \Lambda},
\end{equation}
où $l_4 = 16 \pi^2$ et $l_5 = 24 \pi^3$ sont respectivement les facteurs de boucle à 4D et 5D, et $N$ est le facteur de multiplicité d'états du vertex considéré.

\section{Principales contraintes sur l'échelle de Kaluza-Klein}
Comme nous venons de le voir, autoriser les fermions et les bosons de jauge à avoir accès à la dimension spatiale supplémentaire permet de construire un modèle de saveur robuste. Néanmoins, la présence d'une tour de KK, pour chaque fermion et boson de jauge, implique de nouvelles contraintes expérimentales sur l'échelle de Kaluza-Klein $m_{KK}$, et donc sur l'échelle de coupure sur la brane-IR puisque $\Lambda(L) \gtrsim 10 \; m_{KK}$.

\subsection{FCNCs}
\'Etant donné que les modes zéro des fermions sont localisés à différents endroits dans la dimension spatiale supplémentaire, leurs couplages aux KK-bosons de jauge dépend de la saveur. Ces couplages sont de la forme :
\begin{equation}
g_4^{n00} = \dfrac{g_5}{L \sqrt{L}} \int_0^L dy \; f_A^n(y) \left| f_{C,i}^0(y) \right|^2.
\label{couplage_KK_boson}
\end{equation}
Bien qu'ils soient diagonaux dans la base fermionique des états propres de jauge, le fait qu'ils dépendent du paramètre $c$ implique que des éléments non-diagonaux sont induits dans la base des états propres de masse. Des FCNCs, possiblement associés à une violation de $CP$, pourront alors être générés à l'arbre. Par exemple, un KK-gluon $G_\mu^\prime$ peut participer aux oscillations $K^0 - \bar{K}^0$ via le diagramme de Feynman \cite{Ponton:2012bi}:

\begin{figure}[h]
\begin{center}
\includegraphics[height=1.7cm]{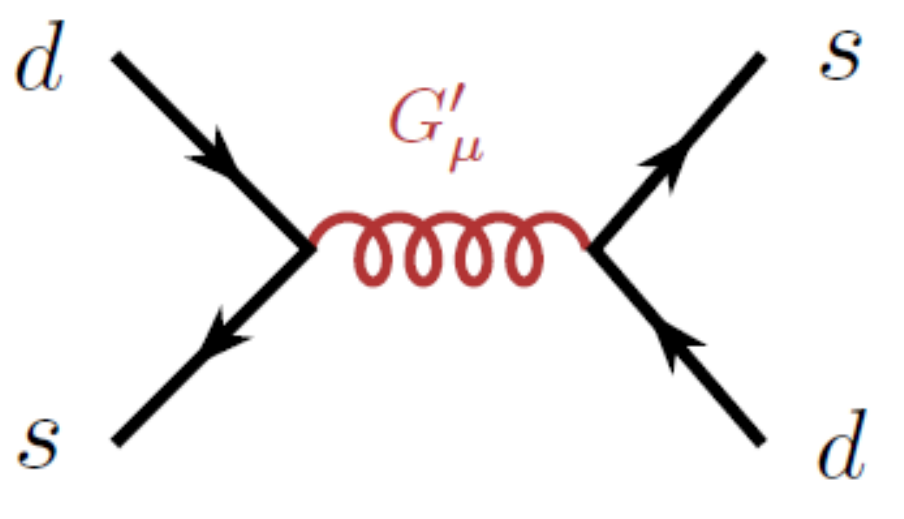}
\end{center}
\end{figure}

\begin{figure}[h]
\begin{center}
\includegraphics[height=6.5cm]{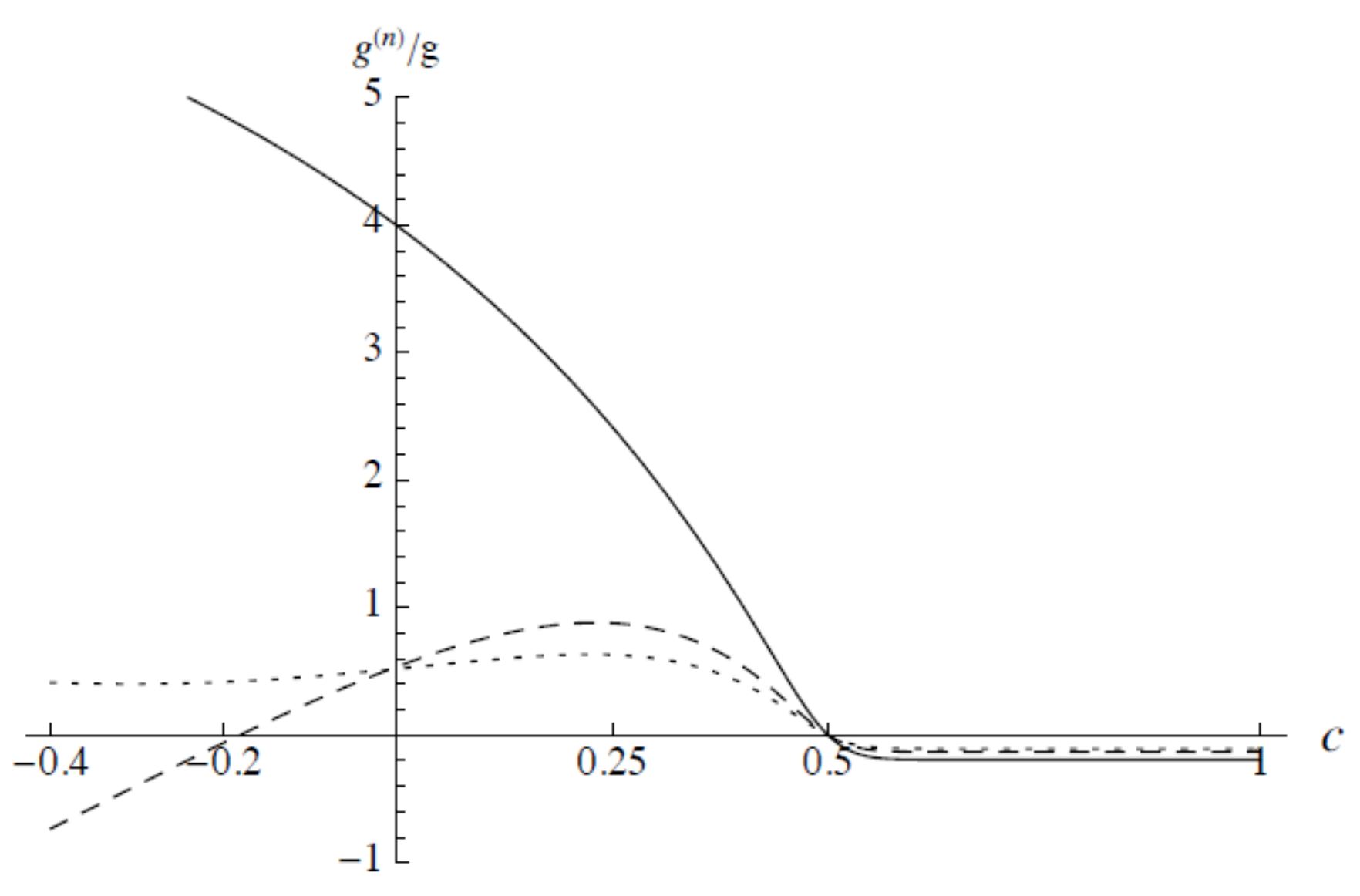}
\end{center}
\caption[Graphique du rapport des couplages $g^{(n)}/g$ en fonction de $c$]{(Adaptée de la Réf.~\cite{Gherghetta:2010cj}.) Graphique du rapport des couplages $g^{(n)}/g$ en fonction de $c$ pour $n=1$ (trait plein), $n=2$ (tirés) et $n=3$ (pointillés).}
\label{KK-gluon_coupling}
\end{figure}

Sur la Fig.~\ref{KK-gluon_coupling}, on a le graphique du rapport $g^{(n)}/g$, où $g^{(n)}$ est le couplage d'un fermion au n-ième KK-boson de jauge, et $g$ est le couplage du même fermion au mode zéro du boson de jauge. Quand $c$ est grand et négatif, les fermions sont quasi-localisés près de la brane IR et le rapport $g^{(n)}/g$ approche une valeur asymptotique $\sqrt{kL}$ correspondant à un fermion localisé sur la brane IR. Pour $c>1/2$, les fermions sont quasi-localisés près de la brane UV, et le couplage devient rapidement universel pour toutes les saveurs de fermions. Ceci est dû au fait que la fonction d'onde des KK-bosons de jauge est presque plate, sauf près de la brane IR. Ainsi dans l'\'Eq.~\eqref{couplage_KK_boson}, on a :
\begin{equation}
\int_0^L dy \; f_A^n(y) \left| f_{C,i}^0(y) \right|^2 \simeq f_A^n(0) \int_0^L dy \left| f_{C,i}^0(y) \right|^2 = f_A^n(0),
\end{equation}
en utilisant la condition de normalisation \eqref{orthonorm_Psi}. Pour un fermion quasi-localisé près de la brane UV, la dépendance en $c$ est donc très faible : les vertex de changement de saveur, impliquant un KK-boson de jauge, sont donc supprimés. Ce résultat est typique des modèles dont la métrique est proche de AdS$_5$ \cite{Gherghetta:2000qt, Agashe:2004ay, Agashe:2004cp}. Pour une métrique AdS$_5$, les Réfs.~\cite{Bauer:2009cf, Cabrer:2011qb} donnent une borne inférieure autour de $m_{KK} \gtrsim 4-8 \ \text{TeV}$. On peut abaisser ces contraintes en ajoutant de nouvelles symétries de saveur \cite{Santiago:2008vq, Bauer:2011ah}.

\subsection{Contraintes électrofaibles}
\subsubsection{Principe}
La présence d'une tour de KK, pour chaque fermion et boson de jauge, impacte aussi le calcul des observables dans le secteur EW, impliquant de nouvelles contraintes sur le modèle. Notamment, la relation $m_W^2 \simeq m_Z^2 \cos^2 \theta_w$, en très bon accord avec les mesures expérimentales, suggère l'existence d'une symétrie carcérale.

Dans le SM, si l'on met les couplages d'hypercharge $g_y$ et de Yukawa à zéro, on peut montrer l'existence d'une symétrie globale :
\begin{equation}
SU(2)_L^{jauge} \times SU(2)_R^{globale} \underset{EWSB}{\longrightarrow} SU(2)^{globale}_{diag} \equiv SU(2)_{carcérale}.
\end{equation}
En pratique, on peut réécrire le lagrangien du SM avec un champ de Higgs sous la forme d'un bidoublet de $SU(2)_L \times SU(2)_R$ :
\begin{equation}
H =
\begin{pmatrix}
H^{0*} & H^+ \\
-H^- & H^0
\end{pmatrix}.
\end{equation}
Sa VEV, $\bra{0} H \ket{0} = v \times I_2$, préserve le sous-groupe diagonal de $SU(2)_L^{jauge} \times SU(2)_R^{globale}$. Comme les trois bosons $W$ se transforment comme un triplet (singulet) sous $SU(2)_L^{jauge}$ \\ ($ SU(2)_R^{globale}$), donc comme un triplet sous $SU(2)_{carcérale}$, les composantes chargées et neutres ont la même masse après EWSB. L'ajout de $g_y \neq 0$ induit un mélange entre $W_\mu^3$ et $B_\mu$, et ainsi une séparation en masse des bosons $Z$ et $W^\pm$ : $m_W^2 = m_Z^2 \cos^2 \theta_w$. Les effets de brisure de la symétrie carcérale par les couplages de Yukawa (dominés par celui du quark top) sont plus petits, puisqu'ils interviennent à l'ordre des boucles.

Dans notre modèle extra-dimensionnel, le lagrangien a la même forme que celui du SM, il y a donc aussi une symétrie carcérale pour $g_y$ et les couplages de Yukawa nuls. Elle est toujours brisée à l'arbre par $g_y$, et par les couplages de Yukawa à l'ordre des boucles. Cependant, les couplages des KK-bosons du champ de jauge $B_M$ au champ de Higgs sont augmentés d'un facteur approximatif $\sqrt{kL} \gg 1$ par rapport aux modes zéro. Même constat pour les couplages des KK-quarks top au champ de Higgs. La violation de la symétrie carcérale par les KK-particules va donc être plus importante que par les modes zéro, via les diagrammes de Feynman suivants, où les lignes internes correspondent aux propagateurs des KK-particules \cite{Ponton:2012bi} :

\begin{figure}[h]
\begin{center}
\includegraphics[height=2cm]{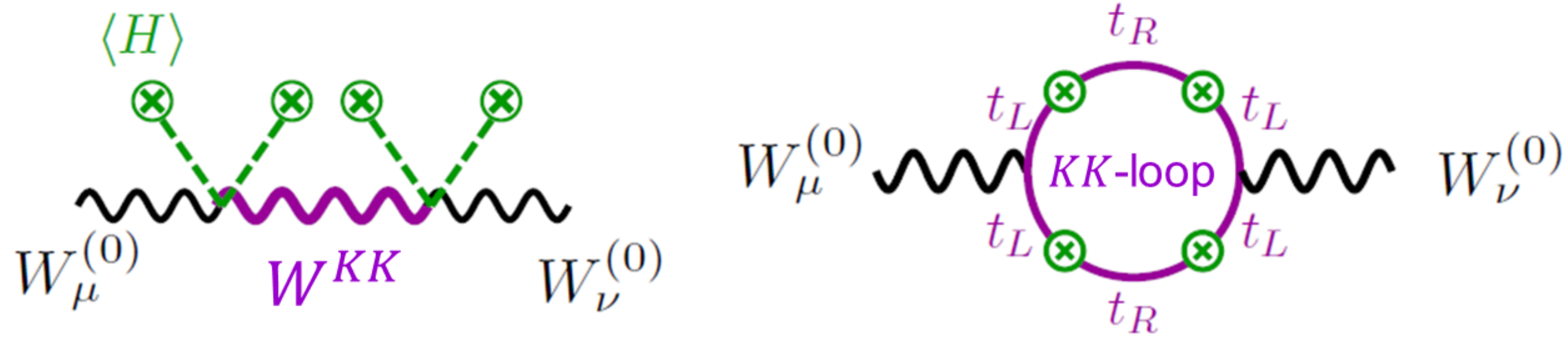}
\end{center}
\end{figure}

La borne inférieure actuelle sur l'échelle de KK, provenant des mesures de précision EW, est \cite{Fichet:2013ola} : $m_{KK} > 6 \ \text{TeV}$. Pour adoucir ces contraintes, une solution populaire est de jauger la symétrie carcérale dans le bulk et sur la brane IR, et de la briser sur la brane UV \cite{Agashe:2003zs}. Cependant, cela nécessite d'ajouter plusieurs champs au modèle, et de fortes contraintes ont été mises en évidence, provenant des mesures expérimentales des couplages du boson de Higgs \cite{Malm:2013jia, Malm:2014gha}. Nous allons présenter ci-dessous une autre solution plus économique.

\subsubsection{Termes cinétiques branaires}
On peut, en toute généralité, ajouter au modèle des termes cinétiques localisés sur les 3-branes (BLKTs -- \textit{Brane Localised Kinetic Terms}), pour les champs de jauge et les fermions dans le bulk, qui n'ont pas de conditions aux bords de Dirichlet sur la brane considérée :
\begin{align}
\mathcal{L}_{BLKT} = \dfrac{\delta(y)}{k} \left[ - \dfrac{1}{4} r_{UV} F_{\mu \nu} F^{\mu \nu} + i \alpha_{UV} \bar{\Psi}_C e^\mu_\alpha \Gamma^\alpha \partial_\mu \Psi_C \right] \nonumber \\
+ \dfrac{\delta(y-L)}{k} \left[ - \dfrac{1}{4} r_{IR} F_{\mu \nu} F^{\mu \nu} + i \alpha_{IR} \bar{\Psi}_C e^\mu_\alpha \Gamma^\alpha \partial_\mu \Psi_C \right].
\end{align}
où $\alpha_{UV, IR}$ et $r_{UV, IR}$ sont les coefficients des BLKTs. Ces termes, autorisés par les symétries du modèle, sont générés par les corrections radiatives \cite{Georgi:2000ks} mais peuvent être aussi présents à l'arbre dans l'EFT. En utilisant les décompositions de KK \eqref{KK_A} et \eqref{KK_Psi}, ils vont modifier les conditions aux bords pour les fonctions d'onde de KK et les conditions d'orthonormalisations \eqref{orthonorm_A} et \eqref{orthonorm_Psi} :
\begin{align}
\dfrac{1}{L} \left[ \int_0^L dy \; f_A^n(y) f_A^m(y) + \dfrac{r_{UV}}{k} f_A^n(0) f_A^m(0) + \dfrac{r_{IR}}{k} f_A^n(L) f_A^m(L) \right] &= \delta^{nm}, \\
\dfrac{1}{L} \left[ \int_0^L dy \; f_C^n(y) f_C^m(y) + \dfrac{\alpha_{UV}}{k} f_C^n(0) f_C^m(0) + \dfrac{\alpha_{IR}}{k} f_C^n(L) f_C^m(L) \right] &= \delta^{nm}.
\end{align}
La relation \eqref{couplage_jauge}, reliant le couplage de jauge effectif à 4D au couplage de jauge à 5D, est également modifiée :
\begin{equation}
\dfrac{1}{g_4^2} = \dfrac{1}{g_5^2} \left[ L + \dfrac{r_{UV}}{k} + \dfrac{r_{IR}}{k} \right].
\end{equation}
Le principal effet des BLKTs est de repousser les fonctions d'onde des KK-particules de la 3-brane où ils sont quasi-localisés. Si on ajoute des BLKTs sur la brane IR où est localisé le champ de Higgs, on va alors diminuer le recouvrement des fonctions d'onde de KK avec le champ de Higgs, et donc diminuer les couplages correspondants. Les contraintes, provenant des mesures de précision EW, sont alors adoucies \cite{Davoudiasl:2002ua, Carena:2002dz, Carena:2003fx, Carena:2004zn, Fichet:2013ola}. La Réf.~\cite{Fichet:2013ola} montre qu'une valeur de $m_{KK} \sim 3 \ \text{TeV}$ reste autorisée pour des coefficients de BLKTs naturels, \textit{i.e.} d'ordre $\mathcal{O}(1)$.

\subsection{Couplages à une boucle du boson de Higgs}
\begin{figure}[h]
\begin{center}
\includegraphics[height=6cm]{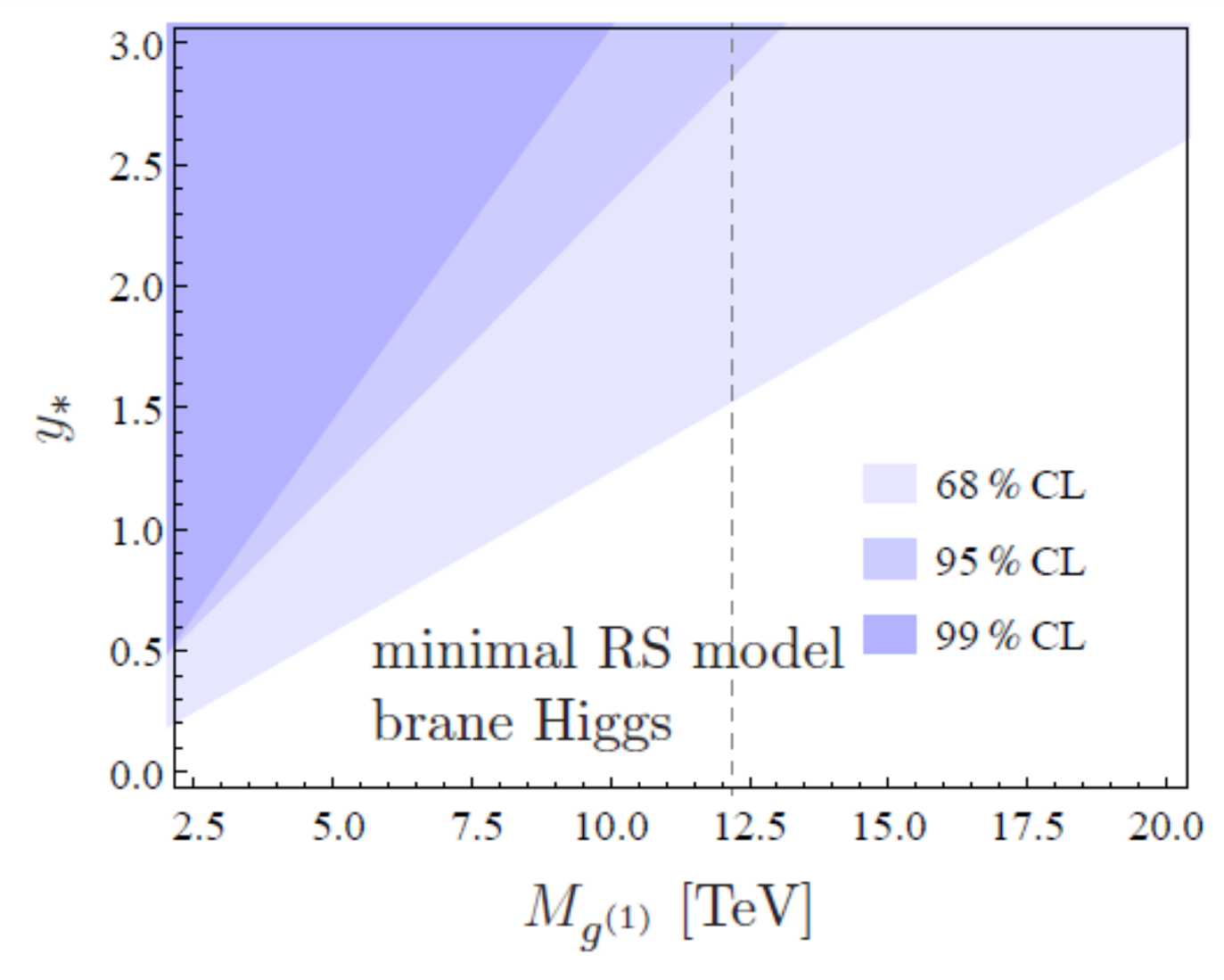}
\end{center}
\caption[Contraintes sur $m_{KK}$ provenant de la mesure de $ggh$]{(Adaptée de la Réf.~\cite{Malm:2013jia}.) Régions d'exclusion des paramètres du modèle en fonction de la masse du premier KK-gluon, $M_{g^{(1)}} = 2.45 \; m_{KK}$, et de $y_*$ défini tel que $\left| Y_{ij} \right| \leqslant y_*$. Les matrices de Yukawa à 5D ont été générées anarchiques et aléatoirement.}
\label{contraintes_ggh}
\end{figure}

Les couplages du boson de Higgs aux gluons et aux photons se font par l'intermédiaire d'une boucle où contribuent les KK-particules \cite{Lillie:2005pt, Djouadi:2007fm, Bhattacharyya:2009nb, Cacciapaglia:2009ky, Bouchart:2009vq, Casagrande:2010si, Carena:2012fk, Malm:2013jia, Hahn:2013nza, Malm:2014gha, Dey:2015pba}. Ainsi, on a le diagramme de Feynman du couplage $hgg$, avec la boucle dans laquelle circulent les quarks du SM et les KK-quarks : \\
\begin{figure}[h]
\begin{center}
\includegraphics[height=2cm]{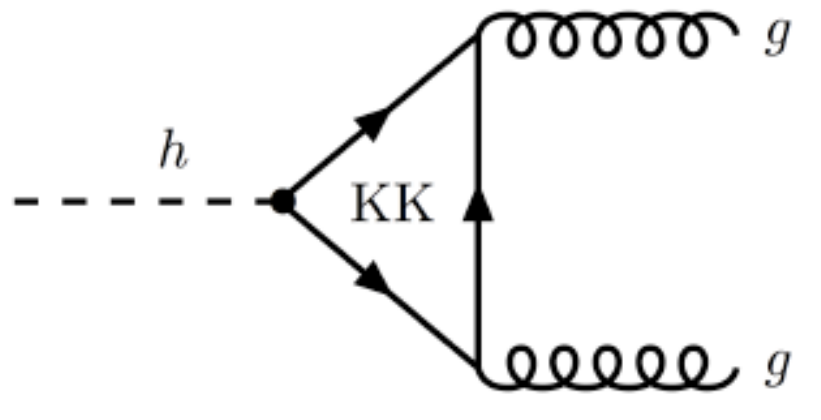}
\end{center}
\end{figure}\\
et les diagrammes du couplages $h \gamma \gamma$, avec les boucles de fermions, KK-fermions, bosons de jauge et KK-bosons de jauge :\\
\begin{figure}[h]
\begin{center}
\includegraphics[height=2cm]{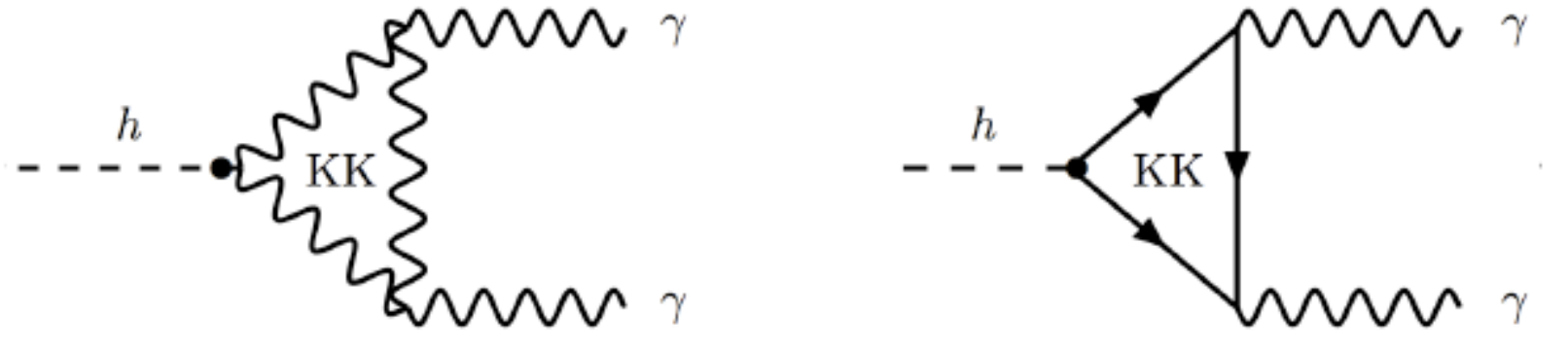}
\end{center}
\end{figure}\\
Les contraintes les plus sévères proviennent de la mesure du couplage $ggh$ \cite{Malm:2013jia}, et la borne inférieure sur $m_{KK}$ augmente avec le couplage de Yukawa $Y_{ij}$ à 5D, puisque les KK-quarks vont alors coupler plus fortement au champ de Higgs. La Fig.~\ref{contraintes_ggh} présente un graphique des régions d'exclusion du modèle en fonction de l'amplitude des $Y_{ij}$ et de l'échelle de KK.

\newpage

\section{Vers un achèvement ultraviolet}
\subsection{Interprétation holographique}
La dualité jauge/gravité, aussi nommée principe holographique, est un cadre pour interpréter une théorie avec une dimension spatiale supplémentaire courbe comme une théorie des champs à 4D fortement couplée. Le point de départ est la correspondance AdS/CFT, proposée en 1997 par Maldacena \cite{Maldacena:1997re} dans le cadre de la théorie des supercordes de type IIB. Cette conjecture établit une dualité entre la théorie des supercordes de type IIB formulée sur l'espace-temps AdS$_5 \times S^5$, et une théorie de jauge $SU(N)$ supersymétrique $\mathcal{N} = 4$ sur l'espace-temps de Minkowski à 4D. Cette théorie de jauge est une théorie des champs conforme (CFT -- \textit{Conformal Field Theory}) : sa constante de couplage $g_{YM}$ ne reçoit pas de correction radiative, et ne varie donc pas avec l'énergie.

Les symétries des deux théories sont reliées :
\begin{itemize}
\item Les isométries de $S^5$ sont les rotations décrites par le groupe $SO(6)$, dont le groupe de recouvrement universel est $SU(4)$, qui est justement le groupe de R-symétrie de la CFT supersymétrique.
\item Le groupe des isométries de AdS$_5$ est précisément celui des transformations conformes de la CFT.
\end{itemize}

La correspondance établit une relation entre les paramètres des deux théories :
\begin{equation}
\left(\dfrac{R_{AdS}}{l_s}\right)^4 = 4 \pi g_{YM}^2 N \, ,
\end{equation}
où $R_{AdS}=1/k$ est la longueur de courbure de l'espace-temps AdS$_5$, et $l_s$ est la longueur d'une corde. On se place dans le régime où la théorie des supercordes se réduit à la SUGRA. On a donc $R_{AdS} \gg l_s$, ce qui implique la condition $g_{YM}^2 N \gg 1$ : la CFT est fortement couplée. On ajoute la condition supplémentaire de perturbativité de la théorie des supercordes, \textit{i.e.} que sa constante de couplage $g_s \ll 1$, de telle manière que les états d'excitation non-perturbatifs de la corde de masse $\sim 1/g_s$ restent lourds.  Puisque $g_s \sim 1/N$, on a la limite de 't Hooft \cite{tHooft:1973alw} $N \rightarrow \infty$. Si on étudie la CFT fortement couplée, on peut alors utiliser la correspondance AdS/CFT et calculer les observables physiques à partir de la théorie gravitationnelle classique faiblement couplée.

\begin{figure}[h]
\begin{center}
\includegraphics[height=8cm]{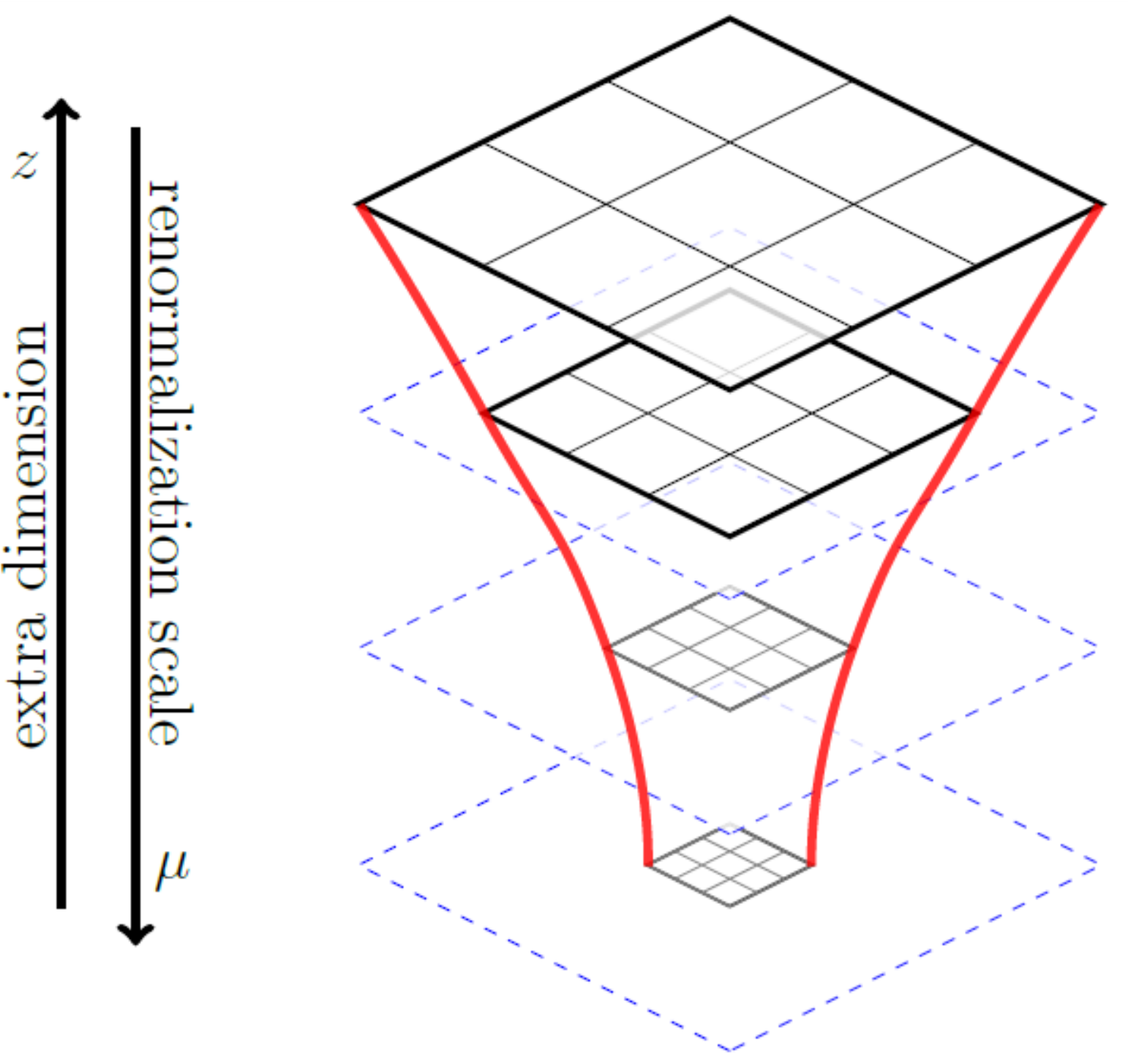}
\end{center}
\caption[Correspondance AdS/CFT]{(Adaptée de la Réf.~\cite{Csaki:2016kln}.) Schéma de la correspondance AdS/CFT. Les isométries de l'espace-temps AdS$_5$ correspondent à la symétrie conforme de la CFT à 4D. Se déplacer dans la direction $z$ revient à une transformation du groupe de renormalisation de la théorie à 4D.}
\label{holo1}
\end{figure}

Dans les modèles à 5D qui nous intéressent, on a simplement une théorie des champs définie sur l'espace-temps AdS$_5$. On conjecture la dualité jauge/gravité qui nous permet d'établir une correspondance entre une théorie gravitationnelle classique définie sur l'espace-temps AdS$_5$ et une CFT fortement couplée.  On peut utiliser l'explication heuristique suivante. Considérons la métrique de l'espace-temps AdS$_5$ (\'Eq.~\eqref{RS1_met}) et effectuons le changement de variable suivant :
\begin{equation}
\dfrac{R_{AdS}}{z} = \text{e}^{-y/R_{AdS}} \, .
\end{equation}
On obtient la métrique :
\begin{equation}
ds^2 = \left( \dfrac{R_{AdS}}{z} \right)^2 \, \left( \eta_{\mu \nu} dx^\mu dx^\nu -dz^2 \right) \, .
\label{z_metric}
\end{equation}
La transformation
\begin{equation}
z \rightarrow \text{e}^\alpha z \, , \ \alpha \in \mathbb{R} \, ,
\end{equation}
d'une tranche à 4D de AdS$_5$ :
\begin{equation}
ds^2 = \left( \dfrac{R_{AdS}}{z} \right)^2 \, \eta_{\mu \nu} dx^\mu dx^\nu \, ,
\end{equation}
se traduit par un changement d'échelle de la métrique effective à 4D :
\begin{equation}
x \rightarrow \text{e}^\alpha x \, .
\end{equation}
Se déplacer vers les $z$ croissants équivaut donc à augmenter les échelles de longueur à 4D, et donc à diminuer les échelles d'énergie, $\mu \sim 1/z$ (\textit{c.f.} la Fig.~\ref{holo1}). Ces considérations mènent à l'interprétation holographique de la dimension spatiale supplémentaire, où la coordonnée $z$ correspond au flot du groupe de renormalisation dans la CFT à 4D. L'évolution d'une fonction d'onde de KK quand $z$ croît correspond à l'évolution du coefficient d'un opérateur de la CFT quand $\mu$ décroît. Pour une description détaillée de la correspondance AdS/CFT dans le cadre de la théorie des champs, voir les Réfs.~\cite{ArkaniHamed:2000ds, Rattazzi:2000hs, Skenderis:2002wp, Strassler:2005qs, Gherghetta:2006ha, Gherghetta:2010cj, Sundrum:2011ic}. Dans la suite, on va donner les éléments principaux du dictionnaire AdS/CFT.

\begin{figure}[h]
\begin{center}
\includegraphics[height=6cm]{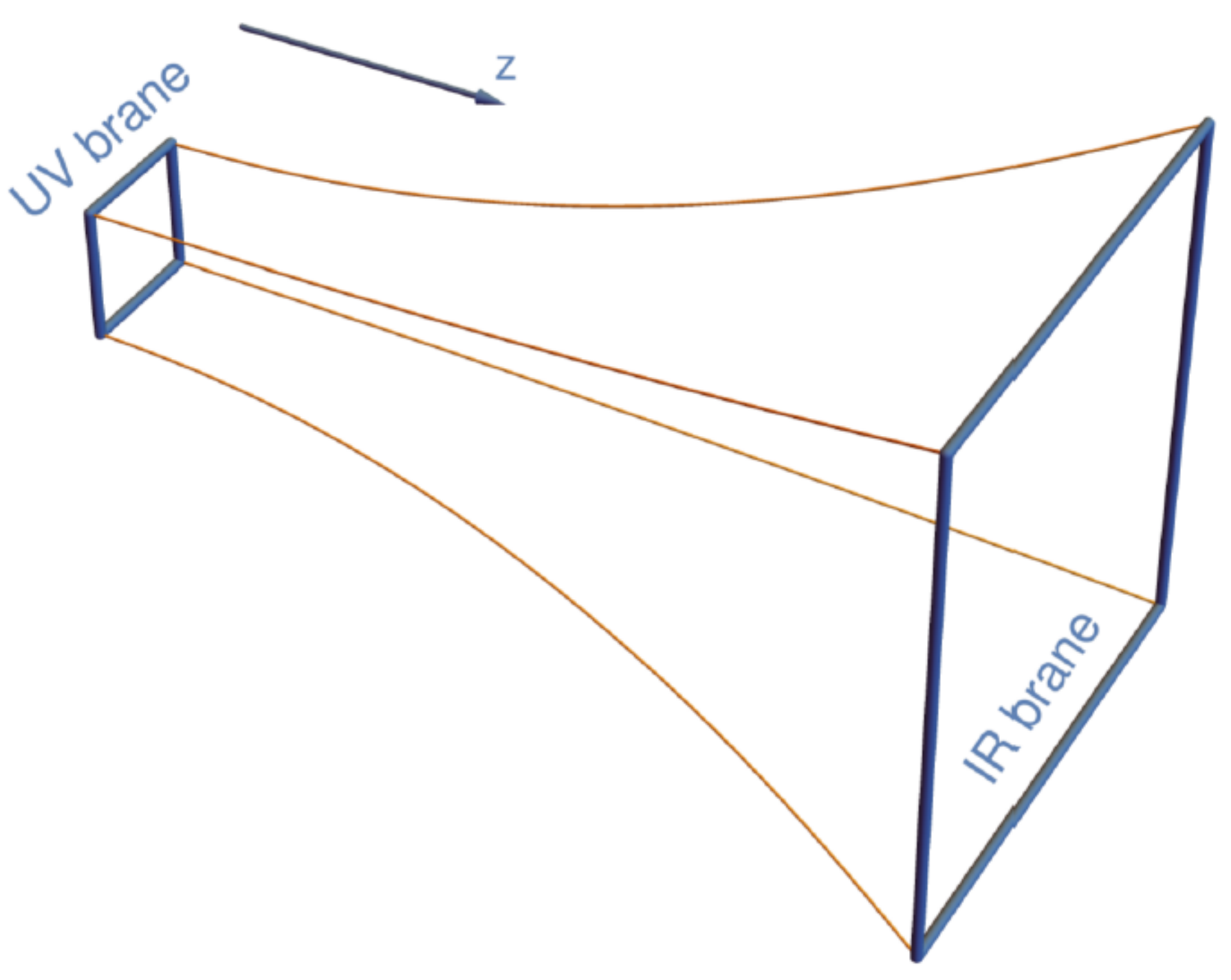}
\end{center}
\caption[Tranche AdS$_5$]{(Adaptée de la Réf.~\cite{Csaki:2018muy}.) Schéma de la tranche AdS$_5$. Les 3-branes coupent l'espace brusquement, introduisant une brisure de la symétrie conforme dans l'UV et l'IR.}
\label{holo2}
\end{figure}

Les modèles type RS1 ne sont pas formulés sur un espace-temps AdS$_5$ entier mais sur une tranche de celui-ci (\textit{c.f.} la Fig.~\ref{holo2}). Quelle est l'interprétation des deux 3-branes ?
\begin{itemize}
\item La brane UV correspond à une échelle de coupure de la CFT à haute énergie, $\Lambda_{UV} \sim 1/R_{AdS}$. Elle introduit une échelle de masse dans les CFT, et est donc une source de brisure de la symétrie conforme. En s'éloignant de la brane UV, on diminue l'énergie dans la CFT, et le bulk devient immédiatement AdS : la CFT devient alors rapidement conforme sous l'échelle de coupure UV. Toute source de brisure de la symétrie conforme, introduite par la coupure UV, doit donc être une déformation irrelevante de la CFT.
\item La brane IR est une coupure brusque de l'espace-temps AdS$_5$ en $z=R'$, et correspond à une déformation relevante de la CFT, impliquant une brisure spontanée de la symétrie conforme, et un phénomène de confinement à l'échelle $\Lambda_{IR} \sim 1/R'$. On peut montrer que la présence de la brane IR est responsable de la quantification du spectre de KK des différents champs dans le bulk, dont les états les plus légers apparaissent près de l'échelle $\Lambda_{IR}$. Pour chaque état de KK de la théorie extra-dimensionnelle, il existe un opérateur composite dans la CFT. La brane IR doit être interprétée comme un modèle simplifié du confinement.
\end{itemize}

\begin{figure}[h]
\begin{center}
\includegraphics[height=6cm]{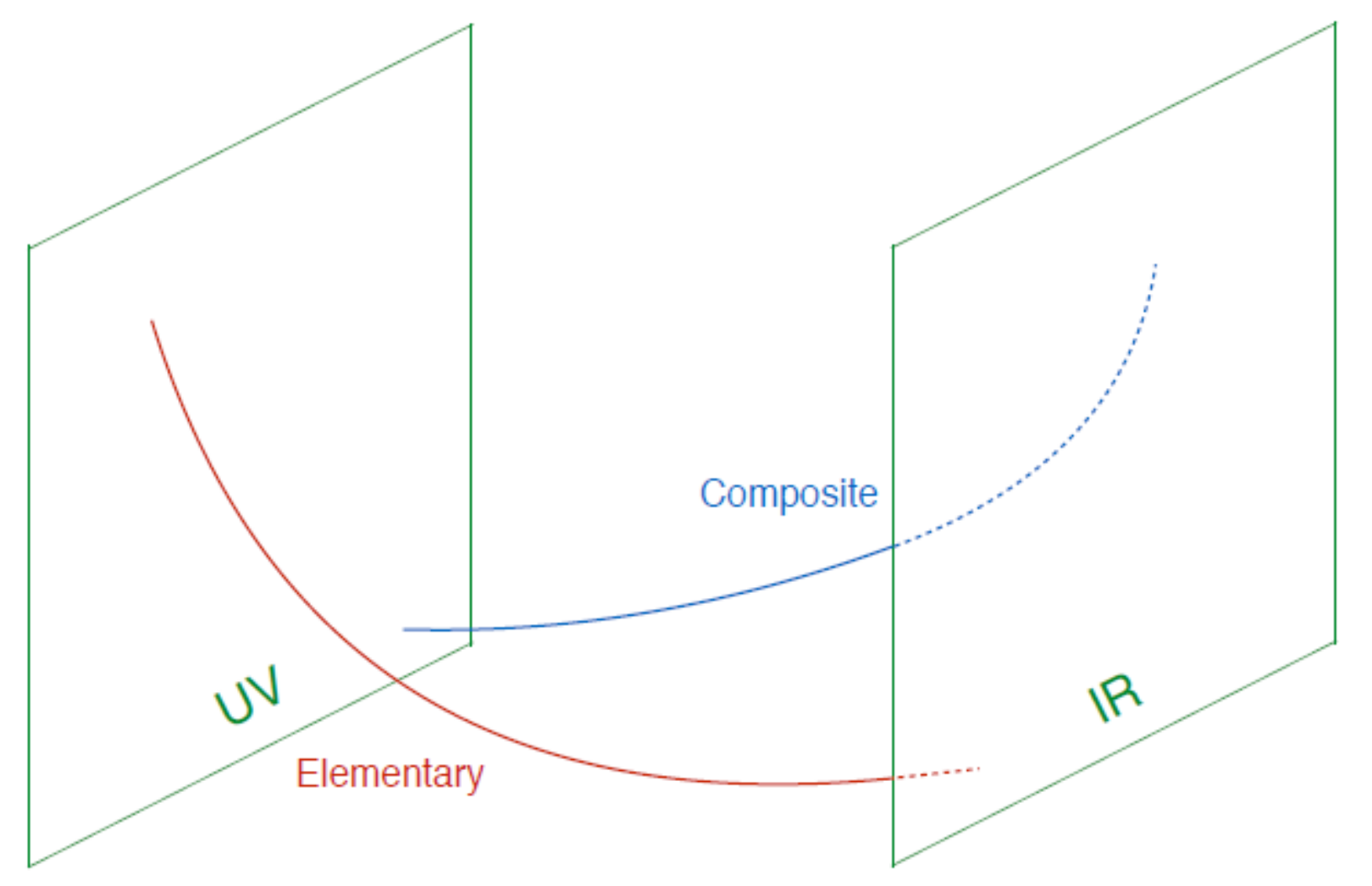}
\end{center}
\caption[Dictionnaire AdS/CFT]{(Adaptée de la Réf.~\cite{Csaki:2018muy}.) Schéma du dictionnaire AdS/CFT. Les états élémentaires (composites) sont localisés près de la brane UV (IR).}
\label{holo3}
\end{figure}

Les états liés existent lorsque l'on considère la CFT à des énergies proches de l'échelle de confinement $\Lambda_{IR}$, ce qui implique que les KK-particules, qui ont leur fonction d'onde localisée près de la brane IR dans la tranche AdS$_5$, sont duaux des états composites. À l'inverse, un état localisé près de la brane UV interagit faiblement avec les états de KK (états composites de la CFT). On en conclut que (\textit{c.f.} la Fig.~\ref{holo3}) :
\begin{itemize}
\item Pour chaque état localisé sur la brane IR, il existe un état composite dans la CFT.
\item Pour chaque état localisé sur la brane UV, il existe un état élémentaire dans la CFT.
\item Pour chaque mode qui se propage dans le bulk, il existe une particule partiellement composite dans la CFT. C'est un mélange d'états élémentaires et composites, à l'instar du mélange entre le méson $\rho$ et le photon en QCD. Pour une particule, dont la fonction d'onde de KK est piquée près de la brane UV (IR), il existe une particule principalement élémentaire (composite) dans la CFT. Pour une particule, dont la fonction d'onde de KK est plate, il existe une particule dans la CFT qui est un mélange équilibré entre états élémentaires et composites.\\
\end{itemize}
Dans le modèle de RS1 originel, tous les champs du SM sont localisés sur la brane IR, il existe donc pour chacun d'entre eux des états composites dans la CFT. Dans les modèles de ce chapitre :
\begin{itemize}
\item Le champ de Higgs est localisé sur la brane IR : il est donc composite dans la CFT. Le problème de hiérarchie de jauge est résolu parce qu'au-dessus de l'échelle de confinement, le boson de Higgs disparaît et les préons\footnote{Si les particules du SM ont une substructure, on appelle parfois préons les particules les constituant.} (non-scalaires), qui le constituent, apparaissent.
\item Les modes zéro du graviton et des fermions légers du SM ont leur fonction d'onde de KK piquée près de la brane UV : ils sont donc partiellement composites dans la CFT, et principalement élémentaires.
\item L'isodoublet et l'isosingulet faibles du quark top, ainsi que les KK-particules, ont leur fonction d'onde de KK piquée près de la brane IR : ils sont donc partiellement composites dans la CFT, et principalement composites.
\item Les modes zéro des bosons de jauge du SM ont leur fonction d'onde de KK plate : ils sont donc partiellement composites dans la CFT, mélanges équilibrés entre états élémentaires et composites. On notera que les bosons de jauge massifs du SM ont une fonction d'onde de KK qui n'est pas tout à fait plate près de la brane IR, ils sont donc légèrement plus composites qu'élémentaires dans la CFT.
\end{itemize}

En conclusion, les modèles de théorie des champs dans une tranche AdS$_5$ peuvent être considérés comme un outil de calcul pour étudier des modèles composites. Dans cette vision, la dimension spatiale supplémentaire n'est pas physique, on peut parler de dimension spatiale supplémentaire auxiliaire : la Nature est décrite par une CFT fortement couplée, et on utilise la dualité jauge/gravité pour calculer les observables dans un modèle extra-dimensionnel. Comme la CFT est une théorie de jauge à 4D, elle est renormalisable. Il est néanmoins très important d'avoir à l'esprit que la correspondance AdS/CFT, dans un modèle phénoménologique à 5D, permet de  faire davantage des discussions qualitatives que quantitatives sur la CFT. Contrairement à la conjecture initiale en théorie des supercordes, le modèle à 5D ne permet pas d'avoir accès à la dynamique préoniques de la CFT. Il n'est d'ailleurs pas garanti qu'il existe une théorie avec confinement dont la dynamique reproduit exactement les valeurs des observables calculés à 5D \cite{Contino:2004vy}. Autrement dit, de la même manière qu'un achèvement dans l'UV n'est pas garanti pour le modèle phénoménologique à 5D, il en va de même pour la CFT. Le principe holographique doit donc être utilisé avec prudence. Les contraintes sur les modèles à 5D ne peuvent pas être, en général, directement appliquées aux modèles composites. Par contre, qualitativement, la découverte des premiers états excités des particules du SM ne permettrait pas de trancher entre un scenario extra-dimensionnel et un scenario composite. Seule la découverte de plusieurs niveaux d'excitation différents, dont on pourrait extraire l'expression du spectre de masse, permettrait de déterminer qui d'une dimension supplémentaire ou d'une théorie de jauge avec confinement décrit correctement ces résonances.

\subsection{Gorges déformées en théorie des supercordes de type IIB}
\subsubsection{Construction}
Si la dimension spatiale supplémentaire est physique, il est nécessaire de trouver un achèvement UV qui soit une théorie extra-dimensionnelle. Celle-ci doit résoudre le problème de non-renormalisabilité des interactions de jauge et de Yukawa dans un espace-temps à plus de 4D. Une candidate naturelle est l'une des théories des supercordes (\textit{c.f.} la Réf.~\cite{Reece:2010xj} pour une revue). Il y a notamment des tentatives intéressantes dans le type IIB. Typiquement, les six dimensions spatiales supplémentaires sont compactifiées, dont la taille est de l'ordre de $\ell_P$. Cependant, en ajoutant des configurations non-triviales de champs de $p$-formes dans l'espace compactifié, on obtient des quanta de flux, lesquels peuvent générer une dimension spatiale supplémentaire plus grande que les autres où la courbure est importante. On obtient une géométrie dite de gorge déformée, jouant un rôle similaire au bulk entre les 3-branes UV et IR du modèle de RS1. La taille de la gorge est stabilisée par les flux, ce que l'on peut modéliser dans l'EFT par le mécanisme de Goldberger-Wise \cite{Brummer:2005sh}.

\begin{figure}[h]
\begin{center}
\includegraphics[height=8cm]{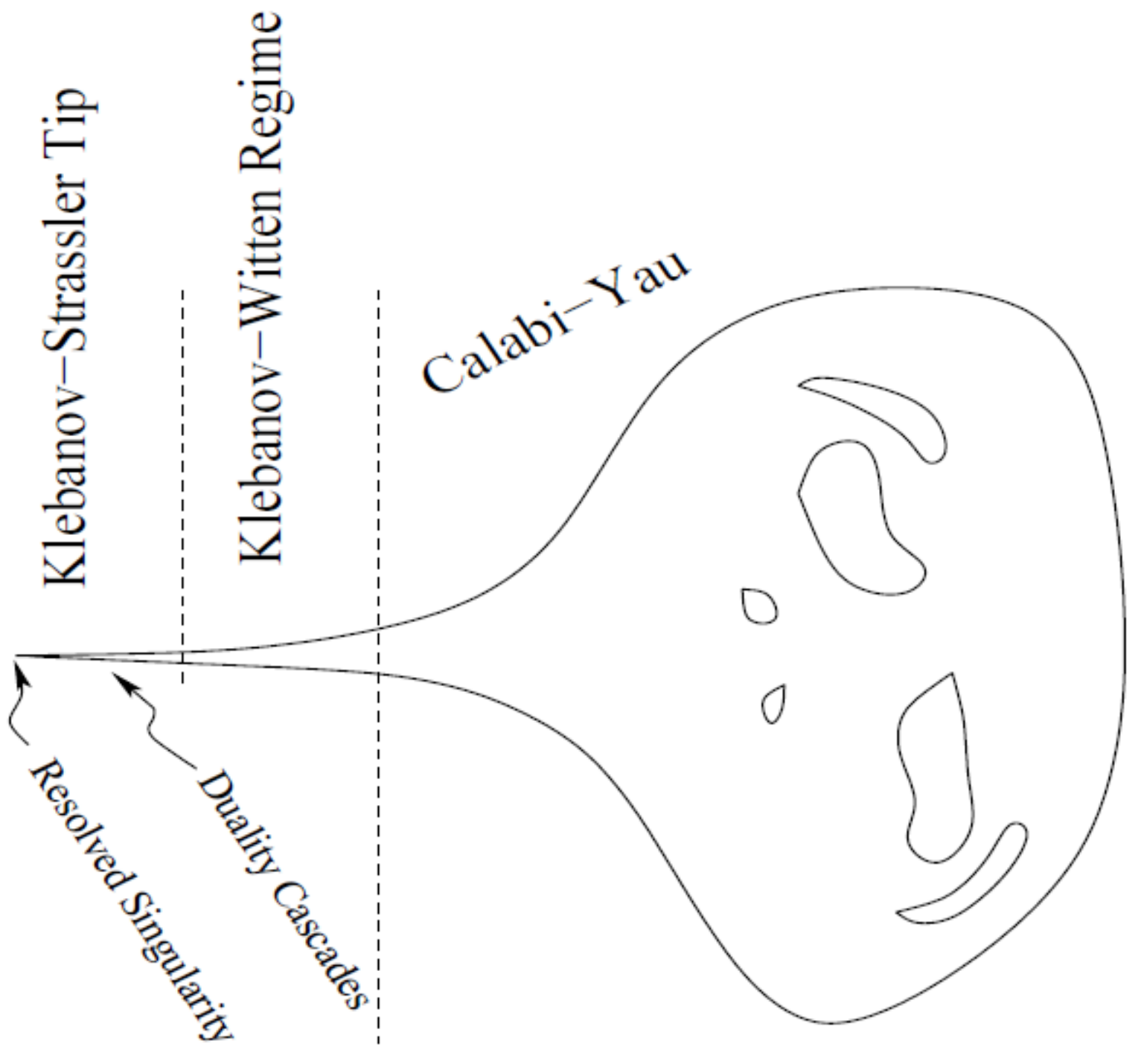}
\end{center}
\caption[Gorge de Klebanov-Strassler collée à un espace de Calabi-Yau]{(Adaptée de la Réf.~\cite{Gherghetta:2006yq}.) Schéma d'une gorge de Klebanov-Strassler collée à un espace de Calabi-Yau. Loin de la pointe IR, la gorge peut être décrite par la solution plus simple de Klebanov-Witten.}
\label{KS_throat}
\end{figure}

Dans ces constructions, l'espace est de la forme AdS$_5 \times X_5$, où $X_5$ est une variété compacte à 5D. AdS$_5$ est la géométrie proche de l'horizon d'une pile de $N$ D3-branes. Il y a alors $N$ unités de flux dans $X_5$. Le point de départ est la construction de Klebanov-Witten \cite{Klebanov:1998hh}, où $X_5 = T^{1,1} \simeq \left( SU(2) \times SU(2) \right) / U(1)$. La théorie n'est pas régulée dans l'IR, \textit{i.e.} la géométrie présente une singularité nue. Pour la résoudre, on utilise la construction de Klebanov-Strassler \cite{Klebanov:2000hb, Strassler:2005qs}, où $X_5 = S^2 \times S^3$, avec $S^2$ se réduisant à un point au bout de la pointe de la gorge. La géométrie, près de la pointe IR, n'est alors plus AdS$_5$. Les constructions de Klebanov-Witten et de Klebanov-Strassler ont des géométries non-compactes, s'étendant à l'infini dans la direction associée au rayon AdS$_5$. Il faut alors compléter la géométrie dans l'UV par une variété compacte de Calabi-Yau. Dans la construction de Klebanov-Strassler, le spectre est alors normalisable et discret \cite{Giddings:2001yu}. Dans le modèle effectif de RS1, la pointe de la gorge et l'espace de Calabi-Yau sont respectivement modélisés par les branes IR et UV. La Fig.~\ref{KS_throat} présente une illustration de la géométrie d'une gorge déformée.

Pour introduire les champs du SM dans le bulk \cite{Karch:2002sh, Gherghetta:2006yq}, on ajoute une pile de D7-branes s'enroulant autour d'un 3-cycle de $X_5$. Les bosons de jauge et les fermions ne se propagent pas alors dans une, mais quatre dimensions spatiales supplémentaires. De plus, la déformation de la géométrie AdS$_5$, près de la pointe IR dans la solution de Klebanov-Strassler, a des conséquences importantes pour la phénoménologie des KK-particules. En effet, les fonctions d'onde de KK de celles-ci sont quasi-localisées près de la brane IR, et leurs couplages sont donc très sensibles à la géométrie de la pointe de Klebanov-Strassler, qui n'est pas AdS$_5$. Dans la littérature, il y a peu d'études des contraintes sur $m_{KK}$ dans ces modèles. Des constructions simplifiées \cite{Shiu:2007tn, McGuirk:2007er, Archer:2010hh} montrent que les KK-particules sont fortement couplées près de la pointe IR. On s'attend donc à avoir une borne inférieure sur $m_{KK}$ dramatiquement plus élevée que dans une tranche AdS$_5$, ce qui remet sérieusement en question le potentiel du modèle pour résoudre le problème de hiérarchie de jauge avec les bosons de jauge et les fermions dans le bulk. Construire un achèvement UV préservant, ou même améliorant, les contraintes expérimentales sur $m_{KK}$ dans un modèle aussi simple que celui de RS1 reste donc un défi important.

\subsubsection{Gorges multiples}
\begin{figure}[h]
\begin{center}
\includegraphics[width=14cm]{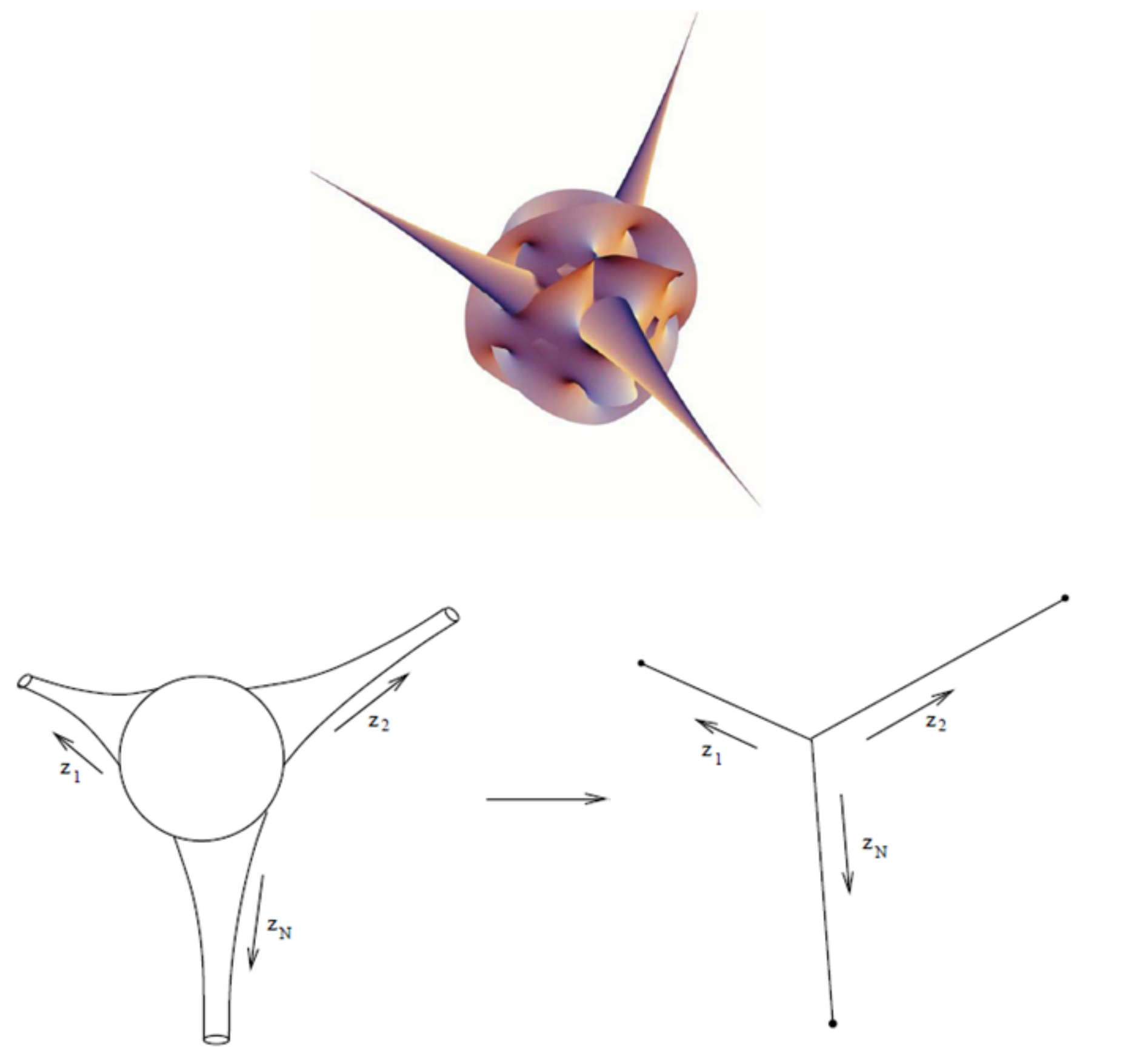}
\end{center}
\caption[Gorges multiples]{(Adaptée de la Réf.~\cite{Cacciapaglia:2006tg}.) En haut : Illustration d'une variété de Calabi-Yau de laquelle naissent les bases de gorges déformées. En bas : Dans l'approche d'EFT de la Réf.~\cite{Cacciapaglia:2006tg}, les dimensions transverses des gorges sont négligées, et ces dernières sont modélisées par une seule dimension spatiale supplémentaire. Les gorges se rejoignent en un point : la brane UV.}
\label{multi-throats}
\end{figure}

En théorie des supercordes de type IIB, les compactifications avec flux peuvent donner naissance, non pas à une, mais à une multitude de gorges déformées, collées au même espace de Calabi-Yau. On obtient ainsi un Univers \og pieuvre \fg{} où la \og tête \fg{} serait l'espace de Calabi-Yau et les \og tentacules \fg{} les gorges (\textit{c.f.} la Fig.~\ref{multi-throats}). On peut alors considérer une EFT où les dimensions transverses des gorges sont négligeables, et où l'espace de Calabi-Yau est réduit à une 3-brane UV. Chaque gorge est alors modélisée par un intervalle, et toutes partagent la même brane UV, comme illustré sur la Fig.~\ref{multi-throats}. Autrement dit, on colle plusieurs tranches AdS$_5$ ensemble, et la physique UV de l'espace de Calabi-Yau est décrite par des opérateurs localisés sur la brane UV. Les champs peuvent alors se propager dans une ou plusieurs gorges, ouvrant la porte vers une large variété de modèles, possibilités pourtant peu exploitées dans la littérature. Les Réfs.~\cite{Kim:2005aa, Cacciapaglia:2006tg} donnent les bases de construction d'une théorie des champs dans cette géométrie, et la Réf.~\cite{Law:2010pv} montre que l'on peut stabiliser la longueur de chaque gorge avec une généralisation du mécanisme de Goldberger-Wise. En ayant des gorges de différentes longueurs, chaque brane IR reçoit un facteur de déformation différent ; on génère ainsi naturellement différentes échelles d'énergie dans le modèle. Quelques applications phénoménologiques ont été proposées \cite{Dimopoulos:2001ui, Dimopoulos:2001qd, Cacciapaglia:2005pa} pour construire notamment des modèles de saveur \cite{Abel:2010kw}, de neutrinos \cite{Gripaios:2006dc}, de leptogénèse \cite{Bechinger:2009qk}, de secteurs cachés \cite{McDonald:2010iq, McDonald:2010fe}, de matière noire \cite{Ahmed:2015ona} et d'axions \cite{Flacke:2006ad, Flacke:2006re}.

\selectlanguage{english}

\part{Original Research Work}
\label{research_work}

\chapter{Beyond Brane-Higgs Regularization}
\label{1_4D perturbative approach}

This chapter is a personal adaptation of Ref.~\cite{Angelescu:2019viv} written in collaboration with Andrei Angelescu, Ruifeng Leng and Grégory Moreau.

\section{Introduction}
The paradigm of scenarii with spacelike extra dimensions (and the composite Higgs models dual via the AdS/CFT correspondance) represents an alternative to SUSY for addressing the gauge hierarchy problem of the SM. Moreover, the warped extra dimension scenarii with SM fields in the bulk allow to generate the SM fermion mass hierarchy from a simple geometrical picture of fermion KK wave functions along the extra dimension. To realize those two hierarchical features, the Higgs field, 
providing masses to fermions and weak gauge bosons of the SM via EWSB, is often stuck exactly on the IR-brane.

In supersymmetric models with a warped geometry (see Ref.~\cite{Gherghetta:2015ssa} for a review), the two Higgs doublets are localized at the boundary of the extra dimension and SUSY is broken on the opposite side. On the one hand, if the Higgs doublets are on the IR-brane \cite{Gherghetta:2003he, Gherghetta:2011wc}, then the big gauge hierarchy problem is still solved by the warp factor, and SUSY protects the masses of the Higgs bosons from radiative corrections between the EW scale and the quantum gravity scale on the IR-brane (indeed the current LHC constraints push the quantum gravity scale on the IR-brane up to $\mathcal{O}(10)$ TeV, implying a little hierarchy problem for the mass of the Higgs boson). On the other hand, if the Higgs doublets are on the UV-brane \cite{Gherghetta:2000kr, Gabella:2007cp, Buyukdag:2018ose, Buyukdag:2018cka}, SUSY solves the entire gauge hierarchy problem. In both cases, the fermion mass hierarchy is generated by allowing the gauge and matter fields to propagate in the bulk. The same mechanism gives a sfermion spectrum compatible with the experimental bounds in the flavor sector.

In the present chapter, we discuss a rigorous 5D treatment of a brane localized Higgs scalar field, interacting with bulk quarks/leptons, 
which presents subtleties that deserve to be looked at more deeply. Let us recall these subtle aspects. The brane localized interactions between the 5D fields and the Higgs boson on the IR-brane involves a Dirac distribution, which may induce an unusual discontinuity in the KK wave function along the 
extra dimension for some of the bulk fermions: the so-called ``jump problem''~\cite{Bagger:2001qi, Csaki:2003sh}. This motivated the introduction~\cite{Csaki:2003sh, Csaki:2005vy, Grojean:2007zz, Casagrande:2008hr, Azatov:2009na, Casagrande:2010si, Barcelo:2014kha} 
of a regularization of the Dirac distribution (smoothing the Dirac distribution or shifting it from the boundary) in the calculation of the KK fermion mass 
spectra and effective four-dimensional (4D) Yukawa couplings. Although there is no profound theoretical reason to apply such a regularization procedure (implying
interaction-free boundary conditions for fermions), nowadays most of the theoretical and phenomenological studies of the warped models with a brane localized SM Higgs field (see {\it e.g.} 
Refs.~\cite{Azatov:2009na, Casagrande:2010si, Azatov:2010pf, Goertz:2011hj, Carena:2012fk, Malm:2013jia, Hahn:2013nza, Malm:2014gha, Barcelo:2014kha}) are relying on this Dirac distribution regularization. 

We first present the mathematical inconsistencies of this regularization procedure used in the literature. Then, instead of regularizing, 
we develop the rigorous determination of the KK wave functions -- taking into account the nature of the Dirac distribution in the Higgs couplings -- which leads to bulk fermion KK wave functions without 
discontinuities on the considered internal space. Alternatively, we suggest to introduce the appropriate formalism of distributions in order to treat properly the Dirac distribution 
as well as all the Lagrangian terms. We conclude from this whole approach that neither a jump of KK wave functions nor a particular problem arises when a proper mathematical 
framework is used, so that there is in fact no motivation to introduce a brane-Higgs regularization.
These methods without regularization are illustrated here in the 
derivation of the KK fermion mass spectrum -- the same ideas apply to the calculation of effective 4D Yukawa couplings. This calculation of the mass spectrum is performed in a simplified 
model with a flat extra dimension, the minimal field content and without gauge symmetry. Nevertheless this toy model already possesses all the key ingredients to study 
the delicate brane-Higgs aspects. Hence our conclusions can be directly extended to the realistic warped models with bulk SM matter addressing the fermion 
flavor and gauge hierarchy.

Several new methods for the calculation of the spectrum are proposed which further allow for confirmations of the analytical results. Those methods apply to alternative uses of 
the 5D or the 4D approaches, employing distributions or functions. Note that from a historical point of view, the method established here arises naturally in the theory of variational calculus, as the Lagrangian boundary terms 
(brane-Higgs couplings to fermions) are included in a new boundary condition \cite{Hilbert}. 
Furthermore, the present method follows the prescription of considering the Dirac function to be a distribution. By the way, the distributions were 
defined with mathematical rigor during the 1940's by L.~Schwartz \cite{Schwartz2, Schwartz1} precisely for the purpose of consistently solving physical problems.  

Besides, the performed exhaustive exploration of the solutions for the KK wave function, in the context of two fermion fields propagating along a flat extra dimension 
with neither bulk mass nor BLKTs, allows to show that while in the free case either Neumann or Dirichlet conditions are possible at both interval boundaries
(with possibly different conditions at the two boundaries), once their brane-Higgs couplings are introduced, only the new type of boundary conditions involving the Yukawa coupling
constants exists at the boundary where the Higgs field is localized.

The rigorous result obtained here for the KK mass spectrum is different from the one derived through the regularization of Dirac distributions, as it is detailed in the 
present work. This difference is physical and 
analytical which is also important because determining the fermion mass formula is part of the precise theoretical understanding of the higher-dimensional setup  
with a brane localized Higgs boson, since the Higgs properties are tested at the LHC with an increasing accuracy. 
Phenomenological studies of the brane localized Higgs interaction within warped models can be found in Refs.~\cite{Lillie:2005pt, Djouadi:2007fm, Bouchart:2009vq, Cacciapaglia:2009ky, Bhattacharyya:2009nb, Casagrande:2010si, Azatov:2010pf, Goertz:2011hj, Carena:2012fk, Frank:2013un, Malm:2013jia, Hahn:2013nza, Malm:2014gha, Dey:2015pba}.

At the level of calculating Higgs production/decay rates, a theoretical debate appeared in the literature \cite{Carena:2012fk} on models with brane-Higgs coupling to bulk matter.
Indeed, a paradoxal non-commutativity arose in the calculation: different physical results were obtained for Higgs processes when taking $\epsilon \to 0$ and then 
$N_{KK} \to \infty$ \cite{Casagrande:2010si, Goertz:2011hj} or the opposite order \cite{Azatov:2010pf}~\footnote{Here $N_{KK}$ is the number of exchanged KK states at the level of the loop amplitude and $\epsilon$ 
is the infinitesimal parameter introduced to regularize the Dirac distribution.}. A consequence of our present work is that the ambiguity is solved since 
our conclusion is that the regularization must not be applied (in turn no $\epsilon$ has to be introduced).

Furthermore, the mass spectrum obtained here allows to point out the necessity for bulk fermions (with or without coupling to a brane localized scalar field) to have certain Bilinear Boundary Terms (BBTs) (studied in a
different context in Refs.~\cite{Henningson:1998cd, Mueck:1998iz, Arutyunov:1998ve, Henneaux:1998ch, Contino:2004vy})
which are mass terms from the point of view of the spinorial structure but do not introduce new contributions to physical masses.  Indeed, such terms guarantee the existence of physical solutions (with correct profile normalizations, Hermitian conjugate boundary conditions and
satisfying the decoupling limit argument) and the (non-trivial) exact matching between the 5D and 4D analytical calculations. These boundary terms are part of the definition of the self-consistent 5D model with bulk fermions 
coupled to a brane localized scalar field, although they are newly introduced in this context.  

At a brane without Yukawa coupling, instead of including the BBTs, 
we find that one can alternatively impose an {\it essential boundary condition} (in contrast to {\it natural boundary conditions} coming from the Lagrangian variations), consistent with the condition that a fermion current along the extra dimension must vanish at the boundaries (exclusively within the 4D approach in case of brane localized Yukawa interactions). Indeed, the generic reason for the presence 
of the BBTs is the consistent and complete geometrical definition of models with an extra dimension compactified on an interval to which fermionic matter is confined. With the BBTs, the chiral nature of the SM at the level of the zero modes (left-handed for the weak isodoublets and right-handed for the weak isosinglets)
is entirely induced by the signs in front of BBTs. This new relation shows how the particular chiral properties of the SM could be explained by an underlying theory, through the signs of the BBTs.

The plan of the chapter is the following. We begin by defining our toy model in Section \ref{1_A_toy_model_of_flat_extra_dimension} and explain how to generalize it to a realistic warped model reproducing the SM at low energy. In Section~\ref{1_free_bulk_fermions}, we discuss in detail the {\it natural} versus {\it essential boundary conditions} for bulk fermions -- at the boundaries of the extra dimension -- 
and also to derive the free fermion mass spectrum (in the absence of Yukawa interactions), which is useful in particular for the 4D approach or more generically
for a solid comprehension of the mass spectrum. In particular the boundary conditions arising from considerations on the probability currents are discussed. We expose the so-called 4D method, where the coupling of the bulk fermions to the Higgs field VEV is introduced with an infinite matrix to bi-diagonalize. In Section~\ref{1_Usual_treatment}, we discuss the main drawbacks of the 5D method with regularization of the Dirac distribution which one can find in the literature. In Section~\ref{1_Yukawa_terms_as boundary_conditions}, we present our 5D method. In Section~\ref{implications}, we apply our results, together with a discussion of the phenomenological impacts. We summarize and conclude in Section~\ref{1_conclusion_4}.

\section{Minimal Consistent Model}
\label{1_A_toy_model_of_flat_extra_dimension}

\subsection{Spacetime Structure}

We consider a 5D toy model with a space-time $\mathcal{E}^5 = \mathcal{M}^4 \times I$.
\begin{itemize}
\item $\mathcal{M}^4$ is the usual 4D Minkowski spacetime. An event in $\mathcal{M}^4$ is characterized by its 4-vector coordinates $x^\mu$ where 
$\mu =0,1,2,3$ is the Lorentz index. 
\item $I$ is the interval $[0, L]$, guaranteeing chiral fermions at low energies (below the first KK mass scale). 
$I$ is parametrized by the extra coordinate $y$ and bounded by two flat 3-branes at $y = 0$ and $y = L$. 
\item A point of the whole 5D spacetime $\mathcal{E}^5$ is labeled by its coordinates $z^M$ with an index $M \in \llbracket 0, 4 \rrbracket$. $z^M$ can be split  
into $\left( x^\mu, y \right)$.
\end{itemize}

\subsection{Bulk Fermions}

We consider the minimal spin-1/2 fermion field content allowing to write down a 4D renormalizable SM Yukawa-like coupling between zero mode
fermions (of different chiralities) and a scalar field (see Subsection~\ref{1_Yukawa}):
a pair of fermions $Q$ and $D$. Both are also propagating along the extra dimension, as we have in mind an extension of the model to a realistic scenario with bulk matter
({\it c.f.} Subsection~\ref{1_extension}) where $Q,D$ will be left and right handed down type quark fields. According to the discussion in Appendix~\ref{app_cont_field}, there are branes only at the boundaries so we take the 5D fields continuous up to the boundaries.
Using the notation, 
$\overleftrightarrow{\partial_M}=\overrightarrow{\partial_M}-\overleftarrow{\partial_M}$, the 5D fields $Q(x^\mu, y)$ and $D(x^\mu, y)$ of mass dimension 2 have thus the following kinetic terms:
\begin{align}
&S_{\Psi} = \int d^4x \int_0^L dy \ \mathcal{L}_\Psi \ , \nonumber \\
&\text{with} \ \ 
\mathcal{L}_\Psi =  \frac{i}{2} \ \left( \bar{Q}\Gamma^M \overleftrightarrow{\partial_M} Q + \bar{D}\Gamma^M \overleftrightarrow{\partial_M} D \right) \ ,
\label{1_eq:actionKin}
\end{align}
where $\Gamma^M$ denote the 5D Dirac matrices. 
In our notations a 5D Dirac spinor, being in the irreducible representation of the Lorentz group, reads
\begin{equation}
Q = \left(\begin{array}{c} Q_L \\ Q_R   \end{array}\right) \,\,\,\, \mbox{and} \,\,\,\,   D = \left(\begin{array}{c} D_L \\ D_R   \end{array}\right)\,,
\label{1_5DDiracSp}
\end{equation}
in terms of the two two-component Weyl spinors. $L/R$ stands for the left/right-handed chirality. Let us rewrite the bulk action of Eq.~\eqref{1_eq:actionKin} in a convenient form:
\begin{align}
&S_{\Psi} = \int d^4x \int_0^L dy \ \mathcal{L}_\Psi \ , \nonumber \\
&\text{with} \ \ 
\mathcal{L}_\Psi = \sum_{F = Q, D} \dfrac{1}{2} \left( i F^\dagger_R \sigma^\mu \overleftrightarrow{\partial_\mu} F_R + i F^\dagger_L \bar{\sigma}^\mu \overleftrightarrow{\partial_\mu} F_L + F^\dagger_R \overleftrightarrow{\partial_y} F_L - F^\dagger_L \overleftrightarrow{\partial_y} F_R \right).
\label{1_L_2}
\end{align}

\subsection{Brane Localized Scalar Field}
\label{1_Brane Localized Scalar Field}

The subtle aspects arise when the fermions couple to a single 
4D real scalar field $H$ (mass dimension 1), confined on a boundary taken here to be at $y=L$ (as inspired by warped scenarii addressing the gauge hierarchy
problem). The action of this scalar field has the generic form
\begin{align}
&S_H  = \int d^4x \int_0^L dy~ \delta(y-L) \ \mathcal{L}_H \, , \nonumber \\
&\text{with} \ \ 
\mathcal{L}_H = \dfrac{1}{2} \, \partial_\mu H \partial^\mu H - V(H) \, ,   
\label{1_eq:actionH}
\end{align}
where the action of a Dirac distribution centered at $y=a$ on a testfunction $\varphi(y)$ is defined as
\begin{equation}
\int_{0}^{L} dy \ \delta(y-a) \, \varphi(y) = \varphi(a) \, .
\end{equation}
The potential $V$ possesses a minimum which generates a non-vanishing 
VEV for the field $H$, which can be expanded as
\begin{equation}
H=\frac{v+h(x^\mu)}{\sqrt{2}} \, ,
\label{1_H_exp}
\end{equation}
in analogy to the SM Higgs field.

\subsection{Yukawa Interactions}
\label{1_Yukawa}

We focus on the following basic interactions in order to study the subtleties induced by the coupling of the brane scalar field to bulk fermions,
\begin{align}
&S_{Y} = \int d^4x \int_0^L dy~ \delta(y-L) \ \mathcal{L}_Y \ , \nonumber \\
&\text{with} \ \ 
\mathcal{L}_Y = - Y_5\ Q^\dagger_LHD_R - Y^\prime_5\ Q^\dagger_RHD_L +  {\rm H.c.} \ .
\label{1_eq:actionYuk}
\end{align}
Considering operators involving $H$, $Q$ and $D$ up to dimension 5 allows to include this Yukawa coupling of interest~\footnote{Notice that for instance a dimension-6
operator of type $\frac{1}{M^2}\delta(y-L)Q^\dagger_{L/R}H^2D_{R/L}$, $M$ being a mass scale, would be treated in a similar way as the couplings in Eq.~\eqref{1_eq:actionYuk} 
(and can contribute to the Yukawa couplings~\eqref{1_eq:actionYuk} through the scalar field VEV).}.
The complex coupling constants $Y_5 = \text{e}^{i \alpha_Y} |Y_5|$ and $Y_5' = \text{e}^{i \alpha_Y'} |Y_5'|$ of Yukawa type (with a mass dimension -1), entering these two distinct terms, are independent (i.e. parameters with possibly different values)  
as the fermion fields on the 3-brane at $y = L$ are strictly chiral (see for instance Ref.~\cite{Azatov:2009na}).


When calculating the KK fermion mass spectrum, we restrict our considerations to the VEV of $H$. Indeed, we have in mind that the interactions of the fluctuation of the Higgs boson with the fermions are treated in perturbation theory, as usual in QFT. Hence, we can decompose the action~\eqref{1_eq:actionYuk} into $S_Y = S_X + S_{hQD}$ such that
\begin{align}
&S_{X} = \int d^4x \int_0^L dy~ \delta(y-L) \ \mathcal{L}_X \ , \nonumber \\
&\text{with} \ \ 
\mathcal{L}_X = - X\ Q^\dagger_L D_R - X^\prime\ Q^\dagger_RD_L +  {\rm H.c.} \ ,
\label{1_L_5}
\end{align}
and the compact notations $X = \dfrac{v}{\sqrt{2}} \, Y_5$ and $X' = \dfrac{v}{\sqrt{2}} \, Y_5'$, and
\begin{align}
&S_{hQD} = \int d^4x \int_0^L dy~ \delta(y-L) \ \mathcal{L}_{hQD} \ , \nonumber \\
&\text{with} \ \ 
\mathcal{L}_{hQD} = - \dfrac{Y_5}{\sqrt{2}} \ h Q^\dagger_L D_R - \dfrac{Y_5'}{\sqrt{2}} \ h Q^\dagger_RD_L +  {\rm H.c.} \ .
\label{1_L_Yuk}
\end{align}

\subsection{Bilinear Boundary Terms}
\label{1_brTermSec}
Interestingly, the presence of certain BBTs for the bulk fermions coupled to the brane Higgs field 
will turn out to be required to recover the SM in the decoupling limit as will be demonstrated in detail in Subsection~\ref{1_Correct_treatment_AFHS_term}.  The necessity and exact form (including numerical coefficients) 
of these terms will be confirmed in Subsection~\ref{1_Correct_treatment_AFHS_term} by the exact analytical matching between 
the 4D and 5D calculations of the KK mass spectrum. These boundary terms read\footnote{Similar terms, 
leading in particular to $\mathcal{L}_{B}=\dfrac{1}{2}(\bar{D}D-\bar{Q}^D Q^D)$, would be present in a model version extended to the EW symmetry of the SM, with the $Q$ field promoted to an $SU(2)_L$ doublet $Q^D$. 
In contrast, terms of the kind $\bar{Q}^U D$ (or $\bar{Q}D$) or $\bar{Q}^U Q^D$ would obviously not belong to a gauge invariant form.}
\begin{align}
&S_B = \int d^4x \int_0^L dy \, \left[ \delta(y) - \delta(y-L) \right] \ \mathcal{L}_{B} \ , \nonumber \\
&\text{with} \ \ 
\mathcal{L}_{B} = \dfrac{1}{2} \left( \bar{D}D - \bar{Q}Q \right) \ , \label{1_eq:actionBound}
\end{align}
They do not involve new parameters (multiplying the operators).

Note that this kind of boundary terms~\eqref{1_eq:actionBound} is a specific case of boundary mass terms that one can add to the model:
\begin{align}
&S_{BBT} = \int d^4x \int_0^L dy \, \left[ \delta(y) + \delta(y-L) \right] \, \mathcal{L}_{BBT} \ , \nonumber \\
&\text{with} \ \ 
\mathcal{L}_{BBT} = - \sum_{F=Q,D} \dfrac{\mu_F(y)}{2} \, \bar{F}F,
\label{1_HS-terms}
\end{align}
when $\mu_Q(0) = \mu_D(L) =  1$ and $\mu_Q(L) = \mu_D(0) = -1$. In Ref.~\cite{Henningson:1998cd}, Henningson and Sfetsos introduced by hand such kind of terms only at the UV-brane ($y=0$) of a warped 5D model with an arbitrary coefficient $\mu_F(0)$, in order to derive the AdS/CFT correspondance in the calculation 
of correlation functions for spinors (see also Ref.~\cite{Mueck:1998iz}). Then, Arutyunov and Frolov justified them in Ref.~\cite{Arutyunov:1998ve} with a fixed coefficient $\mu_F(0) = \pm 1$ in the Lagrangian formulation, 
to obtain a consistent Hamiltonian formulation when one performs the Legendre transformation in the case of a full AdS spacetime. Refs.~\cite{Henneaux:1998ch, Contino:2004vy} showed that they are necessary to apply Hamilton's principle -- see Annexe~\ref{Holo}. Note that one is free to add also such kind of terms at the IR-brane ($y=L$). The AdS/CFT duality can be used to give a holographic interpretation of warped models 
(from which the present simplified scenario is inspired) in terms of composite Higgs models. Refs.~\cite{vonGersdorff:2004eq, vonGersdorff:2004cg} make use of boundary terms to select boundary conditions with Hamilton's principle for 5D symplectic-Majorana fermions.

The necessity of boundary localized terms, in a field theory defined on a manifold with boundaries, is not a surprise as it is known for a long time in the context of gravity. Indeed, Gibbons-Hawking boundary terms \cite{Gibbons:1976ue, Chamblin:1999ya, Lalak:2001fd, Carena:2005gq} are needed in order to cancel the variation of the Ricci tensor at the boundary of the manifold.

\subsection{Model Extension}
\label{1_extension}

The toy model considered is thus characterized by the action
\begin{eqnarray}
S_{5D} = S_{\Psi} + S_H  + S_Y + S_B \ . \label{1_eq:actionTot}
\end{eqnarray}
Nevertheless, the conclusions of the present work can be directly generalized to realistic warped models with bulk SM matter 
solving the fermion mass and gauge hierarchies. Indeed, working with a warped extra dimension instead of a flat one would not affect the conceptual subtleties concerning the 
coupling of bulk fermions to a brane scalar field. The boundaries at $y=0$ and $y=L$ would then become the UV and IR-branes, respectively. 
Similarly, the scalar potential $V(H)$ can be extended to any potential (like the SM Higgs potential breaking the EW symmetry) as long as it still generates a 
VEV for the scalar field as here. In this context the $H$ singlet can be promoted to the Higgs 
doublet under the SM $SU(2)_W$ gauge group, simply by inserting doublets 
in the kinetic term of Eq.~\eqref{1_eq:actionH}. The whole structure of the coupling of Eq.~\eqref{1_L_5} between bulk fermions and the brane localized VEV would remain the same in case of fermions promoted to SM $SU(2)_W$ weak isodoublets: after index contraction of the doublet $(Q^U,Q^D)^t$ with down/up-quark 
singlets $D,U$, one would obtain two replicas of the structure~\eqref{1_L_5} with the forms $\bar Q_C^D D_{C'}$ and $\bar Q_C^U U_{C'}$ where $C(')$ denotes 
the chirality. Hence the procedure described in this work should be applied to both terms separately~\footnote{The action in Eq.~\eqref{1_eq:actionTot} would be trivially generalized as well to a scenario with a gauge symmetry.}. The same comment holds for the SM color triplet contraction 
and the extension of the field content to three generations of quarks and leptons of the SM. Notice that the flavor mixings would be combined with the mixings among
fermion modes of the KK towers, without any impact on the present considerations concerning brane localized couplings.

\section{5D Free Fermions on an Interval}
\label{1_free_bulk_fermions}

In this part, we calculate the fermionic mass spectrum in the basic case where $Y_5=Y_5'=0$ in Eq.~\eqref{1_eq:actionYuk} (Refs~\cite{Csaki:2003sh, Gherghetta:2010cj}), pointing out the correct treatment. We consider the system with a compactification on an interval with only the free bulk action $S_\Psi$ in Eq.~\eqref{1_L_2}.  The main interest of this section is to develop a rigorous procedure for applying the boundary conditions.

\subsection{Natural Boundary Conditions Only}
\label{1_NBC_only}
In order to extract the Euler-Lagrange equations and the boundary conditions for the bulk fermions, we apply Hamilton's principle with the action \eqref{1_L_2} for each fermion $F=Q,D$~\footnote{The Euler-Lagrange equations for the conjugate fields are trivially related.}. According to the discussion in Appendix~\ref{app_cont_field}, we take the variations of the fields continuous in the bulk and at the boundaries since there are no branes away from the boundaries. Assuming the values of the fields at the boundaries $F_{L/R}(x^\mu, y=0, L)$ $\hat = \left . F_{L/R} \right |_{0, L}$ to be initially unknown (unfixed by \textit{ad hoc} boundary conditions), their values should be deduced from the minimization of the action with respect to them, considering thus generic field variations $\left . \delta F_{L/R} \right |_{0, L}$. In other words, $\left . F_{L/R} \right |_{0, L}$ should be obtained from the so-called
natural boundary conditions\footnote{The notion of natural boundary conditions has already been introduced in the treatment of scalars and gauge fields with brane localized mass terms~\cite{Csaki:2003dt}.}~\cite{Hilbert}. The stationary action condition can be split into conditions associated to each field without loss of generality. After an integration by part of each term, we get
\begin{align}
0 = \delta_{F^\dagger_L} S_{\Psi} &= \displaystyle{ \int d^4x \, \int_0^{L} dy \,  
\delta F^\dagger_L \left[ i\bar{\sigma}^\mu \partial_\mu F_L - \partial_y F_R  \right] } 
\nonumber \\
&\displaystyle{ + \ \int d^4x \, \dfrac{1}{2} \left[ \left . \left( \delta F^\dagger_L \, F_R \right) \right |_{L} - \left . \left( \delta F^\dagger_L \, F_R \right) \right |_{0} \, \right] } \ ,
\label{1_HVP_1a}
\end{align}
\begin{align}
0 = \delta_{F^\dagger_R} S_{\Psi} &= \displaystyle{ \int d^4x \, \int_0^{L} dy \,  
\delta F^\dagger_R \left[ i\sigma^\mu \partial_\mu F_R + \partial_y F_L  \right] } 
\nonumber \\
&\displaystyle{ + \ \int d^4x \, \dfrac{1}{2} \left[ - \left . \left( \delta F^\dagger_R \, F_L \right) \right |_{L} + \left . \left( \delta F^\dagger_R \, F_L \right) \right |_{0} \, \right] },
\label{1_HVP_1b}
\end{align}
where we drop the boundary term at infinity of the four usual dimensions.
The bulk and brane variations of Eqs.~\eqref{1_HVP_1a}-\eqref{1_HVP_1b} must vanish separately (respectively the volume and the surface terms) to ensure $\delta_{F_{L/R}^\dagger} S_{\Psi} = 0$ for generic field variations~\footnote{A functional variation reads $\delta F_{L/R}(z^M)=\kappa \  \eta_{L/R} (z^M)$ with a generic function 
$\eta_{L/R} (z^M)$ and an infinitesimal parameter $\kappa \to 0$.}. We get the bulk Euler-Lagrange equations
\begin{equation} 
\left \{
\begin{array}{r c l}
i \sigma^\mu \partial_\mu F_R + \partial_y F_L &=& 0 \, ,
\\ \vspace{-0.2cm} \\
i \bar{\sigma}^\mu \partial_\mu F_L - \partial_y F_R &=& 0 \, ,
\end{array}
\right.
\label{1_ELE_1}
\end{equation}
and the boundary conditions
\begin{equation}
F_L \vert_{0,L} = F_R \vert_{0,L} = 0.
\label{1_BC_1}
\end{equation}
Let us now deduce from these equations involving the 5D fields the relations among the KK wave functions along the extra dimension. \\


To develop a 4D effective picture, let us replace the 5D fields in the relations obtained just above by their standard KK decomposition, 
\begin{equation}
F_{L/R} \left( x^\mu, y \right) = \dfrac{1}{\sqrt{L}} \displaystyle{ \sum_{n} f^n_{L/R}(y) \, F^n_{L/R} \left( x^\mu \right)} \ ,
\label{1_KK_1}
\end{equation}
where $f^n_{L/R}=q_{L/R}^n$ or $d_{L/R}^n$ are the dimensionless KK wave functions along the extra dimension, associated respectively to the 
4D fields $F^n_{L/R}=Q_{L/R}^n$ or $D_{L/R}^n$ ($n$ is the KK level integer index). If they are not vanishing for all $y$, the KK wave functions are orthonormalized such that
\begin{equation}
\forall n \ , \forall m \ , \dfrac{1}{L} \int_0^{L} dy \, f^{n*}_{L/R}(y) \, f^m_{L/R}(y) = \delta^{nm},
\label{1_normalization_1}
\end{equation}
from the requirement of canonically normalized and diagonal kinetic terms for the 4D fields after inserting the KK decomposition~\eqref{1_KK_1} into the 5D field kinetic terms~\eqref{1_L_2}. Each KK-fermion is described by 4D Dirac-Weyl equations,
\begin{equation}
\forall n,
\left\{
\begin{array}{r c l}
i \bar{\sigma}^\mu \partial_\mu F_L^n \left( x^\mu \right) - m_n \, F_R^n \left( x^\mu \right) &=& 0 \, ,
\\ \vspace{-0.2cm} \\
i \sigma^\mu \partial_\mu F_R^n \left( x^\mu \right) - m_n \, F_L^n \left( x^\mu \right) &=& 0 \, ,
\end{array}
\right.
\label{1_Dirac_1}
\end{equation}
where $m_n$ are the mass eigenvalues of the KK excitation tower for the fermions such that, for each KK mode, one has the action
\begin{equation}
S_F^n = \alpha_L^n \, i F_L^{n\dagger} \bar{\sigma}^\mu \partial_\mu F_L^n + \alpha_R^n \,i F_R^{n\dagger} \sigma^\mu \partial_\mu F_R^n - \alpha_L^n \alpha_R^n \, m_n \left( F_R^{n\dagger} F_L^n + F_L^{n\dagger} F_R^n \right) \, ,
\label{1_SKK_1}
\end{equation}
where $\alpha_{L/R}^n =1$ if $f_{L/R}^n(y) \neq 0$, and $\alpha_{L/R}^n =0$ if $f_{L/R}^n(y) = 0$. Inserting the KK decomposition \eqref{1_KK_1} and the Dirac equations \eqref{1_Dirac_1} into the 5D Euler-Lagrange equations \eqref{1_ELE_1}, one can directly extract the differential equations for the KK wave functions,
\begin{equation}
\forall n \, , \ 
\left\{
\begin{array}{r c l}
\partial_y f_R^n(y) - m_n \, f_L^n(y) &=& 0,
\\ \vspace{-0.2cm} \\
\partial_y f_L^n(y) + m_n \, f_R^n(y) &=& 0,
\end{array}
\right.
\label{1_ELE_2}
\end{equation}
whose solutions for the zero modes ($m_0=0$) are
\begin{equation}
f_{L/R}^0(y) = A_{L/R}^{f, 0} \, ,
\label{1_profiles_00}
\end{equation}
where the $A^{f,0}_{L/R}$'s are complex coefficients. For the massive modes ($m_n \neq 0$), the first order differential equations~\eqref{1_ELE_2} can be combined into second order equations,
\begin{equation}
\forall n, \, \left( \partial_y^2 + m_n^2 \right) f_{L/R}^n(y) = 0,
\label{1_ELE_3}
\end{equation}
which are the equations for independent harmonic oscillators whose solutions have the general form
\begin{equation}
\forall n, \ f_{L/R}^n(y) = A^{f,n}_{L/R} \, \cos(m_n \, y) + B^{f,n}_{L/R} \, \sin(m_n \, y) \, ,
\label{1_profiles_0}
\end{equation}
where $A^{f,n}_{L/R}$, $B^{f,n}_{L/R}$ are complex coefficients. Eq.~\eqref{1_ELE_2} is a system of first order coupled equations which imposes the relations $A^{f,n}_{L} = B^{f,n}_{R}$ and $A^{f,n}_{R} = - B^{f,n}_{L}$ for $m_n \neq 0$.


Now, inserting Eq.~\eqref{1_KK_1} into the boundary conditions~\eqref{1_BC_1}, we obtain the following Dirichlet boundary conditions for any KK wave function
\begin{equation}
\forall n, \ f_{L/R}^n(0) =  f_{L/R}^n(L) = 0.
\label{1_profiles_vanish}
\end{equation}
For massive modes, a Dirichlet boundary condition for a KK wave function associated to one chirality, $f_{L/R}^n(0,L) = 0$, combined with Eq.~\eqref{1_ELE_2}, gives a Neumann boundary condition for the KK wave function of the other chirality, $\partial_y f_{R/L}^n(0,L) = 0$. Hence, the KK wave functions have both Dirichlet and Neumann boundary conditions at each boundary. When applied to the expression of the KK wave functions in Eqs.~\eqref{1_profiles_0}-\eqref{1_profiles_0}, the boundary conditions lead to $\forall  n, \, A^{f,n}_{L} = A^{f,n}_{R} = B^{f,n}_{L} = B^{f,n}_{R} = 0$, hence
\begin{equation}
\forall n, \, f_{L/R}^n(y)=0,
\label{1_zero_profiles}
\end{equation} 
which means that there are no KK modes. This problem comes from the fact that the system is overconstrained at the boundaries. Indeed, the KK wave function equations \eqref{1_ELE_2} relate $f_L^n(y)$ and $f_R^n(y)$ on-shell for the massive modes: a boundary condition on $f_L^n(y)$ is also a constraint on $f_R^n(y)$ and vice versa. Therefore, the KK wave functions $f_L^n(y)$ and $f_R^n(y)$ for $m_n \neq 0$ depend on the same three parameters: the mass $m_n$ and the two complex coefficients $B^{f, n}_{L/R}$. For the zero modes, $f_L^0(y)$ and $f_R^0(y)$ depend each on one complex coefficient $A_{L/R}^{f, 0}$. The variation of the action at the boundaries in Eqs.~\eqref{1_HVP_1a}-\eqref{1_HVP_1b} involves the variations of both $F_L$ and $F_R$ so there are two natural boundary conditions for each boundary. The system is thus overconstrained, that is why the KK wave functions vanish everywhere for all modes.

In order to have a non-trivial 5D fermion model interesting for physics, we have to choose another kind of boundary conditions (see the next Subsection~\ref{1_EBC_currents}). In Subsection~\ref{1_NBC_with_AFHS}, we show that if we modify our model by introducing the BBTs \eqref{1_eq:actionBound}, we avoid the above problems.

\subsection{Essential Boundary Conditions from Conserved Currents}
\label{1_EBC_currents}
Regarding the geometrical field configuration within the present model, 
each fermion is defined only along the interval $[0,L]$. This model building hypothesis, that fermions do not propagate outside the interval, translates into the condition of a vanishing conserved current transversal to the branes at both boundaries for each fermion species separately (without possible compensations) -- as 
was discussed in Ref.~\cite{Cheng:2010pt} for a scalar field case.

Formally speaking, there are two conserved currents defined independently for the two bulk fermions, involving the 5D fields $F=Q,D$:
\begin{equation}
j^M_Q = \bar{Q} \, \Gamma^M Q, \  j^M_D = \bar{D} \, \Gamma^M D, \phantom{0} \text{with the local conservation law} \phantom{0} \partial_M j_F^M = 0 \, ,
\label{1_courant_1}
\end{equation}
as predicted by Noether's theorem applied to the global fermionic symmetry $U(1)_Q \times U(1)_D$ of the free action in Eq.~\eqref{1_eq:actionKin}. The symmetries are based on the distinct transformations,
\begin{equation}
Q \mapsto \text{e}^{-i \alpha_q} Q \ , \ D \mapsto \text{e}^{-i \alpha_d} D,
\label{1_symetrie_1} 
\end{equation}
where $\alpha_f = \alpha_q, \, \alpha_d$ ($\in \mathbb{R}$) are the continuous parameters of the transformations. Infinitesimal field varia\-tions give $\delta F = - i \alpha_f F$, $\delta \bar F = i \alpha_f \bar F$. The condition that the component of the conserved currents transverse to the branes vanishes at the boundaries is thus,
\begin{equation}
\left. j_F^4 \right |_{0, L}= \left. \bar{F} \, \Gamma^4 F \right |_{0, L} = i \left. \left( F^\dagger_L F_R - F^\dagger_R F_L \right) \right|_{0, L} = 0 \ ,
\label{1_courant_2}
\end{equation}
It implies the following possible Dirichlet boundary conditions,
\begin{equation}
\left. F_L \right |_{0}=0 \  \text{or} \ \left. F_R \right |_{0}=0  \phantom{0} \text{and} \phantom{0}   \left. F_L \right |_{L}=0 \  \text{or} \ \left. F_R \right |_{L}=0,
\label{1_courant_3}
\end{equation}
which translate into boundary conditions for the KK wave functions after the KK decomposition \eqref{1_KK_1},
\begin{equation}
\forall n, \, f^n_L(0)=0 \  \text{or} \ f^n_R(0)=0 , \phantom{0} \text{and} \phantom{0}  f^n_L(L)=0 \  \text{or} \ f^n_R(L)=0.
\label{1_courant_4}
\end{equation}

One can impose \textit{ad hoc} conditions on the fields at the boundaries from the beginning, instead of using natural boundary conditions derived from the variational calculus \cite{Hilbert}. Such boundary conditions will be called essential boundary conditions. Having fixed fields at boundaries is equivalent to have vanishing functional variations, 
$\left . \delta F_L \right |_{0, L}  = 0$ or $\left . \delta F_R \right |_{0, L}  = 0$. The conditions~\eqref{1_courant_3} of vanishing fields at the boundaries can be imposed as essential boundary conditions: they correspond to some fields initially fixed at the boundaries [geometrical field configuration].
This is in contrast to our first treatment [above], where all the boundary fields $\left . F_{L/R} \right |_{0, L}$ were assumed to be initially arbitrary, and then we obtained the natural boundary conditions through Hamilton's principle. In the literature, Dirichlet boundary conditions are imposed by hand as essential boundary conditions \cite{Csaki:2003sh, Gherghetta:2010cj}. Here, we justify these essential boundary conditions by the requirement of vanishing probability currents at the boundaries.

Keeping these new essential boundary conditions~\eqref{1_courant_3}-\eqref{1_courant_4} in mind, 
let us apply Hamilton's principle to the action \eqref{1_L_2} as in Eqs.~\eqref{1_HVP_1a}-\eqref{1_HVP_1b}. The vanishing of the bulk variations is identical to the previous treatment and leads to the same equations \eqref{1_ELE_1}. However, the boundary variations -- in the two-component notations --
\begin{equation}
\left. \left( \delta F_L^\dagger F_R \right) \right|_{0,L} = \left. \left( \delta F_R^\dagger F_L \right) \right|_{0,L} = 0,
\label{1_boundary_term}
\end{equation}
are vanishing this time by the choice of one of the four possible essential boundary conditions~\eqref{1_courant_3} which read
\begin{eqnarray}
1) \ \left. F_L \right |_{0}=0   \phantom{0} \text{and} \phantom{0}   \left. F_L \right |_{L}=0 ,
& \ \ \ 2) \ \left. F_L \right |_{0}=0  \phantom{0} \text{and} \phantom{0}   \left. F_R \right |_{L}=0 ,
\nonumber  \\ 3) \ \left. F_R \right |_{0}=0  \phantom{0} \text{and} \phantom{0}   \left. F_L \right |_{L}=0 ,
& \ \ \ 4) \ \left. F_R \right |_{0}=0  \phantom{0} \text{and} \phantom{0}   \left. F_R \right |_{L}=0 ,
\label{1_EBC} 
\end{eqnarray}
recalling that $\left. F_{L/R} \right |_{0,L}=0$ implies $\left. \delta F_{L/R} \right |_{0,L}=0$. In other words, for instance, if $\left. F_{L}^{(\dagger)} \right |_{0}=0$ then there is no need to minimize the action with respect to this known field at the boundary so that the corresponding variation term [involving $\left. \delta F^\dagger_{L} \right |_{0}$] in Eq.~\eqref{1_boundary_term} should be absent. Hence to obtain complete boundary conditions, we have to combine only 
the essential boundary conditions~\eqref{1_EBC} with the KK wave functions equations~\eqref{1_ELE_2} [and their solutions~\eqref{1_profiles_0}]. For the fields which are not fixed at the boundaries by essential boundary conditions, their variation must be generic. We obtain the following four possible sets of wave functions and KK mass spectrum
equations ($\forall n$),
\begin{eqnarray}
1) & (--): \ f_{L}^n(y) = - A^{f,n}_{R} \; \sin(m_n \; y), \  (++): \ f_{R}^n(y) = A^{f,n}_{R} \; \cos(m_n \; y) \, ; \  \sin(m_n \; L)=0,
\nonumber  \\ 
2) & (-+): \ f_{L}^n(y) = - A^{f,n}_{R} \; \sin(m_n \; y), \  (+-): \ f_{R}^n(y) = A^{f,n}_{R} \; \cos(m_n \; y) \, ; \  \cos(m_n \; L)=0,
\nonumber  \\ 
3) & (+-): \ f_{L}^n(y) = A^{f,n}_{L} \; \cos(m_n \; y) , \  (-+): \ f_{R}^n(y) = A^{f,n}_{L} \; \sin(m_n \; y) \, ; \  \cos(m_n \; L)=0,
\nonumber  \\ 
4) & (++): \ f_{L}^n(y) = A^{f,n}_{L} \; \cos(m_n \; y) , \  (--): \ f_{R}^n(y) = A^{f,n}_{L} \; \sin(m_n \; y) \, ; \  \sin(m_n \; L)=0,
\nonumber  \\ \label{1_completeEBC} 
\end{eqnarray}
In Eq.~\eqref{1_completeEBC}, we have used the standard boundary condition notations, i.e. $-$ or $+$ for example at $y=0$ stands respectively for the Dirichlet or Neumann KK wave function 
boundary condition: $f_{L/R}^n(0) = 0$ or $\partial_y f_{L/R}^n(0) = 0$. For instance $(-+)$ denotes Dirichlet (Neumann) boundary condition at $y=0$ ($y=L$). For the boundary conditions 1) and 2), we would like to know if the phase of $A_L^{f,n}$ is physical. For that purpose, we perform the transformations:
\begin{equation}
A_L^{f,n} \mapsto \text{e}^{i \theta_n} A_L^{f,n} \ \ \ \Longrightarrow \ \ \ (f_L^n, f_R^n) \mapsto (\text{e}^{i \theta_n} f_L^n, \text{e}^{i \theta_n} f_R^n) \, ,
\end{equation}
which let the KK wave functions equations \eqref{1_ELE_2} and the orthonormalization conditions \eqref{1_normalization_1} invariant, thus the phase of $A_L^{f,n}$ is not physical and one can take $A_L^{f,n} = |A_L^{f,n}|$. For the boundary conditions 3) and 4), the same method is applied to conclude that the phase of $A^{f,n}_{R}$ is not physical. We take $A^{f,n}_{R} = |A^{f,n}_{R}|$.
The boundary conditions 1) and 4) have the following solutions for the KK mass spectrum: 
\begin{eqnarray}
\sin(m_n \, L)=0 \ \Rightarrow \ m_n= n \, \frac{\pi}{L} \ , \ n \in \mathbb{N} \ .
\label{1_PossSolSpctr} 
\end{eqnarray}
As $m_n = - n \pi/L$ is also solution of the mass spectrum equation \eqref{1_PossSolSpctr}, we will show that the sign of $m_n$ is not physical. One can perform the transformations
\begin{equation}
\left\{
\begin{array}{l}
m_n \mapsto -m_n \\
F_R^n \mapsto - F_R^n \ \ \ \text{or} \ \ \ F_L^n \mapsto - F_L^n
\end{array}
\right.
\end{equation}
Then, the 4D actions of each KK modes \eqref{1_SKK_1} (and thus the Dirac equations \eqref{1_Dirac_1}), the KK wave functions equations \eqref{1_ELE_2} and the orthonormalization conditions \eqref{1_normalization_1} (using the parity of the solutions \eqref{1_completeEBC}) are invariant. One can conlude that the sign of $m_n$ is not physical and take $m_n \geq 0$. The boundary conditions 2) and 3) have the solutions:
\begin{equation}
\cos(m_n \, L)=0 \ \Rightarrow \ m_n = (2n+1)\, \frac{\pi}{2L} \ , \ n \in \mathbb{N}.
\label{1_PossSolSpctr2} 
\end{equation}
$m_n = - (2n+1) \pi/(2L)$ is also solution of the mass spectrum equation \eqref{1_PossSolSpctr2}. By using the same method as above, one can show that the sign of $m_n$ is not physical and take $m_n \geq 0$. Therefore, in the case of the first complete boundary conditions above, the 5D field $F$ possesses a zero mode ($m_0=0$) with a KK wave function $f_{R}^0(y)$ associated to a right-handed chirality 4D field $F^0_{R}$ [{\it c.f.} Eq.~\eqref{1_KK_1}], since $f_{L}^0(y)=0$ -- in contrast to the second complete boundary conditions without zero mode ($m_0 \neq 0$). The constants $|A^{f,n}_{L}|$ and $|A^{f,n}_{R}|$ are fixed by the orthonormalization conditions~\eqref{1_normalization_1}.

Now let us impose on the 5D fields $Q,D$ the realistic complete boundary conditions in Eq.~\eqref{1_completeEBC} to fix the model. 
$D,Q$ are associated respectively to the first and fourth complete boundary conditions of Eq.~\eqref{1_completeEBC}, implying the existence of  zero mode 4D fields, which is realistic as,
in standard warped scenarios including Yukawa interactions, the various measured quark masses are expected to originate mainly from the EWSB mechanism.
Furthermore, within an extended realistic warped model (as described in Subsection~\ref{1_extension}) the $Q$ field is promoted to a SM $SU(2)_W$ gauge doublet, which is consistent with
the left-handed chirality of the zero mode 4D field $Q^0_{L}$ imposed by the fourth complete boundary conditions, while $D$ is the $SU(2)_L$ singlet down-quark, compatible with
the right-handed chirality of $D^0_{R}$ driven by the first complete boundary conditions. The KK wave functions associated to these realistic boundary conditions are
\begin{equation}
\left\{
\begin{array}{l}
q_L^0(y) = 1 \, , \ \ \ q_R^0(y) = 0 \, , \\
d_L^0(y) = 0 \, , \ \ \ d_R^0(y) = 1 \, ,
\end{array}
\right.
\label{solfree1}
\end{equation}
for the massless zero modes, and
\begin{equation}
\left\{
\begin{array}{l}
q_L^n(y) = \sqrt{2} \, \cos (m_n \, y) \, , \ \ \ 
q_R^n(y) =  \sqrt{2} \, \sin (m_n \, y) \, , \\
d_L^n(y) = - \sqrt{2} \, \sin (m_n \, y) \, , \ \ \ 
d_R^n(y) =  \sqrt{2} \, \cos (m_n \, y) \, ,
\end{array}
\right.
\label{solfree2}
\end{equation}
for the excited modes with $m_n = n \, (\pi / L)$, $n \in \mathbb{N}^*$.
The second and third complete boundary conditions of Eq.~\eqref{1_completeEBC} are not realistic for a describtion of SM fermions, since they 
predict neither zero mode 4D fields nor chiral states (only vector-like modes). A chiral model 
can thus be recovered at low-energy for a specific choice of complete boundary conditions on the boundaries of the interval.

One can notice that the action $S_\Psi$ \eqref{1_eq:actionKin} has an accidental global $SU(2)$ symmetry whose fundamental representation acts on the doublet $(Q,D)$. However, we do not want this symmetry in our model: our choice of boundary conditions $1)$ and $4)$ in Eq.~\eqref{1_completeEBC} break this $SU(2)$ symmetry explicitely. Indeed, one can check that the component of the associated Noether's current transverse to the 3-branes is not vanishing for these boundary conditions. This is not a problem since the $SU(2)$ symmetry is not in our list of symmetries which define our model.

\subsection{Natural Boundary Conditions with Bilinear Boundary Terms}
\label{1_NBC_with_AFHS}
We propose an alternative solution to the problem of 5D free fermions with the extra dimension compactified on an interval. Instead of imposing essential boundary conditions when performing Hamilton's principle with the action $S_\Psi$ in Eq.~\eqref{1_L_2}, we add the boundary terms \eqref{1_HS-terms} allowing for natural boundary conditions. This is the point of view adopted in Ref.~\cite{Lalak:2001fd} for a 5D scalar field and 5D gravity, and in Refs.~\cite{vonGersdorff:2004eq, vonGersdorff:2004cg} for a 5D symplectic-Majorana fermion. We will see that the two methods give the same physical mass spectrum as in the previous subsection if the boundary terms have fixed coefficients. Essential boundary conditions and natural boundary conditions with BBTs correspond to two ways to describe the same model.

Let us apply Hamilton's principle by considering the sum of the actions~\eqref{1_L_2} and \eqref{1_HS-terms} involving the two-component spinor fields:
\begin{align}
0 = \delta_{F^\dagger_L} (S_{\Psi}+S_{BBT}) &= \displaystyle{ \int d^4x \, \int_0^{L} dy \,  
\delta F^\dagger_L \left[ i\bar{\sigma}^\mu \partial_\mu F_L - \partial_y F_R  \right] } 
\nonumber \\
&\displaystyle{ - \dfrac{1}{2} \int d^4x \left[ 1+\mu_F(0) \right] \left. \left( \delta F^\dagger_L F_R \right) \right|_{0} }
\nonumber \\
&\displaystyle{ + \dfrac{1}{2} \int d^4x \, \left[ 1-\mu_F(L) \right] \left. \left( \delta F^\dagger_L F_R \right) \right|_{L} } \ ,
\label{1_HVP_3a}
\end{align}
\begin{align}
0 = \delta_{F^\dagger_R} (S_{\Psi}+S_{BBT}) &= \displaystyle{ \int d^4x \int_0^{L} dy \,  
\delta F^\dagger_R \left[ i\sigma^\mu \partial_\mu F_R + \partial_y F_L  \right] } 
\nonumber \\
&\displaystyle{ + \dfrac{1}{2} \int d^4x \left[ 1-\mu_F(0) \right] \left. \left( \delta F^\dagger_R F_L \right) \right|_{0} }
\nonumber \\
&\displaystyle{ - \dfrac{1}{2} \int d^4x \, \left[ 1+\mu_F(L) \right] \left. \left( \delta F^\dagger_R F_L \right) \right|_{L} } \ .
\label{1_HVP_3b}
\end{align}
For generic field variations $\left . \delta F^\dagger_{L/R} \right |_{0, L}$, 
the bulk and brane variations in Eqs.~\eqref{1_HVP_3a} and \eqref{1_HVP_3b} have to vanish separately, leading to the same equations as in Eq.~\eqref{1_ELE_1} and in turn, 
via Eqs.~\eqref{1_Dirac_1} and \eqref{1_KK_1}, to the Euler-Lagrange equations~\eqref{1_ELE_2} with solutions~\eqref{1_profiles_00}-\eqref{1_profiles_0}. We are thus left with the natural boundary conditions (generic field variations at the boundaries):
\begin{equation}
\left[ 1+\mu_F(0) \right] \left. F_R \right|_0 = \left[ 1-\mu_F(L) \right] \left. F_R \right|_L = \left[ 1-\mu_F(0) \right] \left. F_L \right|_0 = \left[ 1+\mu_F(L) \right] \left. F_L \right|_L = 0.
\end{equation}
For $\mu_F(0) \neq \pm 1$ and $\mu_F(L) \neq \pm 1$, we get the same boundary conditions as in Eq.~\eqref{1_BC_1} which, after a KK decomposition \eqref{1_KK_1} and solving the KK wave functions equations \eqref{1_ELE_2}, leads to vanishing KK wave functions everywhere \eqref{1_profiles_vanish} in the bulk as in Subsection~\ref{1_NBC_only}. Indeed, the system is still overconstrained by the natural boundary conditions (see discussion below Eq.~\eqref{1_zero_profiles}). In order to obtain interesting natural boundary conditions for physics, one should take $\mu_F(0) = \pm 1$ and $\mu_F(L) = \pm 1$. In these cases, the variation of the action at the boundaries (Eq.~\eqref{1_HVP_3b}) involves only one of the two chiralities: $F_L$ or $F_R$. This implies that there is only one natural boundary condition for each boundary instead of two:
\begin{align}
\mu_F(0) = 1 \ \Rightarrow \ \left. F_R \right|_0 = 0, \ \ \
\mu_F(0) = -1 \ \Rightarrow \ \left. F_L \right|_0 = 0, \nonumber \\
\mu_F(L) = 1 \ \Rightarrow \ \left. F_L \right|_L = 0, \ \ \
\mu_F(L) = -1 \ \Rightarrow \ \left. F_R \right|_L = 0.
\end{align}
Performing a KK decomposition \eqref{1_KK_1} and using Eq.~\eqref{1_ELE_2}, one obtains the following boundary conditions for the KK wave functions:
\begin{equation}
\begin{array}{l l l l l l l l}
1) & \mu_F(0) = - \mu_F(L) = -1 & & \Rightarrow & & (--): \ f_{L}^n(y), & (++): \ f_{R}^n(y), \\ 
2) & \mu_F(0) = \mu_F(L) = -1 & & \Rightarrow & & (-+): \ f_{L}^n(y), & (+-): \ f_{R}^n(y), \\ 
3) & \mu_F(0) = \mu_F(L) = 1 & & \Rightarrow & & (+-): \ f_{L}^n(y), & (-+): \ f_{R}^n(y), \\ 
4) & \mu_F(0) = - \mu_F(L) = 1 & & \Rightarrow & & (++): \ f_{L}^n(y), & (--): \ f_{R}^n(y),
\end{array}
\label{1_AFHS_BCs}
\end{equation}
where we used the same numbering as in Eq.~\eqref{1_completeEBC}. The associated KK wave function solutions and mass spectrum are the same (the system is not overconstraines anymore). As in Subsection~\ref{1_EBC_currents}, we take the boundary conditions 1) and 4) respectively for the fields $D$ and $Q$, so the boundary terms are the ones of Eq.~\eqref{1_eq:actionBound}.

The kinetic terms \eqref{1_eq:actionKin} and the BBTs \eqref{1_HS-terms} are invariant under the global symmetries $U(1)_Q$ and $U(1)_D$ so there are conserved currents $J_Q^M$ and $J_D^M$ (see Appendix~\ref{Hamilton_Noether} for a discussion on Noether's theorem with boundary terms). From Eq.~\eqref{cons_current_app}, the conserved currents are
\begin{equation}
J_F^M = j_F^M \, \Theta_I, \ \ \ \Theta_I(y) = \theta(y) - \theta(y-L)
\label{1_current_3000}
\end{equation}
with $j_F^M$ defined as in Eq.~\eqref{1_courant_1} and where $\theta(y)$ is the Heaviside distribution. $j_F^4$ is given by Eq.~\eqref{1_courant_2}. The 5-divergence of this conserved current vanishes:
\begin{equation}
\partial_M J_F^M = \partial_M j_F^M \, \Theta_I + j_F^4 \, \left[ \delta(y)-\delta(y-L) \right] = 0 \, .
\label{1_current_4000}
\end{equation}
From the conservation of $j_F^M$ in the bulk, we have
\begin{equation}
\partial_M j_Q^M = 0.
\label{1_conserved_current_000}
\end{equation}
The fields are differentiable with continuous derivatives on $[0,L]$, thus Eq.~\eqref{1_conserved_current_000} can be extended to the 3-branes. Using this last result in \eqref{1_current_4000}, we have
\begin{align}
\partial_M J_F^M &= j_F^4 \, \left[ \delta(y)-\delta(y-L) \right] = 0 \, .
\end{align}
We get Eq.~\eqref{1_courant_2}, which is guaranteed by the boundary conditions \eqref{1_AFHS_BCs}. Therefore, using this method, the boundary conditions on the currents are a consequence of Hamilton's principle instead of a requirement imposed by essential boundary conditions.

The global $SU(2)$ symmetry of the action \eqref{1_eq:actionKin} is not a symmetry of the boundary terms \eqref{1_eq:actionBound} and thus of the natural boundary conditions through Hamilton's principle. Here, the symmetry is broken explicitely in the action. In Subsection~\ref{1_EBC_currents}, the action was invariant under the $SU(2)$ symmetry but it was broken by the choice of essential boundary conditions.

\subsection{Summary}
\begin{figure}[!h]
\begin{center}
\includegraphics[width=15cm]{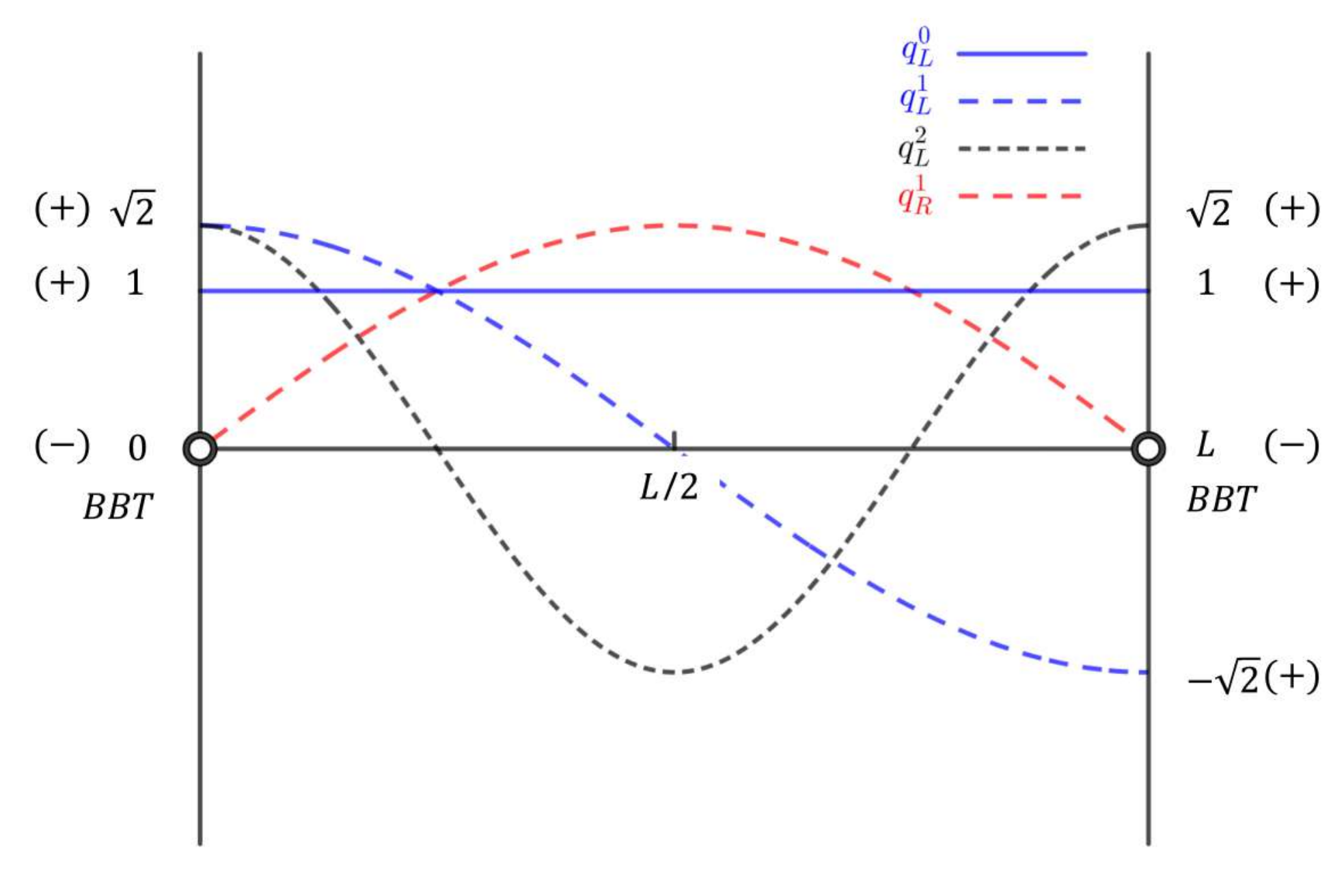}
\end{center}
\caption{Free KK wave functions of the 5D fields $Q$ and $D$.
}\label{1_prof_interval_Free}
\end{figure}

In Fig.~\ref{1_prof_interval_Free}, we give a plot of the free profiles along the extra dimension.

\begin{figure}[!h]
\begin{center}
\includegraphics[height=8cm]{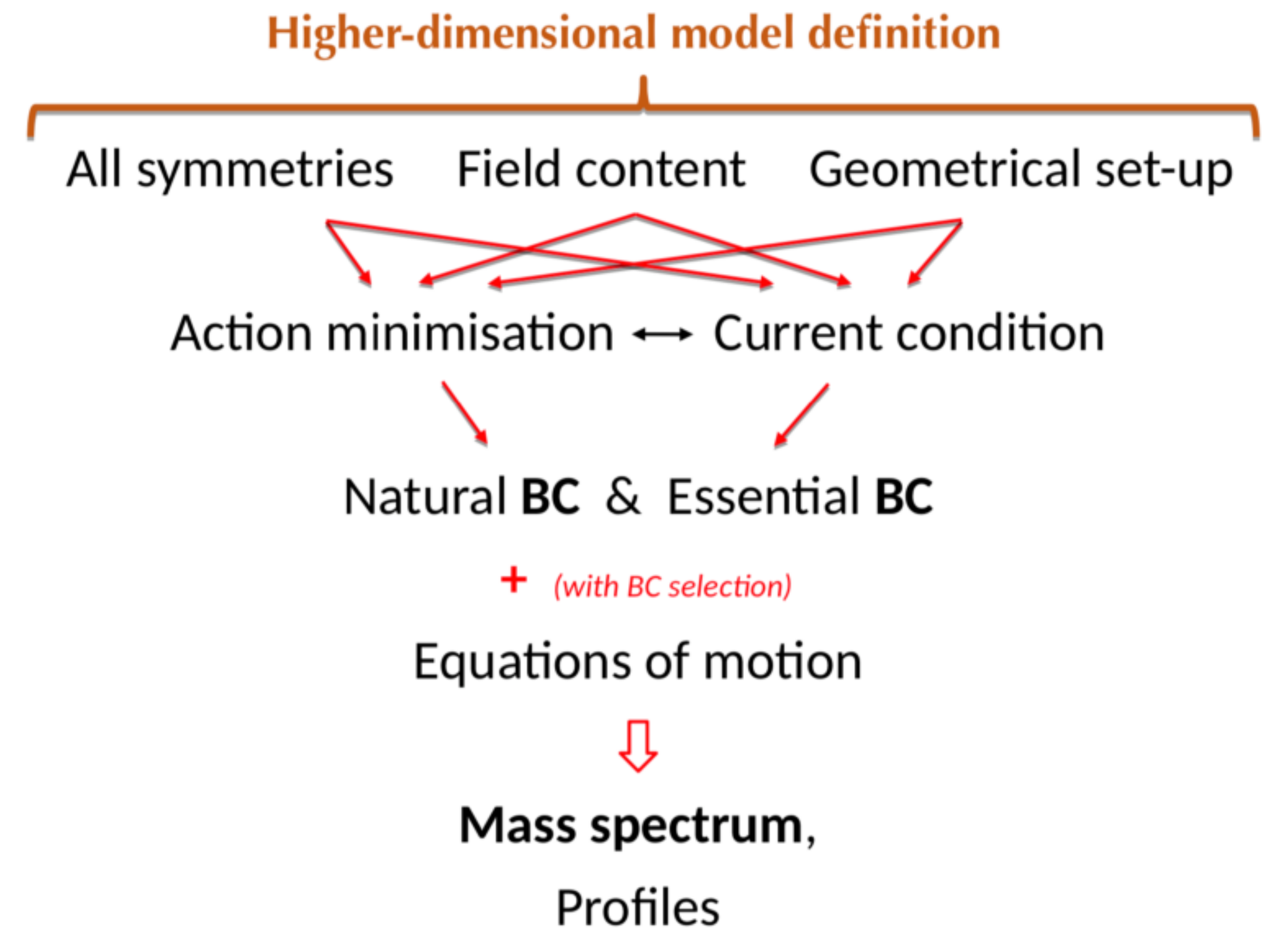}
\end{center}
\caption[Inverse pyramidal picture of the method]{
Inverse pyramidal picture illustrating the general principle for determining KK wave functions and masses within a given model based on extra dimension(s) (BC means ``Boundary Condition'').
}\label{1_fig:Pyramidal}
\end{figure}

We summarize and provide an overview of the method followed in this section to obtain the fermion mass spectrum and KK wave functions along the spacelike extra dimension(s) -- 
allowing to calculate the 4D effective Yukawa couplings which had not been considered here. For this purpose, we present in 
Fig.~\ref{1_fig:Pyramidal} a general schematic description of this method for a well-defined extra-dimensional model.
The figure must be understood as follows. A given extra-dimensional model is defined by its geometrical set-up (spacetime structure and geometrical field configuration), 
its field content and its internal (gauge groups,\dots) 
as well as other kinds of symmetries (restricted to the Poincar\'e group here) of symmetries. These ingredients determine entirely the action
from which Noether's theorem predicts 
the conserved currents. The basic assumptions on the boundaries of the internal space where the fields propagate translate 
into conditions on these currents~\footnote{The geometrical condition imposed 
along the three usual spatial dimensions being that the fields and their derivatives must decrease quickly at large distance and vanish at infinity.}. In turn those geometrical field configurations fix some fields at the boundaries. One can use essential boundary conditions (and hence impose vanishing field variations) or natural boundary conditions with well chosen boundary terms in the action. The KK decomposition allows to separate the Euler-Lagrange equations into the ones for the 4D fields and for the 
KK wave functions along the extra dimension(s). The last step is to solve the KK wave function equation, together with the previously obtained complete boundary conditions, to find the mass spectrum.

\section{4D Fermion Mass Matrix including Yukawa couplings}
\label{1_Mass_matrix_diagonalisation}
In this subsection, we describe the two steps of a first method~\cite{Goertz:2008vr, Barcelo:2014kha} to include the effects of the Yukawa terms~\eqref{1_L_5} on the final fermion mass spectrum. 
First, the free KK wave functions and free spectrum are calculated within a 5D approach whose treatment was exposed in detail in Section~\ref{1_free_bulk_fermions}.
Secondly, one can write a mass matrix for the 4D fermion fields involving the pure KK masses (the free spectrum of the first step) as well as the masses induced by the Higgs 
VEV in the Yukawa terms~\eqref{1_L_5} (with free KK wave functions of first step) which mix the KK modes. The bi-diagonalization of this matrix gives rise to an infinite set of eigenvalues 
constituting the physical masses, as will be presented here.

We focus on the interval compactification with the fermion terms of the 5D action~\eqref{1_eq:actionTot}, with or without $S_B$, in order to work out the mass spectrum. 
This action leads -- after insertion of the free KK decomposition~\eqref{1_KK_1}, the use of the orthonormalization condition~\eqref{1_normalization_1} when one integrates over the fifth dimension -- to the canonical kinetic terms for the 4D fermion fields as well as to the following fermionic 4D effective mass terms in the Lagrangian density (and to independent 4D effective Higgs-fermion couplings not discussed here),
\begin{equation}
- \chi^\dagger_L \, \mathcal{M} \, \chi_R + \text{H.c.}
\end{equation}
where $\chi_L$ and $\chi_R$ are the free KK bases (Eq.~\eqref{1_KK_1}) for the left and right-handed 4D fields\footnote{Notice that there exists only one chirality for the zero modes as explained above with the free solutions~\eqref{1_completeEBC}.}
\begin{equation}
\left\{
\begin{array}{r c l}
\chi_L &=& \left( Q_L^{0}, D_L^{0}, Q_L^{1}, D_L^{1}, Q_L^{2}, D_L^{2}, \cdots \right), \\ \\
\chi_R &=& \left( Q_R^{0}, D_R^{0}, Q_R^{1}, D_R^{1}, Q_R^{2}, D_R^{2}, \cdots \right),
\label{1_basis}
\end{array}
\right.
\end{equation}
and where the infinite mass matrix $\mathcal{M}$ reads (see Ref.~\cite{Barcelo:2014kha})
\begin{equation}
\mathcal{M} =
\begin{pmatrix}
m_0 & \alpha_{00} & 0 & \alpha_{01} & 0 & \alpha_{02} & \cdots \\
\beta_{00} & m_0 & \beta_{01} & 0 & \beta_{02} & 0 & \cdots \\
0 & \alpha_{10} & m_{1} & \alpha_{11} & 0 & \alpha_{12} & \cdots \\
\beta_{10} & 0 & \beta_{11} & m_{1} & \beta_{12} & 0 & \cdots \\
0 & \alpha_{20} & 0 & \alpha_{21} & m_{2} & \alpha_{22} & \cdots \\
\beta_{20} & 0 & \beta_{21} & 0 & \beta_{22} & m_2 & \cdots \\
\vdots & \vdots & \vdots & \vdots & \vdots & \ddots \\
\end{pmatrix}.
\label{1_M}
\end{equation}
$m_n$ is the free spectrum of the first and fourth solutions of Eq.~\eqref{1_completeEBC},
and the off-diagonal terms $\alpha_{ij}$ and $\beta_{ij}$ describe the mixings between the KK modes (obtained in Section~\ref{1_free_bulk_fermions}) induced by the Yukawa couplings. They are given by the overlap of the free KK wave function with the Higgs-brane:
\begin{equation}
\left\{
\begin{array}{l}
\displaystyle{\forall (i, j) \in \mathbb{N}^2, \, \alpha_{ij} =  \dfrac{X}{L} \int dy \ \delta(y-L) \, q_L^i(y) \, d_R^j(y) = \dfrac{X}{L} \, q_L^i(L) \, d_R^j(L)}, \\ \\ \vspace{-0.8cm} \\
\displaystyle{\forall (i, j) \in \mathbb{N}^{\star 2}, \, \beta_{ij} = \dfrac{X'}{L} \int_0^L dy \ \delta(y-L) \, d_L^i(y) \, q_R^j(y) = \dfrac{X'}{L} \, d_L^i(L) \, q_R^j(L)}.
\end{array}
\right.
\label{1_coeff}
\end{equation}
From the first and fourth solutions of Eq.~\eqref{1_completeEBC} respectively for $d^n_{L/R}(y)$ and $q^n_{L/R}(y)$ 
in the studied model, $\beta_{ij}=0$.

The physical fermion mass spectrum is obtained by bi-diagonalizing the mass matrix~\eqref{1_M}. The eigenstate basis is then
\begin{equation}
\left\{
\begin{array}{r c l}
\psi_L &=& \left( \psi_L^{0}, \psi_L^{1},  \psi_L^{2}, \cdots \right), \\ \\
\psi_R &=& \left( \psi_R^{0}, \psi_R^{1},  \psi_R^{2}, \cdots  \right).   
\label{1_basis_2}
\end{array}
\right.
\end{equation}
This method is called the perturbative method in the sense that truncating the mass matrix at a given KK level corresponds to keeping only the dominant contributions
to the lightest mass eigenvalue being the measured fermion mass (higher KK modes tend to mix less with the zero mode due to larger mass differences).

Extracting the mass spectrum equation from the characteristic equation for the Hermi\-tian-squared mass matrix~\eqref{1_M}, in the case of infinite KK towers, is not trivial. 
This useful calculation was addressed analytically in Ref.~\cite{Barcelo:2014kha} for the present toy model with real Yukawa couplings $Y_5$ and $Y_5'$ in Eq.~\eqref{1_eq:actionYuk}. The resulting exact equation, without any approximation is
\begin{equation}
\forall n \in \mathbb{N}, \ \tan^2(|M_n| \, L) = X^2,
\end{equation}
so the mass spectrum $\left\{ |M_n| \right\} = \left\{ |M_k|, |M_l| \right\}$ is given by the two equations:
\begin{align}
\tan(|M_k| \, L) = X \ &\Rightarrow \ |M_k| = \dfrac{1}{L} \left[ k \pi + \arctan (X) \right], \ k \in \mathbb{N} \nonumber \\
\tan(|M_l| \, L) = -X \ &\Rightarrow \ |M_l| = \dfrac{1}{L} \left[ l \pi - \arctan (X) \right], \ l \in \mathbb{N^*}.
\end{align}
One can verify that in the limit of vanishing Yukawa couplings ($X \rightarrow 0$), the two KK spectra ${|M_k|}$ and ${|M_l|}$ converge to the two free spectra \eqref{1_PossSolSpctr} (one for each 5D field $Q$, $D$). Truncating the spectra at a given KK level $N$, one can see that there is a one-to-one correspondance between the KK modes. Rewriting the spectra with the unique label $n$, one gets:
\begin{equation}
|M_n| = \dfrac{1}{L} \left| n \pi + \arctan (X) \right|, \ n \in \mathbb{Z}.
\label{1_4D_spectrum}
\end{equation}

Let us notice that if one includes $S_B$ in Eq.~\eqref{1_eq:actionTot}, the boundary terms are included in the mass matrix \eqref{1_M}. Then the would-be induced 
4D mass matrix elements vanish since $S_B$ involves $(--)$ KK wave functions at $y=L$, like the $\beta_{ij}$ coefficients in Eq.~\eqref{1_coeff}. Hence the mass eigenvalues are unchanged.

\section{Usual 5D Treatment: the Regularization Doom}
\label{1_Usual_treatment}
In the previous Section~\ref{1_free_bulk_fermions}, the fermion wave functions did not satisfy
equations of motion including interactions with the Higgs field localized
on the boundaries. In this part, we work out the fermion mass spectrum in the model defined by the 5D action~\eqref{1_eq:actionTot}, using 
the alternative 5D approach based on the brane-Higgs regularizations~\cite{Casagrande:2008hr, Azatov:2009na, Carena:2012fk, Malm:2013jia, Barcelo:2014kha} 
and we point out non-rigorous features of these methods.

\subsection{Mixed Kaluza-Klein Decomposition}
\label{1_5D_approach}
As we have seen in Eqs.~\eqref{1_basis} and \eqref{1_M}, after EWSB, the infinite towers of $Q^n_L$ and $D^n_L$ fields of the free KK basis \eqref{1_basis} mix (as well as the $Q^n_R$ and $D^n_R$) 
to form 4D fields $\psi^n_L$ (and $\psi^n_R$) of the mass eigenstate basis \eqref{1_basis_2}. In order to take into account this mixing within the 5D approach, we expand the 5D fields in the mass eigenstate basis \eqref{1_basis_2} instead of the free KK basis \eqref{1_basis}. It is called a \textit{mixed} KK 
decomposition [instead of the \textit{free} one in Eq.~\eqref{1_KK_1}]~\cite{Casagrande:2008hr, Azatov:2009na}, defined as follows,
\begin{equation}
\left\{
\begin{array}{r c l}
Q_L \left( x^\mu, y \right) &=& \dfrac{1}{\sqrt{L}} \displaystyle{ \sum_n q^n_L(y) \, \psi^n_L \left( x^\mu \right) },
\\ \vspace{-0.3cm} \\
Q_R \left( x^\mu, y \right) &=& \dfrac{1}{\sqrt{L}} \displaystyle{ \sum_n q^n_R(y) \, \psi^n_R \left( x^\mu \right) },
\\ \vspace{-0.3cm} \\
D_L \left( x^\mu, y \right) &=& \dfrac{1}{\sqrt{L}} \displaystyle{ \sum_n d^n_L(y) \, \psi^n_L \left( x^\mu \right) },
\\ \vspace{-0.3cm} \\
D_R \left( x^\mu, y \right) &=& \dfrac{1}{\sqrt{L}} \displaystyle{ \sum_n d^n_R(y) \, \psi^n_R \left( x^\mu \right) }.
\end{array} 
\right. 
\label{1_KK_2}
\end{equation}
The 4D fields $\psi^n_{L/R}$ ($\forall n$) must satisfy the Dirac-Weyl equations
\begin{equation}
\left\{
\begin{array}{r c l}
i \bar{\sigma}^\mu \partial_\mu \psi^n_L \left( x^\mu \right) &=& M_n \, \psi_R^n \left( x^\mu \right),
\\ \vspace{-0.2cm} \\
i \sigma^\mu \partial_\mu \psi^n_R \left( x^\mu \right) &=& M_n \, \psi_L^n \left( x^\mu \right),
\end{array}
\right.
\label{1_Dirac_2}
\end{equation}
where the spectrum $M_n$ includes contributions to masses from the Yukawa couplings~\eqref{1_L_5}. Note that in contrast to the free case, 
there is a unique mass spectrum $M_n$ for a unique 4D field tower $\psi^n_{L/R} (x^\mu)$. Fo each KK mode, one has the action:
\begin{equation}
S_\psi^n = i \psi_L^{n\dagger} \bar{\sigma}^\mu \partial_\mu \psi_L^n + i \psi_R^{n\dagger} \sigma^\mu \partial_\mu \psi_R^n - M_n \left( \psi_R^{n\dagger} \psi_L^n + \psi_L^{n\dagger} \psi_R^n \right) \, .
\label{1_SKK_2}
\end{equation}
In order to guarantee the existence of diagonal and canonical kinetic terms for the 4D fields $\psi^n_{L/R}$, the associated new KK wave functions must now obey the two following 
orthonormalization conditions,
\begin{equation}
\forall n \ , \forall m \ , \dfrac{1}{L} \int_0^{L} dy \, \left[ q^{n*}_{C}(y) \, q^m_{C}(y) + d^{n*}_{C}(y) \, d^m_{C}(y) \right] = \delta^{nm},
\label{1_normalization_2}
\end{equation}
for a chirality index $C \equiv L$ or $R$. These two conditions are different from the four ones of Eq.~\eqref{1_normalization_1} due to the new mixed KK decomposition.

\subsection{Shift of the Bounday Localized Higgs Field}
\label{1_RegInterval}



Here we highlight the formal problems of the 5D process by shifting the brane-Higgs field away from the boundary~\cite{Csaki:2003sh, Csaki:2005vy, Grojean:2007zz, Barcelo:2014kha} to get the fermion mass 
tower. The considered fermion terms of the 5D 
action~\eqref{1_eq:actionTot} are $S_{\Psi}$ and $S_{X}$ (without $S_{B}$ which was missed in the relevant literature and which will be taken into account 
in Subsection~\ref{1_Yukawa_terms_as boundary_conditions}). The variations of the studied action lead to the same free boundary conditions, $Q_R|_{0,L} = D_L|_{0,L} = 0$, and to the following bulk Euler-Lagrange equations 
including the Yukawa coupling constants present in the mass terms with $X$ and $X'$ [instead of the free ones in Eq.~\eqref{1_ELE_1}], 
\begin{equation}
\left\{
\begin{array}{r c l}
i \bar{\sigma}^\mu \partial_\mu Q_L - \partial_y Q_R - \delta(y-L) \, X \, D_R &=& 0\ ,
\\ \vspace{-0.2cm} \\
i \sigma^\mu \partial_\mu Q_R + \partial_y Q_L - \delta(y-L) \, X' \, D_L &=& 0\ ,
\\ \vspace{-0.2cm} \\
i \bar{\sigma}^\mu \partial_\mu D_L - \partial_y D_R - \delta(y-L) \, X' \, Q_R &=& 0\ ,
\\ \vspace{-0.2cm} \\
i \sigma^\mu \partial_\mu D_R + \partial_y D_L - \delta(y-L) \, X \, Q_L &=& 0 \ ,
\end{array}
\right.
\label{1_ELE_4}
\end{equation}
where $X$ and $X'$ are taken real. Indeed, in view of regularizing the brane-Higgs field, the Yukawa interactions must be included in the bulk Euler-Lagrange equations~\cite{Barcelo:2014kha} as done in the literature.
Inserting the mixed KK decomposition~\eqref{1_KK_2} into the 5D field Euler-Lagrange equations~\eqref{1_ELE_4} allows to factorize out the 4D fields, obeying the 4D Dirac equations~\eqref{1_Dirac_2}, 
and to obtain the KK wave function equations for each mode (instead of the free ones in Eq.~\eqref{1_ELE_2}):
\begin{equation}
\forall n\ , \ 
\left\{
\begin{array}{r c l}
\partial_y q^n_R(y) - M_n \, q^n_L(y) &=& - \ \delta(y-L) \, X \, d^n_R(y) \ ,
\\ \vspace{-0.2cm} \\
\partial_y q^n_L(y) + M_n \, q^n_R(y) &=& \delta(y-L) \, X' \, d^n_L(y) \ ,
\\ \vspace{-0.2cm} \\
\partial_y d^n_R(y) - M_n \, d^n_L(y) &=& - \ \delta(y-L) \, X' \, q^n_R(y) \ ,
\\ \vspace{-0.2cm} \\
\partial_y d^n_L(y) + M_n \, d^n_R(y) &=& \delta(y-L) \, X \, q^n_L(y) \ .
\end{array}
\right.
\label{1_ELE_5}
\end{equation}
Here we underline a first mathematical issue of this usual approach: 
introducing $\delta(y-L)$ Dirac distributions~\footnote{Strictly speaking, a Dirac distribution is a distribution although its historical name is ``Dirac delta function'' \cite{Dirac:1927we}.} 
in these KK wave function equations requires to treat the KK wave functions as distributions~\footnote{Also called ``generalized functions'' in mathematical analysis.}. The right-hand sides of the equations \eqref{1_ELE_5} involve products of the four KK wave functions and $\delta(y-L)$ which are well defined if the KK wave functions are continuous at $y=L$. As we will see, some of the KK wave functions are instead discontinuous at $y=L$.

The apparent ``ambiguity'' noticed in the literature (in the context of a warped extra dimension) was that the Yukawa terms in Eq.~\eqref{1_ELE_5} are present only at the $y=L$ boundary
and might thus affect the fermion boundary conditions. In order to avoid this question, a regularization of the brane-Higgs coupling was suggested 
allowing to maintain the free fermion boundary conditions in the presence of Yukawa interactions. Here is the main drawback of all the regularization methods: when applying Hamilton's principle, the boundary term from the integration by parts of the variation of the bulk action \eqref{1_L_2} has to vanish without the variation of the localized Lagrangian including the Yukawa couplings \eqref{1_L_5}, leading to free boundary conditions. One should instead impose the variation of the complete boundary term at $y=L$ to vanish, as we did in Subsection~\ref{1_Yukawa_terms_as boundary_conditions}.

\subsubsection{Regularization I Drawbacks}
For the type of regularization first applied in the literature~\cite{Csaki:2003sh, Csaki:2005vy, Grojean:2007zz}, called Regularization~I \cite{Barcelo:2014kha}, one begins by using the free boundary conditions $d^n_L(L)=q^n_R(L)=0$ (see respectively the first and fourth solutions in  
Eq.~\eqref{1_completeEBC}) in the system of equations~\eqref{1_ELE_5} which becomes
\begin{equation}
\forall n\ , \ 
\left\{
\begin{array}{r c l}
\partial_y q^n_R(y) - M_n \, q^n_L(y) &=& - \ \delta(y-L) \, X \, d^n_R(y) \ ,
\\ \vspace{-0.2cm} \\
\partial_y q^n_L(y) + M_n \, q^n_R(y) &=& 0 \ ,
\\ \vspace{-0.2cm} \\
\partial_y d^n_R(y) - M_n \, d^n_L(y) &=& 0 \ ,
\\ \vspace{-0.2cm} \\
\partial_y d^n_L(y) + M_n \, d^n_R(y) &=& \delta(y-L) \, X \, q^n_L(y) \ .
\end{array}
\right.
\label{1_ELE_5bis}
\end{equation}
In the next step of this method, the usual trick is to shift the brane-Higgs coupling from the brane at $y=L$  
(TeV-brane in a warped framework) by an amount $\epsilon$:
\begin{equation}
\forall n\ , \ 
\left\{
\begin{array}{r c l}
\partial_y q^n_R(y) - M_n \, q^n_L(y) &=& - \ \delta(y-[L-\epsilon]) \, X \, d^n_R(y) \ ,
\\ \vspace{-0.2cm} \\
\partial_y q^n_L(y) + M_n \, q^n_R(y) &=& 0 \ ,
\\ \vspace{-0.2cm} \\
\partial_y d^n_R(y) - M_n \, d^n_L(y) &=& 0 \ ,
\\ \vspace{-0.2cm} \\
\partial_y d^n_L(y) + M_n \, d^n_R(y) &=& \delta(y-[L-\epsilon]) \, X \, q^n_L(y) \ .
\end{array}
\right.
\label{1_ELE_5ter}
\end{equation}
Then the integration of the four equations~\eqref{1_ELE_5ter} over an infinitesimal range, tending to zero and centered at $y=L-\epsilon$, leads 
to~\footnote{The integration of Eq.~\eqref{1_ELE_5ter} could also be performed over the interval $[L-\epsilon,L]$; this variant of the calculation,
suggested in Refs.~\cite{Csaki:2003sh, Csaki:2005vy, Grojean:2007zz}, represents in fact an equivalent regularization process leading to the same physical results but 
with identical mathematical inconsistencies.}
\begin{equation}
\forall n\ , \ 
\left\{
\begin{array}{r c l}
q^n_R([L-\epsilon]^+)-q^n_R([L-\epsilon]^-) &=& - \, X \, d^n_R(L-\epsilon) \ ,
\\ \vspace{-0.2cm} \\
q^n_L([L-\epsilon]^+)-q^n_L([L-\epsilon]^-)   &=& 0 \ ,
\\ \vspace{-0.2cm} \\
d^n_R([L-\epsilon]^+)-d^n_R([L-\epsilon]^-)   &=& 0 \ ,
\\ \vspace{-0.2cm} \\
d^n_L([L-\epsilon]^+)-d^n_L([L-\epsilon]^-) &=& X \, q^n_L(L-\epsilon) \ .
\end{array}
\right.
\label{1_ELE_5integ}
\end{equation}
We see an inconsistency arising here in the regularization process: the first and fourth relations in Eq.~\eqref{1_ELE_5integ} show that the 
KK wave functions $q^n_R(y)$ and $d^n_L(y)$ possess a discontinuity at $y=L-\epsilon$, and thus at $y=L$ in the limit $\epsilon \to 0$, using the free boundary conditions~\footnote{The KK wave functions $q^n_L(y)$, $d^n_R(y)$ are usually assumed to be continuous at $y=L-\epsilon$ while 
$q^n_R(y)$, $d^n_L(y)$ remain unknown exactly at this point.}. Hence their product with the Dirac distribution centered at $y=L$ is not defined in the original equations \eqref{1_ELE_5}. As a consequence, the action~\eqref{1_L_5} is ill-defined since the Dirac distribution $\delta(y-L)$ multiplies the KK wave functions $q^n_R(y)$ and $d^n_L(y)$ not being continuous
at $y=L$.

In the following steps of this regularization~I, one solves Eq.~\eqref{1_ELE_5ter} first in the interval $[0,L-\epsilon)$ (bulk Euler-Lagrange equations without Yukawa couplings) 
and applies the free boundary conditions $d^n_L(0)=q^n_R(0)=0$. Then one solves Eq.~\eqref{1_ELE_5ter} on $(L-\epsilon,L]$ before applying the jump and continuity  
conditions~\eqref{1_ELE_5integ} at $y=L-\epsilon$ on the resulting KK wave functions. The last step is to apply the free boundary conditions at $y=L$ on these KK wave functions and take the limit $\epsilon \to 0$ 
(to recover the studied brane-Higgs model) of the boundary conditions. The obtained boundary conditions give rise to the equation whose solutions constitute the fermion mass spectrum with real Yukawa couplings \cite{Barcelo:2014kha},
\begin{equation}
\forall n \in \mathbb{N}, \ \tan^2(M_n \, L) = X^2
\label{1_RegulResultI} 
\end{equation}
exactly as the 4D approach result of Eq.~\eqref{1_4D_spectrum}.

\subsubsection{Regularization II Drawbacks}
Within the regularization~II~\cite{Barcelo:2014kha}, the Higgs coupling is first shifted along $y$ in the bulk equations~\eqref{1_ELE_5}, which become
\begin{equation}
\forall n\ , \ 
\left\{
\begin{array}{r c l}
\partial_y q^n_R(y) - M_n \, q^n_L(y) &=& - \ \delta(y-[L-\epsilon]) \, X \, d^n_R(y) \ ,
\\ \vspace{-0.2cm} \\
\partial_y q^n_L(y) + M_n \, q^n_R(y) &=& \delta(y-[L-\epsilon]) \, X' \, d^n_L(y) \ ,
\\ \vspace{-0.2cm} \\
\partial_y d^n_R(y) - M_n \, d^n_L(y) &=& - \ \delta(y-[L-\epsilon]) \, X' \, q^n_R(y) \ ,
\\ \vspace{-0.2cm} \\
\partial_y d^n_L(y) + M_n \, d^n_R(y) &=& \delta(y-[L-\epsilon]) \, X \, q^n_L(y) \ .
\end{array}
\right.
\label{1_ELE_5-shiftII}
\end{equation}
Integrating these four relations over an infinitesimal range centered at $y=L-\epsilon$ gives
\begin{equation}
\forall n\ , \ 
\left\{
\begin{array}{r c l}
q^n_R([L-\epsilon]^+)-q^n_R([L-\epsilon]^-) &=& - \, X \, d^n_R(L-\epsilon) \ ,
\\ \vspace{-0.2cm} \\
q^n_L([L-\epsilon]^+)-q^n_L([L-\epsilon]^-)   &=& X' \, d^n_L(L-\epsilon) \ ,
\\ \vspace{-0.2cm} \\
d^n_R([L-\epsilon]^+)-d^n_R([L-\epsilon]^-)   &=& - \, X' \, q^n_R(L-\epsilon) \ ,
\\ \vspace{-0.2cm} \\
d^n_L([L-\epsilon]^+)-d^n_L([L-\epsilon]^-) &=& X \, q^n_L(L-\epsilon) \ .
\end{array}
\right.
\label{1_ELE_5integII}
\end{equation}
This set of conditions shows that the four KK wave functions undergo a jump at $y=L-\epsilon$, in conflict with the continuity required by the products of the shifted Dirac distributions and the KK wave functions on the right-hand side of the equations \eqref{1_ELE_5-shiftII}. In other terms, the continuity conditions~\eqref{1_ELE_5integII} depend on the right-hand sides of the equations so that one must choose a value for each KK wave function exactly at $y=L-\epsilon$.
Taking a mean value weighted by a real number $c$~\footnote{Different values of $c$ correspond to physically equivalent regularizations based on different input
values of the Yukawa coupling constants (different coupling definitions).}, Eq.~\eqref{1_ELE_5integII} takes the form
\begin{equation}
\forall n\ , \ 
\left\{
\begin{array}{r c l}
q^n_R([L-\epsilon]^+)-q^n_R([L-\epsilon]^-) &=& - \, X \, \frac{d^n_R([L-\epsilon]^-)\, + \, c \ d^n_R([L-\epsilon]^+)}{1+c} \ ,
\\ \vspace{-0.2cm} \\
q^n_L([L-\epsilon]^+)-q^n_L([L-\epsilon]^-)   &=& X' \, \frac{d^n_L([L-\epsilon]^-)\, + \, c\  d^n_L([L-\epsilon]^+)}{1+c}  \ ,
\\ \vspace{-0.2cm} \\
d^n_R([L-\epsilon]^+)-d^n_R([L-\epsilon]^-)   &=& - \, X' \, \frac{q^n_R([L-\epsilon]^-)\, + \, c\  q^n_R([L-\epsilon]^+)}{1+c}  \ ,
\\ \vspace{-0.2cm} \\
d^n_L([L-\epsilon]^+)-d^n_L([L-\epsilon]^-) &=& X \, \frac{q^n_L([L-\epsilon]^-)\, +\, c\  q^n_L([L-\epsilon]^+)}{1+c}  \ .
\end{array}
\right.
\label{1_ELE_5weight}
\end{equation}
Looking at the left-hand sides of those four equations, one observes that jumps may arise at $y=L$ (in the limit $\epsilon \to 0$) for the four KK wave functions [for each 
$n^{th}$ mode]. Determining which KK wave functions are discontinuous requires to consider the free boundary conditions at $y=L$ (before applying the limit $\epsilon \to 0$), 
the various $c$ values (including infinity) and the four KK wave functions simultaneously [as they are related through Eq.~\eqref{1_ELE_5weight}].
The hypothesis that all of the four KK wave functions are continuous at $y=L-\epsilon$ (in the limit $\epsilon \to 0$) corresponds to the same field configuration as in the 
absence of Yukawa interactions and leads thus to a free fermion mass spectrum. This kind of solution was not considered in the literature since it
does not reproduce the SM at low-energies and is thus not realistic. Therefore, there exists at least one KK wave function discontinuous at $y=L$ which in turn involves ill-defined products with $\delta(y-L)$ in the original KK wave function equations \eqref{1_ELE_5}. Furthermore, the obtained discontinuous KK wave function [at $y=L$] multiplies $\delta(y-L)$ in 
Eq.~\eqref{1_L_5}, spoiling the mathematical validity of this action.

In the next steps of regularization~II, Eq.~\eqref{1_ELE_5-shiftII} is first solved over the domain $[0,L-\epsilon)$ (free bulk Euler-Lagrange equations) 
and the free boundary conditions at $y=0$ are applied to the resulting KK wave functions. Eq.~\eqref{1_ELE_5-shiftII} is then solved over $(L-\epsilon,L]$ before the jump/continuity  
conditions~\eqref{1_ELE_5weight} at $y=L-\epsilon$ are applied to the obtained KK wave functions. Finally the free boundary conditions at $y=L$ are implemented on those KK wave functions and one applies 
the limit $\epsilon \to 0$ to the boundary conditions. These boundary conditions lead to the equation giving the fermion mass spectrum with real Yukawa couplings \cite{Barcelo:2014kha},
\begin{equation}
\forall n \in \mathbb{N}, \ \tan^2(M_n \, L) = \left(\dfrac{4 X}{4 + X X'}\right)^2,
\label{1_RegulResultII}
\end{equation}
which can still be shown~\cite{Barcelo:2014kha} to be physically equivalent to the result of regularization~I~\eqref{1_RegulResultI} as the definition of the Yukawa couplings are also different in the two regularizations.

\subsection{Softening of the Boundary Localized Higgs Field}
\label{1_RegSoft}
Another type of regularization used in the literature (for warped models)~\cite{Azatov:2009na, Azatov:2010pf, Casagrande:2010si, Barcelo:2014kha} consists in replacing the
Dirac distribution $\delta(y-L)$ of Eq.~\eqref{1_L_5} by a normalized square function:
\begin{equation}
\delta^\epsilon (y-L) =
\left\{
\begin{array}{rcl}
\dfrac{1}{\epsilon} \, , & y \in [L-\epsilon, L] \, , \\ \\
0 & \text{otherwise,}
\end{array}
\right.
\end{equation}
with $\epsilon > 0$. The limit when $\epsilon \rightarrow 0$ is formaly the Dirac ``function'':
\begin{equation}
\delta (y-L) =
\left\{
\begin{array}{cl}
\infty  & \text{when} \ y = L \, , \\ \\
0 & \text{otherwise}
\end{array}
\right.
\end{equation}
as in Dirac's original article \cite{Dirac:1927we}. Of course, this is not a true function in the mathematical sense. This object has a meaning only in the framework of distribution theory built by L.~Schwartz \cite{Schwartz1, Schwartz2}. The Dirac distribution $\delta(y-L)$ is defined from its action on a function $\varphi(y)$ continuous at $y=L$ such that
\begin{equation}
\int_0^L dy \ \delta(y-L) \, \varphi(y) = \varphi(L) \, .
\end{equation}
What is allowed is to say that $\delta^\epsilon (y-L)$ converges weakly on $\delta (y-L)$ when $\epsilon \rightarrow 0$, which means that
\begin{equation}
\lim_{\epsilon \to 0} \int_0^L dy \ \delta^\epsilon (y-L) \, \varphi(y) = \int_0^L dy \ \delta(y-L) \, \varphi(y) = \varphi(L) \, .
\end{equation}
We insist on the fact that this is allowed because $\varphi(y)$ is continuous at $y=L$.

In Refs.~\cite{Azatov:2009na, Azatov:2010pf, Casagrande:2010si}, they found that by naively integrating the equations in the system \eqref{1_ELE_5} around $y=L$, the KK wave functions are discontinuous and they propose to regularize $\delta (y-L)$ with $\delta^\epsilon (y-L)$ precisely in a case where it is not allowed. Then, they solve the system of equations \eqref{1_ELE_5} in the two regions $[0, L-\epsilon)$ and $(L-\epsilon, L]$. The KK wave functions solutions are connected at $y=L-\epsilon$ by continuity. They get a mass spectrum which depends both of $X$ and $X'$. For giving a meaning to this regularization procedure, $\delta (y-L)$ in Eq.~\eqref{1_ELE_5} should multiply only continuous wave functions. This happens when $X$ and/or $X'$ vanish. We need $X \neq 0$ to give their masses to the zero modes of $Q_L$ and $D_R$ as in the SM. In order to be allowed to soften the brane-Higgs, we have to take $X'=0$ in a regularization procedure which is well defined in distribution theory.

\subsection{Two non-commutativities of calculation limits}
The analytical differences of the mass spectra found in the Regularizations~I and II, as well as via the softened and shifted brane-Higgs, could be compensated by different input values for the Yukawa coupling constants ($Y_5$ and $Y'_5$) to get identical physical mass values \cite{Barcelo:2014kha}. Nevertheless, the Regularizations~I and II are in fact physically different with respect to the existence of measurable flavor violating effective 4D Yukawa couplings at leading order in $v^2/M_{KK}^2$ (with $M_{KK}$ the KK scale) which are generated by the $Y'_5$ couplings~\cite{Azatov:2009na} being present exclusively within Regularization~II (as appears clearly in the 4D approach).
This physical difference between the two schemes of regularization raises the paradoxal question of which one is the correct analytical scheme to use, and represents thus a confirmation 
of the inconsistency of regularizing the Higgs peak.
These two schemes of regularization are obtained~\cite{Barcelo:2014kha} by commuting in the 4D calculation (of masses and couplings) the order of implementation of the two limits $\epsilon \to 0$ 
[the regularization parameter $\epsilon$ defined in Eq.~\eqref{1_ELE_5ter}] and $N \to \infty$ [the upper value $N$ of the KK level $n$ in Eq.~\eqref{1_KK_1}]. 
Therefore, this physical non-commutativity of calculation limits reflects the inconsistency of the Higgs peak regularization.
Another paradoxal non-commutativity of calculation limits arising in the context of regularization of a brane-Higgs coupled to bulk fermions was discussed in Ref.~\cite{Carena:2012fk, Malm:2013jia}: 
different results for Higgs production/decay rates when taking $\epsilon \to 0$ and then $N_{KK} \to \infty$~\footnote{Here $N_{KK}$ 
stands for the number of excited fermion eigenstates exchanged at the loop-level.}~\cite{Casagrande:2010si} or the inverse order~\cite{Azatov:2010pf} in their calculation.
We can thus interpret now this second non-commutativity of calculation limits as another hint for the problematic brane-Higgs regularization (also expected with a warped extra dimension). 
The origin of the two non-commutativities is the mathematically ill-defined (see above) and unnecessary 
(see below) Higgs regularization (introducing $\epsilon$). 

\section{New 5D Treatment}
\label{1_Yukawa_terms_as boundary_conditions}

In this part, we present a method to calculate the fermionic mass spectrum which does not require any kind of regularization. We follow the main lines of the methodology developed for the free case in Section~\ref{1_free_bulk_fermions} and summarized in Fig.~\ref{1_fig:Pyramidal}.

\subsection{Fermionic Currents \& Essential Boundary Conditions}
\label{1_Int_sym}

We begin with considerations of the fermion currents, as justified in Section~\ref{1_free_bulk_fermions}, to define the geometrical field configuration of the considered scenario (see the schematic illustration of Fig.~\ref{1_fig:Pyramidal}). In this scenario, the two 5D fields $Q,D$ propagate only in the interval $[0,L]$. This set-up translates into a condition 
of a vanishing current perpendicular to both boundaries (see Appendix~\ref{Hamilton_Noether} for a discussion on Noether's theorem with boundary terms); this current is now the sum of the individual currents of type~\eqref{1_courant_2} for the two fermion species $Q,D$ since those 
fermions are mixed at $y=L$ through the terms~\eqref{1_L_5}. More precisely, there exists a current involving the two 5D fields:
\begin{equation}
j^M = \bar{Q} \, \Gamma^M Q + \bar{D} \, \Gamma^M D \phantom{0} \text{with the local conservation relation} \phantom{0} \partial_M j^M = 0 \, ,
\label{1_courant_1Y}
\end{equation}
as obtained by Noether's theorem applied to the global fermionic $U(1)$ symmetry of the action made of Eqs.~\eqref{1_L_2}, \eqref{1_eq:actionBound} and
Eq.~\eqref{1_L_5}. The associated transformations,
\begin{equation}
Q_{L/R} \mapsto \text{e}^{-i \alpha} Q_{L/R} \ , \ D_{L/R} \mapsto \text{e}^{-i \alpha} D_{L/R},
\label{1_symetrie_1Y} 
\end{equation}
with $\alpha$ ($\in \mathbb{R}$) a continuous parameter [now forced by the invariant terms~\eqref{1_L_5} to be common to the two fields $Q,D$].
Finally, the conditions for a vanishing current perpendicular to the two boundaries are,
\begin{equation}
\left. j^4 \right |_{0, L}= \left. \left ( \bar{Q} \, \Gamma^4 Q + \bar{D} \, \Gamma^4 D \right ) \right |_{0, L} 
= i \left. \left( Q^\dagger_L Q_R - Q^\dagger_R Q_L + D^\dagger_L D_R - D^\dagger_R D_L \right) \right|_{0, L} = 0 \ .
\label{1_courant_2Y}
\end{equation}
Those relations constitute boundary conditions for the currents.

However, the brane localized Yukawa terms \eqref{1_L_5} are not invariant under the individual $U(1)_Q$ and $U(1)_D$ symmetries so the individual currents $j_Q^M$ and $j_D^M$ defined in Eq.~\eqref{1_courant_1} are separately conserved locally only outside the boundary at $y=L$, which leads to the conditions of vanishing currents at $y=0$ in Eq.~\eqref{1_courant_3}. These are satisfied if one imposes the same essential boundary conditions at $y=0$ on the fermion wave functions, as in the free case, and the 4D mass matrix method in Subsection~\ref{1_4D perturbative approach},
\begin{equation}
Q_R|_0 = D_L|_0 = 0.
\label{1_EBC_0}
\end{equation}

\subsection{Failed Treatment without Bilinear Boundary Terms}
\label{1_Failed_treatment}
We study the mass spectrum considering the action $S_\Psi$~\eqref{1_L_2} for the kinetic terms and the
Yukawa couplings action $S_X$~\eqref{1_L_5}, but without the BBTs $S_B$~\eqref{1_eq:actionBound} for the moment. The boundary fields at $y=0$ are fixed by the essential boundary conditions \eqref{1_EBC_0}, whereas they are initially free at $y=L$ so that their functional varia\-tions will be taken generic and continuous including the boundaries (see Appendix~\ref{app_cont_field}). Without loss of generality, the stationary action condition can be split into the following conditions for each field,
\begin{align}
0 =  \delta_{Q^\dagger_L} (S_{\Psi} + S_X) &= \displaystyle{ \int d^4x \; \int_0^L dy \;  
\delta Q^\dagger_L \left[ i\bar{\sigma}^\mu \partial_\mu Q_L - \partial_y Q_R  \right] } 
\nonumber \\
&\displaystyle{ + \int d^4x \left. \left[ \delta Q^\dagger_L \left( \dfrac{1}{2} Q_R -X D_R \right) \right] \right|_{L} } \ ,
\label{1_HVP_5a}
\end{align}
\begin{align}
0 = \delta_{Q^\dagger_R} (S_{\Psi} + S_X) &= \displaystyle{ \int d^4x \int_0^L dy \;  
\delta Q^\dagger_R \left[ i\sigma^\mu \partial_\mu Q_R + \partial_y F_L  \right] } 
\nonumber \\
&\displaystyle{ + \int d^4x \left. \left[ - \delta Q^\dagger_R \left( \dfrac{1}{2} Q_L + X' D_L \right) \right] \right|_{L} } \ ,
\label{1_HVP_5b}
\end{align}
\begin{align}
0 =  \delta_{D^\dagger_L} (S_{\Psi} + S_X) &= \displaystyle{ \int d^4x \; \int_0^L dy \;  
\delta D^\dagger_L \left[ i\bar{\sigma}^\mu \partial_\mu D_L - \partial_y D_R  \right] } 
\nonumber \\
&\displaystyle{ + \int d^4x  \left. \left[ \delta D^\dagger_L \left( \dfrac{1}{2} D_R -X^{\prime *} Q_R \right) \right] \right|_{L} } \ ,
\label{1_HVP_5c}
\end{align}
\begin{align}
0 = \delta_{D^\dagger_R} (S_{\Psi} + S_X) & = \displaystyle{ \int d^4x \int_0^L dy \;  
\delta D^\dagger_R \left[ i\sigma^\mu \partial_\mu D_R + \partial_y D_L  \right] } 
\nonumber \\
&\displaystyle{ + \int d^4x \left. \left[ -\delta D^\dagger_R \left( \dfrac{1}{2} D_L + X^{*} Q_L \right) \right] \right|_{L} } \ ,
\label{1_HVP_5d}
\end{align}
where the field variations at $y=0$ are zero because of the essential boundary conditions. For generic field variations, the bulk and brane variations of Eqs.~\eqref{1_HVP_5a}, \eqref{1_HVP_5b}, \eqref{1_HVP_5c} and \eqref{1_HVP_5d}, must vanish separately. The bulk part of Eqs.~\eqref{1_HVP_5a}, \eqref{1_HVP_5b}, \eqref{1_HVP_5c} and \eqref{1_HVP_5d} leads to identical bulk Euler-Lagrange equations as in Eq.~\eqref{1_ELE_1}. The brane parts give the natural boundary conditions:
\begin{equation}
\left\{
\begin{array}{l}
\left. \left( Q_R - 2X \; D_R \right) \right|_{L} = 0,  \ \ \ \left. \left( D_L + 2X^{*}  \; Q_L \right) \right|_{L} = 0, \\
\left. \left( Q_L + 2X' \; D_L \right) \right|_{L} = 0,  \ \ \ \left. \left( D_R - 2X^{\prime *}  \; Q_R \right) \right|_{L} = 0, \\
\left. Q_R \right|_0 = 0,   \ \ \ \left. D_L \right|_0 = 0,
\end{array}
\right. \ 
\label{1_BC_3}
\end{equation}
satisfying the boundary current conditions \eqref{1_courant_3} and \eqref{1_courant_2Y}, and leading to consistency relations,
\begin{equation}
4 X X^{\prime *} = 4 X^* X' = 1 \, ,
\label{1_XX'}
\end{equation}
which imply $4|XX'|=1$ and $\alpha_Y' = \alpha_Y \, [\pi]$. Eq.~\eqref{1_XX'} gives conditions which relate different parameters of the model. They appear because our model is overconstrained at the boundaries (similar to the discussion below Eq.~\eqref{1_zero_profiles}). The variation of the action at the boundaries (Eq.~\eqref{1_HVP_5a}-\eqref{1_HVP_5d}) involves the variations of both $F_L$ and $F_R$. There is thus one natural boundary condition for each field at each boundary, while $F_L$ and $F_R$ are related on-shell by the Euler-Lagrange equations \eqref{1_ELE_1}: the system is thus overconstrained.

The Euler-Lagrange equations \eqref{1_ELE_1} lead, via the relevant 
mixed KK decomposition~\eqref{1_KK_2} and 4D Dirac-Weyl equations~\eqref{1_Dirac_2}, to the equations for the KK wave functions along the whole interval $[0,L]$,
\begin{equation}
\forall n \, , \ 
\left\{
\begin{array}{r c l}
\partial_y q_R^n(y) - M_n \, q_L^n(y) &=& 0,
\\ \vspace{-0.2cm} \\
\partial_y q_L^n(y) + M_n \, q_R^n(y) &=& 0,
\\ \vspace{-0.2cm} \\
\partial_y d_R^n(y) - M_n \, d_L^n(y) &=& 0,
\\ \vspace{-0.2cm} \\
\partial_y d_L^n(y) + M_n \, d_R^n(y) &=& 0.
\end{array}
\right. \ 
\label{1_ELE_Y}
\end{equation}
The KK decomposition~\eqref{1_KK_2} in Eq.~\eqref{1_BC_3}, together with Eq.~\eqref{1_ELE_Y}, gives the complete boundary conditions,
\begin{equation}
\forall n \, , \ 
\left\{
\begin{array}{l}
q_R^n(L) - 2X \, d_R^n(L) = 0, \ \ \ \partial_y q_L^n(L) + 2XM_n \, d_R^n(L) = 0, \\
d_L^n(L) + 2X^{*}  \, q_L^n(L) = 0, \ \ \ \partial_y d_R^n(L) + 2X^{*}M_n  \, q_L^n(L) = 0, \\ \\
q_L^n(L) + 2X' \, d_L^n(L) = 0,  \ \ \ \partial_y q_R^n(L) + 2X'M_n \, d_L^n(L) = 0, \\
d_R^n(L) - 2X^{\prime *}  \, q_R^n(L) = 0,  \ \ \ \partial_y d_L^n(L) + 2X^{\prime *}M_n  \, q_R^n(L) = 0, \\ \\
q_R^n(0) = 0,   \ \ \ \partial_y q_L^n(0) = 0,   \\
d_L^n(0) = 0,   \ \ \ \partial_y d_R^n(0) = 0.
\end{array}
\right. \ 
\label{1_BC_4}
\end{equation}
The boundary conditions are not of type $(++)$ or $(--)$ anymore. One can call them $(+ \times)$ and $(- \times)$, where a boundary condition $(\times)$ refers to the dependence on $X$, $X'$ at $y=L$.

We can now solve the bulk equations~\eqref{1_ELE_Y} with the boundary conditions~\eqref{1_BC_4}. This system of coupled first order equations can be decoupled into second order ones so that
\begin{equation}
\forall n, \, \left( \partial_y^2 + M_n^2 \right) f^n_{L/R}(y) = 0,
\label{1_ELE_Y_2}
\end{equation}
whose solutions are given by 
\begin{equation}
\forall n, \ f_{L/R}^n(y) = A^{f,n}_{L/R} \, \cos(M_n \, y) + B^{f,n}_{L/R} \, \sin(M_n \, y),
\label{1_profiles_Y}
\end{equation}
with real constant parameters $A^{f,n}_{L/R}$, $B^{f,n}_{L/R}$. The coupled equations \eqref{1_ELE_Y} impose the relations $A^{f,n}_{L} = B^{f,n}_{R}$ and $A^{f,n}_{R} = - B^{f,n}_{L}$.
The solutions~\eqref{1_profiles_Y} of Eq.~\eqref{1_ELE_Y} inserted into the complete boundary conditions at $y=0$~\eqref{1_BC_4} give rise to the 
following profiles,
\begin{equation}
\forall n \, , \ 
\left\{
\begin{array}{l}
q_{L}^n(y) = A^n_q \, \cos(M_n \, y),  \ \ \ q_{R}^n(y) =  A^n_q \, \sin(M_n \, y), \\
d_{L}^n(y) =  - A^n_d \, \sin(M_n \, y),  \ \ \ d_{R}^n(y) = A^n_d \, \cos(M_n \, y),
\end{array}
\right. \ 
\label{1_prof-BC_Y_1}
\end{equation}
with $A^n_q \equiv A^{q,n}_{L} = B^{q,n}_{R}$ and $A^n_d \equiv A^{d,n}_{R} = - B^{d,n}_{L}$. The normalization conditions \eqref{1_normalization_2} give $|A^n_q| = |A^n_d| = 1$. When these four solutions are injected into the boundary conditions at $y=L$~\eqref{1_BC_3}, one obtains the equations for the mass spectrum,
\begin{equation}
\forall n, \ \tan (M_n \, L) = 2X \, \dfrac{A_d^n}{A_q^n} = 2X^* \, \dfrac{A_q^n}{A_d^n} \ \ \Rightarrow \ \ \tan^2(M_n \, L) = 4|X|^2,
\label{1_syst_BC_1}
\end{equation}
and
\begin{equation}
\forall n, \ \cot (M_n \, L) = 2X' \, \dfrac{A_d^n}{A_q^n} = 2X^{\prime *} \, \dfrac{A_q^n}{A_d^n} \ \ \Rightarrow \ \ \cot^2(M_n \, L) = 4|X^{\prime}|^2.
\label{1_mass_spectrum_2X_2}
\end{equation}
These two mass spectra are equivalent thanks to the relation \eqref{1_XX'} but they do not match the mass spectrum~\eqref{1_4D_spectrum} of the 4D method.

Moreover, the Yukawa coupling $y_{00}$ between two zero modes is obtained by injecting the KK decomposition \eqref{1_KK_2} in Eq.~\eqref{1_L_Yuk} and then by using the BCs at $y=L$ \eqref{1_BC_4} and the relation \eqref{1_XX'}:
\begin{equation}
S_{hQD} = \int d^4x \left( -\dfrac{y_{00}}{\sqrt{2}} \, h \psi_L^{0 \dagger} \psi_R^0 + \text{H.c.} + \text{other terms involving the excited KK modes} \right) \, 
\end{equation}
thus
\begin{equation}
y_{00} = \dfrac{Y_5}{L} \, q_L^0(L) \, d_R^0(L) + \dfrac{Y_5'}{L} \, d_L^0(L) \, q_R^0(L) = 0.
 \label{1_decoupling_Yuk}
\end{equation}
A vanishing Yukawa coupling for the zero modes is not compatible with the SM at low energy. All these problems show that this approach is not consistent.

Another failure of the present treatment can be seen when one takes the limit of vanishing Yukawa couplings ($X=X'=0$). One expects that the 5D and 4D mass matrix methods match, \textit{i.e.} the complete boundary conditions \eqref{1_BC_4} should be the same as the free ones in Subsection~\ref{1_Mass_matrix_diagonalisation}: $Q_L:\, (++)$, $Q_R:\, (--)$, $D_L:\, (--)$, $D_R:\, (++)$. Instead, we have $Q_L:\, (+\pm)$, $Q_R:\, (-\pm)$, $D_L:\, (-\pm)$, $D_R:\, (+\pm)$, where $(\pm)$ means that there is a Neumann and a Dirichlet boundary conditions on the same KK wave function at the same boundary. This implies vanishing KK wave function everywhere along the extra dimension.

\subsection{Treatment with Bilinear Boundary Terms}
\label{1_Correct_treatment_AFHS_term}
In order to overcome the drawbacks discussed in Subsection~\ref{1_Failed_treatment}, we add to the model the BBTs~\eqref{1_eq:actionBound}. The boundary fields are free to vary on the branes. Hamilton's principle applied to $S_{\Psi} + S_X + S_B$ (Eqs.~\eqref{1_L_2}, \eqref{1_L_5}, \eqref{1_eq:actionBound}), insisting on continuous field variations (see Appendix~\ref{app_cont_field}) field variations, gives
\begin{align}
0 =  \delta_{Q^\dagger_L} (S_{\Psi} + S_B + S_X) &= \displaystyle{ \int d^4x \; \int_0^L dy \;  
\delta Q^\dagger_L \left[ i\bar{\sigma}^\mu \partial_\mu Q_L - \partial_y Q_R  \right] }
\nonumber \\
&\displaystyle{ + \int d^4x \left\{ \left. \left[ \delta Q^\dagger_L \left( Q_R -X D_R \right) \right] \right|_{L} - \left. \left[ \delta Q^\dagger_L Q_R \right] \right|_{0} \right\} } \ ,
\label{1_HVP_4a}
\end{align}
\begin{align}
0 = \delta_{Q^\dagger_R} (S_{\Psi} + S_B + S_X) &= \displaystyle{ \int d^4x \int_0^L dy \;  
\delta Q^\dagger_R \left[ i\sigma^\mu \partial_\mu Q_R + \partial_y F_L  \right] }
\nonumber \\
&\displaystyle{ + \int d^4x \left. \left[ \delta Q^\dagger_R \left( -X' D_L \right) \right] \right|_{L} } \ ,
\label{1_HVP_4b}
\end{align}
\begin{align}
0 =  \delta_{D^\dagger_L} (S_{\Psi} + S_B + S_X) &= \displaystyle{ \int d^4x \; \int_0^L dy \;  
\delta D^\dagger_L \left[ i\bar{\sigma}^\mu \partial_\mu D_L - \partial_y D_R  \right] }
\nonumber \\
&\displaystyle{ + \int d^4x  \left. \left[ \delta D^\dagger_L \left( -X^{\prime *} Q_R \right) \right] \right|_{L} } \ ,
\label{1_HVP_4c}
\end{align}
\begin{align}
0 = \delta_{D^\dagger_R} (S_{\Psi} + S_B + S_X) &= \displaystyle{ \int d^4x \int_0^L dy \;  
\delta D^\dagger_R \left[ i\sigma^\mu \partial_\mu D_R + \partial_y D_L  \right] } 
\nonumber \\
&\displaystyle{ + \int d^4x \left\{ \left. \left[ -\delta D^\dagger_R \left( D_L + X^{*} Q_L \right) \right] \right|_{L} + \left. \left[ \delta D^\dagger_R D_L \right]  \right|_{0}  \right\} } \ .
\label{1_HVP_4d}
\end{align}
Variations of the action in the bulk and on the branes must vanish separately. The Euler-Lagrange equations \eqref{1_ELE_1} are still valid and give, after the KK decomposition, the equations for the KK wave functions \eqref{1_ELE_Y}. This time, the natural boundary conditions from the vanishing variations of the action on the branes are
\begin{equation}
\left\{
\begin{array}{l}
\left. \left( Q_R - X \; D_R \right) \right|_{L} = 0,  \ \ \ \left. \left( D_L + X^{*}  \; Q_L \right) \right|_{L} = 0, \\
\left. X' \; D_L \right|_{L} = 0,  \ \ \ \left. X^{\prime *}  \; Q_R \right|_{L} = 0.
\end{array}
\right. \ 
\label{1_BC_13}
\end{equation}
It is straightforward to check that the boundary current conditions \eqref{1_courant_3} and \eqref{1_courant_2Y} are satisfied by the natural boundary conditions \eqref{1_BC_13} which can be further rewritten without loss of generality as
\begin{eqnarray}
& \left. \left( Q_R - X \, D_R \right) \right|_{L} = 0, \ \ \ \left. \left( D_L + X^{*}  \, Q_L \right) \right|_{L} = 0, \ \ \ 
X^{\prime}=0 \ {\rm or} \ ( \left . Q_R \right|_L = 0 \ {\rm and} \ \left. D_L \right|_L = 0 ),
\nonumber \\ 
&   \left. Q_R  \right |_{0}    = 0 , \ \ \ \left .  D_L \right |_{0}  = 0,
\label{1_BC_13cont}
\end{eqnarray}
which leads to
\begin{eqnarray}
{\rm \underline{BCs \ 1:}} & \left. X \, D_R \right|_{L} = 0, \ \ \ \left. X^{*}  \, Q_L \right|_{L} = 0, \ \ \ 
\left . Q_R \right|_L = 0, \ \ \ \ \left. D_L \right|_L = 0 ,
\nonumber \\ 
&   \left. Q_R  \right |_{0}    = 0 , \ \ \ \left .  D_L \right |_{0}  = 0,
\nonumber \\ 
{\rm or, \ \ \underline{BCs \ 2:}} & \left. \left( Q_R - X \, D_R \right) \right|_{L} = 0, \ \ \ \left. \left( D_L + X^{*}  \, Q_L \right) \right|_{L} = 0, \ \ \ 
X^{\prime}=0,
\nonumber \\ 
&   \left. Q_R  \right |_{0}    = 0 , \ \ \ \left .  D_L \right |_{0}  = 0.
\label{1_BC_13ter}
\end{eqnarray}
As shown in Subsection~\ref{1_Failed_treatment} in Eq.~\eqref{1_decoupling_Yuk}, the Yukawa coupling $Y_5$ must be present to allow this scenario to reproduce the SM at EW energies. 
Hence one has $X \neq 0$ so that the BCs~1 reads
\begin{eqnarray}
{\rm \underline{BC \ 1:}} & \left. D_R \right|_{L} = 0, \ \ \ \left. Q_L \right|_{L} = 0, \ \ \ 
\left . Q_R \right|_L = 0, \ \ \ \ \left. D_L \right|_L = 0 ,
\nonumber \\ 
&   \left. Q_R  \right |_{0}    = 0 , \ \ \ \left .  D_L \right |_{0}  = 0.
\label{1_BC_1NBCA}
\end{eqnarray} 
Then, combining these BCs~1 with the bulk Euler-Lagrange equations~\eqref{1_ELE_1} would lead to vanishing 5D fields: the presence of the mass term with $Y_5' \neq 0$ leads to an additional natural boundary condition which overconstrains the system. Let us move to the BCs~2 by taking $Y_5'=0$ (thus $X'=0$ and we recover the condition guessed in Subsection~\ref{1_RegSoft}, which can be expressed in terms of the KK wave functions thanks to the relevant mixed KK 
decomposition~\eqref{1_KK_2}, and then combined with the equations for the KK wave functions \eqref{1_ELE_Y}. We get the complete boundary conditions,
\begin{equation}
\forall n \, , \ 
\left\{
\begin{array}{l}
q_R^n(L) - X \, d_R^n(L) = 0, \ \ \ \partial_y q_L^n(L) + XM_n \, d_R^n(L) = 0, \\
d_L^n(L) + X^{*}  \, q_L^n(L) = 0, \ \ \ \partial_y d_R^n(L) + X^{*}M_n  \, q_L^n(L) = 0, \\ \\
q_R^n(0) = 0,   \ \ \ \partial_y q_L^n(0) = 0,   \\
d_L^n(0) = 0,   \ \ \ \partial_y d_R^n(0) = 0.
\end{array}
\right. \ 
\label{1_BC2}
\end{equation}
One can check that in the limit $X=X'=0$, one recovers the free boundary conditions of the 4D mass matrix approach in Subsection~\ref{1_4D perturbative approach}: $Q_L:\, (++)$, $Q_R:\, (--)$, $D_L:\, (--)$, $D_R:\, (++)$.

\begin{figure}[!h]
\begin{center}
\includegraphics[width=15cm]{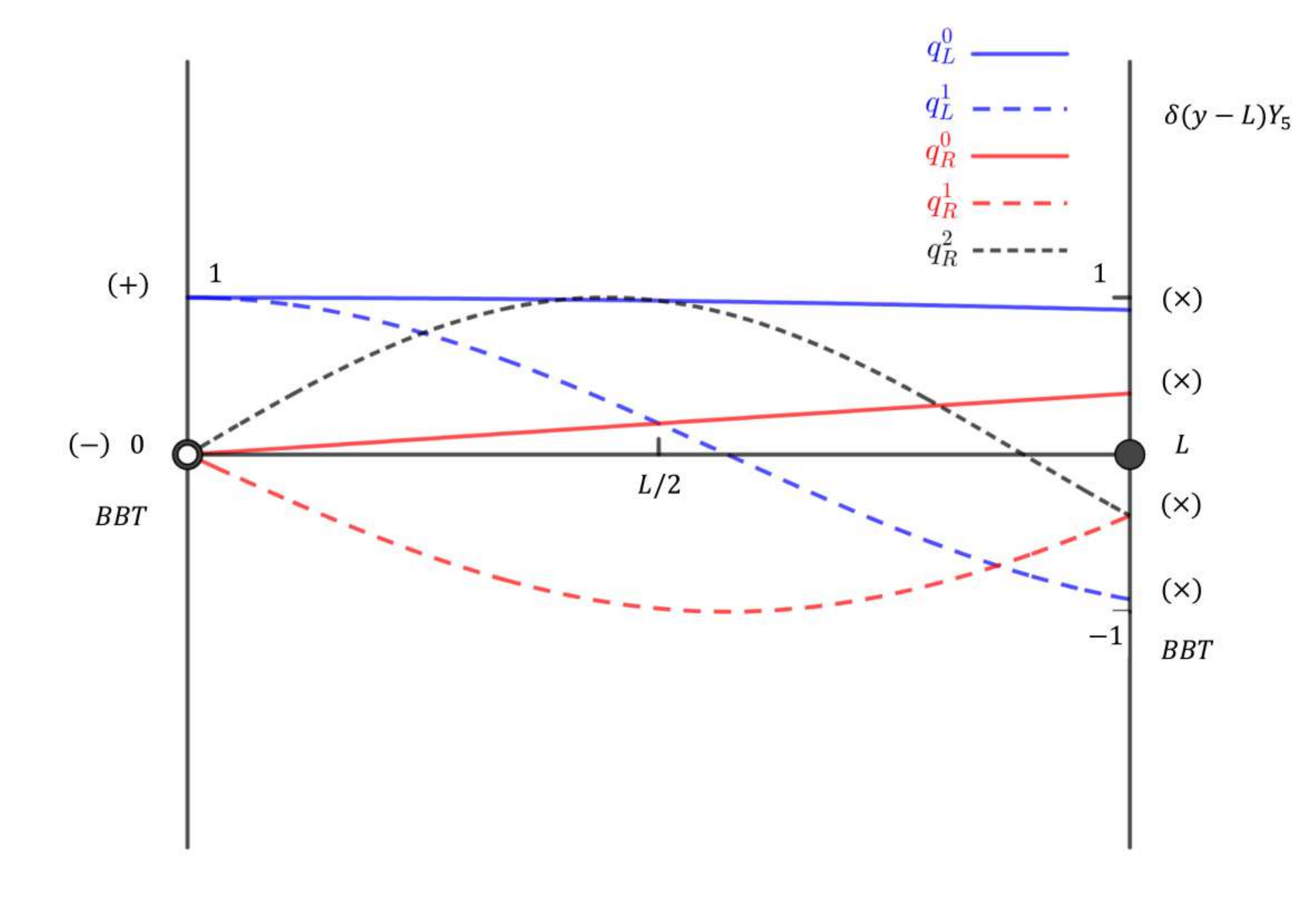}
\end{center}
\caption{KK wave functions of the 5D fields $Q$ and $D$ with Yukawa couplings.
}\label{prof_interval_Yuk}
\end{figure}

Eq.~\eqref{1_ELE_Y} and the boundary conditions at $y=0$ in Eq.~\eqref{1_BC2} lead to the same KK wave functions \eqref{1_prof-BC_Y_1} as in Subsection \ref{1_Failed_treatment}. When these four solutions are injected into the boundary conditions~\eqref{1_BC2} at $y=L$, one obtains
\begin{equation}
\forall n, \ \tan (M_n \, L) = X \, \dfrac{A_d^n}{A_q^n} = X^* \, \dfrac{A_q^n}{A_d^n} \ \ \Rightarrow \ \ \tan^2(M_n \, L) = |X|^2.
\label{1_syst_BC_3}
\end{equation}
The mass spectrum is thus
\begin{equation}
\tan(M_n \, L) = \pm |X| \ \ \Rightarrow \ \ \ M_n = \pm \dfrac{1}{L} \left[ n \pi + \arctan (|X|) \right] \, , n \in \mathbb{Z}
\label{1_spect_final}
\end{equation}
The normalization conditions of Eq.~\eqref{1_normalization_2} ($n=m$) give, with the KK wave functions of Eq.~\eqref{1_prof-BC_Y_1}, $A_q^n = \text{e}^{i \beta_q^n}$ and $A_d^n = \text{e}^{i \beta_d^n}$ with $\beta_q^n$, $\beta_d^n \in \mathbb{R}$. Moreover, Eq.~\eqref{1_syst_BC_3}, together with the mass spectrum equation in Eq.~\eqref{1_spect_final}, leads to
\begin{equation}
\tan(M_n \, L) = |X| \ \ \ \Rightarrow \ \ \ A_q^n = \text{e}^{i (\beta_d^n + \alpha_Y)} \, , \ \ \ A_d^n = \text{e}^{i \beta_d^n} \, ,
\label{1_tower_1}
\end{equation}
\begin{equation}
\tan(M_n \, L) = -|X| \ \ \ \Rightarrow \ \ \ A_q^n = - \text{e}^{i (\beta_d^n + \alpha_Y)} \, , \ \ \ A_d^n = \text{e}^{i \beta_d^n} \, .
\label{1_tower_2}
\end{equation}
When one changes the sign of $M_n$, one goes from the spectrum \eqref{1_tower_1} to the spectrum \eqref{1_tower_2}. When one performs the transformation
\begin{equation}
\left\{
\begin{array}{l}
M_n \mapsto -M_n \, , \\
\psi_R^n \mapsto - \psi_R^n \ \ \ \text{or} \ \ \ \psi_L^n \mapsto - \psi_L^n \, ,
\end{array}
\right.
\end{equation}
the 4D actions of each KK modes \eqref{1_SKK_2} (and thus the Dirac equations \eqref{1_Dirac_2}), the KK wave functions equations \eqref{1_ELE_Y} and the orthonormalization conditions \eqref{1_normalization_2} (using the parity of the solutions \eqref{1_prof-BC_Y_1}) are invariant. One can conlude that the sign of $M_n$ is not physical and consider only the spectrum \eqref{1_tower_1} which is physically equivalent to \eqref{1_tower_2}. The phase $\beta_d^n$ is not fixed yet. To see if it is physical, we perform the shift $\beta_d^n \mapsto \beta_d^n + \theta_n$, and check that the KK wave function equations \eqref{1_ELE_Y} and the orthonormalization conditions \eqref{1_normalization_2} are invariant, so one can take $\beta_d^n=0$ since it is not physical. What about $\alpha_Y$? By performing the shift $\alpha_Y \mapsto \alpha_Y + \theta$, we check also that \eqref{1_ELE_Y} and \eqref{1_normalization_2} are invariant so we fix $\alpha_Y=0$. At the end, we obtain the KK wave functions \eqref{1_prof-BC_Y_1} with $A_q^n=A_d^n=1$ and the mass spectrum:
\begin{equation}
M_n = \dfrac{1}{L} \left[ n \pi + \arctan (|X|) \right] \, , n \in \mathbb{Z}
\label{mass_spect_Y_interval}
\end{equation}
This time, the mass spectrum matches the one obtained in the 4D mass matrix approach (Eq.~\eqref{1_4D_spectrum}). We conclude that this new approach seems consistent. In Fig.~\ref{prof_interval_Yuk}, we give a plot of the KK wave functions along the extra dimension.

The Yukawa couplings between the modes $n$ and $m$ follow from inserting the KK decomposition \eqref{1_KK_2} in Eq.~\eqref{1_L_Yuk} and using the profiles expression \eqref{1_prof-BC_Y_1},
\begin{equation}
S_{hQD} = \int d^4x \ \sum_{n, m} \left( -\dfrac{y_{nm}}{\sqrt{2}} \, h \psi_L^{n \dagger} \psi_R^m + \text{H.c.} \right) \, 
\end{equation}
thus
\begin{align}
y_{nm} &= \dfrac{|Y_5|}{L} \; q_L^n(L) \; d_R^m(L) \nonumber \\
 &= \dfrac{|Y_5|}{L} \; \cos(M_n \; L) \cos(M_m \; L) \nonumber \\
 &= (-1)^{n+m} \dfrac{|Y_5|}{L(1+X^2)},
\end{align}
where we use the mass spectrum \eqref{1_spect_final} and trigonometric identities\footnote{
\begin{equation}
\forall n \in \mathbb{Z}, \cos(x + n \pi) = (-1)^n \cos(x)
\end{equation}
\begin{equation}
\cos(\arctan(x)) = \dfrac{1}{\sqrt{1+x^2}}
\end{equation}
} to get the last equality.

\subsection{Brane Localized Terms \& the Dirac Distribution}
\label{1_Yukawa_terms_and_Dirac_distribution}
In this part, we reformulate the method in Subsection~\ref{1_Correct_treatment_AFHS_term} with the formalism of distributions \cite{Schwartz2, Schwartz1}. As it corresponds just to reexpress the procedure in the language of distributions, the physical results are the same.

\subsubsection{The 5D Action as a Distribution}
\label{1_distrib}
One can treat the 5D fields as distributions in order to handle boundary localized interactions. In this subsection, we choose a different approach where the fields are still functions but the action is built with rectangular and Dirac distributions.

To formulate our model using Dirac distributions for the brane localized terms, we define the 5D fields on $y \in \mathbb{R}$ and use a rectangular distribution $\Theta_{I}(y) = \theta(y) - \theta(y-L)$ to restrict the action to the physical compact space. The 5D action is thus rewritten as
\begin{equation}
S_{\Psi} + S_B + S_X = \int d^4x \int_{-\infty}^{+\infty} dy \left\{ \mathcal{L}_\Psi \, \Theta_I(y) + \mathcal{L}_B  \left[  \delta(y) - \delta(y-L)  \right] + \mathcal{L}_X \, \delta(y-L) \right\}.
\label{1_S_4}
\end{equation}
An important remark is that the fields in the Lagrangians $\mathcal{L}_\Psi$~\eqref{1_L_2}, $\mathcal{L}_X$~\eqref{1_L_5} and $\mathcal{L}_B$\eqref{1_eq:actionBound} must be continuous at the positions of the 3-branes in order to define the product between the Dirac distributions and the Lagrangians $\mathcal{L}_B$, $\mathcal{L}_X$ in Eq.~\eqref{1_S_4}.

Hamilton's principle applied to the action \eqref{1_S_4}, varying each field, with generic variations at the boundaries, leads to
\begin{align}
0 =  \delta_{Q^\dagger_L} (S_{\Psi} + S_B + S_X) &= \displaystyle{ \int d^4x \, \int_{-\infty}^{+\infty} dy \,  
\Theta_I(y) \, \delta Q^\dagger_L \left[ i\bar{\sigma}^\mu \partial_\mu Q_L - \partial_y Q_R  \right] } 
\nonumber \\
&\displaystyle{ + \int d^4x \int_{-\infty}^{+\infty} dy \, \left[ \delta(y-L) \, \delta Q^\dagger_L \left( Q_R -X D_R \right) - \delta(y) \; \delta Q^\dagger_L Q_R \right] } \ ,
\end{align}
\begin{align}
0 = \delta_{Q^\dagger_R} (S_{\Psi} + S_B + S_X) &= \displaystyle{ \int d^4x \int_{-\infty}^{+\infty} dy \,  
\Theta_I(y) \, \delta Q^\dagger_R \left[ i\sigma^\mu \partial_\mu Q_R + \partial_y F_L  \right] } 
\nonumber \\
&\displaystyle{ + \int d^4x \int_{-\infty}^{+\infty} dy \, \delta(y-L) \, \delta Q^\dagger_R \left( -X' D_L \right) } \ ,
\end{align}
\begin{align}
0 =  \delta_{D^\dagger_L} (S_{\Psi} + S_B + S_X) &= \displaystyle{ \int d^4x \, \int_{-\infty}^{+\infty} dy \,  
\Theta_I(y) \, \delta D^\dagger_L \left[ i\bar{\sigma}^\mu \partial_\mu D_L - \partial_y D_R  \right] } 
\nonumber \\
&\displaystyle{ + \int d^4x \int_{-\infty}^{+\infty} dy \, \delta(y-L) \, \delta D^\dagger_L \left( -X^{\prime *} Q_R \right) } \ ,
\end{align}
\begin{align}
0 = \delta_{D^\dagger_R} (S_{\Psi} + S_B + S_X) &= \displaystyle{ \int d^4x \int_{-\infty}^{+\infty} dy \,  
\Theta_I(y) \, \delta D^\dagger_R \left[ i\sigma^\mu \partial_\mu D_R + \partial_y D_L  \right] } 
\nonumber \\
&\displaystyle{ \int d^4x \int_{-\infty}^{+\infty} dy \, \left[ - \delta(y-L) \, \delta D^\dagger_R \left( D_L + X^{*} Q_L \right)  + \delta(y) \; \delta D^\dagger_R D_L \right] } \ ,
\end{align}
where we used the fact that $\partial_y \Theta_I(y) = \delta(y) - \delta(y-L)$ since $\partial_y \theta(y) = \delta(y)$. The generic field variations should be understood as playing the role of testfunctions. One gets the equations including distributions,
\begin{equation}
\left\{
\begin{array}{c c c}
\left( i \bar{\sigma}^\mu \partial_\mu Q_L - \partial_y Q_R \right) \Theta_I(y) + \left( Q_R - X \, D_R \right) \delta(y-L) - Q_R \; \delta(y) &=& 0, \\ \vspace{-0.3cm} \\
\left( i \sigma^\mu \partial_\mu Q_R + \partial_y Q_L \right) \Theta_I(y) - X' \, D_L \, \delta(y-L) &=& 0, \\ \vspace{-0.3cm} \\
\left( i \bar{\sigma}^\mu \partial_\mu D_L - \partial_y D_R \right) \Theta_I(y) - X^{\prime *} \, Q_R \, \delta(y-L) &=& 0, \\ \vspace{-0.3cm} \\
\left( i \sigma^\mu \partial_\mu D_R + \partial_y D_L \right) \Theta_I(y) - \left( D_L + X^* \, Q_L \right) \delta(y-L) + D_L \; \delta(y) &=& 0.
\end{array}
\right.
\label{1_ELE_11}.
\end{equation}
One can notice that, in the formalism of distribution theory, the 5D Euler-Lagrange equations contain the information leading to the natural boundary conditions via the products with $\delta(y-L)$. If one applies the distribution equations \eqref{1_ELE_11} on a testfunction $\eta(y)$, whose support $\text{supp}(\eta)$ is such that $\text{supp}(\eta) \subset ]0,L[$, one gets simply
\begin{equation}
\left\{
\begin{array}{l c r}
\displaystyle{\int_{-\infty}^{+\infty} dy \left[ \left( i \bar{\sigma}^\mu \partial_\mu F_L - \partial_y F_R \right) \Theta_I (y) \, \eta(y) \right]} &=& 0, \\ \vspace{-0.3cm} \\
\displaystyle{\int_{-\infty}^{+\infty} dy \left[ \left( i \sigma^\mu \partial_\mu F_R + \partial_y F_L \right) \Theta_I (y) \, \eta(y) \right]} &=& 0.
\end{array}
\right.
\label{1_ELE_12}
\end{equation}
For a generic testfunction $\eta(y)$, this leads to the bulk Euler-Lagrange equations~\eqref{1_ELE_1} for the 5D fields restricted to $(0,L)$. Assuming that the 5D fields are differentiable with respect to $y$ with continuous derivatives, one can extend the viability of this Euler-Lagrange equations system \eqref{1_ELE_12} to $[0,L]$. Then, using them in the system~\eqref{1_ELE_11}, we obtain
\begin{equation}
\left\{
\begin{array}{c c c}
\left( Q_R - X \, D_R \right) \delta(y-L) &=& 0, \\ \vspace{-0.3cm} \\
X' \, D_L \, \delta(y-L) &=& 0, \\ \vspace{-0.3cm} \\
X^{\prime *} \, Q_R \, \delta(y-L) &=& 0, \\ \vspace{-0.3cm} \\
\left( D_L + X^* \, Q_L \right) \delta(y-L) &=& 0.
\end{array}
\right.
\label{1_BC_18}
\end{equation}
Again, one obtains the same boundary conditions as in Eq.~\eqref{1_BC_13cont}. At the end, we recover the system of 5D Euler-Lagrange equations \eqref{1_ELE_1} and boundary conditions of Subsection~\ref{1_Correct_treatment_AFHS_term}. The solution of the equations is thus the same (see the discussion in Subsection~\ref{1_Correct_treatment_AFHS_term} below Eq.~\eqref{1_BC_13cont}). Therefore, treating the Yukawa terms as distributions is just another mathematical way to treat brane localized couplings of bulk fields.

\subsubsection{Currents and Distributions}
The formalism of distributions is particulary convenient to study the currents associated to the 5D fields. We come back to the two transformations $U(1)_{Q,D}$ in Subsection~\ref{1_EBC_currents}. It is straightforward to generalize Noether's theorem to the case of a Lagrangian distribution as in Eq.~\eqref{1_S_4}. The current associated to the field $F=Q,D$ is
\begin{equation}
J_F^M = j_F^M \, \Theta_I,
\label{1_current_3}
\end{equation}
with $j_F^M$ defined as in Eq.~\eqref{1_courant_1}, and in particular $j_F^4$ is given by the same expression as in Eq.~\eqref{1_courant_2}.
Moreover, Noether's theorem gives the expression for the divergence of these currents ($|\alpha_f| \ll 1$) \cite{Itzykson:1980rh},
\begin{equation}
\partial_M J_F^M = \dfrac{\partial}{\partial \alpha_f} \left( \mathcal{L}_\Psi^* \, \Theta_I(y) + \left( \mathcal{L}_{\rm{B}}^* + \mathcal{L}_{\rm{X}}^* \right) \delta(y-L) \right),
\label{1_div_current}
\end{equation}
where the $^*$ means the transformed Lagrangian under $U(1)_F$. Therefore, even if the currents $J_F^M$ are not conserved on the 3-brane at $y=L$, Eq.~\eqref{1_div_current} can still provide information, linking the bulk currents to the Yukawa couplings on the 3-brane. For the left-hand side of \eqref{1_div_current}, using \eqref{1_current_3}, one gets
\begin{equation}
\partial_M J_F^M = \partial_M j_F^M \, \Theta_I + j_F^4 \, \left[ \delta(y)-\delta(y-L) \right].
\label{1_current_4}
\end{equation}
Restricting our study to the bulk, it is clear from Eq.~\eqref{1_eq:actionKin} that $U(1)_F$ is a symmetry of $\mathcal{L}_\Psi$ so the associated current is conserved outside of the branes,
\begin{equation}
\partial_M j_Q^M = 0.
\label{1_conserved_current}
\end{equation}
Again, we know that the fields are differentiable with continuous derivatives on $[0,L]$, thus Eq.~\eqref{1_conserved_current} can be extended to the 3-branes. Using this last result in \eqref{1_current_4}, we have
\begin{align}
\partial_M J_F^M &= j_F^4 \, \left[ \delta(y)-\delta(y-L) \right], \nonumber \\
&= - j_F^4 \, \delta(y-L),
\label{1_div_current_2}
\end{align}
where we used the fact that $j_F^4$ vanishes at $y=0$ since there is no mixing term between $Q$ and $D$ here\footnote{but the constraint that $j^4$ \eqref{1_courant_2Y} has to vanish at $y=L$ remains.}. Subsequently, one can express the right-hand side of Eq.~\eqref{1_div_current} with \eqref{1_L_5} for the transformation $U(1)_Q$,
\begin{equation}
\dfrac{\partial \mathcal{L}_*}{\partial \alpha_Q} = i \left[ X^* \, D_R^\dagger Q_L - X \, Q_L ^\dagger D_R + X^{\prime *} \, D_L^\dagger Q_R - X' \, Q_R^\dagger D_L \right] \delta(y-L),
\end{equation}
hence \eqref{1_div_current_2} with \eqref{1_courant_2} leads to
\begin{equation}
\left. \left[ \left( Q_R^\dagger Q_L - Q_L^\dagger Q_R \right) - X^* \, D_R^\dagger Q_L + X \, Q_L^\dagger D_R + X' \, Q_R^\dagger D_L - X^{\prime *} \, D_L^\dagger Q_R \right] \right|_{y = L} = 0.
\end{equation}
which is trivially satisfied by the BCs 2 in Eq.~\eqref{1_BC_13ter} that we choose. One can repeat the same exercise for the transformation $U(1)_D$. Our choice of boundary conditions is thus consistent with the conservation laws of the currents.

\section{Implications}
\label{implications}
\subsection{Interpretation of the Analytical Results}
The matching of the fermion mass spectra obtained in the distinct 4D (Subsection~\ref{1_Mass_matrix_diagonalisation}) and 5D (Subsection~\ref{1_Correct_treatment_AFHS_term}) approaches constitutes a confirmation of both the spectra and the feature that regularizing (performed in none of the 4D and 5D approaches) is not required for the calculation. The 5D method is obviously different from the 5D regularizations (of type~I and II) discussed in the literature and whose inconsistencies have been pointed out in Section~\ref{1_Usual_treatment}.

As a corollary, note that there is no other theoretical reason other than a would-be regularization to introduce a temporary shift along $y$ of the Dirac distribution in the spectrum calculation. This shows the uniqueness of the fermion mass spectrum since such a shift could modify the spectrum 
(as in the regularization~II). Indeed, in particular in the 4D approach, the coefficients $\beta_{ij}$ of Eqs.~\eqref{1_M}-\eqref{1_coeff} would tend to zero as shift of the brane on which the Higgs is localized would go to zero, implying a change of the mass spectrum if this limit is taken after (like in the 4D regularization~II) the infinite KK tower 
limit of the mass matrix characteristic equation~\cite{Barcelo:2014kha}.

Besides, the independence of the final fermion mass spectrum~\eqref{1_spect_final} on the parameter $X'$ is an important result since most approaches including brane-Higgs regularization lead to a $X'$ dependeence. The condition $X'=0$ is required in order to have a system of Euler-Lagrange equations with boundary conditions which is not overconstrained.

Let us now interpret physically the absence of the role of the $Y_5'$ coupling (involved in $X'$) in the final spectrum~\eqref{1_spect_final} which
depends only on $X$. In the 4D approach, starting with the free action $S_{\Psi} + S_B$, the KK wave functions $d^n_L(y)$ and $q^n_R(y)$ ($\forall n$), defined in Eq.~\eqref{1_KK_1} and
given respectively by the first and fourth solutions of Eq.~\eqref{1_completeEBC} (see the discussion below this equation), vanish at the boundary 
$y=L$.
Hence the term with a coefficient $X'$ in the action $S_X$ of Eq.~\eqref{1_L_5}, added to the above free action $S_{\Psi} + S_B$, gets multiplied by a vanishing 
factor from the integration over the interval due to the Dirac distribution $\delta (y-L)$. More intuitively, the $Y'_5$ term localized at the boundary $y=L$, and coupling 
the brane localized Higgs boson to the bulk fermions $D_L$, $Q_R$, should not have any effect since the associated free fermionic KK wave functions vanish at this boundary. Moreover, if this term is absent at tree level, it will not be perturbatively generated \cite{Georgi:2000ks, Carena:2004zn}, and thus this situation is technically natural. This feature is confirmed by the rigorous procedure of solving the Euler-Lagrange equations and complete boundary conditions leading to the mass spectrum ~\eqref{1_spect_final}.

\subsection{Phenomenological impacts}
In the appropriate treatment developed in the present chapter without regularization, the obtained mass spectrum and the effective 4D Yukawa coupling depend on $Y_5$ but not on $Y'_5$ coupling constant. 

The results for fermion masses and profiles are also correct when one invokes the brane-Higgs Regularization~I for which the $Y_5'$ dependence cancels.
Hence, the phenomenological analyses in the literature based on such results are still valid: see for instance 
Ref.~\cite{Huber:2000ie,Huber:2003sf,Casagrande:2008hr,Goertz:2008vr,Moreau:2005kz,Moreau:2006np,Bouchart:2009vq}. 
Those arguments apply also to the geometrical background with warped extra dimensions where the spectrum is expected to be independent of $Y_5'$ as well. 

However, if the Regularization~II or the softering of the brane-Higgs are used, the obtained fermion masses and 4D Yukawa couplings depend on both $Y_5$ and $Y'_5$ so that the results differ effectively from the correct ones. 
Hence, the phenomenological studies based on these analytical results (for example Ref.~\cite{Azatov:2009na,Casagrande:2010si,Carena:2012fk,Malm:2013jia,Hahn:2013nza}) should be reconsidered.

In addition, the effective 4D Yukawa couplings to fermions and their KK excitations affect the main Higgs production mechanism at the Large Hadron Collider (LHC): the gluon-gluon fusion via triangular loops of (KK) fermions.
Hence the effect of the realistic limit~\cite{Barcelo:2014kha} of vanishing $Y'_5$ on the constraints on KK masses derived in the studies~\cite{Casagrande:2010si,Carena:2012fk,Malm:2013jia,Hahn:2013nza}, 
within the warped background and based on the Regularization~II, should be considered as well.

Besides, the rotation matrices diagonalizing the 4D fermion mass matrix~\eqref{1_M} do not diagonalize simultaneously the effective 4D Yukawa coupling matrix 
since the latter one does not contain matrix elements due to the pure KK masses. 
The induced flavor violating 4D Yukawa couplings are generated at leading order by $Y'_5$ contributions as can be shown diagrammatically~\cite{Azatov:2009na}.
Hence there exist FCNC effects in measured $\Delta F= 2$ processes such as $\bar K-K$, $\bar B-B$ and $\bar D-D$ mixings, mainly produced by tree-level exchanges of the Higgs boson 
via $Y'_5$ couplings. These lead to considerable lower bounds on the KK boson mass scale found to be around $6-9$~TeV in the analysis~\cite{Azatov:2009na} 
on warped extra dimensions using indeed the Regularization~II. Hence these bounds should be significantly alleviated in the realistic situation where $Y'_5\to 0$;
this limit should be applied since the independence found in the present paper on $Y'_5$ (extended via flavor indices) remains true for the case of three flavors, as well as for fermion bulk masses, as it is clear in the 
4D approach where the $\beta_{ij}$-elements~\eqref{1_coeff} of the mass matrix still vanish. The predictions of Ref~\cite{Azatov:2009na}, based on Regularization~II, that FCNC reactions involving Yukawa couplings, like the rare top 
quark decay $t\to ch$ and exotic Higgs boson decay to charged leptons $h\to \mu \tau$, can be observable at the LHC deserve reconsiderations as well when $Y'_5=0$.

\section{Summary \& Conclusion}
\label{1_conclusion_4}
For bulk fermions coupled to a brane-Higgs boson we have shown that the proper calculation of the fermion masses and effective 4D Yukawa couplings does not rely on brane-Higgs regularizations.
The justifications are the following ones: {\it (i)} There are no jumps of the fermion wave functions at the boundary where the Higgs field is localized so there is no motivation to introduce an arbitrary regularization, {\it (ii)} the regularizations
suffer from several mathematical discrepancies confirmed by two known non-commutativities of calculation limits, {\it (iii)} our method without any regularization is validated in particular by the 
matching between the 4D versus 5D treatments.
 
In the rigorous method developed for both free and brane-coupled bulk fermions, we have also pointed out the necessity to either include BBTs in the Lagrangian, or alternatively impose conditions on vanishing conserved currents at the boundaries of the interval. The arguments go as follows: {\it (i)} the presence of BBTs guarantees the conditions on currents which define the field geometrical configuration of the model, {\it (ii)} the BBTs and the conditions on currents allow to define a model which is not overconstrained and to find physically consistent fermion masses, bulk profiles and effective 4D Yukawa couplings (solutions fulfilling the normalization constraints and the decoupling limit condition), {\it (iii)} the BBTs lead to the expected matching between the 4D and 5D calculation results. We summarize the results in Tab.~\ref{1_summary_table}. The BBTs indicate a possible origin of the chiral nature of the SM as well as of its chirality distribution among quark/lepton $SU(2)_W$ doublets and singlets.

\begin{table}[h]
\begin{center}
\begin{tabular}{c|c|c|c|}
& & & \\
 & \textbf{NBCs only} & \textbf{EBCs} & \textbf{BBTs \& NBCs} \\
& & & \\
\hline 
& & & \\
\textbf{4D approach} & Not physical & BCs ($\pm$) & BCs ($\pm$) \\ 
& & & \\
\hline
& & & \\
\textbf{5D approach} & Not physical & Impossible & BCs ($\times$) \\
& & & \\
\end{tabular} 
\end{center}
\caption[Summary of the results of the 4D and 5D approaches]{Summary of the results of the 4D and 5D approaches. We use the following acronyms: BC for Boundary Condition, EBC for Essential Boundary Condition, NBC for Natural Boundary Condition.}
\label{1_summary_table}
\end{table}

The general methodology worked out reveals that the information regarding the definition of a higher-dimensional model are not necessarily fully contained in the action itself
-- through the deduced Euler-Lagrange equations and the natural boundary conditions -- but might be partly included as well through essential boundary conditions.

We finished the analysis by the phenomenological impacts of the new calculation method which predicts
the independence of the fermion masses and effective 4D Yukawa couplings on the $Y_5'$ parameter of the Lagrangian. This feature, with respect to the Regularization~II or to the softening of the brane-Higgs usually 
applied in the literature, should in particular alleviate significantly the previously obtained severe bounds on KK masses induced by FCNC processes generated via flavor violating couplings of the 
Higgs boson.

\chapter{Generalizations}
\label{Applications}

This chapter is a personal adaptation of a collection of unpublished works in collaboration with Andrei Angelescu, Ruifeng Leng and Grégory Moreau.

\vspace{1cm}

In this chapter, we continue our exploration of the treatment of brane localized terms for 5D fermions. In Section~\ref{One_generation_quarks}, we treat the model of Chapter~\ref{1_4D perturbative approach} on an AdS$_5$ spacetime. In Section~\ref{other_boundary_terms}, we apply our methods to the case of other boundary terms usually studied in brane world literature. In Sections~\ref{1_Orbifold} and \ref{1_Orbifold_2}, we study the model of Chapter~\ref{1_4D perturbative approach} on an orbifold background. Finally, in Section~\ref{H_out_boundary}, we generalize the study of Chapter~\ref{1_4D perturbative approach} to the case of a Higgs field localized on a brane away from a boundary.

\section{Boundary Localized Higgs Field: Generalization to a Slice of AdS$_5$}
\label{One_generation_quarks}

\subsection{Randall-Sundrum 1 Setup}
In this section, we want to generalize the model of Chapter~\ref{1_4D perturbative approach} to the case of a warped extra dimension. We consider the 5D spacetime of the RS1 model. The extra dimension is still compactified on an interval $I=[0, L]$ but now the two 3-branes have opposite tensions. The gravitationnal backreaction of the brane tensions is balanced by the introduction of a negative bulk cosmological constant. At the price of a fine tunning between these quantities, the fifth dimension is warped in order to get a 5D bulk as a slice of an AdS spacetime. Once integration is performed over the compact coordinate, the effective 4D cosmological constant vanishes. The 5D metric, solution of the Einstein equations, is given by
\begin{equation}
ds^2 = e^{-2ky} \eta_{\mu\nu} dx^\mu dx^\nu - dy^2 \equiv g_{MN} dx^M dx^N,
\label{metric}
\end{equation}
where $1/k$ is the AdS curvature. The fünfbein $e^A_M$ is defined by
\begin{equation}
g_{MN} = e^A_M e^B_N \eta_{AB}
\label{metric2}
\end{equation}
and its inverse by $e^M_A e^B_M = \delta^B_A$, with $A \in \left\{ a, 4 \right\}$ and $a \in \left\{ 0, 1, 2, 3 \right\}$. We define also $e = \left|\mathrm{det} \, e_M^A \right|$. It is often more convenient to work in the conformally flat frame by making the coordinate transformation
\begin{equation}
z = \dfrac{\text{e}^{ky}}{k},
\end{equation}
such that the conformally flat fünfbein is
\begin{equation}
e_A^M = kz \, \delta_A^M \ \ \ \text{and} \ \ \ e = (kz)^{-5} \, ,
\label{vielbein}
\end{equation}
so
\begin{equation}
g_{MN} = \left( \dfrac{1}{kz} \right)^2 \eta_{MN} \, .
\label{warped_metric_20}
\end{equation}
In this frame, the UV-brane at $y=0$ and the IR-brane at $y=L$, where the electroweak symmetry breaking occurs, are respectively at $z=1/k$ and $z=1/T$, where $T = k \text{e}^{-kL}$. If $M_*$ is the 5D gravity scale, one should have $M_*/k \gtrsim \mathcal{O}(10)$ such that we are in the classical regime where one can use the Einstein equations: the AdS$_5$ background metric is well defined. The cut-off\footnote{\textit{i.e.} the scale at which gravity becomes strong.} on the UV-brane is usually taken $\Lambda_{UV} \sim M_*$ and the one on the IR-brane is $\Lambda_{IR} = \text{e}^{-kL} \Lambda_{UV}$, because of the gravitational redshift induced by the warp factor. If one takes $\Lambda_{IR} \sim \mathcal{O}(1) \ \text{TeV}$ and a IR-brane localized Higgs field, one can solve the naturalness problem of the Higgs sector in the SM.

In the following, we focus on a toy model containing only bulk fermions and a brane localized Higgs field, and study their resulting Yukawa interactions.

\subsection{Field Content}
\label{model_1}
\subsubsection{Brane Localized Higgs field}
The Higgs field is still confined to the IR-brane. For simplicity, we take just a real scalar without gauge quantum numbers. Its Lagrangian has the generic form
\begin{equation}
\mathcal{L}_H = \dfrac{1}{2} \, \partial_\mu H \, \partial^\mu H - V(H).
\end{equation}
As for the SM, the minimum of the Higgs potential $V(H)$ determines a non-vanishing VEV $v$ for $H$
\begin{equation}
H(x) = \dfrac{k}{T} \cdot \dfrac{v + h(x)}{\sqrt{2}}.
\end{equation}
The factor $k/T$ comes from the induced metric (Eq.~\eqref{warped_metric_20}) on the IR-brane and the usual rescaling of the brane localized Higgs field, in order to define $v$ as the electroweak VEV.

\subsubsection{Fermion Content and Yukawa Interactions}
Our goal is to study a chiral 4D effective field theory with renormalizable Yukawa couplings to the Higgs field, in order to generate Dirac masses for the fermions as, for example, in the quark sector of the SM. For that purpose, the minimal spin-$1/2$ fermion field content is a couple of fermions, $\mathcal{Q}$ and $\mathcal{D}$. In the spirit of RS1 models with bulk matter, the fields $\mathcal{Q}(x, y)$ and $\mathcal{D}(x, y)$ propagate in the extra dimension. We split the fermion Lagrangian into a free part $\mathcal{L}^F_{\Psi}$, the BBTs $\mathcal{L}_B$, and a Yukawa interaction with the Higgs field $\mathcal{L}_Y$, such that the 5D action is
\begin{equation}
S_{5D} = \int d^4x \int_{0}^{L} dy \, e \left\{ \mathcal{L}^Q_{\Psi} + \mathcal{L}^D_{\Psi} + \left[ \delta(y)-\delta(y-L) \right] \mathcal{L}_{B} + \delta(y-L) \left( \mathcal{L}_H +  \mathcal{L}_Y \right) \right\}.
\label{S_5D_2}
\end{equation}
The passage to the conformally flat frame requires the following substitution: $\delta(y) \rightarrow \delta\left(z-k^{-1}\right)$ and $\delta(y-L) \rightarrow\dfrac{k}{T} \delta\left(z-T^{-1}\right)$.

The free Lagrangian for a fermion $\mathcal{F} = \mathcal{Q}, \mathcal{D}$ is
\begin{equation}
\mathcal{L}^F_{\Psi} = \dfrac{i}{2} \left( \bar{\mathcal{F}} e^M_A \Gamma^A \nabla_M \mathcal{F} - \bar{\nabla_M \mathcal{F}} e^M_A \Gamma^A \mathcal{F} \right) - M_F \, \bar{\mathcal{F}}\mathcal{F},
\label{1_L_6}
\end{equation}
with $\bar{\mathcal{F}} = \mathcal{F}^\dagger \Gamma^0$ and $\bar{\nabla_M \mathcal{F}} = \left( \nabla_M \mathcal{F} \right)^\dagger \Gamma^0$. The covariant derivative $\nabla_M = \partial_M + \omega_M$ contains the spin connection $\omega_M$. It is known that in the case of the RS metric, Eq.~\eqref{metric}, the contribution of $\omega_M$ cancels in the action and one can safely replace $\nabla_M \rightarrow \partial_M$ in Eq.~\eqref{1_L_6}. We define the bulk mass $M_F$ in units of the fundamental scale $k$ such that $M_F = c_F \, k$.
The 5D spin-$1/2$ fermions are decomposed as
\begin{equation}
\mathcal{F} =
\begin{pmatrix}
F_\alpha \\
\bar{F}^{\dagger \dot{\alpha}}
\end{pmatrix},
\end{equation}
in terms of two Weyl spinors\footnote{From now on, we use calligraphic letters to denote 5D (Dirac) spinors and typed letters to denote Weyl spinors.} $F_\alpha$ and $\bar{F}^{\dot{\alpha}}$ with $\alpha, \dot{\alpha} \in \{ 1,2 \}$. We use the conventions of Ref.~\cite{Martin:1997ns} which are summarized in Appendix~\ref{conventions}. We point out that the bar on $\bar{F}^{\dot{\alpha}}$ is part of the name of the field, and does not denote the operation of Hermitian conjugation.
After some algebra, Eq.~\eqref{1_L_6} becomes
\begin{equation}
\mathcal{L}^F_\Psi = kz \left[ i \left( F^\dagger \bar{\sigma}^\mu \partial_\mu F + \bar{F} \sigma^\mu \partial_\mu \bar{F}^\dagger \right) + \dfrac{1}{2} \left( \bar{F} \overleftrightarrow{\partial_z} F - F^\dagger \overleftrightarrow{\partial_z} \bar{F}^\dagger \right) \right] - k \, c_F \left( F \bar{F} + F^\dagger \bar{F}^\dagger \right) \, ,
\label{L_7}
\end{equation}
where we have dropped a 4-divergence term from an integration by parts. Below the KK scale (in the decoupling limit), we would like to recover the SM so we would like zero modes for $Q$ and $\bar{D}$ only. For that purpose, one needs the BBTs,
\begin{equation}
\mathcal{L}_B = \dfrac{1}{2} \left( \bar{\mathcal{D}} \mathcal{D} - \bar{\mathcal{Q}} \mathcal{Q} \right) \, ,
\label{1_L_B_2}
\end{equation}
which allow the matching between the 4D and 5D methods.

The Yukawa interactions with the Higgs field are
\begin{equation}
\mathcal{L}_Y = -Y_5 \, \bar{D}HQ -Y_5' \, D^\dagger H \bar{Q}^\dagger + \text{H.c.} \, .
\label{L_8}
\end{equation}
We focus on the VEV of the Higgs field and include the $H$ boson fluctuation only in a later subsection which deals with the Yukawa couplings. $\mathcal{L}^F_Y$ therefore becomes
\begin{equation}
\mathcal{L}^F_X = -X \, \bar{D}Q -X' \, D^\dagger \bar{Q}^\dagger + \text{H.c.} \, ,
\label{L_9}
\end{equation}
with the compact notation $X = Y_5 \, \dfrac{k}{T} \cdot \dfrac{v}{\sqrt{2}}$ and $X' = Y'_5 \, \dfrac{k}{T} \cdot \dfrac{v}{\sqrt{2}}$. 


\subsection{5D Method}
\subsubsection{Euler-Lagrange Equations and Boundary Conditions}
One can generalize Hamilton's principle in Appendix~\ref{HVP_general} to the case of a metric with a warp factor. With the Lagrangian $e ( \mathcal{L}^Q_{\Psi} + \mathcal{L}^D_{\Psi} )$ \eqref{L_7}, the 5D Euler-Lagrange equations \eqref{ELE_5D} give
\begin{equation}
\begin{array}{l c c}
i \bar{\sigma}^\mu \partial_\mu F - \partial_z \bar{F}^\dagger + \dfrac{2-c_F}{z} \bar{F}^\dagger &=& 0, \\ \\
i \sigma^\mu \partial_\mu \bar{F}^\dagger + \partial_z F - \dfrac{2+c_F}{z} F &=& 0.
\end{array}
\label{5D_ELE_3}
\end{equation}
At $z=1/k$, we obtain the natural boundary conditions \eqref{BC_UV}: $\bar{Q}|_{y=0,L} = 0$ and $D|_{y=0,L} = 0$. At $z=1/T$, the Hamilton's principle with the Lagrangians $e \, ( \mathcal{L}^Q_{\Psi} + \mathcal{L}^D_{\Psi} )$ \eqref{L_7}, $e \, \left( \dfrac{1}{T} \right) \,  \dfrac{k}{T} \, \mathcal{L}_B$ \eqref{1_L_B_2} and $e \, \left( \dfrac{1}{T} \right) \, \dfrac{k}{T} \, \mathcal{L}_Y$ \eqref{L_8} give the natural boundary condition \eqref{BC_IR}
\begin{equation}
\left. \bar{Q}  \right|_{z=1/T} = \left. D  \right|_{z=1/T} = 0 \ \ \ \textbf{or else} \ \ \ X' = 0 \, ,
\label{BC_8}
\end{equation}
and
\begin{equation}
\left. \left( \bar{Q}^\dagger - X \, \bar{D}^\dagger \right) \right|_{z=1/T} = \left. \left( D + X^* \, Q \right)\right|_{z=1/T} = 0 \, .
\label{BC_10}
\end{equation}
In order to obtain a mass spectrum which depends on $X$ (we want to recover the SM in the decoupling limit), we choose $X'=0$ so $Y_5^\prime=0$. The only warp factor which affects the boundary conditions at $z=T^{-1}$ is the redshift of the Higgs field VEV in $X$. One can check whether the phase of the 5D Yukawa coupling $Y_5$ has a physical impact. If one performs the following simultaneous transformations:  $Y_5 \rightarrow e^{i \alpha} \, Y_5$, $Q \rightarrow e^{-i \alpha} Q$ and $\bar{Q} \rightarrow e^{i \alpha} \, \bar{Q}$, the Lagrangians \eqref{L_7}, \eqref{1_L_B_2} and \eqref{L_8} are invariant. Since the phase of $Y_5$ can be redefine by a transformation, this phase is not physical, which is similar to the SM with only one generation of fermions. One can thus choose $Y_5=|Y_5|$. In the limit $X \rightarrow \infty$, one needs $\lim_{z \to 1/T} \bar{D} = \lim_{z \to 1/T} Q = 0$ in order to keep $\lim_{z \to 1/T} \bar{Q}$ and $\lim_{z \to 1/T} D$ finite, so an infinite Higgs field VEV is equivalent to a Dirichlet boundary conditions on $\bar{Q}$ and $D$.

\subsubsection{Kaluza-Klein Decomposition \& Equations for the KK Wave Functions}
Because of the mass term $\mathcal{L}_X^F$ localized on the IR-brane, the two KK towers from $Q$ and $\bar{Q}$ will mix respectively with the ones from $D$ and $\bar{D}$. We follow the same method as in Section~\ref{1_Yukawa_terms_as boundary_conditions}, and introduce the \textit{mixed} KK decomposition:
\begin{equation}
\begin{array}{c c c}
Q(x,z) &=& \displaystyle{(kz)^{3/2} \sum_n q_n(z)\, \psi_n(x)}, \\ \\
\bar{Q}^\dagger(x,z) &=& \displaystyle{(kz)^{3/2} \sum_n \bar{q}_n(z)\, \bar{\psi}^{\dagger}_n(x)}, \\ \\
D(x,z) &=& \displaystyle{(kz)^{3/2} \sum_n d_n(z)\, \psi_n(x)}, \\ \\
\bar{D}^\dagger(x,z) &=& \displaystyle{(kz)^{3/2} \sum_n \bar{d}_n(z)\, \bar{\psi}^{\dagger}_n(x)},
\end{array}
\label{KK_decomposition}
\end{equation}
where the Weyl spinors $\psi_n(x)$ and $\bar{\psi}^{\dagger}_n(x)$ obey the usual Dirac-Weyl equations,
\begin{align}
\forall n \, , \ i \bar{\sigma}^\mu \partial_\mu \psi_n \left( x \right) &= m_n \, \bar{\psi}_n^\dagger \left( x\right)\, ,
\nonumber \\
i \sigma^\mu \partial_\mu \bar{\psi}_n^\dagger \left( x \right) &= m_n \, \psi_n \left( x \right)\, ,
\label{Dirac_eq}
\end{align}
with $m_n$ the mass of the $n^{\text{th}}$ KK mode, and where $f_n(z)$ and $\bar{f}_n(z)$ ($f = q, \, d$) are the KK wave functions, orthonormalized so that
\begin{equation}
\begin{array}{c c c}
\displaystyle{\int_{1/k}^{1/T} \dfrac{dz}{kz} \left[ q_n^*(z) \, q_m(z) + d_n^*(z) \, d_m(z) \right]} &=& \delta_{nm}, \\ \\
\displaystyle{\int_{1/k}^{1/T} \dfrac{dz}{kz} \left[ \bar{q}_n^*(z) \, \bar{q}_m(z) + \bar{d}_n^*(z) \, \bar{d}_m(z) \right] } &=& \delta_{nm} \, ,
\end{array}
\label{GOC}
\end{equation}
in order to recover canonical kinetic terms for the KK modes when plugging the mixed KK decomposition into the 5D action from Eq.~\eqref{S_5D_2}.
This, together with the Dirac-Weyl equations, Eq.~\eqref{Dirac_eq}, and the 5D Euler-Lagrange equations \eqref{5D_ELE_3}, gives the equations for the KK wave functions
\begin{equation}
\begin{array}{c c c}
\partial_z \bar{f}_n(z) - m_n \, f_n(z) - \dfrac{1}{z} \left( \dfrac{1}{2} - c_F \right) \bar{f}_n(z) &=& 0, \\ \\
\partial_z f_n(z) + m_n \, \bar{f}_n(z) - \dfrac{1}{z} \left( \dfrac{1}{2} + c_F \right) f_n(z) &=& 0.
\end{array}
\label{1st_order_2}
\end{equation}
One can easily decouple these first order differential equations into second order equations:
\begin{equation}
\begin{array}{c c c}
\left( z^2 \partial_z^2 + z \partial_z + m_n^2 z^2 - \left[ c_F - \dfrac{1}{2} \right]^2 \right) \left( \dfrac{f_n(z)}{kz} \right) = 0, \\ \\
\left( z^2 \partial_z^2 + z \partial_z + m_n^2 z^2 - \left[ c_F + \dfrac{1}{2} \right]^2 \right) \left( \dfrac{\bar{f}(z)}{kz} \right) = 0.
\end{array}
\label{2nd_order_2}
\end{equation}
These are Bessel equations, whose solutions are combinations of Bessel functions of the first and the second kind:
\begin{equation}
\begin{array}{c c c c c c c}
q_n(z) &=& \dfrac{kz}{N_n^q} f^q_n(z) & \ \text{with}\ & f^q_n(z) &=& J_\alpha (m_n z) + b_n^q \, Y_\alpha (m_n z), \\ \\
\bar{q}_n(z) &=& \dfrac{kz}{\bar{N}_n^q} f^{\bar{q}}_n(z) &\ \text{with}\ & f^{\bar{q}}_n(z) &=& J_{\alpha + 1} (m_n z) + \bar{b}_n^q \, Y_{\alpha + 1} (m_n z), \\ \\
d_n(z) &=& \dfrac{kz}{N_n^d} f^d_n(z) &\ \text{with}\ & f^d_n(z) &=& J_{\beta + 1} (m_n z) + b_n^d \, Y_{\beta + 1} (m_n z), \\ \\
\bar{d}_n(z) &=& \dfrac{kz}{\bar{N}_n^d} f^{\bar{d}}_n(z) &\ \text{with}\ & f^{\bar{d}}_n(z) &=& J_\beta (m_n z) + \bar{b}_n^d \, Y_\beta (m_n z),
\end{array}
\label{profiles_2}
\end{equation}
where $\alpha = \pm \left( c_Q - \dfrac{1}{2}  \right)$ and $\beta = \pm \left( c_D + \dfrac{1}{2} \right)$. We choose the prescription $+$ for $\alpha$ and $-$ for $\beta$ and will come back later on the reason for this choice.

\subsubsection{Boundary Conditions \& Mass Spectrum}
The boundary conditions at $z = k^{-1}$, $\bar{Q}|_{z=1/k} = 0$ and $D|_{z=1/k} = 0$, translate into Dirichlet boundary conditions for the KK wave functions of $Q$ and $\bar{D}$,
\begin{equation}
\bar{q}_n \left( k^{-1} \right) = 0 \ \ \ \textbf{and} \ \ \ d_n \left( k^{-1} \right) = 0.
\label{BC_6}
\end{equation}
One can combine them with the first order equations for the KK wave function from Eq.~\eqref{1st_order_2} to get the boundary conditions:
\begin{equation}
\begin{array}{c c c}
\left. \left[ \partial_z q_n(z) - k (\alpha + 1) q_n(z) \right] \right|_{z=1/k} &=& 0, \\ \\
\left. \left[ \partial_z \bar{d}_n(z) - k (\beta + 1) \bar{d}_n(z) \right] \right|_{z=1/k} &=& 0.
\end{array}
\label{BC_16}
\end{equation}
These boundary conditions are consistent with Eq.~\eqref{1st_order_2} and the KK wave functions in Eq.~\eqref{profiles_2} if
\begin{equation}
N_n^q = \bar{N}_n^q \ , \ \ \ N_n^d = - \bar{N}_n^d \ , \ \ \ b_n^q = \bar{b}_n^q = -\dfrac{J_{\alpha + 1}\left( \dfrac{m_n}{k} \right)}{Y_{\alpha + 1}\left( \dfrac{m_n}{k} \right)} \ , \ \ \ b_n^d = \bar{b}_n^d = -\dfrac{J_{\beta + 1}\left( \dfrac{m_n}{k} \right)}{Y_{\beta + 1}\left( \dfrac{m_n}{k} \right)}.
\label{int_const}
\end{equation}

Using the mixed KK decomposition (Eq.~\eqref{KK_decomposition}) in the boundary conditions at $z = T^{-1}$ from Eq.~\eqref{BC_10}, one gets
\begin{equation}
\begin{array}{c c c}
\dfrac{f^{\bar{q}}_n \left( T^{-1} \right)}{N^q_n}  + X \, \dfrac{f^{\bar{d}}_n \left( T^{-1} \right)}{N^d_n} &=& 0, \\ \\
X \, \dfrac{f^q_n \left( T^{-1} \right)}{N^q_n} + \dfrac{f^d_n \left( T^{-1} \right)}{N^d_n} &=& 0,
\end{array}
\end{equation}
where the factorized warp factors in the KK decompositions \eqref{KK_decomposition} simplifies in these boundary conditions. This represents a system of two linear equations for the inverse of the normalization coefficients $N^f_n$. This system has non-trivial solutions (i.e. non-zero) only when the associated determinant vanishes. This gives us the relation from which we can extract the mass spectrum $m_n$:
\begin{equation}
\left. \dfrac{f^{\bar{q}}_n(z) \, f^d_n(z)}{f^q_n(z) \, f^{\bar{d}}_n(z)} \right|_{z=1/T} = X^2,
\label{mass}
\end{equation}
which is independent of the normalization coefficients $N^f_n$.

In the case of equal bulk masses, the left-handed zero mode of $Q$ and the right-handed zero mode of $\bar{D}$ have the same KK wave functions. In our sign convention for $\alpha$ and $\beta$, this is simply the condition $\alpha = \beta$, from which it follows that the common localization parameter is $c \doteq c_Q = -c_D$. This also implies that $b_n \doteq b_n^q = b_n^d$ and $N_n \doteq N_n^q = N_n^d$. Using the KK wave functions from Eq.~\eqref{profiles_2}, the mass spectrum from Eq.~\eqref{mass} is determined by
\begin{equation}
\left. \left( \dfrac{J_{\alpha + 1}(m_n z) + b_n \, Y_{\alpha + 1}(m_n z)}{J_{\alpha}(m_n z) + b_n \, Y_{\alpha}(m_n z)} \right)^2 \right|_{z = T^{-1}} = X^2.
\end{equation}

\subsubsection{Normalization Conditions}
In the same way as in Section~\ref{1_Yukawa_terms_as boundary_conditions}, it is easy to show that one can take the normalization constants $N_n^q$ and $N_n^d$ real. The KK wave functions are thus real. Let us have a closer look at the normalization conditions in Eq.~\eqref{GOC} for the exact KK wave functions. These conditions read for $n=m$.
\begin{align}
\int_{k^{-1}}^{T^{-1}} \frac{\mathrm{d}z}{z} \left[ q_n^2(z) + d_n^2(z) \right] &= k, \label{norm_cdt1} \\ 
\int_{k^{-1}}^{T^{-1}} \frac{\mathrm{d}z}{z} \left[ \bar q_n^2(z) + \bar d_n^2(z) \right] &= k, \label{norm_cdt2}
\end{align}
and in the following we focus on showing that these two conditions are equivalent.

We start by simplifying Eqs.~\eqref{norm_cdt1} and \eqref{norm_cdt2}. To achieve this, we first consider the coupled equations for the KK wave functions $d$ and $\bar{d}$:
\begin{gather}
\partial_z d_n + m_n \bar d_n + \beta \, \frac{d_n}{z} = 0,
\label{d_eom_prime} \\
\partial_z \bar d_n - m_n d_n - (\beta + 1) \, \frac{\partial_z \bar d_n}{z} = 0.
\label{dbar_eom_prime}
\end{gather}
We now perform two operations on the coupled equations: we multiply (i) Eq.~\eqref{d_eom_prime} by $d_n$ and Eq.~\eqref{dbar_eom_prime} by $\bar d_n$ and (ii) Eq.~\eqref{d_eom_prime} by $\bar d_n$ and Eq.~\eqref{dbar_eom_prime} by $d_n$, and in both cases we take the sum of the resulting equations. After a few lines of algebra, it is possible to express the two KK wave functions, $d_n$ and $\bar d_n$, in terms of their derivatives
\begin{align}
\frac{\bar d_n^2}{z} &= \partial_z \left( \frac{d_n^2 + \bar d_n^2}{2}  + \frac{\beta}{m_n} \frac{d_n \bar d_n}{z} \right),
\label{d_expr_deriv} \\
\frac{d_n^2}{z} &= \partial_z \left( \frac{d_n^2 + \bar d_n^2}{2}    + \frac{\beta + 1}{m_n} \frac{d_n \bar d_n}{z} \right).
\label{dbar_expr_deriv}
\end{align}
If one repeats the exercise for the $q$ KK wave functions, the following relations are derived:
\begin{align}
\frac{\bar q_n^2}{z} &= \partial_z \left( \frac{q_n^2 + \bar q_n^2}{2}  - \frac{\alpha+1}{m_n} \frac{q_n \bar q_n}{z} \right),
\label{q_expr_deriv} \\
\frac{q_n^2}{z} &= \partial_z \left( \frac{q_n^2 + \bar q_n^2}{2} - \frac{\alpha}{m_n} \frac{q_n \bar q_n}{z} \right).
\label{qbar_expr_deriv}
\end{align}
Inserting Eqs.~\eqref{d_expr_deriv}-\eqref{qbar_expr_deriv} into the normalization conditions from Eqs.~\eqref{norm_cdt1} and \eqref{norm_cdt2}, we find that
\begin{gather}
\int_{k^{-1}}^{T^{-1}} \mathrm{d}z \, \partial_z \left[ \frac{d_n^2 + \bar d_n^2 + q_n^2 + \bar q_n^2}{2} + \frac{\beta \, d_n \bar d_n - \alpha \, q_n \bar q_n}{m_n z} + \dfrac{d_n \bar{d}_n}{m_n z} \right] = k, \label{refined_norm_cdt1} \\ 
\int_{k^{-1}}^{T^{-1}} \mathrm{d}z \, \partial_z \left[ \frac{d_n^2 + \bar d_n^2 + q_n^2 + \bar q_n^2}{2} + \frac{\beta \, d_n \bar d_n - \alpha \, q_n \bar q_n}{m_n z} - \frac{q_n \bar q_n}{m_n z} \right] = k. \label{refined_norm_cdt2}
\end{gather}
Note that the only difference between the two previous equations is the third terms in Eqs.~\eqref{refined_norm_cdt1} and \eqref{refined_norm_cdt2}. Since the KK wave functions are continuous and differentiable on the closed interval $[k^{-1},T^{-1}]$, we can write the difference between Eqs.~\eqref{refined_norm_cdt1} and \eqref{refined_norm_cdt2} as
\begin{equation}
\int_{k^{-1}}^{T^{-1}} \mathrm{d}z \, \partial_z  \left[ \frac{d_n \bar d_n + q_n \bar q_n}{m_n z} \right] = \frac{d_n (z) \bar d_n (z) + q_n (z) \bar q_n (z)}{m_n z}\bigg|_{k^{-1}}^{T^{-1}}.
\label{third_term}
\end{equation} 
We now use the boundary conditions for the KK wave functions to show that the above expression vanishes. Indeed, as $\bar q_n (k^{-1}) = d_n (k^{-1}) = 0 $ and 
\begin{equation}
\bar d_n (T^{-1}) = \frac{\bar q_n (T^{-1})}{X}, \quad d_n (T^{-1}) = - X q_n (T^{-1}),
\label{boundary_cdt_once_again}
\end{equation}
one immediately realizes that the expression in Eq.~\eqref{third_term} vanishes. As a consequence, Eqs.~\eqref{refined_norm_cdt1} and \eqref{refined_norm_cdt2} are identical, which means that the two normalization conditions in Eqs.~\eqref{norm_cdt1} and \eqref{norm_cdt2} are equivalent. In other words, Eqs.~\eqref{norm_cdt1} and \eqref{norm_cdt2} cannot determine both normalization constants $N_n^d$ and $N_n^q$, but only provide a relationship between the two.

\subsubsection{4D Yukawa Couplings}
\label{Yuk_1}
The 4D Yukawa couplings between the KK modes and the Higgs boson can be obtained by developping the Higgs field around its VEV in Eq.~\eqref{L_8} and performing the KK decompositions of the 5D fields. One simply obtains
\begin{equation}
y_{nm} = Y_5 \, \bar{d}_n \left( T^{-1} \right) q_m \left( T^{-1} \right).
\label{Yukawa_4D}
\end{equation}

\subsection{4D Method}
The 4D method to compute the mass spectrum and Yukawa couplings was done in an unpublished work by our collaborator Andrei Angelescu. We give it in Appendix~\ref{4D_method_appendix} for completeness. As expected, the results of the 5D and 4D methods match.

\subsection{Conclusion}
In this section, we have generalized the treatment of 5D fermions coupled to a boundary localized Higgs field of Chapter~\ref{1_4D perturbative approach} to a warped extra dimension. The lightest mode can be identified with a fermion of the SM which obtain his mass from the boundary localized Higgs field. The method to treat the boundary localized mass terms and Yukawa interactions is still the same: one has to add BBTs in order to have a well defined 5D method and also to get the matching between the 4D and 5D methods.

\section{Various Boundary Localized Terms}
\label{other_boundary_terms}
We are going to study various boundary localized terms without the regularization procedure. We will see the necessity of introducing BBTs for each case.

\subsection{Toy Model: One Fermion in the Bulk}
\label{A_fermion_in_the_bulk}
We consider a 5D flat spacetime. The extra dimension is compactified on an interval $I$ of length $L$, with two 3-branes set at the boundaries. Our goal is to study various brane localized kinetic and mass terms for a bulk fermion field $\mathcal{F}$, using a rigorous mathematical treatment. The corresponding fermionic Lagrangian splits into a free part (without interaction) $\mathcal{L}^F_{K}$ containing the kinetic term, and a brane localized part $\mathcal{L}^F_L$ at $y=L$, such that the 5D action is
\begin{equation}
S_{5D} = \int d^4x \int_{0}^{L} dy \left[ \mathcal{L}^F_{K} + \delta(y-L) \, \mathcal{L}^F_L \right] \, .
\label{S_5D}
\end{equation}

The bulk part of the Lagrangian reads
\begin{equation}
\mathcal{L}^F_{K} = \dfrac{i}{2} \bar{\mathcal{F}} \, \Gamma^M \overleftrightarrow{\partial}_M \mathcal{F},
\label{L_1}
\end{equation}
where the 5D spin-$1/2$ fermions are decomposed as
\begin{equation}
\mathcal{F} =
\begin{pmatrix}
F_\alpha \\
\bar{F}^{\dagger \dot{\alpha}}
\end{pmatrix},
\end{equation}
in terms of two Weyl spinors $F_\alpha$ and $\bar{F}^{\dagger \dot{\alpha}}$ with $\alpha, \dot{\alpha} \in \{ 1,2 \}$ (see Appendix~\ref{conventions}). After some algebra and an integration by parts (we drop the 4-divergence), Eq.~\eqref{L_1} becomes
\begin{equation}
\mathcal{L}^F_K = i \left( F^\dagger \bar{\sigma}^\mu \partial_\mu F + \bar{F} \sigma^\mu \partial_\mu \bar{F}^\dagger \right) + \dfrac{1}{2} \left( \bar{F} \overleftrightarrow{\partial_y} F - F^\dagger \overleftrightarrow{\partial_y} \bar{F}^\dagger \right).
\label{L_2}
\end{equation}

In Appendix~\ref{HVP_general}, we explain how to apply Hamilton's principle for a field theory with an extra dimension compactified on $I$. The 5D Euler-Lagrange equations for $F$ and $\bar{F}$ (Eq.~\eqref{ELE_5D}) originate only from the bulk Lagrangian $\mathcal{L}^F_\Psi$ \eqref{L_2} (Eq.~\eqref{ELE_5D}), and we get
\begin{align}
i \bar{\sigma}^\mu \partial_\mu F - \partial_y \bar{F}^\dagger &= 0 \, ,
\nonumber \\
i \sigma^\mu \partial_\mu \bar{F}^\dagger + \partial_y F &= 0 \, .
\label{5D_ELE}
\end{align}
In absence of a brane localized Lagrangian and a Higgs field, we would like that the 5D fermion field reduces to a massless 4D chiral fermion (SM-like before EWSB) in the limit $L \rightarrow 0$ (decoupling limit). For that purpose, it is convenient to decompose the 5D theory into a 4D theory by performing a KK decomposition: each 5D field becomes an infinite tower of 4D KK modes, each of them has a KK wave function along the extra dimension. In order to conserve only the left-handed or the right-handed zero mode, we impose the so-called chiral boundary conditions. Here, we choose to have a left-handed zero mode so the chiral boundary conditions consist of Dirichlet boundary conditions at $y=0,L$ for $\bar{F}$. In Chapter~\ref{1_4D perturbative approach}, it is shown that the chiral boundary conditions can be essential boundary conditions induced by conditions on the conserved current at the boundaries. Then, one can include the brane localized Lagrangians as an infinite matrix in the KK basis to diagonalize. Here we use the alternative method, where the brane localized operators generate boundary conditions for the 5D fields.


\subsection{Boundary Localized Kinetic Terms} 

We begin by studying Brane Localized Kinetic Terms (BLKTs) at $y=L$ such that
\begin{equation}
\mathcal{L}^F_L = \kappa \, i F^\dagger \bar{\sigma}^\mu \partial_\mu F + \kappa' \, i \bar{F} \sigma^\mu \partial_\mu \bar{F}^\dagger,
\label{BLKT}
\end{equation}
where $\kappa$ and $\kappa'$ are real parameters with mass dimension $-1$: the magnitude of the BLKTs. They only involve the partial dervatives $\partial_\mu$ since terms with $\partial_y$ involve subtleties that we will not consider here \cite{Carena:2002me, delAguila:2003bh, delAguila:2006atw}. In an extra-dimensional model with branes, the BLKTs are allowed by the symmetries in the EFT. The UV completion can imply large BLKTs at tree level leading to important phenomenological consequences \cite{Davoudiasl:2002ua, Carena:2002dz, Carena:2003fx, Carena:2004zn, Fichet:2013ola}. Even if absent at tree level, the BLKTs are still generated radiatively \cite{Georgi:2000ks}. In this case, their magnitude is not large, as they originate from loop corrections.

In order to rewrite the 5D theory as a 4D one, we perform a KK decomposition of the fields,
\begin{align}
F \left( x, y \right) &= \displaystyle{ \dfrac{1}{\sqrt{L}} \sum_n f_n(y) \, \psi_n(x) }\,, \nonumber \\
\bar{F}^\dagger \left( x, y \right) &= \displaystyle{ \dfrac{1}{\sqrt{L}} \sum_n \bar{f}_n(y) \, \bar{\psi}^\dagger_n(x)}\, .
\label{KK_1}
\end{align}
With this KK decomposition, the 5D Euler-Lagrange equations \eqref{5D_ELE} reduce to the first order differential equations for the KK wave functions $f_n(y)$ and $\bar{f}_n(y)$:
\begin{align}
\forall n \, , \ \partial_y f_n(y) + m_n \, \bar{f}_n(y) &= 0 \, ,
\nonumber \\
\partial_y \bar{f}_n(y) - m_n \, f_n(y) &= 0 \, ,
\label{1st_order}
\end{align}
and the two 4D Weyl spinors $\psi_n(x)$ and $\bar{\psi}^{\dagger}_n(x)$ combine into a Dirac spinor of mass $m_n$, which represents the $n$-th fermionic KK level, described by the Dirac-Weyl equations \eqref{Dirac_eq}. In order to get canonical kinetic terms for the 4D KK modes, the wave functions $f_n(y)$ and $\bar{f}_n(y)$ are orthonormalized as
\begin{align}
\forall n \, , \ 
\displaystyle{ \dfrac{1}{L} \int_0^L dy \left[ 1 + \kappa \, \delta (y-L) \right] f^*_n(y) f_m(y) } &= \delta_{nm} \, ,
\nonumber \\
\displaystyle{ \dfrac{1}{L} \int_0^L dy \left[ 1 + \kappa' \, \delta (y-L) \right] \bar{f}^*_n(y) \bar{f}_m(y) } &= \delta_{nm} \, .
\label{norm_1}
\end{align}

The first order equations can be decoupled into second order ones as
\begin{align}
\forall n \, , \ \left( \partial_y^2 + m_n^2 \right) f_n(y) &= 0
\nonumber \\
\left( \partial_y^2 + m_n^2 \right) \bar{f}_n(y) &= 0 \, .
\label{2nd_order}
\end{align}
The general solutions are
\begin{align}
\forall n \, , \ f_n(y) &= A_n \, \cos(m_n \, y) + B_n \, \sin(m_n \, y) \, , \nonumber \\
\bar{f}_n(y) &= \bar{A}_n \, \cos(m_n \, y) + \bar{B}_n \, \sin(m_n \, y) \, .
\label{profile_BLKT_1}
\end{align}
where $A_n$, $B_n$, $\bar{A}_n$ and $\bar{B}_n$ are complex coefficients. The first order equations \eqref{1st_order} impose the relations $A_n = \bar{B}_n$ and $B_n = - \bar{A}_n$. We keep the same boundary conditions at $y=0$ as in the absence of BLKTs, $\bar{F}|_{y=0} = 0$, so after the KK decomposition \eqref{KK_1} and with Eq.~\eqref{1st_order},
\begin{equation}
\forall n \, , \ \bar{f}_n(0) = \partial_y f_n(0) = 0 \, .
\label{BC_BLKT_7}
\end{equation}
When imposed on the KK wave functions \eqref{profile_BLKT_1}, the boundary conditions \eqref{BC_BLKT_7} give
\begin{align}
\forall n \, , \ 
f_n(y) &= A_n \, \cos(m_n \, y) \, ,
\nonumber \\
\bar{f}_n(y) &= A_n \, \sin(m_n \, y) \, .
\label{profile_BLKT_2}
\end{align}

\subsubsection{\textit{\textcolor{black}{Failed treatment without bilinear boundary terms}}}
Using Appendix~\ref{HVP_general}, Hamilton's principle applied to the action \eqref{S_5D} with the Lagrangians \eqref{L_2} and \eqref{BLKT} gives the natural boundary conditions at $y=L$ (Eq.~\eqref{BC_IR}),
\begin{align}
\left. \left( \bar{F}^\dagger + 2 \, \kappa \, i \bar{\sigma}^\mu \partial_\mu F \right) \right|_{y=L} &= 0 \, ,
\nonumber \\
\left. \left( F - 2 \, \kappa' \, i \sigma^\mu \partial_\mu \bar{F}^\dagger \right) \right|_{y=L} &= 0 \, .
\label{BC_BLKT_1}
\end{align}

With the KK decomposition \eqref{KK_1} and the 4D Dirac-Weyl equations \eqref{Dirac_eq}, the natural boundary conditions at $y=L$ \eqref{BC_BLKT_1} become boundary conditions for the KK wave functions,
\begin{align}
\forall n \, , \ 
\bar{f}_n(L) + 2 \, \kappa \, m_n \, f_n(L) &= 0 \, ,
\nonumber \\
f_n(L) - 2 \, \kappa' \, m_n \, \bar{f}_n(L) &= 0 \, ,
\label{BC_BLKT_2}
\end{align}
which lead to
\begin{equation}
f_n(L)=\bar{f}_n(L)=0
\label{kin_DDD}
\end{equation}
or else to the relation:
\begin{equation}
\forall n \, , \ -4 \, \kappa \, \kappa' \, m_n^2 = 1 \, .
\label{bad_eq_BLKT}
\end{equation}
With the KK wave functions of Eq.~\eqref{profile_BLKT_2}, the boundary conditions \eqref{kin_DDD} lead to wave functions (and a 5D field) which vanish everywhere in the extra dimension (not physical), \textit{c.f.} Chapter~\ref{1_4D perturbative approach}. Eqs.~\eqref{profile_BLKT_2} and \eqref{bad_eq_BLKT} give the mass spectrum
\begin{equation}
\tan (m_n \, L) = - 2 \kappa \, m_n \, .
\end{equation}
With Eq.~\eqref{bad_eq_BLKT}, there is thus only one massive KK mode (no KK tower) whose mass fix the parameter $\kappa'$ which is not a free parameter. In fact, Eq.~\eqref{bad_eq_BLKT} results from the fact that the system is overconstrained (see discussion below Eq.~\eqref{1_XX'} where we have a similar problem). Moreover, in the 4D method, $\kappa'$ should not play a role since the matrix element involving $\kappa'$ are proportional to the overlap of wave functions with Dirichlet boundary conditions so they vanish (no 4D/5D matching).

\subsubsection{\textit{\textcolor{black}{Treatment with bilinear boundary terms}}}
To overcome the previous difficulty, we add to the action \eqref{S_5D} BBTs:
\begin{equation}
S_B = \int d^4x \int_{-\infty}^{+\infty} dy \, \left[ \delta(y) - \delta(y-L) \right] \, \mathcal{L}_B^F \, ,
\
\label{S_B}
\end{equation}
with
\begin{equation}
\mathcal{L}_B^F = - \dfrac{1}{2} \bar{\mathcal{F}} \mathcal{F} = - \dfrac{1}{2} \left( \bar{F} F + F^\dagger \bar{F}^\dagger \right) \, ,
\label{BBT}
\end{equation}
and we will take also $\kappa' = 0$. The choice $\kappa' \neq 0$ corresponds to a non-vanishing BLKT leading to a variation of the action at $y=L$ involving $\delta \bar{F}$, while we have also the variation of the BLKT depending on $\kappa$ whose variation implies also $\delta F$: there are two natural boundary conditions at the boundary. As the 5D Euler-Lagrange equations \eqref{5D_ELE} relate $F$ and $\bar{F}$ on-shell, a non-vanishing $\kappa$ overconstrains the system. The absence of BLKT for a 5D field with a Dirichlet boundary condition\footnote{It is a perturbative argument based on the fact that one solves the Euler-Lagrange equations first without the BLKTs and then includes them perturbatively through a 4D matrix to diagonalize.} was already noted in the literature. In Ref~\cite{Carena:2004zn}, it was noticed that the BLKT for a field with a Dirichlet boundary condition vanishes at tree level. In the context of holography, Ref.~\cite{Contino:2004vy} showed that the absence of brane localized terms for fields with Dirichlet boundary conditions is required by Hamilton's principle. Moreover, if a brane localized operator is absent at tree level for fields with Dirichlet boundary conditions, it is not radiatively generated \cite{Georgi:2000ks}.

Hamilton's principle applied to the action \eqref{S_5D}, with the Lagrangians \eqref{L_2}, \eqref{BLKT} and \eqref{BBT}, gives the natural boundary conditions at $y=L$ (Eq.~\eqref{BC_IR}),
\begin{equation}
\left. \left( \bar{F}^\dagger + \kappa \, i \, \bar{\sigma}^\mu \partial_\mu F \right) \right|_{y=L} = 0 \, ,
\label{BC_BLKT_3}
\end{equation}
and
\begin{equation}
\kappa' = 0 \ \ \ \text{or} \ \ \ \left. i \sigma^\mu \partial_\mu \bar{F}^\dagger \right|_{y=L} = 0 \, ,
\label{BC_BLKT_4}
\end{equation}
and the natural boundary conditions at $y=0$, $\bar{F}|_{y=0} = 0$.

As above, we use a KK decomposition \eqref{KK_1}. The KK modes satisfy Eq.~\eqref{Dirac_eq} and their wave functions are normalized with \eqref{norm_1}. The boundary conditions~\eqref{BC_BLKT_3} and \eqref{BC_BLKT_4} give
\begin{equation}
\forall n \, , \ \bar{f}_n(L) + \kappa \, m_n \, f_n(L) = 0 \, ,
\label{BC_BLKT_5}
\end{equation}
and
\begin{equation}
\kappa' = 0 \ \ \ \text{or else} \ \ \ \forall n \, , \ m_n \, \bar{f}_n(L) = 0 \, .
\label{BC_BLKT_6}
\end{equation}
The effect of the BLKTs is due to the boundary conditions. For $m_n \neq 0$ in the boundary condition~\eqref{BC_BLKT_5} with BBTs, the limit $\kappa \rightarrow \infty$ implies $\lim_{y \to L} f_n(y) = 0$ to keep $\lim_{y \to L} \bar{f}_n(y)$ finite, since the KK wave functions are bounded as they must be normalized. One can thus impose Dirichlet boundary condition by a BLKT of infinite magnitude.

The KK wave functions satisfy Eq.~\eqref{1st_order}. This is in contrast to what appeared in the literature. In \cite{Csaki:2003sh}, the BLKT is present in the equations for the KK wave functions multiplied by a Dirac distribution. The mass spectrum is obtained by integrating these equations over a small interval around the boundary. In Chapter~\ref{1_4D perturbative approach}, it was shown that such a treatment leads to mathematical inconsistencies and that a Dirac distribution must be treated more carefully within the theory of distributions. The simplest way is to put the boundary terms in the boundary conditions, as we do here. 

The KK wave functions are still given by Eq.~\eqref{profile_BLKT_2}. With the boundary condition at $y=L$ \eqref{BC_BLKT_5} and the KK wave functions expressions \eqref{profile_BLKT_2}, we get the transcendental equation
\begin{equation}
\tan (m_n \, L) = - \kappa \, m_n \, ,
\label{spetr_BLKT}
\end{equation}
whose solutions give the mass spectrum depending on $\kappa$. We see that the BLKT affect the mass spectrum and the shape of the KK wave functions only for $n \neq 0$. On the other hand, when $n=0$, the mode is massless and $\kappa$ will only play a role via the normalization condition. Indeed, combining Eqs.~\eqref{norm_1}, \eqref{profile_BLKT_2} and \eqref{spetr_BLKT}, we obtain
\begin{equation}
|A_n| = \left( \dfrac{1}{2} + \dfrac{\kappa}{2L (1 + \kappa^2 m_n^2)} \right)^{-1/2} \, .
\end{equation}

\subsection{Boundary Localized Majorana Mass Terms}
In this subsection, we study the case of a Majorana mass term localized on the 3-brane \cite{Csaki:2003sh} at $y=L$, which is described by the following Lagrangian:
\begin{equation}
\mathcal{L}^F_L = -\dfrac{\mu}{2} \, \left( FF + F^\dagger F^\dagger \right) - \dfrac{\mu'}{2} \, \left( \bar{F} \bar{F} + \bar{F}^\dagger \bar{F}^\dagger \right) \, ,
\label{L_Maj}
\end{equation}
with $\mu, \mu'$, dimensionless real parameters. Such terms are possible because physics on the 3-brane is 4D and therefore Majorana spinors exist, contrary to the 5D bulk where only Dirac spinors can be defined. These terms could play an important role in brane world model building with neutrinos or with 5D SUSY.

We define a KK decomposition,
\begin{align}
F \left( x, y \right) &= \displaystyle{ \dfrac{1}{\sqrt{L}} \sum_n f_n(y) \, \nu_n \left( x \right) } \, , \nonumber \\
\bar{F}^\dagger \left( x, y \right) &= \displaystyle{ \dfrac{1}{\sqrt{L}} \sum_n \bar{f}_n(y) \, \nu^\dagger_n \left( x \right) } \, ,
\label{KK_Maj}
\end{align}
where the KK states are Majorana fermions $\nu_n$ of mass $m_n$, obeying the homonymous equations,
\begin{align}
i \bar{\sigma}^\mu \partial_\mu \nu_n \left( x \right) &= m_n \, \nu_n^\dagger \left( x \right) \, , \nonumber \\
i \sigma^\mu \partial_\mu \nu_n^\dagger \left( x \right) &= m_n \, \nu_n \left( x \right) \, .
\label{Majorana_eq}
\end{align}
The KK wave functions are orthonormlized to recover canonical 4D kinetic terms for the KK modes, so
\begin{equation}
\dfrac{2}{L} \int_0^L \left( f_n^*(y) f_m(y) + \bar{f}_n^*(y) \bar{f}_m(y) \right) = \delta_{nm} \, .
\label{norm_3}
\end{equation}
The 5D Euler-Lagrange equations \eqref{ELE_5D} lead, after the KK decomposition \eqref{KK_Maj} and using the Majorana equations \eqref{Majorana_eq}, to the coupled first order differential equations \eqref{1st_order} for the KK wave functions. Their solution is similar to the one below Eq.~\eqref{1st_order} with the same boundary condition at $y=0$ \eqref{BC_BLKT_7}, and we get the KK wave functions in Eq.~\eqref{profile_BLKT_2}.

\subsubsection{\textit{\textcolor{black}{Failed treatment without bilinear boundary terms}}}
We apply Hamilton's principle to the action \eqref{S_5D} with the Lagrangians \eqref{L_2} and \eqref{L_Maj} and get the natural boundary conditions (Appendix~\ref{HVP_general}, Eq.~\eqref{BC_IR}),
\begin{align}
\left. \left( \bar{F}^\dagger - 2 \, \mu \, F^\dagger \right) \right|_{y=L} &= 0 \, , \nonumber \\
\left. \left( F + 2 \, \mu' \, \bar{F} \right) \right|_{y=L} &= 0 \, ,
\label{BC_Maj_1}
\end{align}
so
\begin{equation}
\bar{F}|_{y=L} = F|_{y=L} = 0 \, ,
\label{bad_BC_mu_2000}
\end{equation}
or else
\begin{equation}
-4 \mu \mu' = 1 \, .
\label{mu-tuning}
\end{equation}
The case of Eq.~\eqref{bad_BC_mu_2000} with the KK wave functions \eqref{profile_BLKT_2} gives vanishing wave functions (and 5D field) everywhere in the extra dimension, which is not physical. Now we examine the case of the relation \eqref{mu-tuning}. The KK decomposition \eqref{KK_Maj} in Eq.~\eqref{BC_Maj_1} gives the boundary conditions at $y=L$:
\begin{align}
\bar{f}(L) - 2 \mu \, f(L) &= 0 \, , \nonumber \\
f(L) + 2 \mu^\prime \, \bar{f}(L) &= 0 \, ,
\end{align}
which are applied to the wave functions \eqref{profile_BLKT_2} to get the equation for the mass spectrum:
\begin{equation}
\tan (m_n \, L) = 2 \mu \, .
\end{equation}
$\mu'$ is not an independent parameter, which means that the system is overconstrained. Moreover, in the 4D method, $\mu'$ should not play a role since the matrix element involve the overlap with wave functions which have Dirichlet boundary conditions so they vanish: we do not have the 5D/4D matching. Besides, in the limit $\mu, \mu' \rightarrow 0$, $F|_{y=L} = \bar{F} = 0$. One would expect to recover the free case in this limit, where only $\bar{F}$ has a Dirichlet boundary condition at $y=L$. Indeed, when the brane localized Majorana mass terms \eqref{L_Maj} are included with an infinite matrix in the free KK basis to diagonalize, the limit $\mu, \mu' \rightarrow 0$ reduces to the boundary condition of the free case. Instead, we have Dirichlet boundary conditions for both $F$ and $\bar{F}$ at $y=L$, and this is known to imply vanishing 5D fields at each point in the bulk.

\subsubsection{\textit{\textcolor{black}{Treatment with bilinear boundary terms}}}
In the same way as in the case of the BLKTs, we modify the model by adding the boundary term $S_B$ (Eqs.~\eqref{S_B} and \eqref{BBT}). The effect of the brane localized Majorana term enters via the boundary conditions at $y=L$. In order to give a mass to the zero mode, we would like non-vanishing $\mu$. The presence of the term with $\mu'$ overconstrained the system in the same way as in the discussion with the BLKT involving $\kappa'$ so we will take also $\mu'=0$. Again, the brane-localized term, involving fields with Dirichlet boundary conditions in the free case, is vanishing and is not radiatively generated (Ref.~\cite{Georgi:2000ks}).

Hamilton's principle with the Lagrangians \eqref{L_2}, \eqref{L_Maj} and \eqref{BBT} gives the natural boundary conditions at $y=L$ (Appendix~\ref{HVP_general}, Eq.~\eqref{BC_IR}),
\begin{equation}
\left. \left( \bar{F}^\dagger - \mu \, F^\dagger \right) \right|_{y=L} = 0 \, ,
\label{BC_Maj_2}
\end{equation}
and
\begin{equation}
\mu' = 0 \ \ \ \text{or else} \ \ \ \left. \bar{F} \right|_{y=L} = 0 \, ,
\label{BC_Maj_3}
\end{equation}
and the natural boundary condition at $y=0$, $\bar{F}|_{y=0} = 0$.

The natural boundary conditions \eqref{BC_Maj_2} and \eqref{BC_Maj_3} translate into boundary conditions for the KK wave functions as
\begin{equation}
\forall n \, , \ \bar{f}_n(L) - \mu \, f_n(L) = 0 \, ,
\label{BC_Maj_4}
\end{equation}
and
\begin{equation}
\mu' = 0 \ \ \ \text{or} \ \ \ \forall n \, , \ \bar{f}_n(L) = 0 \, .
\label{BC_Maj_5}
\end{equation}
We chose $\mu'=0$. One can notice that, in the case $\mu \rightarrow \infty$, one needs $\lim_{y \to L} f_n(y) = 0$ in order to maintain $\lim_{y \to L} \bar{f}_n(L)$ finite, since the KK wave functions are bounded. One can thus impose Dirichlet boundary conditions by an infinite brane localized Majorana mass.

The boundary condition at $y=L$ \eqref{BC_Maj_4} gives, with the KK wave functions \eqref{profile_BLKT_2}, the equation for the mass spectrum,
\begin{equation}
\tan(m_n \, L) = \mu \, ,
\end{equation}
whose solutions are
\begin{equation}
m_n = \dfrac{1}{L} \left[ \arctan(\mu) + n \pi \right] \, , \ n \in \mathbb{Z} \, .
\end{equation}
The normalization conditions \eqref{norm_3} applied to the profiles \eqref{profile_BLKT_2} give the normalization coefficients $|A_n| = 1 / \sqrt{2}$.

\subsection{Mixing Terms involving a Boundary Localized Fermion}
We now turn towards the case of a bulk fermion $\mathcal{F}$ and a brane localized Weyl fermion $\bar{W}$. The two fermions can mix on the IR brane via a Dirac mass term, with the relevant piece of Lagrangian being
\begin{equation}
\mathcal{L}^F_L = i \bar{W} \sigma^\mu \partial_\mu \bar{W}^\dagger - M \left( \bar{W}F + F^\dagger \bar{W}^\dagger \right) \, ,
\label{L_loc_ferm}
\end{equation}
where $M$ is a real parameter of mass dimension $1/2$.

We use the KK decomposition \eqref{KK_1}. The wave functions are orthonormalized such that we recover canonical 4D kinetic terms for the KK fields:
\begin{align}
\displaystyle{\dfrac{1}{L} \int_0^L dy\, f^*_n(y) f_m(y)} &= \delta_{nm} \, , \nonumber \\
a^{W*}_n a^W_m + \displaystyle{\dfrac{1}{L} \int_0^L dy\, \bar{f}^*_n(y) \bar{f}_m(y)} &= \delta_{nm} \, .
\label{norm_4}
\end{align}
The boundary field can be decomposed in the basis spanned by $\bar{\psi}_n$,
\begin{equation}
\bar{W}^\dagger \left( x \right) = \displaystyle{\sum_n a^W_n \, \bar{\psi}^\dagger_n \left( x \right)} \, ,
\label{w_decomp}
\end{equation}
where $a^W_n$ are mixing coefficients. The KK decomposition \eqref{KK_1} and the Dirac equations \eqref{Dirac_eq} are used to obtain the system in Eq.~\eqref{1st_order} from \eqref{ELE_5D}. The boundary conditions at $y=0$ are still those in Eq.~\eqref{BC_BLKT_7}. We obtain the KK wave functions of Eq.~\eqref{profile_BLKT_2}.

\subsubsection{\textit{\textcolor{black}{Failed treatment without bilinear boundary terms}}}
The presence of the brane localized term \eqref{L_loc_ferm} requires to use natural boundary condition. Hamilton's principle with the Lagrangians \eqref{L_2} and \eqref{L_loc_ferm} gives (Appendix~\ref{HVP_general}, Eq.~\eqref{BC_IR})
\begin{align}
\left. \bar{F}^\dagger \right|_{y=L} - 2 \, M \, \bar{W}^\dagger &= 0 \, , \nonumber \\
F|_{y=L} &= 0 \, , \nonumber \\
i \sigma^\mu \partial_\mu \bar{W}^\dagger - M \left. F \right|_{y=L} &= 0 \, .
\label{BC_loc_ferm_0}
\end{align}
By using the decompositions \eqref{KK_1} and \eqref{w_decomp}, we get
\begin{align}
\bar{f}(L) - 2 M \, a_n^W &= 0 \, , \nonumber \\
f(L) &= 0 \, , \nonumber \\
m_n \, a_n^W &= 0 \, .
\label{syst_W}
\end{align}
With the KK wave functions of Eq.~\eqref{profile_BLKT_2}, we have wave functions which vanish everywhere in the extra dimension. Indeed, the system \eqref{syst_W} is overconstrained.
Since $F$ and $\bar{F}$ are related on-shell by the Euler-Lagrange equations \eqref{5D_ELE}, there is one natural boundary condition too much which overconstrains the system.

\subsubsection{\textit{\textcolor{black}{Treatment with bilinear boundary terms}}}
In order to treat correctly the brane localized fermion, we add to the action the BBTs (Eqs.~\eqref{S_B} and \eqref{BBT}). When we apply Hamilton's principle to the Lagrangians \eqref{L_2}, \eqref{BBT} and \eqref{L_loc_ferm}, we obtain the natural boundary conditions,
\begin{align}
\left. \bar{F}^\dagger \right|_{y=L} - M \, \bar{W}^\dagger &= 0 \, , \nonumber \\
i \sigma^\mu \partial_\mu \bar{W}^\dagger - M \left. F \right|_{y=L} &= 0 \, ,
\label{BC_loc_ferm_1}
\end{align}
and the natural boundary condition at $y=0$, $\bar{F}|_{y=0} = 0$. 

The two boundary conditions at $y=L$ \eqref{BC_loc_ferm_1} give, with Eqs.~\eqref{KK_1} and \eqref{w_decomp},
\begin{equation}
\begin{array}{c c c}
a_n^W &=& \dfrac{1}{M \sqrt{L}} \, \bar{f}_n(L) \, ,  \\ \\
a_n^W &=& \dfrac{M}{m_n \sqrt{L}} \, f_n(L) \, ,
\end{array}
\end{equation}
from which we get the transcendental equation for $m_n$,
\begin{equation}
m_n \, \dfrac{\bar{f}_n(L)}{f_n(L)} = M^2 \ \ \  \Longrightarrow \ \ \ m_n \, \tan(m_n \, L) = M^2 \, .
\label{mass_ferm_loc}
\end{equation}
The normalization coefficients are given by Eq.~\eqref{norm_4}, using Eqs.~\eqref{profile_BLKT_2} and \eqref{mass_ferm_loc}:
\begin{equation}
|A_n| = \left( \dfrac{1}{2} + \dfrac{M^2}{2L (m_n^2 + M^4)} \right)^{-1/2} \, .
\end{equation}

\subsection{Conclusion \& Summary}
We have studied different scenarii of 5D fermions with a boundary localized term: BLKT, Majorana mass, Dirac mass with a boundary localized 4D fermion. In every cases, the absence of BBTs implies pathologies, which indicates that the model is overconstrained. By adding a BBT and allowing a boundary localized term only for the left or right-handed 5D field, we get models wich are not overconstrained.

\section{Brane-Higgs Field within the Orbifold $S^1/\mathbb{Z}_2$}
\label{1_Orbifold}

\subsection{1D Compactification Classification}
When one builds a field theory with a compactified extra dimension, the existence, the degenerescence and the mass spectrum of the KK modes depend on the compact geometry.

Once we have chosen our 1D compact space (the circle $S^1$ or the interval $I$), one can still impose that the action is symmetric under some field transformations with respect to isometries of this compactified space: physics has to be the same at points related by the isometry. In Fig.~\ref{phyics_id}, we give several examples:
\begin{itemize}
\item The circle $S^1$ of radius $R$ is invariant under a translation of generator $\pi R$ or under a reflection $\mathbb{Z}_2$ with respect to an axis passing through its center. One can thus impose to the Lagrangian: $\mathcal{L}(y+\pi R) = \mathcal{L}(y)$ or $\mathcal{L}(-y) = \mathcal{L}(y)$ for the orbifold $S^1/\mathbb{Z}_2$. One can also wind the infinite real line $\mathbb{R}$ around $S^1$ and define a winding number. The Lagrangian on $\mathbb{R}$ has thus to verify $\mathcal{L}(y+n \, 2 \pi R) = \mathcal{L}(y)$ with the winding number $n \in \mathbb{Z}$.
\item The interval $I = [0, L]$, is invariant under the reflection $\mathbb{Z}_2: \ y \mapsto -y$ with respect to the midpoint of the interval. One can thus impose the Lagrangian to be invariant under this transformation. Like for $S^1$, one can define a winding number on the interval which is the number of return trips.
\end{itemize}

\begin{table}[h]
\begin{center}
\begin{tabular}{c|ccc}
\hline
\multicolumn{1}{c|}{\diagbox{Geometry}{Lagrangian}} & Winding & Translation & Parity \\ 
\hline
& & &  \\
\multicolumn{1}{m{2.5cm}<\centering |}{\includegraphics[width=2.5cm]{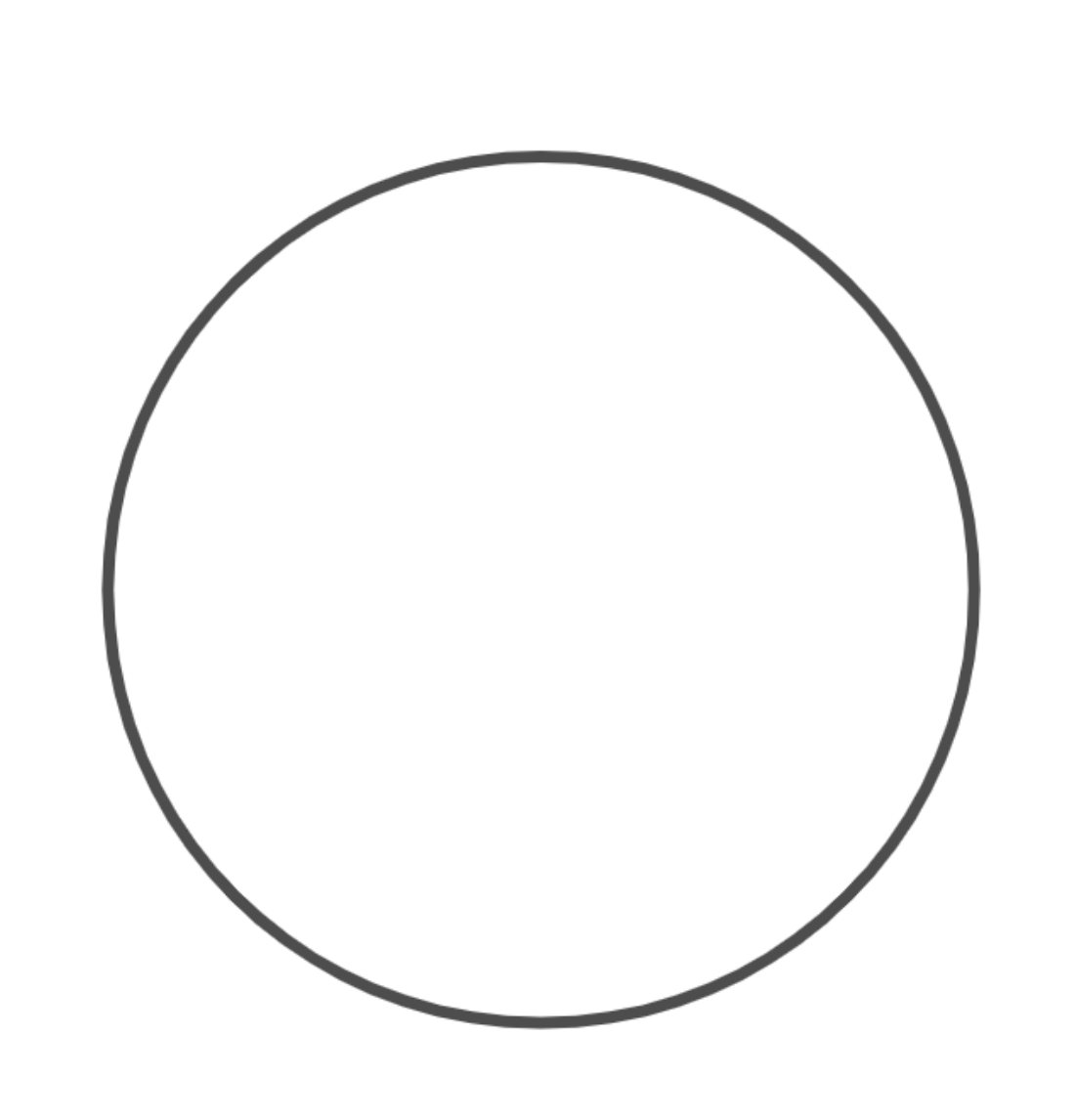}}
& \multicolumn{1}{m{2.5cm}<\centering}{\includegraphics[width=2.5cm]{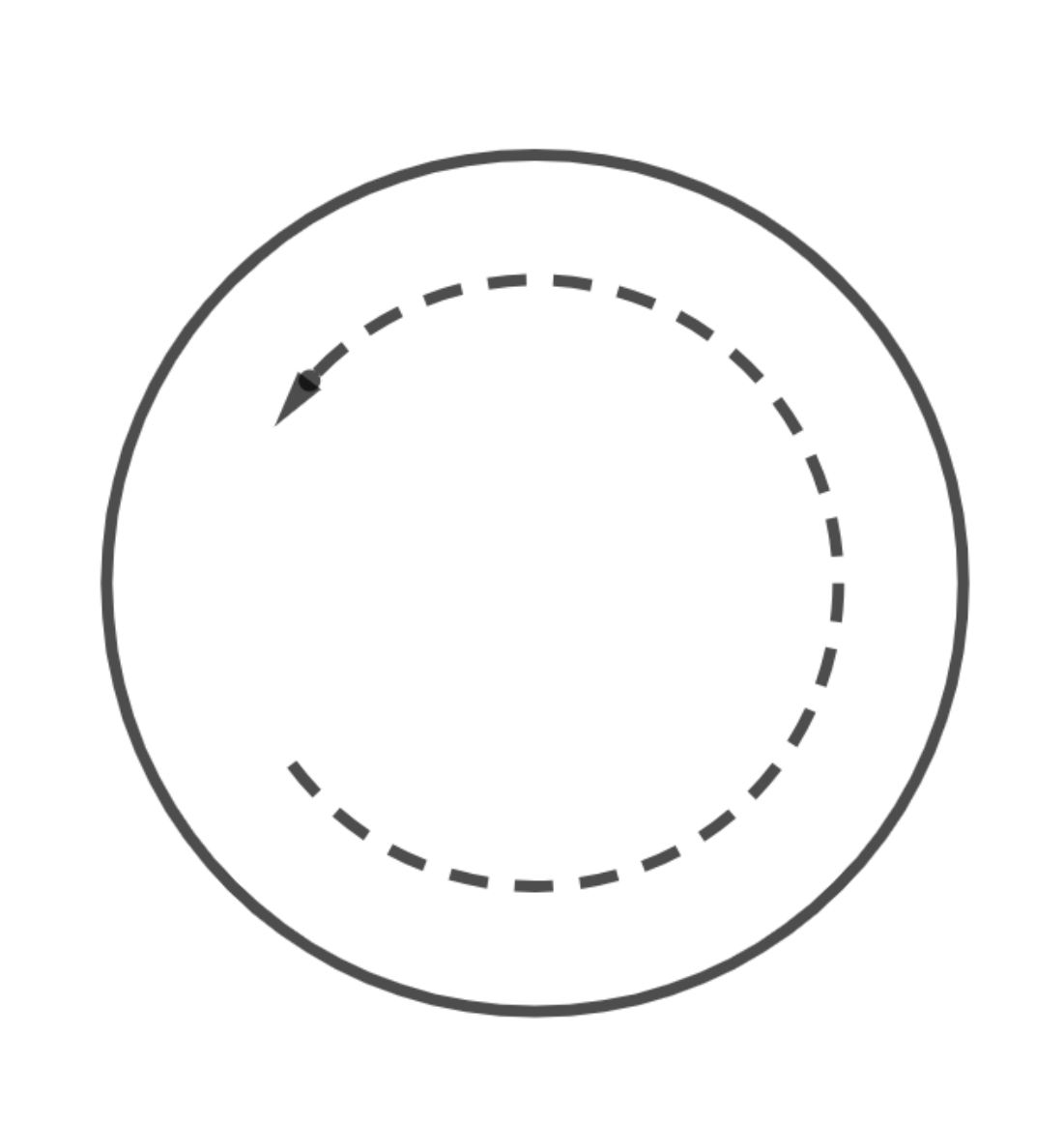}}
& \multicolumn{1}{m{2.5cm}<\centering}{\includegraphics[width=2.5cm]{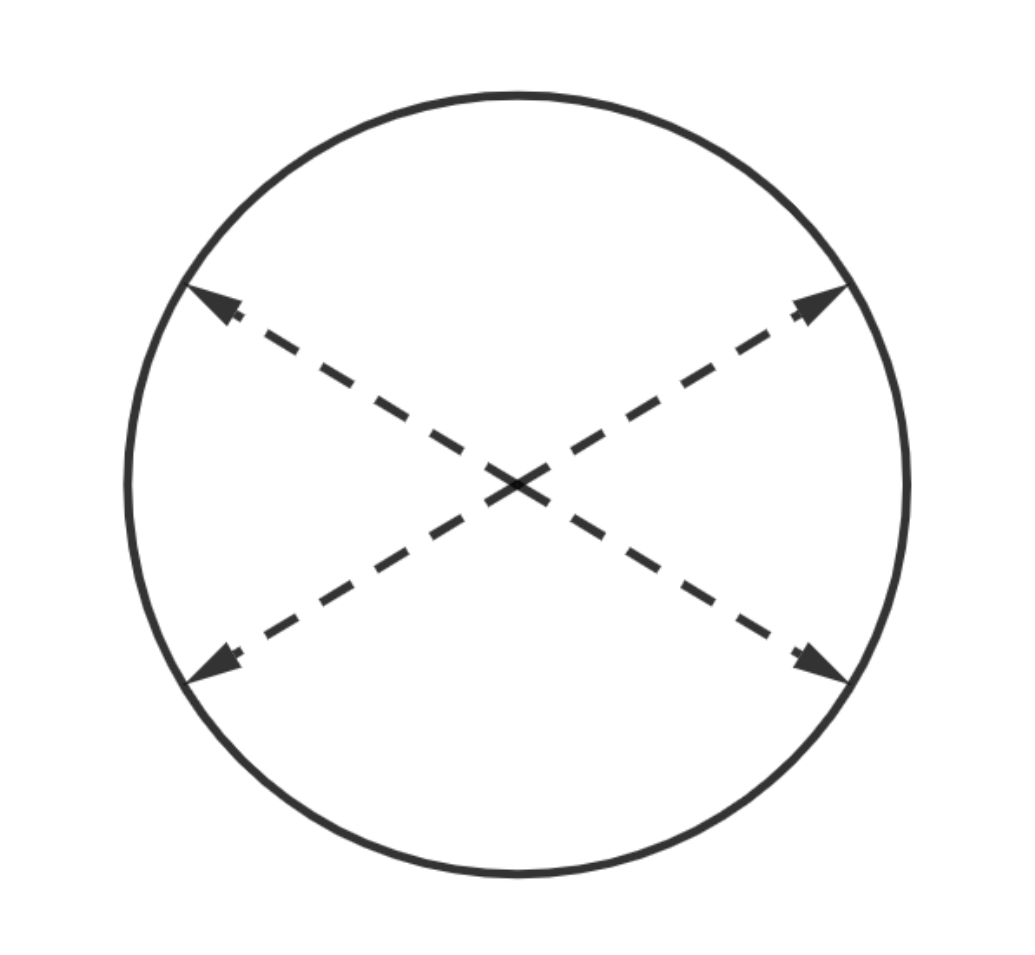}}
& \multicolumn{1}{m{2.5cm}<\centering}{\includegraphics[width=2.5cm]{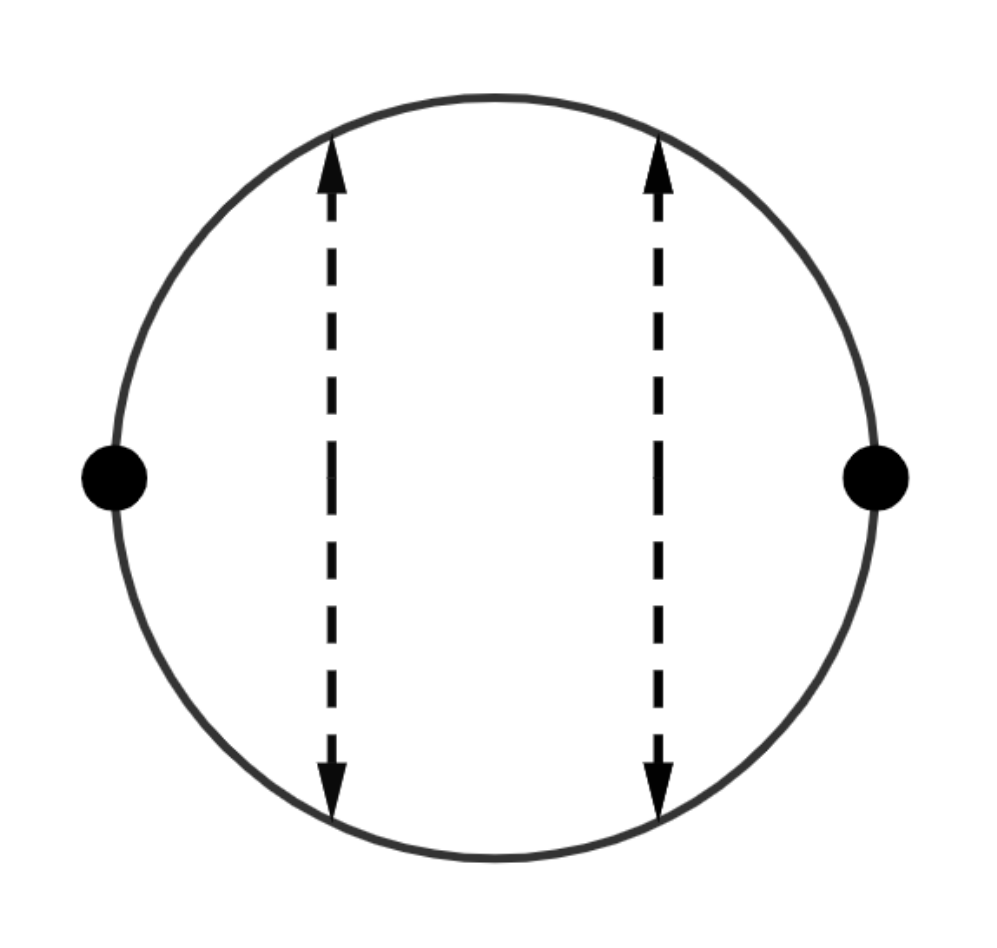}}
\\
& & & \\
\hline
& & &  \\
\multicolumn{1}{m{2.5cm}<\centering |}{\includegraphics[width=2.5cm]{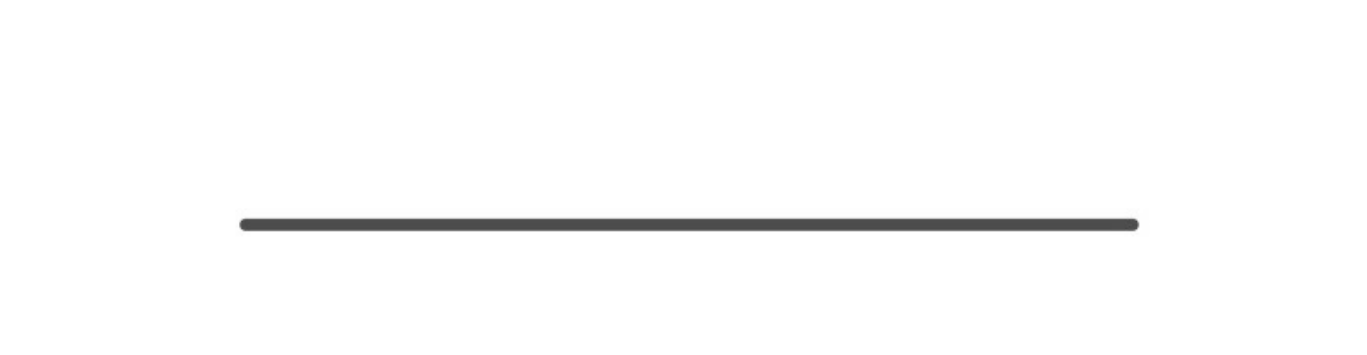}}
& \multicolumn{1}{m{2.5cm}<\centering}{\includegraphics[width=2.5cm]{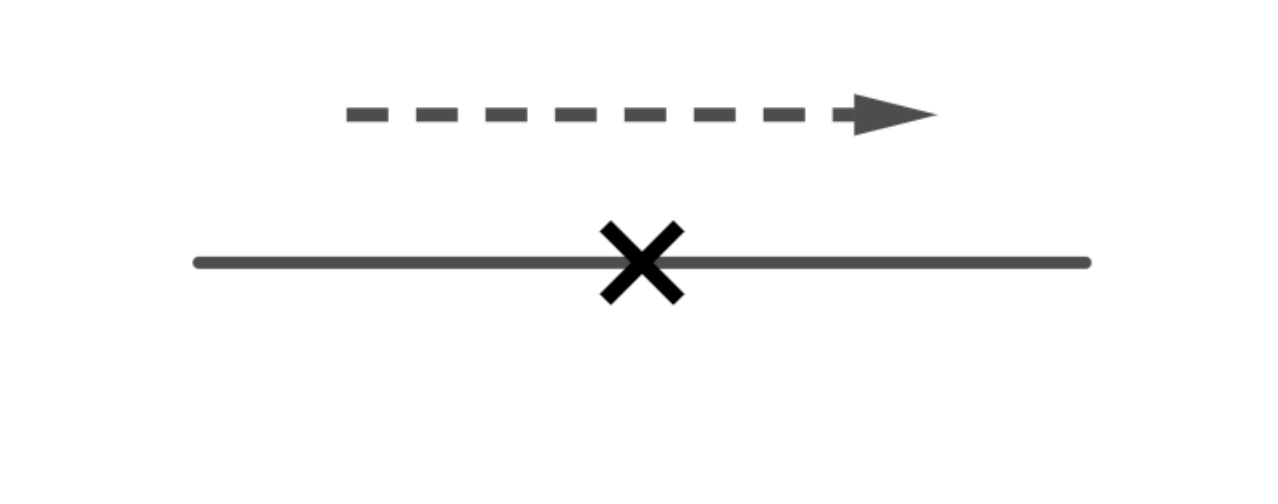}}
&
& \multicolumn{1}{m{2.5cm}<\centering}{\includegraphics[width=2.5cm]{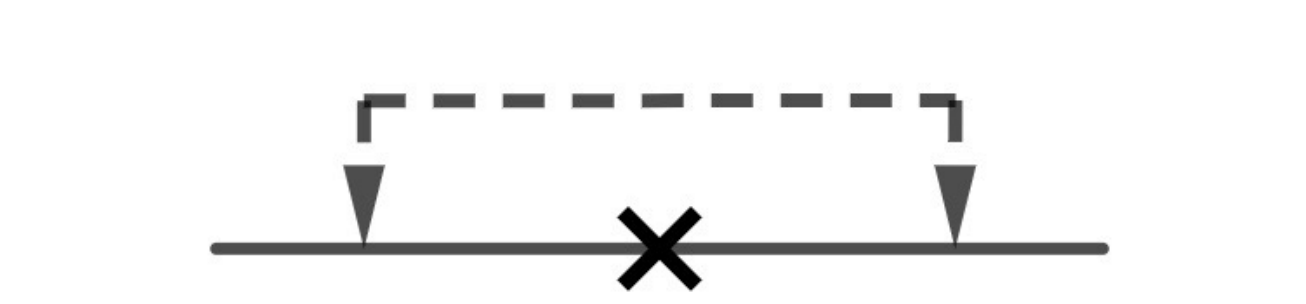}} \\
& & & \\
\hline
\end{tabular} 
\end{center}
\caption{Table of different identifications of the Lagrangian at points related by an isometry of the compact space.
}
\label{phyics_id}
\end{table}

One can then combine these different cases to define more complicated orbifolds like $S^1/(\mathbb{Z}_2 \times \mathbb{Z}_2')$ (see Section~\ref{1_Orbifold_2}). Moreover, the interval and the circle are the smallest building blocks for more complicated 1D geometries: one can glue different circles and/or intervals to obtain metric graphs (see Chapter~\ref{large_star_rose_ED_small_leaves_petals}).


\subsection{The Model on the Orbifold $S^1/\mathbb{Z}_2$}
The traditional way to obtain a 4D chiral theory, at the level of the zero modes, from a 5D theory is by compactifying the extra dimension on an orbifold. In this part, we reformulate our 5D treatment of Section~\ref{1_Yukawa_terms_as boundary_conditions} with an orbifold background. We use the definition of an orbifold given in Subsection~\ref{compact_orbifold}. Let $S^1$ be the circle of radius $R$. We assign a coordinate $y \in (-\pi R, \pi R]$ to the points of $S^1$. The orbifold $S^1/\mathbb{Z}_2$ is obtained by defining an equivalence relation between the points of $S^1$ belonging to the same orbit under $\mathbb{Z}_2$:
\begin{equation}
y \sim -y \, .
\label{1_ident}
\end{equation}
We want to study the model of Chapter~\ref{1_4D perturbative approach} on the geometry $\mathcal{M}^4 \times S^1 / \mathbb{Z}_2$. The standard procedure is to define the field theory on the covering space $S^1$. Physics has to be the same at points belonging to the same equivalence class: the values of the fields at these points have to match up to a symmetry transformation of the fields:
\begin{equation}
\Phi(-y) = Z \, \Phi(y) \, , \ \ \ Z^2 = I \, ,
\end{equation}
and the Lagrangian is the same at these points:
\begin{equation}
\mathcal{L}(-y) = \mathcal{L}(y) \, .
\end{equation}
The orbifold $S^1 / \mathbb{Z}_2$ has two fixed points at $y=0, \pi R$, which have the properties of 3-branes. Its fundamental domain is the interval $[0, \pi R]$.

The model has the same field content as in Subsection~\ref{1_A_toy_model_of_flat_extra_dimension}. The bulk action of Eq.~\eqref{1_eq:actionKin} becomes
\begin{align}
&S_{\Psi} = \int d^4x \oint_{-\pi R}^{+\pi R} dy \ \mathcal{L}_\Psi \, ,
\label{S_Psi_S1/Z2_0}
\end{align}
with $\mathcal{L}_\Psi$ defined as in Eq.~\eqref{1_L_2}.
The transformation laws for the fermionic fields $Q$ and $D$ under $\mathbb{Z}_2$ are chosen as,
\begin{equation} 
\left\{
\begin{array}{c c c}
Q \left( x^\mu, -y \right) = - \gamma^5 \, Q \left( x^\mu, y \right) \, \Longrightarrow \, Q_L \ \text{even}, \ Q_R \ \text{odd},
\\ \vspace{-0.2cm} \\
D \left( x^\mu, -y \right) = \gamma^5 \, D \left( x^\mu, y \right) \, \Longrightarrow \, D_L \ \text{odd}, \ D_R \ \text{even},
\end{array}
\right.
\label{1_Z2}
\end{equation}
under which the action $S_{\Psi}$ \eqref{S_Psi_S1/Z2_0} is invariant. The choice of the sign in front of $\gamma^5$ in Eq.~\eqref{1_Z2} is free and determines the chirality of the zero mode for the 5D Dirac field. Here, we require that the zero modes for the field $Q/D$ are left/right-handed respectively to recover the SM at low energies.

.

\subsection{From Fields as Distributions to Fields as Functions: an Additional Motivation for Bilinear Brane Terms}
The branes at the fixed points are not boundaries of the covering space $S^1$, we have thus to discuss if the fields are continuous or discontinuous across the branes (see Appendix~\ref{app_cont_field}). The even (odd) fields are even (odd) with respect to each fixed point. Even fields are continuous across the branes since a point-like discontinuity has no effect in distribution theory. If the odd fields are continuous at  $y=0, L$, they have to vanish at the fixed points: when one applies Hamilton's principle, their variations have thus to vanish as well at $y=0, L$. If fields are discontinuous at $y=0, L$, their values at the fixed points are not prescribed by the parity of the fields since the jumps are at the fixed points.

We use the distribution theory to define the partial derivatives $\partial_y$ in $\mathcal{L}_\Psi$ \eqref{S_Psi_S1/Z2} as weak derivatives for (discontinuous) odd fields. We treat the fields as distributions\footnote{See Refs.~\cite{Schwartz1, Schwartz2} for an introduction to the theory of distributions by L.~Schwartz. For an application to quantum mechanics, see Ref.~\cite{Messiah1}.}, which act on test functions whose supports are included in the compactified geometry. We define the regular distributions $\widetilde{Q_{L/R}}, \widetilde{D_{L/R}}$. The weak partial derivatives give
\begin{align}
\partial_y \widetilde{Q_L} &= \left\{ \partial_y Q_L \right\} \, , \nonumber \\
\partial_y \widetilde{Q_R} &= \left\{ \partial_y Q_R \right \} + \left( \left. Q_R \right|_{0^+} - \left. Q_R \right|_{0^-} \right) \delta(y) + \left( \left. Q_R \right|_{-\pi R^+} - \left. Q_R \right|_{\pi R^-} \right) \delta(y- \pi R) \, , \nonumber \\
\partial_y \widetilde{D_L} &= \left\{ \partial_y D_L \right\} + \left( \left. D_L \right|_{0^+} - \left. D_L \right|_{0^-} \right) \delta(y) + \left( \left. D_L \right|_{-\pi R^+} - \left. D_L \right|_{\pi R^-} \right) \delta(y- \pi R) \, , \nonumber \\
\partial_y \widetilde{D_R} &= \left\{ \partial_y D_R \right\}
\, ,
\label{weak_dervi_orbi_1}
\end{align}
where $\left\{ \partial_y Q_{L/R} \right\}, \left\{ \partial_y D_{L/R} \right\}$ are the regular distributions associated to the partial derivatives $\partial_y Q_{L/R}$, $\partial_y D_{L/R}$.

The action for the kinetic terms in distribution theory is
\begin{align}
&S_{\Psi} = \int d^4x \oint_{-\pi R}^{+\pi R} dy \ \widetilde{\mathcal{L}_\Psi} \ , \nonumber \\
&\text{with} \ \ 
\widetilde{\mathcal{L}_\Psi} =  \sum_{F=Q,D} \dfrac{1}{2} \left( i \widetilde{F_R}^\dagger \sigma^\mu \overleftrightarrow{\partial_\mu} \widetilde{F_R} + i \widetilde{F_L}^\dagger \bar{\sigma}^\mu \overleftrightarrow{\partial_\mu} \widetilde{F_L} + \widetilde{F_R}^\dagger \overleftrightarrow{\partial_y} \widetilde{F_L} - \widetilde{F_L}^\dagger \overleftrightarrow{\partial_y} \widetilde{F_R} \right) \, .
\end{align}
$F_L$ or $F_R$ have to be continuous at $y=0, L$, such that the products $\widetilde{F_R}^\dagger \overleftrightarrow{\partial_y} \widetilde{F_L}$ and $\widetilde{F_L}^\dagger \overleftrightarrow{\partial_y} \widetilde{F_R}$ in the kinetic terms are well defined. By using \eqref{weak_dervi_orbi_1}, we can thus rewrite the action $S_\Psi$ \eqref{S_Psi_S1/Z2_0} in terms of fields treated as functions:
\begin{align}
&S_{\Psi} = \int d^4x \left\{ \left[\left(\int_{-\pi R^+}^{0-} dy + \int_{0}^{\pi R} dy \right) \mathcal{L}_\Psi \right] -\left. \mathcal{L}_{B} \right|_{0^-} + \left. \mathcal{L}_{B} \right|_{0^+} - \left. \mathcal{L}_{B} \right|_{\pi R^-} + \left. \mathcal{L}_{B} \right|_{-\pi R^+} \right\} \ , \nonumber \\
&\text{with} \ \ 
\mathcal{L}_\Psi =  \sum_{F = Q, D} \dfrac{1}{2} \left( i F^\dagger_R \sigma^\mu \overleftrightarrow{\partial_\mu} F_R + i F^\dagger_L \bar{\sigma}^\mu \overleftrightarrow{\partial_\mu} F_L + F^\dagger_R \overleftrightarrow{\partial_y} F_L - F^\dagger_L \overleftrightarrow{\partial_y} F_R \right) \ , \nonumber \\
&\text{and} \ \
\mathcal{L}_{B} = \dfrac{1}{2} \left( D_L^\dagger D_R - Q_L^\dagger Q_R \right) + \text{H.c.} \ ,
\label{S_Psi_S1/Z2}
\end{align}
where the BBTs $\mathcal{L}_{B}$ (identical to the ones from the compactification on an interval at Eq.~\eqref{1_eq:actionBound}) come from the weak partial derivative of the odd discontinuous fields at the fixed points. We notice that this action is also valid for continuous odd fields at $y=0, L$, where the boundary terms vanish. It is important to see here that the BBTs are not added by hand to the orbifold $S^1 / \mathbb{Z}_2$ description (as in the interval case) but originate from the weak partial derivative of the odd field discontinuous at the fixed points. In the second integral of Eq.~\eqref{S_Psi_S1/Z2}, we have included the points $y=0, \pi R$ to recover the starting point of the method with the fields as functions. Indeed, we require that all the fields have a well defined partial right-hand (left-hand) derivative $\partial_y$ at $y=0$ ($y=\pi R$) in order to be able to include these points in the action.

The boundary mass terms for the fermions of Eq.~\eqref{1_L_5}, using a Dirac distribution, become
\begin{align}
&S_{X} = \int d^4x \oint_{- \pi R}^{+ \pi R} dy~ \delta(y - \pi R) \ \widetilde{\mathcal{L}_X} \ , \nonumber \\
&\text{with} \ \ 
\widetilde{\mathcal{L}_X} = - X \, \widetilde{Q_L}^\dagger \widetilde{D_R} - X^\prime \, \widetilde{Q_R}^\dagger \widetilde{D_L} +  {\rm H.c.} \ ,
\label{S_X_S1/Z2}
\end{align}
where the action of a Dirac distribution centered at $y=a$ on a test function $\varphi(y)$ is defined as
\begin{equation}
\oint_{- \pi R}^{+ \pi R} dy \ \delta(y-a) \, \varphi(y) = \varphi(a) \, .
\end{equation}
As they involve $\delta(y - \pi R)$, the terms involving $X$ and $X'$ require that the products $Q_L^\dagger D_R$ and $Q_R^\dagger D_L$ are continuous at $y=L$ respectively. The $X$-terms give mass to the zero mode of $Q_L$ and $D_R$, which are identified with the 4D fields of the SM, so we need $X \neq 0$ to recover the SM in the decoupling limit: $Q_L^\dagger D_R$ has thus to be continuous. As $Q_L$ and $D_R$ are even fields, we choose them continuous across the branes. The $X'$-terms are not mandatory to recover the SM in the decoupling limit so $Q_R$ and $D_L$ can be discontinous. The fields $Q_R$ and $D_L$ are odd. If they are continuous at the fixed points, they have to vanish at $y=\pi R$ and the terms involving the coupling $X'$ in the Lagrangian $\mathcal{L}_X$ (Eq.~\eqref{S_X_S1/Z2}) vanish. If they are non-vanishing and discontinuous at $y=L$ the product $Q_R^\dagger D_L$ can still be continuous because $Q_R$ and $D_L$ are odd. We have thus to keep the $X'$-terms in general.

\subsection{Method with 5D Fields as Functions}
\label{different_5D_methods_func}
If we want to treat the problem of a brane away from the boundary by keeping the 5D fields as functions, we have to cut the integral of the action in two pieces on both sides of the brane. If a 5D field is discontinuous across a brane, the 5D kinetic term is in general not defined at the brane position. We require that we have a well defined left-hand or right-hand derivative at the brane position which is equivalent to have a 5D field with a well defined left-hand or right-hand derivative at the brane position. For a brane at $y=\ell$ in the extra dimension, we have the action:
\begin{equation}
S = \int d^4x \int_{- \pi R^+}^{+\pi R} dy \ \mathcal{L}_{bulk} + \int d^4x \left. \mathcal{L}_{brane} \right|_{y=\pi R} \, ,
\end{equation}
where $\mathcal{L}_{bulk}$ and $\mathcal{L}_{brane}$ are the Lagrangians in the bulk and on the brane respectively. Then, one uses Hamilton's principle and treats the brane as a boundary in order to get 5D Euler-Lagrange equations in the bulk and junction conditions on the brane.

\subsubsection{5D Free Fermions}
In this part, we revisit our treatment of free fermions of Section~\ref{1_free_bulk_fermions} but with the orbifold $S^1 / \mathbb{Z}_2$ compactification.

\subsubsection{\textcolor{black}{\textit{No Bilinear Brane Terms}}}
If one does not define the fields as distributions from the start but begins with functions, one does not have the argument to include the BBTs. If one tries to treat the free case without BBTs, one finds vanishing 5D fields everywhere as in the interval case in Subsection~\ref{1_NBC_only}, which is a motivation to include the BBTs.

\subsubsection{\textcolor{black}{\textit{Including the Bilinear Brane Terms}}}
We consider $S_\Psi$ \eqref{S_Psi_S1/Z2} with the possibility that odd fields (and their variations) can be discontinuous at the positions of the 3-branes. We apply Hamilton's principle, treating the fixed points as boundaries:
\begin{align}
0 = \delta_{Q^\dagger_L} \left( S_{\Psi} + S_X \right) &= \displaystyle{ \int d^4x \ \left(\int_{-\pi R^+}^{0^-} dy + \int_{0}^{\pi R} dy \right)  
\delta Q^\dagger_L \left[ i\bar{\sigma}^\mu \partial_\mu Q_L - \partial_y Q_R  \right] } 
\nonumber \\
&\displaystyle{ + \ \int d^4x \, \left[ \left . \left( \delta Q^\dagger_L \, Q_R \right) \right |_{0^-} - \left . \left( \delta Q^\dagger_L \, Q_R \right) \right |_{- \pi R^+} \, \right] }
\nonumber \\
&\displaystyle{ + \ \int d^4x \, \left[ \left . \left( \delta Q^\dagger_L \, Q_R \right) \right |_{\pi R^-} - \left . \left( \delta Q^\dagger_L \, Q_R \right) \right |_{0^+} \, \right] } \ ,
\label{orbi_HVP_1a}
\end{align}
\begin{align}
0 = \delta_{Q^\dagger_R} \left( S_{\Psi} + S_X \right) &= \displaystyle{ \int d^4x \left(\int_{-\pi R^+}^{0^-} dy + \int_{0}^{\pi R} dy \right) 
\delta Q^\dagger_R \left[ i\sigma^\mu \partial_\mu Q_R + \partial_y Q_L  \right] } \, ,
\label{orbi_HVP_1b}
\end{align}
\begin{align}
0 = \delta_{D^\dagger_L} \left( S_{\Psi} + S_X \right) &= \displaystyle{ \int d^4x \left(\int_{-\pi R^+}^{0-} dy + \int_{0}^{\pi R} dy \right)
\delta D^\dagger_L \left[ i\bar{\sigma}^\mu \partial_\mu D_L - \partial_y D_R  \right] } \, ,
\label{orbi_HVP_1c}
\end{align}
\begin{align}
0 = \delta_{D^\dagger_R} \left( S_{\Psi} + S_X \right) &= \displaystyle{ \int d^4x \left(\int_{-\pi R^+}^{0-} dy + \int_{0}^{\pi R} dy \right)
\delta D^\dagger_R \left[ i\sigma^\mu \partial_\mu D_R + \partial_y D_L  \right] } 
\nonumber \\
&\displaystyle{ + \ \int d^4x \ \left[ - \left . \left( \delta D^\dagger_R \, D_L \right) \right |_{0^-} + \left . \left( \delta D^\dagger_R \, D_L \right) \right |_{- \pi R^+} \, \right] }
\nonumber \\
&\displaystyle{ + \ \int d^4x \ \left[ - \left . \left( \delta D^\dagger_R \, D_L \right) \right |_{\pi R^-} + \left . \left( \delta D^\dagger_R \, D_L \right) \right |_{0^+} \, \right] },
\label{orbi_HVP_1d}
\end{align}
We vary the action with respect to fields which are eigenstate of the $\mathbb{Z}_2$ symmetry so the variation of a field has the same parity as the field. We can thus use the parity (Eq.~\eqref{1_Z2}) and the right-hand (left-hand) continuity of the fields and their variations at $y=0$ ($y= \pi R$) to rewrite the variations of the action at $y=0^-, -\pi R^+$ as variations at $y=0, \pi R$:
\begin{align}
0 = \delta_{Q^\dagger_L} \left( S_{\Psi} + S_X \right) &= \displaystyle{ \int d^4x \left(\int_{-\pi R^+}^{0-} dy + \int_{0}^{\pi R} dy \right)
\delta Q^\dagger_L \left[ i\bar{\sigma}^\mu \partial_\mu Q_L - \partial_y Q_R  \right] } 
\nonumber \\
&\displaystyle{ + 2 \int d^4x \ \left[ \left . \left( \delta Q^\dagger_L \, Q_R \right) \right |_{\pi R} - \left . \left( \delta Q^\dagger_L \, Q_R \right) \right |_{0} \, \right] } \ ,
\label{orbi_HVP_2a}
\end{align}
\begin{align}
0 = \delta_{D^\dagger_R} \left( S_{\Psi} + S_X \right) &= \displaystyle{ \int d^4x \left(\int_{-\pi R^+}^{0-} dy + \int_{0}^{\pi R} dy \right)
\delta D^\dagger_R \left[ i\sigma^\mu \partial_\mu D_R + \partial_y D_L  \right] }
\nonumber \\
&\displaystyle{ + 2 \int d^4x \ \left[ - \left . \left( \delta D^\dagger_R \, D_L \right) \right |_{\pi R} + \left . \left( \delta D^\dagger_R \, D_L \right) \right |_{0} \, \right] } \, .
\label{orbi_HVP_2d}
\end{align}
The Euler-Lagrange equations are \eqref{1_ELE_1} (with $y \in S^1$) and the natural boundary conditions are
\begin{equation}
Q_R|_{0, \pi R} = D_L|_{0, \pi R} = 0 \, ,
\end{equation}
so we find that the on-shell odd fields are continuous at the fixed points. On the orbifold $S^1 / \mathbb{Z}_2$ the KK decomposition and the orthonormalization of the KK wave functions are defined such that
\begin{equation}
F_{L/R} \left( x^\mu, y \right) = \dfrac{1}{\sqrt{2 \pi R}} \displaystyle{ \sum_{n} f^n_{L/R}(y) \, F^n_{L/R} \left( x^\mu \right)} \ ,
\label{1_KK_3}
\end{equation}
\begin{equation}
\forall n \ , \forall m \ , \dfrac{1}{2 \pi R} \int_{-\pi R}^{+ \pi R} dy \ f^{n*}_{L/R}(y) \, f^m_{L/R}(y) = \delta^{nm} \, ,
\label{1_normalization_3}
\end{equation}
where we normalize on the entire covering space $S^1$. The KK modes satisfy the Dirac-Weyl equations \eqref{1_Dirac_1} and the system of equations for the KK wave function is the one of Eq.~\eqref{1_ELE_2} (with $y \in S^1$). One can solve the equations as in Subsection~\ref{1_EBC_currents}, with the complete boundary conditions
\begin{equation}
\forall n \, , \ q_R^n(0,\pi R) = d_L^n(0,\pi R) = \partial_y q_L^n(0,\pi R) = \partial_y d_R^n(0,\pi R) = 0 \, .
\label{BC_orbi_1}
\end{equation}
The solutions on $[0, \pi R]$ and mass spectrum are still the ones in Eqs.~\eqref{solfree1}-\eqref{solfree2} (with $L = \pi R$). The solutions on $(-\pi R, 0)$ are obtained from the ones on $[0, \pi R]$ with the $\mathbb{Z}_2$ symmetry.

\subsubsection{\textcolor{black}{\textit{Essential Boundary Conditions}}}
We apply Hamilton's principle imposing the continuity of the fields (and their variations) at the fixed points. The odd fieds are thus vanishing at $y=0, \pi R$. This can be used as essential boundary conditions for the Hamilton's principle, so
\begin{equation}
Q_R|_{0, \pi R} = D_L|_{0, \pi R} = 0 \ \ \ \Rightarrow \ \ \ \delta Q_R|_{0, \pi R} = \delta D_L|_{0, \pi R} = 0 \, .
\label{1_EBC_orbi}
\end{equation}
The variations at the fixed points of Eqs.~\eqref{orbi_HVP_1a}-\eqref{orbi_HVP_1d} are thus vanishing. The Euler-Lagrange equations are still \eqref{1_ELE_1}. Then we introduce the KK decomposition \eqref{1_KK_3} and the orthonormalization conditions \eqref{1_normalization_3}. The system of equations for the KK wave function is the one of Eq.~\eqref{1_ELE_2}. One can solve them as in Subsection~\ref{1_EBC_currents}, with the essential boundary conditions \eqref{1_EBC_orbi} leading to the complete boundary conditions \eqref{BC_orbi_1}. The solutions and mass spectrum are still the ones in Eqs.~\eqref{solfree1}-\eqref{solfree2} (with $y \in S^1$ and $L = \pi R$).

\subsubsection{Brane Localized Yukawa Interactions with the 4D Method}
As in Section~\eqref{1_Mass_matrix_diagonalisation}, one can add the brane localized mass term as a matrix in the free KK basis and bi-diagonalize it. In the following, we focus on the 5D method.

\subsubsection{Brane Localized Yukawa Interactions with the 5D Method}
\label{orbi_S1Z2_Yuk}
In this subsection, we use the 5D method to compute the mass spectrum of the fermions coupled to the Higgs VEV with the orbifold $S^1 / \mathbb{Z}_2$ description, parallel to the treatment in Section~\ref{1_Yukawa_terms_as boundary_conditions} in the interval picture.

The regularization procedure in the context of the orbifold $S^1 / \mathbb{Z}_2$ was not considered in detail in the relevant literature for the shift of the Higgs peak \cite{Casagrande:2008hr}. We find similar inconsistencies for the regularizations of the brane-Higgs coupling as on the interval studied in Section~\ref{1_RegInterval} due to products of Dirac distributions with discontinuous functions. The problems of the regularizations relying on the softening of the brane-Higgs coupling discussed in Subsection~\ref{1_RegSoft} 
remain present as well within the orbifold framework.

\subsubsection{\textcolor{black}{\textit{Neither Bilinear Brane Terms nor Essential Boundary Conditions}}}
If we do not include the BBTs, we find the same problems than in Subsection~\ref{1_Failed_treatment} for the interval.

\subsubsection{\textcolor{black}{\textit{Bilinear Brane Terms}}}
We consider the action $S_\Psi + S_X$ (Eqs.~\eqref{S_Psi_S1/Z2} and \eqref{S_X_S1/Z2}).
When we perform Hamilton's principle, we do not impose the continuity of the odd fields through the fixed points:
\begin{align}
0 = \delta_{Q^\dagger_L} S_{\Psi} &= \displaystyle{ \int d^4x \left(\int_{-\pi R^+}^{0-} dy + \int_{0}^{\pi R} dy \right)
\delta Q^\dagger_L \left[ i\bar{\sigma}^\mu \partial_\mu Q_L - \partial_y Q_R  \right] } 
\nonumber \\
&\displaystyle{ + \ \int d^4x \, \left[ \left . \left( \delta Q^\dagger_L \, Q_R \right) \right |_{0^-} - \left . \left( \delta Q^\dagger_L \, Q_R \right) \right |_{- \pi R^+} \, \right] }
\nonumber \\
&\displaystyle{ + \ \int d^4x \, \left[ \left . \left( \delta Q^\dagger_L \, Q_R \right) \right |_{\pi R^-} - \left . \left( \delta Q^\dagger_L \, Q_R \right) \right |_{0^+} \, \right] }
\nonumber \\
&\displaystyle{ \left. - \int d^4x \ X \, \delta Q^\dagger_L D_R \right |_{\pi R} } \ ,
\label{orbi_HVP_3a}
\end{align}
\begin{align}
0 = \delta_{Q^\dagger_R} S_{\Psi} &= \displaystyle{ \int d^4x \left(\int_{-\pi R^+}^{0-} dy + \int_{0}^{\pi R} dy \right)
\delta Q^\dagger_R \left[ i\sigma^\mu \partial_\mu Q_R + \partial_y Q_L  \right] }
\nonumber \\
&\displaystyle{ - \int d^4x \ X' \, \delta Q_R^\dagger D_L }
\, ,
\label{orbi_HVP_3b}
\end{align}
\begin{align}
0 = \delta_{D^\dagger_L} S_{\Psi} &= \displaystyle{ \int d^4x \left(\int_{-\pi R^+}^{0-} dy + \int_{0}^{\pi R} dy \right)
\delta D^\dagger_L \left[ i\bar{\sigma}^\mu \partial_\mu D_L - \partial_y D_R  \right] }
\nonumber \\
&\displaystyle{ - \int d^4x \ X^{\prime *} \, \delta D_L^\dagger Q_R }
\, ,
\label{orbi_HVP_3c}
\end{align}
\begin{align}
0 = \delta_{D^\dagger_R} S_{\Psi} &= \displaystyle{ \int d^4x \left(\int_{-\pi R^+}^{0-} dy + \int_{0}^{\pi R} dy \right)
\delta D^\dagger_R \left[ i\sigma^\mu \partial_\mu D_R + \partial_y D_L  \right] } 
\nonumber \\
&\displaystyle{ + \ \int d^4x \, \left[ - \left . \left( \delta D^\dagger_R \, D_L \right) \right |_{0^-} + \left . \left( \delta D^\dagger_R \, D_L \right) \right |_{- \pi R^+} \, \right] }
\nonumber \\
&\displaystyle{ + \ \int d^4x \, \left[ - \left . \left( \delta D^\dagger_R \, D_L \right) \right |_{\pi R^-} + \left . \left( \delta D^\dagger_R \, D_L \right) \right |_{0^+} \, \right] }
\nonumber \\
&\displaystyle{ \left. - \int d^4x \ X^* \, \delta D^\dagger_R Q_L \right|_{\pi R} } \, .
\label{orbi_HVP_3d}
\end{align}
With the parity of the fields in Eq.~\eqref{1_Z2} and the right-hand (left-hand) continuity at $y=0$ ($y=\pi R$), we rewrite the variations of the action at $y=-\pi R^+, 0^-$ as variations at $y=0, \pi R$:
\begin{align}
0 = \delta_{Q^\dagger_L} S_{\Psi} &= \displaystyle{ \int d^4x \left(\int_{-\pi R^+}^{0-} dy + \int_{0^+}^{\pi R^-} dy \right)
\delta Q^\dagger_L \left[ i\bar{\sigma}^\mu \partial_\mu Q_L - \partial_y Q_R  \right] } 
\nonumber \\
&\displaystyle{ + 2 \int d^4x \ \left[ \left . \left( \delta Q^\dagger_L \, Q_R \right) \right |_{\pi R^-} - \left . \left( \delta Q^\dagger_L \, Q_R \right) \right |_{0^+} \, \right] }
\nonumber \\
&\displaystyle{ \left. - \int d^4x \ X \, \delta Q^\dagger_L D_R \right|_{\pi R} } \ ,
\label{orbi_HVP_4a}
\end{align}
\begin{align}
0 = \delta_{D^\dagger_R} S_{\Psi} &= \displaystyle{ \int d^4x \left(\int_{-\pi R^+}^{0-} dy + \int_{0^+}^{\pi R^-} dy \right)
\delta D^\dagger_R \left[ i\sigma^\mu \partial_\mu D_R + \partial_y D_L  \right] } 
\nonumber \\
&\displaystyle{ + 2 \int d^4x \ \left[ - \left . \left( \delta D^\dagger_R \, D_L \right) \right |_{\pi R^-} + \left . \left( \delta D^\dagger_R \, D_L \right) \right |_{0^+} \, \right] }
\nonumber \\
&\displaystyle{ \left. - \int d^4x \ X^* \, \delta D^\dagger_R Q_L \right|_{\pi R} } \, .
\label{orbi_HVP_4d}
\end{align}
We get the Euler-Lagrange equations \eqref{1_ELE_1} extended on $S^1 \setminus \{ 0, \pi R \}$ and the boundary conditions:
\begin{align}
& \left. Q_R - \dfrac{X}{2}  D_R \right|_{\pi R} = 0 \, , \ \ \ \left. D_L + \dfrac{X^*}{2}  Q_L \right|_{\pi R} = 0 \, , \nonumber\\
& X' \left. D_L \right|_{\pi R} = 0 \, , \ \ \ X^{\prime *} \left. Q_R \right|_{\pi R} = 0 \, ,
\nonumber\\
& \left. Q_R  \right |_{0} = 0 \, , \ \ \ \left .  D_L \right |_{0}  = 0 \, .
\label{1_BC_0_L_orbi}
\end{align}
It is clear that if the odd fields are continuous or $X' \neq 0$ ($Q_R|_{\pi R} = D_L|_{\pi R} = 0$) then, for $X \neq 0$, $D_R|_{\pi R} = Q_L|_{\pi R} = 0$, we get Dirichlet boundary conditions for all fields (and thus vanishing 5D fields everywhere in the bulk). Therefore, in order to have a physical model, one has to take a discontinuity for the odd fields and $X'=0$.

We perform a mixed KK decomposition,
\begin{equation}
\left\{
\begin{array}{r c l}
Q_L \left( x^\mu, y \right) &=& \dfrac{1}{\sqrt{2 \pi R}} \displaystyle{ \sum_n q^n_L(y) \, \psi^n_L \left( x^\mu \right) },
\\ \vspace{-0.3cm} \\
Q_R \left( x^\mu, y \right) &=& \dfrac{1}{\sqrt{2 \pi R}} \displaystyle{ \sum_n q^n_R(y) \, \psi^n_R \left( x^\mu \right) },
\\ \vspace{-0.3cm} \\
D_L \left( x^\mu, y \right) &=& \dfrac{1}{\sqrt{2 \pi R}} \displaystyle{ \sum_n d^n_L(y) \, \psi^n_L \left( x^\mu \right) },
\\ \vspace{-0.3cm} \\
D_R \left( x^\mu, y \right) &=& \dfrac{1}{\sqrt{2 \pi R}} \displaystyle{ \sum_n d^n_R(y) \, \psi^n_R \left( x^\mu \right) },
\end{array} 
\right. 
\label{1_KK_4}
\end{equation}
where the KK modes satisfy the Dirac-Weyl equations \eqref{1_Dirac_2} and the KK wave functions are orthonormalized so that
\begin{equation}
\forall n \ , \forall m \ , \dfrac{1}{2 \pi R} \int_{-\pi R}^{+ \pi R} dy \left[ q^{n*}_{L/R}(y) \, q^m_{L/R}(y) + d^{n*}_{L/R}(y) \, d^m_{L/R}(y) \right] = \delta^{nm}.
\label{1_normalization_4}
\end{equation}
The discussion is the same as in Section~\ref{1_Yukawa_terms_as boundary_conditions}, below Eq.~\eqref{1_BC_13cont}, with the replacements Eq.~\eqref{1_KK_2} $\mapsto$ Eq.~\eqref{1_KK_4}, Eq.~\eqref{1_normalization_2} $\mapsto$ Eq.~\eqref{1_normalization_4} and $Y_5 \mapsto Y_5/2$: we obtain the KK wave function solutions \eqref{1_prof-BC_Y_1} on $[0, \pi R]$ and the mass spectrum \eqref{mass_spect_Y_interval} (with and $L = \pi R$ and $X \mapsto X/2$). The solutions on $(-\pi R, 0)$ are obtained from the ones on $[0, \pi R]$ with the $\mathbb{Z}_2$ symmetry. The mass spectra of the compactification on the orbifold $S^1/\mathbb{Z}_2$ and the interval $[0, \pi R]$ are equivalent if the 5D Yukawa couplings $Y_5$ between the two models differ by a factor 2. $Y_5$ is an input, not an observable, and the two spectra span the same range of values for the observables so one can say that the models are dual (\textit{i.e.} same observables) for values of $Y_5$ which differ by a factor 2.

\subsubsection{\textcolor{black}{\textit{Essential Boundary Conditions}}}
If we do not include the BBTs, and impose a jump to the odd fields as essential boundary conditions, we will find the same problems than without BBTs nor essential boundary conditions.

\newpage

\subsubsection{Summary}
\begin{figure}[!h]
\begin{center}
\includegraphics[width=15cm]{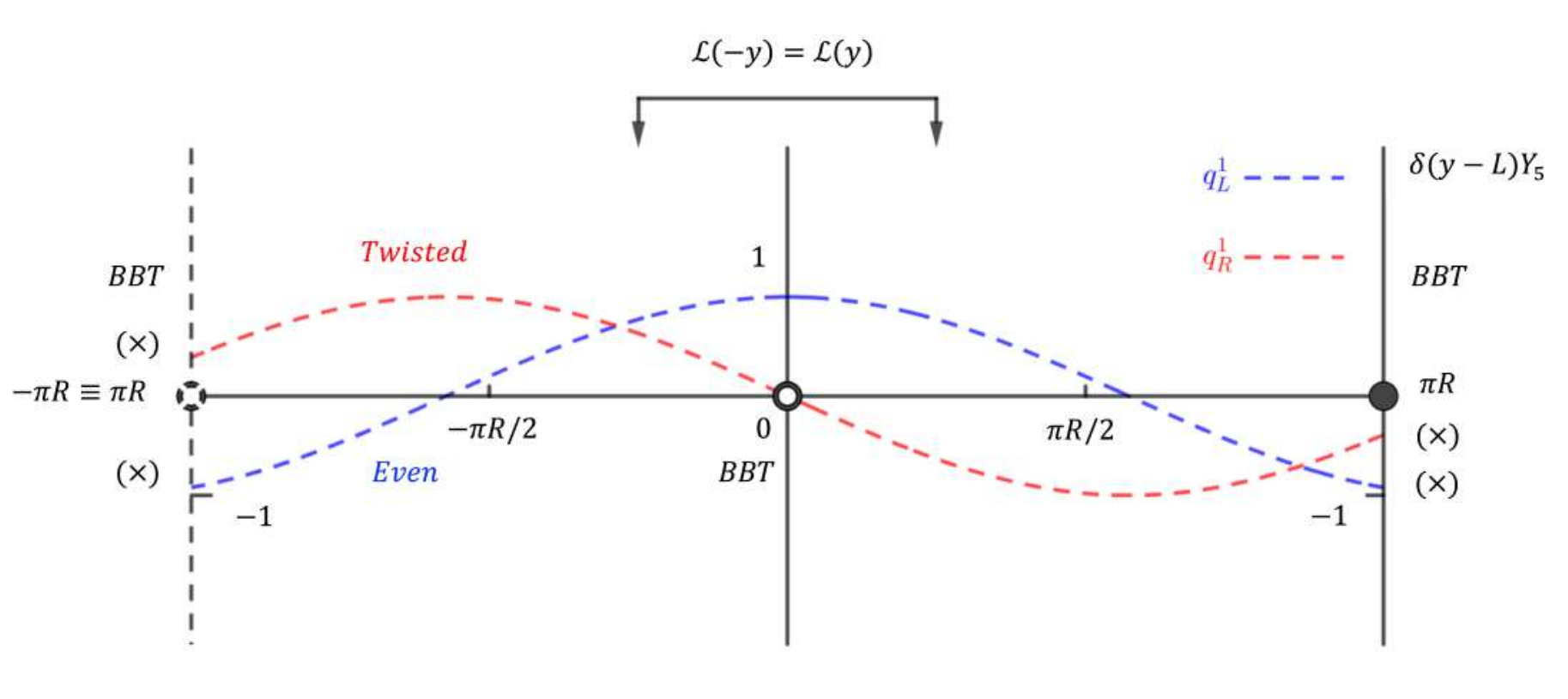}
\end{center}
\caption{KK wave functions of the 5D fields $Q$ and $D$ with Yukawa couplings along the orbifold $S^1/\mathbb{Z}_2$.
}\label{prof_S1_Z2}
\end{figure}

In Fig.~\ref{prof_S1_Z2}, we give a plot of the KK wave functions along the extra dimension.

The summary table is the same as for the interval: Tab.~\ref{1_summary_table}, which illustrates the duality between the interval and the orbifold $S^1/\mathbb{Z}_2$ up to a factor two in the mass spectrum and the normalization of the wave functions.

In the orbifold description, the essential boundary conditions are related to the continuity of the odd fields across the fixed points. In the interval description, they come from the requirement of vanishing transverse conserved currents at the boundaries.

The BBTs are justified by four different ways: (i) there is no physical solutions without them; (ii) They have a geometrical role in the sense that either they take into account the boundary of the interval (vanishing currents) or they specify the field junction in the case of a periodic extra dimension; (iii) they allow the 4D/5D matching; (iv) they appear from the weak derivative of the fields treated as distributions.

As in the case of the interval, in the 4D approach, the $X'$-terms do not affect the mass matrix. In the 5D approach, $X'=0$ is implied by the vanishing of the variation of the action at the fixed points. So there is a 4D/5D matching regarding the independance of the $X'$ parameter. This result hold in both the approaches with functions/distributions and interval/orbifold.

\subsection{Method with 5D Fields as Distributions}
\label{different_5D_methods_distrib}

When a brane is away from a boundary as the fixed point of $S^1 / \mathbb{Z}_2$ on $S^1$, 5D fields can be discontinuous across the brane. It is then practical to associate regular distributions to the 5D fields in order to have 5D Euler-Lagrange equations with the Yukawa couplings in factor of a Dirac distribution. This method is interesting only for the 5D treatment of brane localized Yukawa couplings which involves Dirac distributions (the free case and the 4D method are trivial with distributions since the KK wave functions are continuous).

Once the action is defined with a Lagrangian involving distributions, one can use Hamilton's principle within distribution theory. The variations of the fields are test functions whose supports are included in the compact space. A brane away from the boundaries does not give a boundary term in the action but generates a term proportional to a Dirac distribution in the 5D Euler-Lagrange equations. On the covering space $S^1$, there is no boundary terms since, after the integration by parts,
\begin{equation}
\oint_{-\pi R}^{+ \pi R} dy \ \partial_y \widetilde{A} = 0 \, ,
\end{equation}
where $\widetilde{A}$ is the regular distribution associated to the function $A(x^\mu, y)$ discontinuous at $y=0, \pm \pi R$. To solve the 5D Euler-Lagrange equations, it is convenient to solve them first in the regions outside the brane and then to integrate the equations over the extra dimension in a small neighborhood of the brane to get the junction conditions on the brane. We recover the same boundary conditions as with the method with functions.

\subsection{An alternative Way to Recover the Interval}
In this subsection, we present an alternative treatment of the model of Subsection~\ref{1_Orbifold} with an extra dimension compactified on the orbifold $S^1/\mathbb{Z}_2$, with the 5D fields as functions (but one can do it also with distributions). We start with the action defined on the covering space $S^1$ and reduce it to an action on the interval $[0, \pi R]$ with the $\mathbb{Z}_2$ symmetry at the level of the action. Of course, this is equivalent to the previous methods where we recover the interval solutions by applying the $\mathbb{Z}_2$ symmetry at the level of the KK wave functions.

The action is
\begin{align}
S_{Fermion} = S_{\Psi} + S_{Y} + S_B \, ,
\label{a_S_Psi_S1/Z2_Lagrangian}
\end{align}
with $S_{\Psi}$,  $S_{Y}$ and $S_B$ defined as the actions of the kinetic terms, the brane localized Yukawa interactions and the BBTs respectively ($L = \pi R$):
\begin{align}
&S_{\Psi} = \int d^4x \int_{-L^+}^{L} dy \ \mathcal{L}_\Psi \, , 
\label{a_1_eq:actionKin}
\end{align}
with $\mathcal{L}_\Psi$ defined in Eq.~\eqref{1_L_2} where the left-hand partial derivatives $\partial_y$ are defined at $y=L$,
\begin{align}
&S_{Y} = \int d^4x  \ \mathcal{L}_Y |_{L} \, , 
\label{a_1_eq:actionYuk}
\end{align}
with $\mathcal{L}_Y$ defined in Eq.~\eqref{a_1_eq:actionYuk}, and
\begin{align}
S_B  &=  \int d^4x \  \dfrac{1}{2} \left[ \left. \left( \bar{D}D - \bar{Q}Q \right) \right |_{-L^+}^{0^-} + \left. \left( \bar{D}D - \bar{Q}Q \right)\right |_{0^+}^{ L^-}  \right]  \, .
\label{a_1_eq:actionBound}
\end{align}

When calculating the mass spectrum of the KK fermions, we restrict our attention to the VEV $v$ of $H$ \eqref{1_H_exp}. Hence, we decompose the Yukawa interactions into $S_Y = S_X + S_{hQD}$ and the action involving only the fermions becomes:
\begin{align}
S_{Fermion}^{X} = S_{\Psi} + S_{X} + S_B \, .
\label{a_S_Psi_S1/Z2_Lagrangian_VEV}
\end{align}
\begin{align}
&S_{X} = \int d^4x \ \mathcal{L}_X |_{L} , \nonumber \\
&\text{with} \ \ 
\mathcal{L}_X = - X\ Q^\dagger_L D_R - X^\prime\ Q^\dagger_RD_L +  {\rm H.c.} \ ,
\label{a_1_L_5}
\end{align}
with the compact notations $X = \dfrac{v}{\sqrt{2}} \, Y_5$, $X' = \dfrac{v}{\sqrt{2}} \, Y_5'$, and
\begin{align}
&S_{hQD} = \int d^4x \ \mathcal{L}_{hQD} |_{L}  \ , \nonumber \\
&\text{with} \ \ 
\mathcal{L}_{hQD} = - \dfrac{Y_5}{\sqrt{2}} \ h Q^\dagger_L D_R - \dfrac{Y_5'}{\sqrt{2}} \ h Q^\dagger_RD_L +  {\rm H.c.} \ .
\label{a_1_L_Yuk}
\end{align}

We begin by cutting the integral over $y$ of the kinetic terms \eqref{a_S_Psi_S1/Z2_Lagrangian_VEV} at the brane-Higgs position:
\begin{align}
S_{\Psi} &= \int d^4x \int_{-L^+}^{L} dy \ \mathcal{L}_\Psi  \nonumber \\
&= \int d^4x \ \left \{  \int_{-L^+}^{0^-} dy \ \mathcal{L}_\Psi + \int_{0^+}^{L^-} dy \ \mathcal{L}_\Psi \right \} \, ,
\label{a_1_eq:actionKin_cut}
\end{align}
where the right-hand (left-hand) partial derivatives $\partial_y$ of the 5D fields are defined at $y=0$ ($y=L$). We have thus the right-hand (left-hand) continuity of the 5D fields at $y=0$ ($y=L$).

We choose the transformation laws for the fermionic fields $Q$ and $D$ under $\mathbb{Z}_2$ as:
\begin{equation} 
\left\{
\begin{array}{c c c}
Q \left( x^\mu, -y \right) = - \gamma^5 \, Q \left( x^\mu, y \right) \, \Longrightarrow \, Q_L \ \text{even}, \ Q_R \ \text{odd},
\\ \vspace{-0.2cm} \\
D \left( x^\mu, -y \right) = \gamma^5 \, D \left( x^\mu, y \right) \, \Longrightarrow \, D_L \ \text{odd}, \ D_R \ \text{even}.
\end{array}
\right.
\label{a_1_Z2}
\end{equation}

Considering the Lagrangian symmetry $\mathbb{Z}_2$, and performing a change of variable in the kinetic terms \eqref{a_1_eq:actionKin_cut}:
\begin{align}
S_{\Psi} &\ni \int d^4x \ \int_{-L^+}^{0^-} dy \ \mathcal{L}_\Psi  \nonumber \\
&= \int d^4x \ \int_{0^+}^{L^-} dy' \ \mathcal{L}_\Psi \nonumber \\
&= \int d^4x \ \int_{0^+}^{L^-} dy \ \mathcal{L}_\Psi \ , \nonumber \\
&\text{with $y'=-y$.}
\label{a_1_eq:actionKin_cut_Label}
\end{align}

Then we separate the brane localized mass terms into two identical parts by the left-hand continuity of the 5D fields at $y=L$ and we use the symmetry under $\mathbb{Z}_2$:
\begin{align}
S_{X} &= \int d^4x \ \mathcal{L}_X |_{L} = \int d^4x \ \mathcal{L}_X |_{L^-}  \nonumber \\
&= \int d^4x \ \dfrac{1}{2} \left[  \mathcal{L}_X |_{L^-} + \mathcal{L}_X |_{-L^+} \right]  \ , \nonumber \\
&\text{where $\mathcal{L}_X |_{L^-} = \mathcal{L}_X |_{-L^+}$. }
\label{a_1_L_5_2}
\end{align}

We can also use the transformation laws \eqref{a_1_Z2} in the BBTs \eqref{a_1_eq:actionBound}:
\begin{align}
S_B & \ni  \int d^4x \  \dfrac{1}{2} \left. \left( \bar{D}D - \bar{Q}Q \right) \right |_{-L^+}^{0^-} \nonumber \\
&= \int d^4x \  \dfrac{1}{2} \left\{ \left. \left( D^ \dagger \gamma^5 \gamma^0  \gamma^5 D - Q^ \dagger \gamma^5 \gamma^0  \gamma^5 Q \right) \right |_{L^-}^{0^+} \right\} \nonumber \\
&= \int d^4x \  \dfrac{1}{2} \left. \left( \bar{D}D - \bar{Q}Q \right) \right |_{0^+}^{L^-} \ .
\label{a_1_eq:actionBound_s}
\end{align}

We rewrite $S_{Fermion}^{X}$ as
\begin{align}
S_{Fermion}^{X} &= S_{\Psi} + S_{X} + S_B \nonumber \\
&= \int d^4x  \left \{  \int_{-L^+}^{0^-} dy \ \mathcal{L}_\Psi + \int_{0^+}^{L^-} dy \ \mathcal{L}_\Psi \right \} 
+ \int d^4x \ \dfrac{1}{2} \left[  \mathcal{L}_X |_{L^-} + \mathcal{L}_X |_{-L^+} \right] \nonumber \\
&+  \int d^4x \  \dfrac{1}{2} \left[ \left. \left( \bar{D}D - \bar{Q}Q \right) \right |_{-L^+}^{0^-} + \left. \left( \bar{D}D - \bar{Q}Q \right)\right |_{0^+}^{ L^-}  \right] \nonumber \\
&= S_{Fermion 1}^{X} + S_{Fermion 2}^{X} \ ,\nonumber\\
\text{with} \ \ 
S_{Fermion 1}^{X} &=\int d^4x \ \left[  \int_{-L^+}^{0^-} dy \ \mathcal{L}_\Psi + \left. \dfrac{1}{2} \mathcal{L}_X \right |_{-L^+} + \left. \dfrac{1}{2} \left( \bar{D}D - \bar{Q}Q \right) \right |_{-L^+}^{0^-} \right] \ , \nonumber \\
S_{Fermion 2}^{X} &=\int d^4x \ \left[ \int_{0^+}^{L^-} dy \ \mathcal{L}_\Psi + \left. \dfrac{1}{2} \mathcal{L}_X \right |_{L^-} + \left. \dfrac{1}{2} \left( \bar{D}D - \bar{Q}Q \right)\right |_{0^+}^{ L^-}  \right]  \ .
\label{a_SF_Z2_1&2}
\end{align}

Using \eqref{a_1_eq:actionKin_cut_Label}, \eqref{a_1_L_5_2}, \eqref{a_1_eq:actionBound_s} into \eqref{a_SF_Z2_1&2}, we find the identity:
\begin{align}
S_{Fermion 1}^{X} =S_{Fermion 2}^{X}  \ .
\label{a_SF_1&2}
\end{align}

Finally, we separate the action $S_{Fermion}^{X}$ \eqref{a_S_Psi_S1/Z2_Lagrangian_VEV} into 2 identical parts:
\begin{align}
S_{Fermion}^{X} =2 \ S_{Fermion 2}^{X}  \ .
\label{a_SF_tot}
\end{align}
We can thus treat $S_{Fermion 2}^{X}$ as the basic building block and go back to the interval scenario $[0, L]$ in Subsection~\ref{1_Correct_treatment_AFHS_term}. If we search for the profiles in $(-L, 0)$, we just need to use the transformation laws \eqref{a_1_Z2} for fermionic fields $Q$ and $D$.
\begin{table}[h]
\centering
\begin{tabular}{| c | c | c | c | c | }
\hline
\multirow{2}{*}{\diagbox{Domains}{Fields}} & \multicolumn{2}{| c |}{$Q_{L/R}$} & \multicolumn{2}{| c |}{$D_{L/R}$}  \\
\cline{2-5}
 & $q^n_L(y)$ & $q^n_R(y)$ & $d^n_L(y)$ & $d^n_R(y)$ \\
\hline
$(-L, 0)$ & $A^n_{q} \; \cos(M_n \; y)$ & $A^n_{q} \; \sin(M_n \; y)$ & $- A^n_{d} \; \sin(M_n \; y)$ & $A^n_{d} \; \cos(M_n \; y)$  \\
\hline
$[0, L]$ & $A^n_{q} \; \cos(M_n \; y)$ & $A^n_{q} \; \sin(M_n \; y)$ & $- A^n_{d} \; \sin(M_n \; y)$ & $A^n_{d} \; \cos(M_n \; y)$ \\
\hline
Mass spectrum & \multicolumn{4}{| c |}{$M_n = \dfrac{1}{L} \left[ n \pi + \arctan \left( \dfrac{|X|}{2} \right) \right] \, , n \in \mathbb{Z}$}  \\
\hline
\end{tabular}
\caption{Fermionic fields $Q$ and $D$ on the orbifold $S^1/ \mathbb{Z}_2$.}
\label{a_Q&D_Z2}
\end{table}

\section{Higgs Field Localized at a Fixed Point of the Orbifold $S^1/\left( \mathbb{Z}_2 \times \mathbb{Z}_2' \right)$}
\label{1_Orbifold_2}

\begin{figure}[h]
\begin{center}
\includegraphics[height=8cm]{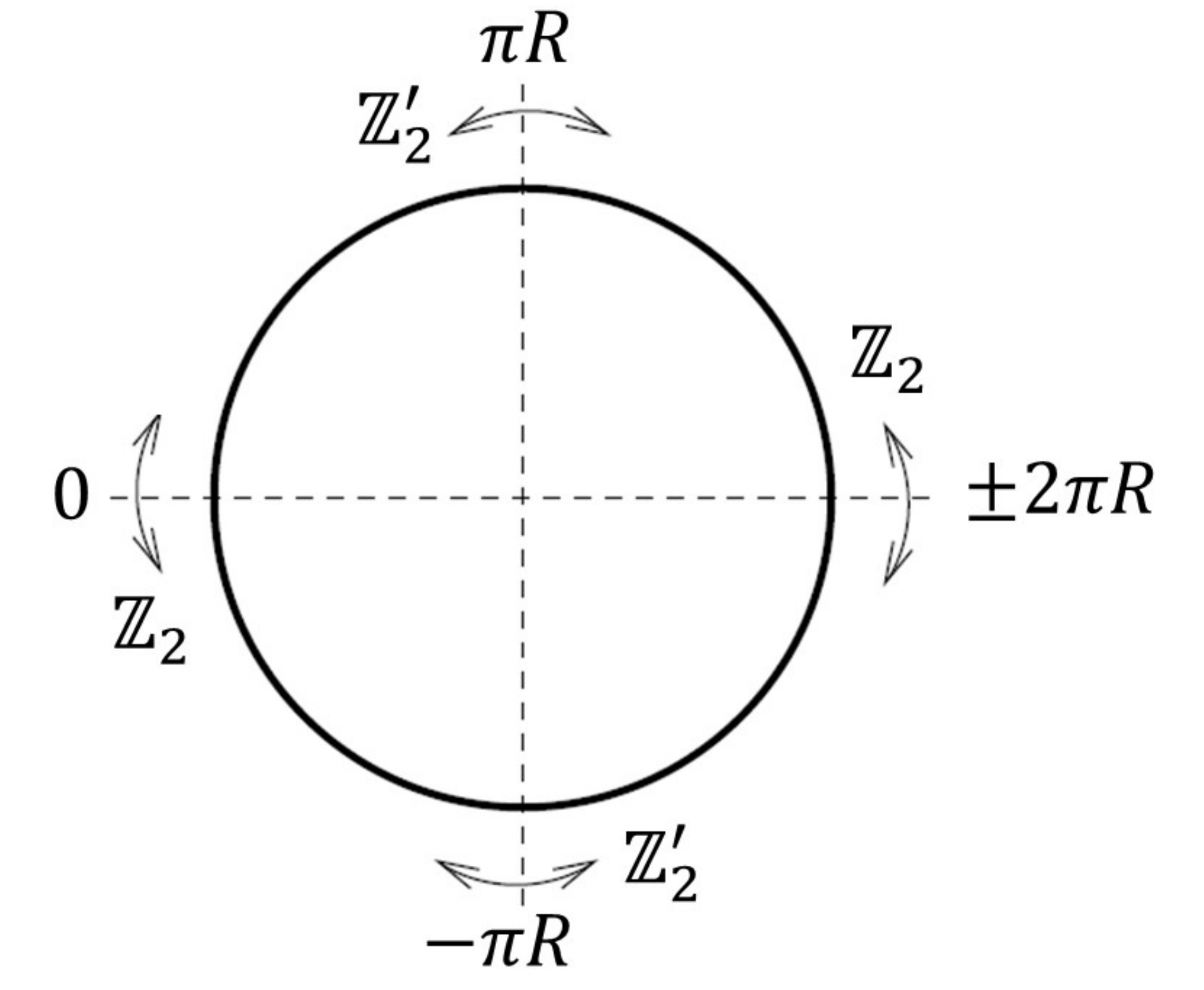}
\end{center}
\caption[Orbifold $S^1/\left( \mathbb{Z}_2 \times \mathbb{Z}_2' \right)$]{(Adapted from Ref.~\cite{Csaki:2005vy}.) The action of the reflections $\mathbb{Z}_2$ and $\mathbb{Z}_2'$ in the extended circle picture. We use the notations of Subsection~\ref{compact_orbifold}.
}
\label{double_S1}
\end{figure}

As we explained in Subsection~\ref{compact_orbifold} (we keep the same definitions and notations), a compactification on $S^1$ allows the possibility of a Scherk-Schwarz twist. If one mods out $S^1$ by a $\mathbb{Z}_2$ symmetry with respect to $y=0$, the Scherk-Schwarz twist is constrained. One can combine it with the $\mathbb{Z}_2$ symmetry in order to define a new reflection symmetry $\mathbb{Z}_2'$ with respect to $y=\pi R$:
\begin{equation}
y \sim 2 \pi R - y \, .
\end{equation}
The two fixed points of the orbifold $S^1/\left( \mathbb{Z}_2 \times \mathbb{Z}_2' \right)$ are still $y=0$ and $y=\pi R$, and the fundamental domain is the interval $[0, \pi R]$. It is convenient to define the field theory on an extended circle of circumference $4 \pi R$ with the coordinate $y \in (-2 \pi R, 2 \pi R]$ (see Fig.~\ref{double_S1}).

\newpage

The model has the same field content as in Subsection~\ref{1_A_toy_model_of_flat_extra_dimension}. Again, we treat the fields as distributions, which act on test functions whose supports are included into the compactified geometry. We will give only the method with functions including the brane localized interaction, and we will motivate the BBTs by defining the fields as distributions in the Lagrangians.

\subsubsection{Fields as Distributions}
The bulk action at the level of distributions is
\begin{align}
&S_{\Psi} = \int d^4x \oint_{-2\pi R}^{+2\pi R} dy \ \widetilde{\mathcal{L}_\Psi} \ , \nonumber \\
&\text{with} \ \ 
\widetilde{\mathcal{L}_\Psi} =  \sum_{F=Q,D} \dfrac{1}{2} \left( i \widetilde{F_R}^\dagger \sigma^\mu \overleftrightarrow{\partial_\mu} \widetilde{F_R} + i \widetilde{F_L}^\dagger \bar{\sigma}^\mu \overleftrightarrow{\partial_\mu} \widetilde{F_L} + \widetilde{F_R}^\dagger \overleftrightarrow{\partial_y} \widetilde{F_L} - \widetilde{F_L}^\dagger \overleftrightarrow{\partial_y} \widetilde{F_R} \right) \, .
\label{bulk_action_orbi_SS}
\end{align}
We choose the transformation laws for the fermionic fields $Q$ and $D$ under $\mathbb{Z}_2$ as in Eq.~\eqref{1_Z2} and under $\mathbb{Z}_2'$ as
\begin{equation} 
\left\{
\begin{array}{c c c}
Q \left( x^\mu, 2 \pi R - y \right) = - \gamma^5 \, Q \left( x^\mu, y \right) \, \Longrightarrow \, Q_L \ \text{even}, \ Q_R \ \text{odd},
\\ \vspace{-0.2cm} \\
D \left( x^\mu, 2 \pi R - y \right) = \gamma^5 \, D \left( x^\mu, y \right) \, \Longrightarrow \, D_L \ \text{odd}, \ D_R \ \text{even},
\end{array}
\right.
\label{1_Z2'}
\end{equation}
under which the action $S_{\Psi}$ \eqref{S_Psi_S1/Z2_0} is invariant (the parities even and odd are considered with respect to the point $y=\pi R$). The choice of the signs in front of $\gamma^5$ in Eqs.~\eqref{1_Z2} and \eqref{1_Z2'} is free. They determine if the 5D field has a zero mode and its chirality. Here, we require that $Q/D$ has a left/right-handed zero mode respectively in order to recover the SM at low energy.

\newpage

On each brane at $y=-\pi R, \pi R$, there is one copy of the same Higgs field. The action for the brane localized Higgs fields $H$ (real scalar fields in this toy model) is thus
\begin{align}
&S_H  = \int d^4x \oint_{-2 \pi R}^{2 \pi R} dy~\left[ \delta(y-\pi R) + \delta(y+\pi R) \right] \mathcal{L}_H \, , \nonumber \\
&\text{with} \ \ 
\mathcal{L}_H = \dfrac{1}{2} \, \partial_\mu H \partial^\mu H - V(H) \, ,   
\label{eq:actionH}
\end{align}
where the action of a Dirac distribution centered at $y=a$ on a test function $\varphi(y)$ is defined as
\begin{equation}
\oint_{- 2\pi R}^{+ 2\pi R} dy \ \delta(y-a) \, \varphi(y) = \varphi(a) \, .
\end{equation}
The brane localized Yukawa interactions are
\begin{align}
&S_{Y} = \int d^4x \int_0^L dy~ \left[ \delta(y-\pi R) + \delta(y+\pi R) \right] \mathcal{L}_Y \ , \nonumber \\
&\text{with} \ \ 
\mathcal{L}_Y = - Y_5\ Q^\dagger_LHD_R - Y^\prime_5\ Q^\dagger_RHD_L +  {\rm H.c.} \ .
\label{eq:actionYuk}
\end{align}
By keeping only the Higgs VEV (Eq.~\eqref{1_H_exp}), we have the brane localized mass terms:
\begin{align}
&S_{X} = \int d^4x \oint_{- 2\pi R}^{+ 2\pi R} dy~ \left[ \delta(y - \pi R) + \delta(y + \pi R) \right] \widetilde{\mathcal{L}_X} \ , \nonumber \\
&\text{with} \ \ 
\widetilde{\mathcal{L}_X} = - X \, \widetilde{Q_L}^\dagger \widetilde{D_R} - X^\prime \, \widetilde{Q_R}^\dagger \widetilde{D_L} +  {\rm H.c.} \ ,
\label{S_X_S1/Z2xz2'}
\end{align}
Again, we need $X \neq 0$ in order to give a mass to the zero modes of $Q_L$ and $D_R$ so these 5D fields have to be continuous at $y=-\pi R, \pi R$. However, the $X'$-terms are not required by the decoupling argument. If the odd fields $Q_R$ and $D_L$ are continuous, they have to vanish on the branes so the $X'$-terms vanish. If the odd fields are discontinuous at the fixed points, their product is continuous so we keep the $X'$-terms.

The weak partial derivatives give
\begin{align}
\partial_y \widetilde{Q_L} &= \left\{ \partial_y Q_L \right\} \, , \nonumber \\
\partial_y \widetilde{D_R} &= \left\{ \partial_y D_R \right\}
\, ,
\end{align}
for the even fields,
\begin{align}
\partial_y \widetilde{Q_R} = \left\{ \partial_y Q_R \right \} 
&+ \left( \left. Q_R \right|_{0^+} - \left. Q_R \right|_{0^-} \right) \delta(y) + \left( \left. Q_R \right|_{\pi R^+} - \left. Q_R \right|_{\pi R^-} \right) \delta(y- \pi R) \nonumber \\
&+ \left( \left. Q_R \right|_{- 2 \pi R^+} - \left. Q_R \right|_{2 \pi R^-} \right) \delta(y - 2 \pi R) + \left( \left. Q_R \right|_{-\pi R^+} - \left. Q_R \right|_{-\pi R^-} \right) \delta(y + \pi R)
\, ,
\end{align}
for the odd field $Q_R$, and
\begin{align}
\partial_y \widetilde{D_L} = \left\{ \partial_y D_L \right\} 
&+ \left( \left. D_L \right|_{0^+} - \left. D_L \right|_{0^-} \right) \delta(y) + \left( \left. D_L \right|_{\pi R^+} - \left. D_L \right|_{\pi R^-} \right) \delta(y- \pi R) \nonumber \\
&+ \left( \left. D_L \right|_{-2 \pi R^+} - \left. D_L \right|_{2 \pi R^-} \right) \delta(y - 2 \pi R) + \left( \left. D_L \right|_{-\pi R^+} - \left. D_L \right|_{-\pi R^-} \right) \delta(y + \pi R)
\, ,
\end{align}
for the odd field $D_L$. After that, we can rewrite the action $S_\Psi$ \eqref{bulk_action_orbi_SS} as
\begin{align}
S_{\Psi}
&= \int d^4x \left\{ \left[\left(\int_{-2\pi R}^{-\pi R} dy + \int_{-\pi R^+}^{0-} dy + \int_{0^+}^{\pi R^-} dy + \int_{\pi R^+}^{2 \pi R^-} dy \right) \mathcal{L}_\Psi \right] \right. \nonumber \\
&\left. -\left. \mathcal{L}_{B} \right|_{0^-}
+ \left. \mathcal{L}_{B} \right|_{0^+} 
- \left. \mathcal{L}_{B} \right|_{\pi R^-} 
+ \left. \mathcal{L}_{B} \right|_{\pi R^+}
- \left. \mathcal{L}_{B} \right|_{2 \pi R^-}
+ \left. \mathcal{L}_{B} \right|_{-2 \pi R^+} 
- \left. \mathcal{L}_{B} \right|_{-\pi R^-} 
+ \left. \mathcal{L}_{B} \right|_{-\pi R^+} \right\} \, ,
\label{action_orbi_SS_50}
\end{align}
As in the case of the orbifold $S^1/\mathbb{Z}_2$ in Section~\ref{1_Orbifold}, BBTs appears on both sides of the branes.

\subsubsection{Mass Spectrum}
In order to obtain the KK mass spectrum and the wave functions, we use the same method as in Subsection~\ref{orbi_S1Z2_Yuk}. We apply Hamilton's principle to the action $S_\Psi + S_X$ Eqs.~\eqref{action_orbi_SS_50} and \eqref{S_X_S1/Z2xz2'}. From the vanishing of the variation of the action in the bulk, we get the Euler-Lagrange equations \eqref{1_ELE_1} extended on $y \in S^1 \setminus \{ - \pi R, 0, \pi R, 2 \pi R \}$. The variation of the action on the branes gives the natural boundary conditions \eqref{1_BC_0_L_orbi}. We give the different steps for the variations of the actions at the two equivalent branes at $y=-\pi R, \pi R$
\begin{align}
\left[ \delta Q_L^\dagger Q_R \right]_{- \pi R^+}^{- \pi R^-} + \left[ \delta Q_L^\dagger Q_R \right]_{\pi R^+}^{\pi R^-} - X \left. \delta Q_L^\dagger D_R \right|_{- \pi R} - X \left. \delta Q_L^\dagger D_R \right|_{\pi R} &= 0 \, , \nonumber \\
- 2 \left. \delta Q_L^\dagger Q_R \right|_{- \pi R^+} + 2 \left. \delta Q_L^\dagger Q_R \right|_{\pi R^-} - X \left. \delta Q_L^\dagger D_R \right|_{- \pi R} - X \left. \delta Q_L^\dagger D_R \right|_{\pi R} &= 0 \, , \nonumber \\
4 \left. \delta Q_L^\dagger Q_R \right|_{\pi R^-} - 2X \left. \delta Q_L^\dagger D_R \right|_{\pi R} &= 0 \, , \nonumber \\
\left. \delta Q_L^\dagger \left( Q_R - \dfrac{X}{2} \, D_R \right) \right|_{\pi R} &= 0 \, ,
\end{align}
where we used the $\mathbb{Z}_2'$ symmetry (Eq.~\eqref{1_Z2'}) to go from the first line to the second one and the $\mathbb{Z}_2$ symmetry (Eq.~\eqref{1_Z2}) to go from the second line to the second one. By using the same method for each field variation, we get the same boundary conditions as for the orbifold $S^1/\mathbb{Z}_2$: Eq.~\eqref{1_BC_0_L_orbi}. As for the orbifold $S^1/\mathbb{Z}_2$, the fields $Q_R$ and $D_L$ are discontinuous across the branes where the Higgs field is localized and $X'=0$. The free case is obtained by taking $X=X'=0$ and we get that every field is continuous at the fixed points.

We perform a mixed KK decomposition,
\begin{equation}
\left\{
\begin{array}{r c l}
Q_L \left( x^\mu, y \right) &=& \dfrac{1}{\sqrt{4 \pi R}} \displaystyle{ \sum_n q^n_L(y) \, \psi^n_L \left( x^\mu \right) },
\\ \vspace{-0.3cm} \\
Q_R \left( x^\mu, y \right) &=& \dfrac{1}{\sqrt{4 \pi R}} \displaystyle{ \sum_n q^n_R(y) \, \psi^n_R \left( x^\mu \right) },
\\ \vspace{-0.3cm} \\
D_L \left( x^\mu, y \right) &=& \dfrac{1}{\sqrt{4 \pi R}} \displaystyle{ \sum_n d^n_L(y) \, \psi^n_L \left( x^\mu \right) },
\\ \vspace{-0.3cm} \\
D_R \left( x^\mu, y \right) &=& \dfrac{1}{\sqrt{4 \pi R}} \displaystyle{ \sum_n d^n_R(y) \, \psi^n_R \left( x^\mu \right) },
\end{array} 
\right. 
\label{1_KK_50}
\end{equation}
where the KK modes satisfy the Dirac-Weyl equations \eqref{1_Dirac_2} and the KK wave functions are orthonormalized so that
\begin{equation}
\forall n \ , \forall m \ , \dfrac{1}{4 \pi R} \int_{-2\pi R}^{+ 2\pi R} dy \left[ q^{n*}_{L/R}(y) \, q^m_{L/R}(y) + d^{n*}_{L/R}(y) \, d^m_{L/R}(y) \right] = \delta^{nm} \, ,
\label{1_normalization_50}
\end{equation}
where we normalize on the whole extended circle. The discussion is the same as in Section~\ref{1_Yukawa_terms_as boundary_conditions}, below Eq.~\eqref{1_BC_13cont}, with the replacements Eq.~\eqref{1_KK_2} $\mapsto$ Eq.~\eqref{1_KK_50}, Eq.~\eqref{1_normalization_2} $\mapsto$ Eq.~\eqref{1_normalization_50} and $Y_5 \mapsto Y_5/2$: we obtain the KK wave function solutions \eqref{1_prof-BC_Y_1} on $[0, \pi R]$ and the mass spectrum \eqref{mass_spect_Y_interval} (with $L = \pi R$ and $X \mapsto X/2$). We obtain the solutions on the whole extended circle by using the symmetries $\mathbb{Z}_2$ and $\mathbb{Z}_2'$. The mass spectrum is the same as in the case of the compactification on the orbifold $S^1/\mathbb{Z}_2$. There is again this difference of a factor 2 with the mass spectrum of the compactification on the interval $[0, \pi R]$ (see below Eq.~\eqref{1_normalization_4} for a discussion). We conclude that all these models are dual.

\subsubsection{Summary and Conclusion}
%
In this section, we have reformulated our model of Chapter~\ref{1_4D perturbative approach} with the orbifold $S^1/(\mathbb{Z}_2 \times \mathbb{Z}_2')$. The BBTs are generated in distribution theory by the odd fields which are in general discontinuous at the fixed points, in contrast to the interval compactification in Chapter~\ref{1_4D perturbative approach} where we need to add the BBTs by hand. We get the mass spectrum with an orbifold geometry from the one with an interval by the replacement $X \mapsto X/2$.

\section{Higgs Field Localized on a Brane Away from a Boundary of an Interval}
\label{H_out_boundary}

\subsection{5D Method with Fields as Distributions}
In this section, we consider the same model as in Section~\ref{1_A_toy_model_of_flat_extra_dimension} but the Higgs boson is now localized on a new 3-brane at $y=\ell$, away from but parallel to a boundary. The reader can follow Section~\ref{1_A_toy_model_of_flat_extra_dimension} for the field content of the model; we keep the same notations. However, one has to substitute $\delta (y-L) \mapsto \delta (y-\ell)$ in Subsections~\ref{1_Brane Localized Scalar Field} and \ref{1_Yukawa}.

Following Appendix~\ref{app_cont_field}, it is important to allow for the possibility that some fields are discontinuous across the brane at $y=\ell$. We choose to treat the fields as distributions (see Appendix~\ref{app_cont_field}) which act on test functions whose supports are part of the compactified geometry. In order to define the partial derivatives $\partial_y Q$ and $\partial_y D$ in the kinetic term, it is useful to define the Lagrangians using distribution theory. Indeed, one can associate regular distributions $\widetilde{Q}$ and $\widetilde{D}$ to the fields $Q$ and $D$ respectively, and thus put a tilde on every field $Q$, $D$, $Q_{L/R}$, $D_{L/R}$ and every Lagrangian of Section~\ref{1_A_toy_model_of_flat_extra_dimension}, which have to be considered as distributions. The action of the system is then the action of Lagrangian on the test function $\mathbf{1}(x^\mu, y): (x^\mu, y) \mapsto 1$. The product of two distributions in the Lagrangian is defined as in Appendix A of Ref.~\cite{Messiah1}. The kinetic term of $S_\Psi$ \eqref{1_L_2} involves the terms $\widetilde{F^\dagger_R} \overleftrightarrow{\partial_y} \widetilde{F_L}$ and $\widetilde{F^\dagger_L} \overleftrightarrow{\partial_y} \widetilde{F_R}$ with $F=Q$ or $D$. They are well defined if at least $F_L$ or $F_R$ is continuous at $y=\ell$ since the weak derivative of a discontinuous function involves a Dirac distribution. We want that the Higgs VEV gives a mass to the zero modes of $Q_L$ and $D_R$ in order to recover the SM in the decoupling limit, so the terms involving $Y_5$ in $\mathcal{L}_Y$ \eqref{1_eq:actionYuk} have to be well defined. As they are proportional to $\delta(y-\ell)$ and involve the fields $Q_L$ and $D_R$, these two fields have to be continuous at $y=\ell$. As for the terms involving $Y_5'$, which are also proportional to $\delta(y-\ell)$, they involve the fields $Q_R$ and $D_L$, and their presence is not necessary for recovering the SM in the decoupling limit. $Q_R$ and/or $D_L$ can be discontinuous at $y=\ell$. The product of two discontinuous function can be continuous but if it is not, the terms proportional to $Y_5'$ are not defined, thus one has to take $Y_5'=0$. With these constraints, we have the weak partial derivatives:
\begin{align}
\partial_y \widetilde{Q_L} &= \left\{ \partial_y Q_L \right\} \, , \nonumber \\
\partial_y \widetilde{Q_R} &= \left\{ \partial_y Q_R \right\} + \left( \left. Q_R \right|_{\ell^+} - \left. Q_R \right|_{\ell^-} \right) \delta(y - \ell) \, , \nonumber \\
\partial_y \widetilde{D_L} &= \left\{ \partial_y D_L \right\} + \left( \left. D_L \right|_{\ell^+} - \left. D_L \right|_{\ell^-} \right) \delta(y - \ell) \, , \nonumber \\
\partial_y \widetilde{D_R} &= \left\{ \partial_y D_R \right\} \, ,
\label{brane_int_weak_deriv}
\end{align}
where the $\left\{ \partial_y F_{L/R} \right\}$'s are the regular distributions associated to the functions $\partial_y F_{L/R}$. One can use Eq.~\eqref{brane_int_weak_deriv} to go from a distributional to a functional treatment of the Lagrangians and fields where BBTs appear from the weak partial derivative of the discontinuous fields. We can thus continue to use the Lagrangians of Section~\ref{1_A_toy_model_of_flat_extra_dimension} with fields as functions if one adds to $S_{5D}$~\eqref{1_eq:actionTot} the action
\begin{equation}
S_B' = \int d^4x \ \left. \mathcal{L}_B \right|_{\ell^+} - \left. \mathcal{L}_B \right|_{\ell^-} \, .
\end{equation}

We apply Hamilton's principle to the action $S_{\Psi} + S_B + S_B' + S_X$ with arbitrary field variations in the bulk and at the boundaries. For the fields continuous (discontinuous) across the brane at $y=\ell$, the variations are continuous (discontinuous) and arbitrary at $y=\ell$. We get
\begin{align}
0 =  \delta_{Q^\dagger_L} (S_{\Psi} + S_B + S_B' + S_X) &= \displaystyle{ \int d^4x \left( \int_0^{\ell^-} dy + \int_{\ell^+}^L dy \right) \delta Q^\dagger_L \left[ i\bar{\sigma}^\mu \partial_\mu Q_L - \partial_y Q_R  \right] }
\nonumber \\
&\displaystyle{ + \int d^4x \left[ \delta Q^\dagger_L Q_R \right]_{0}^L }  \nonumber \\
&\displaystyle{ - \int d^4x \left. \delta Q^\dagger_L \right|_\ell \left( \left[ Q_R \right]_{\ell^-}^{\ell^+} + X \left. D_R \right|_\ell \right) }
\, ,
\label{HVP_20a}
\end{align}
\begin{align}
0 = \delta_{Q^\dagger_R} (S_{\Psi} + S_B + S_B' + S_X) &= \displaystyle{ \int d^4x \left( \int_0^{\ell^-} dy + \int_{\ell^+}^L dy \right) \delta Q^\dagger_R \left[ i\sigma^\mu \partial_\mu Q_R + \partial_y F_L  \right] }
\nonumber \\
&\displaystyle{ - \int d^4x \ X' \left. \left( \delta Q^\dagger_R \, D_L \right) \right|_{\ell} } \ ,
\label{HVP_20b}
\end{align}
\begin{align}
0 =  \delta_{D^\dagger_L} (S_{\Psi} + S_B + S_B' + S_X) &= \displaystyle{ \int d^4x \left( \int_0^{\ell^-} dy + \int_{\ell^+}^L dy \right)  
\delta D^\dagger_L \left[ i\bar{\sigma}^\mu \partial_\mu D_L - \partial_y D_R  \right] }
\nonumber \\
&\displaystyle{ - \int d^4x \ X^{\prime *} \left. \left( \delta D^\dagger_L \, Q_R \right) \right|_{\ell} } \ ,
\label{HVP_20c}
\end{align}
\begin{align}
0 = \delta_{D^\dagger_R} (S_{\Psi} + S_B + S_B' + S_X) &= \displaystyle{ \int d^4x \left( \int_0^{\ell^-} dy + \int_{\ell^+}^L dy \right)  
\delta D^\dagger_R \left[ i\sigma^\mu \partial_\mu D_R + \partial_y D_L  \right] } 
\nonumber \\
&\displaystyle{ - \int d^4x \left[ \delta D^\dagger_R D_L \right]_{0}^L }
\nonumber \\
&\displaystyle{ + \int d^4x \left. \delta D^\dagger_R \right|_\ell \left( \left[ D_L \right]^{\ell^+}_{\ell^-} - X^{*} \left. Q_L \right|_\ell \right) }
\, ,
\label{HVP_20d}
\end{align}
with the notation $\left[ g(y) \right]_a^b = g(b)-g(a)$. We obtain the Euler-Lagrange equations \eqref{1_ELE_1} in the bulk, and the following conditions at the branes at $y=0, L, \ell$ respectively:
\begin{equation}
\left. Q_R \right|_{0, L} = \left. D_L \right|_{0, L} = 0 \, ,
\label{BC_brane_int_1}
\end{equation}
and
\begin{equation}
\left\{
\begin{array}{rcl}
\left[ Q_R \right]_{\ell^-}^{\ell^+} &=& - X \, \left. D_R \right|_\ell \, , \\ \\
\left[ D_L \right]_{\ell^-}^{\ell^+} &=& X^* \, \left. Q_L \right|_\ell \, , \\ \\
X' \, \left. D_L \right|_\ell &=& X^{\prime *} \, \left. Q_R \right|_\ell = 0 \, .
\end{array}
\right.
\label{BC_brane_int_2}
\end{equation}
If $Q_R$ and $D_L$ are continuous at $y=\ell$, with non-vanishing $Y_5$ and $Y_5'$ in general (and thus $X$ and $X'$ are non-vanishing too), we have
\begin{equation}
\left. Q_L \right|_\ell = \left. Q_R \right|_\ell = \left. D_L \right|_\ell = \left. D_R \right|_\ell = 0 \, .
\end{equation}
A Dirichlet condition at $y=\ell$ for both chirality implies in general vanishing KK wave functions and 5D fields everywhere in the extra dimension. In order to have a physically interesting scenario, we need to relax the continuity condition for $Q_R$ and $D_L$ across the brane at $y=\ell$: $Y_5'=0$ so $X'=0$ and the third line of Eq.~\eqref{BC_brane_int_2} is satisfied whatever are the values of $D_L|_\ell$ and $Q_R|_\ell$.

As the fields $Q$ and $D$ mix due to the brane localized mass term involving $X$, we expand the 5D fields into a mixed KK decomposition:
\begin{equation}
\left\{
\begin{array}{r c l}
Q_L \left( x^\mu, y \right) &=& \displaystyle{ \sum_n q^n_L(y) \, \psi^n_L \left( x^\mu \right) },
\\ \vspace{-0.3cm} \\
Q_R \left( x^\mu, y \right) &=& \displaystyle{ \sum_n q^n_R(y) \, \psi^n_R \left( x^\mu \right) },
\\ \vspace{-0.3cm} \\
D_L \left( x^\mu, y \right) &=& \displaystyle{ \sum_n d^n_L(y) \, \psi^n_L \left( x^\mu \right) },
\\ \vspace{-0.3cm} \\
D_R \left( x^\mu, y \right) &=& \displaystyle{ \sum_n d^n_R(y) \, \psi^n_R \left( x^\mu \right) }.
\end{array} 
\right. 
\label{mixed_KK_20}
\end{equation}
The 4D fields $\psi^n_{L/R}$ ($\forall n$) satisfy the Dirac-Weyl equations \eqref{1_Dirac_2}. The mass spectrum $M_n$ includes the contributions whose origin is the Yukawa couplings~\eqref{1_L_5}. The wave functions $f_{L/R}^n$ ($f=q$ or $d$) are orthonormalized with the conditions:
\begin{equation}
\forall n \ , \forall m \ , \int_0^{L} dy \, \left[ q^{n*}_{L/R}(y) \, q^m_{L/R}(y) + d^{n*}_{L/R}(y) \, d^m_{L/R}(y) \right] = \delta^{nm} \, .
\label{normalization_brane_int}
\end{equation}
The mixed KK decompositions \eqref{mixed_KK_20} are inserted into the Euler-Lagrange equations \eqref{1_ELE_1}, and into the conditions at the branes \eqref{BC_brane_int_1} and \eqref{BC_brane_int_2}. We get the equations for the KK wave functions \eqref{1_ELE_Y}, the boundary conditions:
\begin{equation}
q_R^n(0, L) = d_L^n(0, L) = 0 \, ,
\label{BC_brane_int_wave_1}
\end{equation}
and the junction conditions:
\begin{equation}
\left\{
\begin{array}{rcl}
\left[ q_R^n(y) \right]_{\ell^-}^{\ell^+} &=& - X \, d_R^n(\ell) \, , \\ \\
\left[ d_L^n(y) \right]_{\ell^-}^{\ell^+} &=& X^* \, q_L^n(\ell) \, .
\end{array}
\right.
\label{BC_brane_int_wave_2}
\end{equation}

We solve the equations for the KK wave functions \eqref{1_ELE_Y} in the two different regions: $y<\ell$ and $y>\ell$ with the boundary conditions \eqref{BC_brane_int_wave_1}. We obtain:
\begin{equation}
\forall y < \ell \, , \ \left\{
\begin{array}{ll}
q_L^n(y) = A_q \, \cos( M_n \, y ) \, ,
& d_L^n(y) = -A_d \, \sin( M_n \, y ) \, , \\ \\
q_R^n(y) = A_q \, \sin( M_n \, y ) \, ,
& d_L^n(y) = A_d \, \cos( M_n \, y ) \, ,
\end{array}
\right.
\label{profiles_brane_int_1}
\end{equation}
and:
\begin{equation}
\forall y > \ell \, , \ \left\{
\begin{array}{ll}
q_L^n(y) = B_q \, \cos \left[ M_n (y-L) \right] \, ,
& d_L^n(y) = -B_d \, \sin \left[ M_n (y-L) \right] \, , \\ \\
q_R^n(y) = B_q \, \sin \left[ M_n (y-L) \right] \, ,
& d_L^n(y) = B_d \, \cos \left[ M_n (y-L) \right] \, ,
\end{array}
\right.
\label{profiles_brane_int_2}
\end{equation}
with $A_f$, $B_f \in \mathbb{C}^*$. The fields $Q_L$ and $D_R$ are continuous at $y=\ell$ so the wave functions $q_L^n(y)$ and $d_R^n(y)$ are also continuous across this brane, and we get
\begin{equation}
B_f = A_f \, \dfrac{\cos (M_n \, y)}{\cos \left[ M_n (y-L) \right]} \, .
\end{equation}
The junction conditions \eqref{BC_brane_int_wave_2} with the wave functions \eqref{profiles_brane_int_1} and \eqref{profiles_brane_int_2} give
\begin{equation}
\left[ \tan (M_n \, y) \right]_{\ell - L}^\ell = X \, \dfrac{A_d}{A_q} = X^* \, \dfrac{A_q}{A_d} \, ,
\end{equation}
thus we have the equation for the mass spectrum
\begin{equation}
\left( \left[ \tan (M_n \, y) \right]_{\ell - L}^\ell \right)^2 = |X|^2 \, ,
\label{mass_spect_brane_int_1}
\end{equation}
equivalent to
\begin{equation}
\left[ \tan (M_n \, y) \right]_{\ell - L}^\ell = \pm |X| \, ,
\label{mass_spect_brane_int_2}
\end{equation}
and we have $|A_q|=|A_d|$. The KK wave functions are orthonormalized by Eq.~\eqref{normalization_brane_int}, such that
\begin{equation}
|A_q| = |A_d| = \left[ \ell + \left( \dfrac{\cos^2(M_n \, \ell)}{\cos^2 \left[ M_n (\ell - L) \right]} \right) (L-\ell) \right]^{-1/2} \, .
\end{equation}
In a similar way as below Eq.~\eqref{1_tower_2}, one can show that the phases of $A_f$ and $Y_5$, and the sign of $M_n$ are not physical, so one can take $A_f=|A_f|$, $Y_5=|Y_5|$ and $M_n = |M_n|$. Therefore, the two mass spectra of Eq.~\eqref{mass_spect_brane_int_2} are equivalent and correspond to the same KK modes since one can go frome the first to the second by the transformation $M_n \mapsto -M_n$.

The Yukawa couplings between the modes $n$ and $m$ follow from inserting the KK decompositions \eqref{mixed_KK_20} in Eq.~\eqref{1_L_Yuk}:
\begin{equation}
y_{nm} = |Y_5| \, q_L^n(\ell) \, d_R^m(\ell) \, .
\end{equation}

\subsection{Conclusion}
In this section, we have illustrated the differences between a Higgs field localized on a brane away from a boundary and on a brane at a boundary. If the brane is at a boundary, the BBTs are explicitely introduced in the action from the beginning and all fermion fields are continuous along the extra dimension. If the brane is away from a boundary, the brane localized terms imply a jump at the brane position for one of the two chiralities of the 5D fermion field. The distribution theory is very convenient in this case and the BBTs are implicitely generated on both sides of the brane by the weak partial derivative along the extra dimension of the discontinuous fields.

\chapter{Large Star/Rose Extra Dimension with Small Leaves/Petals}
\label{large_star_rose_ED_small_leaves_petals}

This chapter is an adaptation of Ref.~\cite{Nortier:2020lbs} written under the supervision of Ulrich Ellwanger.

\section{Introduction}
The large gauge hierarchy between the 4D Planck scale,
\begin{equation}
\Lambda_P^{(4)} = \sqrt{\frac{1}{8 \pi G_N^{(4)}}} \simeq 2.4 \times 10^{18} \  \mathrm{GeV},
\end{equation}
where $G_N^{(4)}$ is the 4D gravitational Newton constant, and the measured Higgs boson mass $m_h \simeq 125 \ \mathrm{GeV}$, is one of the internal puzzles of the Standard Model (SM) of particle physics \cite{Donoghue2014} when coupled to gravity. If the Higgs boson is described by an elementary scalar field which is not protected by a symmetry, the radiative corrections to $m^2_h$ are quadratically sensitive to mass scales of new degrees of freedom. Such degrees of freedom are expected at least at scales where gravitational self interactions become strong, which is the 4D Planck scale $\Lambda_P^{(4)}$ in the usual setup. Then the measured $m_h$ implies an incredible fine tuning between the Higgs boson bare mass and the radiative corrections: this is the naturalness problem of the Higgs boson mass due to the large hierarchy between the 4D Planck scale $\Lambda_P^{(4)}$ and the ElectroWeak (EW) scale (usually called the gauge hierarchy problem) \cite{Wilson:1970ag,Susskind:1978ms,tHooft:1979rat}. A possibility to solve this issue is to embed the SM into a theory where the true Planck scale, i.e. the scale where gravitational self interactions become strong and new degrees of freedom are expected, is in the $1-10$ TeV range. However, this does not solve the question how a UltraViolet (UV) completion of
quantum gravity is achieved.

\begin{figure}[h]
\begin{center}
\includegraphics[height=8cm]{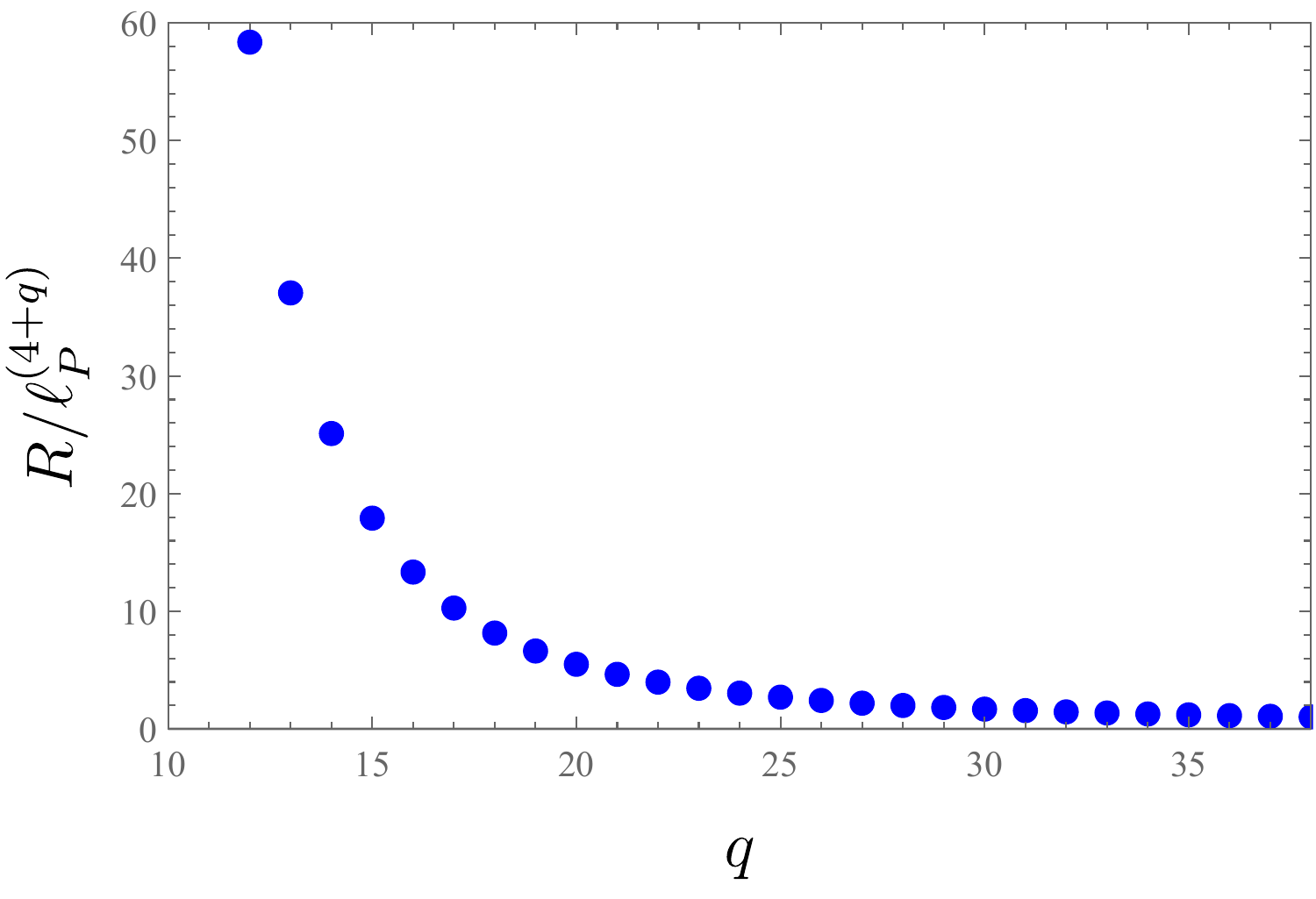}
\end{center}
\caption{Graph of $R/\ell_P^{(4+q)}$ as a function of $q$ for $\Lambda_P^{(4+q)} = 1 \ \mathrm{TeV}$.} 
\label{graph}
\end{figure}

In 1998, N.~Arkani-Hamed, S.~Dimopoulos and G.R.~Dvali (ADD) proposed in Ref.~\cite{ArkaniHamed:1998rs} to compactify $q \in \mathbb{N}^*$ flat spacelike extra dimensions on a $q$-dimensional compact space $\mathcal{C}_q$ of volume $\mathcal{V}_q$ with a factorizable spacetime geometry $\mathcal{M}_4 \times \mathcal{C}_q$, where $\mathcal{M}_4$ is the 4D Minkowski spacetime. The 4D Planck scale $\Lambda_P^{(4)}$ is just an effective scale given by the relation
\begin{equation}
\left[\Lambda_P^{(4)}\right]^2 = \left[ \Lambda_P^{(4+q)} \right]^{q+2} \, \mathcal{V}_q \, ,
\label{ADD_formula}
\end{equation}
involving the $(4+q)$D Planck scale
\begin{equation}
\Lambda_P^{(4+q)} = \left[ \dfrac{1}{8 \pi G^{(4+q)}_N} \right]^{1/(q+2)},
\end{equation}
where $G_N^{(4+q)}$ is the $(4+q)$D gravitational Newton constant. $\Lambda_P^{(4+q)}$ is the real scale at which gravity becomes strongly coupled, so it is the true cut-off of the QFT\footnote{To be more precise, if the UV completion is perturbative, the UV gravitational degrees of freedom appear at a scale $\Lambda_{UV}<\Lambda_P^{(4+q)}$. The cut-off of the EFT is $\Lambda_{UV}$ and not $\Lambda_P^{(4+q)}$. For example, in perturbative string theory, the string scale is lower than the higher-dimensional Planck scale. To simplify the discussion, we ignore this possibility here.}, and this model solves the gauge hierarchy problem if $\Lambda_P^{(4+q)}  \sim \mathcal{O}(1)$ TeV with a large compactified volume $\mathcal{V}_q$. In ADD models, the SM fields must be localized on a 3-brane, in contrast to gravity which is a property of $(4+q)D$ spacetime in general relativity. At large distances between two test masses on the 3-brane, gravity appears as a feebly coupled theory, because gravitational fluxes spread into the large volume $\mathcal{V}_q$ of the bulk. Quickly \cite{ArkaniHamed:1998nn}, it was realized that the Kaluza-Klein (KK) modes of fields propagating into this large volume $\mathcal{V}_q$ have couplings to the SM fields suppressed by $\sqrt{\mathcal{V}_q}$. One can thus build natural models of feebly interacting particles, i.e. particles which have a tiny coupling constant with the SM, like right-handed neutrinos \cite{Dienes:1998sb,ArkaniHamed:1998vp,Dvali:1999cn}, axions \cite{Chang:1999si,Dienes:1999gw}, dark photons \cite{ArkaniHamed:1998nn}, etc. In Ref.~\cite{ArkaniHamed:1998nn}, ADD proposed a simple toroidal compactification $\mathcal{C}_q = \left(\mathcal{R}_1\right)^q$, where $\mathcal{R}_1$ is the circle of radius $R$, and $\mathcal{V}_q = (2 \pi R)^q$. The bulk fields, like the graviton, generate a tower of KK modes with a uniform mass gap of $1/R$. For a benchmark value $\Lambda_P^{(4+q)} = 1$ TeV, the $(4+q)$D Planck length is $\ell_P^{(4+q)} = 1/ \Lambda_P^{(4+q)} \simeq 2 \times 10^{-19} \ \mathrm{m}$, and one gets Tab.~\ref{table} and Fig.~\ref{graph} from Eq.~\eqref{ADD_formula}.

\begin{table}[h]
\begin{center}
\begin{tabular}{c||c|c|c}
$q$ & $R$ (m) & $R/\ell_P^{(4+q)}$ & $M_{KK}$ (eV)  \\
\hline \hline
1 & $2 \times 10^{11}$ & $9 \times 10^{29}$ & $1 \times 10^{-18}$ \\ 
2 & $8 \times 10^{-5}$ & $4 \times 10^{14}$ & $3 \times 10^{-3}$ \\
4 & $2 \times 10^{-12}$ & $8 \times 10^{6}$ & $1 \times 10^5$ \\
6 & $4 \times 10^{-15}$ & $2 \times 10^{4}$ & $5 \times 10^7$ \\
22 & $8 \times 10^{-19}$ & $4$ & $3 \times 10^{11}$
\end{tabular}
\caption{$R$, $R/\ell_P^{(4+q)}$ and $M_{KK}$ as a function of $q$ for $\Lambda_P^{(4+q)} = 1 \ \mathrm{TeV}$.}
\label{table}
\end{center}
\end{table}

Motivated by UV completions in superstring/M-theory \cite{Antoniadis:1998ig,Ibanez:2012zz} requiring 10/11 spacetime dimensions, most of the efforts concentrated on $q \leq 7$. The compactification radius $R$ must be stabilized at a large value compared to $\ell_P^{(4+q)}$, which reintroduces a geometrical hierarchy \cite{ArkaniHamed:1998nn} with a low KK mass gap: too light KK-gravitons are constrained by astrophysics, cosmology and collider physics\footnote{For a review of the constraints on the simplest ADD model with toroidal compactification, c.f. Ref.~\cite{Pomarol:2018oca}. \label{note_const_ADD}}. When one probes gravitational Newton's law at large [small] distances with respect to $R$, gravity appears 4D [$(4+q)$D]. The case $q=1$ is excluded because it leads to a modification of 4D gravitational Newton's law at the scale of the solar system. ADD's proposal is thus often associated with the Large Extra Dimensions (LEDs) paradigm, and is just a reformulation of the gauge hierarchy problem.

In the literature, there are interesting propositions to stabilize such large compactification radii \cite{ArkaniHamed:1998kx, ArkaniHamed:1999dz, Mazumdar:2001ya, Carroll:2001ih, Albrecht:2001cp, Antoniadis:2002gw, Peloso:2003nv}. To circumvent this geometrical hierarchy problem, a solution can be to abandon the framework of UV completions by superstring/M-theory, and to increase the number of extra dimensions. This possibility was mentioned first in Ref.~\cite{ArkaniHamed:1998nn}. Fig.~\ref{graph} shows that, for sufficiently large $q$, the remaining hierarchy disappears: $R \sim \ell_P^{(4+q)}$. A possible trail to UV complete this kind of model could be via Loop Quantum Gravity (LQG) \cite{Rovelli:2014ssa} since, a priori, LQG does not fix the number of spatial dimensions. First attempts to add spacelike extra dimensions to LQG were made in Refs.~\cite{Bodendorfer:2011nv, Bodendorfer:2011nw, Bodendorfer:2011nx, Bodendorfer:2011ny, Bodendorfer:2013jba, Bodendorfer:2011pb, Bodendorfer:2011pc, Bodendorfer:2011hs, Bodendorfer:2011xe, Thurn:2013lko}.

The most popular way to overcome the geometrical hierarchy problem is the warped extra dimension scenario proposed in 1999 by L.~Randall and R.~Sundrum (RS) in Ref.~\cite{Randall:1999ee}, known as the RS1 model. A less known approach is the compactification of $q \geq 2$ spacelike extra dimension on a compact hyperbolic manifold with a large genus (number of holes) proposed in 2000 in Ref.~\cite{Kaloper:2000jb} (see also Ref.~\cite{Orlando:2010kx}).

The goal of the present work is to discuss another compactification geometry which may solve the geometrical hierarchy problem in ADD-like models. In 2005, H.D.~Kim proposed in Ref.~\cite{Kim:2005aa} to realize ADD's idea by compactifying a LED on a 1D singular variety: a metric graph\footnote{Metric graphs have interesting applications in physics, chemistry and mathematics (c.f. Ref.~\cite{Kuchment2002} for a short review). A 2D QFT on a star graph background was developped in Refs.~\cite{Bellazzini:2006jb, Bellazzini:2006kh, Bellazzini:2008mn,Bellazzini:2008cs}.}, like a star or a rose with respectively $N$ leaves/petals of equal length/circumference $\ell$. The reader can find a mathematical introduction to the spectral analysis of differential operators defined on metric graphs, the so-called quantum graphs, in Ref.~\cite{Kuchment2014}. In Ref.~\cite{Kim:2005aa}, it was shown that, for large $N$, one can build a phenomenologically viable model with only a single LED which gives sizable submillimeter deviations from the Newtonian law of gravity in tabletop experiments. The KK mass scale is $M_{KK} = 1/\ell \sim \mathcal{O}(10-100)$ meV. Here, we want to push the concept further and we take $\ell$ close to $\ell_P^{(5)}$ for large $N$, so $M_{KK} = 1/\ell \sim \mathcal{O}(0.1-1)$ TeV which does not reintroduce a scale hierarchy and evade all constrains on traditional ADD models (with a compactification on a low dimensional torus) from submillimeter tests of Newtonian gravity, astrophysics and cosmology. The integer $N$ is radiatively stable so the scenario solves completely the naturalness problem of the Higgs mass under the hypothesis of an exact global permutation symmetry between the leaves/petals. Ref.~\cite{Kim:2005aa} gives no information on the way to embed the SM fields into the proposed geometry. Are they bulk or brane-localized fields? In this work, we will see that the SM fields must be localized on a 3-brane and we find it particulary interesting to localize them on the junction (central vertex) of the star/rose graph.

In the context of the compactification of an extra dimension on a metric graph, the star graph is the most popular \cite{Kim:2005aa, Cacciapaglia:2006tg, Bechinger:2009qk, Abel:2010kw, Law:2010pv, Fonseca:2019aux}, mainly because, with AdS$_5$ leaves, these effective 5D braneworlds capture the low energy behavior of models with warped throats, arising in flux compactification in type IIB superstring theory \cite{Verlinde:1999fy, Klebanov:2000hb, Giddings:2001yu, Dimopoulos:2001ui, Dimopoulos:2001qd, Kachru:2003aw, Cacciapaglia:2006tg}, when one integrates out the modes associated to the transverse dimensions of the throats. In this work, we study a spacelike extra dimension compactified on a flat star/rose graph with identical leaves/petals by adopting a bottom-up approach: we are completely agnostic about the origin of this curious geometry in a UV theory like string theories, LQG, etc.

The authors of Ref.~\cite{Cacciapaglia:2006tg} analyzed a Klein-Gordon field, a Dirac field, a Maxwell field and Einsteinian gravity propagating in an extra dimension compactified on a star with $N$ leaves of different lengths. For that purpose, they define a copy of the same 5D field on each leaf. The copies are connected at the junction of the star through brane-localized interactions and the continuity of the metric. Of course, a different 5D field on each leaf is not equivalent to only one 5D field defined on the whole star graph. In order to recover only one zero mode propagating on the whole star, they add brane-localized mass terms and take the limit of infinite masses such that $N-1$ zero modes decouple from the Effective Field Theory (EFT). However, the meaning of an infinite brane-localized mass term is not clear when the cut-off of the EFT is not so far above the KK scale $M_{KK} = 1/\ell$. That is why we choose in this work the more straigtforward approach of Refs.~\cite{Kim:2005aa, Fujimoto:2019fzb} where a 5D field is defined on the whole metric graph from the start. For that purpose, one needs a distribution theory on the star/rose graph allowing to define a Lagrangian with possible field discontinuities at the junction. Instead of Schwartz's distribution theory \cite{Schwartz1, Schwartz2}, it is more appropriate to use a generalization of Kurasov's one \cite{KURASOV1996297}. In Section~\ref{spacetime_geom}, we give the definitions of a star/rose graph and introduce the elements of the distribution theory we need.

The KK mass spectrum and wave functions of a 5D massless real scalar field on a star/rose graph were studied in Ref.~\cite{Kim:2005aa}. In Section~\ref{KG_field}, we generalize it by adding a 5D mass to the scalar field. Besides, we clarify the method and hypothesises of this previous study, especially the hypothesis of continuity of the scalar field across the cental vertex of the star/rose graph.

Recently, a 5D Dirac field with a compactification on a flat rose graph was considered in Ref.~\cite{Fujimoto:2019fzb}. They took petals of possibly different circumferences and included a 5D Dirac mass for the fermion. In this framework, they considered the rose graph as a master quantum graph since one can reduce it to a star graph by a suitable choice of junction conditions. They studied the general mathematical properties of the junction conditions for the rose graph and classified them. Their work was restricted to the analysis of the zero modes only: the KK mass spectrum and wavefunctions of the excited modes were not considered. They determined the number of zero mode solutions for each type of boundary conditions in their classification. Their work was motivated by the future goals of generating three fermion generations and the features of the flavor sector of the SM fermions from the zero modes of only one generation of 5D fermions. In Section~\ref{Dirac_field}, we study the particular case of a 5D massless Dirac field on a star/rose graph with identical leaves/petals. We use a different approach compared to the one of Ref.~\cite{Fujimoto:2019fzb}. Instead of imposing arbitrary junction conditions, we keep only the natural junction conditions at the vertices, i.e. the junction conditions for which the variation of the action at the vertices vanishes for arbitrary field variations \cite{Hilbert, Csaki:2003dt, Cheng:2010pt, Angelescu:2019viv, Nortier:2020xms}. Indeed, we prefer junction conditions originating from the variation of the action (and thus of the fields) at the vertices. We will see that the natural junction conditions depend only on the hypothesis of the continuity of the fields at the junction. In this approach, we need the bilinear boundary terms for 5D fermions \cite{Henningson:1998cd, Mueck:1998iz, Arutyunov:1998ve, Henneaux:1998ch, Contino:2004vy, vonGersdorff:2004eq, vonGersdorff:2004cg, Angelescu:2019viv, Nortier:2020xms} whose importance was stressed in Chapters~\ref{1_4D perturbative approach} and \ref{Applications}. Besides, we do not restrict ourselves to the study of the zero modes only; we determine the KK mass spectrum and wavefunctions of all KK modes.

In Section~\ref{ADD_star_rose}, we propose a model to reduce the gravity scale to the TeV scale with a large compactified volume, but with EW and KK scales which coincide. The SM fields are localized on the 3-brane at the central vertex of the star/rose, and we compute their couplings to spinless KK-gravitons. We find that the results are very different from standard ADD models in the literature, due to the very specific features of the rose/star graph with identical leaves/petals. We also discuss briefly what kind of physics is expected in the Planckian and trans-Planckian regime of the model, the possibility of a hidden sector made of KK-gravitons and of a dark matter candidate: a stable black hole remnant \cite{Koch:2005ks, Dvali:2010gv, Bellagamba:2012wz, Alberghi:2013hca}, the Planckion \cite{Treder:1985kb, Dvali:2016ovn}.

In Section~\ref{Dirac_Neutrinos}, we revisit the models of Refs.~\cite{Dienes:1998sb, ArkaniHamed:1998vp, Dvali:1999cn}, which generate small Dirac neutrino masses with right-handed neutrinos identified with the zero modes of gauge singlet fermions propagating in a large compactified volume, by adapting this idea to our spacetime geometries. We consider a toy model with only one generation of neutrinos. We use alternatively the zero mode approximation and the exact treatment concerning the brane-localized Yukawa coupling between the SM Higgs field with the 5D neutrino and the 4D left-handed neutrino of the SM particle content. For this exact treatment of a brane-localized Yukawa interaction, we use the 5D method that we developped in Ref.~\cite{Angelescu:2019viv, Nortier:2020xms} with other authors. We find that a large number of KK-neutrinos are sterile and are part of the hidden sector of the proposed models.

We conclude and propose some perspectives in Section~\ref{conclusion_star_rose}. In Appendix~\ref{conventions}, we give our conventions for the 5D Minkowski metric, the Dirac matrices and spinors.

\section{Star \& Rose Graphs}
\label{spacetime_geom}

\subsection{Geometries}
In this subsection, we define the geometries on which we compactify. The reader is refered to Chapter 1 of Ref.~\cite{Kuchment2014} for basic definitions, vocabulary and properties of metric and quantum graphs which we will use in what follows.

\begin{figure}[h]
\begin{center}
\includegraphics[height=7cm]{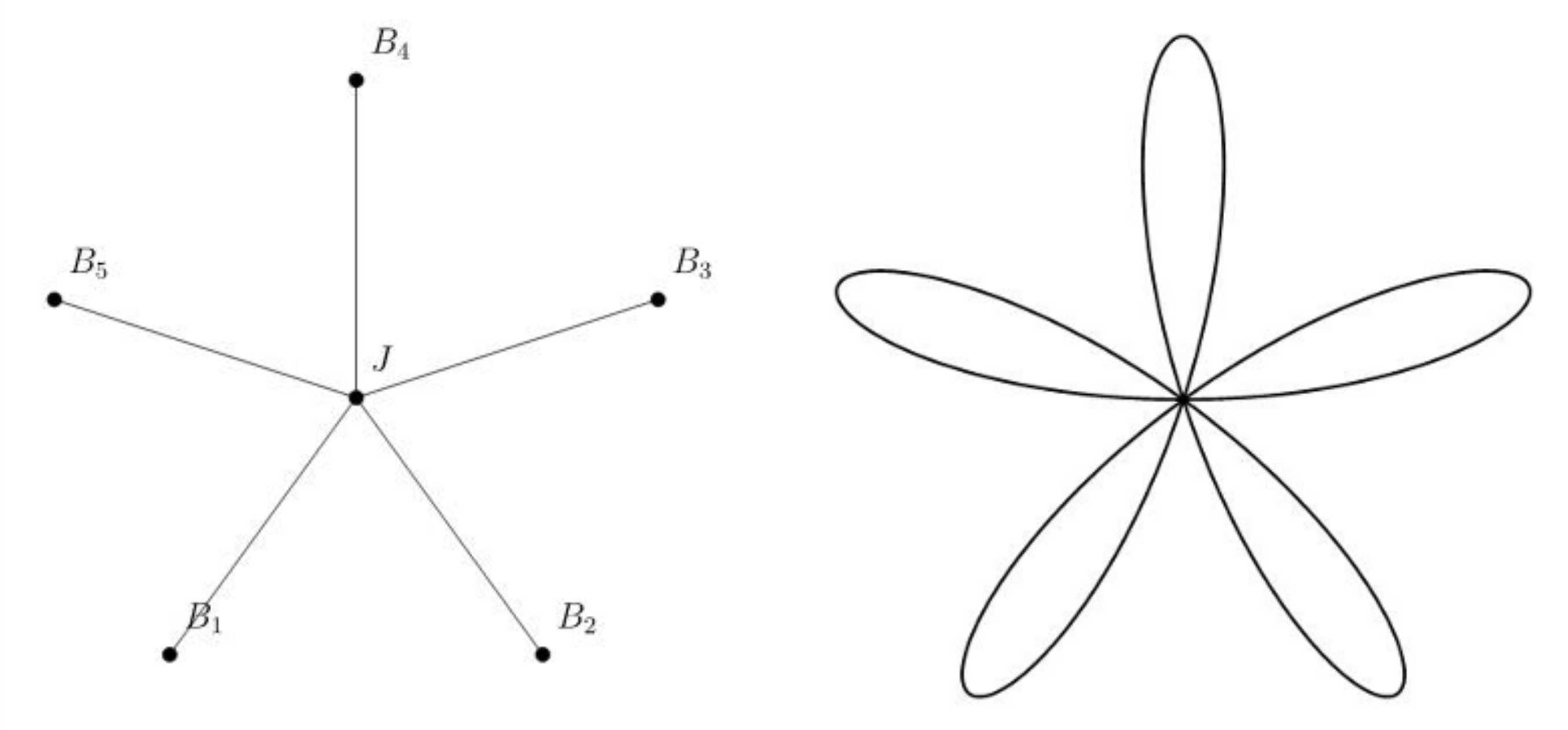}
\end{center}
\caption{Embeddings of a 5-star $\mathcal{S}_5$ (one the left) and of a 5-rose $\mathcal{R}_5$ (on the right) in $\mathbb{R}^2$.}
\label{star_rose_graph}
\end{figure}

\paragraph{$N$-star --}
The $N$-star $\mathcal{S}_N$ (c.f. Fig.~\ref{star_rose_graph}) is defined as the flat equilateral star graph with $N$ bonds directed from 1 vertex of degree $N$ to $N$ vertices of degree 1. It is a flat 1D space of volume $L = N \ell$ obtained by gluing $N$ intervals (the leaves) of length $\ell$ at a common boundary $J$ (the junction). The $i^\text{th}$ leaf ends at the opposite side of the junction $J$: the boundary $B_i$. $\mathcal{S}_N$ is symmetric under the group $\Sigma_N$, which is the set of all permutations of the $N$ leaves. For example, $\mathcal{S}_1$ is the interval of length $\ell$, $\mathcal{S}_2$ the interval of length $2\ell$ symmetric under a reflection ($\Sigma_2 \simeq \mathbb{Z}_2$) with respect to the midpoint $J$, and $\mathcal{S}_3$ is a claw. The couple of coordinates $(y, i) \in [0,\ell] \times \llbracket 1, N \rrbracket$ are assigned to every point of the $i^\text{th}$ leaf with the identification:
\begin{equation}
\forall (i, j) \in \llbracket 1, N \rrbracket^2, \ i \neq j, \ (0, i) \sim (0, j) \, .
\end{equation}

\paragraph{$N$-rose --}
The $N$-rose $\mathcal{R}_N$ (c.f. Fig.~\ref{star_rose_graph}) is defined as the flat equilateral rose graph (also called rhodonea or bouquet of circles), with $N$ directed loops (1 vertex of degree $2N$). It is a flat 1D space of volume $L=N\ell$ obtained by gluing the boundaries of $N$ intervals (the petals), of radius $R$ and circumference $\ell = 2 \pi R$, at a single point $V$ (the vertex/junction). $\mathcal{R}_N$ is symmetric under the group $\Sigma_N$, which is the set of all permutations of the $N$ petals. For example, $\mathcal{R}_1$ is a circle, $\mathcal{R}_2$ a lemniscat, $\mathcal{R}_3$ a trifolium, and $\mathcal{R}_4$ a quadrifolium. The couple of coordinates $(y, i) \in [0, \ell] \times \llbracket 1, N \rrbracket$ is assigned to every point of the $i^\text{th}$ petal, with the identifications:
\begin{equation}
\forall i \in \llbracket 1, N \rrbracket, \ (0, i) \sim (\ell, i) \, ,
\end{equation}
and
\begin{equation}
\forall (i, j) \in \llbracket 1, N \rrbracket^2, \ i \neq j, \, (0, i) \sim (0, j) \, .
\end{equation}

\subsection{Distribution Theory on a Star/Rose Graph}
In order to study a field theory on $\mathcal{K}_N = \mathcal{S}_N$ or $\mathcal{R}_N$ with localized interactions at the vertices, one needs to define a distribution theory on these metric graphs. To be general, we allow the test functions to be discontinuous at the junction. The usual Schwartz's distribution theory \cite{Schwartz1, Schwartz2} is thus not suitable and one should consider instead a generalization on metric graphs of Kurasov's distribution theory \cite{KURASOV1996297}. Up to our knowledge, such a generalization on metric graphs was not considered in the literature. We will thus define in this subsection the objects we need for our study.

\paragraph{Function on $\mathcal{K}_N$ --}
A complex function $f$ on the metric graph $\mathcal{K}_N$ is defined as:
\begin{equation}
f: \left\{
\begin{array}{ccc}
[0, \ell] \times \llbracket 1, N \rrbracket & \rightarrow & \mathbb{C} \, , \\
(y, i) & \mapsto & f(y, i) \, .
\end{array}
\right.
\end{equation}
For each $i \in \llbracket 1, N \rrbracket$, we define a function:
\begin{equation}
f_i: \left\{
\begin{array}{ccc}
[0, \ell] & \rightarrow & \mathbb{C} \, , \\
y & \mapsto & f_i(y) \equiv f(y, i) \, .
\end{array}
\right.
\end{equation}
\begin{itemize}
\item $f$ is continuous/differentiable at $(y,i)=(y_0,i_0)$ if $f_{i_0}$ is continuous/differentiable at $y=y_0$. The derivative of $f$ at $(y,i)=(y_0,i_0)$ is $\partial_y f(y,i) \equiv \partial_y f_i(y)$.
\item $f$ is continuous across the junction if
\begin{equation}
\forall (i,j), \ f(0,i)=f(0,j) \, .
\end{equation}
If it is not the case, $f$ is discontinuous/multivalued at the junction.
\end{itemize}

\paragraph{Test function on $\mathcal{K}_N$ --}
The set of test functions $\mathcal{T}$ is the set of all complex functions $\varphi$ on $\mathcal{K}_N$ such that the functions $\varphi_i$ are infinitely differentiable bounded functions on $[0, \ell]$. We stress that a function $\varphi \in \mathcal{T}$ and/or its derivatives can be discontinuous at the junction.

\paragraph{Distribution --}
A distribution $D \in \mathcal{T}'$ is a linear form on $\mathcal{T}$:
\begin{equation}
\forall \varphi \in \mathcal{T} \, , \ D[\varphi] \equiv \sum_{i=1}^N D_i[\varphi_i] \, ,
\end{equation}
where for every compact set $\mathfrak{B}_i \in [0, \ell]$, there exist constants $C_i$ and $m_i$ such that
\begin{equation}
\forall \varphi \in \mathcal{T} \, , \ \text{supp}(\varphi_i) \in \mathfrak{B}_i \, , \  \left| D_i[\varphi_i] \right| \leq C_i \sum_{\alpha_i\leq m_i} \sup \left| \partial_y^{\alpha_i} \varphi_i (y, i) \right| \, .
\end{equation}

\paragraph{Regular distribution --}
For any integrable complex function  $f$ on $\mathcal{K}_N$, one can define a regular distribution $\widetilde{f} \in \mathcal{T}'$ such that
\begin{equation}
\forall \varphi \in \mathcal{T} \, , \ \widetilde{f}[\varphi] \equiv \sum_{i=1}^N \widetilde{f}_i[\varphi_i] \ \ \ \text{with} \ \ \ \widetilde{f}_i[\varphi_i] \equiv \int_{0}^{\ell} dy \ f(y, i) \, \varphi(y, i) \, .
\end{equation}
A distribution which is not regular is singular.

\paragraph{Product of distributions --}
If $D \in \mathcal{T}'$ and $f \in \mathcal{T}$, one can define the product $f D$ as
\begin{equation}
(fD)[\varphi] \equiv D[f \varphi] \, .
\end{equation}
If $\widetilde{f}$ is the regular distribution associated to $f$, the product $\widetilde{f} D$ is defined as
\begin{equation}
\widetilde{f}D \equiv fD \, .
\end{equation}

\paragraph{Dirac distribution --}
The Dirac distribution on $\mathcal{K}_N$ centered at $(y_0,i_0)$ is the singular distribution $\delta_{y_0,i_0}$ defined as
\begin{equation}
\forall \varphi \in \mathcal{T} \, , \ \delta_{y_0,i_0}[\varphi] \equiv  \varphi(y_0, i_0) \, .
\end{equation}
We want to build a Dirac-like distribution $\delta_{J/V}$ centered at $J/V$ to localize interactions at the junction. Consider the $N$-star $\mathcal{S}_N$. Let $\eta$ be an infinitely differentiable real function on $\mathcal{S}_N$ such that
\begin{equation}
\forall y \in [0, \ell] \, , \ \forall (i, j) \in \llbracket 1, N \rrbracket^2 \, , \ \eta(y,i) = \eta(y,j) \, ,
\label{def_eta}
\end{equation}
and
\begin{equation}
\sum_{i=1}^N \int_0^\ell dy \ \eta(y,i) = 1 \, .
\label{norm_eta}
\end{equation}
We define $\eta_\epsilon$:
\begin{equation}
\eta_\epsilon (y,i) = \dfrac{1}{\epsilon} \, \eta \left( \dfrac{y}{\epsilon}, i \right) \, ,
\label{dilat_eta}
\end{equation}
with $\epsilon > 0$, and we associate the regular distribution $\widetilde{\eta_\epsilon}$ to it. The Dirac distribution $\delta_J$ at the junction $J$ is defined as the weak limit:
\begin{equation}
\delta_J \equiv \lim_{\epsilon \to 0} \eta_\epsilon \, .
\end{equation}
We have
\begin{align}
\forall \varphi \in \mathcal{T} \, , \ \delta_J[\varphi]
&= \lim_{\epsilon \to 0} \sum_{i=1}^N \int_0^\ell dy \ \eta_\epsilon (y,i) \, \varphi (y,i) \, , \nonumber \\
&= \lim_{\epsilon \to 0} \sum_{i=1}^N \int_0^\ell dy \ \eta (y,i) \, \varphi ( \epsilon y,i) \, , \nonumber \\
&= \sum_{i=1}^N \varphi(0,i) \int_0^\ell dy \ \eta(y,i) \, , \nonumber \\
&= \dfrac{1}{N} \, \sum_{i=1}^N \varphi(0,i) \, .
\end{align}
We conclude that
\begin{equation}
\delta_J = \dfrac{1}{N} \, \sum_{i=1}^N \delta_{0,i} \, .
\end{equation}
In the same way, one can build a Dirac distribution $\delta_V$ at the vertx $V$ of the $N$-rose $\mathcal{R}_N$ such that
\begin{equation}
\delta_V \equiv \lim_{\epsilon \to 0} \eta_\epsilon \, ,
\end{equation}
where $\eta_\epsilon$ is defined as in Eq.~\eqref{dilat_eta} and $\eta$ is an infinitely differentiable real function on $\mathcal{R}_N$ such that
\begin{equation}
\forall y \in [0, \ell] \, , \ \forall (i, j) \in \llbracket 1, N \rrbracket^2 \, , \ \eta(y,i)=\eta(y,j) \ \text{and} \ \eta(y-\ell,i)=\eta(y-\ell,j) \, ,
\end{equation}
and normalized as in Eq.~\ref{norm_eta}. Then,
\begin{equation}
\delta_V = \dfrac{1}{2N} \, \sum_{i=1}^N \left( \delta_{0,i} + \delta_{\ell,i} \right) \, .
\end{equation}
We have thus defined a Dirac distribution centered at the junction $J/V$ which acts on test functions possibly discontinuous at $J/V$.

\paragraph{Distributional derivative --} In Kurasov's distribution theory, one defines a distributional derivative in the same way as in Schwartz's distribution theory. The distributional derivative $\partial_y D$ of a distribution $D \in \mathcal{T}'$ is defined by
\begin{equation}
\forall \varphi \in \mathcal{T} \, , \ \partial_y D [\varphi] = - D [\partial_y \varphi] \, .
\end{equation}
The derivative of a regular distribution $\widetilde{f} \in \mathcal{T}'$ is thus
\begin{equation}
\partial_y \widetilde{f} = \left\{ \partial_y f \right\} + \sum_{i=1}^N \left( \delta_{0, i} - \delta_{\ell, i} \right) f \, ,
\end{equation}
where $\left\{ \partial_y f \right\}$ is the regular distribution associated to derivative $\partial_y f$. As in original Kurasov's distribution theory, the distributional derivative does not coincide with the derivative defined in the classical sense. For instance, the distributional derivative of the regular distribution associated to a constant function is not zero. For the unit function $\mathbf{1}: (y, i) \mapsto 1$, we have
\begin{equation}
\partial_y \widetilde{\mathbf{1}} = \sum_{i=1}^N \left( \delta_{0, i} - \delta_{\ell, i} \right) \, .
\end{equation}
Instead, it would be more natural to define the distributional derivative as
\begin{equation}
\forall \varphi \in \mathcal{T} \, , \ \partial_y D [\varphi] = - D [\partial_y \varphi] - \sum_{i=1}^N \left[ \left( \delta_{0, i} - \delta_{\ell, i} \right) D \right] [\varphi] \, .
\end{equation}
However, in this case, a distributional derivative of the Dirac distributions $\delta_{0/\ell, i}$ and $\delta_{J/V}$ would involve Dirac distributions squared which is not defined. Therefore, the price to pay in order to define a usefull distributional derivative is to have extra boundary terms at the vertices, compared to the traditional distributional derivative for a regular distribution in Schwartz's distribution theory. One can thus define the $n^\text{th}$ derivative of the Dirac distribution $\delta_{y_0, i_0}$ as
\begin{equation}
\partial_y^n \delta_{y_0, i_0} [\varphi] =  (-1)^n \, \partial_y^n \varphi(y_0, i_0) \, .
\end{equation}

\paragraph{Moment expansion --}
We will adapt the moment expansion \cite{Estrada:1994} of Schwartz's distribution theory to our case. Consider the $N$-star. The Taylor series of a test function $\varphi \in \mathcal{T}$ is
\begin{equation}
\forall y \in [0, \ell], \forall i \in \llbracket 1, N \rrbracket, \ \varphi(y,i) = \sum_{n=0}^{+\infty} \partial_y^n \varphi(0,i) \, \dfrac{y^n}{n!} \, .
\end{equation}
Then, the action of the previous regular distribution $\widetilde{\eta}$ is
\begin{equation}
\widetilde{\eta} [\varphi] = \sum_{i=1}^N \int_0^\ell dy \ \eta(y,i) \sum_{n=0}^{+\infty} \partial_y^n \varphi(0,i) \, \dfrac{y^n}{n!} \, .
\end{equation}
We define the $n^\text{th}$ moment of the function $\eta$ as
\begin{equation}
\mu_n = \widetilde{\eta} [y^n] = \sum_{i=1}^N \int_0^\ell dy \ \eta(y,i) \, \dfrac{y^n}{n!} \, .
\end{equation}
Thus,
\begin{align}
\widetilde{\eta} [\varphi] &= \left( \sum_{n=0}^{+\infty} \sum_{i=1}^N \dfrac{(-1)^n \, \mu_n}{N n!} \, \partial_y^n \delta_{0,i} \right) [\varphi] \, , \nonumber \\
&= \left( \sum_{n=0}^{+\infty} \dfrac{(-1)^n \, \mu_n}{n!} \, \partial_y^n \delta_J \right) [\varphi] \, .
\end{align}
A similar result is obtained with the $N$-rose. We define the moment expansion of $\widetilde{\eta}$ by
\begin{equation}
\widetilde{\eta} = \sum_{n=0}^{+\infty} \dfrac{(-1)^n \, \mu_n}{n!} \, \partial_y^n \delta_{J/V} \, .
\end{equation}

\subsection{Star/Rose Extra Dimension}
We want to study a field theory on the flat factorizable geometry $\mathcal{M}_4 \times \mathcal{K}_N$, with $\mathcal{K}_N = \mathcal{S}_N$ or $\mathcal{R}_N$. The coordinates can be split as $(z^M, i) = (x^\mu, y, i)$, where $x^\mu$ are the coordinates of $\mathcal{M}_4$. One has $M  \in \llbracket 0, 4 \rrbracket$, and $\mu \in \llbracket 0, 3 \rrbracket$. The junction $J/V$ and the boundaries $B_i$ break explicitely the 5D Lorentz-Poincaré symmetries to the 4D ones, but the 5D symmetries are still preserved locally in the bulk, in the same way as orbifold fixed points. The junction and boundaries are thus 3-branes where one can localize 4D fields and brane-localized kinetic and/or interaction terms for the bulk fields. The 3-branes at the boundaries are called $B_i$-branes, and the 3-brane at the junction is called $J/V$-brane for $\mathcal{K}_N = \mathcal{S}_N/\mathcal{R}_N$. 

One can consider 5D fields which propagate only in one petal/leaf as in Refs.~\cite{Cacciapaglia:2006tg, Bechinger:2009qk, Abel:2010kw, Law:2010pv, Fonseca:2019aux}, or 5D fields which propagate into all the star/rose graph. In this latter case, it is straightforward to generalize our discussion of functions defined on $\mathcal{K}_N$ to the case of 5D fields. The 5D fields can be discontinuous at the junction and thus multivalued at this point. One can interpret it from an EFT point of view (which is always the case in any realistic model with interactions): the value of the field at the point $(x^\mu, 0, i)$ is the one at the neighborhood but outside the core of the $J/V$-brane, since its microscopic description is outside the range of validity of the EFT. One should think the point $(0,i)$ of the graph as $(0^+, i)$. Therefore, the fact that the field is multivalued at the junction is not a problem. It is convenient to define the field theory within a distributional approach with respect to the coordinates $(y, i)$, where the 5D fields are functions of $x^\mu$ but linear forms which act on functions of $(y,i)$.

\section{5D Klein-Gordon Field on a Star/Rose Graph}
\label{KG_field}

\subsection{Klein-Gordon Equation \& Junction/Boundary Conditions}
\label{KG_field_star}
We study a 5D real scalar field $\Phi$ of mass dimension 3/2 and of mass $M_\Phi$ defined on $\mathcal{M}_4 \times \mathcal{K}_N$. The 5D fields $\Phi_i$ are supposed to be smooth functions on the interval $[0, \ell]$. We associate to $\Phi$ a regular distribution $\widetilde{\Phi}$. The Lagrangian $\widetilde{\mathcal{L}_\Phi}$ describing the dynamics of $\Phi$ is defined at the level of distributions. The action is
\begin{equation}
S_\Phi = \int d^4x \ \widetilde{\mathcal{L}_\Phi} [\mathbf{1}] \, ,
\end{equation}
with the unit test function $\mathbf{1}: (y, i) \mapsto 1$, and
\begin{equation}
\widetilde{\mathcal{L}_\Phi} = - \dfrac{1}{2} \, \widetilde{\Phi} \Box_5 \widetilde{\Phi} - \dfrac{1}{4} \, \widetilde{\Phi}^2 \, \partial_y \left( \delta_{\ell, i} - \delta_{0, i} \right) - \dfrac{M_\Phi^2}{2} \, \widetilde{\Phi}^2 \, ,
\label{L_Phi_star}
\end{equation}
with $M_\Phi^2 \geq 0$. We do not include brane-localized kinetic/mass/interaction terms, and the boundary terms are chosen to have Neumann-like conditions at the junction and boundaries. The action reduces to the standard form for a Klein-Gordon field:
\begin{equation}
S_\Phi = \int d^4x \ \sum_{i=1}^N \int_0^\ell dy \left( \dfrac{1}{2} \, \partial^M \Phi \partial_M \Phi - \dfrac{M_\Phi^2}{2} \, \Phi^2 \right) \, ,
\label{S_Phi_star}
\end{equation}
where the boundary terms coming from the distributional derivatives cancel each other.

$\Phi$ can be continuous or discontinuous across the $J/V$-brane. This feature depends on the microscopic structure of the $J/V$-brane in the UV completion. We apply Hamilton's principle to the action $S_\Phi$, with arbitrary variations $\delta \Phi$ of $\Phi$ in the bulk and on the branes. The $\delta \Phi$'s inherite the (dis)continuity properties from $\Phi$ across the $J/V$-brane and we extract the junction and boundary conditions from integrals over total derivatives. We get the Klein-Gordon equation:
\begin{equation}
\left( \partial^M \partial_M + M_\Phi^2 \right) \Phi \left( x^\mu, y, i \right) = 0 \, ,
\label{KG_Phi_star}
\end{equation}
Neumann boundary conditions on the $B_i$-branes:
\begin{equation}
\partial_{y} \Phi \left( x^\mu, \ell, i \right) = 0 \, ,
\label{BCs_Phi_star}
\end{equation}
and the junction condition depends on the (dis)continuity of $\Phi$:
\begin{itemize}
\item If $\Phi$ is allowed to be discontinuous across the junction, we get Neumann junction conditions:
\begin{equation}
\partial_y \Phi(x^\mu, 0/\ell, i) = 0 \, .
\end{equation}
We say that the $J/V$-brane is \textit{airtight} to the field $\Phi$, which means that the spectrum is equivalent to the one obtained by disconnecting the $N$ bonds at the vertex $J/V$ into $N$ disjoined intervals. A brane which is airtight to the field behaves like a boundary for this field.
\item If we impose to $\Phi$ to be continuous across the junction, we get a Neumann-Kirchhoff junction condition:
\begin{equation}
\left\{
\begin{array}{rcl}
\displaystyle{\sum_{i=1}^N \partial_{y} \Phi \left( x^\mu, 0, i \right) = 0} & \text{for} & \mathcal{K}_N = \mathcal{S}_N \, , \\ \\
\displaystyle{\sum_{i=1}^N \left[ \partial_{y} \Phi \left( x^\mu, y, i \right) \right]_{y=0}^{\ell} = 0} & \text{for} & \mathcal{K}_N = \mathcal{R}_N \, ,
\end{array}
\right.
\label{JC_Phi_star}
\end{equation}
with $\left[ g(y) \right]_{y=a}^b = g(b) - g(a)$. When $\Phi$ is continuous across the junction, the leaves/petals communicate through the $J/V$-brane which is thus not airtight.
\end{itemize}

\subsection{Kaluza-Klein Dimensional Reduction}
\label{KK_scalar_20}
We will not study here the KK dimensional reduction when $\Phi$ is allowed to be discontinuous across the junction since it reduces to a 5D scalar field on $N$ disjoined intervals. The case of a 5D scalar field on an interval is very well known in the literature \cite{Dobrescu:2008zz}. In the following, we focus on the continuous case.

\subsubsection{Separation of Variables}
We perform the KK dimensional reduction of the 5D field theory to an effective 4D one in terms of KK degrees of freedom. A general 5D field $\Phi$ can be expanded as
\begin{equation}
\Phi \left( x^\mu, y, i \right) = \sum_b \ \sum_{n_b} \ \sum_{d_b} \phi^{(b, \, n_b, \, d_b)} \left( x^\mu \right) \, f_\phi^{(b, \, n_b, \, d_b)} \left( y, i \right) \, .
\label{KK_Phi_star}
\end{equation}
We label each KK mode solution by a triplet $(b, n_b, d_b)$, where:
\begin{itemize}
\item $b$ labels the different KK towers for the same 5D field which are defined by different mass spectra (see below);
\item $n_b$ labels the levels in the KK tower $b$;
\item $d_b$ labels the degenerate modes for each KK level $(b, n_b)$. We choose the notation $d_b$, instead of the more appropriate one $d(b, n_b)$, for simplifying the notations since we will see that each KK level for a KK tower $b$ has the same degeneracy: there is thus no ambiguity.
\end{itemize}

The 5D equation \eqref{KG_Phi_star} splits into the Klein-Gordon equations for the 4D fields $\phi^{(b, \, n_b, \, d_b)}$:
\begin{equation}
\left( \partial^\mu \partial_\mu + \left[ m_\phi^{(b, \, n_b)} \right]^2 \right) \phi^{(b, \, n_b, \, d_b)} (x^\mu) = 0 \, ,
\label{KG_KK-Phi_star}
\end{equation}
with
\begin{equation}
\left[ m_\phi^{(b, \, n_b)} \right]^2 = M_\Phi^2 + \left[ k_\phi^{(b, \, n_b)} \right]^2 \, ,
\label{m_func_k}
\end{equation}
and the differential equations for the wave functions $f_\phi^{(b, \, n_b, \, d_b)}$:
\begin{equation}
\left( \partial_{y}^2 + \left[ k_\phi^{(b, \, n_b)} \right]^2 \right) f_\phi^{(b, \, n_b, \, d_b)} \left( y, i \right) = 0 \, ,
\label{wave_eq_Phi_star}
\end{equation}
where $\left[ m_\phi^{(b, \, n_b)} \right]^2 \geq 0$ is the mass squared of the KK modes $\phi^{(b, \, n_b, \, d_b)}$, and $\left[ k_\phi^{(b, \, n_b)} \right]^2 \in [0, \ + \infty)$ is an eigenvalue of the operator $\partial_{y}^2$ on $\mathcal{K}_N$ associated to the eigenfunctions $f_\phi^{(b, \, n_b, \, d_b)}$. The orthonormalization conditions for the wave functions $f_\phi^{(b, \, n_b, \, d_b)}$ are
\begin{equation}
\sum_{i=1}^N \int_0^\ell dy \ f_\phi^{(b, \, n_b, \, d_b)}(y, i) \, f_\phi^{(b', \, n'_{b'}, \, d'_{b'})}(y, i) = \delta^{bb'} \, \delta^{n_{b} n'_{b'}} \, \delta^{d_{b} d'_{b'}} \, .
\label{norm_wave_Phi_star}
\end{equation}
The conditions on the 5D field $\Phi$ on the 3-branes are naturally transposed to conditions on the KK wave functions $f_\phi^{(b, \, n_b, \, d_b)}$.

\subsubsection{Zero Modes}
We are looking for zero mode solutions ($b=0$, $n_0 = 0$, $k_\phi^{(0, \, 0)}=0$) of Eq.~\eqref{wave_eq_Phi_star}. For both compactifications on $\mathcal{S}_N$ and $\mathcal{R}_N$, there is only one zero mode ($d_0 \in \{1\}$) whose wave function is flat, such that:
\begin{equation}
f_\phi^{(0, \, 0, \, 1)} (y, i) = \sqrt{\dfrac{1}{N \ell}} \, .
\label{zero_mode_sol_Phi_2}
\end{equation}

\subsubsection{Excited Modes}

\subsubsection*{\boldmath \textcolor{black}{$a)$ $N$-Star}}
The general solutions of Eq.~\eqref{wave_eq_Phi_star}, satisfying the Neumann boundary conditions \eqref{BCs_Phi_star}, are of the form:
\begin{equation}
f_\phi^{(b, \, n_b, \, d_b)} (y, i) = A_i^{(b, \, n_b, \, d_b)} \, \cos \left[ k_\phi^{(b, \, n_b)} \, (y-\ell) \right] \, ,
\end{equation}
with $A_i^{(b, \, n_b, \, d_b)} \in \mathbb{R}$. The continuity condition on the wave functions at the $J$-brane gives
\begin{equation}
\forall (i,j) \, , \ A_i^{(b, \, n_b, \, d_b)} \, \cos \left[ k_\phi^{(b, \, n_b)} \, \ell \right] = A_j^{(b, \, n_b, \, d_b)} \, \cos \left[ k_\phi^{(b, \, n_b)} \, \ell \right] \, ,
\label{cont_phi_star_20}
\end{equation}
which leads to two kinds of excited KK modes: the KK wave functions $f_\phi^{(b, \, n_b, \, d_b)}$ can vanish or not on the $J$-brane.

\subsubsection*{\boldmath \textcolor{black}{\textit{First case: $f_\phi^{(b, \, n_b, \, d_b)}(0,i)=0$}}}
Eq.~\eqref{cont_phi_star_20} gives the KK mass spectrum
\begin{equation}
\cos \left[ k_\phi^{(b, \, n_b)} \, \ell \right] = 0 \ \ \ \underset{b=1}{\Longrightarrow} \ \ \ k_\phi^{(1, \, n_1)} = \left( n_1 + \dfrac{1}{2} \right) \dfrac{\pi}{\ell} \, , \ n_1 \in \mathbb{N} \, ,
\label{mass_spect_Phi_star_2}
\end{equation}
which defines the KK tower $b=1$. The Neumann-Kirchhoff junction condition \eqref{JC_Phi_star} implies
\begin{equation}
\sum_{i=1}^N A_i^{(1, \, n_1, \, d_1)} = 0 \, .
\label{eq_Ai_20}
\end{equation}
Each KK level $(1, n_1)$ is thus $N-1$ times degenerate ($d_1 \in \llbracket 1, N-1 \rrbracket$) and the KK wave functions are
\begin{equation}
f_\phi^{(1, \, n_1, \, d_1)} (y, i) = \epsilon^{(d_1)}_i \sqrt{\dfrac{2}{\ell}} \, \cos \left[ k_\phi^{(1, \, n_1)} \, (y-\ell) \right] \, ,
\end{equation}
with the $(N-1)$-vector basis:
\begin{align}
\overrightarrow{\epsilon^{(1)}} &= \dfrac{1}{\sqrt{2}} \left( 1, -1, 0, \cdots, 0 \right) \, , \nonumber \\
\overrightarrow{\epsilon^{(2)}} &= \dfrac{1}{\sqrt{6}} \left( 1, 1, -2, 0, \cdots, 0 \right) \, , \nonumber \\
\vdots \nonumber \\
\overrightarrow{\epsilon^{(N-1)}} &= \dfrac{1}{\sqrt{N(N-1)}} \left( 1, 1, \cdots, 1, -(N-1) \right) \, .
\label{basis_epsi}
\end{align}

\subsubsection*{\boldmath \textcolor{black}{\textit{Second case: $f_\phi^{(b, \, n_b, \, d_b)}(0,i) \neq 0$}}}
We have $\cos \left[ k_\phi^{(b, \, n_b)} \, \ell \right] \neq 0$ so Eq.~\eqref{cont_phi_star_20} gives
\begin{equation}
\forall (i, j) \, , \ A_i^{(b, \, n_b, \, d_b)} = A_j^{(b, \, n_b, \, d_b)} \equiv A^{(b, \, n_b, \, d_b)} \, .
\label{eq_A_20}
\end{equation}
The Kirchhoff junction condition \eqref{JC_Phi_star} leads to the KK mass spectrum
\begin{equation}
\sin \left[ k_\phi^{(b, \, n_b)} \, \ell \right] = 0 \ \ \ \underset{b=2}{\Longrightarrow} \ \ \ k_\phi^{(2, \, n_2)} = n_2 \, \dfrac{\pi}{\ell} \, , \ n_2 \in \mathbb{N}^* \, ,
\label{mass_spect_Phi_star_1}
\end{equation}
which defines the KK tower with $b=2$ whose KK levels are not degenerate ($d_2 \in \{1\}$). The KK wave functions are
\begin{equation}
f_\phi^{(2, \, n_2, \, 1)} (y, i) = \sqrt{\dfrac{2}{N\ell}} \, \cos \left[ k_\phi^{(2, \, n_2)} \, (y-\ell) \right] \, .
\end{equation}

\subsubsection*{\boldmath \textcolor{black}{$b)$ $N$-Rose}}
Again, to satisfy the continuity condition at the $V$-brane, the KK wave functions $f_\phi^{(b, \, n_b, \, d_b)}$ can vanish or not at the vertex.

\subsubsection*{\boldmath \textcolor{black}{\textit{First case: $f_\phi^{(b, \, n_b, \, d_b)}(0,i)=0$}}}
The general solutions of Eq.~\eqref{wave_eq_Phi_star} with $f_\phi^{(b, \, n_b, \, d_b)}(0,i)=0$ are of the form:
\begin{equation}
f_\phi^{(b, \, n_b, \, d_b)} (y, i) = A_i^{(b, \, n_b, \, d_b)} \, \sin \left[ k_\phi^{(b, \, n_b)} \, y \right] \, ,
\label{gen_prof_rose_50}
\end{equation}
with $A_i^{(b, \, n_b, \, d_b)} \in \mathbb{R}$. The periodicity condition for each petal at the vertex gives $\sin \left[ k_\phi^{(b, \, n_b, \, d_b)} \, \ell \right] = 0$. Moreover, the Neumann-Kirchhoff junction condition \eqref{JC_Phi_star} implies
\begin{equation}
\left( \cos \left[ k_\phi^{(b, \, n_b)} \, \ell \right] - 1 \right) \sum_{i=1}^N A_i^{(b, \, n_b, \, d_b)} = 0 \, .
\label{JC_phi_30}
\end{equation}

There are thus two possibilities:
\begin{itemize}
\item First possibility:\\
The KK mass spectrum is
\begin{equation}
\left\{
\begin{array}{rcl}
\cos \left[ k_\phi^{(b, \, n_b)} \, \ell \right] &\neq& 1 \\
\sin \left[ k_\phi^{(b, \, n_b)} \, \ell \right] &=& 0
\end{array}
\right.
\ \ \ \underset{b=1}{\Longrightarrow} \ \ \ k_\phi^{(1, \, n_1)} = (2n_1+1) \, \dfrac{\pi}{\ell} \, , \ n_1 \in \mathbb{N} \, ,
\label{mass_spect_Phi_rose_3}
\end{equation}
and defines the KK tower $b=1$. From the conditions \eqref{JC_phi_30}, we get Eq.~\eqref{eq_Ai_20} so $d_1 \in \llbracket 1, N-1 \rrbracket$ and the KK wave functions are
\begin{equation}
f_\phi^{(1, \, n_1, \, d_1)} (y, i) = \epsilon^{(d_1)}_i \sqrt{\dfrac{2}{\ell}} \, \sin \left[ k_\phi^{(1, \, n_1)} \, y \right] \, ,
\end{equation}
with the $(N-1)$-vector basis \eqref{basis_epsi}.
\item Second possibility:\\
The KK mass spectrum is
\begin{equation}
\left\{
\begin{array}{rcl}
\cos \left[ k_\phi^{(b, \, n_b)} \, \ell \right] &=& 1 \\
\sin \left[ k_\phi^{(b, \, n_b)} \, \ell \right] &=& 0
\end{array}
\right.
 \ \ \ \underset{b=2}{\Longrightarrow} \ \ \ k_\phi^{(2, \, n_2)} = 2n_2 \, \dfrac{\pi}{\ell} \, , \ n_2 \in \mathbb{N}^* \, ,
\label{mass_spect_Phi_rose_2}
\end{equation}
which satisfies Eq.~\eqref{JC_phi_30} and defines the KK tower $b=2$. We get $d_2 \in \llbracket 1, N \rrbracket$ and the KK wave functions are
\begin{equation}
f_\phi^{(2, \, n_2, \, d_2)} (y, i) = \eta^{(d_2)}_i \sqrt{\dfrac{2}{\ell}} \, \sin \left[ k_\phi^{(2, \, n_2)} \, y \right] \, ,
\label{wave_funct_1}
\end{equation}
with the $N$-vector basis:
\begin{align}
\overrightarrow{\eta^{(1)}} &= \left( 1, 0, \cdots, 0 \right) \, , \nonumber \\
\overrightarrow{\eta^{(2)}} &= \left( 0, 1, 0, \cdots, 0 \right) \, , \nonumber \\
\vdots \nonumber \\
\overrightarrow{\eta^{(N)}} &= \left( 0, \cdots, 0, 1 \right) \, .
\label{basis_eta}
\end{align}
\end{itemize}

\subsubsection*{\boldmath \textcolor{black}{\textit{Second case}: $f_\phi^{(b, \, n_b, \, d_b)}(0,i) \neq 0$}}
The KK mass spectrum is the same as in Eq.~\eqref{mass_spect_Phi_rose_2}. Each KK level $(b, n_b)$ is thus degenerate with the $N$ KK modes of the level $(2, n_2)$ so $d_2 \in \llbracket 1, N+1 \rrbracket$: we label the KK modes with non-vanishing wave functions at the junction by the triplet $(2, n_2, N+1)$. The KK wave functions are
\begin{equation}
f_\phi^{(2, \, n_2, \, N+1)} (y, i) = \sqrt{\dfrac{2}{N \ell}} \, \cos \left( m_\phi^{(2, \, n_2)} \, y \right) \, .
\label{wave_phi_rose_interm}
\end{equation}

\vspace{1cm}
Finally, we insist on the fact that all KK towers labeled by $b$ are present in the spectrum: they do not correspond to different models. The 5D field $\Phi$ has one zero mode of mass $M_\Phi$ in both geometries ($\mathcal{K}_N = \mathcal{S}_N$ or $\mathcal{R}_N$) and excited modes. For a massless 5D field ($M_\Phi=0$), the mass gap between the KK modes is of the order of $1/\ell$. Some of the KK modes have wave functions which vanish at the junction. We will see a physical application of these results in Subsection~\ref{pheno_star_rose_graviton}, where we will study a toy model of a 5D spinless graviton which is just a real scalar field with $M_\Phi = 0$. The zero mode is thus identified with the 4D massless graviton (where we do not take into account the spin).

\section{5D Massless Dirac Field  on a Star/Rose Graph}
\label{Dirac_field}

\subsection{Dirac-Weyl Equations \& Junction/Boundary Conditions}
We study a 5D massless Dirac field
\begin{equation}
\Psi =
\begin{pmatrix}
\Psi_L \\
\Psi_R
\end{pmatrix}
\end{equation}
of mass dimension 2 defined on $\mathcal{M}_4 \times \mathcal{K}_N$, where the fields $\Psi_L$ and $\Psi_R$ describe fermion fields of left and right-handed 4D chirality respectively. To the function $\Psi$, we associate the regular distribution $\widetilde{\Psi}$. The action is
\begin{equation}
S_\Psi = \int d^4x \ \widetilde{\mathcal{L}_\Psi} [\mathbf{1}] \, ,
\end{equation}
with the Lagrangian
\begin{equation}
\widetilde{\mathcal{L}_\Psi}
= \dfrac{i}{2} \, \bar{\widetilde{\Psi}} \Gamma^M \overleftrightarrow{\partial_M} \widetilde{\Psi} + \sum_{i=1}^N \dfrac{s_i}{2} \, \bar{\widetilde{\Psi}} \widetilde{\Psi} \, \left( \delta_{0,i} - \delta_{\ell,i} \right) \, ,
\end{equation}
where $\bar{\widetilde{\Psi}} = \Gamma^0 \widetilde{\Psi}$, $\overleftrightarrow{\partial_M} = \vec{\partial_M} - \overleftarrow{\partial_M}$ and $\Gamma^M = \left( \gamma^\mu, i \gamma^5 \right)$ are the 5D Dirac matrices\footnote{Our conventions for the Dirac algebra is given in Appendix~\ref{conventions}.}
We include the bilinear boundary terms at the vertices with $s_i=\pm 1$. The relative sign between the HS terms at $(y,i)=(0,i)$ and $(y,i)=(\ell,i)$ is chosen in order to allow the existence of zero modes \cite{Angelescu:2019viv, Nortier:2020xms}. If we flip the sign of $s_i$, we exchange the features of the left and right-handed KK modes. In what follows, we choose $s_i = 1$.

The action can be written as
\begin{equation}
S_\Psi = \int d^4x \sum_{i=1}^N \left\{ \left( \int_{0}^{\ell} dy \ \dfrac{i}{2} \, \bar{\Psi} \Gamma^M \overleftrightarrow{\partial_M} \Psi \right) - \left[ \dfrac{1}{2} \, \bar{\Psi} \Psi \right]_{y=0}^{\ell} \right\} \, ,
\label{S_Psi_star}
\end{equation}
where the boundary terms coming from the distributional derivatives cancel each other. The conserved Noether current associated to the symmetry $U(1): \ \Psi \ \mapsto \ e^{-i \alpha} \Psi$, with $\alpha \in \mathbb{R}$, is
\begin{equation}
j_\Psi^M = \bar{\Psi} \Gamma^M \Psi \ \ \ \text{with} \ \ \ \partial_M j_\Psi^M = 0 \, .
\end{equation}
Current conservation requires a Kirchhoff condition for the current at the junction:
\begin{equation}
\left\{
\begin{array}{rcl}
\displaystyle{\sum_{i=1}^N j_\Psi^M \left( x^\mu, 0, i \right) \overset{!}{=} 0} & \text{for} & \mathcal{K}_N = \mathcal{S}_N \, , \\ \\
\displaystyle{\sum_{i=1}^N \left[ j_\Psi^M \left( x^\mu, y, i \right) \right]_{y=0}^{\ell} \overset{!}{=} 0} & \text{for} & \mathcal{K}_N = \mathcal{R}_N \, .
\end{array}
\right.
\end{equation}
For the component $M = 4$ one gets at the junction:
\begin{equation}
\left\{
\begin{array}{rcl}
\displaystyle{\sum_{i=1}^N \left. \left( \Psi_L^\dagger \Psi_R - \Psi_R^\dagger \Psi_L \right) \right|_{y=0} \overset{!}{=} 0} & \text{for} & \mathcal{K}_N = \mathcal{S}_N \, , \\ \\
\displaystyle{\sum_{i=1}^N \left[ \Psi_L^\dagger \Psi_R - \Psi_R^\dagger \Psi_L \right]_{y=0}^{\ell} \overset{!}{=} 0} & \text{for} & \mathcal{K}_N = \mathcal{R}_N \, .
\end{array}
\right.
\label{Kir_current_psi}
\end{equation}

We apply Hamilton's principle to the action $S_\Psi$, with arbitrary variations of the fields $\delta \Psi_{L/R}$ in the bulk and on the branes. We get the massless Dirac-Weyl equations for the 5D fields $\Psi_{L/R}$:
\begin{equation}
\left \{
\begin{array}{r c l}
i \sigma^\mu \partial_\mu \Psi_R  (x^\mu, y, i) + \partial_y \Psi_L  (x^\mu, y, i) &=& 0 \, ,
\\ \vspace{-0.2cm} \\
i \bar{\sigma}^\mu \partial_\mu \Psi_L  (x^\mu, y, i) - \partial_y \Psi_R  (x^\mu, y, i) &=& 0 \, .
\end{array}
\right.
\label{Dirac_Psi_star}
\end{equation}
Therefore, when the fields are on-shell, $\Psi_L$ and $\Psi_R$ are not independent so the junction/boundary conditions must not overconstrain $\Psi_L$ and $\Psi_R$ at the same point \cite{Henneaux:1998ch, Contino:2004vy, Nortier:2020xms}. The addition of the HS terms guarantee that only $\Psi_L$ is constrained on the branes by the minimization of the action \cite{Henneaux:1998ch, Contino:2004vy, Angelescu:2019viv, Nortier:2020xms}. We get the Dirichlet boundary conditions at the $B_i$-branes:
\begin{equation}
\Psi_L \left( x^\mu, \ell, i \right) = 0 \ \text{(for $\mathcal{K}_N = \mathcal{S}_N)$} \, .
\label{D_psi_B}
\end{equation}
The fields $\Psi_L$ and $\Psi_R$ can be taken independently (dis)continuous across the junction. $\delta \Psi_{L/R}$ is (dis)continuous as $\Psi_{L/R}$. One can explore the different possibilities of junction conditions for $\Psi_L$, depending on the (dis)continuity of the fields here, summarized in Tab.~\ref{table_junction}:
\begin{itemize}
\item Case 1: If both $\Psi_L$ and $\Psi_R$ are allowed to be discontinuous, one gets Dirichlet junction conditions:
\begin{equation}
\left\{
\begin{array}{l}
\Psi_L(x^\mu, 0, i) = 0 \ \text{for $\mathcal{K}_N = \mathcal{S}_N$ or $\mathcal{R}_N$,} \\
\Psi_L(x^\mu, \ell, i) = 0 \ \text{for $\mathcal{K}_N = \mathcal{R}_N$,}
\end{array}
\right.
\label{D_junction_psi}
\end{equation}
which correspond to a 5D field $\Psi$ defined on $N$ disjoined intervals. The spectrum of a 5D fermion on an interval is well known in the literature \cite{Csaki:2003sh}. The $J/V$-brane is airtight, which is illustrated by the fact that each incoming or outcoming current at the $J/V$-brane vanishes like at a boundary \eqref{Kir_current_psi}. There is a chiral zero mode (here right-handed) in each leaf/petal and a KK tower of vector-like fermions with a mass gap of $\pi/\ell$. If one generation of the SM fermion sector propagates on a $3$-star/$3$-rose it is possible to generate three generations at the level of zero modes with a airtight $J/V$-brane. The mechanism is the same as Refs.~\cite{Fujimoto:2012wv, Fujimoto:2013ki, Fujimoto:2014fka, Fujimoto:2014pra, Fujimoto:2017lln, Fujimoto:2019lbo} with point interactions along an interval/circle to generate several zero modes from a unique discontinuous 5D fermion field.

\item Case 2: If we impose that both $\Psi_L$ and $\Psi_R$ are continuous, we obtain no additional boundary condition at the $V$-brane, but a Dirichlet condition \eqref{D_junction_psi} for $\Psi_L$ at the $J$-brane.

\item Case 3: If we impose only the continuity on $\Psi_L$, there is a Dirichlet condition \eqref{D_junction_psi} for $\Psi_L$ at the $J/V$-brane.

\item Case 4: If we impose only the continuity on $\Psi_R$, Hamilton's principle gives a Kirchhoff junction condition for $\Psi_L$ at the $J/V$-brane:
\begin{equation}
\left\{
\begin{array}{rcl}
\displaystyle{\sum_{i=1}^N \Psi_L \left( x^\mu, 0, i \right) = 0} & \text{for} & \mathcal{K}_N = \mathcal{S}_N \, , \\ \\
\displaystyle{\sum_{i=1}^N \left[ \Psi_L \left( x^\mu, y, i \right) \right]_{y=0}^{\ell} = 0} & \text{for} & \mathcal{K}_N = \mathcal{R}_N \, ,
\end{array}
\right.
\label{JC_Psi_star}
\end{equation}
which solves the current condition \eqref{Kir_current_psi}.
\end{itemize}
For the physical applications in this article, we will add brane-localized terms at the junction for $\Psi_R$, which are in general are incompatible with the continuity of $\Psi_L$ \cite{Angelescu:2019viv, Nortier:2020xms}. In this article, we will not consider airtight branes so we impose only that $\Psi_R$ is continuous at the junction in what follows.

\begin{table}[h]
\begin{center}
\begin{tabular}{l||c|c}
$N$-star & $\Psi_L$ continuous & $\Psi_L$ discontinuous  \\
\hline \hline
$\Psi_R$ continuous & Dirichlet & Kirchhoff   \\ 
$\Psi_R$ discontinuous & Dirichlet & Dirichlet
\end{tabular}

\vspace{0.5cm}

\begin{tabular}{l||c|c}
$N$-rose & $\Psi_L$ continuous & $\Psi_L$ discontinuous  \\
\hline \hline
$\Psi_R$ continuous &   & Kirchhoff   \\ 
$\Psi_R$ discontinuous & Dirichlet & Dirichlet
\end{tabular}
\caption{The different possibilities of junction conditions.}
\label{table_junction}
\end{center}
\end{table}

\subsection{Kaluza-Klein Dimensional Reduction}
\label{KK_Drac_20}

\subsubsection{Separation of Variables}
In order to perform the KK dimensional reduction of the 5D field theory, we use the same method as in Subsection~\ref{KK_scalar_20} for the scalar field, with the same system of labels for the KK modes. We expand the 5D fields $\Psi_{L/R}$ as
\begin{equation}
\left\{
\begin{array}{rcl}
\Psi_{L} \left( x^\mu, y, i \right) &=& \displaystyle{\sum_b \ \sum_{n_b} \ \sum_{d_b} \psi_L^{(b, \, n_b, \, d_b)} \left( x^\mu \right) \, f_L^{(b, \, n_b, \, d_b)} \left( y, i \right)} \, ,  \\
\Psi_{R} \left( x^\mu, y, i \right) &=& \displaystyle{\sum_b \ \sum_{n_b} \ \sum_{d_b} \psi_R^{(b, \, n_b, \, d_b)} \left( x^\mu \right) \, f_R^{(b, \, n_b, \, d_b)} \left( y, i \right)} \, ,
\end{array}
\right.
\label{KK_Psi_star_1}
\end{equation}
where $\psi_{L/R}^{(b, \, n_b, \, d_b)}$ are 4D Weyl fields and $f_{L/R}^{(b, \, n_b, \, d_b)}$ are wave functions defined on $\mathcal{K}_N$. The 5D equations \eqref{Dirac_Psi_star} split into Dirac-Weyl equations for the 4D fields $\psi_{L/R}^{(b, \, n_b, \, d_b)}$:
\begin{equation}
\left \{
\begin{array}{r c l}
i \sigma^\mu \partial_\mu \psi_R^{(b, \, n_b, \, d_b)} (x^\mu) - m_\psi^{(b, \, n_b)} \, \psi_L^{(b, \, n_b, \, d_b)} (x^\mu) &=& 0 \, ,
\\ \vspace{-0.2cm} \\
i \bar{\sigma}^\mu \partial_\mu \psi_L^{(b, \, n_b, \, d_b)} (x^\mu) - m_\psi^{(b, \, n_b)} \, \psi_R^{(b, \, n_b, \, d_b)} (x^\mu) &=& 0 \, ,
\end{array}
\right.
\label{Dirac_KK-Psi_star}
\end{equation}
and the differential equation for the wave functions $f_{L/R}^{(b, \, n_b, \, d_b)}$:
\begin{equation}
\forall y \neq 0 \, , \ \left \{
\begin{array}{r c l}
\partial_y f_R^{(b, \, n_b, \, d_b)} (y, i) - m_\psi^{(b, \, n_b)} \, f_L^{(b, \, n_b, \, d_b)} (y, i) &=& 0 \, ,
\\ \vspace{-0.2cm} \\
\partial_y f_L^{(b, \, n_b, \, d_b)} (y, i) + m_\psi^{(b, \, n_b)} \, f_R^{(b, \, n_b, \, d_b)} (y, i) &=& 0 \, ,
\end{array}
\right.
\label{wave_eq_Psi_star}
\end{equation}
where $m_\psi^{(b, \, n_b)}$ is the mass of the KK modes $(b, \, n_b, \, d_b)$. The wave functions $f_{L/R}^{(b, \, n_b, \, d_b)}$ are orthonormalized with the conditions
\begin{equation}
\sum_{i=1}^N \int_0^\ell dy \ \left[ f_{L/R}^{(b, \, n_b, \, d_b)}(y, i) \right]^* \, f_{L/R}^{(b', \, n'_{b'}, \, d'_{b'})}(y, i) = \delta^{bb'} \, \delta^{n_{b} n'_{b'}} \, \delta^{d_{b} d'_{b'}} \, ,
\label{orthonorm_KK-Phi_star}
\end{equation}
The conditions on the 5D field $\Psi_{L/R}$ at the vertices are naturally transposed to conditions on the KK wave functions $f_{L/R}^{(b, \, n_b, \, d_b)}$.

\subsubsection{Zero Modes}
We are looking for zero mode solutions ($b=0$, $n_0=0$, $m_\psi^{(0, \, 0)}=0$) of Eq.~\eqref{wave_eq_Psi_star} for which the first order differential equations are decoupled. For both compactifications on $\mathcal{S}_N$ and $\mathcal{R}_N$, there is only one right-handed zero mode ($d_0 \in \{1\}$). Its wave function is continuous across the $J/V$-brane and flat:
\begin{equation}
f_R^{(0, \, 0, \, 1)} (y, i) = \sqrt{\dfrac{1}{N \ell}} \, .
\label{zero_mode_star_psi_20}
\end{equation}
For the left-handed zero modes, it is necessary to distinguish between the compactification on $\mathcal{S}_N$ and $\mathcal{R}_N$.

\subsubsection*{\boldmath \textcolor{black}{$a)$ $N$-Star}}
There is no left-handed zero mode for $\mathcal{K}_N = \mathcal{S}_N$.
The theory is thus chiral at the level of the zero mode, which generalizes the well known result of the particular case of a compactification on the interval $\mathcal{S}_1$. The compactification on a star graph is thus very interesting since it allows to build models where the SM fields propagate in the extra dimension: the SM particles are identified with the zero modes of the 5D fields. In Section~\ref{Dirac_Neutrinos}, we will identify the right-handed neutrinos with the zero modes of 5D Dirac fields coupled to brane-localized 4D left-handed neutrinos. The goal is to propose a toy model to obtain small Dirac neutrino masses.

\subsubsection*{\boldmath \textcolor{black}{$b)$ $N$-Rose}}
For $\mathcal{K}_N = \mathcal{R}_N$, we have $N$ degenerate left-handed zero modes ($d_0 \in \llbracket 1, N \rrbracket$). The theory is vector-like at the level of the zero modes, which generalizes the result of the compactification on a circle $\mathcal{R}_1$ in the literature. Therefore, with the compactification on a rose graph and without an airtight $V$-brane, one cannot build models with the SM fields propagating in the extra dimension except if one is able to propose a mechanism which generates chirality by giving a mass to the mirror partners of the SM fermions. If this is possible, one recovers the three SM generations by taking $N=3$. The KK wave functions $f_L^{(0, \, 0, \, d_0)}$ are flat in each petal and discontinuous across the $V$-brane (except for $\mathcal{K}_N = \mathcal{R}_1$, the circle, where they can be taken continuous):
\begin{equation}
f_L^{(0, \, 0, \, d_0)} (y, i) = \eta_i^{(d_0)} \sqrt{\dfrac{1}{\ell}} \, ,
\label{zero_mode_f_L}
\end{equation}
with the $N$-vector basis \eqref{basis_eta}.

\subsubsection{Excited Modes}
\label{excited_modes_fermion}
We are looking for massive KK modes ($m_\psi^{(b, \, n_b)} \neq 0$). The coupled first order differential equations \eqref{wave_eq_Psi_star} can be decoupled into second order ones:
\begin{equation}
\left( \partial_{y}^2 + \left[m_\psi^{(b, \, n_b)}\right]^2 \right) f_{L/R}^{(b, \, n_b, \, d_b)} \left( y, i \right) = 0 \, .
\label{eq_Psi_wave_2nd}
\end{equation}
The KK wave functions $f_{R}^{(b, \, n_b, \, d_b)}$ are continuous across the junction. In the same way as in the case of the scalar field, it is necessary to distinguish between the cases where the $f_{R}^{(b, \, n_b, \, d_b)}$'s vanish or not at the junction. One can follow the same method as in Subsection~\ref{KK_scalar_20}. We will not give again all the details here since there is no major technical difference. We summarize the results in what follows.

\subsubsection*{\boldmath \textcolor{black}{$a)$ $N$-Star}}

\subsubsection*{\boldmath \textcolor{black}{\textit{First case: $f_R^{(b, \, n_b, \, d_b)}(0,i)=0$}}}
\label{1st_case_psi_star}
The KK mass spectrum is
\begin{equation}
m_\psi^{(1, \, n_1)} = \left( n_1 + \dfrac{1}{2} \right) \dfrac{\pi}{\ell} \, , \ n_1 \in \mathbb{N} \, ,
\label{mass_spect_Psi_star_2}
\end{equation}
and defines the KK tower $b=1$. Each KK level is $N-1$ times degenerate ($d_1 \in \llbracket 1, N-1 \rrbracket$) and the KK wave functions are
\begin{equation}
\left\{
\begin{array}{rcl}
f_L^{(1, \, n_1, \, d_1)} (y, i) & = & - \epsilon^{(d_1)}_i \sqrt{\dfrac{2}{\ell}} \, \sin \left[ m_\psi^{(1, \, n_1)} \, (y-\ell) \right] \, , \\
f_R^{(1, \, n_1, \, d_1)} (y, i) & = & \epsilon^{(d_1)}_i \sqrt{\dfrac{2}{\ell}} \, \cos \left[ m_\psi^{(1, \, n_1)} \, (y-\ell) \right] \, ,
\end{array}
\right.
\label{Psi_free_fR=0}
\end{equation}
with the $(N-1)$-vector basis \eqref{basis_epsi}. The $f_L^{(1, \, n_1, \, d_1)}$'s are discontinuous across the $J$-brane (except for $\mathcal{K}_N = \mathcal{S}_1$, the interval, where they are taken continuous).

\subsubsection*{\boldmath \textcolor{black}{\textit{Second case: $f_R^{(b, \, n_b, \, d_b)}(0,i) \neq 0$}}}
The KK mass spectrum is
\begin{equation}
m_\psi^{(2, \, n_2)} = n_2 \, \dfrac{\pi}{\ell} \, , \ n_2 \in \mathbb{N}^* \, ,
\label{mass_spect_Psi_star_1}
\end{equation}
which is not degenerate ($d_2 \in \{1\}$) and defines the KK tower $b=2$. The KK wave functions are
\begin{equation}
\left\{
\begin{array}{rcl}
f_L^{(2, \, n_2, \, d_2)} (y, i) &=& - \sqrt{\dfrac{2}{N \ell}} \, \sin \left[ m_\psi^{(2, \, n_2)} \, (y-\ell) \right] \, , \\
f_R^{(2, \, n_2, \, d_2)} (y, i) &=& \sqrt{\dfrac{2}{N \ell}} \, \cos \left[ m_\psi^{(2, \, n_2)} \, (y-\ell) \right] \, ,
\end{array}
\right.
\end{equation}
where the $f_L^{(2, \, n_2, \, d_2)}$'s can be taken continuous across the $J$-brane.

\subsubsection*{\boldmath \textcolor{black}{$b)$ $N$-Rose}}

\subsubsection*{\boldmath \textcolor{black}{\textit{First case: $f_R^{(b, \, n_b, \, d_b)}(0,i)=0$}}}
\label{1st_case_psi_rose}
There are two cases:
\begin{itemize}
\item First case: \\
The KK mass spectrum is
\begin{equation}
m_\psi^{(1, \, n_1)} = (2n_1+1) \, \dfrac{\pi}{\ell} \, , \ n_1 \in \mathbb{N} \, ,
\label{mass_spect_Psi_rose_3}
\end{equation}
and defines the KK tower $b=1$ with $d_1 \in \llbracket 1, N-1 \rrbracket$. The KK wave functions are
\begin{align}
\left\{
\begin{array}{rcl}
f_L^{(1, \, n_1, \, d_1)} (y, i) &=& \epsilon^{(d_1)}_i \sqrt{\dfrac{2}{\ell}} \, \cos \left[ m_\psi^{(1, \, n_1)} \, y \right] \, , \\
f_R^{(1, \, n_1, \, d_1)} (y, i) &=& \epsilon^{(d_1)}_i \sqrt{\dfrac{2}{\ell}} \, \sin \left[ m_\psi^{(1, \, n_1)} \, y \right] \, ,
\end{array}
\right.
\label{wave_funct_Psi_rose_fR(0)=0}
\end{align}
with the $(N-1)$-vector basis \eqref{basis_epsi}. The $f_L^{(1, \, n_1, \, d_1)}$'s are discontinuous across the $V$-brane.
\item Second possibility:\\
The KK mass spectrum is
\begin{equation}
m_\psi^{(2, \, n_2)} = 2n_2 \, \dfrac{\pi}{\ell} \, , \ n_2 \in \mathbb{N}^* \, ,
\label{mass_spect_Psi_rose_bis}
\end{equation}
and defines the KK tower $b=2$ with $d_2 \in \llbracket 1, N \rrbracket$. The KK wave functions are
\begin{equation}
\left\{
\begin{array}{rcl}
f_L^{(2, \, n_2, \, d_2)} (y, i) &=& \eta^{(d_2)}_i \sqrt{\dfrac{2}{\ell}} \, \cos \left[ m_\psi^{(2, \, n_2)} \, y \right] \, , \\
f_R^{(2, \, n_2, \, d_2)} (y, i) &=& \eta^{(d_2)}_i \sqrt{\dfrac{2}{\ell}} \, \sin \left[ m_\psi^{(2, \, n_2)} \, y \right] \, ,
\end{array}
\right.
\label{wave_funct_psi_1_rose}
\end{equation}
with the $N$-vector basis \eqref{basis_eta}. The $f_L^{(2, \, n_2, \, d_2)}$'s are discontinuous across the $V$-brane (except for $\mathcal{K}_N = \mathcal{R}_1$, the circle, where they can be taken continuous).
\end{itemize}

\subsubsection*{\boldmath \textcolor{black}{\textit{Second case}: $f_R^{(b, \, n_b, \, d_b)}(0,i) \neq 0$}}
The KK mass spectrum is the same as in Eq.~\eqref{mass_spect_Psi_rose_bis}. Each KK level $(b, n_b)$ is thus degenerate with the $N$ KK modes of the level $(2, n_2)$ so $d_2 \in \llbracket 1, N+1 \rrbracket$: we label the KK modes with non-vanishing wave functions at the junction by the triplet $(2, n_2, N+1)$. The KK wave functions are
\begin{equation}
\left\{
\begin{array}{rcl}
f_L^{(2, \, n_2, \, N+1)} (y, i) &=& - \sqrt{\dfrac{2}{N \ell}} \, \sin \left[ m_\psi^{(2, n_2)} \, y \right] \, , \\
f_R^{(2, \, n_2, \, N+1)} (y, i) &=& \sqrt{\dfrac{2}{N \ell}} \, \cos \left[ m_\psi^{(2, n_2)} \, y \right] \, ,
\end{array}
\right.
\label{wave_func_psi_rose_period}
\end{equation}
where the $f_L^{(2, \, n_2, \, N+1)}$'s can be taken continuous across the $V$-brane.

\vspace{1cm}
Like for the scalar field, all KK towers labeled by $b$ are present in the spectrum. Each excited KK level is vector-like for both compactifications, and the mass gap between the KK modes is of the order of $1/\ell$.

\section{A Low 5D Planck Scale with a Star/Rose Extra Dimension}
\label{ADD_star_rose}
In this section, we propose an ADD model with brane-localized 4D SM fields where gravity propagates in a large star/rose extra dimension with large $N$ and a natural value for $\ell$.

\subsection{Lowering the Gravity Scale}
With a LED compactified on a metric graph $\mathcal{K}_N = \mathcal{S}_N$ or $\mathcal{R}_N$, one can obtain a low 5D Planck scale. In this case, Eq.~\eqref{ADD_formula} gives
\begin{equation}
\left[\Lambda_P^{(4)}\right]^2 = L \, \left[ \Lambda_P^{(5)} \right]^{3} \, , \ \ \  L = N \ell \, .
\label{ADD_formula_star}
\end{equation}
To solve the gauge hierarchy problem with the hypothesis of an exact global $\Sigma_N$ symmetry (see Section~\ref{conclusion_star_rose} for a discussion when this hypothesis is relaxed), we choose $\Lambda_P^{(5)} \simeq 1$ TeV, obtained with $L \simeq 1 \times 10^{12}$ m. In order to be in the EFT regime, i.e. below the 5D Planck scale, we need $\ell > \ell_P^{(5)}$, with $\ell_P^{(5)}=1/\Lambda_P^{(5)} \simeq 2 \times 10^{-19}$ m. In practice, $\ell/\ell_P^{(5)} \simeq 10$ with a large $N \simeq 6 \times 10^{29}$ should be enough, and thus a KK mass scale near the EW scale: $M_{KK} = 1/\ell \simeq 100$ GeV. Such heavy KK-gravitons evade completely the constrains from submillimeter tests of 4D gravitational Newtons's law, stellar physics and cosmology (c.f. footnote \ref{note_const_ADD} p.~\pageref{note_const_ADD}). If one allows for $1\%$ of fine tuning for $m_h$ by pushing $\Lambda_P^{(5)}$ up to 10 TeV with $N \simeq 6 \times 10^{27}$, one can even allow for $M_{KK} \simeq 1$ TeV. Moreover, if the concepts of space and volume still make sense at the Planck scale $\Lambda_P^{(5)}$, by taking $\ell \simeq \ell_P^{(5)}$ and $N \simeq 6 \times 10^{30}$ to get $\Lambda_P^{(5)} \simeq 1$ TeV, there is no tower of KK-gravitons in the EFT, instead the first experimental hints for a low 5D Planck scale are strongly coupled quantum gravity phenomena near $\Lambda_P^{(5)}$. Such a large $N$ seems puzzling at first glance: the reader would wonder whether our proposition is just a reformulation of the gauge hierarchy into why $N$ is large. However, in the EFT, $N$ is a conserved integer so it is stable under radiative corrections, it does not need to be dynamically stabilized, and has no preferred value. When $\ell \gg \ell_P^{(5)}$, the models proposed in this article are formulated in the context of EFTs defined on a classical background spacetime $\mathcal{M}_4 \times \mathcal{K}_N$, where the number of leaves/petals $N$ is fixed by the definition of the model, even in presence of gravity. Possibly, $N$ becomes a dynamical quantity in a theory of Planckian gravity involving a quantum spacetime. The situation is somewhat similar to other beyond SM scenarii in the literature to solve the hierarchy problem:
\begin{itemize}
\item The model of Ref.~\cite{Kaloper:2000jb} where $q \geq 2$ spacelike dimension are compactified on a compact hyperbolic manifold of genus $g$ (the number of holes) and of volume $\mathcal{V}_q$. A compact hyperbolic manifold has two length scales: a curvature radius $R_c$ and a linear size $L \sim R_c \log (g)$. For $L \gg R_c/2$, we have
\begin{equation}
\mathcal{V}_q \sim R_c^q \exp \left[ (q-1) \dfrac{L}{R_c} \right] \, .
\end{equation}
For $q=3$, $\Lambda_P^{(7)} \simeq 1$ TeV, and $R_c \sim 1/\Lambda_P^{(7)}$, the formula gives $L \sim 35 R_c$ so the number of holes is very large: $g \sim e^{35} \sim 10^{15}$.
\item The model of Refs.~\cite{ArkaniHamed:1998kx, Corley:2001rt}, where $q$ spacelike extra dimensions are stabilized by a large number $N$ of branes with inter-brane forces, forming a brane crystal.
\begin{equation}
N \sim \dfrac{1}{\alpha^q} \left[ \dfrac{\Lambda_P^{(4)}}{\Lambda_P^{(4+q)}} \right]^2 \simeq \dfrac{10^{30}}{\alpha^q} \, ,
\label{crystal_brane}
\end{equation}
for $\Lambda_P^{(4+q)} \simeq 1$ TeV, where $\alpha$ is a parameter which controls the inter-brane distance. One should have $\alpha \simeq 10$ in order to be in the regime where general relativity is valid and also to avoid a new fine tuning.
\item The model proposed by G.R.~Dvali~et al. in Refs.~\cite{Dvali:2007hz,Dvali:2007wp,Dvali:2008fd,Dvali:2008jb,Dvali:2009fw,Dvali:2009ne,Dvali:2019ewm} involving a large number $N_p$ of particle species. This is a 4D model where the scale at which gravity becomes strongly coupled $\Lambda_G$ is given by
\begin{equation}
\Lambda_G = \dfrac{\Lambda_P^{(4)}}{\sqrt{N_p}} \, .
\label{N-species}
\end{equation}
This effect can be understood perturbatively as being the result of the radiative corrections of $N_p$ particle species to the graviton propagator. Then, $\Lambda_G \simeq 1$ TeV for $N_p \simeq 6 \times 10^{30}$. From Eqs.~\eqref{ADD_formula_star} and \eqref{N-species} it is a curious coincidence that $N_p=N=\left(\Lambda_P^{(4)}/\Lambda_P^{(5)}\right)^2 \simeq 6 \times 10^{30}$ when $\ell=\ell_P^{(5)}$. One can make the same remarks for the number of branes in the brane crystal model when $\alpha_q = 1$ in Eq.~\eqref{crystal_brane}, which means that the inter-brane distance is the fundamental Planck length.
\item The model of $N$-naturalness proposed by N.~Arkani-Hamed~et al. in Ref.~\cite{Arkani-Hamed:2016rle}. There are $N$ SM-like sectors which are mutually non-interacting. The Higgs mass parameter squared $\mu_H^2$ takes values between $- \Lambda_H^2$ and $\Lambda_H^2$, where $\Lambda_H^2$ is the scale common to the $N$ sectors that cuts the quadratic divergences to $\mu_H^2$. Then, for a wide range of $\mu_H^2$ distributions, one expects that some sectors are accidentally tuned at the $1/N$ level, such that $|\mu_H^2| \sim \Lambda_H^2/N$. The sector with the smallest non-zero Vacuum Expectation Value (VEV) is identified with our sector. When $\Lambda_H \gg \Lambda_{EW}$, $N$ has thus to be large in order to have $|\mu_H| \sim 100$ GeV. There is no need for new physics at the TeV scale!
\end{itemize}

The only quantity which needs to be dynamically stabilized is the leaf (petal) length (circumference) $\ell$, otherwise the radion, i.e. the scalar field which represents the fluctuations of $\ell$, remains massless and conflicts with the null result from the search for a new force of infinite range. Moreover, if only the graviton propagates into the extra dimension, the bosonic quantum loops are known to make the extra dimension unstable, which shrinks to a point \cite{Appelquist:1982zs, Appelquist:1983vs}. On the one hand, if $\ell \gtrsim \mathcal{O}(10) \times \ell_P^{(5)}$, the corrections of Planckian gravity can safely be neglected (EFT regime), and it is important to add a field theoretical mechanism to the model to stabilize $\ell$. On the other hand, if $\ell \sim \ell_P^{(5)}$, one expects $\mathcal{O}(1)$ corrections from Planckian gravity, and one needs a complete theory of gravity to formulate the model and to address its stabilization.
 
In the EFT regime, we have supposed the existence of an exact global $\Sigma_N$ symmetry of the $N$-star/rose, which is not realistic since gravity is supposed to break global symmetries (see Section~\ref{conclusion_star_rose} for a discussion on the impact on the resolution of the gauge hierarchy problem). This is reminiscent of the 5D models with a Universal Extra Dimension (UED) compactified on an interval symmetric under an exact $\mathbb{Z}_2$ reflection with respect to its midpoint \cite{Appelquist:2000nn}. One can also mention the model of Ref.~\cite{Agashe:2007jb} where two identical slices of AdS$_5$ are glued to a common UV-brane. The extra dimension of these models can be stabilized by the dynamics of additonal bulk fields at the quantum level by a balance between the contribution of bosonic and fermionic loops \cite{Ponton:2001hq}, or at the classical level by a potential for a scalar field as in the Goldberger-Wise mechanism originally proposed for the RS1 model \cite{Goldberger:1999uk, DeWolfe:1999cp, Csaki:2000zn}. In Ref.~\cite{Law:2010pv}, it is shown how to stabilize a $N$-star $\mathcal{S}_N$ with the potential of a different scalar field in each leaf, these $N$ scalar fields are related by the $\Sigma_N$ symmetry. One can apply these mechanisms here with $N$ large. Other stabilization mechanisms were proposed in Ref.~\cite{ArkaniHamed:1998kx}, in particular it is possible to stabilize an extra dimension compactified on a circle with the help of a complex scalar field with a topologically conserved winding number. As it is possible to stabilize one petal by this mechanism, one can repeat it with a different scalar field in each petal of the $N$-rose $\mathcal{R}_N$. Both for $\mathcal{S}_N$ and $\mathcal{R}_N$, the $N$ scalar fields meet only at the junction $J/V$ and we assume that they interact only through gravity, like the $N$ copies of the SM in Ref.~\cite{Dvali:2007wp, Dvali:2009ne}. Therefore, the picture reduces to stabilize $N$ independent leaves/petals, which is a simplified version of the mechanism in Ref.~\cite{Law:2010pv}. If the bulk fields in all leaves/petals have an exact global $\Sigma_N$ symmetry, the geometrical $\Sigma_N$ symmetry is preserved (see Section~\ref{conclusion_star_rose} for a discussion when it is not the case).

\subsection{Embedding the Standard Model Fields}
Up to this subsection we did not mention how we embed the SM fields into the proposed spacetime geometries. On the one hand, in the ADD-like models in the literature, the SM fields must be localized on a 3-brane, while gravity and possibly other exotic fields propagate in the bulk \cite{ArkaniHamed:1998rs,Antoniadis:1998ig,ArkaniHamed:1998nn,ArkaniHamed:1998vp,ArkaniHamed:1998kx,ArkaniHamed:1998sj,Berezhiani:1998wt}. On the other hand, in the RS1-like models, one can allow some or all SM fields to propagate in the extra dimension \cite{Davoudiasl:1999tf, Pomarol:1999ad, Grossman:1999ra, Chang:1999nh, Gherghetta:2000qt, Davoudiasl:2000wi, Luty:2004ye, Davoudiasl:2005uu, Cacciapaglia:2006mz}. One reason is that the KK scale in the ADD and RS1 models are respectively below and above the TeV scale. The absence of discovery of low mass KK excitations for the SM particles rules out an ADD model with bulk SM fields. In the case of an ADD model with a star extra dimension, where it is possible to embed a chiral model at the zero mode level, one can naively think that it allows the SM fields to propagate in the bulk with a KK mass scale $M_{KK} = 1/\ell \sim \mathcal{O}(1)$ TeV. However, one should also consider the magnitude of the couplings of the zero mode gauge bosons. In the RS1 model with a 5D gauge field, the 5D gauge coupling $g_{RS}^{(5)}$ (of mass dimension $-1/2$) is related to the zero mode gauge coupling $g_{RS}^{(4)}$ by the relation \cite{Pomarol:1999ad}:
\begin{equation}
g_{RS}^{(4)} = \dfrac{g_{RS}^{(5)}}{\sqrt{L_{RS}}} \, ,
\label{gauge_RS}
\end{equation}
where $L_{RS}$ is the proper length of the warped extra dimension. $L_{RS}$ is not large compared to the 5D Planck length $\ell_P^{(5)}$, so for a natural gauge coupling
\begin{equation}
g_{RS}^{(5)} \sim \sqrt{\ell_P^{(5)}} \Rightarrow g_{RS}^{(4)} \sim \mathcal{O}(1) \, ,
\end{equation}
which is the good magnitude for a SM gauge coupling. However, in the ADD model with a $(4+q)$D gauge field, the higher dimensional gauge coupling $g_{ADD}^{(4+q)}$ (of mass dimension $-q/2$) is related to the zero mode gauge coupling $g_{ADD}^{(4)}$ by the relation \cite{Csaki:2004ay}:
\begin{equation}
g_{ADD}^{(4)} = \dfrac{g_{ADD}^{(4+q)}}{\sqrt{\mathcal{V}_q}} \, .
\label{gauge_ADD}
\end{equation}
With a natural value for the gauge coupling
\begin{equation}
g_{ADD}^{(4+q)} \sim \left[ \Lambda_P^{(4+q)} \right]^{-q/2} \, ,
\end{equation}
one obtains with Eq.~\eqref{ADD_formula}:
\begin{equation}
g_{ADD}^{(4)} \sim \dfrac{1}{\sqrt{\mathcal{V}_q \left[ \Lambda_P^{(4+q)}\right]^q}} = \dfrac{\Lambda_P^{(4+q)}}{\Lambda_P^{(4)}} \sim 10^{-16} \, ,
\end{equation}
so $g_{ADD}^{(4)}$ is a very tiny coupling and cannot be identified with a SM gauge coupling. This result depends only on the volume of the compactified space, i.e. on the hierarchy between the 4D and $(4+q)$D Planck scales. It is still valid for the geometries considered in this article, where $\mathcal{V}_1 = L$. Therefore, the gauge coupling argument is much stronger than the one of the KK mass scale to rule out bulk SM fields: it applies to every compactified geometry one can imagine to realize an ADD model.

After this discussion, it is clear that in the case of a star/rose extra dimension with a large $N$, even if $M_{KK} = 1/\ell \gtrsim 1$ TeV, the SM fields must be localized on a 3-brane, like in the other ADD models in the literature. Consider a 5D EFT with a brane. The cut-off in the bulk and the 3-brane thickness are noted $\Lambda$ and $\epsilon$ respectively. There are two cases~\cite{delAguila:2006atw}:
\begin{itemize}
\item The fat brane ($\epsilon > 1/\Lambda$): its microscopic description is in the range of validity of the 5D EFT. Usually, a fat brane is a topological defect \cite{Akama:1982jy, Rubakov:1983bb} and it is necessary to provide a field theoretical mechanism to trap the zero modes of 5D fields of various spins in the neighborhood of the brane \cite{Visser:1985qm, Jackiw:1975fn, Dvali:1996xe, Dubovsky:2001pe, Ohta:2010fu}. The topological defects are the first prototypes of braneworlds in the literature and are chosen by ADD to trap the SM fields in their first article on LEDs \cite{ArkaniHamed:1998rs}. It is also possible to localize the zero modes of 5D fields towards orbifold fixed points or spacetime boundaries with large 5D masses of brane-localized kinetic terms \cite{Dvali:2000rx, Dubovsky:2001pe, Fichet:2019owx}. Irrespectively of the trapping mechanism of the fat brane, we speak about quasi-localized 5D fields.
\item The thin brane ($\epsilon \leq 1/\Lambda$): its microscopic description is outside the range of validity of the 5D EFT. A thin brane is described in the EFT by an infinitely thin hypersurface where 4D fields are strictly localized \cite{Sundrum:1998sj, Csaki:2004ay, Fichet:2019owx}. The trapping mechanism of the fields is relegated to the UV completion. This case became popular when it was relalized that 4D fields can live in the worldvolume of solitonic objects in some UV completions, like D-brane stacks in superstring theories where matter fields are described by open strings attached to them \cite{Polchinski:1996na, Bachas:1998rg, Johnson:2003gi}. In EFTs, orbifold fixed points, spacetime boundaries or metric graph vertices are perfect candidates for thin branes. One can also obtain a thin brane by integrating out the width of a fat brane: one gets an EFT with a cut-off equal to the the inverse of the brane width, and 4D fields (the zero modes of the quasi-localized 5D fields of the UV completion) strictly localized on the thin brane (depending on the quasi-localization mechanism, the excited KK-modes do not necessarily decouple \cite{Fichet:2019owx}). Quickly, after the theoretical discovery of D-branes, physicists explored the new possibilities offered by thin branes \cite{Antoniadis:1998ig, ArkaniHamed:1998nn, Randall:1999ee, Randall:1999vf, Lykken:1999nb, Kogan:1999wc, Gregory:2000jc, Dvali:2000hr}. 
\end{itemize}
For the models studied in this article, if one considers quasi-localized 5D SM fields on the $J/V$-branes (fat branes), one has a problem. Indeed, consider the $N$-star $\mathcal{S}_N$, the fat brane has a thickness $\epsilon > \ell_P^{(5)}$ extended into each leaf, so a zero mode gauge coupling $g_4$ is related to the 5D gauge coupling $g_5 \sim \sqrt{\ell_P^{(5)}}$ as
\begin{equation}
g_4 \sim \dfrac{g_5}{\sqrt{N \epsilon}} \lesssim \mathcal{O} \left( \dfrac{1}{\sqrt{N}} \right) \, ,
\label{gauge_fat_J}
\end{equation}
similar to Eqs.~\eqref{gauge_RS}-\eqref{gauge_ADD}. As $N$ is large, the model will suffer from the same problem as for bulk SM fields in ADD models: the gauge couplings of the zero modes are too suppressed to match the values measured in experiments. The same problem arises with a fat $V$-brane in the case of the $N$-rose $\mathcal{R}_N$. However, there is no problem with quasi-localized SM fields on a fat $B_i$-brane where
\begin{equation}
g_4 \sim \dfrac{g_5}{\sqrt{\epsilon}} \lesssim \mathcal{O} (1) \, .
\end{equation}
Therefore, we will consider only 4D SM fields localized on thin $J/V$-branes\footnote{See Ref.~\cite{Fichet:2019owx} for a recent discussion on the consistency of a 5D EFT coupled to gravity with 4D fields strictly localized on a 3-brane. Only the argument in Subsection~5.1 in Ref.~\cite{Fichet:2019owx} applies to our model (see discussion in Appendix~\ref{Brane_Thickness}).}: the SM gauge fields do not arise from the limit of quasi-localized 5D fields which propagate into the leaves/petals, so the gauge couplings are not suppressed by $\sqrt{N}$. Moreover, spacetime symmetries allow us to localize 4D degrees of freedom exactly on the 3-branes located at the vertices. However, there are various arguments that gravity implies the existence of a minimal length scale in Nature of the order of the fundamental Planck length (see Ref.~\cite{Hossenfelder:2012jw} for a review). Therefore, in a UV completion including gravity, the singular behavior the singular feature of the junction should be regularized. In Ref.~\cite{EXNER200577}, it is shown that the spectrum of a quantum graph can arise in the thin limit of a ``graph-like manifold''. Therefore, the metric graph structure of our extra dimension could emerge from a UV completion where the internal space is a $q$D graph-like manifold with $q-1$ transverse dimensions of 5D Planck size, and where the vertex at the junction is regularized \cite{Cacciapaglia:2006tg}. After integrating out these transverse dimensions, one is left with only one extra dimension compactified on a metric graph. Concerning the UV origin of the brane-localized 4D SM fields, we adopt a bottom-up approach where we do not assume a specific UV completion. We stress that it is crucial that it does not rely on quasi-localized higher-dimensional fields propagating into the leaves/petals. In fact, one can imagine a UV-completion with a $q$D graph-like manifold and a fat $J/V$-brane made of $q$D quasi-localized fields. However, the wave functions of the zero modes must be highly peaked inside the protrusion \cite{Kuchment2002} at the vertex $J/V$, i.e. they must decrease quickly inside the vertex protrusion such that they are suppressed at least by $1/\sqrt{N}$ at the entrance of a leaf/petal to avoid the problem of Eq.~\eqref{gauge_fat_J}. One can also imagine another UV-completion: if it is possible to generate graph-like 6D manifolds in superstring theories, the 4D SM fields may live in the worldvolume of a D-brane stack at the regularized $J/V$ vertex. 

It is interesting to notice that if one localizes the SM fields on the $B_i$-branes of a star extra dimension, one has $N$ copies of the SM fields, if one does not want to break explicitly the $\Sigma_N$ symmetry of $\mathcal{S}_N$. Then, if one softly breaks this symmetry only by the mass term of the Higgs field, one can hope to be able to realize the $N$-naturalness idea proposed in Ref.~\cite{Arkani-Hamed:2016rle}, with the reheaton (the field which populates the Universe after inflation by decaying into SM particles and possibly other fields) as a bulk field. The Higgs mass parameter is
\begin{equation}
|\mu_H| \sim \dfrac{\Lambda_P^{(5)}}{N} \, ,
\end{equation}
where a first $1/\sqrt{N}$ factor comes from the effect of the $N$ SM fields copies coupled to gravity (c.f.~\eqref{N-species}), and a second one comes from the uniform distribution of the parameters $\mu_H^2$ discussed in Ref.~\cite{Arkani-Hamed:2016rle}. Although there are $N$ copies of the SM fields, Ref.~\cite{Arkani-Hamed:2016rle} gives two explicit models where the reheaton decays preferencially into our SM sector, which allows a large number of SM sectors subject to the constraints from cosmology. Applied to our setup, it is thus possible to have a higher $\Lambda_P^{(5)}$ with no new physics at the TeV scale, which could explain the null result from the search for beyond SM particles at the Large Hadron Collider (LHC, $\sqrt{s}=13$~TeV). This idea needs to be investigated deeper in the future. Here, we will consider only one copy of the SM fields, and localize them on the $J/V$-brane. The gauge hierarchy is generated by the volume of the compactified extra dimension, with strongly coupled quantum gravity effects accessible to the LHC ($\sqrt{s}=14$~TeV) or a future hadronic collider.

\subsection{Phenomenology}
\label{pheno_star_rose_graviton}

\subsubsection{Kaluza-Klein Gravitons}
\label{KK-graviton}
In an ADD model, gravity and possibly other exotic fields propagate into the extra dimensions. It is crucial for our proposition to have an idea of the implication of gravitons propagating into the bulk. The KK dimensional reduction of a 5D graviton \cite{Csaki:2004ay} leads to a tower of KK-gravitons with a zero mode, and one massless graviscalar (the radion). A massless graviphoton is also present if there is no boundary for the extra dimension (present for the $N$-rose $\mathcal{R}_N$ and absent for the $N$-star $\mathcal{S}_N$). As the existence of KK-gravitons can have important phenomenological effects, one has to extract their KK mass spectrum and their couplings to the SM fields. By a suitable gauge choice, the Euler-Lagrange equations for a 5D massless graviton reduce to Klein-Gordon equations. One can thus study a 5D massless real scalar field coupled minimaly to the energy momentum tensor of the SM to obtain the KK mass spectrum and the couplings of spinless KK-gravitons. This 5D spinless graviton $\Phi$ couples to the energy momentum sources through the effective metric:
\begin{equation}
g_{\mu \nu} = \left( 1 + \dfrac{\Phi}{2\left[ \Lambda_P^{(5)} \right]^{3/2}} \right) \eta_{\mu \nu} \, .
\end{equation}
The metric $g_{\mu \nu}$ and thus $\Phi$ have to be continuous at the junction. We focus on the compactification on $\mathcal{S}_N$. The case of $\mathcal{R}_N$ is very similar. The coupling of the 5D spinless graviton to the energy momentum tensor 
\begin{equation}
T^{\mu \nu} = \left. \dfrac{2}{\sqrt{|g|}} \dfrac{\delta S_{SM}}{\delta g_{\mu \nu}} \right|_{g_{\mu\nu} = \eta_{\mu\nu}}
\end{equation}
of the 4D SM fields (of action $S_{SM}$) localized on the $J$-brane is
\begin{equation}
\left(\dfrac{1}{2\left[ \Lambda_P^{(5)} \right]^{3/2}} \, \widetilde{\Phi} \, T^\mu_\mu \, \delta_J\right)[1] = \int d^4x \sum_{i=1}^N \ \dfrac{1}{2N\left[ \Lambda_P^{(5)} \right]^{3/2}} \,  \Phi (x^\mu, 0, i) \, T^\mu_\mu \, .
\label{int_Phi-SM_star_1}
\end{equation}
One can use the KK decomposition of Subsection~\ref{KK_scalar_20} with $M_\Phi = 0$ and treat the brane-localized interactions with the SM fields as a perturbation. The zero mode is identified with the 4D massless graviton. We note $n_*$ the number of KK-modes which couple to the SM fields below the cut-off $\Lambda_P^{(5)}$. Only the KK-gravitons with a wave function which does not vanish on the $J$-brane couple to the SM fields. The interaction term \eqref{int_Phi-SM_star_1} gives
\begin{equation}
\int d^4x \left[ \dfrac{1}{2\Lambda_P^{(4)}} \, \phi^{(0, \, 0, \, 1)} (x^\mu) \, T^\mu_\mu + \dfrac{\sqrt{2}}{2\Lambda_P^{(4)}} \, \sum_{n_2 = 1}^{n_*} (-1)^{n_2} \, \phi^{(2, \, n_2, \, 1)}(x^\mu) \, T^\mu_\mu \right] \, ,
\label{int_Phi-SM_star_2}
\end{equation}
where
\begin{equation}
n_* \sim \dfrac{\Lambda_P^{(5)} \, \ell}{\pi} \, .
\label{n_max}
\end{equation}
The KK modes whose wave functions do not vanish at $y=0$ couple individually to the energy momentum tensor of the SM with a coupling suppressed by $\Lambda_P^{(4)}$: they are thus very feebly coupled, and the probability $P_1$ to emit a single KK-graviton is proportional to its coupling squared:
\begin{equation}
P_1 \propto \left[ \dfrac{E}{\Lambda_P^{(4)}} \right]^2 \, ,
\label{P_1}
\end{equation}
where $E$ is the energy of matter originating from $T_\mu^\mu$ in Eq.~\eqref{int_Phi-SM_star_2}. We compare two benchmark scenarii with $\Lambda_P^{(5)} \simeq 1$ TeV:

\paragraph{Benchmark scenario \#1 -- $N = 1$.}
This case is the traditional situation of ADD models in the literature with only one extra dimension. From Eq.~\eqref{ADD_formula_star}, we have $M_{KK} = 1/\ell \sim \mathcal{O}(10^{-18})$ eV, which is excluded by the success of 4D gravitational Newton's law at the scale of the solar system. Eqs.~\eqref{ADD_formula_star} and \eqref{n_max} give
\begin{equation}
n_* \sim \dfrac{1}{\pi} \left[ \dfrac{\Lambda_P^{(4)}}{\Lambda_P^{(5)}} \right]^2 \sim 10^{30} \, ,
\end{equation}
so we have a large number of KK-gravitons below the cut-off. At colliders with a center of mass energy which reaches $\Lambda_P^{(5)}$, the probability to produce one out of $n_*$ gravitons becomes then
\begin{equation}
P_* = n_* \, P_1 \propto \left[ \dfrac{E}{\Lambda_P^{(5)}} \right]^2 \, .
\label{P*ADD}
\end{equation}
where we used Eqs.~\eqref{ADD_formula_star} , \eqref{n_max} and \eqref{P_1}.
This last result is also valid in more realistic models with more than one extra dimension which can pass with success the submillimeter tests of 4D gravitational Newton's law. The KK tower can thus be probed and constrained at the LHC ($\sqrt{s}=13$~TeV), c.f. Ref.~\cite{Pomarol:2018oca}.

\paragraph{Benchmark scenario \#2 -- $N \simeq 6 \times 10^{29}$.} In this case, the large volume in Eq.~\eqref{ADD_formula_star} is generated by a large $N$ and $M_{KK} = 1/\ell \simeq 100$ GeV. Thus there are few KK modes which couple to the SM fields: $n_* \simeq 3$ from Eq.~\eqref{n_max}, and $P_* \sim P_1$ at the LHC. So the KK tower is completely invisible in current experiments. The compactification on $\mathcal{S}_N$ can thus circumvent the current LHC constraints on the KK-gravitons of traditionnal ADD models. 

However, these results follow from the zero-thickness brane hypothesis. How are they modified by a brane width in the UV completion? Indeed, we have already discussed that one expects that the singular behavior of the junction is soften in a UV-completion including gravity. After integrating out the UV degrees of freedom, one is left with an effective brane form factor as in Ref.~\cite{Kiritsis:2001bc} to modelize the brane width\footnote{We stress that this effective brane form factor has nothing to do with the wave function of the zero mode of quasi-localized 5D fields on the $J$-brane but is related to the UV description of the brane. We have already discussed that the brane-localized SM fields are 4D degrees of freedom in the EFT.}. It is a function $\mathcal{B}_J(y)$ rapidly decreasing over a distance $\ell_P^{(5)}$ and normalized such that
\begin{equation}
\sum_{i=1}^N \int_0^\ell dy \ \mathcal{B}_J(y) = 1 \, .
\end{equation}
One can perform a moment expansion of $\mathcal{B}_J(y)=\Lambda_P^{(5)} b \left( \Lambda_P^{(5)} y \right)$, where $b(y)$ is an intermediate function defined for convenience, such that
\begin{equation}
\widetilde{\mathcal{B}_J} =  \sum_{n=0}^{+ \infty} \dfrac{b_n}{\left[\Lambda_P^{(5)}\right]^{n}} \, \partial_y^n \delta_J \, ,
\label{asymp_exp}
\end{equation}
with
\begin{equation}
b_n = \dfrac{(-1)^n}{n!} \int_0^\ell dy \ y^n \, b(y) \, .
\end{equation}
The action describing the interaction between the spinless graviton and the SM fields is
\begin{align}
&\int d^4x \left( \dfrac{1}{2 \left[ \Lambda_P^{(5)} \right]^{n + 3/2}} \, \widetilde{\Phi} \, T^\mu_\mu \, \widetilde{\mathcal{B}_J} \right) [\mathbf{1}] \nonumber \\
&=\int d^4x \ \left(\sum_{n=1}^{+\infty}\dfrac{b_n}{2 \left[ \Lambda_P^{(5)} \right]^{n + 3/2}} \, \widetilde{\Phi} \, T^\mu_\mu \, \partial_y^n \delta_J\right)[\mathbf{1}] \nonumber \\
&= \int d^4x \sum_{n=1}^{+\infty} \sum_{i=1}^N \ \dfrac{(-1)^n \, b_n}{2N\left[ \Lambda_P^{(5)} \right]^{n+3/2}} \, \partial_y^n \Phi (x^\mu, 0, i) \, T^\mu_\mu \, .
\end{align}
One can naively think that the large number of KK-gravitons which do not couple to the SM fields through the operator \eqref{int_Phi-SM_star_1} will have non-vanishing couplings to the SM via the higher-dimensional operators. Then, one expects that $P_*$ is less suppressed than in Eq.~\eqref{P_1}. However, this is not the case. Indeed, by using the equations for the wave functions \eqref{wave_eq_Phi_star}, one can show that
\begin{equation}
\forall l \geq 1 \left\{
\begin{array}{l c l}
\partial_y^{2l} f_\phi^{(b, \, n_b, \, d_b)} (0, i) &=& (-1)^l \, \left[ k_\phi^{(b, \, n_b)} \right]^{2l} f_\phi^{(b, \, n_b, \, d_b)} (0, i) \, , \\ \\
\partial_y^{2l+1} f_\phi^{(b, \, n_b, \, d_b)} (0, i) &=& (-1)^l \, \left[ k_\phi^{(b, \, n_b)} \right]^{2l} \partial_y f_\phi^{(b, \, n_b, \, d_b)} (0, i) \, .
\end{array}
\right.
\end{equation}
Therefore, for $n$ even, again only the tower $b=2$ contributes, with an extra suppression factor $\left[ M_{KK}/\Lambda_P^{(5)} \right]^n$. For $n$ odd, the Neumann-Kirchhoff junction conditions \eqref{JC_Phi_star} imply that these operators vanish. We conclude that even when we take into account the brane width in the UV, the KK-graviton towers are still invisible at the LHC ($\sqrt{s}=14$~TeV). The KK-gravitons with $b \neq 2$ constitute a hidden sector.

One notices also that this important feature of the $N$-star compactification is valid only if the SM fields are localized on the $J$-brane. If one localizes them on one of the $B_i$-branes instead, they couple also to the KK-gravitons whose wave functions vanish at the $J$-brane. One can easily show that this crucial difference implies that $P_*$ is of the same form as in Eq.~\eqref{P*ADD}, as in the case of standard ADD models, and one will be able to constrain this scenario at the LHC.

\subsubsection{Ultraviolet Gravitational Objects}
Black holes are expected to appear near the cut-off scale $\Lambda_P^{(5)}$, when the coupling to 5D gravitons becomes non-perturbative. However, in the case of the benchmark scenario \#2, we saw that the coupling of the energy-momentum tensor of the SM to the linear superposition of KK-gravitons is suppressed by $\Lambda_P^{(4)}$ instead of $\Lambda_P^{(5)}$, so one expects that the couplings of the brane-localized SM fields to the tower of KK-gravitons remains perturbative well above $\Lambda_P^{(5)}$, questioning the possibility of producing black holes in trans-Planckian collisions of SM particles. However, once the linear superposition of KK-gravitons with a trans-Planckian energy leaves the $J/V$-brane, where it was perturbatively produced through SM fields in a trans-Planckian collision, it will interact with all the KK-gravitons, including those whose wave functions vanish on the $J/V$-brane. This last process is non-perturbative above $\Lambda_P^{(5)}$ and will produce a black hole. Near this threshold, the black holes are dominated by quantum corrections, we speak about Quantum Black Holes (QBHs) \cite{Rizzo:2006zb, Alberghi:2006km, Meade:2007sz, Casadio:2008qy, Calmet:2008dg, Gingrich:2009hj, Gingrich:2010ed, Dvali:2010gv, Nicolini:2011nz, Mureika:2011hg, Calmet:2012fv, Kilic:2012wp, Belyaev:2014ljc, Arsene:2016kvf} which need a complete theory of quantum gravity to be described. Besides, the lightest QBH, the Planckion \cite{Treder:1985kb, Dvali:2016ovn}, is the last stage of the evaporation of a semi-classical black hole by Hawking radiation. In some models, this black hole remnant \cite{Koch:2005ks, Dvali:2010gv, Bellagamba:2012wz, Alberghi:2013hca} is stable and one can speculate that it can constitute a part of dark matter \cite{Conley:2006jg, Dvali:2010gv, Nakama:2018lwy}. There are also a large number of KK-gravitons below the TeV scale whose wave functions vanish on the $J/V$-brane, where the SM fields are localized (c.f. Subsection~\ref{KK_scalar_20}): these KK-gravitons interact only with gravity in the bulk, and constitute a natural hidden sector which could be populated by black hole evaporation during the early Universe.


\section{Toy Model of Small Dirac Neutrino Masses}
\label{Dirac_Neutrinos}

\subsection{Zero Mode Approximation}
It is known from Refs.~\cite{Dienes:1998sb, ArkaniHamed:1998vp, Dvali:1999cn} that if the left-handed neutrinos, localized on the SM brane, interact with gauge singlet neutrinos, propagating in the bulk in the form of an internal torus $\left( \mathcal{R}_1 \right)^q$ of radius $R$ and large volume $\mathcal{V}_q$, one can get small Dirac masses for the neutrinos. With one left-handed neutrino and one gauge singlet neutrino (without BLKTs), the Dirac mass is \cite{ArkaniHamed:1998vp}:
\begin{equation}
m_\nu \simeq \dfrac{\left| y_\nu^{(4+q)} \right| v}{\sqrt{2 \mathcal{V}_q}} \, ,
\ \ \ \mathcal{V}_q = (2 \pi R)^q \, ,
\label{m_nu_1}
\end{equation}
where $v$ is the SM Higgs field VEV, and $y_\nu^{(4+q)}$ is the $(4+q)$D Yukawa coupling of mass dimension $-q/2$. Eq.~\eqref{m_nu_1} is valid if one can use the zero mode approximation, i.e. neglect the mixing between the zero mode and the KK-excitations of the bulk gauge singlet neutrino:
\begin{equation}
\dfrac{\left| y_\nu^{(4+q)} \right| v}{\sqrt{2\mathcal{V}_q}} \ll M_{KK} \equiv \dfrac{1}{R} \, .
\end{equation}
For a natural value
\begin{equation}
\left| y_\nu^{(4+q)} \right| \sim \left[ \ell_P^{(4+q)} \right]^{q/2} \, ,
\end{equation}
with $\Lambda_P^{(4+q)} \sim \mathcal{O}(1) \ \rm{TeV}$, one has, with Eq.~\eqref{ADD_formula},
\begin{equation}
m_\nu \sim \dfrac{v}{\sqrt{2\mathcal{V}_q \left[\Lambda_P^{(4+q)}\right]^q}} = \dfrac{v \Lambda_P^{(4+q)}}{\sqrt{2}\Lambda_P^{(4)}} \sim \mathcal{O}(0.1) \ \rm{meV} \, ,
\end{equation}
which is a good order of magnitude for the neutrino masses.

We want to see if it is possible to build such a model for a LED compactified on the metric graph $\mathcal{K}_N$. For the compactification on a star/rose graph, one takes a 4D left-handed neutrino $\nu_L$ of mass dimension 3/2 localized on the $J/V$-brane, and a 5D gauge singlet neutrino $\Psi$ of mass dimension 2 propagating into the bulk. The action of the model is
\begin{align}
S_{\nu} &= S_\Psi + \int d^4x \left( \widetilde{\mathcal{L}_\nu} + \widetilde{\mathcal{L}_{\Psi \nu}} \right) \delta_{J/V}[\mathbf{1}] \, , \nonumber \\
&= S_\Psi + \int d^4x \left( \mathcal{L}_\nu + \left. \widetilde{\mathcal{L}_{\Psi \nu}} \right|_{y=0} \right) \, .
\label{S_nu}
\end{align}
The free action $S_\Psi$ is given by Eq.~\eqref{S_Psi_star}, and
\begin{equation}
\mathcal{L}_\nu = \dfrac{i}{2} \, \nu_L^\dagger \bar{\sigma}^\mu \overleftrightarrow{\partial_\mu} \nu_L \, ,
\end{equation}
The brane-localized mass term is
\begin{equation}
\widetilde{\mathcal{L}_{\psi \nu}} = - \dfrac{y_\nu^{(5)} v}{\sqrt{2}} \, \nu_L^\dagger \widetilde{\Psi_R} + \rm{H.c.} \, ,
\label{nu_Yuk}
\end{equation}
where $y_\nu^{(5)}$ can be taken real since a phase shift of the Yukawa coupling can be compensated by a phase shift of the field $\nu_L$. We have imposed that the leptonic number $L$ is conserved, so $U(1)_L$ is a symmetry of the model: in this way, bulk and brane-localized Majorana mass terms for the neutrino fields are not allowed. We have also assumed the absence of a bulk Dirac mass term to simplify the discussion. By adopting a perturbative approach, where $\mathcal{L}_{\psi \nu}$ is treated as a perturbation, one can perform the KK dimensional reduction of Section~\ref{Dirac_field}. In the regime where we can use the zero mode approximation, i.e. when
\begin{equation}
\dfrac{y_\nu^{(5)} v}{\sqrt{2L}} \ll M_{KK} \equiv \dfrac{1}{\ell} \, ,
\end{equation}
we get a mass term for the zero mode neutrino:
\begin{equation}
-m_\nu \, \nu_L^\dagger \psi_R^{(0, \, 0, \, 1)} + \rm{H.c.} \, ,
\end{equation}
with
\begin{equation}
m_\nu = \dfrac{y_\nu^{(5)} v}{\sqrt{2}} \, f_R^{(0, \, 0, \, 1)}(0)
= \dfrac{y_\nu^{(5)} v}{\sqrt{2L}} \, .
\label{m_nu_2}
\end{equation}
For a natural value
\begin{equation}
y_\nu^{(5)} \sim \sqrt{\ell_P^{(5)}} \, ,
\end{equation}
with $\Lambda_P^{(5)} \simeq 1 \ \rm{TeV}$, one has, from Eqs.~\eqref{ADD_formula_star}, \eqref{m_nu_1} and \eqref{m_nu_2},
\begin{equation}
m_\nu \sim \dfrac{v}{\sqrt{2L \Lambda_P^{(5)}}} = \dfrac{v \Lambda_P^{(5)}}{\sqrt{2}\Lambda_P^{(4)}} \sim \mathcal{O}(0.1) \ \rm{meV} \, .
\end{equation}
As $m_\nu \ll M_{KK} \equiv 1/\ell$ for the benchmark scenario \#2, the zero mode approximation is thus justified.

However, within the perturbative approach, we find that the $N$ left-handed zero modes in Section~\ref{Dirac_field} for the $N$-rose $\mathcal{R}_N$ do not get masses from the brane-localized mass term \eqref{nu_Yuk}. They remain massless and do not mix with the left-handed neutrino localized on the $V$-brane: they are sterile neutrinos which do not participate in neutrino oscillations. However, they are coupled to gravity and may have an impact on the cosmological history. As our model requires a large $N$, it appears to be ruled out by cosmological constraints which are sensible to the number of light fermionic degrees of freedom. Even with a brane-localized reheaton, which does not couple to the modes with discontinuous wave functions like the $N$ left-handed zero modes, mini-black hole evaporation should produce them in the early Universe. A solution could be to add a new ingredient to the model to give a mass to these $N$ zero modes. A priori, our toy model is thus interesting only for the compactification on a star graph.

\subsection{Exact Treatment}
\label{exact_treatment}

\subsubsection{Euler-Lagrange Equations \& Junction/Boundary Conditions}
In this subsection, we take the effect of the brane-localized mass term $\widetilde{\mathcal{L}_{\Psi \nu}}$ on the KK mass spectrum exactly into account with the 5D method of Ref.~\cite{Angelescu:2019viv, Nortier:2020xms}. From Hamilton's principle applied to the action $S_\nu$ \eqref{S_nu}, we get the Euler-Lagrange equations: Eq.~\eqref{Dirac_Psi_star} and
\begin{equation}
i \bar{\sigma}^\mu \partial_\mu \nu_L(x^\mu) - M \, \Psi_R(x^\mu, 0, i) = 0 \, ,
\label{nu_ELE_5D}
\end{equation}
with
\begin{equation}
M = \dfrac{y^{(5)}_\nu v}{\sqrt{2}} \, .
\end{equation}
We get also a Kirchhoff junction condition for the left-handed field on the $J/V$-brane:
\begin{equation}
\left\{
\begin{array}{rcl}
\displaystyle{\sum_{i=1}^N \Psi_L (x^\mu, 0, i) = M \, \nu_L(x^\mu)} & \text{for} & \mathcal{K}_N = \mathcal{S}_N \, ,   \\
\displaystyle{\sum_{i=1}^N \left[ \Psi_L (x^\mu, y, i) \right]_{y=0}^{\ell} = M \, \nu_L(x^\mu)} & \text{for} & \mathcal{K}_N = \mathcal{R}_N \, ,
\end{array}
\right.
\label{Kir_junc_mix_Psi}
\end{equation}
and Dirichlet boundary conditions \eqref{D_psi_B} for the left-handed field on the $B_i$-branes.

\subsubsection{Separation of Variables}
We want to solve the field equations by separation of variables and sum over all linearly independant solutions. We thus write the KK decomposition \eqref{KK_Psi_star_1} and expand $\nu_L$ as a linear superposition of the left-handed KK modes which are mass eigenstates:
\begin{equation}
\nu_L(x^\mu) = \sum_{b} \ \sum_{n_b} \sum_{d_b} a^{(b, \, n_b, \, d_b)} \, \psi_L^{(b, \, n_b, \, d_b)} \left( x^\mu \right) \, ,
\label{nu_L_sep}
\end{equation}
with $a^{(b, \, n_b, \, d_b)} \in \mathbb{C}$. Indeed, the brane-localized mass term induces a mixing between the field $\nu_L$ and the KK modes of $\Psi_L$ obtained in Section~\ref{Dirac_field}. Here, we expand the fields $\nu_L$ and $\Psi_L$ in the same KK basis spanned by the $\psi_L^{(b, \, n_b, \, d_b)}$'s (the basis of the mass eigenstates). The reader can follow the discussion between Eqs.~\eqref{KK_Psi_star_1} and \eqref{orthonorm_KK-Phi_star}, we will use the same notations but we stress that, in Section~\ref{Dirac_field}, $\psi_L^{(b, \, n_b, \, d_b)}$ is an element of the KK basis without brane-localized mass term but here it is an element of the KK basis including the effect of the brane-localized mass term. Besides, the orthonormalization conditions \eqref{orthonorm_KK-Phi_star} for the functions $f_{L/R}^{(b, \, n_b, \, d_b)} \neq 0$ are replaced by
\begin{equation}
\left\{
\begin{array}{rcl}
\displaystyle{\left[a^{(b, \, n_b, \, d_b)}\right]^* a^{(b', \, n'_{b'}, \, d'_{b'})} + \sum_{i=1}^N \int_0^\ell dy \ \left[ f_{L}^{(b, \, n_b, \, d_b)}(y, i) \right]^* \, f_{L}^{(b', \, n'_{b'}, \, d'_{b'})}(y, i)} &=& \delta^{bb'} \, \delta^{n_{b} n'_{b'}} \, \delta^{d_{b} d'_{b'}} \, , \\
\displaystyle{\sum_{i=1}^N \int_0^\ell dy \ \left[ f_{R}^{(b, \, n_b, \, d_b)}(y, i) \right]^* \, f_{R}^{(b', \, n'_{b'}, \, d'_{b'})}(y, i)} &=& \delta^{bb'} \, \delta^{n_{b} n'_{b'}} \, \delta^{d_{b} d'_{b'}} \, .
\end{array}
\right.
\label{norm_wave_Psi_star_nu}
\end{equation}
The conditions on the 5D fields $\Psi_{L/R}$ become conditions on the KK wave functions $f_{L/R}^{(b, \, n_b, \, d_b)}$ by using Eq.~\eqref{KK_Psi_star_1}. There is a new Kirchhoff junction condition on the $J/V$-brane from Eq.~\eqref{Kir_junc_mix_Psi}:
\begin{equation}
\left\{
\begin{array}{rcl}
\displaystyle{\sum_{i=1}^N f_L^{(b, \, n_b, \, d_b)} (0, i) = a^{(b, \, n_b, \, d_b)} \, M} & \text{for} & \mathcal{K}_N = \mathcal{S}_N \, ,   \\
\displaystyle{\sum_{i=1}^N \left[ f_L^{(b, \, n_b, \, d_b)} (y, i) \right]_{y=0}^{\ell} = a^{(b, \, n_b, \, d_b)} \, M } & \text{for} & \mathcal{K}_N = \mathcal{R}_N \, .
\end{array}
\right.
\label{JC_nu_Psi_wave_1}
\end{equation}
Moreover, Eq.~\eqref{nu_ELE_5D}, with Eqs.~\eqref{Dirac_KK-Psi_star}, \eqref{KK_Psi_star_1} and \eqref{nu_L_sep}, gives:
\begin{equation}
a^{(b, \, n_b, \, d_b)} \, m_\psi^{(b, \, n_b)} = M \, f_R^{(b, \, n_b, \, d_b)}(0, i) \, .
\label{eq_a}
\end{equation}
For $m_\psi^{(b, \, n_b)} \neq 0$, Eqs.~\eqref{JC_nu_Psi_wave_1} and \eqref{eq_a} together lead to:
\begin{equation}
\left\{
\begin{array}{rcl}
\displaystyle{\sum_{i=1}^N f_L^{(b, \, n_b, \, d_b)} (0, i) = \dfrac{M^2}{m_\psi^{(b, \, n_b)}} \, f_R^{(b, \, n_b, \, d_b)}(0, i)} & \text{for} & \mathcal{K}_N = \mathcal{S}_N \, ,   \\
\displaystyle{\sum_{i=1}^N \left[ f_L^{(b, \, n_b, \, d_b)} (y, i) \right]_{y=0}^{\ell} = \dfrac{M^2}{m_\psi^{(b, \, n_b)}} \, f_R^{(b, \, n_b, \, d_b)}(0, i) } & \text{for} & \mathcal{K}_N = \mathcal{R}_N \, .
\end{array}
\right.
\label{JC_nu_Psi_wave_2}
\end{equation}

\subsubsection{Kaluza-Klein Mode Analysis on the Star Graph}
We give here only the KK mode analysis of the $N$-star $\mathcal{S}_N$, since the $N$-rose $\mathcal{R}_N$ compactification should be incompatible with cosmology without additional assumptions. For completeness, we give the KK-mode analysis on $\mathcal{R}_N$ in Appendix~\ref{neutrino_rose_app}.

\vspace{1cm}

There are no zero modes ($b=0$, $n_0=0$, $m_\psi^{(0, \, 0)}=0$) with $M \neq 0$ for $\mathcal{K}_N = \mathcal{S}_N$. Let us look at massive KK modes ($m_\psi^{(b, \, n_b)} \neq 0$). The coupled first order differential equations \eqref{wave_eq_Psi_star} can be decoupled into second order ones: Eq.~\eqref{eq_Psi_wave_2nd}. The KK wave functions $f_{R}^{(b, \, n_b, \, d_b)}$ are still continuous across the junction. We will use the same method as in Subsections~\ref{KK_scalar_20} and \ref{KK_Drac_20}. We give the results in what follows.

\subsubsection*{\boldmath \textcolor{black}{\textit{First case: $f_R^{(b, \, n_b, \, d_b)}(0,i)=0$}}}
The results are identical to the ones in Subsection~\ref{excited_modes_fermion}, Paragraph ``First case: $f_R^{(b, \, n_b, \, d_b)}(0, i)=0$'' of a) p.~\pageref{1st_case_psi_star}. This condition gives the mass spectrum \eqref{mass_spect_Psi_star_2} which defines the KK tower $b=1$. We have $a^{(1, \, n_1, \, d_1)} = 0$ from Eq.~\eqref{eq_a} so the left-handed modes do not mix with $\nu_L$: they are completely sterile, interact only with gravity, and are thus part of the hidden sector of the model.

\subsubsection*{\boldmath \textcolor{black}{\textit{Second case: $f_R^{(b, \, n_b, \, d_b)}(0,i) \neq 0$}}}
The KK mass spectrum is given by the transcendental equation:
\begin{equation}
m_\psi^{(2, \, n_2, \, d_2)} \, \tan \left[ m_\psi^{(2, \, n_2)} \, \ell \right] = \dfrac{M^2}{N} \, , \ \ \ n_2 \in \mathbb{N} \, ,
\end{equation}
whose solutions $m_\psi^{(2, \, n_2)}$ define the KK tower $b=2$ and are not degenerate ($d_2 \in \{1\}$). We have $a^{(2, \, n_2, \, 1)} \neq 0$ from Eq.~\eqref{eq_a} so the left-handed modes mix with $\nu_L$. The lightest mode $(2, 0, 1)$ is identified with the neutrino we observe in Nature\footnote{Of course, we observe three generations of neutrinos in Nature and here we consider a toy model with only one generation.}. In the decoupling limit $\ell \rightarrow 0$, the mass of this mode is given by Eq.~\eqref{m_nu_2} in the zero mode approximation. Indeed, in this limit the excited KK modes decouple and their mixing with the lightest massive mode $(2, 0, 1)$ goes to zero. The KK wave functions are
\begin{equation}
\left\{
\begin{array}{rcl}
f_L^{(2, \, n_2, \, 1)} (y, i) &=& - \left[ \dfrac{N\ell}{2} + \dfrac{M^2(2N-1)}{2 \left(\left[m_\psi^{(2, \, n_2)}\right]^2 + \left[ \dfrac{M^2}{N} \right]^2 \right)} \right]^{-1/2} \, \sin \left[ m_\psi^{(2, \, n_2)} \, (y-\ell) \right] \, , \\ \\
f_R^{(2, \, n_2, \, 1)} (y, i) &=& \left[ \dfrac{N\ell}{2} + \dfrac{M^2(2N-1)}{2 \left(\left[m_\psi^{(2, \, n_2)}\right]^2 + \left[ \dfrac{M^2}{N} \right]^2 \right)} \right]^{-1/2} \, \cos \left[ m_\psi^{(2, \, n_2)} \, (y-\ell) \right] \, ,
\end{array}
\right.
\end{equation}
where the $f_L^{(2, \, n_2, \, 1)}$'s are discontinuous across the $J$-brane (except for $\mathcal{K}_N = \mathcal{S}_1$, the interval, where they are taken continuous). This discontinuity is sourced by the brane-localized interaction.

\vspace{1cm}

In a nutshell, the massive KK modes have still a mass gap of order $1/\ell$. Only the KK modes, whose right-handed Weyl spinors have non-vanishing wave functions at the junction without brane-localized Yukawa couplings (c.f. Subsection~\ref{KK_Drac_20}), are affected by the addition of the brane-localized SM left-handed neutrino $\nu_L$ and Higgs field. The KK masses are shifted and the wave functions of the left-handed Weyl spinors become discontinuous at the junction\footnote{In the literature, it is known that an interaction localized on a brane away from a boundary or at a fixed point of the orbifolds $\mathcal{R}_1/\mathbb{Z}_2$ and $\mathcal{R}_1/(\mathbb{Z}_2 \times \mathbb{Z}_2')$ implies a discontinuity for a 5D fermion field \cite{Bagger:2001qi, Csaki:2003sh, Casagrande:2008hr, Casagrande:2010si, Carena:2012fk, Malm:2013jia, Barcelo:2014kha}.}. This result is easy to understand when the Yukawa interaction with the VEV of the Higgs field is treated as a perturbation: only non-vanishing wave functions at the junction in Subsection~\ref{KK_Drac_20} will have non-vanishing matrix elements. These modes mix with the SM left-handed neutrino. The other ones are completely sterile and interact only through gravity.

\section{Conclusion \& Perspectives}
\label{conclusion_star_rose}
In this work, we have studied the possibily of compactifying a large spacelike extra dimension on a star/rose graph with identical leaves/petals. In Section~\ref{spacetime_geom}, we have adapted Kurasov's distribution theory to a star/rose graph. In this way, we have defined a rigorous framework to build a field theory on these geometries.

In Sections~\ref{KG_field} and \ref{Dirac_field}, we have worked out the KK decomposition of a Klein-Gordon and Dirac field respectively. Our main contributions, compared to the previous articles \cite{Kim:2005aa, Fujimoto:2019fzb}, are discussions concerning the different possibilities for the continuity of the fields at the junction and the impact on the KK-spectrum. In particular, we have pointed out the case of an airtight brane (when the off-shell fields are allowed to be discontinuous), which is equivalent to $N$ disconnected bonds. Moreover, we have studied for the first time the KK-modes of a massless 5D Dirac fermion propagating into the whole star/rose graph. We have also discussed the chirality of the zero modes. For both bosonic and fermionic massless fields, the KK scale is given by the inverse of the leaf/petal length/circumference.

One can realize a large compactified volume with a high KK scale for a large number of small leaves/petals. This possibility has been investigated in Section~\ref{ADD_star_rose} in order to lower the gravity scale to the TeV scale and solve the gauge hierarchy problem under the hypothesis of an exact global $\Sigma_N$ symmetry. Moreover, we have shown that if the SM fields are localized on the 3-brane at the junction of the star/rose graph, they couple only to few modes of the whole tower of KK-gravitons, even when a UV brane thickness is taken into account. The couplings of the SM fields to this KK tower are suppressed by the large 4D Planck scale instead of the 5D one at the TeV scale: the KK-gravitons are thus completely invisible at the LHC ($\sqrt{s}=14$~TeV). This result is in sharp contrast to standard ADD models in the literature with compactification on a torus or its orbifolds, where the SM fields couple to the whole tower of KK-gravitons, which implies couplings suppressed only by the 5D Planck scale (near a TeV) which one can constrain at the LHC. 

Besides KK-gravitons, our proposition can still be probed at hadronic colliders through the search for strongly coupled phenomena induced by gravity like QBHs at the LHC ($\sqrt{s}=14$~TeV) or semi-classical black holes at the Future Circular Collider proton-proton (FCC pp, $\sqrt{s}=100$~TeV). The absence of a theory of Planckian gravity renders difficult to make precise predictions concerning the production and decay of QBHs or other exotic states near the Planck scale. It is thus delicate to translate the LHC data ($\sqrt{s}=13$~TeV) into constraints on the 5D Planck scale of our model and to estimate the degree of fine tuning which remains to accomodate an EW scale at 100 GeV.

Finally, in Section~\ref{Dirac_Neutrinos} we have proposed to realize in our scenario a toy model of small Dirac neutrino masses. For that purpose, we have considered only one generation of neutrinos coupled to one gauge singlet fermion in the bulk. The large compactified volume suppresses the 5D Yukawa coupling of order unity, and we have been able to reproduce the good order of magnitude for the mass of the neutrinos, with a model which accomodates also a 5D Planck scale at the TeV scale. This kind of model was discussed previously only with a toroidal/orbifold compactification, and the adaptation to a star/rose graph is new. The model is realistic only for the compactification on a star graph since the rose graph predicts a large number of massless left-handed sterile neutrinos incompatible with cosmology. Moreover, we have found that our models have a hidden sector consisting in secluded KK-gravitons and sterile KK-neutrinos possibly populated during the early Universe by the decays of mini-black holes in the bulk. The Planckion could also be a candidate to dark matter.

Our effective model (as well as the traditionnal UED models) has a global $\Sigma_N$ symmetry which acts on the geometry. One of the Swampland conjectures is the absence of global symmetries in a complete theory of quantum gravity (see Refs.~\cite{Brennan:2017rbf, Palti:2019pca} for reviews on the Swampland program). If true, there is no global $\Sigma_N$ symmetry and one has a different leaf/petal size $\ell_i$ for each $i$. One can define the average of the leaf/petal size as
\begin{equation}
\langle \ell \rangle = \dfrac{1}{N} \sum_i \ell_i \, ,
\end{equation}
such that Eq.~\eqref{ADD_formula_star} is still valid but with $L = N \langle \ell \rangle$. The mass spectrum must be studied numerically. In general, the mass scale of the lightest KK-mode is given by the inverse of the largest $\ell_i$. If the $\ell_i$'s are incommensurate (i.e. $\forall i \neq j, \ \ell_i/\ell_j \notin \mathbb{Q}$), the KK-spectrum is chaotic \cite{PhysRevLett.79.4794, Kottos_1999, Gaspard2000a, Dabaghian2001a, PhysRevLett.88.044101, kottos2005quantum, Cacciapaglia:2006tg}. Moreover, there are more KK-modes below the cut-off $\Lambda_P^{(5)}$ which couple to the $J$-brane for different $\ell_i$'s: $P_*$ is thus less suppressed than for identical $\ell_i$'s such that one is in an intermediate situation between the benchmark scenarii \#1 and \#2. If the $\ell_i$'s are incommensurate, there are no KK modes whose wavefunctions vanish at the $J$-brane and the KK-spectrum is chaotic.

In a realistic model including gravitationnal effects, instead of an exact global $\Sigma_N$ symmetry, one should consider a geometry with an approximate one. In case of GW mechanisms within each leaf, the $\ell_i$'s would depend on the mass parameter for the GW scalar field within each leaf. Identical leaf lengths (exact $\Sigma_N$ symmetry) corresponds to $\Sigma_N$ symmetric mass parameters within all leafs. The question is then in how far quantum gravity effects affect classically $\Sigma_N$ symmetric mass parameters. If these modifications remain within $\sim 10\%$ for all leafs, i.e. the effects of gravity can be considered as a perturbation of the geometry with an exact $\Sigma_N$ symmetry, the hierarchy problem could be considered as solved, with the common scale being given by $1/\langle \ell \rangle$. On the other hand, if the modified leaf lengths follow a statistical distribution (like a Gaussian) around a central value $\langle \ell \rangle$, the hierarchy problem can be considered as solved only if this distribution is extremely narrow such that the number of very light KK-gravitons remains compatible with present constraints, see subsection~\ref{KK-graviton}. In the absence of a concrete theory of quantum gravity it is impossible to answer these questions. Therefore, the toy model of this article with an exact global $\Sigma_N$ symmetry should not be considered as a viable solution of the gauge hierarchy problem but as a scenario within which (among others) a solution of the hierarchy problem may be possible.

We also want to discuss some perspectives for future investigations. As a follow-up of the present work, it would be important to study the unitarity constraints on our models, since a low gravity scale is known to need a UV completion at a scale lower than the higher-dimensional Planck scale in standard ADD models with a toroidal compactification \cite{Atkins:2010re, Antoniadis:2011bi}. It is also important to see how the mechanism to produce small neutrino masses is influenced by adding bulk and brane-localized Majorana mass terms, in the way of Ref.~\cite{Dienes:1998sb}. Moreover, strongly coupled gravity at the TeV scale may generate dangerous brane-localized higher-dimensional operators inducing proton decay, large Majorana neutrino masses and Flavor Changing Neutral Currents (FCNCs) \cite{Antoniadis:1998ig}. Without knowledge of the UV completion, we cannot know if these operators are naturally suppressed. If this is not the case, the natural value of their Wilson coefficients are suppressed only by the TeV scale and one has to add new ingredients to the scenarii to forbid them, like gauging some global symmetries of the SM as the baryon and lepton numbers \cite{Perez:2015rza} and other flavor symmetries \cite{Berezhiani:1998wt, ArkaniHamed:1998sj, ArkaniHamed:1999yy}.

Beyond the motivations for the present work, we stress that the 5D background geometries that we studied here can be used in general to generate feebly coupled interactions. Indeed, the couplings of the whole KK tower of a 5D field, coupled to SM fields localized at the junction of the star/rose graph, are in general suppressed by the square root of the compactified volume. One can easily imagine how it can be used to build consistent models of axions and dark matter with order one 5D couplings. Moreover, a 5D field has KK modes whose wave functions vanish at the junction where the SM is localized: they are thus good candidates for a hidden sector. The star graph compactification with a small number of leaves can also be used to build models of 5D SM fields as the theory is chiral at the level of zero modes for 5D fermions. Generating $N$ fermion zero modes from only one 5D fermion propagating into a star/rose graph with $N$ leaves/petals and an airtight brane  is interesting from the point of view of flavor physics. Moreover, it would be interesting to see if one can implement a 5D supersymmetric field theory or 5D supergravity on the star/rose background. In every scenario, it is important to investigate different possibilities of field theoretical mechanism to stabilize the leaf/petal length scale.

\chapter*{Conclusion}
\addcontentsline{toc}{chapter}{Conclusion}

In modern particle physics, the gauge and flavor hierarchies are one of the main motivations for believing that new mass scales, at which new degrees of freedom pop up, exist well above the TeV scale. This belief motivates model building to stabilize the EW scale $\Lambda_{EW} \sim 100$ GeV against radiative corrections. The paradigm of spacelike extra dimension offers the possibility that all these new scales, including the gravity scale, are in reality in the $1-10$ TeV range.

\vspace{1cm}

In the RS1 model, spacetime is a slice of AdS$_5$ and the warp factor redshifts the scale at which gravity becomes strongly coupled from the 4D Planck scale to the TeV scale on the IR-brane where the SM Higgs boson is localized. If gauge bosons and fermions propagate into the bulk, one can address the flavor hierarchy by quasi-localizing the fermion zero modes along the warped extra dimension. In this picture, it is important to understand the field theoretical treatment of 5D fermions coupled to a Higgs field localized on a 3-brane, which is treated in the literature with an unnecessary and incorrect brane regularization procedure. This last point is at the heart of the first part of this PhD thesis. We have shown that it is required to add BBTs to the action for the 5D fermions in order to be able to employ natural boundary conditions without overconstraining the system. These terms are used for both free and interacting 5D fields. When the brane is a boundary of the extra dimension, all 5D fields are continuous on this brane and the BBTs must be added by hand from the start. Moreover, only one of the two chiralities can have boundary localized kinetic and/or interaction terms. However, when the Higgs field is localized on a brane away from a boundary, the left or the right-handed field is discontinuous across the brane. This jump generates BBTs at both sides of the brane through the 5D kinetic term defined in distribution theory. In this framework, only the field which is continuous across the brane is allowed to have brane localized interactions with the Higgs field. These results can be extended to any kind of brane localized kinetic or interaction terms involving 5D fermions. They require to reexamine the phenomenology of warped models with a brane localized Higgs field coupled to 5D fermion and gauge fields when they were treated by the brane-Higgs regularization as in previous studies.

\vspace{1cm}

In the ADD models, one adds $q$ LEDs compactified on a $q$D torus with a factorizable geometry. The large compactified volume dilutes gravity which appears feebly coupled to an observer localized on a 3-brane. Nevertheless, gravity becomes strongly coupled at the TeV scale. Within a UV completion in superstring/M-theory, one has $q \leq 7$, and the gauge hierarchy problem is reformulated as a geometrical hierarchy problem: the compactification radii need to be stabilized at large values compared to the $(4+q)$D Planck length, which is difficult to achieve in practice. In the second part of this PhD thesis, we have proposed to compactify a single LED on a star/rose graph with a large number of identical leaves/petals of length/circumference of $\mathcal{O}(1/\Lambda_{EW})$, where the 5D Planck scale is of $\mathcal{O}(1)$ TeV (without a large geometrical hierarchy to stabilize). The 4D SM fields are localized at the central vertex of the star/rose graph. We predict a feebly coupled tower of KK-gravitons invisible in current experiments with a KK scale of $\mathcal{O}(\Lambda_{EW})$. Our scenarii lead also to TeV scale strongly coupled gravitational phenomena and to an infinite tower of trans-TeV semi-classical black holes. If there is a stable black hole remnant after evaporation, the Planckion, it could constitute a part of dark matter. Moreover, we propose a toy model to generate light Dirac neutrinos, where the right-handed neutrinos are KK modes of gauge singlet fermions propagating in the extra dimension. In our models, a large number of KK-gravitons and of sterile KK-neutrinos interact only with gravity and consititute a dark sector. This part was also the opportunity to clarify the field theory and KK decomposition of bosons and fermions propagating into an extra dimension compactified on a metric graph. For that purpose, we developped a distribution theory on these geometries. If the brane is not airtight, the bosonic fields are continuous across the vertices. For the fermions, the left and/or right-handed chirality must be continuous across the vertices. Only the continuous field can have brane localized kinetic and/or interaction terms at the vertices, which generalizes the result of the first part.

\vspace{1cm}

Compactification of an extra dimension on metric graphs is an interesting new area to build new 5D models. The traditional compactifications on a circle, an interval and their orbifolds are the smallest building blocks since one can glue them as we want in order to get any metric graph. It would be interesting to explorate the various possibilities that such new geometries allow in future works. Having KK wave functions which vanish at the vertex where SM fields are localized is a new portal for dark matter. Moreover, the compactification on a rose graph with $N$ petals allows to have $N$ zero modes for one chirality of the 5D field, and we have seen that a brane localized mass term at the vertex gives a mass to the zero mode of opposite chirality. It is thus tempting to imagine that one can generate the three generations of SM fermions from only one 5D generation. In this manuscript, we have sketched a model for small Dirac neutrino masses. It is important to investigate the possibility of adding Majorana mass terms and to take into account the mixings among the three generations of neutrinos.

\appendix

\part*{Appendices}
\addcontentsline{toc}{part}{Appendices}
\label{Appendices}

\chapter{Conventions}
\label{conventions}

We use the natural units where we have the reduced Planck constant $\hslash = 1$ and the speed of light $c=1$.

\vspace{1cm}

The flat metric is
\begin{equation}
\eta_{\mu \nu} = \text{diag}(+1, -1, -1, -1, -1).
\label{metric_1}
\end{equation}
The 4D Dirac matrices (in the local Lorentz frame of each point of the manifold) are taken in the Weyl representation
\begin{equation}
\gamma^a =
\begin{pmatrix}
0 & \sigma^a \\
\bar{\sigma}^a & 0
\end{pmatrix}
\phantom{000} \text{with} \phantom{000}
\left\{
\begin{array}{r c l}
\sigma^a &=& \left( I, \sigma^i \right), \\
\bar{\sigma}^a &=& \left( I, -\sigma^i \right).
\end{array}
\right.
\label{gamma_1}
\end{equation}
where $\left( \sigma^i \right)_{i \in\llbracket 1, 3 \rrbracket}$ are the 3 Pauli matrices. We define $\gamma^\mu = \delta^\mu_a \gamma^a$ in order to symplify the notations.
\begin{equation}
\sigma^1 =
\begin{pmatrix}
0 & 1 \\
1 & 0
\end{pmatrix},
\phantom{000}
\sigma^2 =
\begin{pmatrix}
0 & -i \\
i & 0
\end{pmatrix},
\phantom{000}
\sigma^3 =
\begin{pmatrix}
1 & 0 \\
0 & -1
\end{pmatrix},
\end{equation}
and thus a 4D Dirac spinor $\Psi$ can be decompose in its chiral components (see Ref.~\cite{Martin:1997ns}):
\begin{equation}
\Psi = \Psi_L + \Psi_R \ \ \ \text{with} \ \ \
\Psi_L \doteq
\begin{pmatrix}
\psi_L \\
0
\end{pmatrix}
=
\begin{pmatrix}
\psi_\alpha \\
0
\end{pmatrix}
\ \ \text{and} \ \ 
\Psi_R \doteq
\begin{pmatrix}
0 \\
\psi_R
\end{pmatrix}
=
\begin{pmatrix}
0 \\
\bar{\psi}^{\dagger \dot{\alpha}}
\end{pmatrix}.
\end{equation}
The spinor indices are raised and lowered with the antisymmetric symbol
\begin{equation}
\epsilon^{12} = - \epsilon^{21} = \epsilon^{21} = - \epsilon^{12} = 1 \ \ \ \text{and}\ \ \ \epsilon^{11} = \epsilon^{22} = \epsilon^{11} = \epsilon^{22} = 0,
\end{equation}
with $\epsilon_{\alpha\beta}\epsilon^{\beta\gamma} = \epsilon^{\gamma\beta}\epsilon_{\beta\alpha} = \delta_\alpha^\gamma$ and $\epsilon_{\dot{\alpha}\dot{\beta}}\epsilon^{\dot{\beta}\dot{\gamma}} = \epsilon^{\dot{\gamma}\dot{\beta}}\epsilon_{\dot{\beta}\dot{\alpha}} = \delta_{\dot{\alpha}}^{\dot{\gamma}}$. Repeated indices contracted like
\begin{equation}
\phantom{0}^{\alpha}_{\ \alpha} \ \ \ \text{or}\ \ \ \phantom{0}^{\ \dot{\alpha}}_{\dot{\alpha}}
\end{equation}
can be omitted for clarity. \newline \newline
We have also the 4D chirality operator
\begin{equation}
\gamma^5 = i \prod_{a = 0}^3 \gamma^a =
\begin{pmatrix}
-I & 0 \\
0 & I
\end{pmatrix},
\label{gamma_2}
\end{equation}
which defines the projectors on 4D chirality
\begin{equation}
P_{L, R} = \dfrac{I \mp \gamma^5}{2},
\label{chirality_projectors_1}
\end{equation}
such that for the 4D Dirac spinor $\Psi$
\begin{equation}
\left\{
\begin{array}{r c l}
\Psi_{L, R} &=& \mp \gamma^5 \, \Psi_{L, R}, \\
\bar{\Psi}_{L, R} &=& \pm \bar{\Psi}_{L, R} \, \gamma^5.
\end{array}
\right.
\end{equation}
With our conventions, the 5D Dirac matrices are
\begin{equation}
\Gamma^A = \left( \gamma^a, i \gamma^5 \right).
\label{gamma_3}
\end{equation}

\chapter{Supplement Material for 5D Models with Branes}
\label{Hamilton_Principle_5D_Field_Theory}

\section{Holography for Fermions}
\label{Holo}
In this appendix, we recall the way of performing Hamilton's principle in a holographic context following Ref~\cite{Contino:2004vy}, to show the effect of the bilinear boundary terms. We want to describe a free fermion field $F$ in a 5D bulk with the extra dimension compactified on an interval $[0,L]$. Usually, because of the AdS/CFT correspondence, the bulk is a slice of AdS$_5$, but this is not essential for our discussion, so we consider a flat extra dimension. The bulk part of the action is similar to the one in Eq.~\eqref{1_L_2},
\begin{equation}
S_\Psi = \int d^4x \int_0^L dy \, \dfrac{1}{2} \left( i F^\dagger_R \sigma^\mu \overleftrightarrow{\partial_\mu} F_R + i F^\dagger_L \bar{\sigma}^\mu \overleftrightarrow{\partial_\mu} F_L + F^\dagger_R \overleftrightarrow{\partial_y} F_L - F^\dagger_L \overleftrightarrow{\partial_y} F_R \right).
\label{L_6}
\end{equation}
In the holographic context, one has to choose a source field which has to be fixed at one boundary, and allows the remaining degrees of freedom to vary. One cannot fix simultaneously $F_L$ and $F_R$ on the brane at $y=0$ because a 5D Lagrangian for a fermion contains only first order derivatives. Here we choose $F_L$ as our source field,
\begin{equation}
F_L(x^\mu, y=0) = F_L^0 \ \ \ (\delta F_L = 0 \ \text{at $y=0$}),
\end{equation}
and $F_R$ is free to vary. Hamilton's principle for each field leads to
\begin{align}
0 = \delta_{F^\dagger_L} S_{\Psi} &= \displaystyle{ \int d^4x \, \int_0^{L} dy \,  
\delta F^\dagger_L \left[ i\bar{\sigma}^\mu \partial_\mu F_L - \partial_y F_R  \right] } 
\nonumber \\
&\displaystyle{ + \ \int d^4x \, \dfrac{1}{2} \, \left[ \left . \left( \delta F^\dagger_L \, F_R \right) \right |_{L} - \left . \left( \delta F^\dagger_L \, F_R \right) \right |_{0} \, \right] } \ ,
\label{HVP_10a}
\end{align}
\begin{align}
0 = \delta_{F^\dagger_R} S_{\Psi} &= \displaystyle{ \int d^4x \, \int_0^{L} dy \,  
\delta F^\dagger_R \left[ i\sigma^\mu \partial_\mu F_R + \partial_y F_L  \right] } 
\nonumber \\
&\displaystyle{ + \ \int d^4x \, \dfrac{1}{2} \,  \left[ - \left . \left( \delta F^\dagger_R \, F_L \right) \right |_{L} + \left . \left( \delta F^\dagger_R \, F_L \right) \right |_{0} \, \right] } \ .
\label{HVP_10b}
\end{align}
$\left. \delta F_L \right|_0 = 0$, but $\left. \delta F_R \right|_0$ is generic so the boundary variations at $y=0$ vanish only if one adds a bilinear boundary term at $y=0$,
\begin{equation}
S_B = - \int d^4x \, dy \, \delta(y) \, \dfrac{1}{2} \left( F_L^\dagger F_R + F_R^\dagger F_L \right).
\label{L_B_2}
\end{equation}
This bilinear boundary term at $y=0$ is thus essential to cancel the brane variations in the holographic context. It is important to notice that other brane terms (as brane kinetic terms for example) can be localized at $y=0$ if they are functions of $F_L$ only ($\left. \delta F_L \right|_0 = 0$ so the brane variations vanish). In the holographic approach, the value of the field at $y=0$ is determined by the equations of motion for $F_L^0 \neq 0$ after the bulk has been integrated out. The choice of $F_L$ as a source field is equivalent to a Neumann boundary condition for it (and thus a Dirichlet boundary condition for $F_R$). Brane localized terms for a field are hence only consistent with a $(+)$ boundary condition. In the same way, if $F_R$ is chosen as the source field ($F_R$ is $(+)$ and $F_L$ is $(-)$), the boundary variations vanish with a brane term as in Eq.~\eqref{L_B_2} but with the opposite sign.

Since in the holographic approach the source fields are chosen at one boundary so there is no similar motivation to introduce a bilinear boundary term at $y=L$. In the literature, the brane variations at $y=L$ vanish by imposing by hand a Dirichlet boundary condition for $F_L$ or $F_R$ at $y=L$.

\section{Note on the Continuity of the Fields on a Brane}
\label{app_cont_field}
\textit{A priori}, a 5D field can be discontinuous across a brane. It is thus important to investigate whether continuity is required by the definition of the Lagrangian in order to know if one has to take continuous or discontinuous variations of the field across the brane when applying Hamilton's principle. In order to deal with a discontinuous field, it is convenient to use distribution theory. One can associate a regular distributions $\widetilde{\Phi}(x^\mu, y) = \phi(x^\mu) \, \widetilde{f}(y)$ to a 5D field\footnote{In general, a 5D field is a sum of products of 4D fields with wave functions along the extra dimension.} $\Phi(x^\mu, y) = \phi(x^\mu) \, f(y)$. There are two cases to distinguish for the brane position:

1) If the brane is at $y = \ell$ away from a boundary of the extra dimension, the weak partial derivative with respect to $y$ is
\begin{equation}
\partial_y \widetilde{\Phi} = \left\{ \partial_y \Phi \right\} + \left( \left. \Phi \right|_{y=\ell^+} - \left. \Phi \right|_{y=\ell^-} \right) \delta(y-\ell) \, ,
\label{app_jump_formula}
\end{equation}
where $\left\{ \partial_y \Phi \right\} (x^\mu, y) = \phi(x^\mu) \, \widetilde{\partial_y f}(y)$ is the regular distribution associated to the partial derivative $\partial_y \Phi (x^\mu, y) = \phi(x^\mu) \, \partial_y f(y)$. We distinguish two cases:
\begin{itemize}
\item $f(y=\ell^-) \neq f(y=\ell^+)$ so the discontinuity implies non-vanishing brane localized terms proportional to $\delta(y-\ell)$ from Eq.~\eqref{app_jump_formula}. When one applies Hamilton's principle, one has to take discontinuous field variations allowing for $\left. \delta \Phi \right|_{y=\ell^-} \neq \left. \delta \Phi \right|_{y=\ell^+}$.
\item $f(y=\ell^\pm) \neq f(y=\ell)$ but $f(y=\ell^-) = f(y=\ell^+)$ so the term proportional to $\delta(y-\ell)$ vanishes in Eq.~\eqref{app_jump_formula}. One can define a continuous field $\Phi'(x^\mu, y) = \phi(x^\mu) \, f'(y)$ such that $f'(y \neq \ell) = f(y \neq \ell)$ and $f'(\ell) = f(\ell^\pm)$. According to distribution theory \cite{Schwartz1, Schwartz2}, $\Phi'(x^\mu, y)$ and $\Phi(x^\mu, y)$ are associated to the same distribution $\widetilde{\Phi}(x^\mu, y)$ since they are equal everywhere except on the hypersurface at $y=\ell$. From the point of view of distribution theory, $\Phi(x^\mu, y)$ and $\Phi'(x^\mu, y)$ belong to the same equivalence class so one can take a field $\Phi(x^\mu, y)$ continuous at $y=\ell$ from the beginning. In this case, when one applies Hamilton's principle, the variations of the field $\delta \Phi(x^\mu, y)$ are taken continous at $y=\ell$ too.
\end{itemize}

2) If the brane is the boundary at $y=L$ of the extra dimension, the field $\Phi(x^\mu, y) = \phi(x^\mu) \, f(y)$ can be discontinuous on this hypersurface: $f(L^-) \neq f(L)$. However, if we define the continuous field $\Phi'(x^\mu, y) = \phi(x^\mu) \, f'(y)$ such that $f'(y \neq L) = f(y \neq L)$ and $f'(L) = f(L^-)$, it belongs to the same equivalence class as $\Phi(x^\mu, y)$ from the point of vue of distribution theory, since the two fields are equal everywhere except on the boundary at $y=L$. As in the second case of 1), the weak partial derivative with respect to $y$ does not have a singular term at $y=L$:
\begin{equation}
\partial_y \widetilde{\Phi} = \left\{ \partial_y \Phi \right\} \, .
\end{equation}
One can take a field $\Phi(x^\mu, y)$ continuous at $y=L$ as starting point. In this case, when one applies Hamilton's principle, the variations of the field $\delta \Phi(x^\mu, y)$ are taken continuous at $y=L$ too.

\section{Hamilton's Principle \& Noether Theorem with Distributions}
\label{Hamilton_Noether}
In this Appendix, we want to generalize Hamilton's principle and Noether's theorem with the Lagrangian as a distribution. We have a 5D field $\phi(x^\mu, y)$ propagating into an extra dimension compactified on an interval $I=[0, L]$. We call $\mathcal{L}_{bulk} (\phi, \partial_M \phi)$, $\mathcal{L}_{0} (\phi, \partial_\mu \phi)$ and $\mathcal{L}_{L} (\phi, \partial_\mu \phi)$ the Lagrangians in the bulk, on the boundary at $y=0$ and on the boundary at $y=L$ respectively. The action is
\begin{equation}
S = \int d^4x \int_{-\infty}^{+\infty} dy \  \widetilde{\mathcal{L}} \, ,
\end{equation}
with the Lagrangian distribution 
\begin{equation}
\widetilde{\mathcal{L}} = \Theta_{I}(y) \, \mathcal{L}_{bulk} + \delta(y) \, \mathcal{L}_0 + \delta(y-L) \, \mathcal{L}_L \, ,
\end{equation}
where $\Theta_{I}(y) = \theta(y) - \theta(y-L)$, with the Heaviside distribution $\theta(y)$.

Hamilton's principle gives:
\begin{align}
0 = \delta S &= \int d^4x \int_{-\infty}^{+\infty} dy  \left\{ \delta \phi \, \dfrac{\partial \widetilde{\mathcal{L}}}{\partial \phi} + \delta (\partial_M \phi) \, \dfrac{\partial \widetilde{\mathcal{L}}}{\partial (\partial_M \phi)} \right\} \, ,
\nonumber \\
&=\int d^4x \int_{-\infty}^{+\infty} dy  \left\{ \delta \phi \left[ \dfrac{\partial \widetilde{\mathcal{L}}}{\partial \phi} - \partial_M \dfrac{\partial \widetilde{\mathcal{L}}}{\partial (\partial_M \phi)} \right] \right\} + \int d^4x \left[ \delta \phi \, \dfrac{\partial \widetilde{\mathcal{L}}}{\partial (\partial_M \phi)} \right]_{y=-\infty}^{+\infty} \, ,
\end{align}
where we have used the fact that the integral of the 4-divergence give a surface term at the infinity which vanishes. The field variations $\delta \phi$ are test functions whose supports are included into $\mathbb{R}$ so $\delta \phi (x^\mu, \pm \infty) = 0$, and we get the 5D Euler-Lagrange equations with distributions:
\begin{equation}
\dfrac{\partial \widetilde{\mathcal{L}}}{\partial \phi} - \partial_M \dfrac{\partial \widetilde{\mathcal{L}}}{\partial (\partial_M \phi)} = 0 \, .
\label{ELE_6000}
\end{equation}

Let the field transform under an infinitesimal global continuous symmetry:
\begin{equation}
\phi(x^\mu, y) \mapsto \phi(x^\mu, y) + \alpha \Delta \phi(x^\mu, y) \, .
\end{equation}
If the symmetry is internal, the Lagrangian is invariant:
\begin{equation}
\widetilde{\mathcal{L}}(x^\mu, y) \mapsto \widetilde{\mathcal{L}}(x^\mu, y)
\end{equation}
so
\begin{align}
0 &= \alpha \Delta \widetilde{\mathcal{L}} = (\alpha \Delta \phi) \, \dfrac{\partial \widetilde{\mathcal{L}}}{\partial \phi} + \partial_M (\alpha \Delta \phi) \, \dfrac{\partial \widetilde{\mathcal{L}}}{\partial (\partial_M \phi)}
\nonumber \\
&= \alpha \Delta \phi \left[ \dfrac{\partial \widetilde{\mathcal{L}}}{\partial \phi} - \partial_M \dfrac{\partial \widetilde{\mathcal{L}}}{\partial (\partial_M \phi)} \right] + \alpha \partial_M \left( \Delta \phi \, \dfrac{\partial \widetilde{\mathcal{L}}}{\partial (\partial_M \phi)} \right) \, .
\end{align}
\newpage
When the Euler-Lagrange equations \eqref{ELE_6000} are satisfied, the current 
\begin{align}
J^M &= \Delta \phi \, \dfrac{\partial \widetilde{\mathcal{L}}}{\partial (\partial_M \phi)}
\nonumber \\
&= \Theta_I(y) \, \Delta \phi \, \dfrac{\partial \mathcal{L}_{bulk}}{\partial (\partial_M \phi)}
+ \delta(y) \, \Delta \phi \, \dfrac{\partial \mathcal{L}_{0}}{\partial (\partial_\mu \phi)}
+ \delta(y-L) \, \Delta \phi \, \dfrac{\partial \mathcal{L}_{L}}{\partial (\partial_\mu \phi)}
\label{cons_current_app}
\end{align}
is conserved:
\begin{equation}
\partial_M J^M = 0\, ,
\end{equation}
and we recover Noether's theorem.

\section{Hamilton's Principle for 5D Fields on an Interval}
\label{HVP_general}
In this appendix, we formulate the Hamilton's principle for a field theory with an extra dimension compactified an interval $I=[0, L]$. We consider a set of complex fields $\phi_i(x^\mu, y)$. The bulk Lagrangian is
$\mathcal{L}_B (\phi_i, \, \partial_M \phi_i)$. In general, there are boundary localized terms so we add a brane Lagrangian at $y=0$, $\mathcal{L}_0 (\phi_i, \, \partial_\mu \phi_i)$, and at $y=\ell$, $\mathcal{L}_L (\phi_i, \, \partial_\mu \phi_i)$. The 5D action is
\begin{equation}
S = \int d^4x \int_{-\infty}^{+\infty} dy \ \left[ \Theta_I(y) \, \mathcal{L}_B + \delta(y) \, \mathcal{L}_0 + \delta(y-L) \, \mathcal{L}_L \right].
\end{equation}
The Hamilton's principle applied to $S$, varying the fields $\phi_i$, reads
\begin{align}
0 &= \delta S \nonumber \\
  &= \int d^4x \int_0^L dy \, \sum_i \left\{ \delta \phi_i \, \dfrac{\partial \mathcal{L}_B}{\partial \phi_i} + \delta(\partial_M \phi_i) \, \dfrac{\partial \mathcal{L}_B}{\partial \left( \partial_M \phi_i \right)} \right\} \nonumber \\
  &\ \ \ + \int d^4x \, \left. \sum_i \left\{ \delta \phi_i \, \dfrac{\partial \mathcal{L}_0}{\partial \phi_i} + \delta(\partial_\mu \phi_i) \, \dfrac{\partial \mathcal{L}_0}{\partial \left( \partial_\mu \phi_i \right)} \right\} \right|_{y=0} \nonumber \\
  &\ \ \ + \int d^4x \, \sum_i \left. \left\{ \delta \phi_i \, \dfrac{\partial \mathcal{L}_L}{\partial \phi_i} + \delta(\partial_\mu \phi_i) \, \dfrac{\partial \mathcal{L}_L}{\partial \left( \partial_\mu \phi_i \right)} \right\} \right|_{y=L} \nonumber \\
  &= \int d^4x \int_0^L dy \, \sum_i \left\{ \delta \phi_i \, \dfrac{\partial \mathcal{L}_B}{\partial \phi_i} - \delta \phi_i \, \partial_M \left( \dfrac{\partial \mathcal{L}_B}{\partial \left( \partial_M \phi_i \right)} \right) + \partial_y \left( \delta \phi_i \, \dfrac{\partial \mathcal{L}_0}{\partial \left( \partial_y \phi_i \right)} \right) \right\} \nonumber \\
     &\ \ \ + \int d^4x \, \sum_i \left. \left\{ \delta \phi_i \, \dfrac{\partial \mathcal{L}_0}{\partial \phi_i} - \delta \phi_i \, \partial_\mu \left( \dfrac{\partial \mathcal{L}_0}{\partial \left( \partial_\mu \phi_i \right)} \right) \right\} \right|_{y=0} \nonumber \\
     &\ \ \ + \int d^4x \, \sum_i \left. \left\{ \delta \phi_i \, \dfrac{\partial \mathcal{L}_L}{\partial \phi_i} - \delta \phi_i \, \partial_\mu \left( \dfrac{\partial \mathcal{L}_L}{\partial \left( \partial_\mu \phi_i \right)} \right) \right\} \right|_{y=L} \nonumber \\
  &= \int d^4x \int_0^L dy \, \sum_i \left\{ \delta \phi_i \left[ \dfrac{\partial \mathcal{L}_B}{\partial \phi_i} - \partial_M \left( \dfrac{\partial \mathcal{L}_B}{\partial \left( \partial_M \phi_i \right)} \right) \right] \right\} \nonumber \\
     &\ \ \ + \int d^4x \, \sum_i \left\{ \delta \phi_i \left. \left[ \dfrac{\partial \mathcal{L}_0}{\partial \phi_i} - \partial_\mu \left( \dfrac{\partial \mathcal{L}_0}{\partial \left( \partial_\mu \phi_i \right)} \right) - \dfrac{\partial \mathcal{L}_B}{\partial \left( \partial_y \phi_i \right)} \right] \right\} \right|_{y=0} \nonumber \\
     &\ \ \ + \int d^4x \, \sum_i \left\{ \delta \phi_i \left. \left[ \dfrac{\partial \mathcal{L}_L}{\partial \phi_i} - \partial_\mu \left( \dfrac{\partial \mathcal{L}_L}{\partial \left( \partial_\mu \phi_i \right)} \right) + \dfrac{\partial \mathcal{L}_B}{\partial \left( \partial_y \phi_i \right)} \right] \right\} \right|_{y=L} \, .
\end{align}
In the third equality we use an integration by parts, dropping as usual the 4-divergence which gives boundary terms at infinity. For generic field variations, the action variations in the bulk and at the boundaries vanish separately, giving the bulk Euler-Lagrange equations,
\begin{equation}
\forall i \, , \ 
\dfrac{\partial \mathcal{L}_B}{\partial \phi_i} - \partial_M \left( \dfrac{\partial \mathcal{L}_B}{\partial \left( \partial_M \phi_i \right)} \right) = 0 \, ,
\label{ELE_5D}
\end{equation}
and the brane variations,
\begin{equation}
\forall i \, , \ 
\int d^4x \left\{ \delta \phi_i \left. \left[ \dfrac{\partial \mathcal{L}_0}{\partial \phi_i} - \partial_\mu \left( \dfrac{\partial \mathcal{L}_0}{\partial \left( \partial_\mu \phi_i \right)} \right) - \dfrac{\partial \mathcal{L}_B}{\partial \left( \partial_y \phi_i \right)} \right] \right\} \right|_{y=0} = 0 \, ,
\end{equation}
\begin{equation}
\forall i \, , \ 
\int d^4x \left\{ \delta \phi_i \left. \left[ \dfrac{\partial \mathcal{L}_L}{\partial \phi_i} - \partial_\mu \left( \dfrac{\partial \mathcal{L}_L}{\partial \left( \partial_\mu \phi_i \right)} \right) + \dfrac{\partial \mathcal{L}_B}{\partial \left( \partial_y \phi_i \right)} \right] \right\} \right|_{y=L} = 0 \, .
\end{equation}
On the one hand, if the field values at the boundaries are prescribed by a constraint, then the brane variations are zero so Eqs.~\eqref{BC_UV} and \eqref{BC_IR} are satisfied. On the other hand, for unconstrained fields at the boundaries, the field variations have to be generic and Eqs.~\eqref{BC_UV} and \eqref{BC_IR} imply the so-called natural boundary conditions,
\begin{equation}
\forall i \, , \ 
\left. \dfrac{\partial \mathcal{L}_0}{\partial \phi_i} - \partial_\mu \left( \dfrac{\partial \mathcal{L}_0}{\partial \left( \partial_\mu \phi_i \right)} \right) - \dfrac{\partial \mathcal{L}_B}{\partial \left( \partial_y \phi_i \right)} \right|_{y=0} = 0 \, ,
\label{BC_UV}
\end{equation}
and
\begin{equation}
\forall i \, , \ 
\left. \dfrac{\partial \mathcal{L}_L}{\partial \phi_i} - \partial_\mu \left( \dfrac{\partial \mathcal{L}_L}{\partial \left( \partial_\mu \phi_i \right)} \right) + \dfrac{\partial \mathcal{L}_B}{\partial \left( \partial_y \phi_i \right)} \right|_{y=L} = 0 \, .
\label{BC_IR}
\end{equation}

\section{Ultraviolet Brane Thickness}
\label{Brane_Thickness}
In Ref.~\cite{Fichet:2019owx}, it is argued that a brane treated as an infinitely thin hypersurface in an EFT coupled to gravity should have a thickness in the UV completion. Indeed, if there is a minimal length scale in Nature as argued in several approaches to quantize gravity, a brane should have a thickness at least of the order of this minimal length. In the EFT where the brane is treated as a hypersurface, the brane thickness in the UV completion can manifest itself by the presence of higher-dimensional operators such that the brane action is \cite{Fichet:2019owx}
\begin{equation}
S_{brane} = \int d^4x \int dy \sum_{n=0}^{+ \infty} \beta^{(n)} \, \mathcal{L}^{(n)} \, \partial_y^n \delta(y-\ell) \, ,
\end{equation}
where the coefficients $\beta^{(n>0)}$ vanish in the limit of an infinite Planck scale. Ref.~\cite{Fichet:2019owx} claims that the fields localized on the brane must be quasi-localized 5D fields (so they must have a $y$-dependence) in order to have non-vanishing higher-dimensional operators involving derivatives of $\delta(y-\ell)$ required by the UV completion. We find that this argument can be easily circumvented if the Lagrangians $\mathcal{L}^{(n>0)}$ involve the 5D graviton field which depends on $y$. Indeed, this is enough to guarantee that the action of the products $\mathcal{L}^{(n)} \, \partial_y^n \delta(y-\ell)$ ($n>0$) on the test function $\mathbf{1}(x^\mu, y)$ are non-vanishing. In this case, the other brane-localized fields can be 4D fields coupled to the 5D graviton without any problem to UV complete the EFT with a theory involving a brane thickness.

\section{Kaluza-Klein Mode Analysis of a Neutrino Model on the Rose Graph}
\label{neutrino_rose_app}

This appendix refers to the model of Section~\ref{Dirac_Neutrinos}. We give the KK mode analysis on the $N$-rose $\mathcal{R}_N$ of the exact treatment of Subsection~\ref{exact_treatment}.

\subsection{Zero Modes}
We are looking for zero modes ($b=0$, $n_0=0$, $m_\psi^{(0, \, 0)}=0$) with $M \neq 0$ for which the first order differential equations \eqref{wave_eq_Psi_star} are decoupled. For $\mathcal{K}_N = \mathcal{R}_N$, there is no right-handed zero mode. However, we have $N$ degenerate left-handed zero modes described by the wave functions of Eq.~\eqref{zero_mode_f_L}: the theory is chiral at the level of the zero modes. Therefore, the brane-localized mass term generates chirality in the rose graph compactification by lifting the degeneracy between the right and left-handed zero modes which exists in absence of brane-localized Yukawa couplings (c.f. Subsection~\ref{KK_Drac_20}). Moreover, we have $a^{(0, \, 0, \, d_0)} = 0$, which means that the left-handed zero modes do not mix with the left-handed neutrino localized on the $V$-brane: they are sterile neutrinos which do not participate in neutrino oscillations and they interact only through gravity. As discussed previously, this scenario with large $N$ should be ruled out by cosmological constraints on the number of light fermion species.

\subsection{Massive Modes}

\subsubsection*{\boldmath \textcolor{black}{\textit{First case}: $f_R^{(b, \, n_b, \, d_b)}(0,i) = 0$}}
The results are identical to the ones in Subsection~\ref{excited_modes_fermion}, Paragraph ``First case: $f_R^{(b, \, n_b, \, d_b)}(0, i)=0$'' of b) p.~\pageref{1st_case_psi_rose}. By following this discussion, we get two different mass spectra \eqref{mass_spect_Psi_rose_3} and \eqref{mass_spect_Psi_rose_bis} which defines the KK towers $b=1, 2$ respectively. We have $a^{(b, \, n_b, \, d_b)} = 0$ from Eq.~\eqref{eq_a} with $b = 1, 2$ so the left-handed modes do not mix with $\nu_L$: they are completely sterile (hidden sector).

\subsubsection*{\boldmath \textcolor{black}{\textit{Second case}: $f_R^{(b, \, n_b, \, d_b)}(0,i) \neq 0$}}
The KK mass spectrum is given by the transcendental equation:
\begin{equation}
m_\psi^{(3, \, n_3)} \, \tan \left[ m_\psi^{(3, \, n_3)} \, \dfrac{\ell}{2} \right] = \dfrac{M^2}{2N} \, , \ \ \ n_3 \in \mathbb{N} \, ,
\end{equation}
whose solutions $m_\psi^{(3, \, n_3)}$ define the KK tower $b=3$ and are not degenerate ($d_3 \in \{1\}$). We have $a^{(3, \, n_3, \, 1)} \neq 0$ from Eq.~\eqref{eq_a} so the left-handed modes mix with $\nu_L$. The lightest massive mode $(3, 0, 1)$ is identified with the observed neutrino. In the decoupling limit $\ell \rightarrow 0$, we recover that the mass of this mode is given by Eq.~\eqref{m_nu_2} of the zero mode approximation. The KK wave functions are
\begin{equation}
\left\{
\begin{array}{rcl}
f_L^{(3, \, n_3, \, 1)} (y, i) &=& - \left[ \dfrac{N\ell}{2} + \dfrac{M^2(2N-1)}{2 \left( \left[ m_\psi^{(3, \, n_3)}\right]^2 + \left[ \dfrac{M^2}{2N} \right]^2 \right)} \right]^{-1/2} \, \sin \left[ m_\psi^{(3, \, n_3)} \left( y - \dfrac{\ell}{2} \right) \right] \, , \\ \\
f_R^{(3, \, n_3, \, 1)} (y, i) &=& \left[ \dfrac{N\ell}{2} + \dfrac{M^2(2N-1)}{2 \left( \left[ m_\psi^{(3, \, n_3)}\right]^2 + \left[ \dfrac{M^2}{2N} \right]^2 \right)} \right]^{-1/2} \, \cos \left[ m_\psi^{(3, \, n_3)} \left( y - \dfrac{\ell}{2} \right) \right] \, ,
\end{array}
\right.
\end{equation}
where the $f_L^{(3, \, n_3, \, 1)}$'s have a discontinuity across the $V$-brane sourced by the brane-localized interaction.

\chapter{4D Method with a Warped Extra Dimension}
\label{4D_method_appendix}

Unpublished work by Andrei Angelescu.

\vspace{1cm}

In this appendix, we calculate the fermionic mass spectrum and the Yukawa couplings of the model in Section~\ref{One_generation_quarks} by using the so-called 4D method. The idea of this method is to treat the brane-localized Yukawa interactions as perturbations. Thus, one solves the free Euler-Lagrange equations to obtain the free fermionic KK wave functions, and then uses these KK wave functions to calculate their overlaps with the brane localized Higgs field, which in turn gives the fermionic mass and Yukawa matrices in the interaction basis. To pass to the mass or physical basis, the infinite mass matrix has to be diagonalized, and the Yukawa matrix rotated to the mass basis. In the following, we describe this calculation in detail.

The relevant part of the Lagrangian is the effective 4D mass Lagrangian, given by
\begin{equation}
\mathcal{L}_{mass} = -\Psi \cdot \left[ M \right] \cdot \bar{\Psi}  + \mathrm{h.c.} \,.
\end{equation}
By choosing to work in the basis spanned by the 4D spinors $Q_i$ and $D_i$, 
\begin{align}
\Psi &= (Q_0, Q_1, Q_2, \cdots, D_0, D_1, D_2,\cdots) , \\
\bar{\Psi} &= (\bar{Q}_0,\bar{Q}_1,\bar{Q}_2, \cdots, \bar{D}_0, \bar{D}_1, \bar{D}_2,\cdots),
\end{align}
the mass matrix reads, in block matrix notation, as
\begin{equation}
\label{mass_matrix}
\left[ M \right] = \begin{pmatrix}
 M_q & A \\ 0 &  M_d
\end{pmatrix}.
\end{equation}
The elements of the block matrices can be written as
\begin{equation}
(M_q)_{ij} = M_{q_i} \delta_{ij}, \quad (M_d)_{ij} = M_{d_i} \delta_{ij}, \quad (A)_{ij} = a_{q_i} a_{d_j},
\end{equation}
where \(i,j\)\footnote{Throughout this section, we will not use the Einstein summation convention. Sums over indices will always be explicit.} go from \(0\) to \(N\), $N$ being the number of KK levels taken into account (which shall be sent to infinity at the end). $M_{q_i}$ ($M_{d_i}$) represent the free KK masses of the $Q$ ($D$) 5D fermion KK tower. Similarly, $a_{q_i}$ ($a_{d_i}$) are the $(++)$ free fermion KK wave functions evaluated on the IR brane and multiplied by a Yukawa coupling. For the moment, we do not need the explicit expressions of the masses and KK wave functions. We will specify them later on, when needed. 

In order to obtain the 0-mode and KK fermion eigenmasses, one has to diagonalize either one of the two "squared" mass matrices,
\begin{equation}
[M]^{\dagger} [M] = \begin{pmatrix} M_q^2  & M_q A \\ A^{\dagger} M_q & M_d^2 + A^{\dagger} A \end{pmatrix}, \quad 
[M] [M]^{\dagger} = \begin{pmatrix} M_q^2 + A A^{\dagger}  & A M_d \\  M_d A^{\dagger} & M_d^2 \end{pmatrix}.
\label{mass_matrix_block}
\end{equation}
Given the high (or infinite, when $N \to \infty$) dimensionality, the usual methods of finding the eigenvalues of $[M]^{\dagger} [M]$ or $ [M] [M]^{\dagger}$, such as calculating the characteristic polynomial, would be cumbersome, if not impossible. However, as we show in the following, there is a way of circumventing this technical difficulty by closely inspecting the unitary matrices that diagonalize $[M]^{\dagger} [M]$ and $ [M] [M]^{\dagger}$.

We thus consider the two unitary matrices \(U_L\) and \(U_R\), which diagonalize \([M][M]^{\dagger}\) and \([M]^{\dagger}[M]\) respectively:
\begin{equation}
U_L  \left([M] [M]^{\dagger}\right) U_L^{\dagger} = U_R  \left([M]^{\dagger} [M]\right) U_R^{\dagger} = D^2, \quad \left(D \right)_{mn} = m_n \delta_{mn}. \label{diagonalization1}
\end{equation}
At the same time, the two unitary matrices bi-diagonalize the mass-matrix,
\begin{equation}
U_L \, [M] \, U_R^{\dagger} = D.
\label{bidiagonalization}
\end{equation}
Using the unitarity of \(U_L\) and \(U_R\), $U^{\dagger}_{L,R}=U^{-1}_{L,R}$, one derives the following matrix equations:
\begin{equation}
\left([M]^{\dagger} [M]\right) U_R^{\dagger} = U_R^{\dagger} D^2, \quad
\left([M] [M]^{\dagger}\right) U_L^{\dagger} = U_L^{\dagger} D^2, \quad
[M] \, U_R^{\dagger} = U_L^{\dagger} D,
\end{equation}
which represent the starting point for our calculation of $U_{L,R}$.

We start by inspecting $U_R$. Borrowing the bra-ket notation from Quantum Mechanics (and slightly abusing it), we write $U_R$ as
\begin{equation}
U_R^{\dagger} = \begin{pmatrix} \ket{\mathbf{V}^0} & \ket{\mathbf{V}^1} & \ket{\mathbf{V}^2} & \cdots \end{pmatrix},
\end{equation}
with $\ket{\mathbf{V}^n}$ denoting column vectors. By construction, $\ket{\mathbf{V}^n}$ are the eigenvectors of $[M]^{\dagger} [M]$, with the physical squared masses $m_n^2$ as eigenvalues:
\begin{equation}
[M]^{\dagger} [M] \, \ket{\mathbf{V}^n} = m_n^2 \, \ket{\mathbf{V}^n}.
\label{eigenvect_eq_R}
\end{equation}
Similarly to the block decomposition used in Eq.~\eqref{mass_matrix_block}, we write the eigenvector $\ket{\mathbf{V}^n}$ as
\begin{equation}
\ket{\mathbf{V}^n} = \begin{pmatrix}
\ket{\varphi^n} \\ \ket{\chi^n}
\end{pmatrix},
\end{equation}
where
\begin{equation}
\ket{\varphi^n} = \sum_i \scp{i}{\varphi^n} \ket{i} \to \begin{pmatrix} \varphi^n_0 \\ \varphi^n_1 \\ \vdots  \end{pmatrix}, \quad \ket{\chi^n} \to \begin{pmatrix} \chi^n_0 \\ \chi^n_1 \\ \vdots  \end{pmatrix}.
\end{equation}
Therefore, in block matrix notation, the eigenvector and eigenvalue equation, Eq.~\eqref{eigenvect_eq_R}, becomes:
\begin{equation}
\begin{pmatrix} M_q^2  & M_q A \\ A^{\dagger} M_q & M_d^2 + A^{\dagger} A \end{pmatrix} \begin{pmatrix} \ket{\varphi^n} \\ \ket{\chi^n} \end{pmatrix} = m_n^2 \begin{pmatrix} \ket{\varphi^n} \\ \ket{\chi^n} \end{pmatrix}.
\end{equation}
The matrix equation above is equivalent to the system of equations
\begin{align}
\left(M_{d_i}^2 - m_n^2\right) \chi_i^n + a_{d_i} \left(\mathbf{a}_q^2 \sum_{j=0}^{N} a_{d_j} \chi_j^n + \sum_{j=0}^{N} M_{q_j} a_{q_j} \varphi_j^n\right) &= 0, \\
\left(M_{q_i}^2 - m_n^2\right) \varphi_i^n + M_{q_i} a_{q_i} \sum_{j=0}^{N} a_{d_j} \chi_j^n &= 0,
\end{align}
where $\varphi_i^n$ and $\chi_i^n$ are the unknowns, and $ \mathbf{a}_q^2 \equiv \sum_{j=0}^{N} a_{q_j}^2 $. Solving for the unknowns, one finds
\begin{equation}
\varphi_i^n = c_n \frac{M_{q_i} a_{q_i}}{m_n^2 - M_{q_i}^2}, \quad
\chi_i^n = d_n \frac{a_{d_i}}{m_n^2 - M_{d_i}^2}, \label{phi_chi}
\end{equation}
where we used the notation
\begin{equation}
c_n \equiv \sum_{j=0}^{N} a_{d_j} \chi_j^n, \quad
d_n \equiv \mathbf{a}_q^2 \sum_{j=0}^{N} a_{d_j} \chi_j^n + \sum_{j=0}^{N} M_{q_j} a_{q_j} \varphi_j^n. \label{cn_dn}
\end{equation}
Although the two terms in $d_n$ diverge when $N\to\infty$, it turns out that their sum, and hence $d_n$, are finite, as we shall see below. Before continuing, we define the following sums:
\begin{align}
S_q (x) &\equiv \sum_{i=0}^N \frac{a_{q_i}^2}{x^2-M_{q_i}^2}, \quad S_q (m_n) \equiv S_{q_n}, \label{sum_q} \\
S_d (x) &\equiv \sum_{i=0}^N \frac{a_{d_i}^2}{x^2-M_{d_i}^2}, \quad S_d (m_n) \equiv S_{d_n}. \label{sum_d}
\end{align}
With the help of the two definitions above, and inserting Eq.~\eqref{phi_chi} in Eq.~\eqref{cn_dn}, we find the following system of equations:
\begin{equation}
c_n = d_n S_{d_n} , \quad
d_n = \mathbf{a}_q^2 \left( d_n S_{d_n} - c_n \right) + c_n m_n^2 S_{q_n}. 
\end{equation}
Clearly, this system admits a solution only if
\begin{equation}
m_n^2 S_{q_n} S_{d_n} = 1,
\label{mass_spectrum}
\end{equation}
which represents the equation that determines the physical mass spectrum, i.e. the eigenvalues $m_n$. Thus, the matrix elements of $U_R$  become:
\begin{equation}
\varphi_i^n = c_n \frac{M_{q_i} a_{q_i}}{m_n^2 - M_{q_i}^2}, \quad
\chi_i^n = \frac{c_n}{S_{d_n}} \frac{a_{d_i}}{m_n^2 - M_{d_i}^2}. \label{eigenvectors_R}
\end{equation}


Computing the elements of $U_L$ proceeds in a similar fashion as for the case of $U_R$ described above. Writing $U_L$ as
\begin{equation}
U_L^{\dagger} = \begin{pmatrix} \ket{\tilde{\mathbf{V}}^0} & \ket{\tilde{\mathbf{V}}^1} & \ket{\tilde{\mathbf{V}}^2} & \cdots \end{pmatrix},
\end{equation}
and decomposing its columns as 
\begin{equation}
\ket{\tilde{\mathbf{V}}^n} = \begin{pmatrix}
\ket{\tilde\varphi^n} \\ \ket{\tilde\chi^n}
\end{pmatrix},
\end{equation}
we repeat the steps taken earlier for $U_R^{\dagger}$ and find the same equation for the mass spectrum, $m_n^2 S_{q_n} S_{d_n} = 1$. We also find that the elements of $U_L$ are given by
\begin{equation}
\tilde\varphi_i^n = \frac{\tilde c_n}{S_{q_n}}\frac{a_{q_i}}{m_n^2 - M_{q_i}^2}, \quad
\tilde\chi_i^n =  \tilde c_n  \frac{M_{d_i} a_{d_i}}{m_n^2 - M_{d_i}^2}. \label{eigenvectors_L}
\end{equation}
Moreover, the relation between $c_n$ and $\tilde c_n$ can be determined by using the matrix equation $[M] \, U_R^{\dagger} = U_L^{\dagger} D $, and reads
\begin{equation}
\tilde c_n = m_n S_{q_n} c_n = \sqrt{\frac{S_{q_n}}{S_{d_n}}} c_n.
\end{equation}
The expression of $c_n$ can be found through a multitude of ways, such as enforcing unitarity for $U_{L/R}$ or inserting Eqs.~\eqref{eigenvectors_L} and \eqref{eigenvectors_R} into Eqs.~\eqref{diagonalization1} or \eqref{bidiagonalization}. Either way, $c_n$ is given by the equation
\begin{equation}
c_n^2 \, S_{q_n} \, F(m_n) = 1,
\label{cn_formula}
\end{equation} 
with 
\begin{equation}
-F(m_n) = 1 + \frac{m_n^2}{S_{q_n}} \frac{{\rm d} S_q (x)}{{\rm d} x^2} \bigg |_{x=m_n} + \frac{m_n^2}{S_{d_n}} \frac{{\rm d} S_d (x)}{{\rm d} x^2} \bigg |_{x=m_n} = \frac{1}{2} \frac{{\rm d} \log \left[ x^2 S_q(x) S_d(x)\right]}{{\rm d} \log x} \bigg |_{x=m_n}.
\label{F_mn}
\end{equation}
The expression of $c_n$ is important for calculating the mass basis Yukawa couplings $y_{m n}$, which are given by
\begin{equation}
v \, y_{mn} = \begin{pmatrix} \bra{\tilde\varphi^m} & \bra{\tilde\chi^m} \, \end{pmatrix}  \begin{pmatrix} 0 & A \\ 0 &  0 \end{pmatrix} \begin{pmatrix} \ket{\varphi^n} \\ \ket{\chi^n} \end{pmatrix} = \bra{\tilde\varphi^m} A \ket{\chi^n} = m_m S_{q_m} c_m c_n.
\label{yukawas}
\end{equation}
It is now clear that, in order to explicitly calculate the eigenmasses, given by Eq.~\eqref{mass_spectrum}, and the Yukawa couplings, given by Eq.~\eqref{yukawas}, we need to evaluate the infinite sums $S_q (x)$ and $S_d (x)$ and their first derivatives, both for $x = m_n^2$. We detail this procedure in the following.

To start with, we list the expressions of the free fermionic KK wave functions, evaluated at $z=1/T$, i.e. on the IR brane:
\begin{align}
q_{i}(T^{-1}) &= \frac{\lambda}{N_i(q)} \left[ Y_{\alpha+1} (\gamma_{q_i}) J_{\alpha} (\lambda \gamma_{q_i}) - J_{\alpha+1} (\gamma_{q_i}) Y_{\alpha} (\lambda \gamma_{q_i}) \right], \label{qi}\\
\bar q_{i}(T^{-1}) &= \frac{\lambda}{N_i(q)} \left[ Y_{\alpha+1} (\gamma_{q_i}) J_{\alpha+1} (\lambda \gamma_{q_i}) - J_{\alpha+1} (\gamma_{q_i}) Y_{\alpha+1} (\lambda \gamma_{q_i}) \right], \label{qbari} \\
d_{i}(T^{-1}) &= \frac{-\lambda}{N_i(d)} \left[ Y_{\beta} (\gamma_{d_i}) J_{\beta} (\lambda \gamma_{d_i}) - J_{\beta} (\gamma_{d_i}) Y_{\beta} (\lambda \gamma_{d_i}) \right], \label{di} \\
\bar d_{i}(T^{-1}) &= \frac{\lambda}{N_i(d)} \left[ Y_{\beta} (\gamma_{d_i}) J_{\beta-1} (\lambda \gamma_{d_i}) - J_{\beta} (\gamma_{d_i}) Y_{\beta-1} (\lambda \gamma_{d_i}) \right], \label{dbari}
\end{align}
where $i>0$. For $i=0$ (zero-mode KK wave functions), which correspond to $\gamma_0 = 0$, only the $(++)$ fields have non-vanishing KK wave functions, whose values on the IR brane read:
\begin{equation}
q_0 (T^{-1}) = \sqrt{\frac{2(\alpha+1) k}{1-\lambda^{-2(\alpha+1)}}}, \quad \bar d_0 (T^{-1}) = \sqrt{\frac{2\beta k}{1-\lambda^{-2\beta}}}.
\end{equation} 
These KK wave functions listed above are necessary for writing down the expressions of $a_{q_i}$ and $a_{d_i}$, which are given by
\begin{equation}
a_{q_i}^2 = \frac{X}{\lambda} q_i^2 (T^{-1}), \quad a_{d_i}^2 = \frac{X}{\lambda} \bar{d}_i^2 (T^{-1}),
\end{equation}
for all $i \geq 0$. In the previous equations, we used several simplifying notations:
\begin{equation}
\lambda \equiv \frac{k}{T} = \exp (kL), \quad \gamma_{q_i,d_i} \equiv \frac{M_{q_i,d_i}}{k}.
\end{equation}
Because of their similarity, the derivation of both $S_{d_n}$ and $S_{q_n}$ proceeds along the same lines. Thus, we concentrate on the evaluation of $S_{d_n}$, defined in Eq.~\eqref{sum_d}, for which we rely on several mathematical formulas presented in Ref.~\cite{Saharian:2000xx}, which provide closed-form expressions for infinite sums involving Bessel functions. Also, to avoid notation clutter, we write in the following $\gamma_i \equiv \gamma_{d_i}$ and $a_i \equiv a_{d_i}$. Thus, $S_{d_n}$ is given by
\begin{equation}
S_{d_n} = \frac{1}{k^2} \left( \frac{a_0^2}{x_n^2} + \sum_{i=1}^{\infty} \frac{a_i^2}{x_n^2 - \gamma_i^2} \right),
\label{sum_intermediate_1}
\end{equation}
where $x_n \equiv m_n / k$. While $a_0^2$ is straightforward to compute and reads
\begin{equation}
a_0^2 = \frac{2 \beta X k}{\lambda(1-\lambda^{-2\beta})},
\label{a0}
\end{equation}
$a_i$ are more challenging, as they involve the calculation of the normalization constant $N_i(d)$. This constant can be obtained by imposing the normalization condition for $d_i (z)$, namely
\begin{equation}
\int_{k^{-1}}^{T^{-1}} \!\! {\rm d} z \frac{d_i^2 (z)}{k z} = 1.
\end{equation}
The integral above can be computed with the help of Eq.~(4.5) of Ref.~\cite{Saharian:2000xx}, and gives the following value for the normalization constant:
\begin{equation}
N_i^2(d) = \frac{2}{\pi^2 \gamma_i^2 k} \left[ \frac{J_{\beta}^2 (\gamma_i)}{J_{\beta}^2 (\lambda \gamma_i)} - 1 \right]\!, \, i>0,
\end{equation}
which in turn allows us to express $a_i^2$ in closed form for $i>0$:
\begin{equation}
a_i^2 = \frac{\lambda \pi^2 X k \, \gamma_i^2}{2} \times \frac{\left[ Y_{\beta} (\gamma_i) J_{\beta-1} (\lambda \gamma_i) - J_{\beta} (\gamma_i) Y_{\beta-1} (\lambda \gamma_i) \right]^2}{ J_{\beta}^2 (\gamma_i)/J_{\beta}^2 (\lambda \gamma_i) - 1 }. \label{ai}
\end{equation} 
Inserting Eqs.~\eqref{a0} and \eqref{ai} into Eq.~\eqref{sum_intermediate_1}, one can write $S_{d_n}$ as
\begin{equation}
S_{d_n}= \frac{X}{\lambda \, k} \left[ \frac{2 \beta}{1-\lambda^{-2\beta}} \frac{1}{x_n^2} + \frac{\pi^2 \lambda^2}{2} \sum_{i=1}^{\infty} \frac{\gamma_i^2}{x_n^2-\gamma_i^2}  \frac{\left[ Y_{\beta} (\gamma_i) J_{\beta-1} (\lambda \gamma_i) - J_{\beta} (\gamma_i) Y_{\beta-1} (\lambda \gamma_i) \right]^2}{ J_{\beta}^2 (\gamma_i)/J_{\beta}^2 (\lambda \gamma_i) - 1 }  \right].
\label{sum_intermediate_2}
\end{equation}
Clearly, the challenge is to evaluate the infinite sum in the equation above. Fortunately, the infinite sum we are interested in can be evaluated with the help of Eq.~(4.19) from Ref.~\cite{Saharian:2000xx}. We take a short break from our computation to briefly explain this equation.

Let $\gamma_i$ be defined by 
\begin{equation}
Y_{\beta} (\gamma_{i}) J_{\beta} (\lambda \gamma_{i}) - J_{\beta} (\gamma_{i}) Y_{\beta} (\lambda \gamma_{i})  = 0,
\label{gamma_i_def}
\end{equation}
and let $F(z)$\footnote{Here and in the following $z$ denotes a complex variable and not the conformal coordinate along the extra dimension.} be a complex function satisfying certain properties listed in Ref.~\cite{Saharian:2000xx} (all the cases of $F(z)$ we are considering satisfy these properties). Denoting as $z_n$ the poles of $F(z)$ lying complex right half-plane (including the imaginary axis and thus the origin), Eq.~(4.19) from Ref.~\cite{Saharian:2000xx} states that:
\begin{align}
&\sum_{i=1}^{\infty} \frac{\gamma_i \left[ J_{\beta} (\gamma_i) Y_{\beta-1} (\lambda \gamma_i) - Y_{\beta} (\gamma_i) J_{\beta-1} (\lambda \gamma_i) \right]}{J_{\beta}^2 (\gamma_i)/J_{\beta}^2 (\lambda \gamma_i) - 1} F(\gamma_i) = \notag \\ 
&= \frac{1}{\pi} \sum_{\eta = 0,z_n} \left( 2 - \delta_{0 \eta} \right) {\rm Res}_{z=\eta} \left[ \frac{Y_{\beta} (\lambda z) J_{\beta-1} (\lambda z) - J_{\beta} (\lambda z) Y_{\beta-1} (\lambda z)}{J_{\beta} (z) Y_{\beta} (\lambda z) - Y_{\beta} (z) J_{\beta} (\lambda z) }  F(z)\right],
\label{central_eq}
\end{align}
with Res standing for residue. In our case, $\gamma_i$ are given by the equation $d_i(T^{-1})=0$, which coincides with the condition in Eq.~\eqref{gamma_i_def}. This means that indeed we can use Eq.~\eqref{central_eq} to evaluate the infinite sum arising in Eq.~\eqref{sum_intermediate_2}. The function $F(z) $ needed for our purpose is given by
\begin{equation}
F(z) = \frac{z \left[ J_{\beta} (z) Y_{\beta-1} (\lambda z) - Y_{\beta} (z) J_{\beta-1} (\lambda z) \right]}{x_n^2 - z^2},
\label{F(z)}
\end{equation}
whose only pole in the complex right half-plane is at $z = x_n$ (for a given $n$), which is a simple pole. Using the expression of $F(z)$ and the Bessel function identity
\begin{equation}
Y_{\beta} (\lambda z) J_{\beta-1} (\lambda z) - J_{\beta} (\lambda z) Y_{\beta-1} (\lambda z) = -\frac{2}{\pi \lambda z},
\end{equation}
the sum in Eq.~\eqref{central_eq} becomes
\begin{equation}
-\frac{2}{\pi^2 \lambda} \sum_{\eta = 0,z_n} \left( 2 - \delta_{0 \eta} \right) {\rm Res}_{z=\eta} \frac{1}{x_n^2 - z^2} \frac{J_{\beta} (z) Y_{\beta-1} (\lambda z) - Y_{\beta} (z) J_{\beta-1} (\lambda z)}{J_{\beta} (z) Y_{\beta} (\lambda z) - Y_{\beta} (z) J_{\beta} (\lambda z)}.
\label{sum_intermediate_3}
\end{equation}
Let us denote the function in above equation, whose residues have to be evaluated, as $G(z)$. In the right half of the complex plane, $G(z)$ has a simple pole at $z=x_n$ (inherited from $F(z)$) and another simple pole at $z=0$. The residue at the origin is given by
\begin{equation}
{\rm Res}_{z=0} G(z) = \frac{2\beta}{\lambda(1-\lambda^{-2\beta})}\frac{1}{x_n^2},
\label{residue_0}
\end{equation}
with the residue at $z=x_n$ being
\begin{equation}
{\rm Res}_{z=x_n} G(z) = -\frac{1}{2 x_n} \frac{J_{\beta} (x_n) Y_{\beta-1} (\lambda x_n) - Y_{\beta} (x_n) J_{\beta-1} (\lambda x_n)}{J_{\beta} (x_n) Y_{\beta} (\lambda x_n) - Y_{\beta} (x_n) J_{\beta} (\lambda x_n)}=\frac{1}{2 x_n} \frac{\bar{d}_n(T^{-1})}{d_n(T^{-1})},
\label{residue_xn}
\end{equation}
with $d_n$ and $\bar d_n$ being the exact KK wave functions obtained in the so-called 5D approach. Gluing all the pieces together, we find that the infinite sum from Eq.~\eqref{sum_intermediate_2} evaluates to
\begin{equation}
\frac{\pi^2 \lambda^2}{2} \sum_{i=1}^{\infty} \frac{\gamma_i^2}{x_n^2-\gamma_i^2}  \frac{\left[ Y_{\beta} (\gamma_i) J_{\beta-1} (\lambda \gamma_i) - J_{\beta} (\gamma_i) Y_{\beta-1} (\lambda \gamma_i) \right]^2}{ J_{\beta}^2 (\gamma_i)/J_{\beta}^2 (\lambda \gamma_i) - 1 } = - \frac{2\beta}{1-\lambda^{-2\beta}}
\frac{1}{x_n^2} - \frac{\lambda}{x_n} \frac{\bar{d}_n(T^{-1})}{d_n(T^{-1})}.
\end{equation}
Inserting the above equality into the expression of $S_{d_n}$ from Eq.~\eqref{sum_intermediate_2}, we note that the contribution from the residue in $z=0$ exactly cancels the contribution from the term corresponding to $a_0^2$, leaving a simple final expression for $S_{d_n}$:
\begin{equation}
S_{d_n} = -\frac{X}{m_n} \frac{\bar{d}_n(T^{-1})}{d_n(T^{-1})}.
\label{sum_final}
\end{equation} 
Using the same techniques as above, one can compute $S_{q_n}$ as well, whose expression is
\begin{equation}
S_{q_n} = \frac{X}{m_n} \frac{q_n(T^{-1})}{\bar{q}_n(T^{-1})}.
\label{other_sum_final}
\end{equation}
The sign difference between the two sums comes from the negative sign in front of $d_i$, cf Eq.~\eqref{di}. With the expressions of the two sums, Eq.~\eqref{mass_spectrum}, which determines the physical mass spectrum, translates to
\begin{equation}
-\frac{\bar{q}_n(T^{-1}) d_n(T^{-1})}{q_n(T^{-1}) \bar{d}_n(T^{-1})} = X^2,
\end{equation}
which indeed coincides with the mass spectrum \eqref{mass} obtained in the 5D approach.

Having derived the mass spectrum, we now focus on the calculation of the 4D Yukawa couplings in the so-called "4D approach". In the following, we work only with the exact KK wave functions, which we write below in a way that will be convenient later on:
\begin{align}
q_{n}(z) \equiv \frac{k z}{N_{q_n}} f_{q} (x_n,z)  &= \frac{k z}{N_{q_n}} \left[ Y_{\alpha+1} (x_n) J_{\alpha} (x_n k z) - J_{\alpha+1} (x_n) Y_{\alpha} (x_n k z) \right] , \label{qn}\\
\bar q_{n}(z) \equiv \frac{k z}{N_{q_n}} f_{\bar q} (x_n,z) &= \frac{k z}{N_{q_n}} \left[ Y_{\alpha+1} (x_n) J_{\alpha+1} (x_n k z) - J_{\alpha+1} (x_n) Y_{\alpha+1} (x_n k z) \right], \label{qbarn} \\
-d_{n}(z) \equiv \frac{k z}{N_{d_n}} f_{d} (x_n,z) &= \frac{k z}{N_{d_n}} \left[ Y_{\beta} (x_n) J_{\beta} (x_n k z) - J_{\beta} (x_n) Y_{\beta} (x_n k z) \right], \label{dn} \\
\bar d_{n}(z) \equiv \frac{k z}{N_{d_n}} f_{\bar d} (x_n,z) &= \frac{k z}{N_{d_n}} \left[ Y_{\beta} (x_n) J_{\beta-1} (x_n k z) - J_{\beta} (x_n) Y_{\beta-1} (x_n k z) \right]. \label{dbarn}
\end{align}
Combining Eqs.~\eqref{cn_dn} and \eqref{yukawas}, the expression for the effective Yukawa couplings becomes
\begin{equation}
v \, y_{mn}=  \sqrt{\frac{S_{q_m}}{S_{q_n}}} \frac{m_m}{\sqrt{F(m_m)F(m_n)}} =\frac{1}{\sqrt{S_{d_m} F(m_m) S_{q_n}F(m_n)}},
\label{yukawas_2}
\end{equation} 
where we have use the mass spectrum equation~\eqref{mass_spectrum} for the second step. In order to obtain the expressions for the Yukawa couplings,  one needs to calculate $F(m_n)$, defined in Eq.~\eqref{F_mn}:
\begin{align}
F(m_n) &= \frac{1}{2} \frac{\rm d}{{\rm d} \log x} \left[ \log \frac{f_q (x,T^{-1}) f_{\bar{d}} (x,T^{-1}) }{f_{\bar{q}} (x,T^{-1}) f_d (x,T^{-1})} \right]_{x=x_n} \notag \\
 &= \frac{x_n}{2} \frac{\rm d}{{\rm d} x} \left[ \log \frac{f_q (x,T^{-1}) }{f_{\bar{q}} (x,T^{-1})} + \log \frac{f_{\bar{d}} (x,T^{-1})}{f_d (x,T^{-1})} \right]_{x=x_n}.
\label{F_mn_2}
\end{align}
Note that in the above equation we took advantage of the logarithmic derivative from Eq.~\eqref{F_mn} to substitute $m_n$ by $x_n = m_n / k$.

In tackling Eq.~\eqref{F_mn_2}, we first evaluate the part containing the $d$-KK wave functions:
\begin{equation}
\frac{\rm d}{{\rm d} x} \left[ \log \frac{f_{\bar{d}} (x,T^{-1})}{f_d (x,T^{-1})} \right]_{x=x_n} = \frac{f_d (x_n,T^{-1}) \partial_x f_{\bar{d}} (x,T^{-1}) - f_{\bar{d}} (x_n,T^{-1}) \partial_x f_{d} (x,T^{-1})}{f_d (x_n,T^{-1}) f_{\bar{d}} (x_n,T^{-1})}\bigg|_{x=x_n}.
\label{d_profile_deriv}
\end{equation}
Using the following two relations between Bessel functions,
\begin{gather}
Z^{\prime}_a (x)= Z_{a-1} (x)  - \frac{a}{x} Z_a (x) = \frac{a}{x} Z_a(x) - Z_{a+1}(x), \\
Y_a (x) J_{a-1} (x) - J_a (x) Y_{a-1} (x) = -\frac{2}{\pi x}, 
\end{gather}
where $Z$ denotes either first ($J$) or second order ($Y$) Bessel functions, we find that
\begin{align}
\partial_x \left[ f_d(x,T^{-1}) \right]_{x=x_n} &= \lambda f_{\bar{d}} (x_n,T^{-1}) - f_{d} (x_n,T^{-1}) \frac{J_{\beta+1}(x_n)}{J_{\beta} (x_n)} + f_{\bar{d}} (x_n,k^{-1}) \frac{J_{\beta}(\lambda x_n)}{J_{\beta} (x_n)}, \\
\partial_x \left[ f_{\bar{d}} (x,T^{-1}) \right]_{x=x_n} &= -\lambda f_{d} (x_n,T^{-1}) + f_{\bar{d}} (x_n,T^{-1}) \frac{J_{\beta-1}(x_n)}{J_{\beta} (x_n)} \notag \\
& + f_{\bar{d}} (x_n,k^{-1}) \frac{J_{\beta-1}(\lambda x_n)}{J_{\beta} (x_n)} - \frac{f_{\bar{d}} (x_n,T^{-1})}{x_n}.
\end{align}
Using the two equations above, the Bessel function recurrence relation
\begin{equation}
J_{\beta-1} (x) + J_{\beta+1} = \frac{2 \beta}{x} J_{\beta} (x),
\end{equation}
and the relation
\begin{equation}
\frac{J_{\beta} (\lambda x_n) f_{\bar{d}} (x,T^{-1}) - J_{\beta-1} (\lambda x_n) f_{d} (x,T^{-1})}{J_{\beta} (x_n)} = - \frac{f_{\bar{d}} (x_n,k^{-1})}{\lambda},
\end{equation}
we obtain from Eq.~\eqref{d_profile_deriv}
\begin{equation}
\frac{\rm d}{{\rm d} x} \left[ \log \frac{f_{\bar{d}} (x,T^{-1})}{f_d (x,T^{-1})} \right]_{x=x_n} = -\frac{1}{x_n} + \frac{\lambda}{\bar{d}_n (T^{-1}) d_n (T^{-1})} \left[ \frac{2 \beta}{\lambda x_n} \bar{d}_n (z) d_n (z) + \bar{d}^2_n (z) + d_n^2 (z) \right] \Bigg|^{T^{-1}}_{k^{-1}}.
\label{d_profile_deriv_final}
\end{equation} 
Note that, at an intermediate step, we replaced $f_{q,d}$ and $f_{\bar{q},\bar{d}}$ by the exact KK wave functions, using Eqs.~\eqref{qn} - \eqref{dbarn}. We have also used the simplifying notation
\begin{equation}
f(z) \big|^b_a \equiv f(b) - f(a).
\end{equation}

In a similar way, one can obtain the piece containing the $q$-KK wave functions from Eq.~\eqref{F_mn_2}:
\begin{equation}
\frac{\rm d}{{\rm d} x} \left[ \log \frac{f_{q} (x,T^{-1})}{f_{\bar{q}} (x,T^{-1})} \right]_{x=x_n} = \frac{1}{x_n} - \frac{\lambda}{\bar{q}_n (T^{-1}) q_n (T^{-1})} \left[ -\frac{2 \alpha}{\lambda x_n} \bar{q}_n (z) q_n (z) + \bar{q}^2_n (z) + q_n^2 (z) \right] \Bigg|^{T^{-1}}_{k^{-1}}.
\label{q_profile_deriv_final}
\end{equation}
Taking another step towards the final expression for $F(m_n)$, we make use of the boundary conditions in $z=T^{-1}$ of the exact KK wave functions, $\bar{q}_n (T^{-1}) = X d_n (T^{-1})$ and $d_n (T^{-1}) = - X q_n (T^{-1})$, and insert Eqs.~\eqref{d_profile_deriv_final} and \eqref{q_profile_deriv_final} in Eq.~\eqref{F_mn_2} to get
\begin{align}
F(m_n) &= \frac{\lambda x_n}{2} \frac{1}{\bar{d}_n (T^{-1}) d_n (T^{-1})} \times \notag \\   
&\times \left\lbrace \bar{q}^2_n (z) + q_n^2 (z) + \bar{d}^2_n (z) + d_n^2 (z)  + \frac{2}{\lambda x_n} \left[ \beta \bar{d}_n (z) d_n (z) - \alpha \bar{q}_n (z) q_n (z) \right] 
\right\rbrace \bigg|^{T^{-1}}_{k^{-1}}.
\end{align}
Note that, in the equation above, the quantity which is to be evaluated between $T^{-1}$ and $k^{-1}$ is exactly the same quantity appearing in the normalization condition for the KK wave functions, as described in the section dealing with the 5D approach:
\begin{equation}
\left\lbrace \bar{q}^2_n (z) + q_n^2 (z) + \bar{d}^2_n (z) + d_n^2 (z)  + \frac{2}{\lambda x_n} \left[ \beta \bar{d}_n (z) d_n (z) - \alpha \bar{q}_n (z) q_n (z) \right] 
\right\rbrace \bigg|^{T^{-1}}_{k^{-1}} = 2 k.
\end{equation}
With this observation, we arrive at the final expression for $F(m_n)$:
\begin{equation}
F(m_n) = \frac{\lambda m_n}{\bar{d}_n (T^{-1}) d_n (T^{-1})},
\label{F_mn_final}
\end{equation}
where we have used the definition of $x_n$, $m_n = k x_n$. Inserting Eq.~\eqref{F_mn_final} into Eq.~\eqref{yukawas_2}, and using the expression of $S_{d_n}$ and $S_{q_n}$ form Eqs.~\eqref{sum_d} and \eqref{sum_q}, together with the boundary conditions in $z=T^{-1}$, we find that the effective 4D Yukawa couplings are given by \eqref{Yukawa_4D} which indeed matches the relation derived using the 5D approach.

\selectlanguage{french}

\chapter{Résumé en français des chapitres en anglais}
\label{summary_french}

\section{Introduction}

Le modèle standard de la physique des particules connait un succès expérimental indéniable. Son secteur électrofaible est basé sur le mécanisme de brisure spontanée de symétrie de jauge dont la pierre angulaire est le champ de Higgs. Cependant, le boson de Higgs est traité dans le modèle standard comme une particule élémentaire de spin 0 qui n'est pas protégée par une symétrie, donc sa masse est quadratiquement sensible à toute échelle de nouvelle physique (critère de naturalité technique de 't Hooft) et les paramètres de la théorie UltraViolette (UV) doivent être finement ajustés : c'est le problème de hiérarchie de jauge. Si une telle nouvelle physique existe au-dessus du TeV, un mécanisme devrait protéger la masse du Higgs des corrections radiatives non loin de l'échelle électrofaible. L'existence d'au moins l'échelle de Planck à $10^{18}$ GeV, où la plupart des théoriciens s'attendent à ce que de nouveaux degrés de liberté apparaissent pour réguler le comportement de la gravité dans l'UV, rend le problème de hiérarchie de jauge bien réel.

L'hypothèse, que des dimensions spatiales supplémentaires compactifiées et des branes existent, permet de construire des modèles où l'échelle de gravité est ramenée au TeV sur la brane où se trouve le champ de Higgs : elle agit comme une échelle de coupure sur les corrections radiatives, et un boson de Higgs léger est naturel. Il existe deux classes de tels modèles : ADD (Arkani-Hamed - Dimopoulos - Dvali) et RS (Randall - Sundrum). Les modèles ADD utilisent des dimensions spatiales supplémentaires compactifiées sur des espaces avec un grand volume, où les champs du modèle standard sont localisés sur une 3-brane, et où la gravité est diluée par le grand volume et apparaît faiblement couplée sur la brane. En revanche, elle devient fortement couplée au TeV. Pour un petit nombre de dimensions supplémentaires compactifiées sur un tore, les rayons de compactification sont grands devant la longueur de Planck dans le bulk. Cette nouvelle hiérarchie géométrique est en général difficile à stabiliser dans un modèle réaliste. Quant au modèle RS, il nécessite une seule dimension supplémentaire courbe compactifiée sur un intervalle (ou de manière équivalente l'orbifold $S^1/\mathbb{Z}_2$). La courbure entraine que les échelles de gravité aux deux bords de l'intervalle soient exponentiellement séparées : l'une est à $10^{18}$ GeV, et correspond à l'échelle de Planck 4D de la relativité générale, et l'autre est au TeV. Si le boson de Higgs est localisé sur une 3-brane au bord où l'échelle de gravité est au TeV, il n'y a plus de problème de hiérarchie de jauge. Les bosons de jauge et les fermions du modèle standard sont libres de se propager dans le bulk. Si les fermions ont des masses de Dirac 5D différentes, il est possible de localiser différemment les modes zéro le long de la dimension supplémentaire, et on peut reproduire la hiérarchie de masse des fermions du modèle standard : la fonction d'onde d'un fermion léger (lourd) aura un petit (grand) recouvrement avec la brane où est localisé le champ de Higgs. Il est alors crucial de comprendre comment traiter en théorie des champs l'interaction de fermions 5D avec un champ de Higgs branaire.

\section{Au-delà de la régularisation du champ de Higgs branaire}
Au Chapitre~\ref{1_4D perturbative approach}, on étudie un modèle jouet avec une dimension spatiale supplémentaire compactifiée sur un intervalle. Deux champs de Dirac 5D sont couplés à un scalaire localisé sur une 3-brane à un bord de l'intervalle. Le champ scalaire acquiert une valeur moyenne dans le vide, ce qui donne une masse localisée aux fermions. Ce modèle jouet est ainsi inspiré du modèle de génération des masses des fermions dans le modèle RS, avec les bosons de jauge et les fermions du modèle standard dans le bulk.

On réexamine tout d'abord le cas d'un fermion 5D libre. Si on applique naïvement le principe de Hamilton sur l'action du bulk, et que l'on prend des conditions aux bords naturelles, on obtient des solutions pour les champs nulles partout : le système est sur-contraint aux bords. Pour avoir des solutions physiques, il faut imposer des conditions aux bords essentielles de Dirichlet, compatibles avec l'annulation sur les bords de la composante transverse du courant de Dirac conservé. Une autre possibilité est d'ajouter des termes de bord qui sont des termes de masse avec un coefficient fixé à $\pm 1/2$, analogues aux termes de Gibbons-Hawking en gravité. Ces termes de bord vont compenser la variation sur le bord de l'une des deux chiralités du champ 5D, évitant ainsi au système d'être sur-contraint.

Ensuite, on critique la procédure de régularisation du champ de Higgs branaire de la littérature qui est en fait inutile et qui manque de rigueur mathématique. Deux types de régularisations sont proposés dans la littérature : déplacer la brane sur laquelle est localisé le champ de Higgs dans le bulk ou lui donner une épaisseur. Dans les deux cas, les auteurs incluent des termes de masse localisés qui impliquent le produit d'une distribution de Dirac avec des champs discontinus en ce point, ce qui n'est pas défini en théorie des distributions.

On propose alors une méthode 5D où les termes de masse localisés sont pris en compte dans les conditions aux bords en résolvant les équations différentielles donnant les fonctions d'onde dans le bulk de chaque mode de Kaluza-Klein. Les termes de bord optionnels dans le cas libre sont alors essentiels pour obtenir des solutions physiques viables. Le principe de Hamilton entraîne que certains termes de masse localisés sur les bords doivent être nuls pour ne pas que le système soit sur-contraint : les termes de masse localisés ne peuvent faire intervenir que l'une des deux chiralités des champs 5D. Le spectre de masse est alors identique à celui de la méthode dite 4D, où on résout d'abord le spectre de Kaluza-Klein libre, et où on inclut les termes de masse localisés comme une matrice infinie à bi-diagonaliser. Dans cette méthode, les opérateurs localisés, qui doivent être nuls dans la méthode 5D, ne contribuent pas car ils impliquent les champs libres qui s'annulent sur les bords.

\section{Généralisations}
Au Chapitre~\ref{Applications}, on présente diverses applications et extensions de notre méthode 5D pour des fermions avec des termes branaires.

Tout d'abord, on reprend le modèle du Chapitre~\ref{1_4D perturbative approach} mais avec une dimension spatiale supplémentaire courbe. On prend l'exemple d'une tranche AdS$_5$. Ceci ne change pas le traitement des conditions aux bords de l'intervalle et les conclusions sont similaires à celles dans le cas d'une dimension plate.

On s'intéresse ensuite à différents termes localisés sur un bord que l'on peut rencontrer dans la littérature sur les fermions 5D : termes cinétiques branaires, masse de Majorana branaire, mélange avec un fermion branaire. On se limite à une dimension spatiale supplémentaire plate et on calcule le spectre de masse. \`A chaque fois, on montre qu'il est nécessaire d'inclure les termes de bord, discutés plus haut, pour avoir des solutions consistantes où le système n'est pas sur-contraint (relations entre les paramètres indépendants du modèle).

On étudie également le modèle du Chapitre~\ref{1_4D perturbative approach} avec les orbifolds $S^1/\mathbb{Z}_2$ et $S^1/\left( \mathbb{Z}_2 \times \mathbb{Z}_2' \right)$. Les champs de chiralités opposées, provenant d'un même champ 5D, ont une parité différente sous les symétries $\mathbb{Z}_2$ et $\mathbb{Z}_2'$. On définit la théorie des champs sur les espaces de recouvrement : les branes ne sont alors plus sur les bords de l'espace. Les termes de masse induisent des discontinuités à travers les branes pour les champs impairs. Comme pour l'intervalle, le principe de Hamilton implique que les termes de masse branaires pour les champs impairs doivent être nuls. Quand les champs sont définis au sens des distributions, on peut passer au formalisme des fonctions, et les champs discontinus génèrent ainsi des termes de bord identiques à ceux que l'on doit ajouter \textit{ab initio} dans le cas de l'intervalle. Le spectre de masse est identique à celui du modèle avec un intervalle (à un facteur 2 près qui peut être éliminé par une redéfinition du couplage de Yukawa 5D). Ces modèles sont donc duaux.

Pour terminer, on reprend le modèle du Chapitre~\ref{1_4D perturbative approach} où on a déplacé la brane sur laquelle est localisé le champ de Higgs dans le bulk. Les termes de masse impliquent que certains champs sont discontinus à travers la brane. La forme des termes cinétiques impose que l'une des deux chiralités d'un champ 5D doit être continue. De la même manière que pour les orbifolds, des termes de bord sur la brane sont générés par la dérivée au sens des distributions des champs discontinus.

\section{Grande dimension supplémentaire en étoile/rose avec de petit(e)s branches/pétales}
Au Chapitre~\ref{large_star_rose_ED_small_leaves_petals}, on propose un type de modèle ADD qui ne souffre pas d'une grande hiérarchie géométrique à stabiliser. L'idée est de compactifier une dimension spatiale supplémentaire sur un graphe étoile/rose avec un grand nombre $N$ de petit(e)s branches/pétales identiques. Avec une échelle de Planck 5D $\Lambda_P^{(5)} \simeq 1$ TeV et une longueur/circonférence des branches/pétales $\ell \simeq 100$ GeV, on obtient $N \simeq 6 \times 10^{29}$. $N$ est fixé par la géométrie, il est donc conservé et stable sous les corrections radiatives dans la théorie effective. L'origine de la grande valeur de $N$ se trouve dans la théorie UV de la gravité où on s'attend à ce que $N$ soit déterminé dynamiquement.

On étudie la décomposition de Kaluza-Klein d'un champ 5D scalaire réel massif et d'un champ 5D de Dirac sans masse. Certains modes de Kaluza-Klein ont leurs fonctions d'onde (celles qui sont continues) qui s'annulent à la jonction de l'étoile/rose. On trouve que l'échelle de Kaluza-Klein est donnée par $1/\ell$. Pour un champ scalaire, il n'y a qu'un seul mode zéro. Pour un champ de Dirac sur l'étoile, il n'y a un mode zéro que pour le champ chiral continu à travers la jonction. Sur la rose, le champ chiral continu n'a qu'un seul mode zéro alors que le champ de chiralité opposée, qui peut être discontinu, en a $N$. On peut ainsi générer $N$ générations au niveau des modes zéro à partir d'une seule génération de champ 5D.

En supposant les champs du modèle standard localisés sur une 3-brane à la jonction de l'étoile/rose, on montre avec un modèle jouet de graviton 5D sans spin que les couplages de la tour de KK-gravitons au modèle standard sont très supprimés comparés aux modèles ADD avec une compactification toroïdale, car seul le petit nombre de KK-gravitons, dont la fonction d'onde ne s'annule pas à la jonction, couple au modèle standard. Les KK-gravitons, qui ne couplent pas au modèle standard, constituent un secteur caché. On s'attend donc seulement à observer des phénomènes de gravité quantique fortement couplés aux LHC à 14 TeV. Un FCC-pp à 100 TeV permettrait d'observer la base de la tour de trous noirs semi-classiques trans-planckiens. Un trou noir quantique au TeV, le Planckion, est peut-être stable et constituerait ainsi une partie de la matière noire.

Pour terminer, on étudie également un modèle jouet où une génération de neutrino gauche est localisée à la jonction d'une étoile/rose, et est couplée via le champ de Higgs à un neutrino 5D singulet de jauge. La masse de Dirac du mode le plus léger est supprimée par la racine carrée du volume du graphe, ce qui permet de générer le bon ordre de grandeur pour la masse des neutrinos observés tout en ayant l'échelle de gravité au TeV. Seul le graphe étoile est phénoménologiquement viable car, sur la rose, on a $N$ neutrinos gauches stériles sans masse couplés seulement à la gravité qui posent un problème pour la cosmologie du modèle. Une partie des KK-neutrinos sont stériles et constituent un secteur caché.

\section{Perspectives}
On a vu que certains opérateurs localisés doivent être nuls pour obtenir des solutions physiquement intéressantes. On sait néanmoins que les corrections radiatives génèrent des opérateurs localisés sur les branes. En traitant les termes localisés comme des perturbations des solutions libres, on voit que les opérateurs qui doivent être nuls à l'arbre ne sont pas générés radiativement. Il paraît intéressant de vérifier cela avec les solutions exactes.

Dans la littérature sur les modèles à 5D, la dimension spatiale supplémentaire est généralement compactifiée sur un cercle ou un intervalle (ainsi que sur les orbifolds équivalents). Quelques auteurs ont généralisé la compactification à des graphes métriques. Le cercle et l'intervalle sont alors vus comme les briques élémentaires que l'on assemble pour obtenir n'importe quel graphe.

Il serait intéressant d'utiliser les propriétés géométriques des graphes pour construire des modèles de matière noire où une KK-particule est stable. Les exemples de l'étoile/rose, avec un grand nombre de branches/pétales, permettent d'avoir des KK-particules très faiblement couplées, ce qui est intéressant pour les modèles de matière noire, d'axions, de neutrinos, etc.

On a vu également comment générer $N$ modes zéro pour un champ de Dirac 5D sur une rose, et comment donner une masse au mode zéro miroir en localisant une interaction sur le vertex. Il serait alors intéressant de construire un modèle de saveur où les trois générations de fermions du modèle standard sont les modes zéro d'une seule génération de fermions 5D se propageant sur une rose à trois pétales. On pourrait aussi rechercher d'autres graphes métriques qui ont la propriété de générer plusieurs modes zéro fermioniques.

Pour terminer, il serait intéressant d'étudier aussi la construction de théorie des champs 5D supersymétriques sur des graphes métriques.

\selectlanguage{english}

\chapter*{Glossary}
\addcontentsline{toc}{chapter}{Glossary}
\label{Glossary}

\selectlanguage{english}

\begin{tabular}{ll}
\textbf{ADD} : & \textbf{A}rkani-Hamed, \textbf{D}imopoulos and \textbf{D}vali \\
\\
\textbf{AdS} : & \textbf{A}nti-\textbf{d}e  \textbf{S}itter \\
\\
\textbf{BBN} : & \textbf{B}ig \textbf{B}ang \textbf{N}ucleosynthesis \\
\\
\textbf{BBT} : & \textbf{B}ilinear \textbf{B}oundary \textbf{T}erm \\ \\
\textbf{BLKT} : & \textbf{B}rane \textbf{L}ocalized \textbf{K}inetic \textbf{T}erm \\
\\
\textbf{BSM} : & \textbf{B}eyond the \textbf{S}tandard \textbf{M}odel \\
\\
\textbf{CDM} : & \textbf{C}old \textbf{D}ark \textbf{M}atter \\
\\
\textbf{CFT} : & \textbf{C}onformal \textbf{F}ield \textbf{T}heory \\
\\
\textbf{CMB} : & \textbf{C}osmic \textbf{M}icrowave \textbf{B}ackground \\
\\
\textbf{DGP} : & \textbf{D}vali, \textbf{G}abadadze and \textbf{P}orrati \\
\\
\textbf{EFT} : & \textbf{E}ffective \textbf{F}ield \textbf{T}heory \\
\\
\textbf{EWSB} : & \textbf{E}lecto\textbf{W}eak \textbf{S}ymmetry \textbf{B}reaking \\
\\
\textbf{FCC pp} : & \textbf{F}uture \textbf{C}ircular \textbf{C}ollider \textbf{p}roton-\textbf{p}roton \\
\\
\textbf{FCNC} : & \textbf{F}lavor \textbf{C}hanging \textbf{N}eutral \textbf{C}urrent \\
\\
\textbf{GUT} : & \textbf{G}rand \textbf{U}nified \textbf{T}heory \\
\\
\textbf{GSW} : & \textbf{G}lashow-\textbf{S}alam-\textbf{W}einberg \\
\\
\textbf{IR} : & \textbf{I}nfra\textbf{R}ed \\
\\
\textbf{KK} : & \textbf{K}aluza-\textbf{K}lein \\
\\
\textbf{LED} : & \textbf{L}arge \textbf{E}xtra \textbf{D}imension
\end{tabular}

\newpage

\begin{tabular}{ll}
\textbf{LHC} : & \textbf{L}arge \textbf{H}adron \textbf{C}ollider \\
\\
\textbf{LKP} : & \textbf{L}ightest \textbf{K}aluza-Klein \textbf{P}article \\
\\
\textbf{LQG} : & \textbf{L}oop \textbf{Q}uantum \textbf{G}ravity \\
\\
\textbf{NPGO} : & \textbf{N}on-\textbf{P}erturbative \textbf{G}ravitationnal \textbf{O}bject \\
\\
\textbf{PGO} : & \textbf{P}erturbative \textbf{G}ravitational \textbf{O}bject \\
\\
\textbf{QBH} : & \textbf{Q}uantum \textbf{B}lack \textbf{H}ole \\
\\
\textbf{QCD} : & \textbf{Q}uantum \textbf{C}hromo\textbf{D}ynamics \\
\\
\textbf{QFT} : & \textbf{Q}uantum \textbf{F}ield \textbf{T}heory \\
\\
\textbf{RS} : & \textbf{R}andall and \textbf{S}undrum \\
\\
\textbf{SM} : & \textbf{S}tandard \textbf{M}odel \\
\\
\textbf{SUGRA} : & \textbf{SU}per\textbf{GRA}vity \\
\\
\textbf{SUSY} : & \textbf{SU}per\textbf{SY}mmetry \\
\\
\textbf{UED} : & \textbf{U}niversal \textbf{E}xtra \textbf{D}imension \\
\\
\textbf{UV} : & \textbf{U}ltra\textbf{V}iolet \\
\\
\textbf{VEV} : & \textbf{V}acuum \textbf{E}xpectation \textbf{V}alue \\
\\
\end{tabular}

\selectlanguage{english}

\bibliographystyle{JHEP}

\providecommand{\href}[2]{#2}\begingroup\raggedright\endgroup

\newpage
\blankpage
\includepdf[pages={-}]{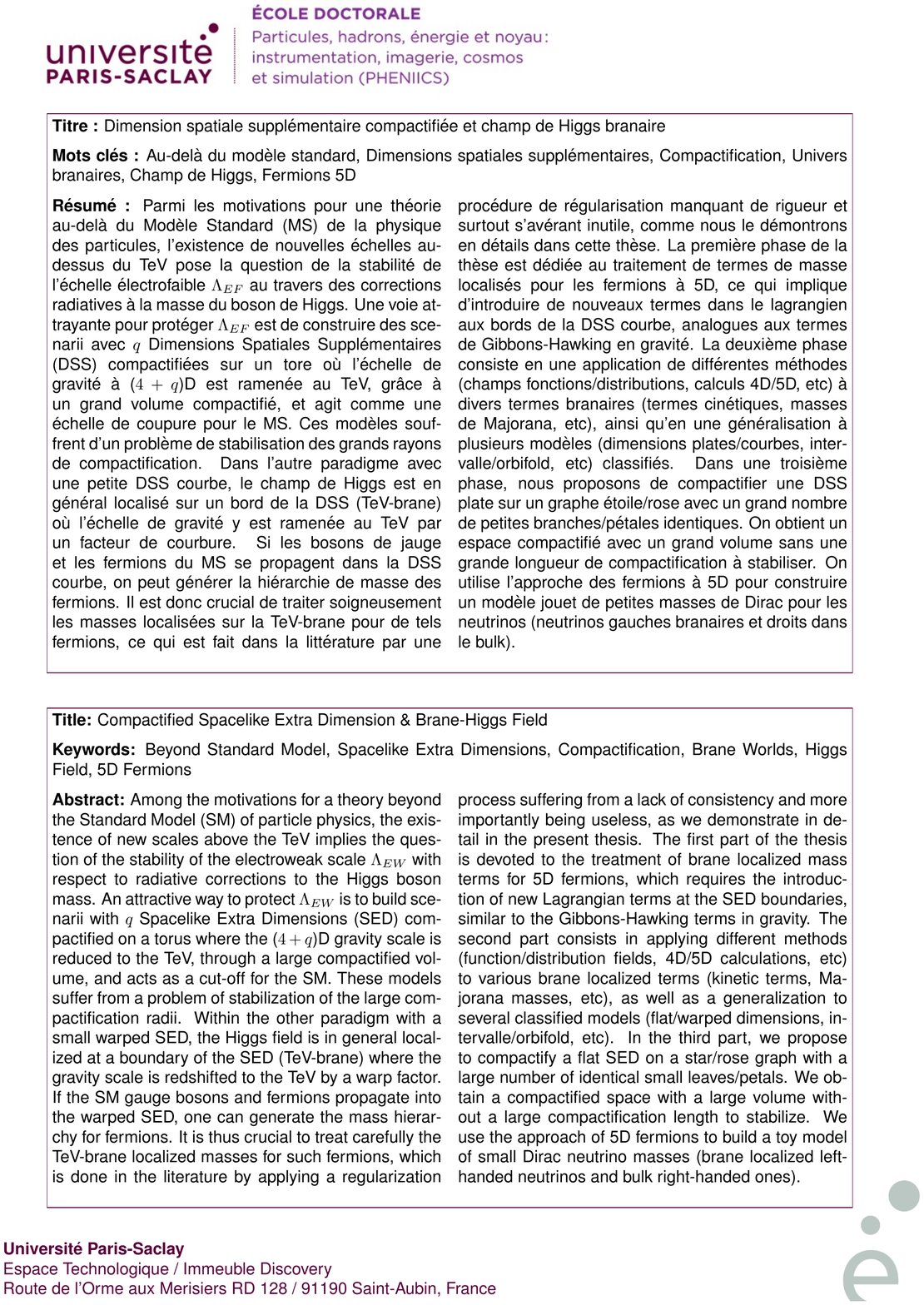}

\end{document}